\pdfoutput=1
\documentclass[graphicx,amssymb,amsmath,enumerate,11pt]{report}
\usepackage[a4paper,inner=3.5cm,outer=2.5cm,top=2.5cm,bottom=2.5cm,pdftex]{geometry}
\usepackage{graphicx}
\usepackage{epsfig}
\usepackage{amsmath}
\usepackage{amssymb}
\usepackage{enumerate}
\usepackage{subfigure}
\usepackage{multirow}
\usepackage{url}
\usepackage{color}
\begin{document}
\vskip 3cm
\begin{center}
{\Large\bf\sc
Proceedings of the Workshop \\
on Monte Carlo's, Physics \\
and Simulations at the LHC \\
\vskip 4cm
PART II}
\end{center}
\newpage
{\it
\begin{center}
Authors
\end{center}
\vskip 0.5cm
\noindent
F. Ambroglini~$^{\ref{UniPG}}$,
R. Armillis~$^{\ref{Lecce}}$,
P. Azzi~$^{\ref{INFNPD}}$,
G. Bagliesi~$^{\ref{INFNPI}}$,
A. Ballestrero~$^{\ref{INFNTO}}$,
G. Balossini~$^{\ref{UniPV}}$,
A. Banfi~$^{\ref{UniMIB}}$,
P. Bartalini~$^{\ref{Taiwan}}$,
D. Benedetti~$^{\ref{Northeastern}}$,
G. Bevilacqua~$^{\ref{Demokritos}}$,
S. Bolognesi~$^{\ref{UniTO}}$,
A. Cafarella~$^{{\ref{Demokritos},\ref{Lecce}}}$,
C.M. Carloni Calame~$^{\ref{Southampton}}$,
L. Carminati~$^{\ref{UniMI}}$,
M. Cobal~$^{\ref{Udine}}$,
G. Corcella~$^{{\ref{FermiZ},\ref{Normale}}}$,
C. Corian\`{o}~$^{\ref{Lecce}}$,
A. Dainese~$^{\ref{Legnaro}}$,
V. Del Duca~$^{\ref{LNF}}$,
F. Fabbri~$^{\ref{INFNBO}}$,
M. Fabbrichesi~$^{\ref{INFNTS}}$,
L. Fan\`{o}~$^{\ref{INFNPG}}$,
Alon E. Faraggi~$^{\ref{Liverpool}}$,
S. Frixione~$^{{\ref{CERNTH},\ref{EPFL}}}$,
L. Garbini~$^{\ref{UniPG}}$,
A. Giammanco~$^{\ref{Louvain}}$,
M. Grazzini~$^{\ref{INFNFI}}$,
M. Guzzi~$^{\ref{Lecce}}$,
N. Irges~$^{\ref{Crete}}$,
E. Maina~$^{\ref{UniTO}}$,
C. Mariotti~$^{\ref{INFNTO}}$,
G. Masetti~$^{\ref{UniBO}}$,
B. Mele~$^{\ref{INFNRM1}}$,
E. Migliore~$^{\ref{UniTO}}$,
G. Montagna~$^{\ref{UniPV}}$,
M. Monteno~$^{\ref{INFNTO}}$,
M. Moretti~$^{\ref{Ferrara}}$,
P. Nason~$^{\ref{INFNMIB}}$,
O. Nicrosini~$^{\ref{INFNPV}}$,
A. Nisati~$^{\ref{INFNRM1}}$,
A. Perrotta~$^{\ref{INFNBO}}$,
F. Piccinini~$^{\ref{INFNPV}}$,
G. Polesello~$^{\ref{INFNPV}}$,
D. Rebuzzi~$^{\ref{INFNPV}}$,
A. Rizzi~$^{\ref{ETH}}$,
S. Rolli~$^{\ref{Tufts}}$,
C. Roda~$^{\ref{INFNPI}}$,
S. Rosati~$^{\ref{INFNRM1}}$,
A. Santocchia~$^{\ref{UniPG}}$,
D. Stocco~$^{{\ref{Subatech},\ref{UniTO}}}$,
F. Tartarelli~$^{\ref{UniMI}}$,
R. Tenchini~$^{\ref{INFNPI}}$,
A. Tonero~$^{\ref{SISSA}}$,
M. Treccani~$^{{\ref{granada},\ref{Ferrara}}}$,
D. Treleani~$^{\ref{UniTS}}$,
A. Tricoli~$^{\ref{Rutherford}}$,
D. Trocino~$^{\ref{UniTO}}$,
L. Vecchi~$^{\ref{SISSA}}$,
A. Vicini~$^{\ref{UniMI}}$,
I. Vivarelli~$^{\ref{INFNPI}}$.

{\scriptsize\it
\begin{enumerate}\addtolength{\itemsep}{-0.7\baselineskip}

\item \label{Lecce}       University of Salento and INFN, Lecce, Italy

\item \label{INFN} INFN, Frascati, Italy

\item \label{UniTO}      University of Torino and INFN, Torino, Italy

\item \label{INFNTO}      INFN, Sezione di Torino, Torino, Italy

\item \label{UniMIB}     University of Milano Bicocca and Sezione INFN, Milano, Italy

\item \label{UniMI}      University of Milano and Sezione INFN, Milano, Italy

\item \label{UniTS}      University of Trieste and Sezione INFN, Trieste, Italy

\item \label{UniBO}      University of Bologna and Sezione INFN, Bologna, Italy

\item \label{INFNBO}     INFN, Sezione di Bologna, Bologna, Italy

\item \label{INFNMIB}    INFN, Sezione di Milano Bicocca, Milano, Italy

\item \label{INFNMI}     INFN, Sezione di Milano, Milano, Italy

\item \label{INFNRM1}     INFN, Sezione di Roma, Roma, Italy

\item \label{UniPV}      University of Pavia and INFN, Pavia, Italy

\item \label{INFNPV}     INFN, Sezione di Pavia, Pavia, Italy

\item \label{INFNTS}     INFN, Sezione di Trieste, Trieste, Italy

\item \label{INFNPG}     INFN, Sezione di Perugia, Perugia, Italy

\item \label{SISSA}      SISSA/ISAS, Trieste, Italy

\item \label{Ferrara}     University of Ferrara and INFN, Ferrara, Italy

\item \label{INFNFI}      INFN, Sezione di Firenze, Firenze, Italy

\item \label{UniPG}       University of Perugia and INFN, Perugia, Italy

\item \label{INFNPD}      INFN, Sezione di Padova, Padova, Italy

\item \label{UniPA}       University of Padova and INFN, Padova, Italy

\item \label{UniPI}       University of Pisa and INFN, Pisa, Italy

\item \label{INFNPI}      INFN, Sezione di Pisa, Pisa, Italy

\item \label{Normale} Scuola Normale Superiore and INFN, Pisa, Italy

\item \label{Crete}       Department of Physics and Institute of Plasma Physics,
                         University of Crete, Heraklion, Greece

\item \label{Taiwan}     National Taiwan University, Taipei, Taiwan.

\item \label{Udine}       INFN Gruppo Collegato di Udine, Udine, Italy

\item \label{FermiZ}      Museo Storico della Fisica e Centro Studi e Ricerche E. Fermi, Roma, Italy

\item \label{Legnaro}     INFN, Laboratori Nazionali di Legnaro, Padova, Italy

\item \label{Demokritos}  Institute of Nuclear Physics, NCSR ``Demokritos'', Athens, Greece

\item \label{Liverpool}   Department of Mathematical Sciences,
             University of Liverpool, Liverpool, United Kingdom

\item \label{granada} Departamento de F\'{i}sica Te\'orica y del Cosmos, University of Granada, Granada, Spain

\item \label{Southampton} INFN and School of Physics and Astronomy, University of Southampton, 
      Highfield, Southampton, UK

\item \label{Subatech} Subatech (Universit\'e de Nantes, Ecole des Mines and CNRS/IN2P3),
Nantes, France

\item \label{Northeastern} Northeastern University, Department of Physics, Boston, MA, USA

\item \label{Louvain}     Universit\'e Catholique de Louvain, Louvain-la-Neuve, Belgium

\item \label{ETH}   Institute for Particle Physics, ETH Zurich, Zurich, Switzerland

\item \label{Rutherford}
    Rutherford Appleton Laboratory, Science and Technology Facilities
Council, Harwell Science and Innovation Campus, Didcot OX11 0QX,
United Kingdom

\item \label{CERNTH} PH Department, TH Unit, CERN, Geneva, Switzerland

\item \label{EPFL} ITPP, EPFL, Lausanne, Switzerland

\item \label{LNF} INFN, Laboratori Nazionali di Frascati, Frascati, Italy

\item\label{Tufts} Tufts University, Medford, Massachusetts, USA

\end{enumerate}
}
}

\newpage
\tableofcontents
\newcommand\mchapter[2]{\chapter*{#1}
\vskip -0.5cm
\noindent
{\it #2}
\vskip 0.3cm
\addcontentsline{toc}{chapter}{#1\\{\normalsize\it #2}}}

\newpage
\addtocounter{chapter}{1}
%\documentclass[a4paper,12pt,twoside]{report}
%\usepackage{epsfig}
%\usepackage{amssymb}
%\usepackage{cite}
% \usepackage{lineno}
%\usepackage{setspace}
%%%%%%%%%%%%%%%%%%%%%%%%%%%%%%%%%%%%%%%%%%%%%%%%%%%%%%%%%%
%\begin{document}
%
%%%%%%%%%%%%%%%%%%%%%%%%%%%%%%%%%%%%%%%%%%%%%%%
% Toggle line numbering
% Won't work with the PRD revtex4 !
% \pagewiselinenumbers
% uncomment if you want doublespace
%\doublespace
%%%%%%%%%%%%%%%%%%%%%%%%%%%%%%%%%%%%%%%%%%%%%%%
%\chapter{Introduction} {\it C. Mariotti, E. Migliore and P. Nason}
%%%%%%%%%%%%%%%%%%%%%%%%%%%%%%%%%%%%%%%%%%%%%%%

\mchapter{ALICE and its pp physics programme}{M. Monteno for the ALICE Collaboration}
%-------------- NEWCOMMANDS ----------------------
\newcommand\pt{p_{\scriptscriptstyle T}}
\newcommand\Et{E_{\scriptscriptstyle T}}
\newcommand\alphas{\alpha_{\scriptscriptstyle S}}

\section{Introduction}

ALICE (A Large Ion Collider Experiment) is the dedicated heavy-ion experiment 
designed to measure the properties of the strongly interacting matter created 
in nucleus-nucleus interactions at the LHC energies ($\sqrt{s} = 5.5$ TeV 
per nucleon pair for \mbox{Pb--Pb} collisions) 
\cite{ALICE-PPRVol1,ALICE-PPRVol2}.

In addition, with its system of detectors, ALICE will also allow to perform 
interesting measurements during the proton--proton LHC runs at $\sqrt{s} = 14$ 
TeV \cite{ppDay1}. Special strength of ALICE is the low $\pt$ cut-off 
($\sim$100~MeV/$c$) due to the low magnetic field, the small amount of 
material in the tracking detectors, and the excellent capabilities in 
particle identification over a large momentum range (up to $\sim$100~GeV/$c$). 
The above features of its conceptual design as soft-particle (low $\pt$) 
tracker make ALICE suitable to explore very effectively the global properties 
of minimum--bias proton--proton collisions (such as the distributions of 
charged tracks in multiplicity, pseudorapidity and transverse momentum) 
in the new domain of the LHC energies.

In addition, these measurements will provide also an indispensable complement 
to those performed in the other pp experiments, ATLAS and CMS, where the  
superposition of minimum-bias collisions at the highest LHC luminosity will be 
the main source of background to the search for rare signals (Higgs boson,  
SUSY particles, '{\it new physics}'). On the other hand also the Underlying 
Event (i.e. the softer component accompanying a hard QCD process) must be 
carefully understood, since it accounts for a large fraction of the event 
activity in terms of the observed transverse energy or charged particle 
multiplicity and momenta.

Furthermore, as it will be shown in the following, the ALICE proton--proton 
programme will include also cross-section measurements of strange particles, 
baryons, resonances, heavy-flavoured mesons, heavy quarkonia, photons, and 
also jet studies.

Another motivation for studying pp events with ALICE is the necessity to 
provide a reference, in the same detector, to measurements performed with 
nucleus--nucleus (and proton--nucleus) collisions. 
The latter could be done via interpolation to $\sqrt{s} = 5.5$ TeV (the centre-of-mass 
energy for Pb–-Pb runs) between the Tevatron and the maximum LHC energy. However, 
since this interpolation will be affected by rather large uncertainties, additional 
dedicated runs at the same centre-of-mass energy as measured in heavy-ion collisions 
could be necessary to obtain a more reliable reference. Indeed, as it was shown by 
the past experiments at the SPS and at the RHIC, such comparison is important 
in order to disentangle genuine collective phenomena and to be more sensitive 
to any signatures of critical behaviour at the largest energy densities reached 
in head-on heavy-ion collisions.

Last, a more technical motivation of pp studies is that pp collisions are  
optimal for the commissioning of the detector, since most of the calibration 
and alignment tasks can be performed most efficiently with low multiplicity 
events. 

For all the above reasons, from the Technical Proposal onwards the 
proton--proton programme has been considered an integral part of the ALICE 
experiment. This programme is going to be started at the commissioning of 
the LHC, which will happen with proton beams at low luminosities .

This review paper is organized as follows. After a section describing the ALICE
detector, we will review the features of ALICE operations with pp collisions 
at the LHC, and the statistics and triggers required to accomplish its physics 
programme. Several physics topics will be addressed, but special emphasis 
will be given to the soft physics programme (event characterization, strange 
particle and resonance production, particle correlations, event-by-event 
fluctuations and baryon asymmetries measurements). Then, the ALICE 
capabilities in measuring some diffractive processes and its potentialities 
in the study of hard processes (jet and photon physics) will be presented, 
to conclude with some hints of possible studies of exotic processes 
(like mini black holes eventually produced by large extra dimensions). 

The physics programme will include also measurements of heavy-flavoured mesons 
(open charm and beauty) and of quarkonia states, both in the central detector 
and in the forward muon spectrometer. However these studies will not be 
discussed here, since they are already reported in other contributions 
included in these proceedings (\cite{Dainese_MCWS,Stocco_MCWS}).

\section{ALICE detector overview}

\begin{figure}[htb]
\begin{center} 
\includegraphics[width=.92\textwidth]{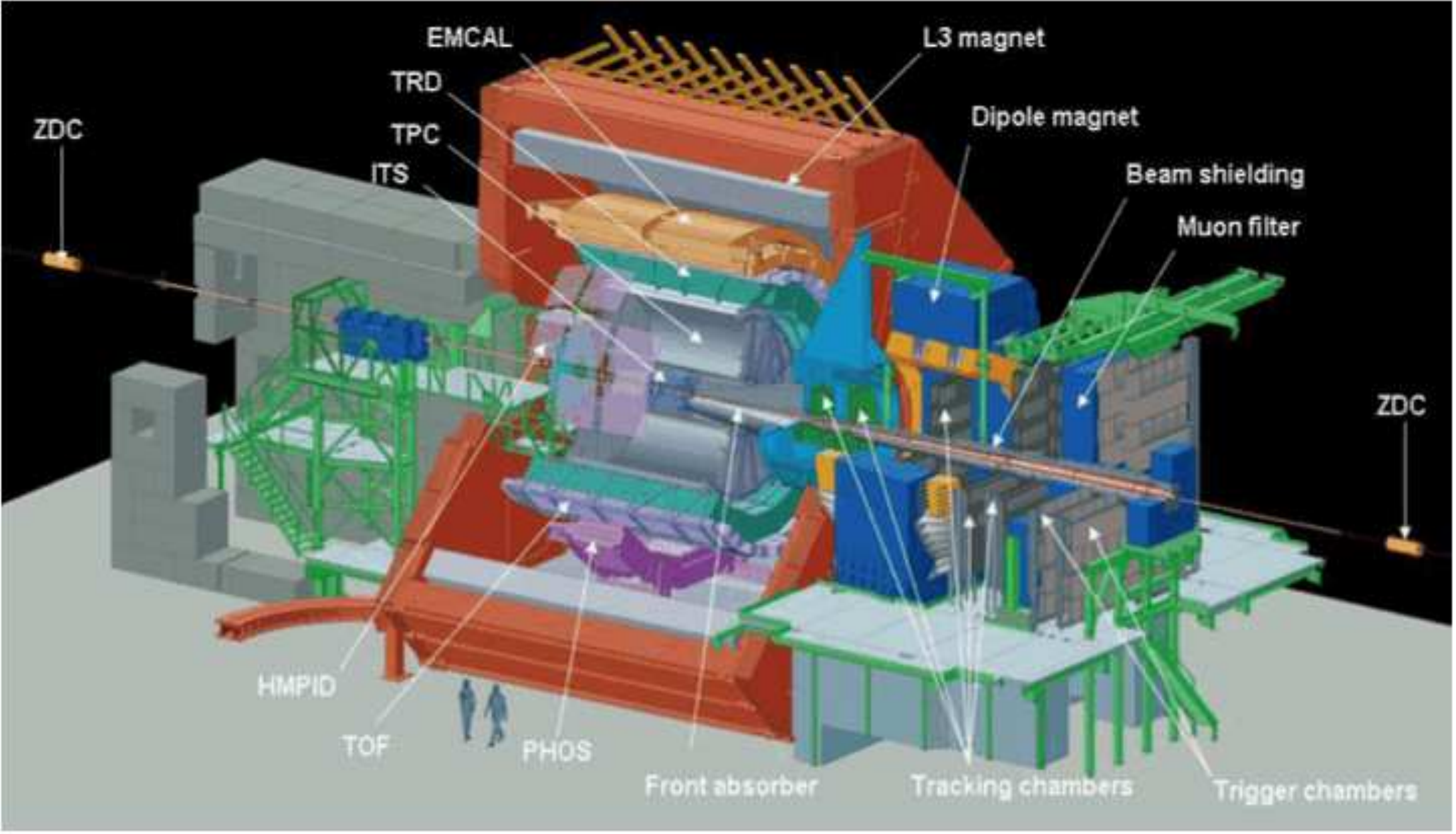}
\caption{
The ALICE experiment in its final layout.
}
\vskip -0.7 cm 
\label{fig:ALICE_layout}
\end{center} 
\end{figure}

ALICE, whose setup is shown in Fig.~\ref{fig:ALICE_layout}, is a 
general-purpose experiment whose detectors measure and identify mid-rapidity 
hadrons, leptons and photons produced in an interaction. A unique 
design, with very different optimisation than the one selected for the 
dedicated pp experiments at LHC, has been adopted for ALICE.

This results from the requirements to track and identify particles from very 
low ($\sim$100~MeV/$c$) up to fairly high ($\sim$100~GeV/$c$) $\pt$, to 
reconstruct short-lived particles such as hyperons, D and B mesons, and to 
perform these tasks even in a heavy-ion collision environment, with large 
charged-particle multiplicities.

Theoretically founded predictions for the multiplicity in central 
\mbox{Pb--Pb} collisions at the LHC range at present from 1000 to 
4000 charged particles per rapidity unit at mid-rapidity, while 
extrapolations from RHIC data point at values of about 1500. The ALICE 
detectors are designed to cope with multiplicities up to 8000 charged 
particles per rapidity unit, a value which ensures a comfortable safety 
margin. 

The detection and identification of muons are performed with a dedicated 
spectrometer, including a large warm dipole magnet and covering a domain 
of large rapidities\footnote{In ALICE the $z$-axis is 
parallel to the mean beam direction, pointing in the direction opposite to the 
muon spectrometer} ($-4.0 \leq \eta \leq -2.4$). 

Hadrons, electrons and photons are detected and identified 
inside the central barrel, a complex system of detectors immersed in 
a moderate (0.5~T) magnetic field provided by the solenoid of the former 
L3 experiment.

Tracking of charged particles is performed by a set of four concentric 
detectors: the Inner Tracking System (ITS), consisting of six cylindrical 
layers of silicon detectors, a large-volume Time-Projection Chamber (TPC), 
a high-granularity Transition-Radiation Detector (TRD), and a high-resolution 
array of Multi-gap Resistive Plate Chambers (TOF). 
These detectors allow global reconstruction of particle momenta in the central 
pseudorapidity range $|\eta|<0.9$ (with good momentum resolution up to 
$\pt \sim 100$~GeV/c), and particle identification is performed by 
measuring energy loss in the ITS and in the TPC, transition radiation in the 
TRD, and time of flight with the TOF.

However, in the case of pp collisions, the lower particle density allows to 
increase the TPC acceptance by considering also tracks with only a partial 
path through the TPC, i.e. ending in the readout chambers; in that case the 
pseudorapidity coverage can be enlarged up to $|\eta| \le 1.5$, with a lower 
momentum resolution.

Two additional detectors provide particle identification at central rapidity 
over a limited acceptance: the High-Momentum Particle Identification Detector 
(HMPID), that is an array of Ring Imaging Cherenkov counters dedicated to the 
identification of hadrons with $\pt > 1$ GeV/c, and a crystal Photon 
Spectrometer (PHOS) to detect electromagnetic particles and provide photon 
and neutral meson identification.

Additional detectors located at large rapidities, on both sides of the central 
barrel, complete the central detection system to characterise the event on a 
wider rapidity range or to provide interaction triggers. The measurement of 
charged particle and photon multiplicity is performed respectively by the 
Forward Multiplicity Detector (FMD) (over the intervals 
$-3.4 \leq \eta \leq -1.7$ and $1.7 \leq \eta \leq 5.1$) and by the 
Photon Multiplicity Detector (PMD) (over the range $2.3 \leq \eta \leq 3.5$). 
The V0 and T0 detectors, designed for triggering purposes, have an acceptance 
covering a rather narrow domain at large rapidities, whereas a set of four 
Zero-Degree Calorimeters (ZDC) will measure spectator nucleons in heavy-ion 
collisions and leading particles in pp collisions around beams' rapidity.

Finally, in order to complete the ALICE capabilities in jet studies, a large 
lead-scintillator electromagnetic calorimeter (EMCal) 
\cite{ALICE_EMCal,ALICE_EMCal_TDR} will be located between the TOF and the L3 
magnetic coils, adjacent to HMPID and opposite to PHOS. 
In its final configuration, the EMCal will have a central acceptance in 
pseudorapidity of $|\eta| < 0.7$, with a coverage of 
$180 \mbox{$^\circ$}$ in azimuth, and an energy resolution of 
$\Delta E / E = 10\%/ \sqrt{E}$. It will be optimized for the detection of 
high-$\pt$ photons, neutral pions and electrons and, together with the 
central tracking detectors, it will improve the jet energy resolution.

The charged-particle multiplicity and the 
${\mathrm d}N_{\mathrm ch}/{\mathrm d}\eta$ distribution will constitute 
the first basic observable which will be measured in ALICE, both for 
pp and \mbox{Pb--Pb} collisions at the LHC. In the central region the best 
performance in these measurements will be obtained with the Silicon Pixel 
Detector (SPD), the first two layers of the ITS, with approximate radii 
of 3.9 and 7.6 cm. A simple 
algorithm can be used in the SPD to measure multiplicity in a robust way 
by using `tracklets', defined by the association of clusters of hits in 
two different SPD layers through a straight line pointing to the primary 
interaction vertex, assumed to be known \cite{MultipSPD}. 
The limits of the geometrical acceptance for an event with primary vertex at 
the center of the detector are $|\eta|<2$ for clusters measured on the first 
SPD layer and $|\eta|<1.5$ for tracklets measured with both SPD layers. 
However, the effective acceptance is larger, due to the longitudinal 
spread of the interaction vertex position, and its limits extend up to about 
$|\eta|<2$ for the multiplicity estimate with tracklets \cite{MultipSPD2006}.

Therefore, by considering the partial overlap between the $\eta$ ranges 
covered by the SPD ($-2 < \eta <2$) 
and by the FMD ($-3.4 \leq \eta \leq -1.7$ and $1.7 \leq \eta \leq 5.1$), 
it follows that the pseudorapidity range covered by the ALICE experiment for 
the charged-particle multiplicity and the 
${\mathrm d}N_{\mathrm ch}/{\mathrm d}\eta$ measurements 
spans over about 8 pseudorapidity units.

\section{ALICE operation with pp collisions}

The proton--proton programme of ALICE will start already during the phase of
commissioning of the LHC, when the luminosity will be low 
($L < 10^{29}$ cm$^{-2}$s$^{-1}$). This time will be a privileged period for 
ALICE to measure pp collisions, because there will be only a small pile-up in 
its slowest detectors and a low level of beam background \cite{ppDay1}.

However, when higher luminosities will be delivered by the LHC, a limiting 
factor for ALICE will be given by the readout of its detectors, essentially by 
the TPC, which is the slowest detector with its drift time of 88 $\mu$s, 
during which additional collisions may occur, causing several superimposed 
events (pile-up).

From the point of view of track reconstruction this would not be a problem, 
since the piled-up interactions in the TPC will keep a regular pattern with 
virtual vertices shifted along the drift direction. This can be tolerated, 
although at the price of heavier tracking and larger data volume for the same 
physics information, at least up to $L = 3 \times 10^{30}$ cm$^{-2}$s$^{-1}$. 

At this luminosity the interaction rate amounts to about 200 kHz, assuming 
that the total inelastic pp cross section is 70 mb. 
The TPC records tracks from interactions which have occurred during the time 
interval 88 $\mu$s before and after the triggered bunch crossing. 
Hence on average 40 events will pile-up during the drift time of the TPC, 
before and after the trigger. However, on average only half of the tracks will 
be recorded, due to the fact that the other half will be emitted outside the 
acceptance. Therefore the total data volume will correspond only to the equivalent 
of 20 complete events. The charged-particle density at mid-rapidity in pp 
collisions at the nominal LHC centre-of-mass energy of $\sqrt{s}=14$ TeV is 
expected to be about 7 particles per unit of pseudorapidity, resulting in a 
total of  $\sim$ 250(400) charged tracks within the TPC acceptance 
$|\eta| < 0.9$ (or within the extended acceptance $|\eta| < 1.5$, when 
including also tracks with only a short path through the TPC). Clearly, 
tracking under such pile-up conditions is still feasible, since the occupancy 
is more than an order of magnitude below the design value of the TPC.

For higher luminosity pile-up becomes progressively more difficult to handle, 
since events start to pile-up also in other detectors (silicon drift and 
silicon strip detectors, and then the HMPID). 
Therefore the luminosity $L = 3 \times 10^{30}$ cm$^{-2}$s$^{-1}$ is the 
maximum that can be tolerated, and in the following we will consider
it as a benchmark for ALICE. When the LHC will reach its design luminosity 
($L = 10^{34}$ cm$^{-2}$s$^{-1}$) some strategies will be needed to record 
meaningful pp data by reducing the luminosity at the ALICE interaction point 
(e.g. beam defocussing and displacement).

For the benchmark luminosity $L = 3 \times 10^{30}$ cm$^{-2}$s$^{-1}$ the 
total pp event size (including pile-up and possible electronics noise) is 
estimated to be of the order of 2.5 MB, without any data compression. Thus, 
running at the foreseen maximum TPC rate of 1 kHz would lead to a total data 
rate of 2.5 GB/s. 
However, the online tracking of the High Level Trigger will select only tracks 
belonging to the interesting interaction. This pile-up suppression will 
reduce the event size by at least a factor 10.

According to current estimates of event sizes and trigger rates, 
a maximum data rate (bandwidth) of the Data Acquistion (DAQ) system of 
1.25 GB/s to mass storage, consistent with the constraints imposed by 
technology, cost and storage capacity, would provide adequate statistics 
for the full physics programme. This will be possible by using a combination 
of increased trigger selectivity, data compression and partial readout.

The above needs of ALICE for data acquisition are well within the limits of 
bandwidth to mass storage provided by the central computing facility (TIER-0) 
of the LHC Computing GRID project, that will be installed at CERN. 

\section{Required statistics and triggers}

An extensive soft hadronic physics programme will be feasible in ALICE 
using LHC proton beams since the machine commissioning phase, when the low 
luminosity will limit the experimental programme to the measurement of 
large-cross-section processes. This programme will include:

\begin{itemize}
\item event characterization, with the measurement of charged particle 
multiplicity, pseudorapidity and momentum spectra, and of the 
$\langle \pt \rangle$-multiplicity correlation; 
\item particle production measurements, i.e. yields and spectra of various 
identified particles, like strange particles ($\Lambda$, $\Xi$, $\Omega$, etc) 
and resonances (i.e. $\rho$, $K^*$ and $\Phi$), and baryon-antibaryon 
asymmetries; 
\item particle correlations (i.e HBT interferometry and forward-backward 
correlations) and event-by-event fluctuations.
\end{itemize}

The soft hadronic physics programme will rely on data samples of minimum-bias 
triggered events. 

The statistics needed depends on the observable under study and spans the 
range from a few $10^5$ to a few $10^8$ events. For a multiplicity 
measurement, a few $10^5$ events will give a meaningful data sample; an order 
of magnitude more is needed for particle spectra; to study rare hadronic 
observables (e.g. $\Omega$ production) we will need a few times $10^8$ pp 
events. Therefore, a statistics of $10^9$ minimum-bias triggered events will 
fulfill the whole soft hadronic physics programme.
 
Since the readout rate of ALICE is limited to 1 kHz by the TPC gating 
frequency, the requirement is to be able to collect the data at the maximum 
possible rate: 1000 events/s, at an average of 100 events/s. 
In this way, at an average acquisition rate of 100 Hz, the required 
statistics can be collected during one typical year of operation ($10^7$ s) 

However, at the same acquisition rate, a reasonable statistics for different
physics topics can be collected already in the first few hours, days, or 
weeks of data taking. For example a few minutes will be sufficient to 
measure pseudorapidity density with $\sim 10^4$ events, while a few hours 
will allow to collect sufficient event statistics for multiplicity studies.
 
For all the above outlined soft physics programme ALICE will need a simple 
minimum-bias trigger for inelastic interactions, that will be provided by 
two of its sub-detectors: the Silicon Pixel Detector (the two innermost
layers of the ITS), and the V0.

The basic building blocks of the Silicon Pixel Detector (SPD) are ladders, 
arranged in two concentric layers covering the central pseudorapidity region, 
and consisting of a 200 $\mu$m thick silicon sensor bump-bonded to 5 front-end 
chips. The signals produced by each chip are logically combined to form the 
global fast-OR ({\tt GLOB.FO}) trigger element.
 
The V0 detector is composed of two independent 
arrays of fast scintillator counters located along the beam pipe on each side 
of the nominal interaction point and at forward/backward rapidities. 
Two different trigger elements are built with the logical combination of the 
signals from counters on the two sides: {\tt V0.OR} requires at least one 
hit in one counter on one side, while {\tt V0.AND} requires at least 
one hit in one counter on both sides.
 
The main background to minimum-bias events are beam--gas and beam--halo
interactions. The rate of beam--gas collisions is expected to be much smaller 
than the rate of beam--halo collisions, whose magnitude should be of the same
order as proton--proton collisions. It has been shown that the structure of 
beam--halo events is similar to that of beam--gas events, the difference being 
that beam--halo events happen at greater distances to the nominal interaction 
point (more than 20 m). 

The proposed proton--proton minimum bias triggers, that use logical 
combinations of the above outlined trigger elements, result to be sensitive to 
interactions corresponding to $\sim 90$ \% of the total inelastic cross
section (and $\sim 99$ \% of the non-diffractive cross section), and still
reject the majority of beam--gas interactions \cite{MBtrigger}.

On the other hand the SPD global fast-OR ({\tt GLOB.FO}) trigger element can 
also be used to provide a high multiplicity trigger, that will allow to 
collect enriched statistics in the tail of multiplicity distributions. 

As regards the other physics topics (open heavy flavour mesons and quarkonia 
production; diffractive processes studies; jet and photon physics) they 
require separate high statistics data samples that would need high rates and 
bandwidth. Dedicated trigger and HLT algorithms 
will significantly improve the event selection and data reduction, and will
allow to collect data samples of adequate statistics already in one year 
of data taking.

Some details on triggers for diffractive processes and jets will be 
given in following dedicated sections.   

\section{Event characterization}

For the first physics measurements, shortly after the LHC start-up, in order 
to minimize the uncertainty stemming from non-optimal alignment and 
calibration, a few detectors systems will be sufficient: the two inner layers 
of the ITS (the Silicon Pixel Detector), the TPC, and the minimum-bias trigger 
detectors (V0 and T0). Indeed, particle identification will have a limited scope 
during initial runs, since it requires a precise calibration and a very good understanding 
of the detectors. Four measurements of soft hadronic physics which can be 
addressed during the first days of data taking will be outlined in this 
section: 1) the pseudorapidity density of primary charged particles, 2) the 
charged particle multiplicity distribution, 3) transverse momentum spectra 
and 4) the correlation of $\langle \pt \rangle$ with multiplicity. 

These measurements of global event properties will be discussed in the context 
of previous collider measurements at lower energies and of their theoretical 
interpretations. 

\subsection{Pseudorapidity density}
The pseudorapidity density of primary charged particles at mid-rapidity 
${{\mathrm d}N_{\mathrm ch}/{\mathrm d}\eta |}_{\eta \approx 0} =  1/\sigma^{inel} \cdot ({\mathrm d} \sigma_{\mathrm ch}/{\mathrm d}\eta )_{\eta \approx 0}$ 
has been traditionally among the first measurements performed by experiments 
exploring a new energy domain. Indeed this measurement is important since 
it gives general indications on the interplay between hard and soft processes 
in the overall particle production mechanisms, and furthermore it brings 
important information for the tuning of Monte Carlo models. A simple scaling 
law ($\sim \ln s$) for the energy dependence of particle production at 
mid-rapidity was predicted by Feynman \cite{Feynman}, but it appeared clearly 
broken in collisions at the SPS \cite{UA5_dNdeta,UA1_dNdeta}. Indeed, the best 
fit to the pp and $\mathrm{p}$$\bar{\mathrm{p}}$ data, including that from SPS 
and Tevatron colliders \cite{CDF_dNdeta}, follows a ln$^2 s$ dependence, whose 
extrapolation at $\sqrt{s}$=14 TeV gives about 6 particles per rapidity unit 
for non-single diffractive interactions. 
 
A reasonable description of the energy dependence of the charged particle 
density is obtained within the framework of the Quark Gluon String Model 
(QGSM) \cite{QGSM}, a phenomenological model that makes use of very few 
parameters to describe high-energy hadronic interactions. In this model, 
based on the ideas of Regge theory, the inclusive cross sections 
${\mathrm d} \sigma_{\mathrm ch}/{\mathrm d}\eta$ increase at very high 
energies and at $\eta \approx 0$ as a power-law $\sim ({s/s_{0}})^\Delta$, 
where $s_{0} = 1$ GeV$^2$ and $\Delta = \alpha_{P} -1$ is related to the 
intercept $\alpha_P$ of a Pomeron (Regge) trajectory. Indeed, with the value 
$\alpha_{P} = 0.12 \pm 0.02$ found from the analysis of $\sigma^{tot}(s)$
\cite{sigmatot} it results that the QGSM model reproduces successfully the 
observed growth of pseudorapidity distributions with energy \cite{Kaidalov}. 

Furthermore, the increase with energy of the charged particle density 
as well as the bulk properties of minimum bias events and of underlying event 
in hard processes are successfully reproduced (up to Tevatron energy) by 
models assuming the occurrence of multiple parton interactions in the same 
pp collision \cite{CDF_tuning, ATLAS_tuning, CMS_tuning}. 
Examples of such models, extending the QCD perturbative picture to the soft 
regime, are implemented in the general purpose Monte Carlo programmes PYTHIA
\cite{PYTHIA}, JIMMY \cite{JIMMY}, SHERPA \cite{SHERPA} and HERWIG++
\cite{HERWIG++}, all of them containing several parameters that must 
be tuned by comparison against available experimental data. On the other hand, 
another successful description of the available data is provided by the Monte 
Carlo model PHOJET \cite{PHOJET} which is based on both perturbative QCD and 
Dual Parton Model. However, the growth of particle density predicted by 
PHOJET is slower than in multiple parton interaction models, and so the 
charged particle density at LHC energy results to be $\sim$ 30\% smaller. 

ALICE will measure the ${\mathrm d}N_{\mathrm ch}/{\mathrm d}\eta$ distribution
 around mid-rapidity by counting correlated clusters (tracklets) in the two 
layers of the SPD (${|{\eta}|<2}$), and/or by counting tracks in the TPC (up 
to ${|{\eta}|=1.5}$). At the low multiplicity typical for proton--proton 
events, the occupancy in the highly segmented detectors will be very low, and 
corrections for geometrical acceptance, detector inefficiency and background 
contamination (from secondary interactions and feed-down decays) will be 
applied on track level. A second correction, taking into account the bias 
introduced by the vertex reconstruction inefficiency, will be applied on a 
event-by-event level (see Ref.\cite{JFGO1} for more details).  

The measurement can be done with very few events ($10^4$ events will 
give a statistical error of ~$\sim$ 2 \% for bins of $\Delta \eta = 0.2$, 
assuming ${{\mathrm d}N_{\mathrm ch}/{\mathrm d}\eta |}_{\eta \approx 0} = 6$).

In addition, the measurement of the pseudorapidity distribution can also be 
performed in the forward region (on the pseudorapidity intervals 
$-3.4 \leq \eta \leq -1.7$ and $1.7 \leq \eta \leq 5.1$), with the Forward 
Multiplicity Detector, but a complete understanding of secondary processes, 
which are dominant at low angles, is required.

\subsection{Multiplicity distribution}
The multiplicity distribution is the probability $P_n$ to produce $n$ 
primary charged particles in a collision. 

At energies below $\sqrt{s}$=63 GeV (up to the ISR domain), the multiplicity 
distributions still scale with the mean multiplicity \cite{ISR_mult}, 
following an universal function 
($P_{n} = \langle n \rangle ^{-1} \Phi (n/<n>)$)\cite{KNO}. For higher 
energies, starting from the SPS, the KNO-scaling appears clearly broken 
\cite{UA5_multip}. 
The peculiarities of the measured multiplicity distributions (as the shoulder
structure in their shape) have been explained in a multi-component scenario, 
by assuming an increased contribution to particle production from hard 
processes (jets and minijets). Multiplicity distributions are fitted to a 
weighted superposition of negative binomial distributions corresponding to 
different classes of events (soft and semi-hard) 
\cite{GiovannUgocc_1,GiovannUgocc_2,two-comp}. 
In alternative approaches, the violation of the KNO-scaling is understood as an
effect of the occurrence of multiple parton interactions \cite{Walker}, or in 
terms of multi-Pomeron exchanges \cite{Matinyan}.

However, the general behaviour of multiplicity distributions in pp collisions 
in full phase space is quite uncertain. For example data at $\sqrt{s}$=546 GeV 
from E735 and UA5 differ by more than a factor of two above 
$N_{\mathrm ch} \approx 80$ \cite{E735_98}.
Therefore extrapolations to higher energies or to full phase space of 
distributions measured within limited rapidity intervals are affected by 
rather big inaccuracies. 

Experimentally the multiplicity distribution is not straightforward to extract.
 The detector response matrix, i.e. the probability that a certain true 
multiplicity gives a certain measured multiplicity, can be obtained from 
detector simulation studies. Using this, the true multiplicity spectrum can 
be estimated from the measured spectrum using different unfolding techniques 
\cite{Anykeev,Dagostini,UA5_multfit}. The procedure of measuring the 
multiplicity distribution with the ALICE detector (using the Silicon Pixel 
Detector of the ITS, as well as the full tracking based on the TPC), is 
thoroughly described in Ref.\cite{JFGO2}

ALICE reach in multiplicity with the statistics foreseen for the first physics 
run ($\geq 10^7$ minimum-bias triggered events) is about 125 ($|\eta|< 0.9$). 
However, a large statistics of high-multiplicity events, with charged-particle 
rapidity densities at mid-rapidity in the range $60$--$70$ (i.e.\, ten times 
the mean multiplicity) can be collected by using a high-multiplicity trigger 
based on the SPD Fast-OR trigger circuit. This class of events may give 
access to initial states where new physics such as high-density effects and
saturation phenomena set in. 

Also, local fluctuations of multiplicity distributions in momentum space and 
related scaling properties (intermittent behaviour) might be a possible 
signature of a phase transition to QGP\cite{VanHove}. This makes it 
interesting to study such multiplicity fluctuations in pp collisions.

\subsection{Transverse momentum spectra}

Collider data on charged-particle $\pt$ spectra have shown that the high 
$\pt$ yield rises dramatically with the collision energy, due to the increase 
of the hard processes cross sections \cite{collider_pt}.

At high $\pt$ the transverse momentum spectra are well described by LO or NLO 
pQCD calculations, but involving several phenomenological parameters and 
functions (K-factor, parton distribution functions and fragmentation functions)
which need an experimental input to be determined. 
At lower $\pt$, where perturbative QCD calculations cannot be performed, 
theoretical foundations of different models are even more insecure. 
Therefore, early measurements of $\pt$ spectrum are important for the tuning 
of the model parameters and for the understanding of the background in the 
experimental study of rare processes. Also, the measurement of the $\pt$ 
spectrum is important to perform high-$\pt$ hadron suppression studies in 
in heavy-ion collisions, where the proton--proton data is used as reference.

In ALICE the track reconstruction is performed within the pseudorapidity 
interval $|\eta| < 0.9$ through several steps (see section 5.1.2 of Ref. 
\cite{ALICE-PPRVol2} for a detailed description of the procedure).  
Firstly, track finding and fitting in the TPC are performed from outside 
inward by means of a Kalman filtering algorithm \cite{Kalman}.  
In the next step, tracks reconstructed in the TPC are matched to the 
outermost ITS layer and followed in the ITS down to the innermost pixel layer.
As a last step, reconstructed tracks can be back-propagated outward 
in the ITS and in the TPC up to the TRD innermost layer and then followed 
in the six TRD layers, in order to improve the momentum resolution.

As it was already said before, the TPC acceptance covers the pseudorapidity 
region  $|\eta| < 0.9$, but this range can be extended up to 
$|\eta| \simeq 1.5$ when analyzing tracks with reduced track length and 
momentum resolution. 

The $\pt$ spectrum is measured by counting the number of tracks in each $\pt$ 
bin and then correcting for the detector and reconstruction inefficiencies 
(as a function of $z$, $\eta$ and $\pt$). Finally, the $\pt$ distribution is 
normalized to the number of collisions and corrected for the effect of vertex
reconstruction inefficiency and trigger bias. 

With an event sample of $\geq 10^7$ event that could be collected in the first 
runs ALICE could reach $\pt > 40$ GeV/c. 

\subsection{Mean transverse momentum versus multiplicity}

The correlation between charged-track $\langle \pt \rangle$ and multiplicity, 
describing the balance between particle production and transverse energy, is 
known since its first observation by UA1~\cite{UA1_avept}, and it has been 
successively studied at the ISR \cite{ISR_avept} and Tevatron 
\cite{E735_avept,CDF_soft} energies. The increase of $\langle \pt \rangle$ as 
a function of multiplicity has been also suggested by cosmic ray 
measurements \cite{Lat80}.

This correlation between $\langle \pt \rangle$ and multiplicity is generally 
attributed to the onset of gluon radiation, and explained in terms of the 
jet and minijet production increasing with energy~\cite{Wan87}. However, CDF 
data \cite{CDF_soft} have shown that the rise of the mean $\pt$ with 
multiplicity is also present in events with no jets (soft events). 
This behaviour is not yet satisfactorily explained by any models or 
Monte Carlo generators (as PYTHIA \cite{PYTHIA} and HERWIG \cite{HERWIG}).
 
In ALICE it will be relatively straightforward to obtain the correlation 
between the mean $\pt$ and the charged particle multiplicity, once the 
multiplicity distribution and the $\pt$ spectra have been measured. The 
$\pt$ cut-off imposed by the detector ($\sim$ 100 MeV/c for pions and 
$\sim$ 300 MeV/c for protons) introduces a rather large systematic uncertainty 
on the $\langle \pt \rangle$ estimate.
   
Detailed measurements of the $\langle \pt \rangle$ versus multiplicity 
(eventually in different regions in $\eta$-$\phi$ relative to leading-jet 
direction, as in CDF analyses \cite{CDF_UE}) will give an insight 
to jet fragmentation processes and to the general Underlying Event structure. 

Another interesting subject for ALICE, due to its powerful particle 
identification system at low and high $\pt$, will be the correlation between 
$\langle \pt \rangle$ and 
multiplicity studied separately for pions, kaons and proton/antiprotons. The 
data collected at Tevatron by the E735 experiment \cite{E735_avept} indicate 
that the correlation has rather different behaviour for the three types of 
particles, especially as regards the proton and antiproton 
$\langle \pt \rangle$, that do not appear to saturate at high multiplicity as 
pions (and maybe also kaons, within experimental uncertainties). 
This is not yet understood in terms of the available hadronic models.

\section{Strange particle measurements}

There are basically two main motivations for ALICE to measure strange 
particle production in pp collisions at the LHC centre-of-mass energy 
of 14 TeV: 1) extending the range where strange quark production has been 
probed in pp collisions; 2) providing a reference for the measurement of 
strangeness production in heavy ion collisions, in view of the strangeness 
enhancement which was observed to set in at the SPS centre-of-mass energy
($\sqrt{s_{\mathrm{NN}}} = 17$ GeV). 

Strange and light quark production rates are usually compared by means 
of several observables. The most simple ones are measured particle 
ratios, like the widespreadly used $K/\pi$, that features a slight and 
remarkably stable increase in pp and $\mathrm{p\overline{p}}$ collisions, 
between $\sqrt{s} = 27$ and 1800 GeV.

\begin{figure}[ht]
\begin{center}
\includegraphics[width=0.7\textwidth]{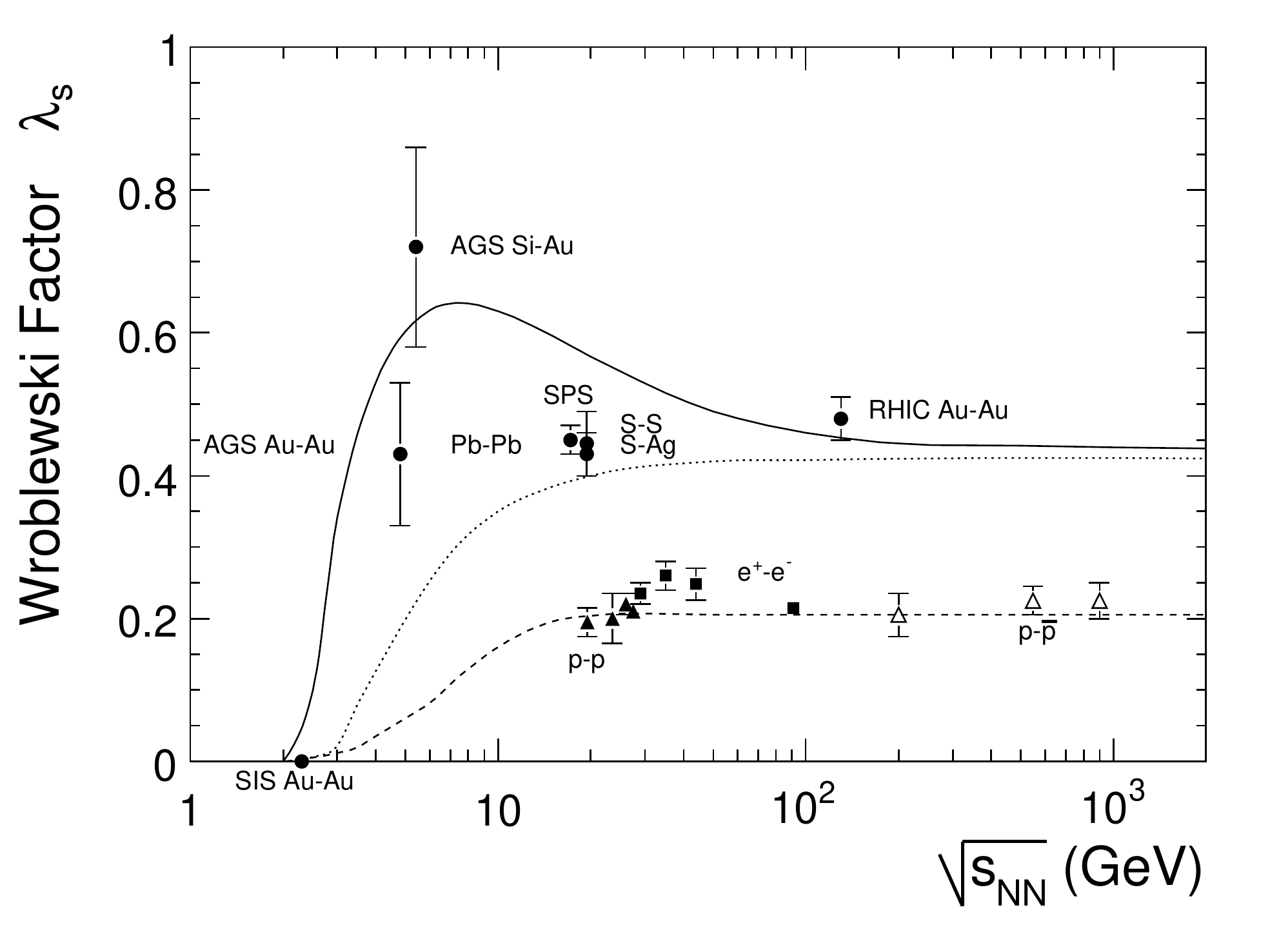}
\caption{The Wroblewski factor $\lambda_\mathrm{s}$ as a function of $\sqrt{s}$.}
\label{fig:TmdWrobBeamNrg}
\end{center}
\end{figure}

However, ideally one would like to extract directly from data the ratio 
between newly produced $s$ and $u$, $d$ quarks at hadronization, before 
hadronic decays take place. A useful way to measure such strangeness content 
is the so-called ``Wroblewski ratio'', defined as: 
$\lambda_S =  \frac{2 s\bar s}{ u\bar u + d\bar d }$. 
The earliest attempts to determine $\lambda_S$ were done in \cite{Wroblewski}, 
on the basis of the models in Refs.~\cite{lambda_s_models}. 
In pp collisions and in the $\sqrt{s}$ range from 10 GeV to 900 GeV 
a fairly constant value of $\lambda_\mathrm{s}\sim 0.2$ has been extracted 
from the data (see Fig.~\ref{fig:TmdWrobBeamNrg}) and there is no evidence 
of a rise \cite{Becattini_strange} with increasing energy. This figure also 
summarizes the results from heavy-ion collisions.

On the other hand, more recent analyses based on statistical models  
of hadronization \cite{Becattini_analyses} have had great success 
in describing experimental data. 
This is shown in Fig.~\ref{fig:TmdWrobBeamNrg} by the dashed
line, which comes from a canonical description using a
correlation volume of two protons. This correlation volume causes
a strangeness reduction as compared to heavy-ion collisions, which
have a $\lambda_\mathrm{s}$ around 0.43. 

In the case of heavy-ion collisions the parameter of the statistical models
are interpreted thermodynamically, ascribing a ``temperature'' and some 
``thermodynamic potentials'' to the system. However, it remains unclear 
as to how such models can successfully describe particle production in 
systems of small volume like those occurring during pp collisions. 
On the other hand it must be remarked that a pp system does not have 
to be thermal on a macroscopic scale to follow statistical emission. The 
apparently statistical nature of particle production observed in pp data 
could be simply a reflection of the statistical features of underlying 
jet fragmentation, or a result of phase space dominance considerations.

From Fig.~\ref{fig:TmdWrobBeamNrg} one might conclude that the
strangeness content in elementary collisions will hardly increase
with incident energy. However, this is far from being clear. The
number of produced particles in pp collisions will increase up to
values similar to those observed in heavy-ion collisions at the
SPS. Hence, the volume parameter in the canonical description may
have to be increased to account for the higher multiplicity. This
would result in an increasing strangeness content. 

Furthermore, at the LHC energies, the dominance of jet and minijet production 
will raise new questions since the final hadronic yields will originate 
from two different sources: a source reflecting the equilibrated (grand)
canonical ensemble (soft physics) and on the other hand the fragmentation 
of jets (hard physics), that differs from the behavior of an equilibrated 
ensemble. 
By triggering on events with one, two or more jets, a
`chemical analysis' of these collisions will be possible. This very 
new opportunity would allow us to study whether the occurrence of hard 
processes influences the Underlying Event distributions. Particularly 
interesting in this context is the behaviour of strange and multi-strange
particles, e.g.~the K/$\pi$ or $\Omega/\pi$ ratio, in combination
with extremely hard processes.

Possible effects of strangeness enhancement might be amplified when
selecting events with high multiplicity. In this respect ALICE, thanks 
to the high-multiplicity trigger provided by the Silicon Pixel Detector
can collect samples enriched in high multiplicity events, and so it 
will reach multiplicities 10 times the mean multiplicity. This study has 
gained special interest recently, when arguments for `deconfinement' have 
been advocated in $\rm{\bar p}p$ collisions at Tevatron energies 
\cite{Alexopoulos_2002}.

Moreover, several kinematical properties of strange particle production, like 
the multiplicity density dependence of their yield and their $\pt$ spectra, 
have been measured in the past, up to Tevatron energies, and still await a
full theoretical explanation. 
For example the $K^{0}_{S}$ and $\Lambda$ $\pt$ spectra recently measured 
with high statistics and rather large $\pt$ coverage by the STAR experiment 
\cite{STAR_ptspectra} in $\sqrt{s} = 200$ GeV pp collisions at the RHIC  
collider have been compared to NLO pQCD calculations with varied factorization 
scales and fragmentation functions (taken from \cite{KKP} for $K^{0}_{S}$ and 
from \cite{DSV} for $\Lambda$). Although for the $K^{0}_{S}$ a reasonable 
agreement is achieved between the STAR data and the NLO pQCD calculations, 
the comparison is much less favorable for the $\Lambda$. 
A better agreement is obtained by Albino, Kniehl and Kramer in \cite{AKK}
when using a new set of fragmentation functions constrained by light-quark 
flavour-tagged $e^{+}e^{-}$ data from the OPAL experiment \cite{OPAL_flavFF}.
It will prove helpful to perform similar comparisons at LHC energies. 

It was shown by the E735 Collaboration at Tevatron 
\cite{Alexopoulos_1993} that kaons $\langle \pt \rangle$ has stronger 
correlations with the charge multiplicity per unit rapidity than the pions 
$\langle \pt \rangle$: while the latter shows a saturation at 
$\langle dN/d\eta \rangle \approx 10$, the former continues to grow, although 
slightly decreasing its slope; the same behaviour is seen for antiprotons. 
Since $\langle dN/d\eta \rangle$ can be related to the energy 
density or entropy density \cite{VanHove_ptQGP}, this behaviour
is certainly relevant for quark-gluon plasma searches, besides providing 
constraints on models attempting to describe hadron production processes. 
ALICE can test this behaviour at much higher multiplicity densities
and for other identified mesons and baryons carrying strangeness quantum 
numbers.

Finally, the measurement of higher resonances in pp will be important to
obtain the respective population. This can be useful as input for the 
statistical models, but also for comparison with what is found in heavy-ion 
collisions, though in the latter case the yields are likely to be changed by 
the destruction of the resonances following the rescattering in the medium.

For all the reasons discussed above it appears very important to measure 
strange particles over a broad range of transverse momentum in the new regime 
of LHC energy. The ALICE experiment will face this challenge, for both pp
and Pb--Pb collisions, thanks to the large acceptance and high precision of 
its tracking apparatus and particle identification methods.

Strange particle can be identified over a wide range in $\pt$ from the 
topology of their decays (``kinks'' for charged kaons and secondary vertices 
for $K^{0}_{S}$, $\Lambda$, $\Xi$ and $\Omega$ decays) or otherwise from 
invariant mass analyses (for resonance decays). 

The decay pattern of charged kaons into the muonic channel, with one charged 
daughter track (a muon) and one neutral daughter (a $\nu_{\mu}$) which is not 
observable in the tracking detectors, is known as a ``kink'', as the track of 
the charged parent (the $K^{\pm}$ candidate) appears to have a discontinuity
at the point of the parent decay. The kink-finding software loops on all 
charged tracks by applying to them some cuts to look for pairs of tracks 
compatible with the kink topology described above.
The reconstruction of the kink topology is a key technique for identifying 
charged kaons over a momentum range much wider than that achieved by combining 
signals from different detectors (ITS, TPC, TOF and HMPID). Simulation 
studies have shown that for a total sample of $10^9$ pp events, in a full 
year of pp data taking at the LHC, a usable statistics of kaons can be 
obtained up to 14 GeV/c. However, when exploiting the relativistic rise of 
the energy loss signal in the TPC, the momentum reach can be further on  
enlarged up to 50 GeV/c.

Strange particles as $K^{0}_{S}$, $\Lambda$, $\Xi$ and $\Omega$ decay via 
weak interactions a few centimeters away from the primary vertex, and 
therefore they can be identified by using topological selections.

In the case of $K^{0}_{S}$ and $\Lambda$ the dominant decay channels are 
$K^{0}_{S}\rightarrow \pi^{+} \pi^{-}$ and $\Lambda \rightarrow p \pi^{-}$. 
The charged tracks of the daughter particles form a characteristic 
V-shaped pattern known as a ``V0'', whose identification is performed
by pairing oppositely charged particle tracks to form V0 candidates. Then,   
a set of geometrical cuts is applied, for example to the distance of closest 
approach (DCA) between the daughter track candidates and to the V0 pointing  
angle, in order to reduce the background and to maximize the signal-to-noise 
ratio. More efficient algorithms for V0 reconstruction, named ``on-the-fly'', 
i.e. performed during track finding, are also under study. 

The identification of the so-called ``cascades'' 
($\Xi^{-} \rightarrow \Lambda \pi^{-}$ and 
$\Omega^{-} \rightarrow \Lambda K^{-}$), 
goes through pairing V0 candidates with a single charged track, referred to as
the ``bachelor'', and then using selections on the V0 mass and impact 
parameter, the DCA between the V0 and the bachelor, the bachelor impact
parameter and the cascade pointing angle. 

The reconstruction of secondary vertices (that has been thoroughly investigated
in section 6.2 of \cite{ALICE-PPRVol2}), relies on the primary vertex 
reconstruction. ALICE shows good performances for the identification of 
secondary vertices from strange hadrons decays, both in Pb--Pb and pp 
collisions. In the latter case particle multiplicities are low, and 
combinatorial background is even lower, so that topological cuts can be 
loosened in order to gather more signal. However, the reconstruction of 
the primary vertex position in the low-multiplicity events produced by pp 
collisions is affected by a large error, which substantially alter the 
reconstruction efficiency. 

\begin{figure}[!ht]
\includegraphics[width=0.52\textwidth]{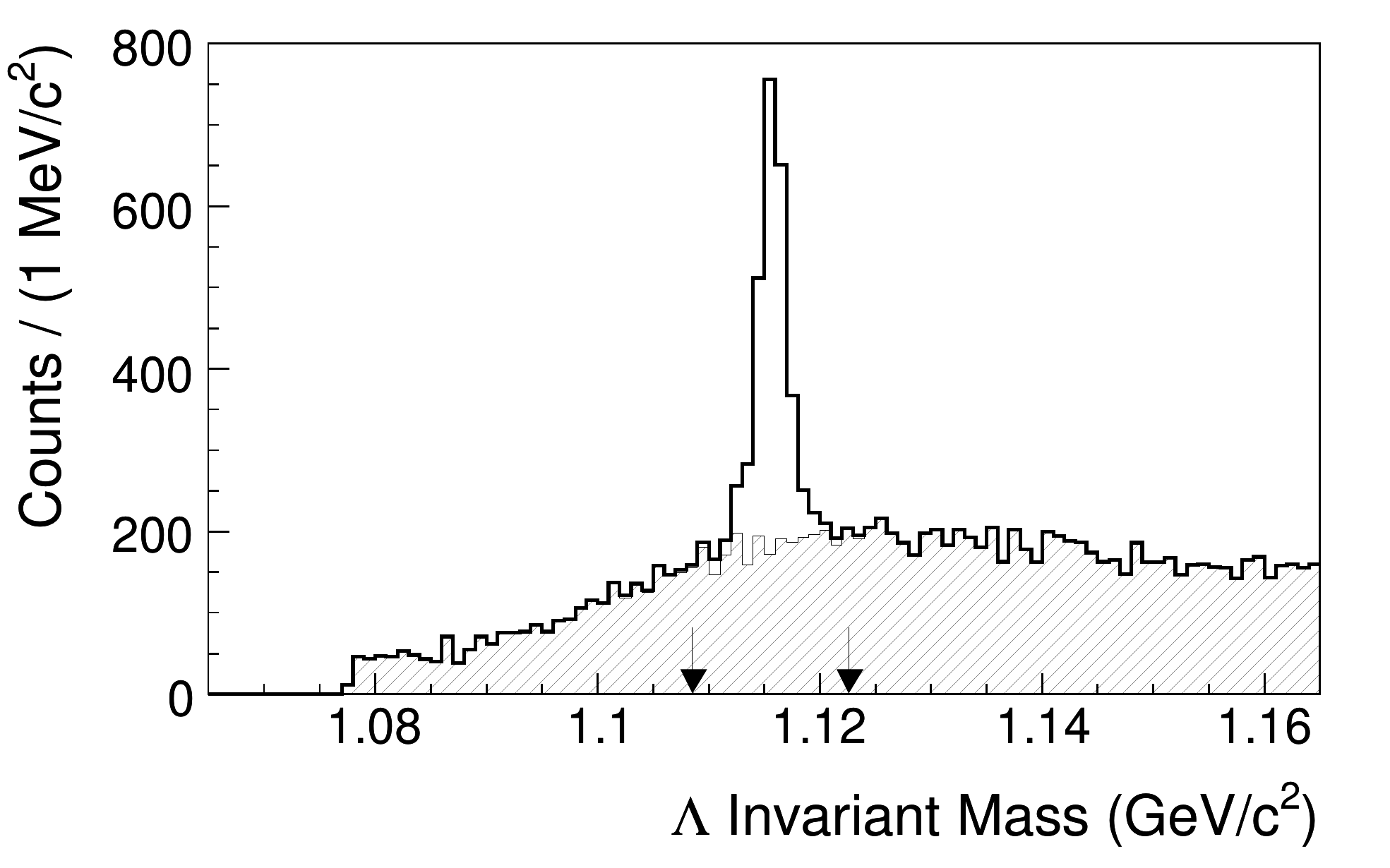}
\includegraphics[width=0.52\textwidth]{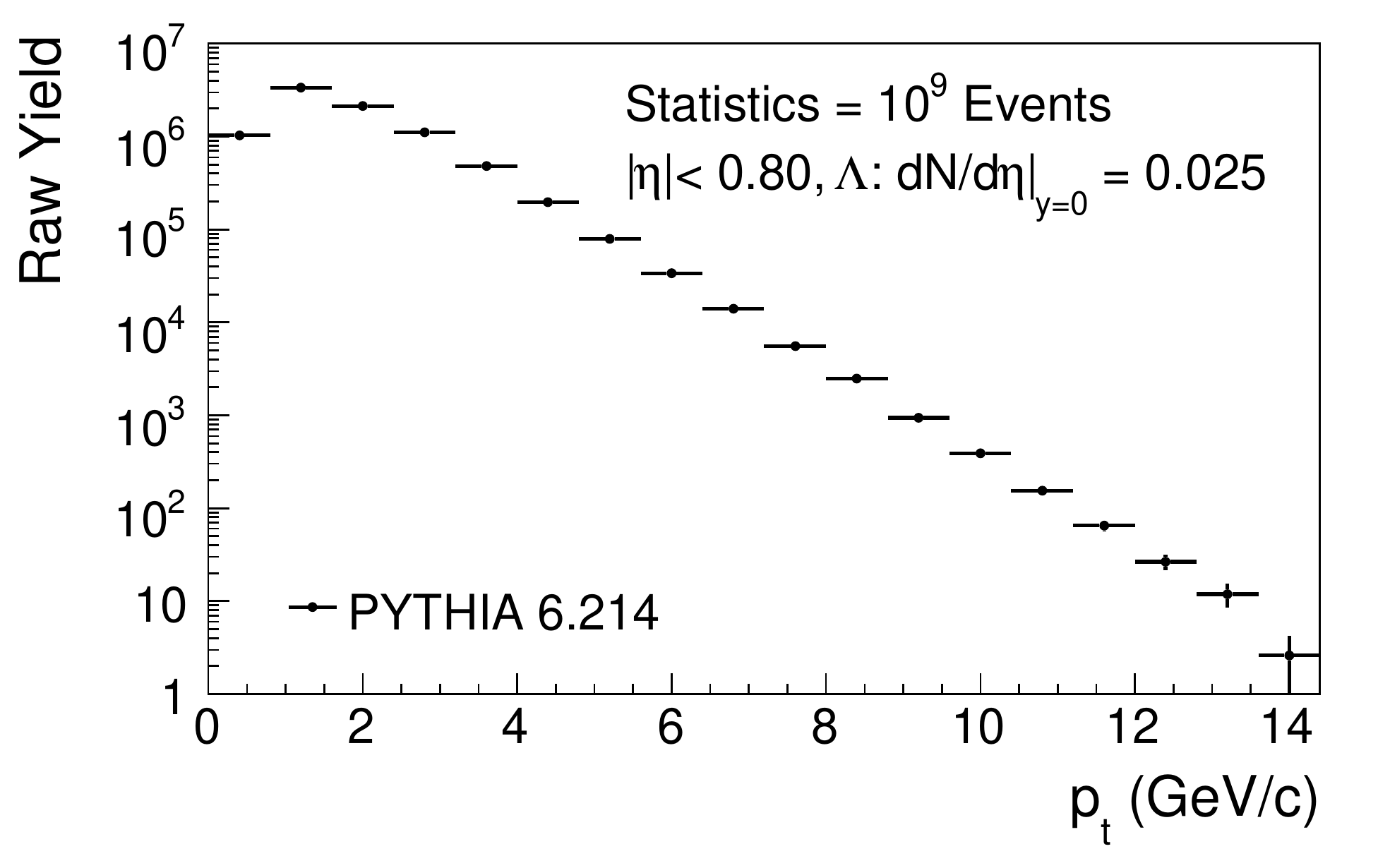}
\caption{Invariant mass distributions of $\Lambda$ reconstructed in pp events 
for tight selections (left); distribution of reconstructed $\Lambda$ as a 
function of $\pt$ for $10^{9}$ pp events at $\sqrt{s}=14$~TeV (right).}
\label{Lambda_PP}
\end{figure}

In any case a clear signal is obtained, as can bee seen in Fig.\ref{Lambda_PP} 
that shows in the left panel, for the case of the $\Lambda$ reconstruction,
the signal and background invariant mass spectra obtained for pp events 
generated with PYTHIA 6.214. The right panel shows the estimated distribution 
of reconstructed $\Lambda$ versus $\pt$ in the central rapidity range 
$|y|<0.8$, for one year of LHC running ($10^9$ events). 
  
Therefore the simulation (presented in \cite{ALICE-PPRVol2} and in 
\cite{Vernet})
shows that transverse mass spectra for $\Lambda$ should be measurable up 
to $\sim 11$~GeV/c in a full year of pp data taking at the LHC. On the other 
hand the maximum $\pt$ reachable for $K^{0}_{S}$ and the multi-strange 
hyperons $\Xi$ and $\Omega$ are 12, 8 and 7 GeV/c respectively. 

It must be remarked that in all the topological methods described above the 
single particle identification is not required, which makes them especially 
efficient at intermediate and high $\pt$, where particle identification based 
on energy loss (in ITS or TPC) and time-of-flight (in TOF) measurement fail. 
Therefore these studies can be performed by using only the basic ALICE 
tracking devices (the TPC and the ITS). 

As regards strange resonance identification, as for example 
the $K^{*}(892)$ and the $\Phi(1020)$ since they decay very early, their 
daughters are not discernible from other primary particles. 

Their main decay modes are $K^{*} \rightarrow K^{+} \pi^{-}$ and 
$\Phi \rightarrow K^{+} K^{-}$.
Therefore these resonances are identified via invariant mass reconstruction
methods that combine all possible pairs of primary daughter candidates. 
The background is very high since no selection other 
than particle identification or track quality is applied, and can be 
accurately estimated by means of 'like-sign' or 'event mixing' procedures.
The kaon and pion identification is obtained from the energy loss in the TPC 
and from the time-of-flight measured in the TOF.

It has been found by preliminary analyses that for both resonances
reaching $\pt$ as high as 4 GeV/c is not problematic. However, the 
identification of higher-$\pt$ resonances should be done without using 
particle identification.

All the tools to identify strange secondary vertices and resonances 
will provide first-physics observables, and a rather large statistics 
can be detected within the very first hours of LHC run. However, within 
a larger time scale (like the first full year of pp LHC run), the statistics
of strange particles reconstructed with ALICE will by far overstep that of the 
previous pp  and $\mathrm{p\overline{p}}$ experiments, and will allow several 
new studies that were barely achievable up to now because of statistics, such 
as the properties of $\pt$ spectra in a range of $\pt$ covering soft, 
intermediate and hard regimes, as wide as possible to understand the 
underlying QCD processes; and to be compared with the phenomena observed 
in nucleus--nucleus collisions at comparable centre-of-mass energy. 

\section{Baryon measurements}

Studies of baryon production in the central rapidity region of 
high energy pp collisions provide a crucial possibility to test 
the baryon structure and to establish how the baryon number is 
distributed among the baryon constituents: valence quarks and 
sea quarks and gluons.

Hadronic processes are described by several models (DPM\cite{DPM}, QGSM 
\cite{QGSM}, PYTHIA \cite{PYTHIA}) 
in terms of color strings stretched between the constituents of the colliding 
hadrons. In the framework of such models the dominant contribution to particle 
production in pp collisions involves diquark--quark string excitations 
followed by string breaking.
The unbroken diquark system, playing the role of carrier of baryon 
number, will take large part of the original proton momentum and 
subsequently fragment into leading baryons, concentrated in the fragmentation 
region of the colliding protons. Such approach, where baryon number transfer 
over wide rapidity intervals is strongly suppressed, describes successfully 
the bulk of data on leading baryon production. 

However, the observed high yield of protons in central rapidity region 
observed in experiments at the ISR pp collider (in the energy interval 
$\sqrt{s} = 23 - 63$~GeV), cannot be explained in the framework of 
such models, that assume an indivisible diquark. These measurements indicate 
that the baryon number can be transported with high probability over a rather
large rapidity gap ($\Delta y \le$~4)\cite{ISR_baryons}. An appreciable 
{\it baryon stopping} is observed, with baryons exceeding antibaryons, 
and in association with higher hadron multiplicities.

To describe such data other mechanisms of baryon number transfer have been 
suggested, following the approach originally introduced by Rossi and 
Veneziano~\cite{rv}. They have shown how it is possible to generalize to 
baryons the successful schemes employed to unify gauge, dual and 
Regge-Gribov theories of mesons. Their results on the topological 
structure of diagrams of processes involving baryons can be rephrased 
in a dual string picture in which the baryon (for $N_{c} = 3$) is a 
Y-shaped object with valence quarks sitting at the ends and with a string 
junction in the middle. Then, it can be assumed that the baryon number is 
carried by valence quarks, or otherwise by the string junction itself, which 
is a non-perturbative configuration of gluon fields.

In a first approach~\cite{kz89} the baryon number of the incident proton is 
assumed to be transferred to a more central rapidity region through a mechanism
 by which a valence quark is slowed down to the central rapidity region, while 
a fast spectator diquark is destroyed. The cross section of the baryon number 
flow has been estimated using perturbative QCD calculations: it has been found 
to depend on the rapidity gap $\Delta \mathrm{y}$ approximately as 
$\exp^{-\Delta \mathrm{y}}$ and nicely agrees with the data at ISR energies. 
Another estimate \cite{kharz96}, based on the topological approach and Regge 
phenomenology, and considering also the stopping of string junctions in the 
central rapidity region, finds a similar dependence of single baryon stopping 
cross section on energy and rapidity, in agreement with the ISR data.

In an alternative approach~\cite{kp97} the baryon number is assumed to be 
transferred dominantly by gluons. This mechanism does not attenuate baryon 
number transfer over large rapidity gaps, since the transfer probability is 
independent of rapidity. 

The HERA data on high-energy photon-proton collisions have offered a unique 
opportunity to study the mechanisms of baryon number transfer. 
The asymmetry in the e-p beam energies made it possible to study baryon 
production in the photon hemisphere up to 8 units of rapidity distance from 
the leading baryon production region. It has been shown in \cite{kp97} that 
at such large rapidity intervals the gluonic mechanism give a dominant 
contribution to the baryon number transfer. An experimental observable that 
is useful to distinguish between different baryon production models is the 
proton to anti-proton yield asymmetry 
$A_p = 2 \frac{N_{p}-N_{\bar{p}}}{N_{p}+N_{\bar{p}}}$, where $N_{p}$ and 
$N_{\bar{p}}$ are the number of protons and anti-protons produced in a 
given rapidity interval. The calculations made in \cite{kp97} predicted the 
asymmetry to be as big as about 7 \%, which appeared to be in reach with the 
statistics collected by the experiments at the HERA $ep$ collider 
\cite{H1baryons}. However, both the gluonic and valence quark exchange 
mechanisms were estimated in \cite{kp97} to give about the same asymmetry 
at $\eta = 0$, and appeared to explain the HERA data within experimental 
and theoretical uncertainties. 
On the other hand, it was shown in \cite{kp99} that the two mechanisms can be 
discriminated by studying the dependence of the baryon asymmetry on the 
multiplicity of the produced hadrons. Comparison with HERA data from 
\cite{H1baryons} strongly supports the assumption that the baryon asymmetry 
is dominated by the gluonic mechanism, and excludes a large contribution of 
baryon number transfer by valence quarks. Such asymmetry reflects the baryon  
asymmetry of the sea partons in the proton at the very low $x$ values, that are 
reached (down to $x \sim 10^{-5}$) at HERA.

More recently, the $R=\bar{p}/p$ ratio has been measured at the RHIC 
collider by the BRAHMS experiment \cite{BRAHMS_ratios} in pp collisions 
at $\sqrt{s} = 200$~GeV. The introduction of a string junction scheme appears 
to provide a good description of their data over the full coverage 
of $0<y<2.9$.

The ALICE detector at the LHC, with its particle identification capabilities 
and abundant baryon statistics  in the central-rapidity region, 
($7 \cdot 10^{8}$ $\overline{\rm p}$, $10^{7} $ $\Lambda$, $2 \cdot 10^{5}$ 
$\Xi$ and $10^{4}$ $ \Omega$ will be recorded with $10^9$ minimum bias 
events), is ideally suited to perform baryon flow studies.

Experimental observables that are useful to distinguish between such 
different models are the proton to anti-proton yield ratio $R=p/\bar{p}$ 
and their asymmetry $A_p = 2 \frac{N_{p}-N_{\bar{p}}}{N_{p}+N_{\bar{p}}}$. 
Similar observables can be defined also for $\Lambda$ and other identified 
hyperons, and can be studied as a function of particle multiplicity. 

At the LHC energies, the rapidity gap between incoming protons and central 
rapidity will be 9.6. That would allow the contribution from valence quarks to 
be probably negligible in comparison to that from gluons. On the other hand, 
within the limited acceptance of the ALICE central detectors ($|\eta|<0.9$)  
the proton-antiproton asymmetry predicted by different baryon flow models 
(being on the order of 5 \% at the LHC), would differ only slightly (a few \%).
 
A detailed studied \cite{Christag_ppbar} has been performed of the 
systematic errors affecting the asymmetry measurement, coming from transport 
code and material uncertainties, contamination from beam--gas events, and 
from the different quality cuts imposed at the event and track level. 
The estimated upper limit of the systematic error in the anti-proton to 
proton ratio and in the asymmetry $A_p$ is below 1\%, that is sufficient 
to keep these measurements at the level of accuracy required at the LHC. 

Such measurements will be also relevant for comparison to heavy ion 
collisions where baryon stopping should be dramatically enhanced.

Finally, ALICE can also study heavy flavour baryons ($\Lambda_b$, $\Xi_b$, 
$\Omega_b$...) which are poorly known. Due to the branching ratio 
$(4.7 \pm 2.8) \cdot 10^{-4}$ of the decay channel $\Lambda_{b} \rightarrow J/\psi \, \Lambda $, $10^9$ events triggered on J/$\psi$ using the TRD detector 
should produce a few thousands $\Lambda_b$.

\section{Correlations and fluctuations}

The study of the correlations among particles emitted in hadronic collisions 
is important in order to unveil the properties of the underlying production 
mechanisms. 

First, the analysis of the two-hadron momentum correlations provides valuable
information to constrain the space-time description of the particle production 
processes. These measurements are of great interest both for nucleus--nucleus 
collisions, where collective effects in nuclear matter are studied, and for pp 
collisions, where they provide clues about the nature of hadronization. 

Momentum correlations can be analysed using an interferometric technique that 
extracts space-time information on the particle emitting source by means of 
a Fourier transformation of the measured two-particle correlation function. 
Such technique was initially developed in astronomy by Hanbury-Brown and Twiss 
(HBT) to infer star radii from the measurement of a two-photon correlation 
function.

Particle correlations arise mainly from quantum statistics effects for 
identical particles and from final state interactions (Coulomb interactions 
for charged particles and strong interactions for all hadrons).
The two-particle correlation function C($\vec{p_1}$,$\vec{p_2}$) is defined as 
the ratio of the differential two-particle production cross section to a 
reference cross section which would be observed in the absence of the effects 
of quantum statistics and final state interactions. Therefore, experimentally 
the two-particle correlation function can be obtained from the ratio 
C($\vec{q}$,$\vec{K}$)= A($\vec{q}$,$\vec{K}$)/B($\vec{q}$,$\vec{K}$), 
normalized to unity at large $\vec{q}$, where $\vec{q}$ is the relative 
momentum of a pair, and $\vec{K}$ is the average pair momentum: the numerator 
A($\vec{q}$,$\vec{K}$) is the the distribution of the relative momentum for 
pairs of particles in the same event, whereas the denominator is the same 
distribution for pairs of particles in different events.

In order to extract information from the measured correlation function about 
the space-time geometry of the particle emitting source, it is generally 
assumed that the source distribution can be parameterised as a Gaussian. 
Simple analyses generally reconstruct source size in one dimension, thus 
providing a correlation function of the form:
\begin{displaymath}
C = 1 + \lambda \exp ( -R^{2} q_{t}^{2}) 
\end{displaymath}

\noindent where $q_t$ is the component of $\vec{q}=\vec{p_1} -\vec{p_2}$ 
normal to $\vec{p_1} + \vec{p_2}$, and $R$ and $\lambda$ (chaoticity) are the 
parameters related to the source size and to the strength of the correlation 
effect, respectively. 

Most recent analyses have done a major effort in reconstructing the 
3-dimensional source shape using the so-called Pratt-Bertsch cartesian 
parameterization to decompose the relative momentum vector of a pair $\vec{q}$ 
into a longitudinal direction $q_l$ along the beam axis, an outward direction 
$q_o$ transverse to the pair direction, and a sideward direction $q_s$ 
perpendicular to those two. Then, according to some given assumptions, the 
correlation function takes the simple form:

\begin{displaymath}
C(\vec{q},\vec{K})= 1 + \lambda \exp ( - R_{o}^{2}(\vec{K}) q_{o}^{2} - R_{s}^{2}(\vec{K}) q_{s}^{2} - R_{l}^{2}(\vec{K}) q_{l}^{2})
\end{displaymath}

\noindent Thus, three HBT parameters ($R_{o}$, $R_{s}$ and $R_{l}$) are 
extracted from the data, containing information about the space-time extent 
of the particle emitting source in the {\it out}, {\it side} and {\it long} 
directions. 

A pronounced dependence of HBT parameters on charged particle multiplicity 
in hadron--hadron collisions has been observed by several experiments: UA1 
\cite{UA1_HBT} and E735 \cite{E735_HBT} in p$\bar{\mathrm p}$ collisions at 
respectively $\sqrt{s} = $630 GeV and 1.8 TeV, and more recently STAR 
\cite{STAR_HBT} in pp collisions at $\sqrt{s} = $200 GeV. Furthermore pion HBT 
results from the STAR experiment \cite{STAR_HBT} have shown a transverse 
mass dependence ($m_{T} = \sqrt{ k^{2}_{t} + m^{2}_{\pi}}$) of the HBT radii
which is surprisingly independent of collision system (pp or nucleus--nucleus 
collisions), and very similar to the $m_T$ dependence measured by NA22 
\cite{NA22_HBT} in hadron--hadron reactions at the lower CERN SPS energies.  
Since the $m_T$ dependence of the HBT radii in heavy-ion collisions is usually 
attributed to the collective flow of a bulk system, results observed for 
hadron--hadron collisions could suggest that also in this case a thermalized 
bulk system undergoing hydrodynamical expansion is generated 
\cite{BudaLund_STAR}.
However, alternative scenarios have been proposed to explain the observed 
$m_T$ dependence, and the question is still open. 

As shown in sect. 6.3 of Ref.\cite{ALICE-PPRVol2}, all such studies of 
particle interferometry can be performed with good accuracy also in ALICE, 
thanks to its accurate tracking devices and its low $\pt$ cutoff, in order 
to test different theoretical models of particle production in pp collisions 
in the TeV region.

Since the expected source sizes in pp collisions are of the order of 1-2 fm, 
two-particle correlation functions are much wider than those obtained with  
nucleus--nucleus collisions. This, together with the smaller track density, 
makes in principle the momentum correlation analysis easier in pp than in 
heavy-ion collisions. On the other hand in pp collisions additional 
correlations come from the fact that at LHC energies a substantial fraction 
of the particles is produced inside jets. Therefore, additional analysis cuts 
are needed to prevent the merging of close track pairs.
Predictions for two pion correlations in $\sqrt{s} = $14 TeV collisions are 
provided in Ref.\cite{Humanic_HBT}, where it is shown how it might be possible 
to obtain information on the hadronization time in these collisions. 

Besides momentum correlations, other kinds of correlations among final state 
particles are important, in order to reveal the properties of the underlying 
production mechanisms, 

First, we can consider two-particle correlations in rapidity: if
 $C_{n}(\eta_{1},\eta_{2}) = \rho^{n}_{2}(\eta_{1},\eta_{2}) - 
  \rho^{n}_{1}(\eta_{1}) \rho^{n}_{1}(\eta_{2}) $
is the semi-inclusive two-particle correlation function for events with a 
fixed multiplicity $n$, written in terms of the single and two-particle 
densities, then we can define an inclusive correlation function 
$C_{s}(\eta_{1},\eta_{2})$ in terms of the $C_{n}(\eta_{1},\eta_{2})$ as:
\begin{displaymath}
C_{s}(\eta_{1},\eta_{2}) = \sum_{n} P_{n} C_{n}(\eta_{1},\eta_{2})
\end{displaymath}
with $P_n$ the probability to find an event with the multiplicity $n$.
As it was shown by the UA5 data at $\sqrt{s}$ = 200 and 900 GeV,  
$C_{s}(\eta_{1},\eta_{2})$ is sharply peaked at $\eta_{1}=\eta_{2}$, 
and for this reason it is usually referred to as a ``short-range'' correlation 
function. The qualitative shape of such correlation is well reproduced by a 
model \cite{Kawrakow} where the equation of the perturbative Pomeron results 
from the summation the of all orders of pQCD in the Leading Log Approximation 
(LLA).

On the other hand in hadron-hadron collisions clear evidence exists for strong
long-range correlations between the charged particles produced into opposite 
(forward and backward) c.m.s. hemispheres of a collision, and also between the 
particles produced in two rapidity bins separated by a wide rapidity gap 
$\Delta \eta$. 
For pp and p$\bar{\mathrm p}$ collisions the forward-backward multiplicity 
correlation coefficient increases logarithmically with energy over a large 
energy interval (from ISR to Tevatron energies). On the other hand, such 
dynamical correlations are absent or quite small in $e^{+}e^{-}$ collisions,  
up to LEP energies. Several attempts have been made to explain such 
correlations within the framework of hadronic string models, or by assuming 
that particles are produced through the decay of ancestors bodies named 
clusters (or clans) \cite{Berger,Ranft_clusters,GiovanniniVanhove,ChouYang,Lim}
, but the exact dynamical origin of such correlations still seems unclear. 
Therefore the study of the forward-backward multiplicity correlations 
represents a useful tool to test any model of hadron production, also 
in the LHC energy domain \cite{GiovanniniUgoccioni_FB}.
On such respect the ALICE experiment is well designed for such studies, since 
its Forward Multiplicity Detector extends the charged particle multiplicity 
measurement from the pseudorapidity interval $-2 < \eta < 2$ covered by the 
SPD to the range $-3.4 \leq \eta \leq 5.1$, thus allowing to study 
the multiplicity correlation between largely separated rapidity bins. 

Another interesting subject is the study of two-particle correlations in 
azimuthal angle $\phi$, initially proposed by Wang \cite{Wang} as a method to 
understand the role of minijets in high energy hadronic interactions. It was 
argued that calculating $C(\phi_{1},\phi_{2})$ for samples of particles with 
$\pt$ above a given $p_{T}^{cut}$, the influence of the underlying soft 
processes could be reduced: the higher the $p_{T}^{cut}$, the more the 
correlation should look like the profile of high-$\pt$ jets.

New analysis approaches have been developed recently by STAR collaboration 
\cite{STAR_whitebook, PorterTrainor} 
to study two-particle correlations in 200 GeV pp (and nucleus--nucleus) 
collisions at RHIC. By looking at the two-particle correlations on transverse 
rapidity $y_T = \ln [(m_{T}+p_{T})/m_\pi ]$, pseudorapidity $\eta$, azimuth 
$\phi$ and on the angular difference variables 
$\eta_{\Delta}= \eta_{1}-\eta_{2}$ and $\phi_{\Delta}=\phi_{1}-\phi_{2}$,
they found that low-$Q^2$ parton fragments (minijets) dominate the correlation 
structure observed both in pp and in nucleus--nucleus collisions. In 
particular they found that at low $Q^2$ the fragmentation process in pp 
differs markedly from the pQCD factorization picture, the 'jet cone' being 
strongly elongated in the azimuth direction.

Additional valuable information on the collision dynamics may be obtained 
in the event-by-event studies of the correlations between various observables 
measured in separated rapidity intervals. 
Model-independent detailed experimental information on long-range correlations
between such observables as charge, net charge, strangeness, multiplicity 
and transverse momentum of specific type particles could be a powerful 
tool to discriminate theoretical reaction mechanisms. 

On the other hand the experimental studies of the correlations in small 
domains of the phase space have to cope with the problem of the local 
fluctuation of the produced hadrons and, more generally, of the experimental 
observables. Indeed, large concentrations of particles in small pseudorapidity 
intervals for single events have been seen in JACEE cosmic ray experiment 
\cite{JACEE}, and in the fixed-target experiment NA22 \cite{NA22}. A possible 
explanation of these spikes was related to an underlying intermittent 
behaviour, i.e. to the guess that there exists a correlation at all scales 
which implies a power-law dependence of the so-called ``normalized factorial moments'' of the multiplicity distribution on the size of the phase-space bins. 
If the above mentioned scaling law will be confirmed in the LHC energy 
domain a new horizon will be opened on self-similar cascading structure and 
fractal properties of hadron-hadron collisions.

The ALICE experiment is well designed for correlation studies, as well for 
the event-by-event measurement of several observables. Charged particle 
measurement in the central region is given by the combination of the ITS, TPC  
and TOF detectors, that provides momenta and particle identification of 
hadrons. 
The charged particle multiplicity measurement in the pixel-detector of the ITS
can be measured up to $\eta=\pm$2 , and the FMD extends this range to 
$-3.4 \leq \eta \leq 5.1$. 
In the central rapidity region the calorimeter PHOS with a
rather limited coverage provides photon multiplicity and photon momenta, 
whereas PMD is designed for photon multiplicity in the high particle 
density region of forward rapidity ($2.3<\eta<3.5$).
Therefore the combination of the information coming from these detectors 
provides an excellent opportunity to study particle correlations as well 
event-by-event physics and fluctuation phenomena at the LHC energies.
More details can be found in sect. 6.5 of Ref. \cite{ALICE-PPRVol2}.

\section{Diffractive physics}

Diffractive reactions in proton--proton collisions are characterised by 
the presence of rapidity gaps and by forward scattered protons. 
Experimentally, a diffractive trigger can therefore be defined by the 
tagging of the forward proton or by the detection of rapidity gaps.

In ALICE, in absence of Roman pot detectors for proton tagging, a diffractive 
double-gap Level-0 trigger can be defined by requiring little or no activity 
in the forward detectors (as the V0), and a low multiplicity in the Silicon 
Pixel Detector (SPD) of the central barrel \cite{Schicker}. 
However, in defining a L0 diffractive trigger, also the signals of other fast  
detectors of the central barrel must be used, especially the TRD, that is put 
in sleep-mode after the readout of an event. Therefore, since the SPD signal 
would not be in time for the wake-up call of the TRD, the V0 signals are  
firstly trasferred to the TRD pre-trigger system, where a wake-up call signal 
is generated by using the information provided by the time-of-flight (TOF) 
array. The output of such a trigger unit is fast enough to reach the ALICE 
central trigger processor well before the time limit for L0 decision.

The acceptance and segmentation in pseudorapidity of the V0 detectors allow 
to select a gap width of approximately 3 and 4 pseudorapidity units beyond 
$|\eta| = 2$ on the two sides, in steps of half a unit. Then, the high-level 
software trigger (HLT), having access to the full information coming from the 
central tracking detectors, can enlarge the rapidity gap to the range 
$-3.7 \leq \eta \leq -0.9$ and $0.9 \leq \eta \leq 5$. 

Furthermore, the information of the zero-degree calorimeter (ZDC) can be used 
in the high-level trigger to identify different diffractive event classes. 
Events of the type $pp \rightarrow p N^{*} X$ (where X denotes a centrally 
produced diffractive state), are characterised by a signal in either the two 
ZDC calorimeters, whereas events  $pp\rightarrow N^{*} N^{*} X$ present a 
signal in the calorimeters of both sides. 

Therefore the geometry of the ALICE experiment is suited for measuring a 
centrally produced diffractive state with a rapidity gap on either side. 
Such topology results from double-Pomeron exchange with subsequent 
hadronization of the central state. It is expected that such events show 
markedly different characteristics as compared to inelastic minimum bias 
events. For example mean transverse momenta of secondary particles are expected 
to be larger, and also the $K/\pi$ ratio is expected to be enhanced. 

A soft/hard scale can also be defined according to whether the $\pt$ of the 
secondaries is smaller or larger than some threshold value $p_{thr}$.
The invariant-mass differential cross section is thought to follow a 
power law: $\frac{d\sigma}{dM^{2}}\sim \frac{1}{M^{\lambda}}$. 
A study of the exponent $\lambda$ as a function of the threshold value 
$p_{thr}$ can reveal the contribution from soft/hard exchanges.
Such analysis can be carried out as a function of rapidity gap width. 

Signatures of Odderon exchanges can be searched for in exclusive reactions 
where, besides a photon, an Odderon (a color singlet with negative C-parity), 
can alternatively be exchanged. For example, diffractively produced C-odd 
states such as vector mesons $\phi$, $J/\psi$, $\Upsilon$ can result from 
photon-Pomeron or Odderon-Pomeron exchanges. Any excess beyond the photon 
contribution would be an indication of Odderon exchange. Estimates of cross 
sections for diffractively produced  $J/\psi$ in pp collisions at LHC energies 
\cite{Bzdak} result to be at a level that in $10^6$ s of ALICE data taking
the $J/\psi$ could be measured in its $e^{+}e^{-}$ decay channel at a level 
of 4\% statistical uncertainty (see Section 6.7.5 of \cite{ALICE-PPRVol2}
for more information on quarkonia detection in the dielectron channel in the 
ALICE central barrel). Furthermore, a transverse momentum analysis of 
the  $J/\psi$ might allow to disentagle the Odderon and photon contributions, 
following their different t-dependence.

Finally, diffractive heavy quark photoproduction, characterised by two 
rapidity gaps in the final state, represents an interesting probe to look for 
gluon saturation effects at the LHC \cite{Goncalves1}, where the cross 
sections for diffractive charm and bottom photoproduction amount respectively 
to 6 nb and 0.014 nb \cite{Goncalves2}. Heavy quarks with two rapidity gaps in 
the final state can, however, also be produced by central exclusive production,
 i.e. two-Pomeron fusion.  However, since the two production mechanisms have 
a different t-dependence, a careful analysis of the $\pt$ dependence of the 
$\mathrm{Q}\bar{\mathrm{Q}}$ pair might allow to disentangle the two 
contributions.

\section{Jet physics}

The measurement of jet production in pp collisions is an important 
benchmark for understanding the same phenomenon in nucleus--nucleus 
collisions. The energy loss experienced by fast partons in the nuclear medium 
(through both radiative \cite{Gyulassy,BDMS,BSZ,Kovner} and collisional 
\cite{Bjorken_1982,Mustafa,Dutt,Djordjevic,Djordjevic_2} mechanisms) is 
expected to induce modifications of the properties of the produced jets.  
This so-called {\it \mbox{jet quenching}} has been suggested to behave very 
differently in cold nuclear matter and in QGP, and has been 
postulated as a tool to probe the properties of this new state of the matter.
The strategy is to identify these medium-induced modifications that 
characterise the hot and dense matter in the initial stage of a 
nucleus--nucleus collision, by comparing the cross sections for some jet 
observables in benchmark pp collisions at the same centre-of-mass energy. 

An accurate understanding of jet and individual hadron inclusive production 
in pp collisions is therefore quite important in order that this strategy be
successful. In this respect, the LHC will open a new kinematic
regime, in which the pp collisions involve features which are not
well understood yet. Therefore, the ALICE experimental programme
will also involve specific studies on jet and high-$\pt$ particle 
production in pp collisions.

In its original design ALICE can only study charged-particle jets by using the 
tracking detectors of the central barrel part of the experiment, covering 
the region $|\eta| < 0.9$. Their high-$\pt$ capabilities, with a momentum 
resolution better than 10\% at $\pt$~= 100 ${\mathrm GeV}$, are sufficient 
for jet identification and reconstruction up to $\Et \simeq$~200 GeV.

However, the strength of ALICE consists in the possibility of combining these 
features with low-$\pt$ tracking and particle identification capabilities, 
to perform detailed studies of jet-structure observables over a wide range 
of momenta and particle species \cite{Morsch:2005sv}. 

Furthermore, since the charged-jet energy resolution is severely limited by  
the amount of charged to neutral fluctuations ($\simeq 30\%$), an 
electromagnetic calorimeter (EMCal) has been designed \cite{ALICE_EMCal,ALICE_EMCal_TDR} 
to complete the ALICE capabilities at high $\Et$. The EMCal covers the region 
$|\eta| < 0.7$, $60 \mbox{$^\circ$} < \varphi < 180 \mbox{$^\circ$}$ 
and has a design energy resolution of $\Delta E / E = 10\%/ \sqrt{E}$. 
The EMCal will improve the jet energy resolution, increase the selection 
efficiency and further reduce the bias on the jet fragmentation through the 
measurement of the neutral portion of the jet energy. Furthermore, it will 
add the jet trigger capabilities which are needed to record jet enriched data 
at high $\Et$.

The low and high transverse momentum tracking capabilities combined with 
electromagnetic calorimetry will represent an ideal tool for jet structure 
studies at the LHC over a wide kinematic region of jet energy and associated 
particle momenta, from the hardest down to very soft hadronic fragments.
A similar strategy has also been used by the STAR collaboration at the RHIC 
collider to reconstruct jets with an electromagnetic calorimeter and a TPC, 
and then to perform systematic studies of fragmentation functions in inclusive 
jets from pp collisions at $\sqrt{s} = 200$ GeV \cite{Heinz_STAR_HP08}. 

ALICE will study jet production on a large $\Et$ range, from minijet
region ($\Et >$2~GeV) up to high-$\Et$ jets of several hundred GeV.
However, the event-by-event jet reconstruction will be restricted to 
relatively high-energy jets, approximately $\Et >$30--40~GeV, whereas 
leading-particle correlation studies will play an important role 
at low-$\Et$. 

Observables of interest for jet studies will include: 1) the semi-hard  
cross sections, measured by counting all events with at least one jet 
produced above some given $\Et$; 2) the relative rates of production of 1, 2 
and 3 jets as a function of the lower $\Et$ cutoff; 3) the double-parton 
collision cross-section and their distinction from the leading QCD $2\to 4$ 
process; 4) the properties of the Underlying Event (UE) in jet events, as it 
has been done extensively by the CDF Collaboration at the Tevatron 
\cite{CDF_UE} by examining the multiplicity and the $\pt$ spectra of 
charged tracks in the ``transverse'' region in $\eta$-$\phi$ space with 
respect to the direction of the leading charged particle jet. 

\begin{figure}[htb]
\begin{center} 
\includegraphics[width=.80\textwidth]{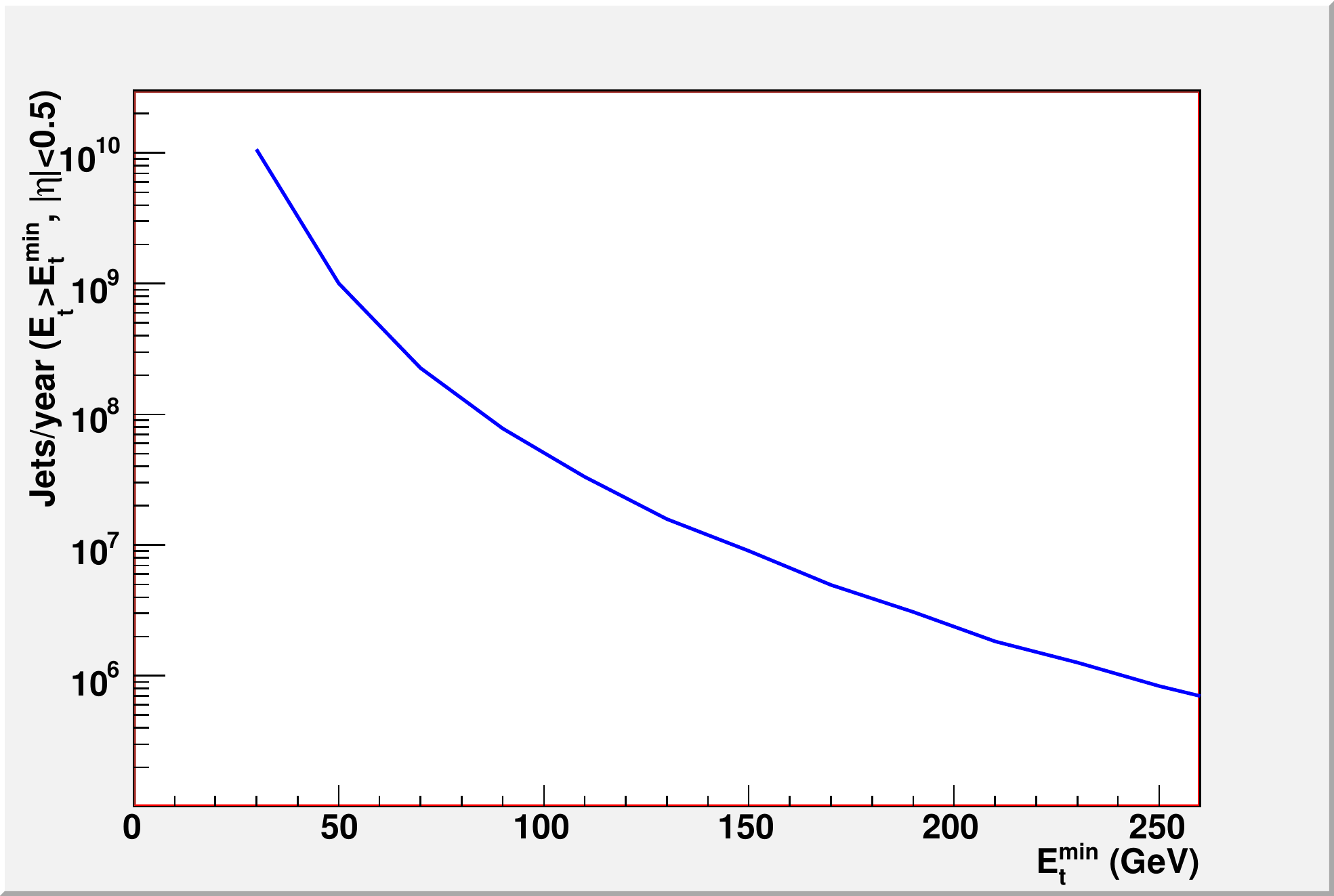}
\caption{
Number of jets with $\Et > {\Et}^{min}$ produced per year within 
$|\eta| < 0.5$ (ALICE central barrel fiducial region) in minimum bias pp 
collisions at $\sqrt{s}$=14 TeV with luminosity 
$5 \times 10^{30}$ cm$^{-2}$s$^{-1}$. 
}
\label{fig:JetYields_ALICE_pp}
\end{center} 
\end{figure}

The jet yield that can be measured with ALICE in a running year ($10^7$~s)
has been estimated by using the hadronic cross sections calculated at NLO 
\cite{Armesto_jets} for a cone algorithm with R=0.7, and using CTEQ5M p.d.f.
and factorization and renormalization scales equal to $\mu=E_{T}/2$. 
Fig.~\ref{fig:JetYields_ALICE_pp} shows the annual jet yield for inclusive 
jets with $\Et > {\Et}^{min}$ produced within the ALICE central barrel 
fiducial region $|\eta| < 0.5$ for minimum bias pp collisions at the nominal 
luminosity in the ALICE interaction point 
$L = 5 \times 10^{30}$ cm$^{-2}$s$^{-1}$ 
(or $L_{int} = 50$ pb$^{-1}$ per year).

However, the rates estimated as above are production rates, which could only 
be exploited by fast dedicated hardware triggers. The EMCal will provide 
$\gamma$, $\pi^0$ and electrons triggers, that can be considered to be jet 
triggers of a sort, but the resulting sample will 
be dominated by relatively low-$\Et$ jets that fragment hard. A more refined 
selection of high-$\Et$ jets requires a jet trigger which sums energy over a 
finite area of phase space and finds the location of the patch with the 
highest integrated EMCal energy. The expected enhancement in statistics due to 
the EMCal trigger can be estimated by comparing the rates to tape of 
EMCal-triggered observables and equivalent observables using only charged 
tracks in the TPC and simple interaction (`minimum bias') triggers. 

It has been estimated (see Sect.~6.8 in Ref.~\cite{ALICE-PPRVol2} and 
Sect.~7.1 in Ref.~\cite{ALICE_EMCal}) that the EMCal trigger will 
significantly increase the statistics, by a factor $\sim$~70 for the $\pi^0$ 
trigger (relative to untriggered charged pion measurements), and by a factor 
$\sim$~50 for finite-area jets of trigger patch-size 
$\Delta \eta\times\Delta \varphi = 0.4 \times 0.4$. The enhancement will be 
limited by the EMCal acceptance ($\sim25$\% of the TPC acceptance) and by its 
reduced effective value for jet triggers of finite extent in phase space 
relatively to small-area triggers ($\gamma$, $\pi^0$ and electrons). However, 
jet measurements incorporating both the EMCal and tracking have significantly 
better resolution and less bias than jet measurements based solely on charged 
particles. Thus, the EMCal-triggered jets provide more robust measurements 
even for modest trigger enhancements.

Depending on the setup of the Level-1 (L1) triggering detectors, the software 
High-Level-Trigger (HLT) will be used to either verify the L1 hypothesis or 
to solely inspect events at L2. A very simple online algorithm can run on the 
nodes of the HLT system to online search for jets using the full event 
information from the central tracking detectors. This algorithm is supposed 
to trigger if it finds at least one charged particle jet with more than 
$m$~GeV in a cone with $R$=0.7. 

Trigger simulations \cite{Loizides_2005PhDThesis} show that for $m$=30 GeV 
data rates in pp can be reduced by a factor of 100 relative to L1 rates, while 
keeping 1/5 of the events where ${\Et}^{min}>$~50 GeV and slightly more than 
half of the events with ${\Et}^{min}>$~100 GeV. 

In case of a HLT running without the help of a jet trigger at L1 
(as in the running scenario before the installation of the EMCal, and 
neglecting the possibility of a trigger provided by the TRD), the yields 
will drop by a factor of $\sim$ 350. The inspection rate of the HLT will be 
limited to the TPC maximum gating frequency of 1 kHz. 
The expected jet yield accumulated in one year for ${\Et}^{min}$= 100 GeV 
when ALICE is running in this configuration, is on the level of $10^4$ events, 
that is at the statistical limit for the analysis of jet fragmentation 
function at high-$z$.
 
Therefore a trigger with EMCal will be necessary to collect jet enriched data 
at $\Et >$~100 GeV and extend the kinematic reach for inclusive jets to above 
$200$~GeV. For di-jets, with a trigger jet in the EMCal and the recoiling jet 
in the TPC acceptance, the kinematic reach will be about $170$~GeV.

\section{Photons}

The study of prompt photons processes, in which a real photon is created in 
the hard scattering of partons, offers possibilities for quantitative and 
clean tests of perturbative QCD (pQCD).
 
Lowest-order QCD predicts that prompt photons can be produced directly at a 
parton interaction vertex mainly by two processes: quark-antiquark annihilation
 ($\mathrm{q}+\bar{\mathrm{q}}\rightarrow\gamma+\mathrm{g}$) and quark-gluon 
Compton scattering ($\mathrm{g}+\mathrm{q}\rightarrow\gamma+\mathrm{q}$). 
Because of the latter process, which dominates the photon production in pp
collisions, the measurement of prompt photons provides a sensitive means to 
extract information on the gluon momentum distribution inside the proton.

However, an additional source of high-$\pt$ prompt photons is due to the hard 
bremmstrahlung of final state partons (fragmentation photons). The latter is a 
long-distance process which is not perturbatively calculable, since it emerges 
from the collinear singularities occurring when a high-$\pt$ parton undergoes a
 cascade of successive splittings ending up with a photon. These singularities 
can be factorised and absorbed into a parton-to-photon fragmentation function 
which has to be determined experimentally and then included in the theoretical 
calculations.

The calculations of the production cross section of prompt photons at large 
$\pt$ have been carried out in the framework of perturbative QCD up to 
next-to-leading order (NLO) accuracy in $\alphas$. Their results have been
found to describe rather well, within experimental errors and theoretical 
uncertainties, all the prompt photon data collected in pp and 
$\mathrm{p}$$\bar{\mathrm{p}}$ collisions over the last 25 years, both 
at fixed-target experiments ($\sqrt{s}$=20--40~GeV) and at colliders 
($\sqrt{s}$=63--1800~GeV) \cite{Tornbull,Aurenche_1999,Aurenche_2000}.  

Corrections for bremmstrahlung processes, for higher-order QCD diagrams and 
for higher-twist processes have been applied to the theory in recent years. 
Furthermore, full QCD calculations have been implemented up to NLO accuracy  
in more flexible Monte Carlo programmes at the partonic level, that allow 
to account easily for any kind of experimental cut 
\cite{Binoth2000,Catani2002}. 
This is particularly important for the analyses of data collected in collider 
experiments, that require isolation criteria on photon candidates in order  
to suppress the huge background of secondary photons coming from hadron 
decays (mainly $\pi^0$, $\eta$) and to reduce the fragmentation component 
of prompt photon production.

It has been found that NLO pQCD predictions agree very well also with the most 
recent data collected by D0 \cite{D0_2006} at the Tevatron Run II 
($\sqrt{s}$=1.96~TeV) and by PHENIX \cite{PHENIX_2006} at the RHIC 
($\sqrt{s}$=200~GeV). D0 has measured isolated prompt photons in the range 
$23 < \pt < 300$ GeV/c, whereas PHENIX has collected both inclusive and 
isolated photon data in the range $4 < \pt < 16$ GeV/c. 

In a recent phenomenological analysis~\cite{Aurenche_2006} all available 
prompt photon cross section data, including the most recent data from D0 and 
PHENIX, have been compared with NLO pQCD theoretical predictions evaluated at 
the common scale $\mu$=$\pt$/2. The data span two orders of magnitude in 
energy and there is an agreement over nine orders of magnitude in the cross 
sections between theory and experimental data.

An exception still comes from the fixed-target experiment E706 
\cite{E706_2004}, at the Fermilab, that measured cross sections several 
times above theoretical predictions based on NLO pQCD calculations, with data 
and theory differing both in magnitude and shape. Although resummed 
calculations \cite{recoil} accounting for recoil effects due to soft gluon 
radiation have reduced the theoretical uncertainties, E706 data still suggest 
large non perturbative parameters (i.e. an intrinsic $k_T$ \cite{kT}) not 
required by any other data sets. Data in the small-$x$ domain probed by LHC 
may contribute to clarify this issue, and its relation to the recoil 
resummation.

Predictions of production rates at LHC, obtained from calculations performed 
at next-to-leading order, still suffer from rather large uncertainties. These 
uncertainties are associated with the choice of renormalization, factorization 
and fragmentation scales~\cite{Aurenche_1999,Gordon93}(of the order of 30\%). 
As for the uncertainties associated to the structure functions they 
are expected to be relatively small (almost 10\% in the lowest $\pt$ region).

However, for photon production at the LHC energies, a new kinematic region of 
small $x$ values ($x \approx  x_\mathrm{T}= 2 \pt /\sqrt{s}$ will be 
explored, especially at low transverse momenta (for $\pt = 2$~GeV/c, 
this corresponds to $x = 3\times 10^{-4}$ at $\sqrt s = 14$~TeV). Therefore 
one may question the reliability of straightforward NLO pQCD calculations in a 
kinematic domain where they have never been tested before, and where recoil 
corrections and resummed calculations may be required. 

Furthermore, a specific feature of photon production at very high energy 
is related to the fact that the bremmstrahlung component becomes large and 
dominant at small $x$. The bremmstrahlung from a gluon is dominant at 
not too high $\pt$ (up to 20~GeV), whereas the production of photons from 
final-state quarks still remains important at higher $\pt$ (about 40\% of the 
prompt photon yield at $\pt \sim$~50~GeV/c \cite{GTagged}). However, the 
bremmstrahlung component is not well under control: in particular the gluon 
fragmentation function into a photon is not sufficiently constrained by 
previous data, and that results in a factor two uncertainty in the prompt 
photon rate for $\pt < 20$ GeV. On the other hand the uncertainties in the 
quark fragmentation component might introduce difficulties in the calibration 
of the energy of a jet through the measurement of the energy of the recoiling 
photon in $\gamma$--jet events. 

\begin{figure}[htb]
\begin{center} 
\includegraphics[width=.80\textwidth]{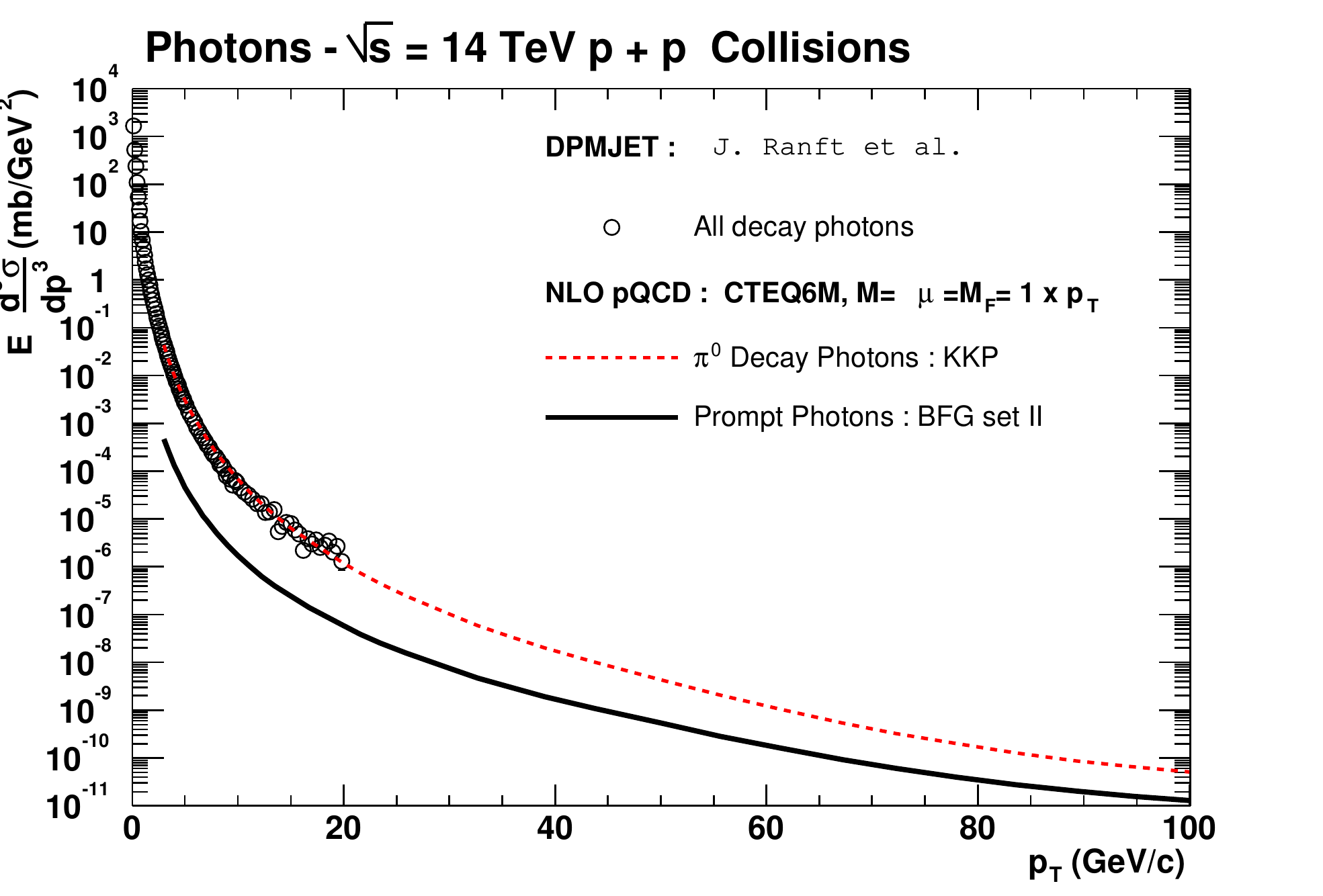}
% Fig. 29b of Yellow Report
\caption{
Comparison of prompt and decay photon spectra in pp collisions at
$\sqrt{s}$=14~TeV.
}
\label{fig:Photons_pp_14TeV}
\end{center} 
\end{figure}

Figure~\ref{fig:Photons_pp_14TeV} shows the predictions at the LHC energies 
of prompt and decay photon spectra \cite{YellowReport_2003_photons}. Prompt 
photon spectra are calculated with NLO pQCD, while for decay photons the NLO 
pQCD estimates, extracted from calculated $\pi^0$ spectra, are compared to the 
predictions of PHOJET/DPMJET \cite{DPMJET}, an implementation of Dual Parton 
Model which includes soft physics (pomeron exchanges) and semi-hard dynamics. 

The results show that the prompt photon spectrum is dominated by more than 
an order of magnitude by the decay photon spectrum, for which there is an 
excellent agreeement between the predictions of DPMJET and NLO pQCD. 

NLO pQCD predictions for the ratio $\gamma_{prompt}/\pi^0$ are in the range 
$5 \times 10^{-3}$--$10^{-2}$ for $\pt < 10$ GeV/c, and in the range 
$10^{-2}$--$10^{-1}$ for $10 < \pt < 100$ GeV/c.
The ratios, slowly increasing with $\pt$, are rather small at LHC energies, 
and isolation cuts will certainly be necessary to reduce the amount of the 
huge background due to decay photons.

At low $\pt$ the uncertainty in the photon spectrum is largely due to the 
choice of the gluon fragmentation function into photons, which is hardly 
constrained by the present data \cite{frag-phot}, and it goes from a factor 
up to 2.5 at 3 GeV/c to a factor 10\% or less for $\pt > 20$ GeV/c. The 
uncertainties associated to the structure functions are much smaller, of the 
order of 10\% and 2.5\% at low and high $\pt$, respectively.

Photons will be detected in ALICE by the Photon Spectrometer PHOS (see 
\cite{SCINT} and Sect.~3.9 in Ref.~\cite{ALICE-PPRVol1} for details), 
an electromagnetic calorimeter with high resolution but limited acceptance 
($\eta < 0.12$, $\Delta \Phi = 120 \mbox{$^\circ$}$). 
The identification power of prompt photons in PHOS is limited by the background
created by decay photons (mainly, $\pi^{0},\eta \rightarrow \gamma+\gamma$),
and is optimal for photons with energy larger than 20~GeV.
Below this value, decay and prompt photons cannot be efficiently distinguished
on an event-by-event basis, and statistical methods are needed. 

Photon measurements can be performed also with the EMCal, the second 
electromagnetic calorimeter in ALICE, which has coarser granularity compared 
to PHOS, but a factor eight larger phase space coverage, and so it may allow 
to extend ALICE photon measurements to higher $\pt$. However, the EMCal 
capabilities in the photon measurements will not be discussed here.

Two different procedures to select (with PHOS) prompt photons among inclusive 
photons are possible: Shower Shape Analysis (SSA), and Isolation Cut Method 
(ICM) (see \cite{SCINT} and Section 6.9 of Ref.~\cite{ALICE-PPRVol2}).
The former identifies photons by analysing the shape of the shower in PHOS, 
and the latter tags and identifies a photon as prompt if it appears isolated, 
i.e., without charged particles emitted in the same direction. 

To optimize the number of parameters conveying the maximal information about 
the shower topology, a principal component analysis was performed.
The values of the two principal components, corresponding to the largest 
eigenvalues, have a Gaussian distribution. Low, medium and high purity photon 
samples can be defined by cutting at three, two and one standard deviations,
respectively. For medium purity photon samples prompt photon identification 
efficiency is about 85\% for pp collisions. 
The contamination from misidentified neutral pions ranges from 0\% at 
$E_\gamma=40$~GeV to 40\% at $E_\gamma=100$~GeV and the contamination from 
misidentified charged hadrons and neutrons ranges from 5\% at $E_\gamma=20$~GeV
to 15\% at $E_\gamma=100$~GeV. Requiring higher purity photons the rejection 
improves at the cost of an important identification efficiency reduction. 
To improve the situation, additional identification procedures are required.

The main source of background to the prompt-photon spectrum at high $\pt$ 
is due to $\pi^0$ which at $\pt >40$ GeV/c produce single clusters in PHOS.
Therefore isolation algorithms can be devised, that search for hadrons inside 
a cone centred around the direction ($\eta_{0},\:\varphi_{0}$) of 
high-$\pt$~photon candidates ($\pt > 20$ GeV/c) identified by PHOS with the 
SSA method.

In the case of pp collisions a prompt photon identification probability of 
100\% and a $\pi^0$ misidentification probability of 3\% were estimated 
from simulations with cone radius $R=0.2$.  

Finally, our Monte Carlo simulations indicate that with PHOS the photon 
spectrum in pp (and also in \mbox{Pb--Pb}) collisions can be measured with 
the statistics of one standard year up to about 80--100 GeV/c, with a 
total systematic error of the order of 20\%.  

As already explained in another section of the present report, because of the 
large cross sections available for hard processes at LHC, exclusive jet 
measurements will be within reach. In particular, the measurement of jet 
topology (jet shape, jet \emph{heating}, fragmentation functions, etc.) will 
require the identification of jets and the measurement of the parton (or jet) 
energy. A very attractive method of performing these studies is to tag jets 
with prompt photons emitted in the opposite direction to the jet direction.

A $\gamma$-tagging algorithm was developed~\cite{GAMMAJET} in ALICE 
to identify $\gamma$--jet events and to reconstruct the hadronic jet features. 
The algorithm was tuned for two experimental configurations of ALICE: 
({\it i}) Charged particles are detected in the central tracking system and 
neutral particles in EMCal (this configuration is labelled as `TPC+EMCal'); 
({\it ii}) Only the central tracking system is available and consequently only 
charged particles can be detected (this configuration is labeled as `TPC').

The jet selection efficiency, defined as the ratio of the number of identified 
$\gamma$-tagged jets to the number of prompt photons found in PHOS was 
calculated using a Monte Carlo simulation. The efficiency for the 
configuration with EMCal is about 30\%. 
For the configuration without EMCal we obtained an efficiency of 40--50\%,
because of: ({\it i}) the wider selection range, implying a lower
identification quality; and ({\it ii}) the larger acceptance in azimuth of
the central tracking system as compared to that of EMCal.

Jet fragmentation functions to be measured in a standard year of LHC running 
(for both pp and \mbox{Pb--Pb} collisions) were studied for identified 
$\gamma$--jet events in the $\pt$ range from 20 to 100 GeV/c. 
The fragmentation functions obtained for jet--jet events misidentified as
$\gamma$--jet events were also studied. 
For pp collisions, we obtained a signal ($\gamma$--jet) to background 
(jet--jet) ratio of about 20 in the configuration without EMCal, and near   
to 100\% background rejection for the setup with EMCal.

It is not advisable to use PHOS as a detector of jet neutral particles, because
of its reduced acceptance. However, we may still consider another approach in 
which the prompt photon is detected in EMCal and jets are detected by the 
central tracking system. In such a setup, considering similar prompt photon 
identification features in PHOS and EMCal and the larger acceptance of the 
EMCal, the prompt photon detection would be enhanced by a factor 7 and 
consequently the statistical errors would be reduced by a factor 2.6.
The EMCal granularity provides $\gamma/\pi^0$ discrimination via shower shape 
in the range $\pt \sim$ 10--30 GeV/c. Due to the low $\gamma/\pi^0$ ratio, 
however, a robust $\gamma$-jet measurement requires additional hadron 
rejection from isolation cuts. The $\pt$-reach up to $\pt \sim 30$ GeV/c 
matches well the statistical reach in a standard year for $\gamma$-jet 
analysis \cite{ALICE_EMCal}.

Therefore an experimental study of the fragmentation function of photon-tagged 
jets (i.e. the distribution of charged hadrons within jets as a function of 
the variable \emph{z}, defined as $z=\pt/E_{\gamma}$) will be feasible with 
EMCal in one year up to $\pt \sim 30$ GeV/c.

%          TWO-PHOTON PROCESSES
Furthermore, in addition to single photon production and photon--jet 
(or photon--hadron) correlations, photon--photon correlations can be studied 
as well. The LO contributions to di-photon production are quark-antiquark 
annihilation ($\mathrm{q}+\bar{\mathrm{q}}\rightarrow\gamma+\gamma$) and 
gluon-gluon scattering ($\mathrm{g}+\mathrm{g}\rightarrow\gamma+\gamma$).

Processes where both photons originate from parton fragmentation or where 
one photon is prompt and the other photon comes from the fragmentation of a 
recoiling parton also contribute in LO. In this way, many of the inputs 
entering the theoretical calculations, in particular the fragmentation 
functions \cite{kniehl}, can be tested. 

However, di-photon final states are not only interesting to perform tests of
pQCD but they are also signatures for many new physics processes, such as 
Higgs production at the LHC or large extra dimensions. 

%     PP AS A BENCHMARK OF PB-PB
A further incentive to study prompt photons in pp collisions in ALICE
comes from the need to provide a baseline against which medium effects 
observed in the measurements of prompt photons in heavy-ion collisions 
can be disentangled.

Medium effects in nucleus--nucleus collisions modify the vacuum production 
cross sections of prompt photons as measured in pp collisions: nuclear 
shadowing and in-medium parton energy loss lead to a suppression of the 
yield~\cite{Jalilian00}, whereas the intrinsic transverse momentum 
distribution of the partons~\cite{Wong98,Papp99} and medium-induced photon 
radiation from quark jets~\cite{Zakharov04} enhance the yield.

On the other hand an expected signature of Quark Gluon Plasma formation in 
central heavy-ion collisions is an increased production of thermal photons,  
emitted in radiation processes, roughly in the $\pt$ range $1<\pt < 10$ GeV/c; 
therefore it is important to understand non-thermal production mechanisms 
in pp collisions in the same energy range.
 
Furthermore, in the case of \mbox{Pb--Pb} collisions, photons emerge almost 
unaltered from the dense strongly-interacting medium and provide a measurement 
of the original energy and direction of the parton emitted in the opposite 
direction. Therefore medium effects will be also identified through 
modifications of the jet fragmentation function, i.e. by the redistribution 
of the jet energy rather than by reduction of jet rate. A broadening of the 
distribution of the jet-particle momenta perpendicular to jet axis ($j_{T}$), 
directly related to the colour density of the medium, is also expected 
\cite{SalgadoWiedemann}.

\section{Exotica: mini black holes from large extra dimensions}

The concept of large extra dimensions provides a way of solving the 
hierarchy problem which concerns the weakness of gravity compared with strong 
and electro-weak forces. The extra space-dimensions, beyond the usual 
three dimensions, are assumed to be compactified. i.e. finite, so they are
too small to be normally detected. A consequence of large extra dimensions is that 
mini black holes (BH) could exist at the greatly reduced Planck mass of around 1 TeV, 
and thus might be produced at the LHC in pp collisions.

Quantitative calculations for BH production and detection in the ALICE 
experiment at the LHC have been presented in Ref.~\cite{Humanic,Humanic2}.
In this study the BH event generator code CHARYBDIS \cite{CHARYBDIS} 
has been used, that is coupled to PYTHIA code for parton evolution and 
hadronization. Taking advantage of the large-acceptance and high-precision
tracking detectors available in ALICE, namely the Inner Tracking System 
(ITS), the Time Projection Chamber (TPC) and the Transition Radiation Detector 
(TRD), two event-by-event hadronic observables were used for BH studies: 
charged particle multiplicity and summed-$\pt$. 

The conclusions drawn from this study are that under the standard running 
conditions, with a minimum-bias trigger running for four months at the LHC 
initial luminosity, and with a maximum data acquisition rate in ALICE of 
100 Hz, only a few BH events could be visible above the QCD background and 
only for a Planck mass $M_{P}=1$~ TeV, occurring for multiplicity above 200 
and summed-$\pt$ above 0.5 TeV/c.

However it is possible to improve this situation, when applying a simple 
charged particle multiplicity trigger to ALICE events, which is expected to
greatly reduce the QCD background allowing for significant BH signals to 
be detected. 

For charged multiplicity, the sensitivity to $M_P$ is raised to 2 TeV and 
hundreds of BH events above background corresponding to this case are 
expected for multiplicity greater than 250. An even better situation occurs
for summed $\pt$ distribution, since ten of thousands of BH events above 
background are expected for the $M_{P}=1$~ TeV case, and tens of BH events
even for the $M_{P}=5$~ TeV case. 

The signature for BH creation from these simple distributions is seen to be 
an abrupt flattening of their slope, as the transition from QCD to 
BH-dominated charged particle production takes place. 

\section{Concluding remarks}

We have presented in this document the potentialities of the ALICE detector 
in the field of pp physics. A special emphasis has been given to the 
minimum-bias pp physics programme, that is expected to dominate the 
start-up of the LHC operation. 
The importance of such a programme has been pointed out both for 
its intrinsic interest and also as a reference system for comparison with the 
nucleus--nucleus and proton--nucleus studies in ALICE. However, it has been 
shown that significant contributions can be given by ALICE also in other pp 
physics topics. 

The complete ALICE detector has significant advantages compared to other 
LHC detectors in pp physics attainable at the low luminosity stage of the LHC, 
mainly because of its low momentum thereshold, good momentum resolution and 
unique capacity to measure and identify a large spectrum of particles, 
including baryons and strange particles. On the other hand the ALICE momentum 
and angular resolution is at least comparable to the one of the other LHC 
experiments up to 10 GeV/c. Moreover ALICE has the capability to measure the 
transverse momentum of charged particles in the range $|\eta| \le 1.5$ and,  
by exploiting both the Silicon Pixel Detector and the Forward Multiplicity 
Detector, charged track multiplicity in the range $-3.4 \leq \eta \leq 5.1$. 
Therefore in many essential ways the ALICE pp programme described here is 
complementary to those possible with other LHC experiments. And especially 
at the early stage of the LHC operation, ALICE will be able to provide a 
significant contribution to this field. 

\section{Acknowledgements}

The author of the present contribution expresses his thanks to 
Andreas Morsch, Panos Christakoglou, Rainer Schicker, Jan Fiete
Grosse-Oetringhaus, Alberto Giovannini and Roberto Ugoccioni 
for useful help and suggestions, and he gratefully acknowledges 
Nestor Armesto for having provided his NLO calculations of jet 
rates at the LHC.

%######## BIBLIOGRAPHY ########################################

%\end{document}

\addtocounter{chapter}{1}
%\documentclass[a4paper,12pt,twoside]{report}
%\usepackage{epsfig}
%\usepackage{amssymb}
%\usepackage{lineno}
%\usepackage{setspace}
%%%%%%%%%%%%%%%%%%%%%%%%%%%%%%%%%%%%%%%%%%%%%%%%%%%%%%%%%%
%%%
%%%  Aliases
%%%
\newcommand{\lsim}{\,{\buildrel < \over {_\sim}}\,}
\newcommand{\gsim}{\,{\buildrel > \over {_\sim}}\,}
\newcommand{\sqrtsNN}{\sqrt{s_{\scriptscriptstyle{{\rm NN}}}}}
\newcommand{\av}[1]{\left\langle #1 \right\rangle}
\newcommand{\eV}{\mathrm{eV}}
\newcommand{\kev}{\mathrm{keV}}
\newcommand{\mev}{\mathrm{MeV}}
\newcommand{\gev}{\mathrm{GeV}}
\newcommand{\tev}{\mathrm{TeV}}
\newcommand{\fm}{\mathrm{fm}}
\newcommand{\mm}{\mathrm{mm}}
\newcommand{\cm}{\mathrm{cm}}
\newcommand{\m}{\mathrm{m}}
\newcommand{\mum}{\mathrm{\mu m}}
\newcommand{\s}{\mathrm{s}}
\newcommand{\ns}{\mathrm{ns}}
\newcommand{\mrad}{\mathrm{mrad}}
\newcommand{\mb}{\mathrm{mb}}
\newcommand{\de}{{\rm d}}
\newcommand{\PbPb}{\mbox{Pb--Pb}}
\newcommand{\pPb}{\mbox{p--Pb}}
\newcommand{\Pbp}{\mbox{Pb--p}}
\newcommand{\NN}{\mbox{nucleon--nucleon}}
\newcommand{\pp}{\mbox{proton--proton}}
\newcommand{\pA}{\mbox{proton--nucleus}}
\renewcommand{\AA}{\mbox{nucleus--nucleus}}
\newcommand{\RAA}{R_{\rm AA}}
\newcommand{\RDh}{R_{{\rm D}/h}}
\renewcommand{\pt}{p_{\rm t}}
\renewcommand{\d}{{\rm d}}
\newcommand{\dEdx}{{\rm d}E/{\rm d}x}
\newcommand{\dNdy}{{\rm d}N_{\rm ch}/{\rm d}y}
\newcommand{\dNdeta}{{\rm d}N_{\rm ch}/{\rm d}\eta}
\newcommand{\qqbar}{\mbox{$\mathrm {q\overline{q}}$}}
\newcommand{\QQbar}{\mbox{$\mathrm {Q\overline{Q}}$}}
\newcommand{\ppbar}{\mbox{$\mathrm {p\overline{p}}$}}
\newcommand{\ccbar}{\mbox{$\mathrm {c\overline{c}}$}}
\newcommand{\bbbar}{\mbox{$\mathrm {b\overline{b}}$}}
\newcommand{\sccbar}{\mbox{$\scriptstyle\mathrm {c\overline{c}}$}}
\newcommand{\sbbbar}{\mbox{$\scriptstyle\mathrm {b\overline{b}}$}}
\newcommand{\Dz}{\mbox{$\mathrm {D^0}$}}
\newcommand{\DtoKpi}{\mbox{${\rm D^0\to K^-\pi^+}$}}
\newcommand{\sDz}{\mbox{$\sigma({\rm D^0})$}}
\newcommand{\dsDzdpt}{\mbox{${\rm d}\sigma({\rm D^0})/{\rm d}\pt$}}
\newcommand{\Bz}{\mbox{$\mathrm {B^0}$}}
\newcommand{\Bp}{\mbox{$\mathrm {B^+}$}}
\newcommand{\Dzbar}{\mbox{$\mathrm \overline{D^0}$}}
\newcommand{\DDbar}{\mbox{$\mathrm {D \overline{D}}$}}
\newcommand{\BBbar}{\mbox{$\mathrm {B \overline{B}}$}}
\newcommand{\pizero}{\mbox{$\mathrm {\pi^0}$}}
\newcommand{\K}{\mbox{$\mathrm {K}$}}
\newcommand{\Kzs}{\mbox{$\mathrm {K^0_S}$}}
\newcommand{\KzS}{\mbox{$\mathrm {K^0_S}$}}
\newcommand{\KzL}{\mbox{$\mathrm {K^0_L}$}}
\newcommand{\Jpsi} {\mbox{J\kern-0.05em /\kern-0.05em$\psi$}\xspace}
\newcommand{\psip} {\mbox{$\psi^\prime$}\xspace}
\newcommand{\Ups} {\mbox{$\Upsilon$}\xspace}
\newcommand{\Upsp} {\mbox{$\Upsilon^\prime$}\xspace}
\newcommand{\Upspp} {\mbox{$\Upsilon^{\prime\prime}$}\xspace}
%%%%%%%%%%%%%%%%%%%%%%%%%%%%%%%%%%%%%%%%%%%%%%%%%%%%%%%%%%
%\begin{document}
%%%%%%%%%%%%%%%%%%%%%%%%%%%%%%%%%%%%%%%%%%%%%%%
% Toggle line numbering
% Won't work with the PRD revtex4 !
%\pagewiselinenumbers
% uncomment if you want doublespace
%\doublespace
%%%%%%%%%%%%%%%%%%%%%%%%%%%%%%%%%%%%%%%%%%%%%%%
%\chapter{ALICE} {\it A. Dainese, M. Monteno and D. Stocco}
%%%%%%%%%%%%%%%%%%%%%%%%%%%%%%%%%%%%%%%%%%%%%%%%%%%%%%%%%
\mchapter{Measurement of heavy-flavour production with ALICE}
{A. Dainese for the ALICE Collaboration.}

\section{Introduction}
\label{intro}

The ALICE experiment~\cite{alicePPR1} will study nucleus--nucleus
collisions at the LHC, with a centre-of-mass energy per
nucleon--nucleon collision $\sqrtsNN=5.5~\tev$ for the Pb--Pb system, 
in order to investigate the properties of QCD matter at energy densities of 
up to several hundred times the density of atomic nuclei. Under 
these conditions
a deconfined state of quarks and gluons is expected to be formed.

The measurement of open charm and open beauty production allows to 
investigate the mechanisms of heavy-quark production, propagation and
 ha\-dro\-ni\-za\-tion
in the hot and dense medium formed in high-energy nucleus--nucleus collisions.
Of particular interest is the study of the effects of parton energy loss
on c and b quarks.
Believed to be at the origin of the jet quenching phenomena observed in 
\mbox{Au--Au} collisions at the Relativistic Heavy Ion Collider (RHIC), 
energy loss is expected to depend on the properties of the medium 
(gluon density and volume) and on the properties of the `probe'
(colour charge and mass).  
The open charm and open 
beauty cross sections are also needed as a reference to measure the effect of
the transition to a deconfined phase on the production of quarkonia.
Heavy-quark production measurements in proton--proton and 
proton--nucleus collisions at the LHC, 
besides providing the necessary baseline for
the study of medium effects in nucleus--nucleus collisions, are interesting 
{\it per se}, as a test
of QCD in a new energy domain.

\section{Heavy-flavour production from pp to Pb--Pb}

Heavy-quark pairs ($\rm Q\overline Q$) 
are expected to be produced in 
primary partonic scatterings with large virtuality $Q^2>(2m_{\rm Q})^2$ and, 
thus, on small temporal and 
spatial scales, $\Delta t\sim \Delta r\sim 1/Q \lsim 0.1~\fm$
for $m_{\rm c}=1.2~\gev$.
In nucleus--nucleus reactions, this implies that
the initial production process is not 
affected by the presence of the dense medium formed in the collision.
Given the large virtualities, 
the baseline production cross sections in nucleon--nucleon collisions can be 
calculated in the framework of perturbative QCD
(pQCD). 
For the estimate of baseline production yields in nuclear collisions
(to be used for performance studies and preparation of the analysis
strategies),
scaling of the yields with the average number $\av{N_{\rm coll}}$ 
of inelastic nucleon--nucleon collisions (binary scaling) is usually assumed:
\begin{equation} 
\d^2 N^{\scriptstyle\rm Q}_{\rm AA(pA)}/\d\pt\d y =
\av{N_{\rm coll}}\times\d^2 N^{\scriptstyle\rm Q}_{\rm pp}/\d\pt\d y\,.
\end{equation}

The expected $\ccbar$ and $\bbbar$ production yields for different collision
systems at the LHC are reported in the first line of 
Table~\ref{tab:xsec}~\cite{notehvq}.
These numbers, assumed as the baseline for ALICE simulation studies, 
are obtained from pQCD calculations at NLO~\cite{hvqmnr}, including 
the nuclear modification of the parton distribution functions 
(PDFs)~\cite{EKS98}
in the Pb nucleus 
(details on the choice of pQCD parameter values and PDF sets can be found 
in~\cite{notehvq}).
Note that the predicted yields have large uncertainties, of about a factor 2,
estimated by varying the values of the calculation parameters.
An illustration of 
the theoretical uncertainty bands for the D and B 
meson cross sections
will be shown in section~\ref{exp}, along with the expected 
sensitivity of the ALICE experiment.

\begin{table}[!t]
\caption{Expected $\rm Q\overline Q$ yields per event at the LHC, 
         from NLO pQCD calculations~\cite{notehvq}. 
         For \pPb~and \mbox{Pb--Pb}, 
         Modification of the PDFs in nuclei is taken into account
          and $N_{\rm coll}$ scaling is assumed.}
\label{tab:xsec}
\begin{center}
\begin{tabular}{cccc}
\hline
colliding system & pp & p--Pb & Pb--Pb \\
$\sqrtsNN$ & 14 TeV & 8.8 TeV & 5.5 TeV\\
centrality & -- & min. bias & 0--5\% $\sigma^{\rm inel}$\\
\hline
$\ccbar$ pairs & 0.16 & 0.78 & 115 \\
$\bbbar$ pairs & 0.0072 & 0.029 & 4.6  \\
\hline
\end{tabular}
\end{center}
\end{table}

\section{Heavy-flavour detection in ALICE}
\label{exp}

The ALICE experimental setup, described in detail in~\cite{alicePPR1,monteno}, 
was designed in order to allow the detection
of ${\rm D}$ and ${\rm B}$ mesons in the high-multiplicity environment 
of central \PbPb~collisions at LHC energy, where a few thousand 
charged particles might be produced per unit of rapidity. 
The heavy-flavour capability of the ALICE detector is provided by:
\begin{itemize}
\item Tracking system; the Inner Tracking System (ITS), 
the Time Projection Chamber (TPC) and the Transition Radiation Detector (TRD),
embedded in a magnetic field of $0.5$~T, allow track reconstruction in 
the pseudorapidity range $-0.9<\eta<0.9$ 
with a momentum resolution better than
2\% for $\pt<20~\gev/c$ 
and a transverse impact parameter\footnote{The transverse impact parameter,
$d_0$, is defined as the distance of closest approach of the track to the 
interaction vertex, in the plane transverse to the beam direction.} 
resolution better than 
$60~\mum$ for $\pt>1~\gev/c$ 
(the two innermost layers of the ITS are equipped with silicon pixel 
detectors)\footnote{Note that, for pp collisions, the 
impact parameter resolution may be slightly worse, due to the 
larger transverse size of the beam at the ALICE interaction point.
This is taken into account in the studies presented in the following.}.
\item Particle identification system; charged hadrons are separated via 
$\dEdx$ in the TPC and in the ITS and via time-of-flight measurement in the 
Time Of Flight (TOF) detector; electrons are separated from charged 
hadrons in the dedicated
Transition Radiation Detector (TRD), and in the TPC; 
muons are identified in the muon 
spectrometer covering the pseudo-rapidity range $-4<\eta<-2.5$~\cite{stocco}. 
\end{itemize}

Detailed analyses~\cite{alicePPR2}, 
based on full simulation of the detector and of the 
background sources, have shown that ALICE has a good potential to carry out
a rich heavy-flavour Physics programme. 
In section~\ref{D0} we describe the expected performance for the exclusive
reconstruction of ${\rm D^0\to K^-\pi^+}$ 
decays in pp, p--Pb and Pb--Pb collisions, 
and the estimated sensitivity for the comparison with pQCD predictions,
for the pp case. 
In section~\ref{Btoe} we present the perspectives
for the measurement of beauty production in the semi-electronic 
channel in pp collisions. The expected performance for beauty 
production measurement using muons is described in section~\ref{Btomu}.

For all studies a multiplicity of $\dNdy=6000$
was assumed for central \PbPb~collisions\footnote{This value of the 
multiplicity can be taken as a conservative assumption, since 
extrapolations based on RHIC data predict $\dNdy\simeq 2000$--$3000$.}.
We report the results corresponding to the 
expected statistics collected by ALICE per LHC year: 
$10^7$ central (0--5\% $\sigma^{\rm inel}$) \PbPb~events at luminosity
$\mathcal{L}_{\rm Pb-Pb}=5\times 10^{26}~\cm^{-2}{\rm s}^{-1}$
and $10^9$ pp events at 
$\mathcal{L}_{\rm pp}^{\rm ALICE}=5\times 10^{30}~\cm^{-2}{\rm s}^{-1}$,
in the barrel detectors; the muon spectrometer will collect
about 40 times larger samples (i.e.\, $4\times 10^8$ central Pb--Pb events).

\section{Measurement of charm production in the $\Dz\to K^-\pi^+$ channel}
\label{D0}

One of the most promising channels for open charm detection is the 
${\rm D^0 \to K^-\pi^+}$ decay (and its charge conjugate) which 
has a branching ratio (BR) of about $3.8\%$.
The expected production yields (${\rm BR}\times\d N/\d y$ at $y=0$) 
for ${\rm D^0}$ (and ${\rm \overline{D^0}}$) 
mesons decaying in a ${\rm K^\mp\pi^\pm}$ pair 
in central 
Pb--Pb (0--$5\%~\sigma^{\rm inel}$) at $\sqrtsNN=5.5~{\rm TeV}$,
in minimum-bias p--Pb collisions at $\sqrtsNN=8.8~{\rm TeV}$ and in pp 
collisions at $\sqrt{s}=14~{\rm TeV}$ are, in the order, 
$5.3\times 10^{-1}$, $3.7\times 10^{-3}$ and $7.5\times 10^{-4}$ per event.

\begin{figure}[!b]
  \begin{center}
  \includegraphics[width=\textwidth]{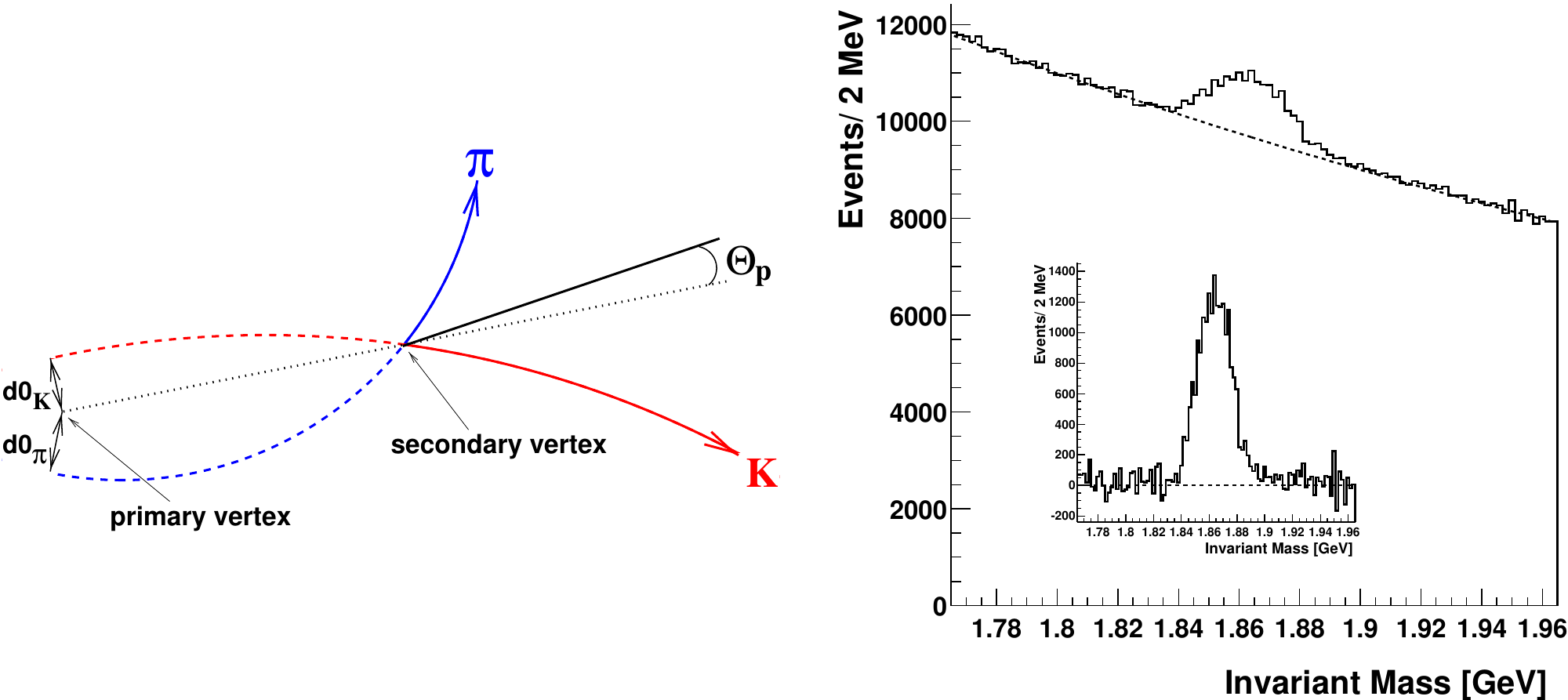}
  \caption{Schematic representation of the ${\rm D^0 \to K^-\pi^+}$ 
           decay (left). 
           ${\rm K\pi}$ 
           invariant-mass distribution corresponding to $10^7$ central Pb--Pb 
           events (right); the background-subtracted distribution is shown 
           in the insert.}
\label{fig:D0combined}
\end{center}
\end{figure}

Figure~\ref{fig:D0combined} 
(left) shows a sketch of the decay: the main feature 
of this topology is the presence of two tracks with impact parameters 
$d_0\sim 100~\mum$. The detection strategy to cope with
the large combinatorial background from the underlying event is based on:
\begin{enumerate}
\item selection of displaced-vertex topologies, i.e. two tracks with 
large impact parameters
and small pointing angle $\Theta_{\rm p}$ 
between the ${\rm D^0}$ momentum and flight-line
(see sketch in Fig.~\ref{fig:D0combined});
\item identification of the K track in the TOF detector;
\item invariant-mass analysis (see $\pt$-integrated
distribution in \PbPb~after selections in Fig.~\ref{fig:D0combined}).
\end{enumerate}
This strategy was optimized separately for pp, p--Pb 
and \PbPb~collisions, as a 
function of the ${\rm D^0}$ transverse momentum~\cite{alicePPR2}. 

\begin{figure}[!t]
  \begin{center}
  \includegraphics[width=0.7\textwidth]{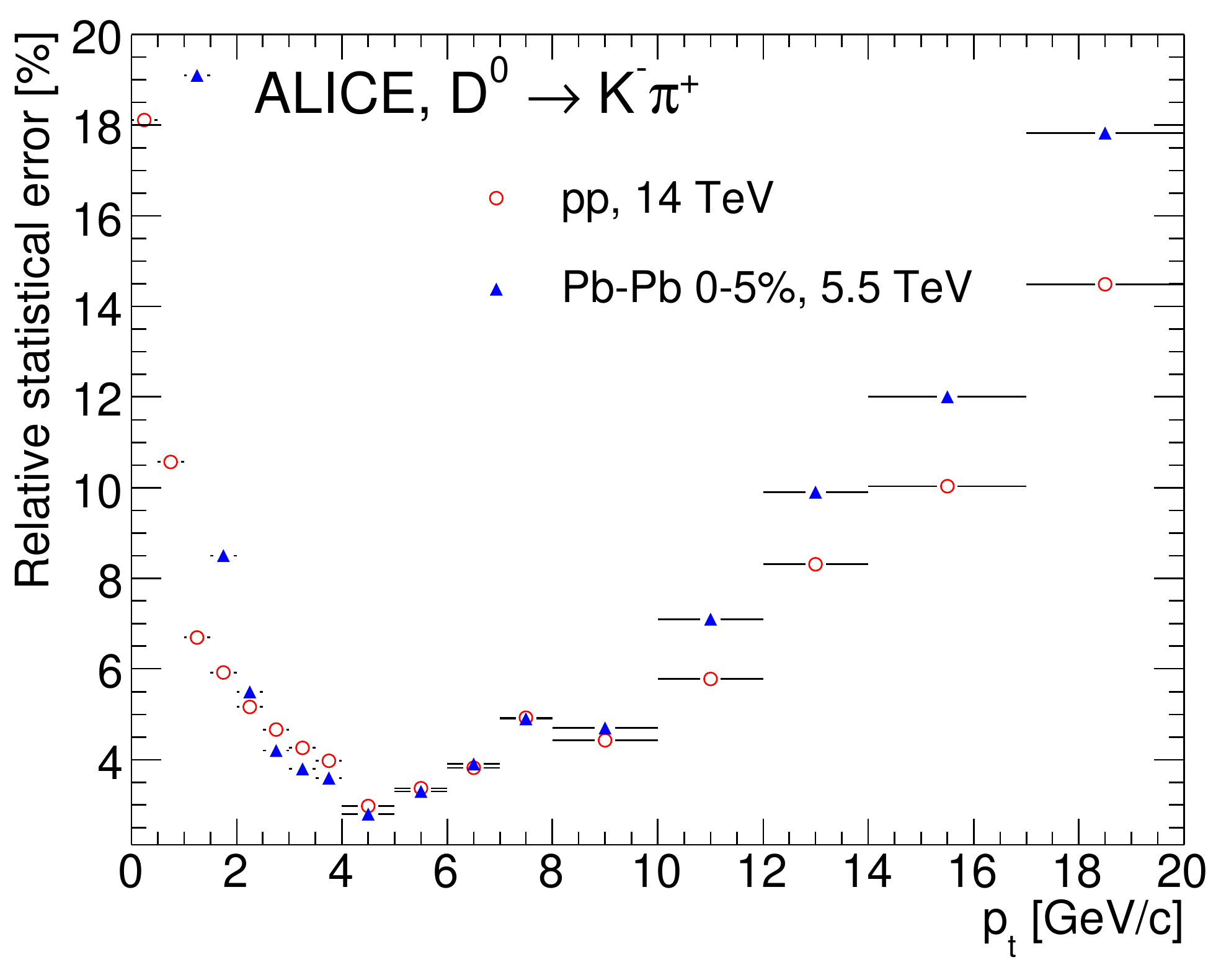}
  \caption{Expected relative statistical errors for the measurement in ALICE
           of the production cross section of $D^0$ mesons in 0--5\%
           central Pb--Pb and in pp collisions.}
\label{fig:D0errors}
\end{center}
\end{figure}

Figure~\ref{fig:D0errors} shows the expected relative statistical errors
on the measured ${\rm D^0}$ $\pt$ distribution 
for pp collisions at 14~TeV ($10^9$ events, i.e. 7 months at nominal pp 
luminosity for ALICE) and central Pb--Pb collisions at 5.5~TeV ($10^7$ events, 
i.e. 1 month at nominal Pb--Pb luminosity).
The accessible $\pt$ range is $1$--$20~\gev/c$ for Pb--Pb and 
$0.5$--$20~\gev/c$ for pp (and p--Pb, not shown), 
with a point-by-point statistical error better than 15--20\%.
The statistical error on the cross section for $\pt>\pt^{\rm min}$ is 
estimated to be of about 3\% in pp and p--Pb ($\pt^{\rm min}=0.5~\gev/c$)
and of about 7\% in central Pb--Pb ($\pt^{\rm min}=1~\gev/c$). 
The systematic error 
(acceptance and efficiency corrections, 
centrality selection for \PbPb) is expected to be smaller than 20\%. More details 
are given in~\cite{alicePPR2}.

\begin{figure}[!t]
  \begin{center}
  \includegraphics[width=0.7\textwidth]{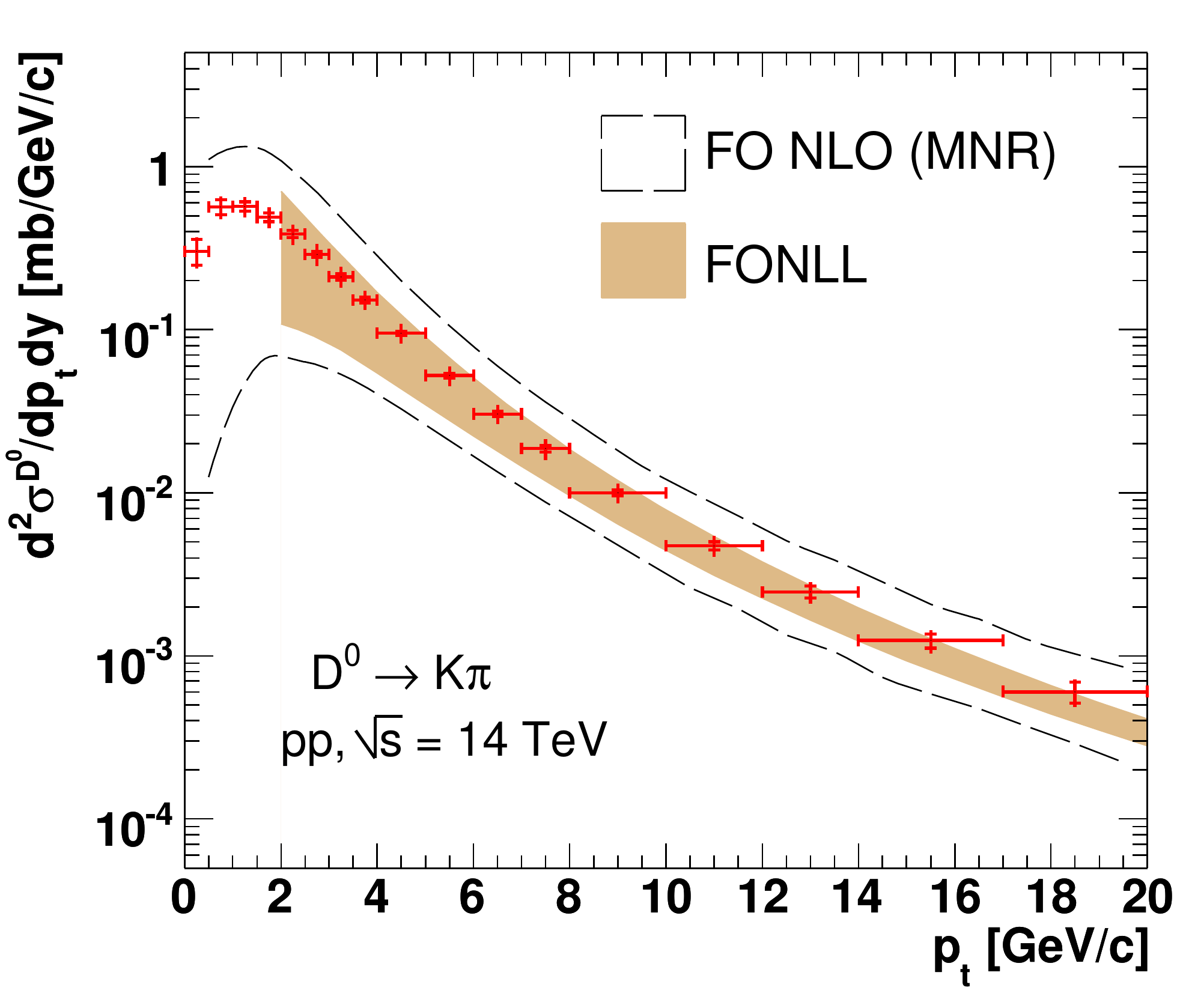}
  \caption{Sensitivity on d$^2\sigma^{\rm D^0}/$d$\pt$d$y$,
           in pp at 14~TeV, compared to 
             NLO pQCD predictions from the MNR~\cite{hvqmnr} and 
             FONLL~\cite{fonll} calculations.
             The inner error bars represent the statistical errors,
             the outer error bars represent the quadratic sum of 
             statistical and $\pt$-dependent systematic errors.
              A normalization error of 5\% is not shown.}
\label{fig:D0ptcmp}
\end{center}
\end{figure}

For the case of pp collisions, the experimental errors on the 
$\pt$-differential cross section are
expected to be significantly smaller than the current theoretical uncertainty 
from perturbative QCD calculations. 
In Fig.~\ref{fig:D0ptcmp} we superimpose the simulated ALICE measurement 
points to the prediction bands from the MNR fixed-order massive 
calculation~\cite{hvqmnr} and from the FONLL fixed-order next-to-leading log
calculation~\cite{fonll,cacciari}.
The perturbative uncertainty bands were estimated by varying the values of the 
charm quark mass and of the factorization and renormalization scales.
The comparison shows that ALICE will be able  to perform 
a sensitive test of the pQCD predictions for charm production at LHC energy.

\section{Measurement of beauty production in the semi-electronic 
decay channel}
\label{Btoe}

The production of open beauty can be studied by detecting the 
semi-electronic decays of beauty hadrons, mostly B mesons. 
Such decays have a branching ratio of $\simeq 10\%$ 
(plus 10\% from cascade decays ${\rm b\to c} \to e$, that only populate 
the low-$\pt$ region in the electron spectrum).
The expected yields (${\rm BR}\times\d N/\d y$ at $y=0$)  
for ${\rm b}\to e+X$ plus ${\rm b}\to {\rm c}\,(\to e +X)+X'$ 
in central \PbPb ($0$--$5\%~\sigma^{\rm inel}$) at $\sqrtsNN=5.5~{\rm TeV}$
and in in pp collisions at $\sqrt{s}=14~{\rm TeV}$ 
are $1.8\times 10^{-1}$  and $2.8\times 10^{-4}$ per event, 
respectively.

\begin{figure}[!t]
  \begin{center}
  \includegraphics[width=0.7\textwidth]{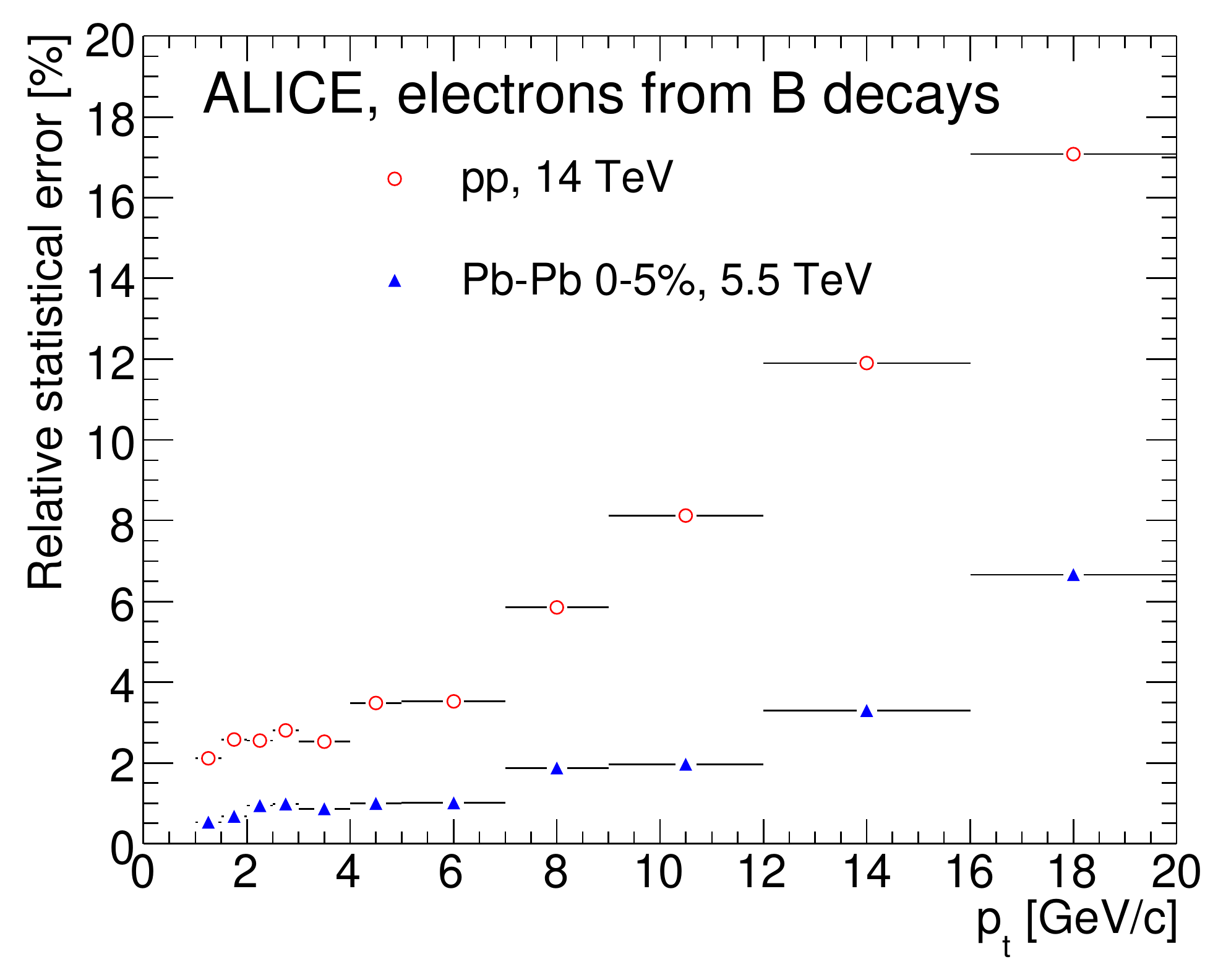}
  \caption{Expected relative statistical errors for the measurement in ALICE
           of the production cross section of B-decay electrons in 0--5\%
           central Pb--Pb and in pp collisions.}
\label{fig:Berrors}
\end{center}
\end{figure}

The main sources of background for the signal of beauty-decay electrons are:
decays of primary D mesons, which have a branching ratio 
  of $\approx10\%$ in the semi-electronic channels, and
  an expected production yield about 20 times larger 
  than B mesons (see Table~\ref{tab:xsec}); 
decays of light mesons (mainly $\rho$, $\omega$, K) and neutral pion Dalitz 
    decays ($\pi^0\to \gamma e^+e^-$);
conversions of photons in the beam pipe or in the inner layers of the ITS; 
charged pions misidentified as electrons.
Given that electrons from beauty have an average 
impact parameter $d_0\simeq 500~\mum$
and a hard momentum spectrum, it is possible to 
obtain a high-purity sample with a strategy that relies on:
\begin{enumerate}
\item Electron identification with a combined $\dEdx$ (TPC) and transition
radiation selection, which is expected to reduce the pion contamination 
by a factor of about $10^4$ at low $\pt$.
\item Impact parameter ($d_0$) cut to reject misidentified $\pi^\pm$ and $e^{\pm}$
from Dalitz decays and $\gamma$ conversions 
(the latter have small impact parameter for $\pt\gsim 1~\gev/c$) and to 
reduce the contribution of electrons from charm decays.
   We have optimized the value of the impact parameter
 cut as a function of the transverse momentum 
   in order to minimize the total errors (statistical + systematic).
The typical value of the cut is $d_0>200~\mu m$.
\end{enumerate}
The residual contamination of about 10\%, mainly 
accumulated in the low-$\pt$ region, 
of electrons from prompt charm decays, from misidentified charged pions
and $\gamma$-conversion electrons 
can be evaluated and subtracted using a Monte Carlo simulation tuned 
to reproduce the measured cross sections for pions and 
$\rm D^0$ mesons.
Figure~\ref{fig:Berrors} shows the expected relative statistical errors
for pp collisions at 14~TeV ($10^9$ events, i.e. 7 months at nominal LHC 
luminosity) and central Pb--Pb collisions at 5.5~TeV ($10^7$ events, 
i.e. 1 month at nominal LHC luminosity).

\begin{figure}[!t]
  \begin{center}
  \includegraphics[width=0.7\textwidth]{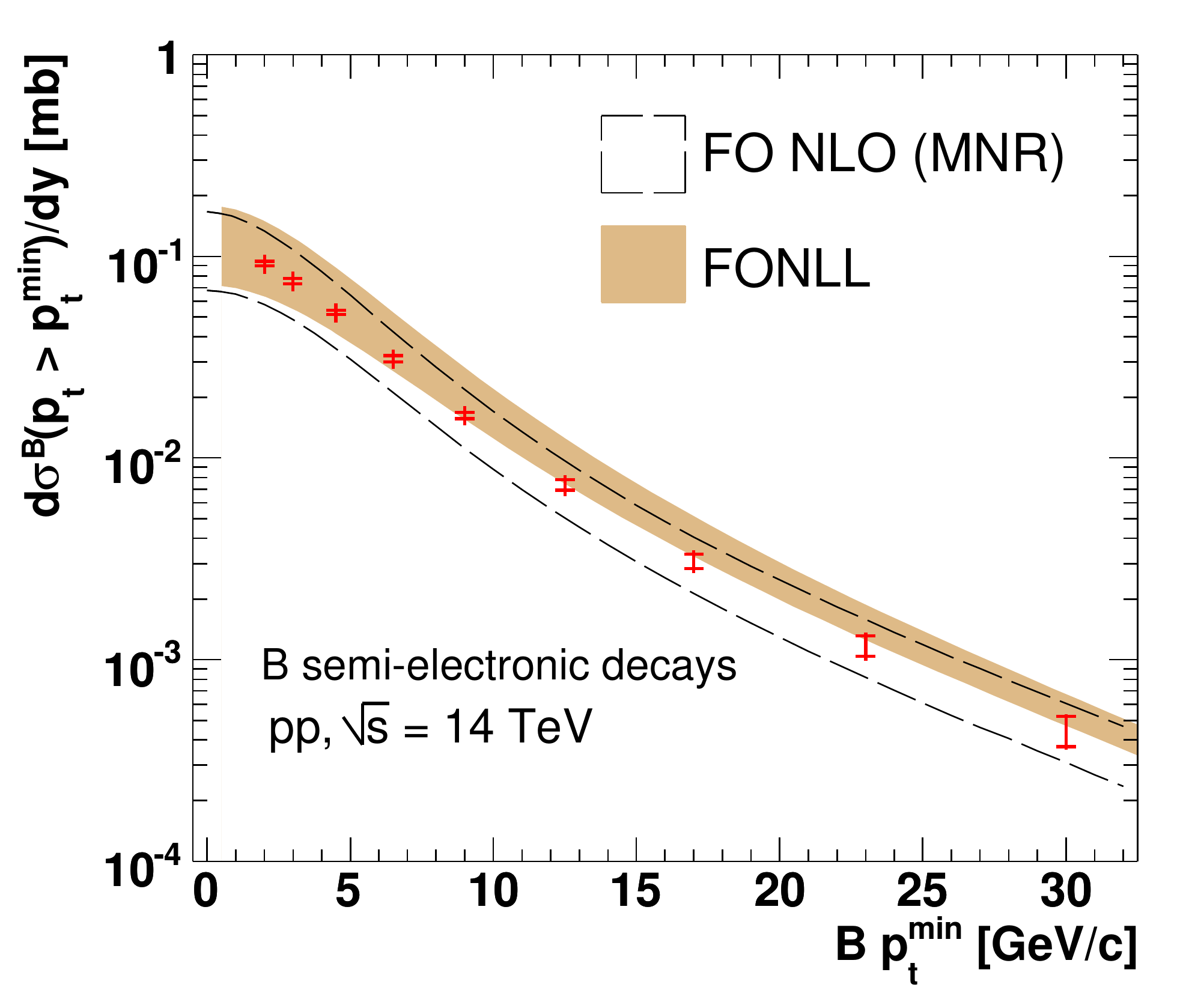}
  \caption{Sensitivity on d$\sigma^{\rm B}(\pt>\pt^{\rm min})/$d$y$,
           in pp at 14~TeV, compared to 
             NLO pQCD predictions from the MNR~\cite{hvqmnr} and 
             FONLL~\cite{fonll} calculations.
             Error bars are defined as in Fig.~\ref{fig:D0ptcmp}.}
\label{fig:sigmaB_cmpTh}
\end{center}
\end{figure}

We tested the possibility to 
infer the $\pt^{\rm min}$-differential cross section for beauty mesons,
${\rm d}\sigma^{B}(\pt>\pt^{min})/{\rm d}y$, from the electron-level cross 
section using a procedure similar to that developed 
by the UA1 Collaboration~\cite{ua1Bextraction}.
The method, described in detail in Ref.~\cite{alicePPR2}, 
is based on Monte Carlo simulation and it relies on the fact that the B meson decay kinematics, 
measured and studied in several experiments, is well understood.
It has been shown~\cite{alicePPR2} for the Pb--Pb case that, if electrons with $\pt>2~\gev/c$ are used
(below this limit, the correlation between the electron and B meson momenta is very poor),
the additional systematic error 
is negligible with respect to the systematic uncertainties already present at the electron level.

Figure~\ref{fig:sigmaB_cmpTh} presents the expected ALICE performance for the measurement of the $\pt^{min}$-differential
cross section of B mesons, ${\rm d}\sigma^{B}(\pt>\pt^{min})/{\rm d}y$ vs. $\pt^{min}$
averaged in the range $|y| < 1$.
For illustration of the sensitivity in the comparison to pQCD calculations, we 
report in the same figure the predictions and the theoretical uncertainty bands from 
the perturbative calculations in the MNR~\cite{hvqmnr} and FONLL~\cite{fonll,cacciari} approaches.   
It can be seen that the expected ALICE performance for $10^9$ events will provide a meaningful comparison with pQCD 
predictions.

\section{Measurement of beauty production in the semi-muonic decay channel}
\label{Btomu}

Beauty production can be measured also in the ALICE muon 
spectrometer, $-4<\eta<-2.5$, analyzing the single-muon $\pt$ distribution
and the opposite-sign di-muons invariant mass 
distribution~\cite{alicePPR2}.

\begin{figure}[!t]
  \begin{center}
    \includegraphics[width=.7\textwidth]{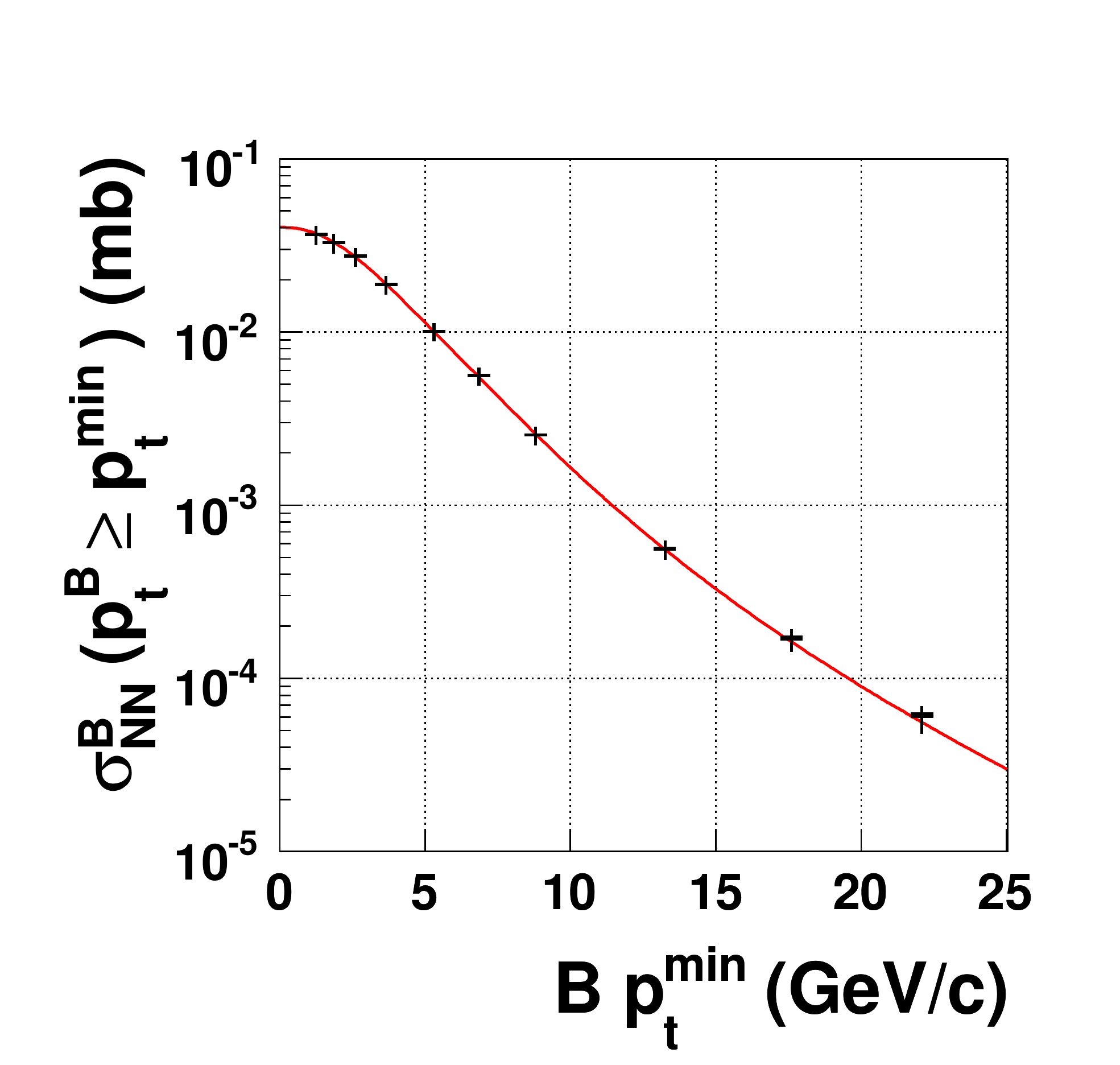}
  \caption{Minimum-$\pt$-differential production 
          cross section per nucleon--nucleon collision for B mesons
          with $-4<y<-2.5$ in central Pb--Pb 
           collisions, as expected to be measured 
           from the single-muon data set. 
           Statistical errors (represented by the thickness of the horizontal 
             bars) corresponding to 
            $4\times 10^8$ events
            and $\pt$-dependent systematic errors (vertical bars) 
            are shown. A normalization error of 10\% is not shown.
           The line indicates the input cross section.}   
  \label{fig:Btomu}
  \end{center} 
\end{figure}

The main backgrounds to the `beauty muon' signal are $\pi^\pm$, 
$\rm K^\pm$ and charm decays. The cut $\pt>1.5~\gev/c$ is applied to all
reconstructed muons in order to increase the signal-to-background ratio.
For the opposite-sign di-muons, the residual combinatorial background is
subtracted using the technique of event-mixing and the resulting distribution
is subdivided into two samples: the low-mass region, $M_{\mu^+\mu^-}<5~\gev$,
dominated by di-muons originating from a single b quark decay through
$\rm b\to c(\to \mu^+)\mu^-$ ($\rm BD_{\rm same}$), and the high-mass region,  
$5<M_{\mu^+\mu^-}<20~\gev$, dominated by $\bbbar\to\mu^-\mu^+$, with each muon
coming from a different quark in the pair ($\rm BB_{\rm diff}$). 
Both samples have a background 
from $\ccbar\to \mu^+\mu^-$ and a fit is performed to extract the charm- and 
beauty-component yields. The single-muon $\pt$ distribution has three
components with different slopes: K and $\pi$, charm, and beauty decays. 
The first component is subtracted on the basis of the identified hadron spectra
measured in the central barrel. Then, a fit technique allows to 
extract a $\pt$ distribution of muons from beauty decays.
A Monte Carlo procedure, similar to that used for semi-electronic decays, 
allows to extract 
B-level cross sections for the data sets (low-mass $\mu^+\mu^-$, 
high-mass $\mu^+\mu^-$, 
and $\pt$-binned single-muon distribution), 
each set covering a different B-meson $\pt>\pt^{\rm min}$ region. 
The results for central Pb--Pb collisions at 
$\sqrtsNN=5.5~\tev$ using only the single muons are 
shown in Fig.~\ref{fig:Btomu}. 
Since only minimal cuts are applied, the reported statistical errors 
(represented by the thickness of the horizontal bars) are very 
small and the high-$\pt$ reach is excellent. Similar performance,
in terms of $\pt$ coverage, is 
expected for pp collisions at $\sqrt{s}=14~\tev$.
The main sources of systematic errors (vertical bars) are: corrections for 
acceptance and efficiency, subtraction of the background 
muons from charged pion and kaon decays, and fit procedure to 
separate the beauty and charm components.

\section{Conclusions}

We presented the performance of ALICE for the measurement 
of charm and beauty production in $\pp$ collisions at $\sqrt{s}=14~\tev$
and central Pb--Pb collisions at $\sqrtsNN=5.5~\tev$. 
For the pp case, these measurements will 
provide sensitive tests for perturbative QCD in
a new energy domain. They will also be essential for a comparison with the corresponding measurements
in $\PbPb$ collisions, for example for the investigation of 
c and b quark in-medium energy loss~\cite{hotquarks06}.

\begin{figure}[!t]
\begin{center}
\includegraphics[width=.7\textwidth]{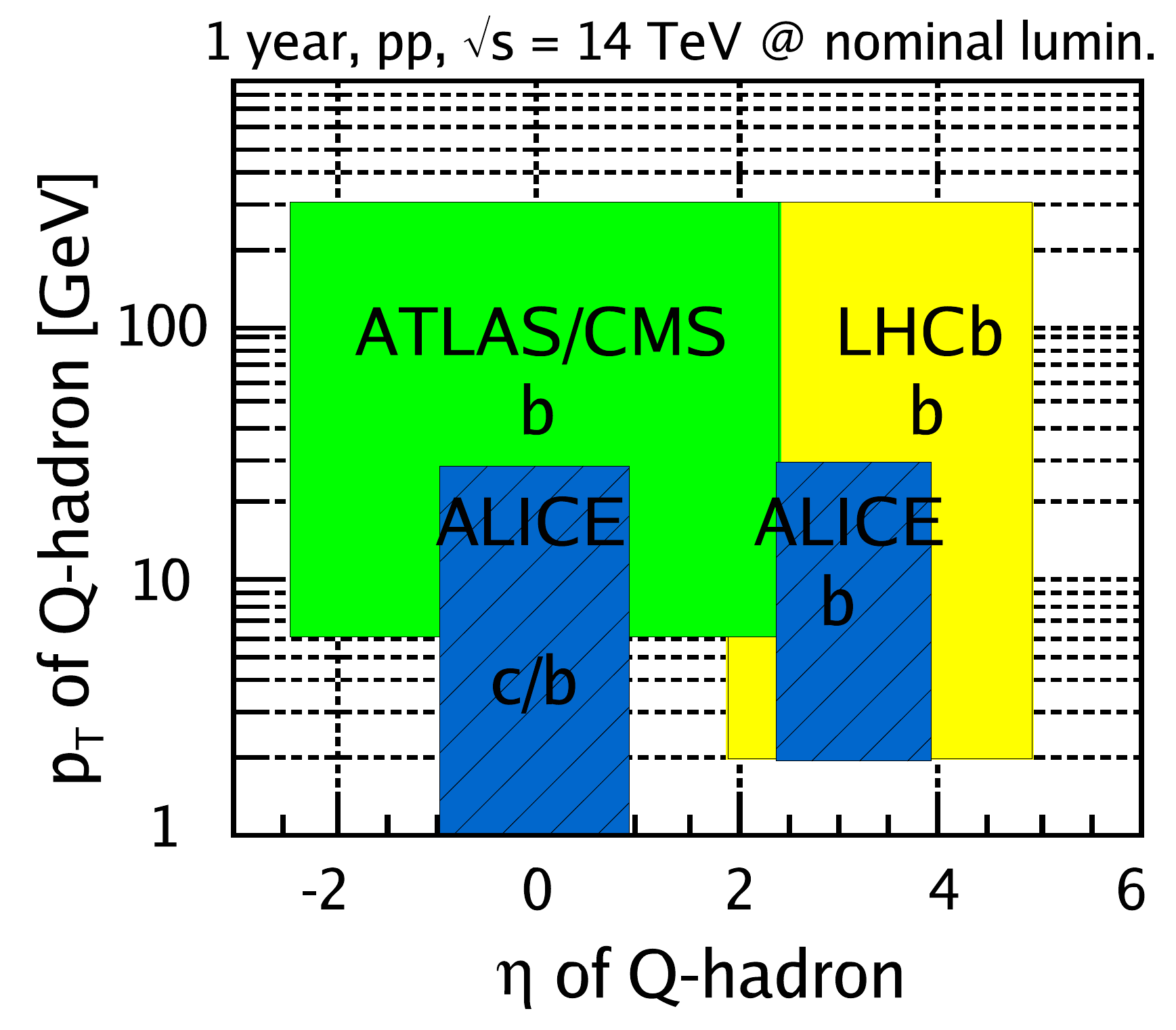}
\caption{Schematic acceptances in transverse momentum and pseudorapidity 
         for open heavy flavour hadrons (indicated as ``Q-hadrons'')
         in the four LHC experiments~\cite{heralhc}. 
         The high-$\pt$ coverages correspond
         to one year (i.e.\,7 months) of running at nominal luminosity
         (see text). [Note that the acceptance of the ALICE muon 
         spectrometer, indicated as $-4<\eta<-2.5$ in the ALICE coordinate
         system, is reported in the figure as $2.5<\eta<4$ to display the
         overlap with the acceptance of LHCb.]}
\label{fig:accHQ}
\end{center}
\end{figure}

We conclude by comparing, for pp collisions at $\sqrt{s}=14~\tev$,
 the ALICE envisaged acceptance for heavy-flavour 
production measurements to that of the other LHC experiments.
Figure~\ref{fig:accHQ} (from the proceedings of the workshop 
``HERA and the LHC''~\cite{heralhc}) 
shows schematically the $\pt$ vs. $\eta$ acceptances
for charm (c) and beauty (b) 
hadrons in the four experiments, as expected for one year of 
running at nominal luminosity (note that the value of the luminosity is 
different for each experiment: $\mathrm{10^{34}~cm^{-2}s^{-1}}$ for 
ATLAS and CMS,
$(2$--$5) \times \mathrm{10^{32}~cm^{-2}s^{-1}}$ for LHCb, and
$\mathrm{ 5 \times 10^{30}~cm^{-2}s^{-1}}$
for ALICE).
ATLAS and CMS have similar acceptances for beauty measurements; the 
minimum accessible $\pt$ is larger than for ALICE because of the strong 
magnetic fields, which in turn, together with the high luminosity, 
allow to cover transverse momenta up to 200--300~$\gev/c$, 
and because of the larger material budget (amount of material) 
in the inner tracking detectors.
In terms of acceptance for beauty measurements, 
ALICE overlaps with ATLAS and CMS at central 
rapidity and with LHCb at forward rapidity. The moderate magnetic field
allows measurements down to transverse momenta of about $2~\gev/c$ for B mesons
in the muon arm and in the barrel, and down to less than 
$1~\gev/c$ for D mesons in the barrel.

%%%%%%%%%%%%%%%%%%%%%%%%%%%%%%%%%%%%%%%%%%%%%%%%%%%%%%%%%%%%%%%%%%%%%%%%

%%%%%%%%%%%%%%%%%%%%%%%%%%%%%%%%%%%%%%%%%%%%%%%%%%%%%%%%%%%%%%%%%%%%%%%%%%

%\end{document}

\addtocounter{chapter}{1}
\mchapter{Quarkonia detection with the ALICE Muon Spectrometer in pp collisions at 14 TeV and PDF sensitivity in the low \textit{x} region
}{D. Stocco for the ALICE Collaboration}

\newcommand{\xspace}{\ }
\newcommand{\dd}{\mathrm{d}}

\renewcommand{\pt}{\ensuremath{p_{\mathrm{t}}}}
\newcommand {\gmom} {\mbox{\rm GeV$\kern-0.15em /\kern-0.12em c$}}
\renewcommand {\tev} {\mbox{${\rm TeV}$}}
\newcommand {\lum} {\, \mbox{${\rm cm}^{-2} {\rm s}^{-1}$}}

\renewcommand{\ccbar}{\mbox{$\mathrm {c\overline{c}}$}}
\renewcommand{\bbbar}{\mbox{$\mathrm {b\overline{b}}$}}
\renewcommand{\qqbar} {\mbox{$q\bar{q}$}\xspace}
\renewcommand{\QQbar} {\mbox{$Q\bar{Q}$}\xspace}

\renewcommand{\Jpsi} {\mbox{J\kern-0.05em /\kern-0.05em$\psi$}\xspace}
\newcommand{\jpsi} {\mbox{J\kern-0.05em /\kern-0.05em$\psi$}}
\newcommand{\Psip} {\mbox{$\psi^\prime$}\xspace}
\renewcommand{\psip} {\mbox{$\psi^\prime$}}
\renewcommand{\Ups} {\mbox{$\Upsilon$}\xspace}
\renewcommand{\Upsp} {\mbox{$\Upsilon^\prime$}\xspace}
\renewcommand{\Upspp} {\mbox{$\Upsilon^{\prime\prime}$}\xspace}

\def\PL#1#2#3{Phys. Lett. {\bf #1}\ (#2)\ #3}
\def\ALI#1#2{ALICE Internal Note {\sc #1--#2}}

The ALICE Muon Spectrometer~\cite{TDR} is a forward detector, with acceptance in the polar angle interval $171^{\circ} < \theta < 178^{\circ}$.
It consists of a composite absorber ($\sim 10 \lambda_{int}$), made with layers of both high and low Z materials, starting 90 cm from the interaction vertex, a large dipole magnet with a 0.7 T magnetic field and 10 planes of high-granularity tracking stations.
A second absorber ($\sim 7 \lambda_{int}$ of iron) at the end of the spectrometer and four more detector planes are used for muon identification and triggering.
The spectrometer is shielded throughout its length by a dense absorber tube surrounding the beam pipe.
The spectrometer was designed in order to detect quarkonia down to $\pt \sim 0$ in the rapidity region $-4.0 < y < -2.5$.

The study of quarkonia production in pp collisions presents a twofold interest.
On the one hand, pp measurements represent a baseline for quarkonia production in heavy-ion collisions.
On the other, they have an intrinsic interest since they are expected to shed light on quarkonia production mechanisms by testing the existing theoretical models in an unexplored energy regime.
In this respect, the relevant observables are quarkonia cross sections and {\pt} distributions.
In addition, the rapidity acceptance of the Muon Spectrometer for quarkonia 
will allow access to PDFs at very small $x$.

The results of simulation studies of the ALICE Muon Spectrometer physics performance for quarkonia detection in proton-proton collisions at $\sqrt{s} = 14$~{\tev} are presented.\\

The simulation input is provided by the Color Evaporation Model (CEM)~\cite{CEM,CEMvogt} predictions.
In this model the quarkonium production cross section is a measurable fraction ($F_{C}$) of all \QQbar pairs below the $H\bar{H}$ threshold (where $H$ is the lowest mass heavy flavor hadron) without any constraints on the color or spin of the final state.
The \QQbar pair then neutralizes its color by interaction with the collision-induced color field.
At leading order, the production cross section of quarkonium state $C$ in an $AB$ collision is:
\begin{equation}\label{eqn:quarkoniaXsec} 
\sigma_C^{\scriptscriptstyle{CEM}} = F_C \sum_{i,j}{\int_{4 m_Q^2}^{4 m_H^2}{\dd \hat{s} \int{\dd x_1 \dd x_2 \, f_{i/A}(x_1,\mu^2)
f_{j/B}(x_2,\mu^2) \hat{\sigma}_{ij}(\hat{s}) \delta(\hat{s}-x_1 x_2 s)}}}
\end{equation}
where $A$ and $B$ can be any hadron or nucleus, $ij = $\qqbar or $gg$, $\hat{\sigma}_{ij}(\hat{s})$ is the $ij \rightarrow$ \QQbar subprocess cross section and $f_{i/A}(x_1,\mu^2)$ is the parton density in the hadron or nucleus.
Finally, $s$ and $\hat{s}$ are respectively the hadronic and partonic center of mass energies.
The results presented here have been obtained with the set of parameters (from~\cite{CEMvogt}) listed in Table~\ref{tab:CEMxSec}.

The resulting cross sections in pp collisions, which will be referred to as \textit{prompt} (and include direct production and feed-down from higher mass resonances within the same family), are shown in the same table.
The values take into account branching ratios in the $\mu^+ \mu^-$ channel as well.

\begin{table}[!ht]
\centering
\begin{tabular}{|*{6}{c|}}
\cline{2-6}
\multicolumn{1}{c|}{} & \Jpsi & \Psip &  \Ups & \Upsp & \Upspp\\
\hline
$\sigma$ ($\mu$b) & 3.18  & 0.057 & 0.028 & 0.0069 & 0.0041\\
\hline
$F_C$ & 0.0144  & 0.0021 & 0.0201 & 0.00636 & 0.00335\\
\hline
PDF      & \multicolumn{2}{|c|}{MRST98 NLO} & \multicolumn{3}{|c|}{MRST98 NLO}\\
\hline
$m_{q}$  & \multicolumn{2}{|c|}{1.2}     & \multicolumn{3}{|c|}{4.5}\\
\hline
$\mu/m_{q}$ & \multicolumn{2}{|c|}{2}    & \multicolumn{3}{|c|}{2}\\
\hline
\end{tabular}
\caption{CEM parameters and resulting cross sections for quarkonia production in pp collisions at 14 {\tev}. Cross sections include feed-down from higher mass resonances and branching ratios in muon pairs. The adopted PDF comes from calculations at NLO precision by Martin-Roberts-Stirling-Thorne in 1998 (see~\cite{Martin:1998sq}).}
\label{tab:CEMxSec} 
\end{table}

In addition to prompt \Jpsi and \psip, also those from B decay are taken into account in this study.
These cross sections have been obtained from the open beauty cross section using the $B \rightarrow \jpsi + X $ and $B \rightarrow \psip + X$ branching ratios.

The rapidity distributions for prompt production of the different quarkonia states are a parameterization of CEM predictions, while the {\pt} distributions are obtained by extrapolating to LHC energies those measured by the CDF experiment at $\sqrt{s} \sim 2$~{\tev}~\cite{Acosta:2004yw,Acosta:2001gv}.

The invariant mass continuum from semileptonic decay of beauty and charm hadrons and from weak decay of pions and kaons was produced with PYTHIA.
{\ccbar} and {\bbbar} pairs were produced with a cross section of 11.2 and 0.51~mb, respectively~\cite{carrerDainese, PPR2}.

The so obtained  {\bbbar} pairs were then used in order to get the \Jpsi and \Psip from B decay.
Since the Muon Spectrometer will not be able to distinguish in a direct way between the two main sources of charmonia (prompt and coming from B decay), both contributions were summed together to evaluate the expected yields.

Since the full simulation of a sufficient number of events would require long computing times, a fast simulation was performed.
Such a method is based on the parameterization of the whole spectrometer response at the single muon level.
Given a muon of momentum $p$ generated at the interaction point with polar and azimuthal angles $\theta$ and $\varphi$, the fast simulation applies the smearing of the apparatus and gives the reconstructed $p^\prime$, $\theta^\prime$ and $\varphi^\prime$ together with the detection probability for that muon.

At the trigger level, loose cuts on single muon {\pt} are applied: a low cut for muons from charmonia resonances and a high cut for muons from bottomonia ones.
Such trigger cuts are not sharp but they roughly correspond to {\pt} $\sim$ 1~{\gmom} and {\pt} $\sim$ 2~{\gmom}, respectively.\\

\begin{figure}[!ht]
  \centering
  \includegraphics[width=.47\textwidth, height=.44\textwidth,trim=30 40 90 10]{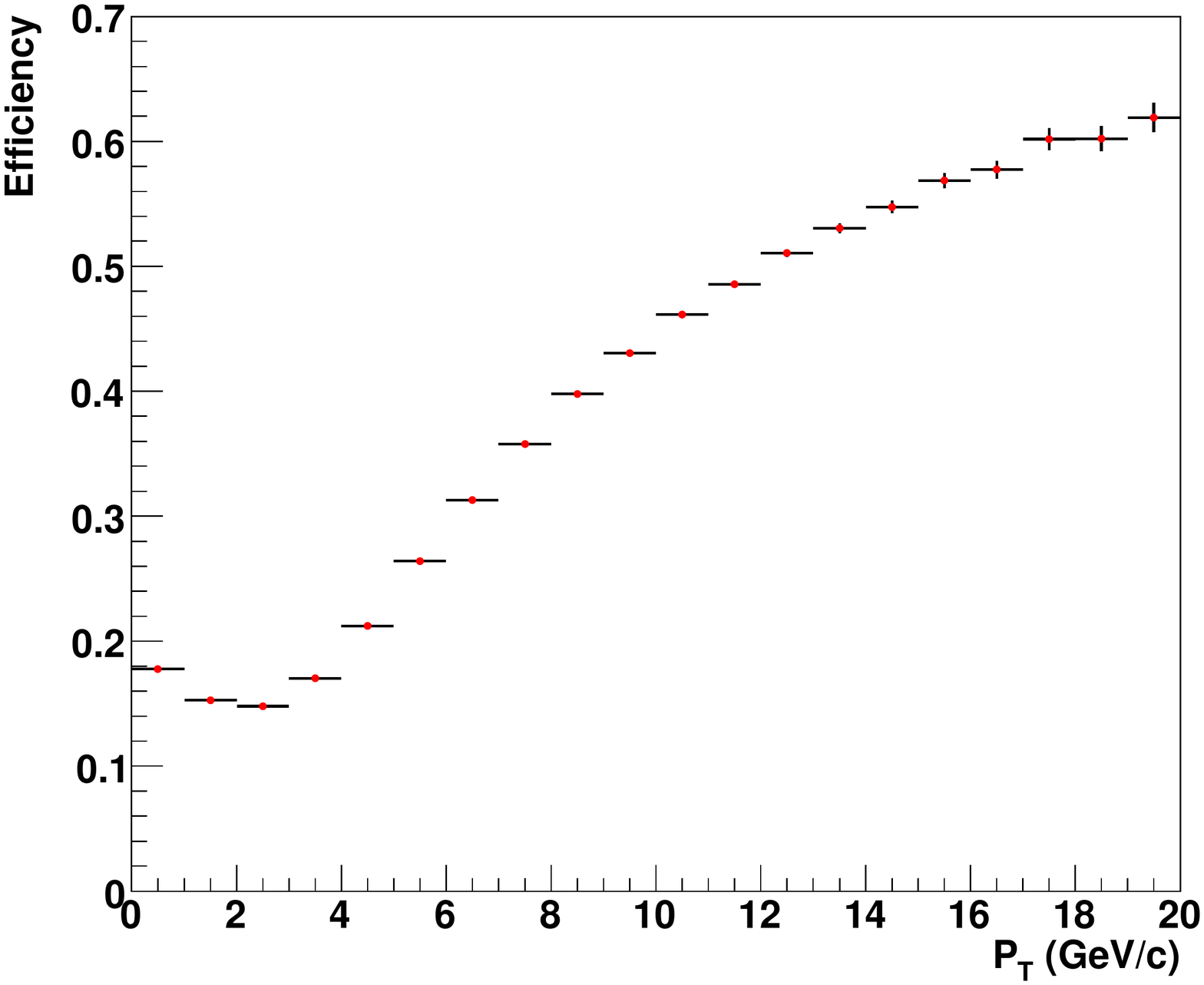}
  \includegraphics[width=.47\textwidth, height=.44\textwidth,trim=30 40 90 10]{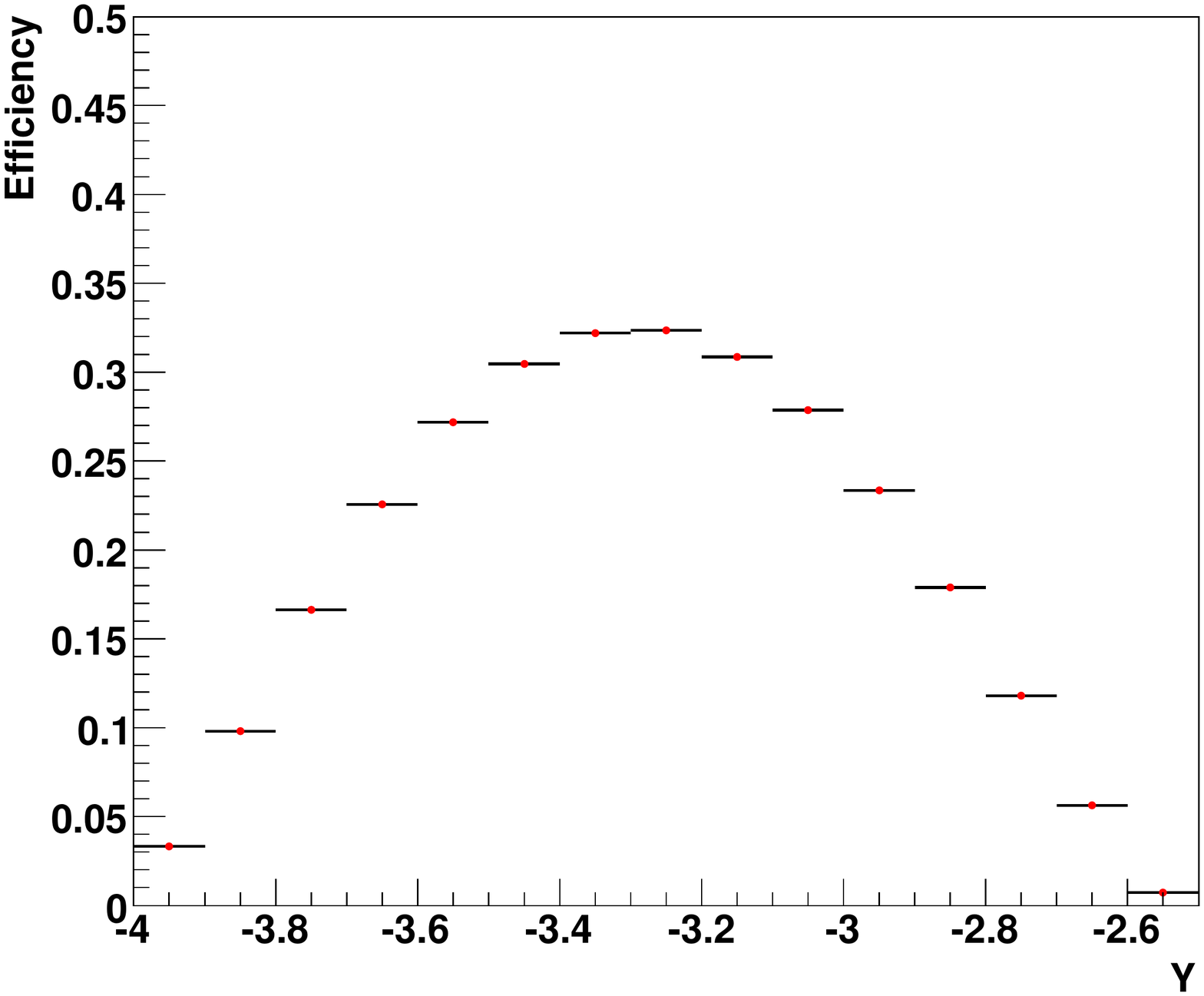}
  \includegraphics[width=.47\textwidth, height=.44\textwidth,trim=30 40 90 10]{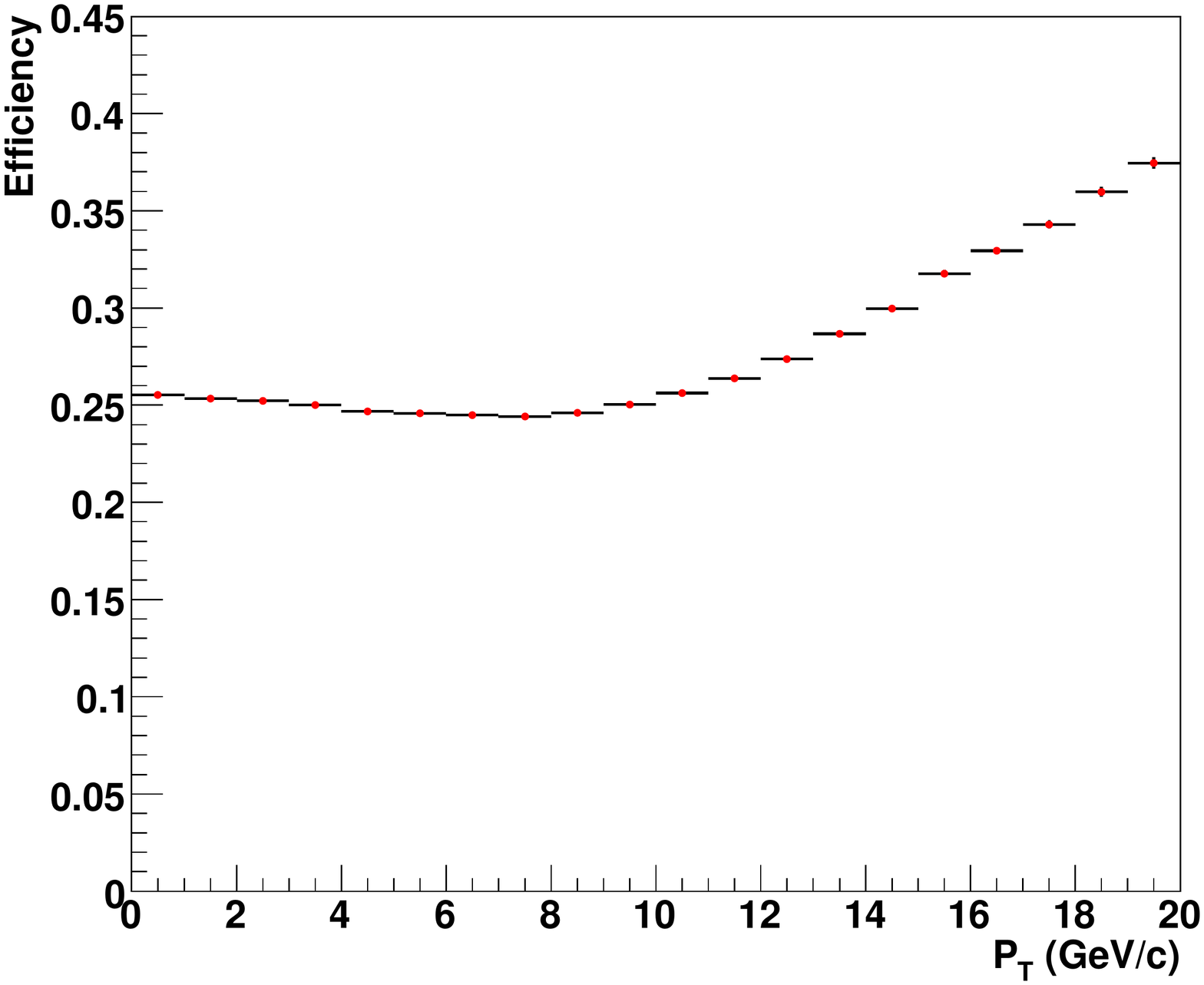}
  \includegraphics[width=.47\textwidth, height=.44\textwidth,trim=30 40 90 10]{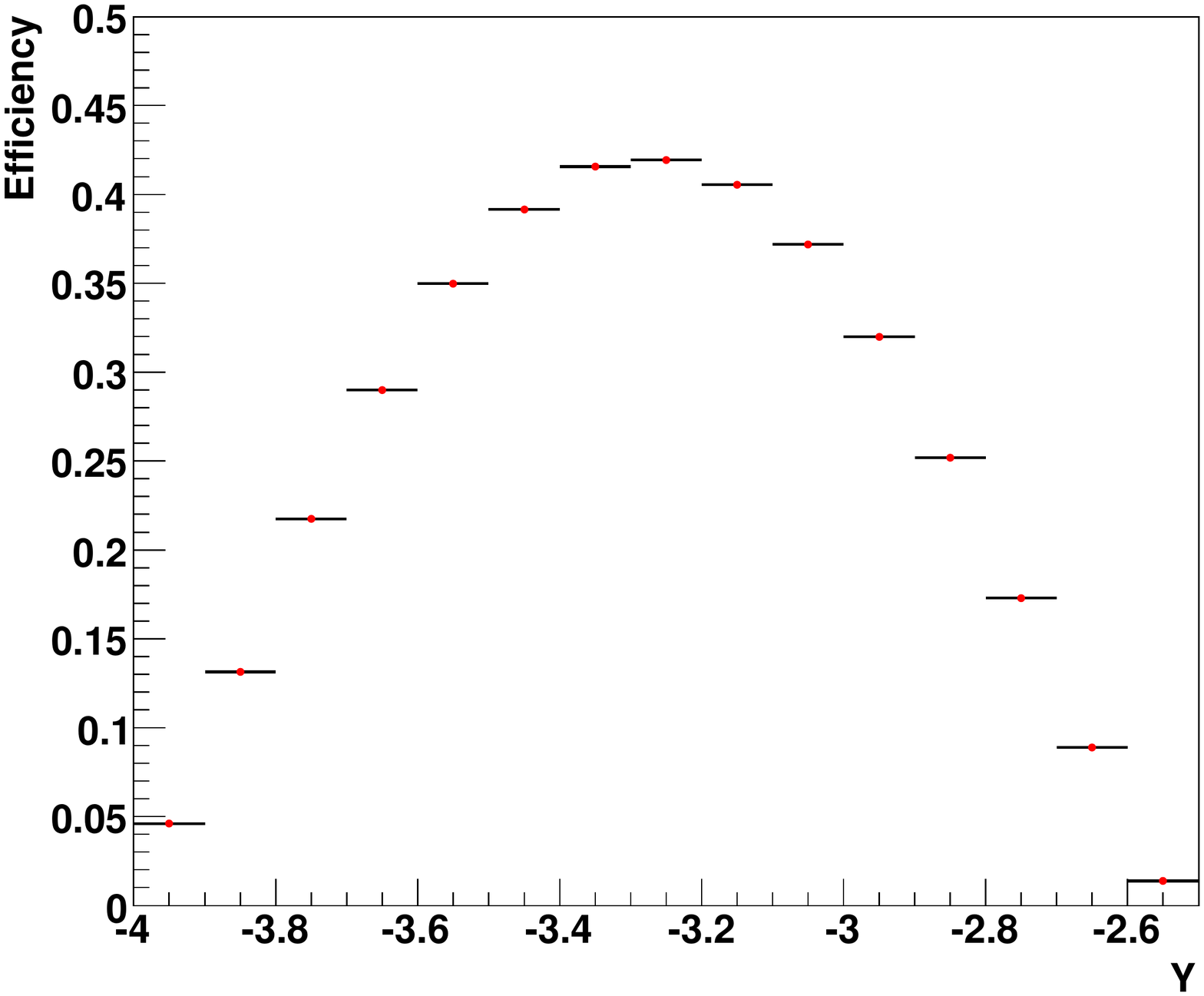}
\vspace{-4mm}

  \caption{{\pt} (left) and $y$ (right) detection probabilities for \Jpsi (top) and \Ups (bottom). The {\pt} detection probabilities are calculated in the Muon Spectrometer rapidity acceptance ($-4.0 < y < -2.5$).}
  \label{fig:detectionProb}
\end{figure}

The simulations allow to calculate the global geometrical acceptances for quarkonia, integrated over the whole phase space, which are found to be of the order of 4\% for both \Jpsi and $\Upsilon$.
The detection probabilities for the different onium states were computed by applying the trigger and tracking response to each muon of the pair.
The rapidity (transverse momentum) dependencies of the detection probabilities for \Jpsi and \Ups are shown in the right (left) panels of Fig.~\ref{fig:detectionProb}.
These were computed as the ratio between the number of detected and generated quarkonia at given $y$ ({\pt}).\footnote{The former ratio was computed by generating quarkonia on the whole {\pt} range, while the latter was actually computed by generating quarkonia only in the rapidity interval $-4.0 < y < -2.5$ covered by the spectrometer.}
The depletion in the {\pt} detection probabilities at low {\pt} is related to the trigger cuts applied.

Quarkonia yields were computed for a data taking scenario of one year of pp data taking (assumed to be equivalent to $10^7$~s) at a luminosity of $3 \times 10^{30} \lum$ (cf.~\cite{PPR1}).

\begin{figure}[!ht]
\centering
\includegraphics[width=0.61\textwidth,trim=30 40 100 10]{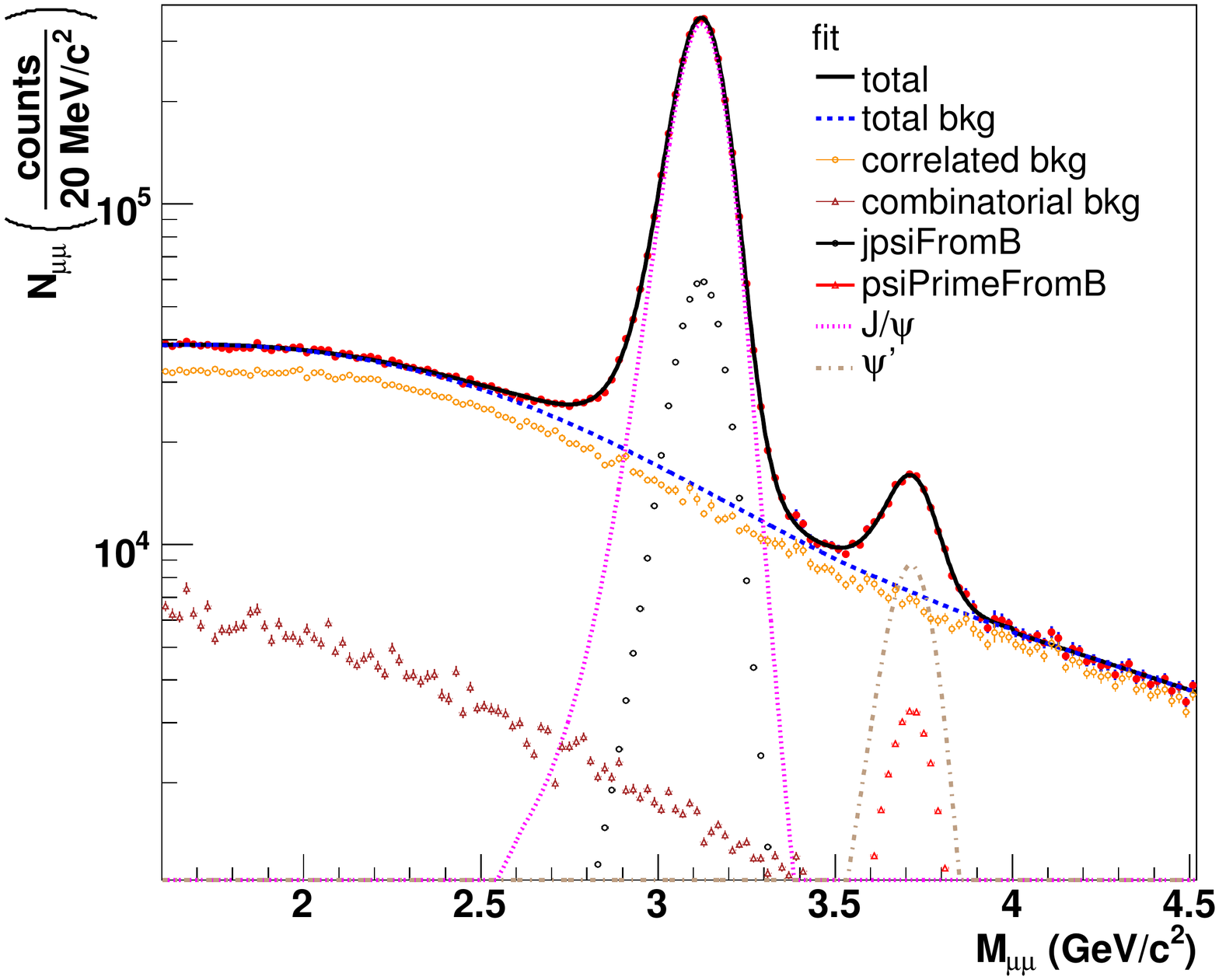}\\
\includegraphics[width=0.61\textwidth,trim=30 40 100 10]{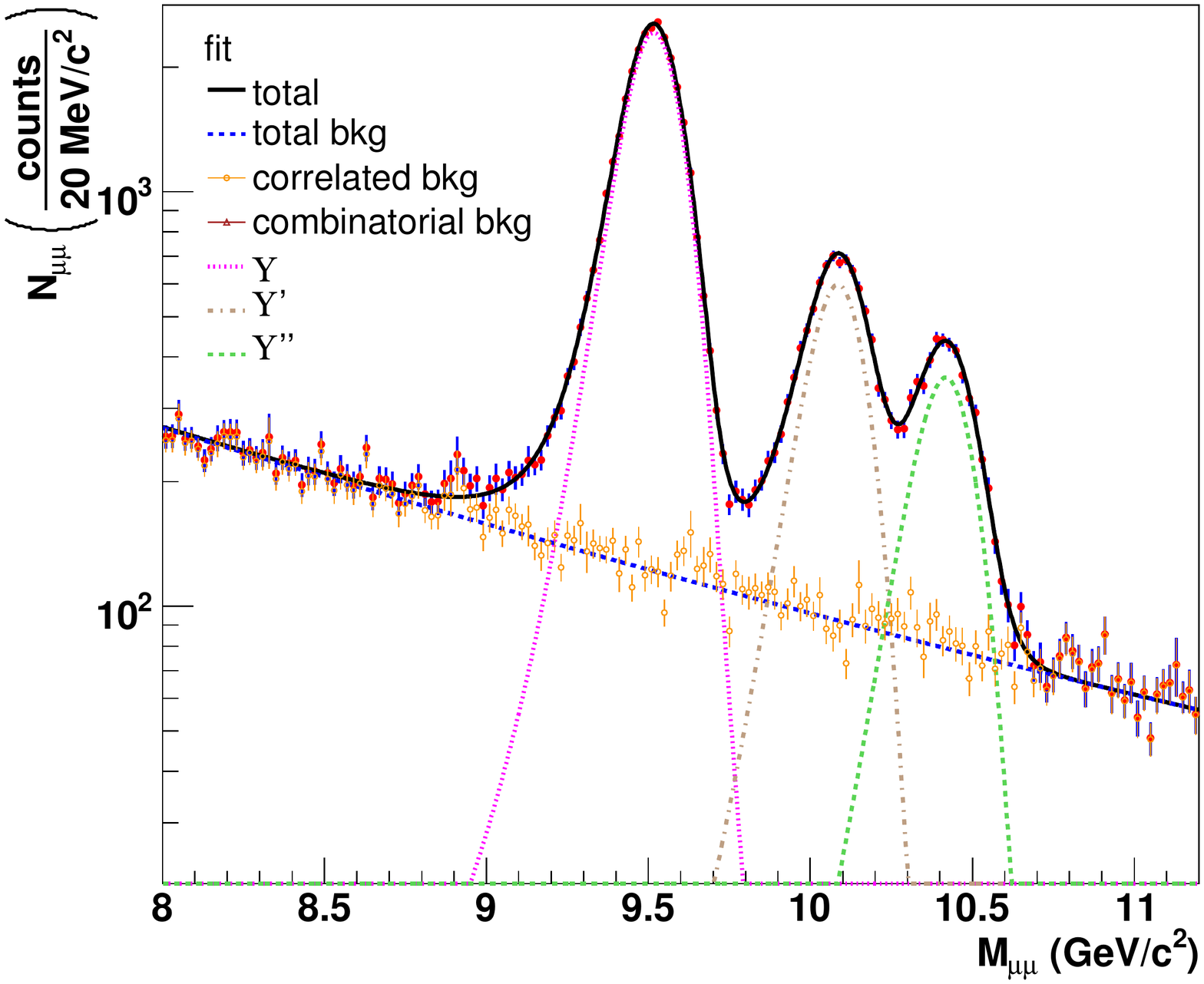}
\caption{Opposite-sign dimuon mass spectra in pp collisions at $\sqrt{s} = 14$~{\tev} for a running time of $10^7$~s at a luminosity of $3 \times 10^{30} \lum$. The \Jpsi and \Ups mass regions are shown in the top and bottom panel respectively.}
\label{fig:massYield}
\end{figure}

The resulting $\mu^+\mu^-$ invariant mass distributions are shown in Fig.~\ref{fig:massYield}, both for the \Jpsi and \Ups regions.
As seen, all charmonium and bottomonium states are clearly resolved.

Together with quarkonia, all sources contributing to the $\mu^+\mu^-$ invariant mass continuum were taken into account, including muons from correlated and uncorrelated decay of {\ccbar} and {\bbbar} pairs and from the decay of $\pi$ and $K$.
The invariant mass dimuon continuum is dominated by correlated sources.

The quarkonia signal is extracted from the total distribution by means of an interpolation.
For each onium state, a Gaussian function was used for the central part of the peak and two more Gaussian with variable width were added to describe the tails.
The correlated continuum was parametrized with two Gaussian functions with variable width, describing the low and high invariant mass regions, respectively.

The total number of detected \Jpsi is of the order of $3 \times 10^6$, while the statistics expected for \Ups is about two orders of magnitude smaller ($3 \times 10^4$). The yields for all quarkonia states are summarized in Table~\ref{tab:totalYields}, where the corresponding signal to background ratios and significances are also given.

\begin{table}[!ht]
  \centering
\begin{tabular}{|c||*{5}{c|}}
    \hline
    state  & S ($\times 10^3$) & B ($\times 10^3$) & S/B   & S/$\sqrt{S+B}$ \\
    \hline
    \Jpsi  &              2807 &               235 &  12.0 &            1610 \\
    \hline
    \Psip  &                75 &               120 &  0.62 &             170 \\
    \hline
    \Ups   &              27.1 &               2.6 &  10.4 &             157 \\
    \hline
    \Upsp  &               6.8 &               2.0 &   3.4 &             73  \\
    \hline
    \Upspp &               4.2 &               1.8 &   2.4 &             55  \\
    \hline
  \end{tabular}

  \caption{Expected quarkonia signal and background yields. Numbers refer to an interval corresponding to $\pm$ 1 FWHM around the resonance mass peak. Signal to background ratios and significances are also listed. All yields and significances are for a $10^7$~s running time with a luminosity of $3 \times 10^{30} \lum$.}
  \label{tab:totalYields}
\end{table}

\begin{figure}[!ht]
\centering
\includegraphics[width=0.49\textwidth,trim=30 40 90 10]{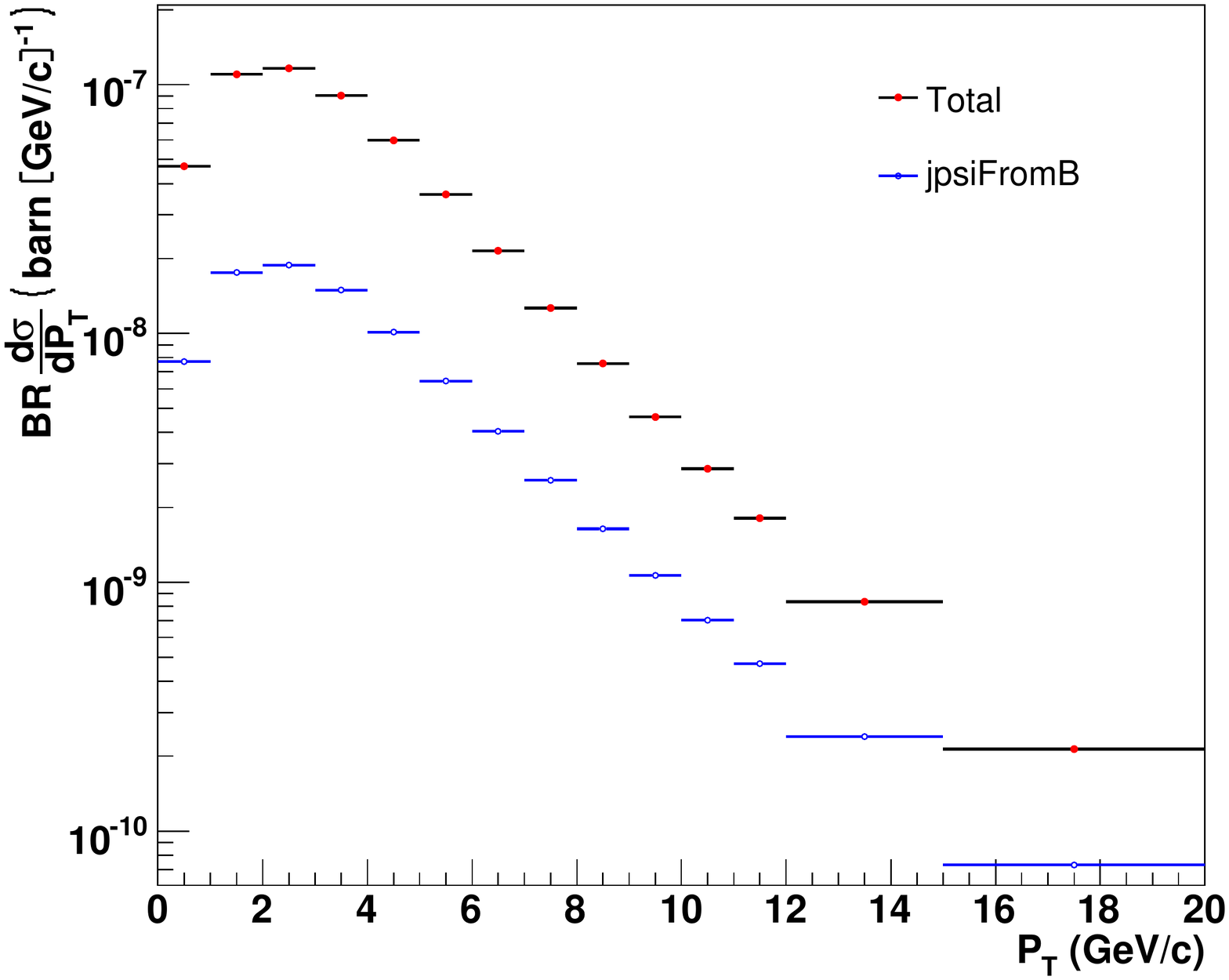}
\includegraphics[width=0.49\textwidth,trim=30 40 90 10]{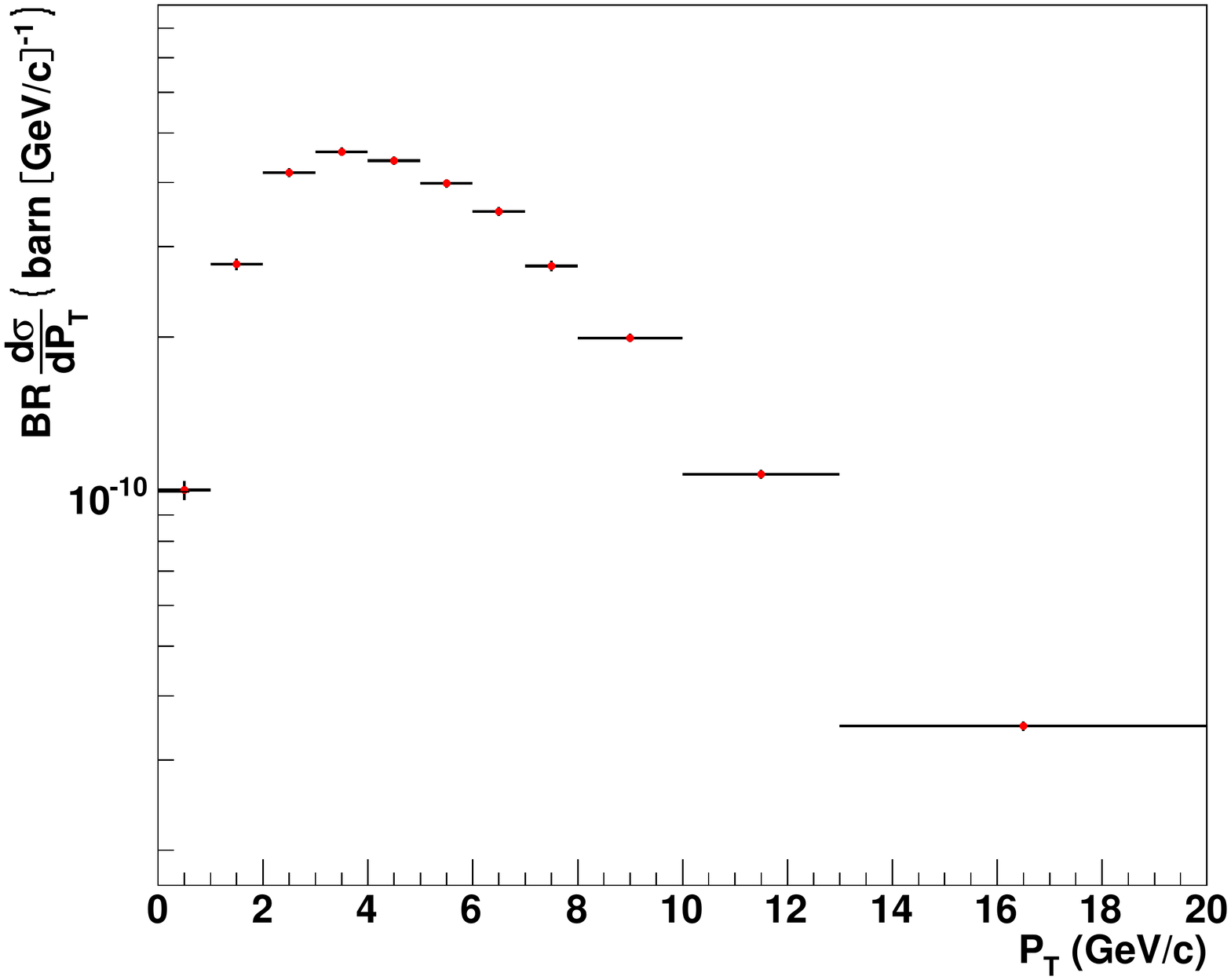}
\vspace{-0.3cm}
\caption{$BR_{\mu\mu} \dd\sigma/\dd\pt$ for \Jpsi (left) and \Ups (right) expected to be measured in a data taking time of $10^7$~s at a pp luminosity of $3 \times 10^{30} \lum$. The cross section is referred to the rapidity window $-4.0 < y < -2.5$ covered by the Muon Spectrometer. The contribution of \Jpsi from B decay is also shown in the left panel.}
\label{fig:dSigmadPt}
\end{figure}

\begin{figure}[!ht]
\centering
\includegraphics[width=0.49\textwidth,trim=30 40 90 10]{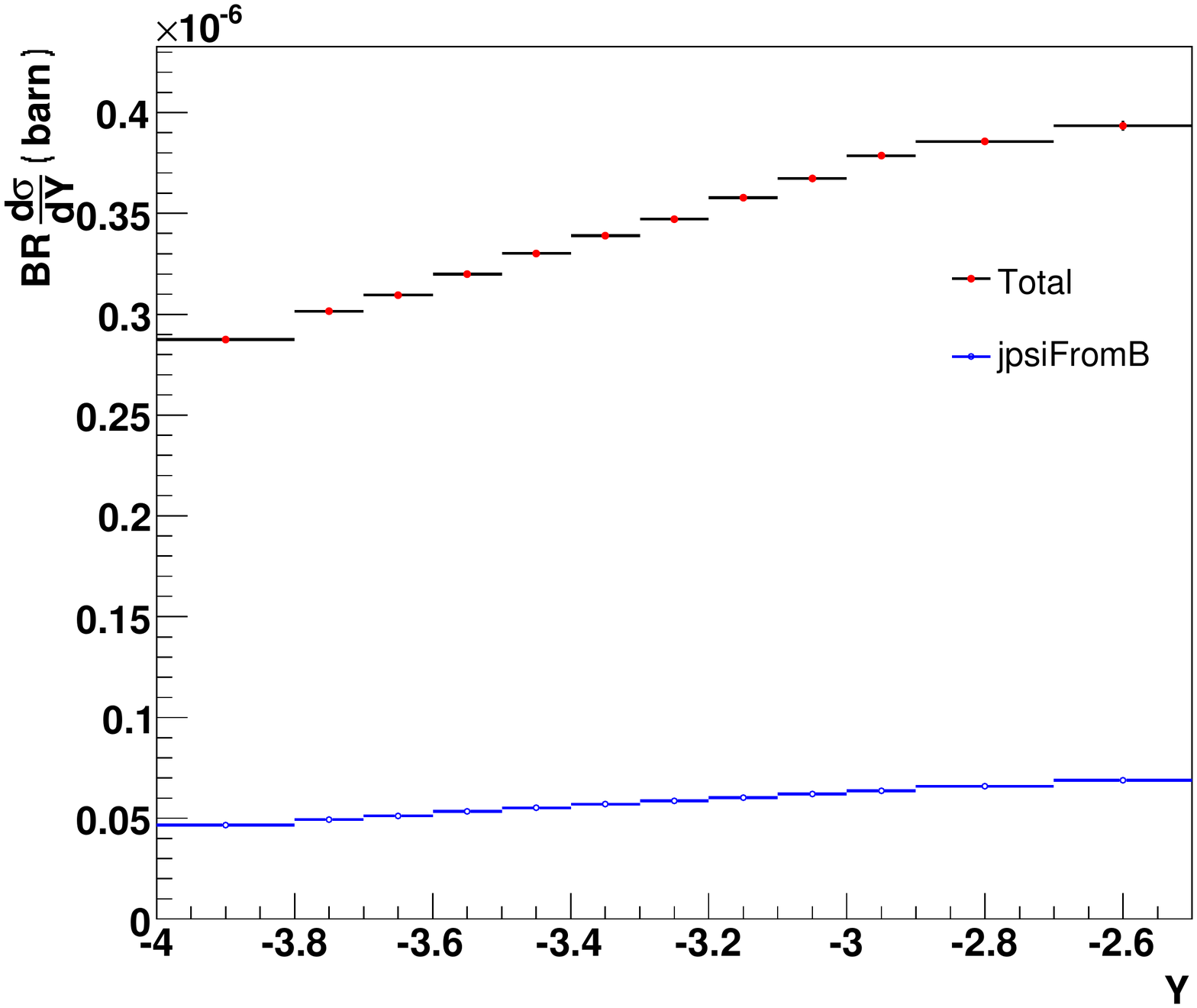}
\includegraphics[width=0.49\textwidth,trim=30 40 90 10]{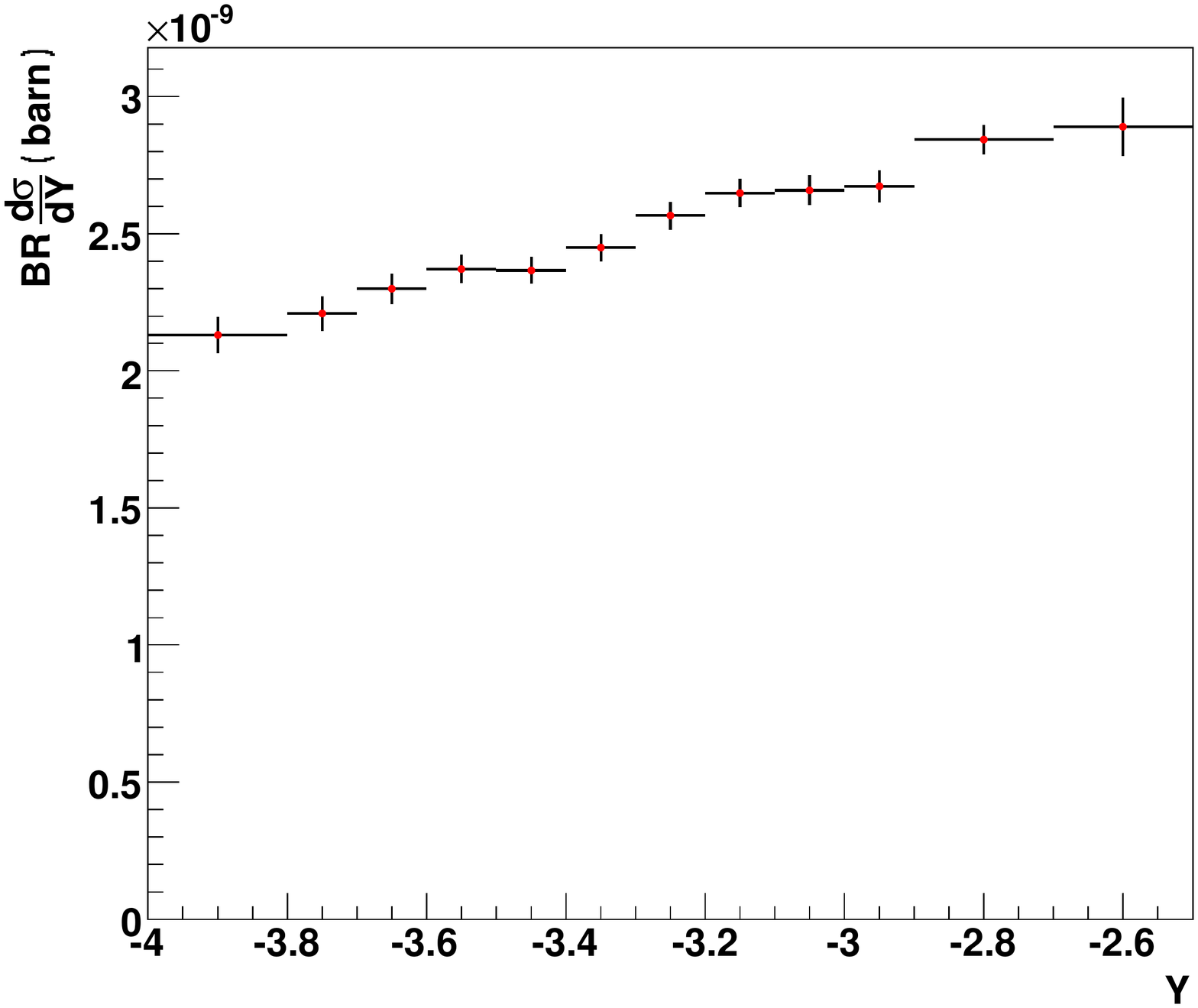}
\vspace{-0.3cm}
\caption{$BR_{\mu\mu} \dd\sigma/{\dd}y$ for \Jpsi (left) and \Ups (right) measured in a data taking time of $10^7$~s at a pp luminosity of $3 \times 10^{30} \lum$. The contribution of \Jpsi from B decay is also shown in the left panel.}
\label{fig:dSigmadY}
\end{figure}

The obtained statistics will be high enough to allow extracting the dimuon yields per bin of transverse momentum and rapidity.
To this aim the transverse momentum (rapidity) of the detected opposite-sign muon pairs was computed and the complete sample of events was divided in several bins.
For each {\pt} ($y$) bin, the corresponding dimuon invariant mass distribution was produced.
From each of these distributions, the \Jpsi and \Ups signals were extracted by fitting the invariant mass spectrum with the fitting procedure previously described.
Then the raw number of detected resonances was corrected for the detection probability to obtain the differential cross section $\dd\sigma/\dd\pt$ ($\dd\sigma/{\dd}y$).
The obtained $\dd\sigma/\dd\pt$ is normalized to the rapidity interval ($-4.0 < y < -2.5$) covered by the Muon Spectrometer.
The results in Fig.~\ref{fig:dSigmadPt} (Fig.~\ref{fig:dSigmadY}) show that the statistical error bars on the measured differential cross sections are small, in particular for the \jpsi, due to the high expected statistics.

For completeness the contribution of \Jpsi from B decay, though not directly measurable, is also shown in the left panels of both figures.\\

Leading order calculations show that in pp collisions at $\sqrt{s} = 14$~\tev, {\jpsi}'s with rapidity higher than 3.0 are produced by gluons\footnote{At high energy the gluon fusion becomes dominant in \QQbar production, while mechanisms involving quarks can be neglected. Hence, in the following we will speak about gluon distribution functions instead of parton ones.} carrying a fraction $x$ of the proton momentum lower than $10^{-5}$.

\begin{figure}[!ht]
\centering
\includegraphics[width=0.59\textwidth,trim=30 40 90 10]{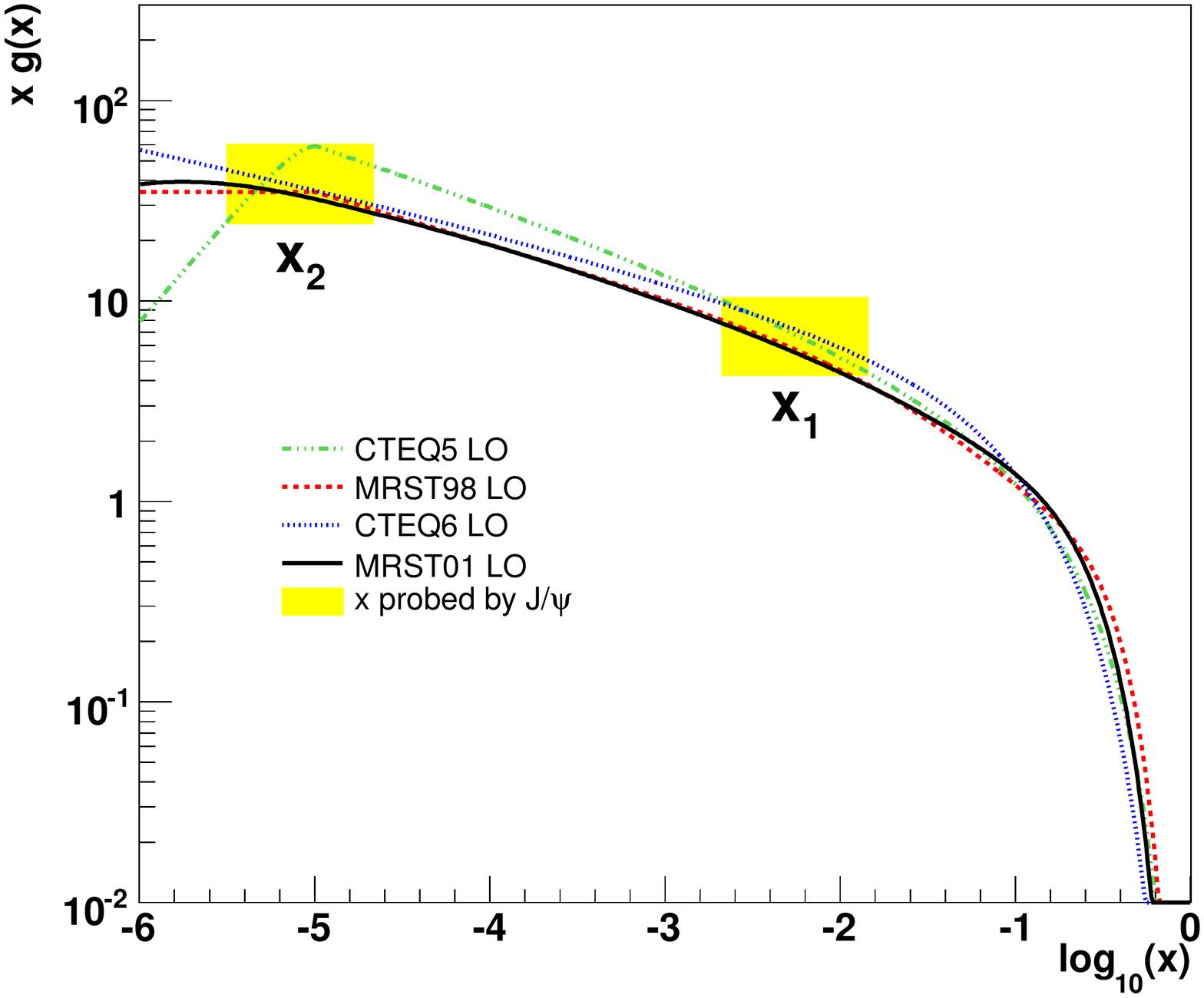}
\includegraphics[width=0.4\textwidth,trim=30 40 90 10]{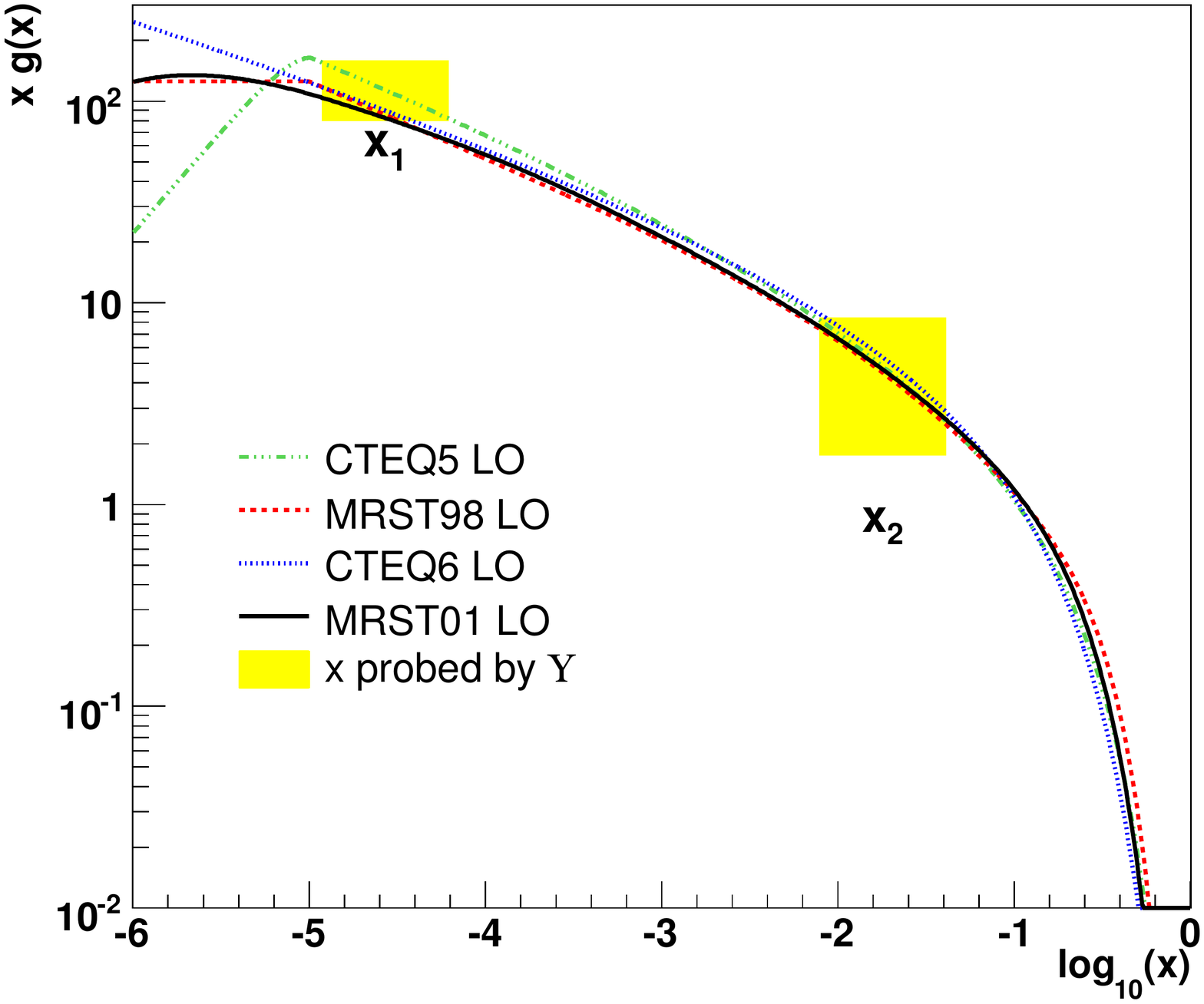}
\caption{Comparison of the gluon distributions from MRST and CTEQ. $x$ regions probed by J/$\psi$ and $\Upsilon$ produced in pp collisions at $\sqrt{s} = 14\tev$ in the rapidity region $-4.0 < y < -2.5$ are shown.}
\label{fig:PDFs}
\end{figure}

In such region, due to a lack of experimental data, the available gluon distribution functions rely on extrapolations, thus manifesting a significant disagreement.
The feature is depicted in Fig.~\ref{fig:PDFs}, showing a comparison between PDF sets calculated at Leading Order (LO) precision by different collaborations, namely Martin-Roberts-Stirling-Thorne (MRST98~\cite{Martin:1998sq} and MRST01~\cite{Martin:2002dr}) and the Coordinated Theoretical-Experimental Project on QCD (CTEQ5~\cite{Lai:1999wy} and CTEQ6~\cite{Pumplin:2002vw}): the $x$-values explored by \Jpsi in the ALICE Muon Spectrometer acceptance (in yellow) partially sits in the region of extrapolation.
It is worth noting that, due to its larger mass, \Ups production is sensitive to $x$-values larger than $10^{-5}$.

Starting from the Color Evaporation Model as a guideline, it is possible to show that the \Jpsi rapidity distribution,
\begin{equation}\label{eqn:dSigmadY}
  \frac{\dd\sigma_{\scriptscriptstyle{J/\psi}}^{\scriptscriptstyle{CEM}}}{{\dd}y} = \frac{F_{\scriptscriptstyle{J/\psi}}}{s} \sum_{i,j}{\int_{4 m_Q^2}^{4 m_H^2}{\dd \hat{s} \, \hat{\sigma}_{ij}(\hat{s}) f_{i/A}(\sqrt{\frac{\hat{s}}{s}}e^y,\mu^2) f_{j/B}(\sqrt{\frac{\hat{s}}{s}}e^{-y},\mu^2)}}
\end{equation}
is an observable sensitive to the PDF variation at low $x$.

This is done by making the following two approximations:
\begin{itemize}
\item in the elementary cross section $\hat{\sigma}_{ij}$, only the dominant contribution $gg \rightarrow \,$\QQbar is taken into account
\item both the elementary cross section $\hat{\sigma}_{gg}$ and the gluon distributions are taken at leading order
\end{itemize}

\begin{figure}[!ht]
  \centering
  \includegraphics[width=.7\textwidth,trim=30 40 90 10]{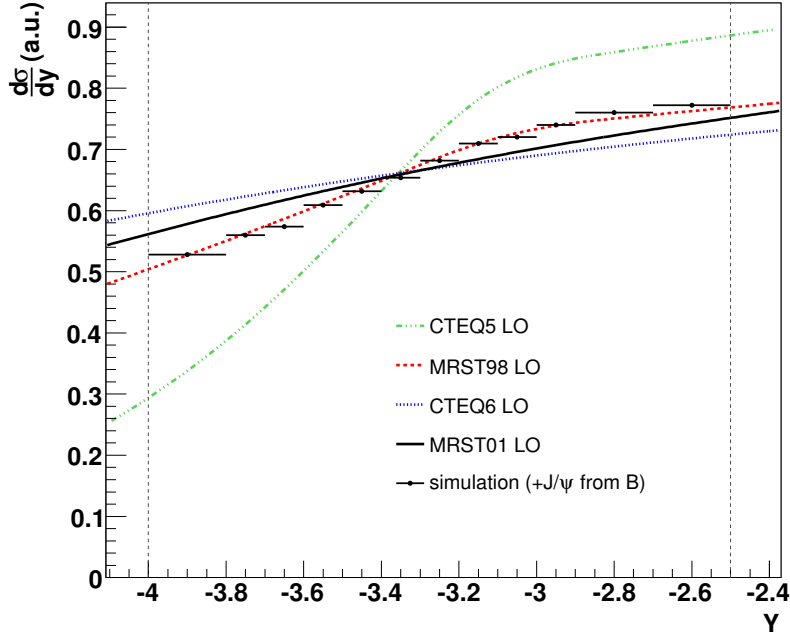}
  \caption{Comparison between \Jpsi rapidity distributions obtained with four different PDF sets (area in detector acceptance is normalized to 1). Simulation results are also shown.}
  \label{fig:compWithSimulation}
\end{figure}

The calculation was carried out with different PDFs.
The results are summarized by the curves in Fig.~\ref{fig:compWithSimulation}.
We note that the rapidity distributions shown in this figure are normalized by setting equal to unit their integral from -4.0 to -2.5 rapidity units.
This way of presenting the results emphasizes the fact that different behaviors of the gluon distribution functions at low $x$ (see Fig.~\ref{fig:PDFs}) lead to different \textit{shapes} of the \Jpsi rapidity distribution
(see Fig.~\ref{fig:compWithSimulation}) in the interval covered by the muon spectrometer.

Such behavior can be easily understood by analyzing Eq.~\ref{eqn:dSigmadY}.
The rapidity distribution depends on:
\begin{itemize}
 \item the product of parton distribution functions
 \item the elementary cross section $\hat{\sigma}_{gg}(\hat{s})$
 \item the ratio $F_{\jpsi}/s$ (assumed to be constant in the model)
\end{itemize}
The last term only affects normalization.
Hence the \textit{shape} of the distribution is related only to the first two quantities.
However, since the elementary cross section doesn't depend explicitly on $y$ and it is not expected to give rise to large variation of magnitude in the small range of integration ($4m_Q^2 < \hat{s} < 4m_H^2$), one can conclude that the shape of the distribution is mainly dependent on PDFs.

The possibility of focusing the study on the \textit{shape} of the rapidity distribution, disregarding the absolute normalization, is extremely favorable from the experimental point of view, since lots of systematic errors (e.g. on luminosity, global acceptance effects, etc.) do not enter.

The next (and last) step of the study presented here is aimed to show that the accuracy of data collected with the Muon Spectrometer will be good enough in order to resolve the different rapidity distributions.

A direct comparison with the simulation results in Fig.~\ref{fig:dSigmadY} is not possible, since such data were obtained by adopting the MRST98 set computed at NLO, while the calculations presented in this section concern LO quantities.
Hence the \Jpsi rapidity distribution was re-obtained from simulation after adopting MRST98 LO as input.
Such distribution (the simulation points in Fig.~\ref{fig:compWithSimulation}), is then compared with the calculations with different PDF sets (curves in Fig.~\ref{fig:compWithSimulation}).
It is worth noting that, differently from calculations, the simulated data already include contribution from B decay, which slightly change the shape of the distribution.
However the change is limited as it can be seen by comparing simulated data and the calculation of prompt \Jpsi distribution obtained with the same PDF set (red curve of Fig.~\ref{fig:compWithSimulation}).
This figure shows that, due to the high statistics, the accuracy of the data that are expected to be collected by the ALICE Muon Spectrometer will be good enough to allow to discriminate among different gluon distribution functions in the region of $x < 10^{-5}$ (at least in the frame of a leading order analysis).

\addtocounter{chapter}{1}
%\documentclass[a4paper,12pt,twoside]{report}
%\usepackage{epsfig}
%\usepackage{amssymb}
%\usepackage{lineno}
%\usepackage{setspace}
%%%%%%%%%%%%%%%%%%%%%%%%%%%%%%%%%%%%%%%%%%%%%%%%%%%%%%%%%%
%\begin{document}
%%%%%%%%%%%%%%%%%%%%%%%%%%%%%%%%%%%%%%%%%%%%%%%
% Toggle line numbering
% Won't work with the PRD revtex4 !
%\pagewiselinenumbers
% uncomment if you want doublespace
%\doublespace

\renewcommand\pt{p_{\scriptscriptstyle T}}
\newcommand\kt{k_{\scriptscriptstyle T}}
\renewcommand\Et{E_{\scriptscriptstyle T}}
\newcommand\mt{m_{\scriptscriptstyle T}}
\newcommand\Ecm{E_{\scriptscriptstyle CM}}
\newcommand\muF{\mu_{\scriptscriptstyle F}}
\newcommand\muR{\mu_{\scriptscriptstyle R}}
\renewcommand\alphas{\alpha_{\scriptscriptstyle S}}
\newcommand\nf{n_f}

\def\etmiss{\slashchar{E}_T}
 
%  \slashchar puts a slash through a character (i.e. ETmiss or Dirac matrices etc.)
 \def\slashchar#1{\setbox0=\hbox{$#1$}           % set a box for #1
    \dimen0=\wd0                                 % and get its size
    \setbox1=\hbox{/} \dimen1=\wd1               % get size of /
    \ifdim\dimen0>\dimen1                        % #1 is bigger
       \rlap{\hbox to \dimen0{\hfil/\hfil}}      % so center / in box
       #1                                        % and print #1
    \else                                        % / is bigger
       \rlap{\hbox to \dimen1{\hfil$#1$\hfil}}   % so center #1
       /                                         % and print /
    \fi}

%%%%%%%%%%%%%%%%%%%%%%%%%%%%%%%%%%%%%%%%%%%%%%%
%\chapter{Parton Distribution Functions at the LHC} {\it A. Tricoli}
\mchapter{Parton Densities at the LHC}{A. Tricoli}
%%%%%%%%%%%%%%%%%%%%%%%%%%%%%%%%%%%%%%%%%%%%%%%%%%%%%%%%%
\section{Introduction}

%In the past decade or so a great amount of work has been done by theorists to develop new theories that extend the Standard Model (SM) and refine calculations of SM predictions, by experimentalists to build the detectors and enhance their capability of detecting new physics signals against backgrounds. 
The start up of the LHC machine is now imminent and theorists and experimentalists are converging their efforts to enhance the LHC discovery potential. 
%experiments successfully reach their goals. 
This implies
%prefiguring more and more realistic physics scenarios and 
 minimising theoretical and experimental uncertainties.
Among the theoretical uncertainties the knowledge of the proton structure plays a major role: the accurate evaluation of parton density functions (PDF's) is vital to provide reliable predictions of new physics signals (i.e. Higgs, Supersymmetry, Extra Dimensions etc.) and their background cross sections at the LHC. 
As shown in the contribution by C. Mariotti, E. Migliore and P. Nason
%, eq. (\ref{eq:HadCrossSect})
, at hadron colliders the inclusive cross section for hard production processes is the convolution of the cross section at parton level, calculable at fixed order in perturbation theory, and the parton densities
%, $f_{i}^{H}(x, \mu)$, 
of the two interacting partons.

\begin{figure}[!t]
\begin{center}
\includegraphics[scale=0.25]{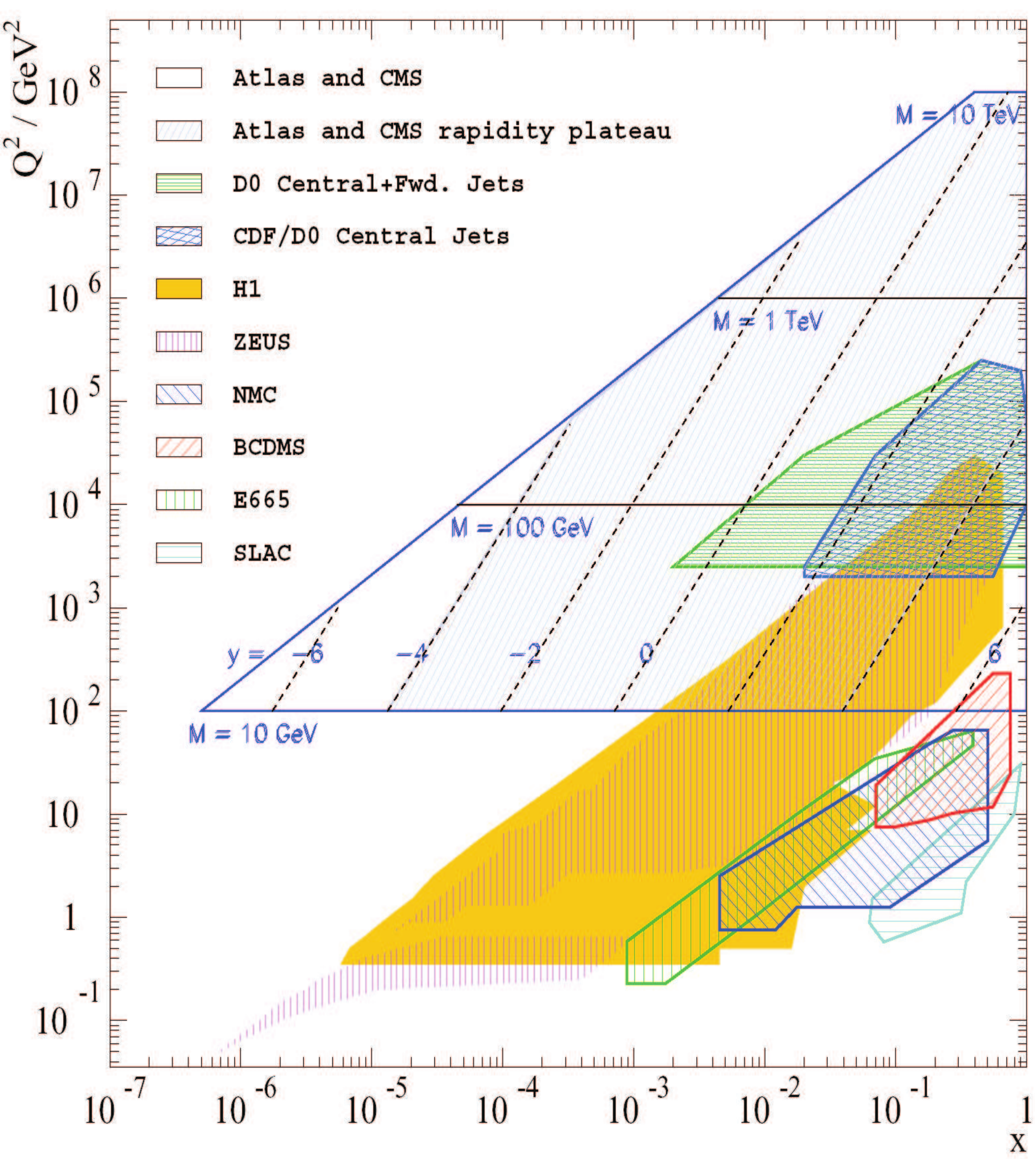}
\end{center}
\caption[LHC Kinematic Regime]{\label{fig:LHCKinRegime}}{\emph{The $Q^2$-$x$ kinematic plane for the LHC and previous experiments, showing the mass ($M$) and rapidity ($y$) dependence.}}
\end{figure}

Our knowledge of the proton structure is improving fast thanks to more experimental data being available and thanks to more precise and sophisticated theoretical calculations: PDF's are nowadays available up to the next-to-next-to leading order ({\em NNLO}) in perturbative QCD and in recent years they have been also providing uncertainties which take into account experimental systematic errors and the correlations between data points that enter the global fits.
Despite the great improvement on PDF's in recent years, their uncertainty dominates many cross section calculations for the LHC. As visible in fig.~\ref{fig:LHCKinRegime}, the LHC will probe kinematic regions in $x$ (parton momentum fraction) and $Q^2$ (hard scattering scale) never explored before, such as the $very ~high$-$Q^2$ and the $very ~low$-$x$ regions. At low-$x$ the current theoretical formalism (DGLAP) is at the edge of its supposed applicability. 
For the production of $Z$ and $W$ bosons the participating partons have small momentum fractions at central rapidity, $x \sim 10^{-3}$, and in the whole measurable rapidity region, $|y|<2.5$, they are within the range $10^{-4}<x<0.1$. Thus, at the electro-weak scale the theoretical predictions for the LHC cross sections are dominated by low-$x$ PDF uncertainty. At the ${\rm TeV}$ scale, where we expect new physics, the interacting partons have higher momentum fractions and very high $Q^2$ ($\geq 10^6~{\rm GeV^2}$).
%: at such high scale the interacting partons are mostly generated by the $g \rightarrow q\bar{q}$ splitting process. 
Thus, at the ${\rm TeV}$ scale the cross section predictions are dominated by high-$x$ PDF uncertainty and rely on the extrapolation of the DGLAP equations. 
In both kinematic regimes the gluon density, which is in most regions the less well constrained density function, plays a major part: at low $x$ the gluon density dominates the quark and anti-quark densities, at high $Q^2$ the interacting partons get an important contribution from the sea, which is driven by the gluon density, via the $g \rightarrow q\bar{q}$ splitting process. For a review on hard interactions of quarks and gluons at the LHC refer to~\cite{LHCPrimer}.

Past and running experiments, such as HERA, have been providing vital information to improve our knowledge of the parton densities, however the broad kinematic region of the LHC forces (and offers a unique opportunity to) ATLAS and CMS experiments to use their own data to constrain the parton densities, in particular the gluon, in the kinematic regions where they are not sufficiently well determined.
In section~\ref{sec:PDFconstraints} it will be shown that significant improvement on PDF fits can be made with LHC data.

%%%%%%%%%%%%%%%%%%%%%%%%%%%
\section{Global fits and error analysis}
%%%%%%%%%%%%%%%%%%%%%%%%%%%%%%%%%

Perturbative QCD provides the evolution equations for the parton densities, DGLAP equations, but does not provide us with their analytic forms as function of $x$.
The most common approach to extrapolate PDF's as function of $x$ and $Q^2$ consists in solving the DGLAP equations by parameterising the parton densities $q_i(x)$ at a fixed scale $Q^2_0= 1 - 7 ~{\rm GeV^2}$, applying assumptions and constraints derived from theory and measurements. Then, with the DGLAP equations, we numerically extrapolate the values of $q_i(x,Q^2)$ to different values of $Q^2$ and a global fit of experimental data is performed.
For valence quarks the parameterisations have usually this behaviour $q_V \approx x^\lambda (1-x)^\eta $, whereas for the gluon and sea quarks they are of this kind $q_S(g) \approx x^{- \lambda} (1-x)^\eta $.
However there is no unanimous agreement on the parametric functions to use and on the number of free parameters. For a review refer to~\cite{Devenish_Cooper-Sarkar}.

%Global fits take advantage of available experimental results to constrain the parton distributions in the widest possible kinematic regions. 
Different regions in the $x, Q^2$ plane and also different partonic components are probed by the available world experimental data. These include DIS data from fixed target experiments and HERA, Drell-Yan data, inclusive jet production and $W$ charge asymmetry from Tevatron.

There are various groups who are fitting the proton structure function data, among them CTEQ and MRST. Recent PDF sets include in their analyses up-to-date experimental data and attempt to provide coherent estimates of the uncertainties, including experimental correlated systematic errors. The differences between these PDF sets can be summarised in three categories: different choices of input data sets, different theoretical model assumptions and different error analyses.

There are many sources of uncertainty which contribute to a global fit uncertainty. These can have experimental and theoretical origins. The former are related to the data errors which enter the fit, the latter are due to the model uncertainties of the theoretical framework.
The theoretical uncertainties concern both the non-perturbative (parameterisations) and perturbative parts of the calculations: assumptions imposed to limit the number of free parameters, higher order truncations in the DGLAP formalism etc.

The treatment of the experimental uncertainties, especially the correlated systematic uncertainties, is a complex subject which is partly still under debate. A modified version of the standard $\chi^2$ method is used to take into account non-Gaussian systematic errors and their correlations: $\chi^2\rightarrow \tilde\chi^2+\Delta T^2$, where $\Delta T$ is the so-called ``tolerance'', a complicated mathematical expression that includes correlated systematic terms~\cite{Devenish_Cooper-Sarkar}.
There are then two methods to compute the central values of the theoretical PDF parameters and their uncertainties: the {\em offset} and the {\em Hessian} method. In the offset method the correlated systematic errors affect only the determination of the PDF uncertainty, not the best fit. This method is used for ZEUS PDF's.
% with $\Delta T^2\approx 49$. 
Conversely in the Hessian method, used by CTEQ and MRST groups, the collective effect of the correlated systematic errors has also an impact on the best fit.
% with  $\Delta T^2\approx 100$ and %with $\Delta T^2\approx 50$ respectively.
%what about CTEQ6.5E?????

\begin{figure}[!tb]
  \begin{minipage}[t]{.40\textwidth} 
  	\begin{center}  
	 \includegraphics[width=7.5cm, height=8.3cm]{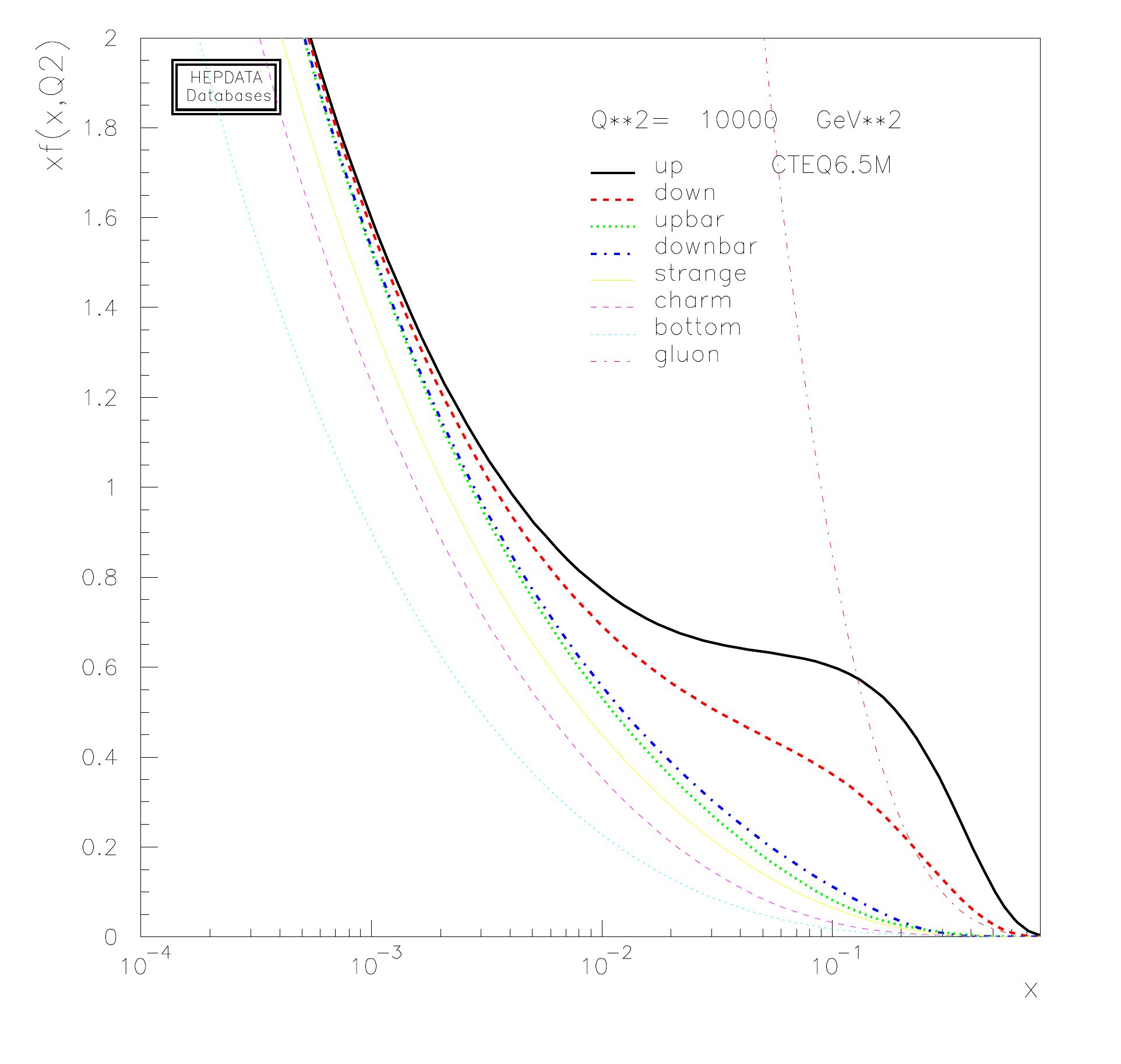}   
	\end{center}
  \end{minipage}
  \hspace{15mm}
  \begin{minipage}[t]{.40\textwidth}
    \begin{center}  
      \includegraphics[width=7.0cm, height=8.cm]{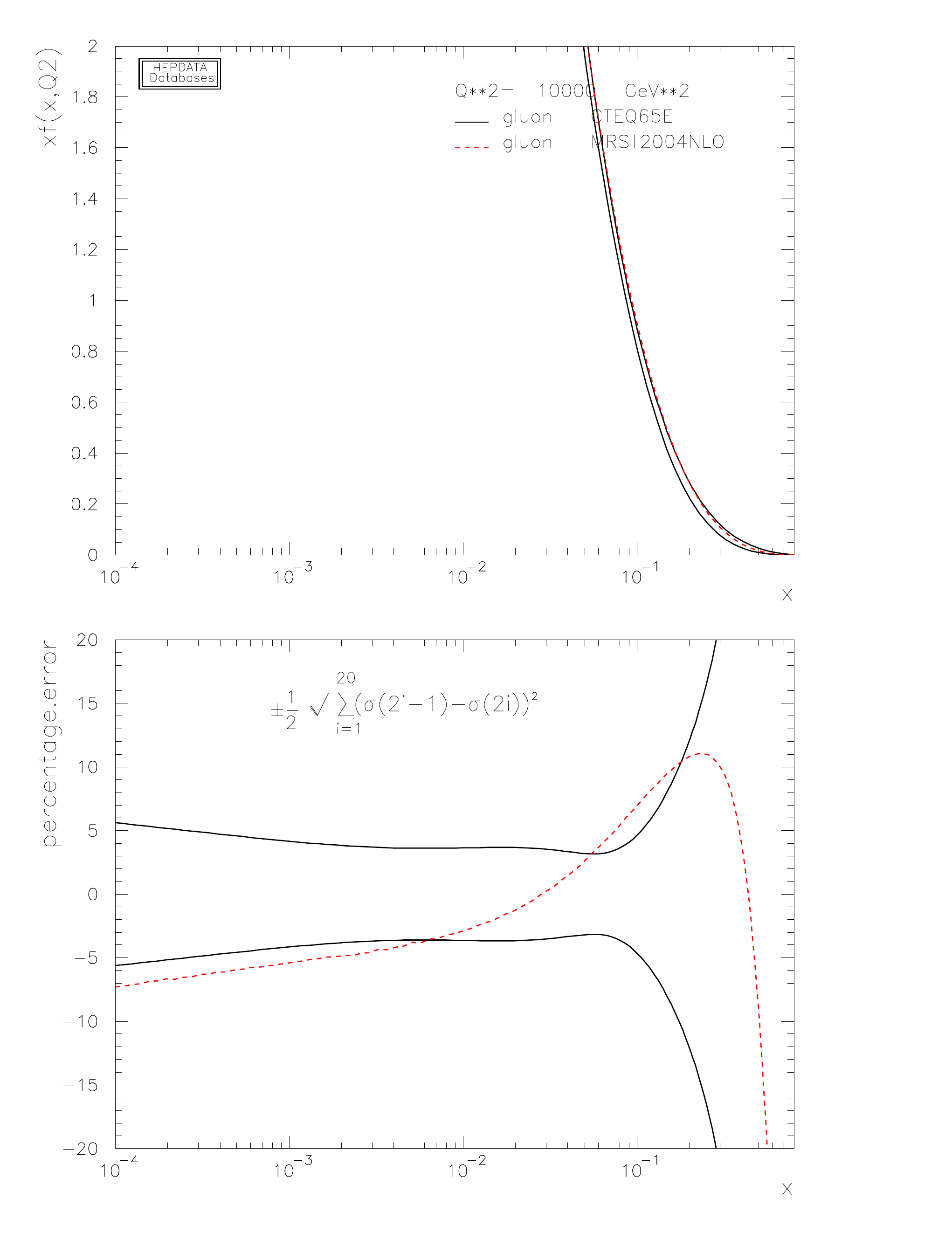} 
    \end{center}
  \end{minipage}
  \caption[CTEQ65 and MRST2004NLO PDF]{\label{fig:CTEQ65vsMRST2004NLO}}{\emph{Left: CTEQ6.5M set at $Q^2\sim M_W^2$. Right: comparison between CTEQ6.5M (black) and MRST2004NLO (red) gluon PDF's and their uncertainties.}} 
\end{figure}

For both, the offset and the Hessian methods, the PDF uncertainty is conventionally computed along the eigenvectors of the diagonalised covariance or Hessian matrices.
The number of eigenvectors corresponds to the number of free parameters in the parton density parameterisations.
Contemporary PDF sets provide a central value PDF set, corresponding to the best data fit, and two PDF sets for each uncertainty eigenvector, giving the upper and lower limit on the uncertainty. Given a PDF set, the upper limit of the PDF uncertainty is calculated for a physical observable by adding in quadrature the upward displacement eigenvectors, whereas the lower limit by adding in quadrature the downward displacement eigenvectors. 
MRST group has chosen 15 free parameters, leading to 30 error sets; CTEQ6 has 20 free parameters and 40 error sets.
Fig.~\ref{fig:CTEQ65vsMRST2004NLO} shows CTEQ6.5 fit for all parton densities at the scale $Q^2\sim M_W^2$ and its gluon uncertainty compared to the MRST2004NLO gluon best fit.

%%%%%%%%%%%%%%%%%%%%%%%%%%%%%%%%%%%%
\section{Impact of PDF uncertainty on LHC physics}
%%%%%%%%%%%%%%%%%%%%%%%%%%%%%%%%%%%%
The experience from previous experiments teaches that the PDF uncertainties must be properly taken into account or features of the SM physics can be misinterpreted as evidence of new physics. For example an unexplained discrepancy between data and theory was originally found in the Tevatron Run-I jet data, which was subsequently reabsorbed within the theoretical uncertainty when a more accurate PDF error analysis was performed.

G. Polesello's contribution on inclusive jet cross-section has shown that the PDF uncertainty is dominating for high $E_T$ jets over the renormalisation/factorisation scale and the experimental energy scale uncertainties: $10\%$ at $1~{\rm TeV}$, $25\%$ at $2~{\rm TeV}$, $60\%$ at $5~{\rm TeV}$.

\subsubsection{Extra dimensions}
In extra dimensions models, if the compactification scale $M_C$ is about few TeV\footnote{In this context the compactification scale is defined as $M_C = 1/R_C$ where $R_C$ is the compactification radius of the extra dimensions on a hypersphere.}, it is possible to observe the production of gravitons and Kaluza Klein (KK) excitations at the LHC. If gauge bosons can propagate in the extra dimensions, we also expect a violation of the SM logarithmic behaviour of the running couplings.
In this scenario, if we consider the CTEQ6M PDF uncertainty on the di-jet cross-section, we see the extra dimensions prediction being absorbed within the SM prediction zone: the high-$x$ gluon uncertainty can cause a decrease of the discovery reach from $M_C = 5~(10)~{\rm TeV}$ to $M_C < 2~(3)~{\rm TeV}$, depending on the number of extra dimensions~\cite{PDFError_ED}.

\subsubsection{Higgs}
The accurate measurements of the Higgs production cross sections and decay branching ratios are crucial to explore all Higgs boson fundamental properties. At the same time, we need very precise estimates of the various theoretical uncertainties.

It is found that the PDF uncertainty can be of the same order of magnitude or even higher than the other theoretical uncertainties. In fact the perturbative calculations of Higgs production cross section are becoming more stable as higher orders are included, leaving the PDF uncertainty as one of the largest contributions to the total theoretical uncertainty.
For example for the dominant Higgs production channel, $gg\rightarrow H$, the PDF uncertainty on $gg$ luminosity, can be larger than the factorisation and renormalisation scale uncertainty: in fact the differences in the $gg$ luminosity prediction between MRST2002 and Alekhin2002 can be higher than $10\%$ for low Higgs mass scenarios. 
Furthermore, studying the effect of three different PDF sets (i.e. CTEQ6M, MRST2001E and Alekhin2002) with their quoted uncertainties, on various Higgs productions channels, we see that the PDF uncertainty can be of the order of $\sim 10-15\%$ on the production cross-section~\cite{PDFError_Higgs}. 

\subsubsection{High mass Drell-Yan}

Several new physics models predict events with two charged leptons originating from the decay of a massive object. A peak in the $d\sigma/dM$ distribution is a clean signature of a new resonance: the identification and reconstruction of high-mass di-lepton final states can be done with high efficiency and the SM background can be small.
However the shape and normalisation of the predicted observable distributions depend on PDF and its uncertainty.

%\begin{figure}[!tb]
% \hspace{-15mm} 
%  \begin{minipage}[t]{.40\textwidth} 
%  	\begin{center} 
%	  \includegraphics[width=11cm, height=6.5cm]{./DrellYanPDFUncertaintyAsymmetric.pdf}  
%	\end{center}
%  \end{minipage}
%  \hspace{55mm}
%  \begin{minipage}[t]{.40\textwidth}
%    \begin{center}  
%      \includegraphics[width=5.5cm, height=6.5cm, clip=true, viewport=280pt 0pt 567pt 414pt]{./DrellYanPDFYUncertaintyAsymmetric.pdf}
%    \end{center}
%  \end{minipage}
%  \caption[High Mass Drell-Yan]{\label{fig:HighMassDilepton}}{\emph{CTEQ6.1 uncertainty on distributions of the high-mass di-electron $M_{ll}$ (left and centre) and rapidity $y$ (right). Herwig+Jimmy generation and ATLAS full simulation~\cite{Florian}.
%%, $Z\rightarrow e^+e^-$ events with $M_{ll}>150~{\rm GeV}$. 
%N.B.: the drop in the low $M_{ll}$ spectrum is an artifact of the event selection in the Monte Carlo.
%}} 
%\end{figure}

\begin{figure}[!tb]
% \hspace{-15mm} 
  \begin{minipage}[t]{.60\textwidth} 
  	\begin{center} 
	  \includegraphics[width=0.98\textwidth]{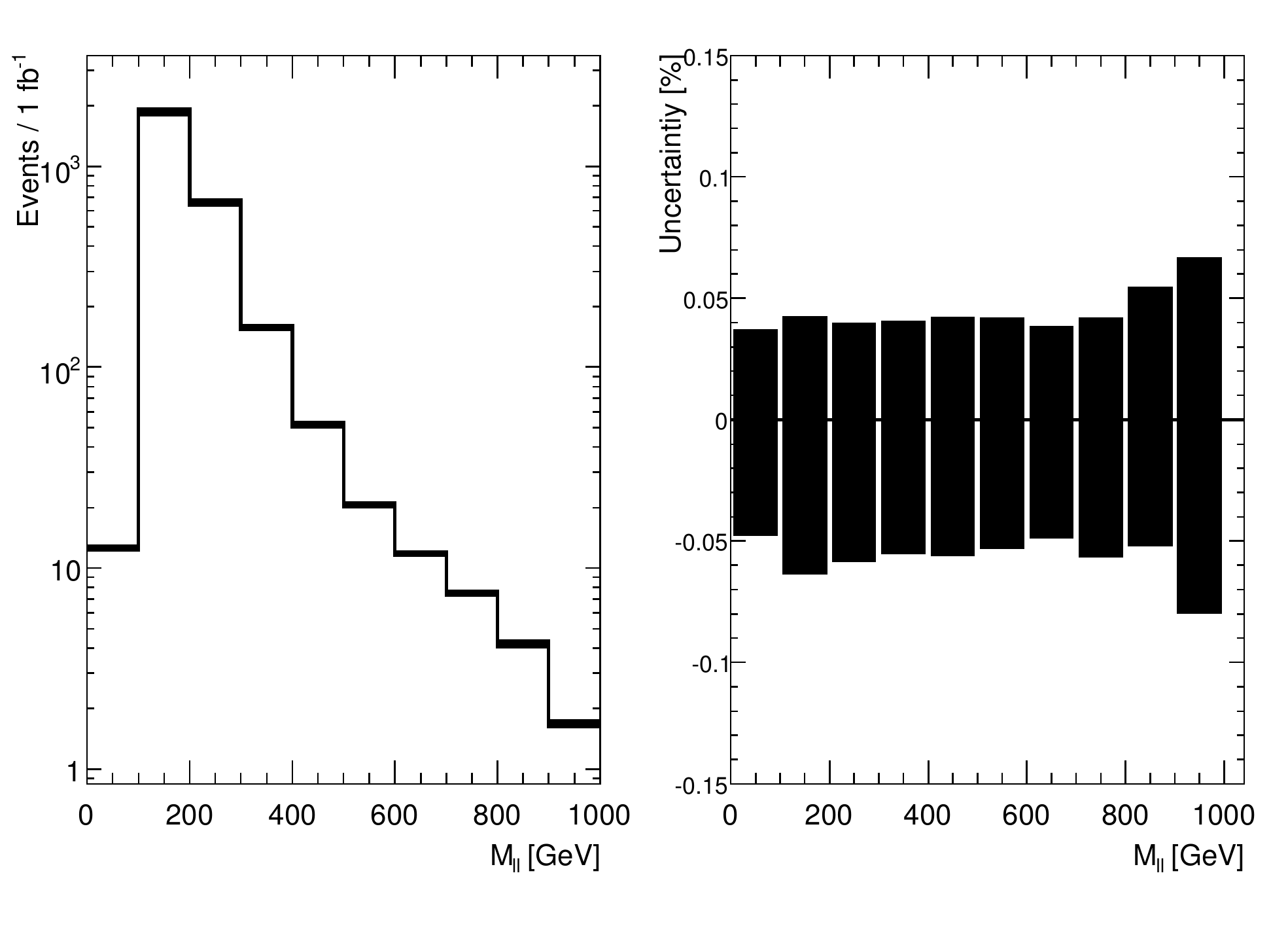}  
	\end{center}
  \end{minipage}
  \hspace{55mm}
  \begin{minipage}[t]{.35\textwidth}
     \hspace{-1.1\textwidth}
      \includegraphics[width=0.85\textwidth,clip=true, viewport=280pt 0pt 567pt 414pt]{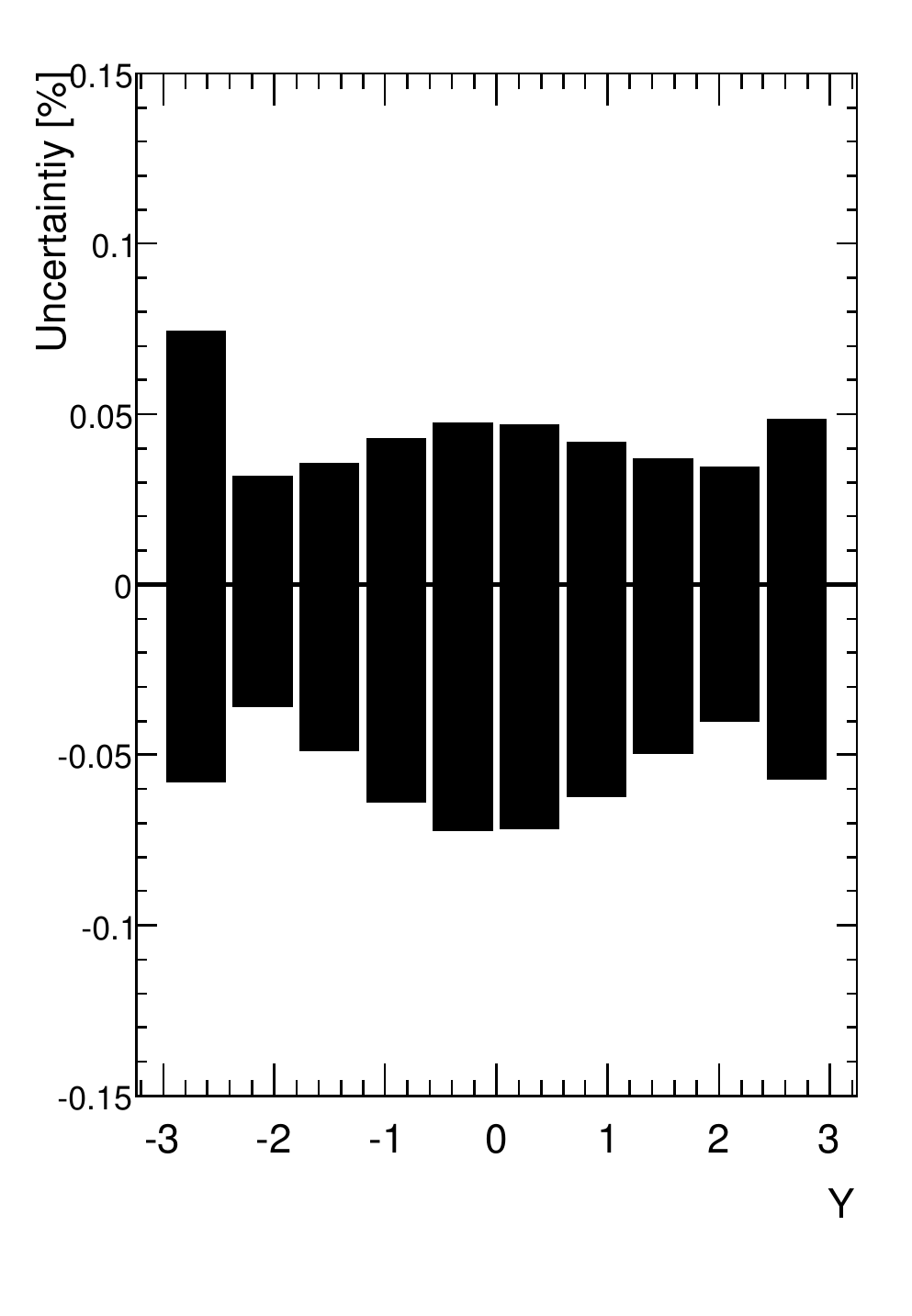}
  \end{minipage}
  \caption[High Mass Drell-Yan]{\label{fig:HighMassDilepton}}{\emph{CTEQ6.1 uncertainty on distributions of the high-mass di-electron $M_{ll}$ (left and centre) and rapidity $y$ (right). Herwig+Jimmy generation and ATLAS full simulation~\cite{Florian}.
%, $Z\rightarrow e^+e^-$ events with $M_{ll}>150~{\rm GeV}$. 
N.B.: the drop in the low $M_{ll}$ spectrum is an artifact of the event selection in the Monte Carlo.
}}
\end{figure}

In fig.~\ref{fig:HighMassDilepton} we see the total CTEQ6.1 uncertainty on the distributions of the reconstructed rapidity $y$ and invariant mass $M_{ll}$ of the lepton pair: 40 CTEQ error sets have been accounted for, applying the PDF reweighting technique (see sec.~\ref{sec:PDFreweighting}). The uncertainty is in the range $4-7\%$ on both $y$ and $M_{ll}$ up to $1~{\rm TeV}$. Excluding the bins at the edge of the rapidity distributions, where statistical fluctuations are present, we see that the largest PDF uncertainty is at $y\sim 0$.
As explained in~\cite{Florian}, a study shows that NLO QCD corrections, applied on Monte Carlo (MC) and on PDF, enhance the cross section with respect to the LO prediction by $24-36\%$, with the largest NLO corrections at $y\sim 0$. A discrepancy of about $6\%$ is found between MRST-NLO and CTEQ-NLO PDF's.%~\cite{Florian}.
%, depending on the chosen PDF set: MRST-NLO and ZEUS-NLO PDF sets are in good agreement with each other and predict a cross section about $6\%$ higher than CTEQ-NLO. The largest NLO corrections are at $y\sim 0$.
%say which versions of PDF sets!!!!!!!!!!!!!!!!!!!!!!!!

%\subsection{SUSY} i.e. W+jets

%%%%%%%%%%%%%%%%%%%%%%%%%%%%%%%%%%%%%%%%%%%%%%%%%%%%%%%%%%%%%%%%%%%%%%%%%%
\section{How to constrain PDF at LHC}\label{sec:PDFconstraints}

Several Standard Model processes are under study to constrain parton densities: $\gamma$, $W$ and $Z$ boson and inclusive jet production processes are equally important to constrain the parton densities and in particular the gluon density in complementary kinematic regions (see~\cite{Tricoli_Photon05}).

In G. Polesello's contribution we appreciate how the LHC jet data can be used to better constrain PDF fits: if the experimental systematic uncertainty is under control to $\leq 10\%$ level, LHC jet data can significantly contribute to constraining the high-$x$ gluon density with $1~fb^{-1}$ luminosity.  
Other studies~\cite{Hollins} have also shown that the prompt photon production process is extremely sensitive to PDF differences and can probe the perturbative theory of the gluon at high-$x$: the discrepancy between MRST2004-NLO, CTEQ6.1M and older PDF sets can be of the order of $16-18\%$ on the photon $\eta$ and $\pt$ distributions. \\
Furthermore, the $bg \rightarrow Zb$ process is sensitive to the $b$-quark content of the proton and the LHC predictions for the $Z+b$ cross-section, using different PDF sets, are $\pm 5-10\%$~\cite{Zb_VerducciEtAl}.

%\subsection{Prompt photon production}
%At LO the direct photon production will take place via the Compton scattering ($\sim 90\%$ rate) $qg\rightarrow \gamma q$ and annihilation ($\sim 10\%$ rate) $q \bar{q}\rightarrow \gamma g $ processes. The typical event topology in the ATLAS detector will be a photon and jet with a back-to-back configuration in the $r-\phi$ plane. 
%At low $\pt$, i.e. $\pt \sim 5 - 10~{\rm GeV}$, the differential cross section  $d^2\sigma=dEd\pt$ is sensitive to non-perturbative information about the intrinsic $\kt$ of the gluons in the proton, to resummation of threshold logarithms, i.e. $\ln(1 - x_T )$ where $x_T=2\pt/\sqrt{s}$, and to the interplay between the two.
%However, far cleaner experimental measurements are performed at the LHC at much higher $\pt$, i.e. $\pt\geq 330~GeV$, where we probe the perturbative theory of the gluon at high-$x$, $x \sim 0.05$ at $\eta \sim 0$ for a photon $\pt = 350~{\rm GeV}$.  

%In~\cite{Hollins} it is shown that the photon $p_T$ and $\eta$ distributions are extremely sensitive to PDF differences: the discrepancy between MRST2004nlo, CTEQ6.1M and older PDF sets can be of the order of $16-18\%$. 

\subsection{W rapidity distributions}

%When LHC is up and running, the number of $W$ bosons produced will be so large that soonthe statistical uncertainty will become negligible with respect to the systematic uncertainties on $W$ cross section. 
%After a collected integrated luminosity of ..., corresponding to ... days
A few days of LHC running at the nominal low luminosity ($10^{33} {\rm cm^{-2} s^{-1}}$) are sufficient to make the statistical uncertainty negligible with respect to the systematic uncertainties on $W$ cross section.
Among the systematic uncertainties there are experimental and theoretical contributions.

%\begin{figure}[!t] 
%\centerline{
% \epsfig{figure=./mt_missEt_sigBkg_isEMcut.pdf,width=5cm,height=5cm,angle=90}
% \epsfig{figure=./mt_missEt_sigBkg_missEtcut.pdf,width=5cm,height=5cm,angle=90}
% \epsfig{figure=./eta_sigBkg_missETcut.pdf,width=5cm,height=5cm,angle=90} 
%}
%\caption[$W\rightarrow e \nu_e$ signal and backgrounds]{\label{fig:WSigBkg}}{\emph{ATLAS fully simulated $W\rightarrow e \nu_e$ signal events (white) and cumulative backgrounds: $W\rightarrow \tau \nu_\tau$ (red), QCD (yellow), $Z\rightarrow e^+ e^-$ (green) and $Z\rightarrow \tau^+ \tau^-$ (blue). On the left, the $W$ boson $M_T$ after the trigger and off-line electron identification only; in the middle and right, the $W$ boson $M_T$ and the electron $\eta$ distributions respectively, after all the inclusive $W$ selection cuts (trigger and off-line included). Pythia events corresponding to $1.3~{\rm pb^{-1}}$ luminosity.}}
%\end{figure}

The ATLAS strategy for selecting $W$ bosons consist of identifying an isolated and highly energetic lepton, $\Et>25~{\rm GeV}$, and requiring a large amount of missing energy in the event due to the neutrino escaping detection, $\etmiss>25~{\rm GeV}$.  
%Fig.~\ref{fig:WSigBkg} shows the fully simulated $W\rightarrow e \nu_e$ signal with respect to the main backgrounds. 
The analysis of $W\rightarrow e \nu_e$ events fully simulated in the ATLAS detector, in the early data scenario, shows that the $W$ boson is a very clean signature: the trigger and the electron off-line identification with the electron $\Et$ and $\etmiss$ cuts leave a background contamination dominated by QCD events (less than $5\%$) and $W\rightarrow \tau \nu_\tau$ (about $0.5\%$). If a jet veto cut is added, the QCD background can be further reduced to a level of $\le 1\%$~\cite{CSCNote}.
Therefore the $W$ sector is an ideal environment to study and constrain theoretical and experimental systematics.

\subsubsection{Higher order corrections}
%A great amount of progress has been made on theoretical uncertainties in the recent years. 
The differential cross section $d\sigma/dy$ for $W$ production has been calculated to the NNLO order in QCD with an energy scale uncertainty of  $\leq 1\%$~\cite{NNLO_WZ_y}.
With this level of precision in perturbative QCD calculations, the electro-weak (EW) contributions are no more negligible. As presented in this workshop, leading order electro-weak contributions with multi-photon radiation introduce corrections of the order of few percent on $W$ boson cross-sections.
%Fig.~\ref{fig:MuPlus_ew_nlo_corr_eta} shows the differential cross-section $d\sigma/d\eta$ for $\mu^+$ decaying from $W^+$, including NLO perturbative QCD corrections and $O(\alpha)$ electro-weak corrections.
%\begin{figure}[!t]
%\begin{center}
%\includegraphics[width=6cm,height=6cm]{./wplep_mrst04_alfew_yw.pdf}
%\end{center}
%\caption[The $W^+$ rapidity distributions with NLO-QCD and LO-EW corrections]{\label{fig:MuPlus_ew_nlo_corr_eta}}{\emph{Analytic calculation for the $W^+$ rapidity distribution, including NLO-QCD and LO-EW corrections with MRST2004QED PDF.}}
%\end{figure}
The EW corrections, computed by the program HORACE interfaced to HERWIG in the $\alpha(0)$ scheme in the muon channel~\cite{CarloniPolesello}, are constant in rapidity and are about $3.5\%$ for a cut on the muon transverse momentum of $\pt>25~{\rm GeV}$ and can be up to $5.2\%$ for loser $\pt$ cuts. The dependence on the muon charge is negligible (up to $0.4\%$ for lose $\pt$ cuts)~\cite{TricoliCooper-Sarkar}.
Considering that these corrections in the muon channel are flat in rapidity and negligible on the muon-charge asymmetry, we can state that they do not have an impact on the PDF extraction, however they are relevant for luminosity measurements in order to achieve a precision of $6\%$ or better.
The electron channel needs further investigation. 

\subsubsection{PDF uncertainty on $W^{\pm}$ rapidity distribution.}

\begin{figure}[!tb]
 \hspace{-5mm} 
  \begin{minipage}[t]{.40\textwidth} 
  	\begin{center} 
       \includegraphics[width=7cm, height=6.5cm]{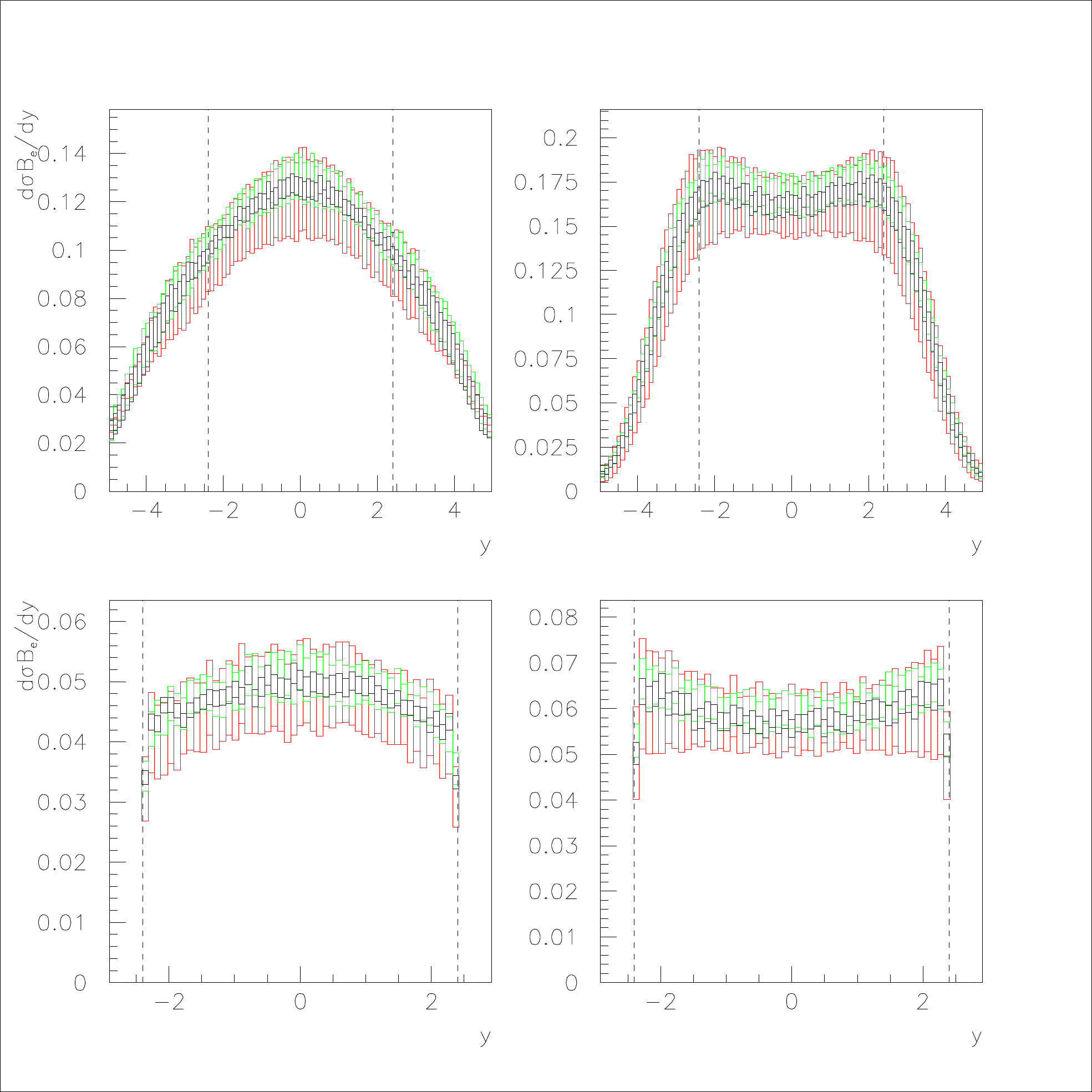}  
	\end{center}
  \end{minipage}
  \hspace{15mm}
  \begin{minipage}[t]{.40\textwidth}
    \begin{center}  
      \includegraphics[width=7cm, height=6.5cm]{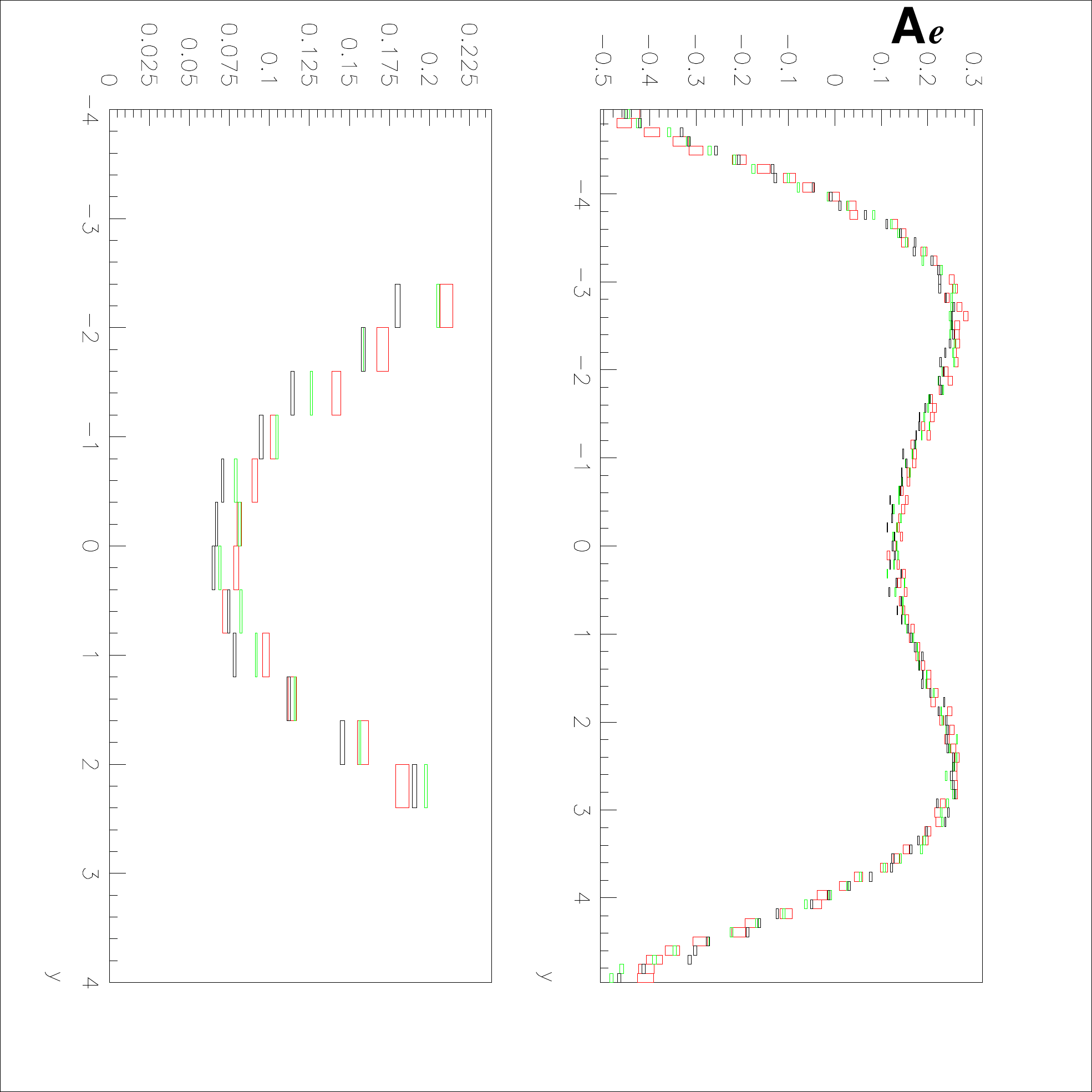}
    \end{center}
  \end{minipage}
  \caption[MC simulated $y$-spectra for $e^{\pm}$ and $A_l$ with PDF errors]{\label{fig:eplus_emin_rap_MC}}{\emph{HERWIG simulations of $e^{\pm}$ from $W^{\pm}$ decay, with CTEQ6.1M (red), MRST2001 (black) and ZEUS-S (green) PDF's and their quoted uncertainties (estimated with the PDF reweighting technique as in sec.~\ref{sec:PDFreweighting}). The top plots are at generator level, the bottom plots at ATLFAST detector level. Left fig: $e^-$ (left plots) and $e^+$ (right plots) rapidity spectra with NLO-QCD corrections.  Right fig: electron-charge asymmetry~\cite{Tricoli}.}}
\end{figure}

From fig.~\ref{fig:eplus_emin_rap_MC} we can see the full PDF uncertainties for three different PDF analyses, on the rapidity distribution of $e^{\pm}$ originating from $W^{\pm}$ decays. Their predictions are compatible within their uncertainties, which are in the range $4\%-12\%$, and are dominated by the gluon density. 
%\begin{figure}[!tb]
%\begin{center}
%\includegraphics[width=10cm,height=6cm]{./emin_eplus_gen_allpdf_nocuts.pdf}
%\end{center}
%\caption[MC simulated $y$-spectra for $e^{\pm}$ with PDF errors]{\label{fig:eplus_emin_rap_MC}}{\emph{Monte Carlo simulations for $e^-$ (top plot) and $e^+$ (bottom plot) rapidity spectra to NLO-QCD, according to CTEQ6.1M (red), MRST2001 (black) and ZEUS-S (green) PDF's with their quoted uncertainties.}}
%\end{figure}

In a previous paper~\cite{HERAtoLHC} it is demonstrated that the LHC can improve the current constraint on the low-$x$ gluon parameter $\lambda_g$ ($xg(x)\approx x^{-\lambda_g}$) by more than $41\%$ by fitting the $e^+$ and $e^-$ rapidity distributions, if their experimental systematic uncertainties are kept under $5\%$ level.  

In the lepton-charge asymmetry $A_l=(\frac{d\sigma}{d\eta}^{l^+} -\frac{d\sigma}{d\eta}^{l^-} )/(\frac{d\sigma}{d\eta}^{l^+} + \frac{d\sigma}{d\eta}^{l^-})$ most of the gluon uncertainty cancel out leaving the valence up ($u_V$) and down ($d_V$) densities as main contributions to the total PDF uncertainty, which is reduced to $\sim 5\%$ at $\eta\approx 0$. However a discrepancy of $\sim 15\%$ is present at $\eta\approx 0$ between the MRST2002 and other two PDF's, CTEQ6.1M and ZEUS-S~\cite{Mandy}. In fact the MRST PDF's prediction for $u_V-d_V$ valence density is different from the other PDF's and is outside the quoted PDF uncertainty bands. This difference in current PDF fits comes from the lack of data on valence quantities at such low-$x$.
The LHC can be the first experiment to perform such measurement in the kinematic region $x\approx 10^{-3}$ and $Q^2=M_W^2$. %maybe figure????   

%\subsection{ $Z + b$-$jet$ production}
%The $bg \rightarrow Zb$ process is sensitive to the $b$-quark content of the proton and is also a background to the Higgs boson search~\cite{Campbell_1,Campbell_2}. The accurate determination of the $b$-PDF is important for the precise measurement of the $Z$ boson production cross section: 
%in order to measure $\sigma_Z$ to about $1\%$, a $b$-PDF precision of $\sim 20\%$ is required.
%For ATLAS, the prediction of the $Z+b$ cross-section using different PDF sets is $\pm 5-10\%$.
%In the $Z\rightarrow \mu^+ \mu^-$ channel we can obtain a clear signal over the background, with a $Z+b$ event selection efficiency of $\sim 15\%$ and sample purity $53\pm 10\%$~\cite{VerducciEtAl}. 

\subsubsection{A posteriori inclusion of PDF's in NLO calculations.}
The MC computation of QCD final state observables to NLO is a lengthy process. In order to study the impact of PDF uncertainties on QCD cross section measurements in a faster way and allow for PDF fitting of these quantities, the technique of ``a posteriori'' inclusion of PDF's in NLO calculations has been developed for LHC processes~\cite{Carli-Salam-Siegert}~\cite{Clements-et-al}. A MC run is used to generate a grid (in $x_1$, $x_2$ and $Q$) of cross section weights that can subsequently be combined with an arbitrary PDF set. This enables the decoupling of the lengthy calculation of perturbative MC weights from the convolution with the parton densities.
Perturbative coefficients for jet (using NLOJET++), W and Z boson (using MCFM) production processes can be collected on a grid with an accuracy better than $0.02\%$.

%%%%%%%%%%%%%%%%%%%%%%%%%%%%%%%%%%%%%%%%%%%%%%%%%%%%%%%%%%%%%%%%%%%%%%%%%%
\section{PDF reweighting of Monte Carlo events}\label{sec:PDFreweighting}

The computation of the full PDF uncertainty on a physics process is a cumbersome procedure. Given one PDF set, such as CTEQ or MRST, it requires the generation of twice as many MC samples as the number of free parameters in the global fit.
Furthermore one error analysis might not be sufficient since, as seen above, there can be large discrepancies between the results of different error analyses. 

A PDF reweighting technique has been studied and tested, requiring only one Monte Carlo generation with one conventional PDF set
% and consists of applying event weights in order to give predictions for any other arbitrary, but conventional, PDF sets
~\footnote{This techniques is not as reliable if the PDF set is as ``unconventional'' as MRST2003, i.e. the validity of its kinematic space is smaller than the one available to the LHC. }~\cite{Tricoli}\cite{HERAtoLHC}.

This technique has been implemented using hard process parameters of the MC generation: flavours ($flav_{1}$ and $flav_{2}$) and  momentum fractions of the interacting partons $x_{flav_{1}}$, $x_{flav_{2}}$ and  the energy scale $Q$. The PDF set used for the MC generation is named $PDF_{1}$.
%the event weights are calculated only from hard-process variables, such as the momentum fractions of the interacting partons ($x_1$ and $x_2$)\footnote{The two momentum fractions $x_1$ and $x_2$ are taken before the parton shower is applied in the MC with the backward evolution.} and the energy scale $Q^2$.
%The invariant mass $M$ and, in first approximation, the rapidity $y$ of the $W$ boson are determined by the hard process calculation.

%Each MC event is generated with one specific PDF set and happens to have: one energy scale $Q$ at the hard process, i.e. $M_W$, two interacting partons of flavours $flav_{1}$, $flav_{2}$ and momentum fractions $x_{flav_1}$, $x_{flav_2}$.
%The probability of having such flavoured partons in the hard process at the energy scale $Q$ is determined according to the chosen PDF set ($PDF_{1}$): $f_{PDF_{1}}(x_{flav_{1}},Q)$ and $f_{PDF_{1}}(x_{flav_{2}},Q)$.

The PDF reweighting technique consists of evaluating, on the event-by-event basis, the probability of picking up the same flavoured partons with the same momentum fractions $x_{flav_{1}}$, $x_{flav_{2}}$, according to a second PDF set, $PDF_{2}$, at the same energy scale $Q$%,\footnote{In order to evaluate the probability $f(x_{flav},Q)$ the PDF Library LHAPDF~\cite{LHAPDF} is used since it contains a large and up-to-date sample of PDF sets.} $f_{PDF_{2}}(x_{flav_{1}},Q)$ and $f_{PDF_{2}}(x_{flav_{2}},Q)$
, then evaluating the following ratio

\begin{equation}
Event~Weight = \frac{f_{PDF_{2}}(x_{flav_{1}},Q)}{f_{PDF_{1}}(x_{flav_{1}},Q)}\cdot %%@
\frac{f_{PDF_{2}}(x_{flav_{2}},Q)}{f_{PDF_{1}}(x_{flav_{2}},Q)}~.
\end{equation} 

After the $Event~Weight$ is applied on MC events generated with $PDF_{1}$, they will effectively be distributed according to $PDF_{2}$. 

This technique has been tested using HERWIG (for inclusive W production) and ALPGEN interfaced to HERWIG (for W+jets production) as Monte Carlo generators and with various recent PDF sets. Similar results have been obtained with these two MC generators and with different PDF sets, as discussed below.
% and with MRST2002, CTEQ6.1 and ZEUS-S PDF sets. 
\begin{figure}[tb]
  \begin{minipage}[t]{.40\textwidth} 
  	\begin{center} 
       \includegraphics[width=7.0cm, height=6.cm]{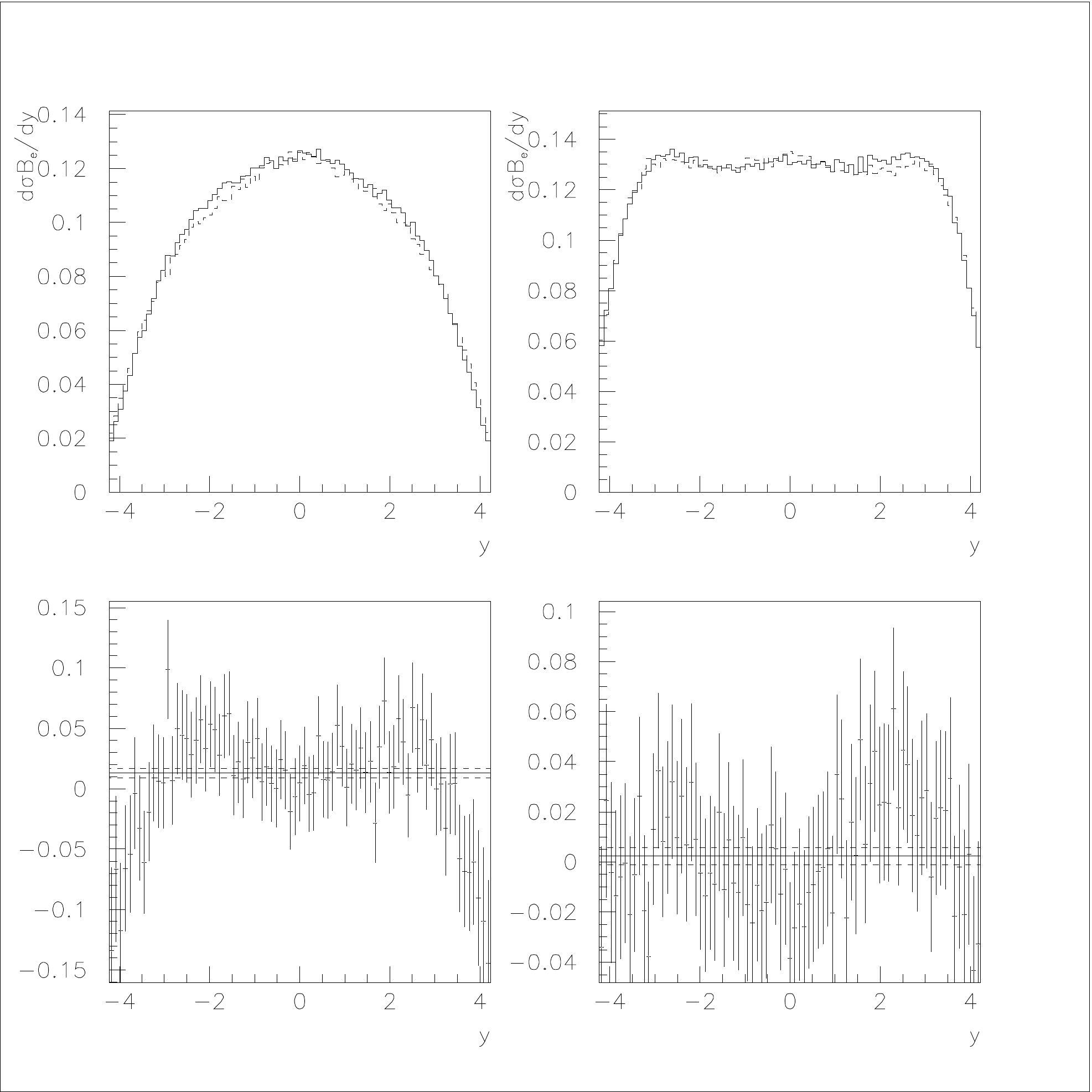}  
	\end{center}
  \end{minipage}
  \hspace{15mm}
  \begin{minipage}[t]{.40\textwidth}
    \begin{center}  
      \includegraphics[width=7.0cm, height=6.cm]{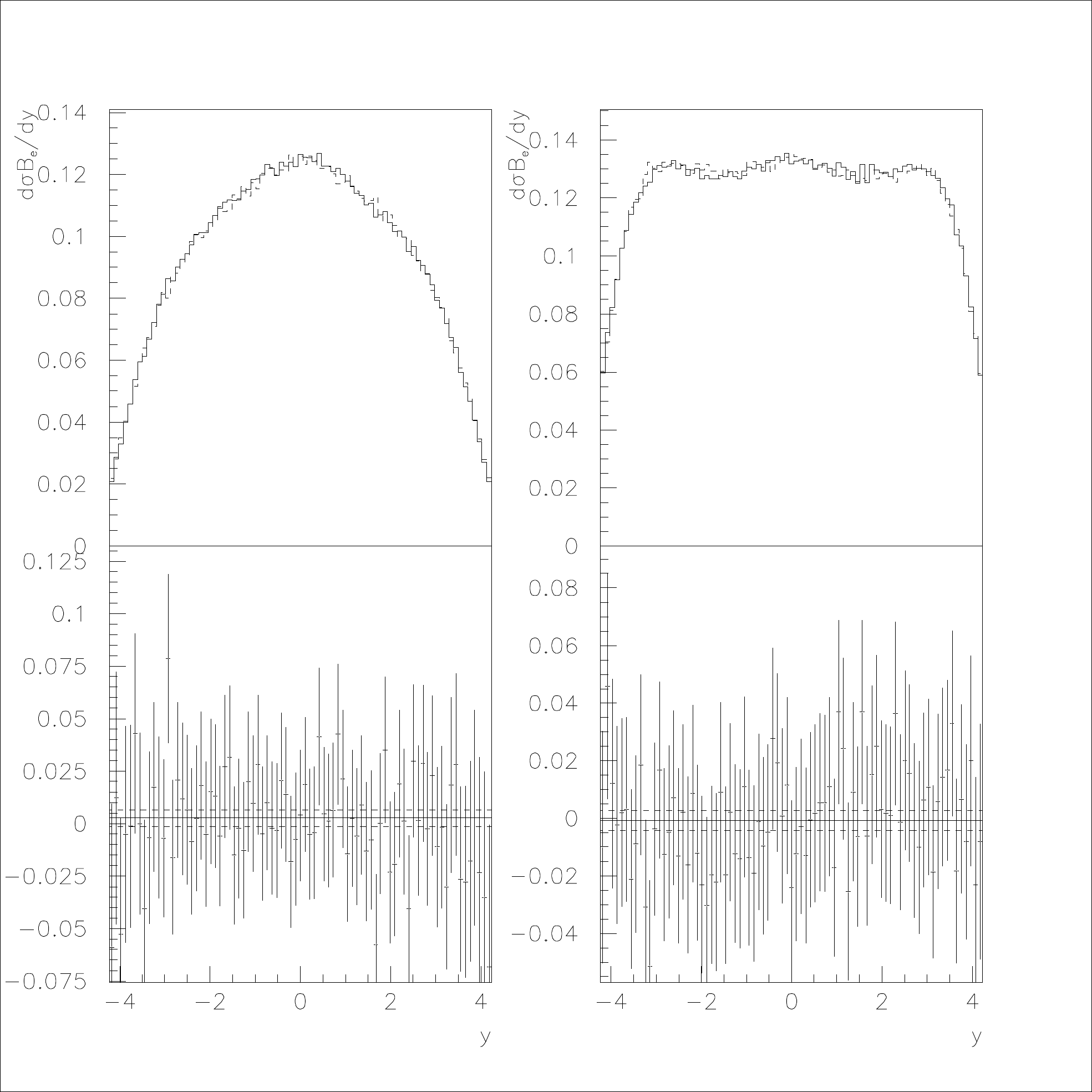}
    \end{center}
  \end{minipage}
  \caption[PDF reweighting]{\label{fig:PDFreweighting}}{\emph{Left fig: $W^{-}$ and $W^{+}$ rapidity distributions at HERWIG generator level for events generated with CTEQ6.1M (dashed lines) and for events generated with MRST2002 (solid lines) and their relative differences (at the bottom).
%: $(MRST2002_{generated} - CTEQ6.1M_{generated})/CTEQ6.1M_{generated}$. 
The straight lines are the means of the points with uncertainty bands. Right fig: same as left hand side plots for events generated with CTEQ6.1M (dashed lines) and for events generated with MRST2002 and PDF-reweighted with CTEQ6.1 (solid lines). Similar results have been obtained reweighting between MRST2002 and ZEUS-S PDF's.}}  
\end{figure}
%The left hand side of fig.~\ref{fig:PDFreweighting} compares the $W^+$ and $W^-$ spectra for events generated using MRST02 and events generated with CTEQ6.1.
%The right hand side of fig.~\ref{fig:PDFreweighting} compares the $W^+$ and $W^-$ spectra for the events generated using MRST02 as $PDF_{1}$ and reweighted with CTEQ6.1 as $PDF_{2}$, with the events directly generatd with CTEQ6.1. Beneath the spectra the fractional difference between these distributions is shown. 
Fig.~\ref{fig:PDFreweighting} shows the accuracy of this technique using HERWIG: the bias over the all $y$ range is of the order of $0.5\%$ or less and there is no evidence of $y$ dependence. 
%Similar results have been obtained reweighting between MRST2002 and ZEUS-S PDF's. 
Comparing the bottom plots on the right and left hand sides of fig.~\ref{fig:PDFreweighting} we see that the PDF reweighting technique corrects for the difference in normalisation between $PDF_1$ and $PDF_2$
%(shifting the means down by more than $1\%$ for $W^-$, $0.3\%$ for $W^+$) 
 and corrects for the $y$ modulation.
%, making the difference between the $y$ distributions of $CTEQ6.1M_{reweighted}$ and $CTEQ6.1M_{generated}$ essentially flat over the whole kinematic regime. 

This technique can be used to estimate the full PDF uncertainty, starting from one sample of MC generated events, for distributions that are determined by the MC hard process.
% such as $y$ and the invariant mass $M$ in HERWIG or PYTHIA. 

%%%%%%%%%%%%%%%%%%%%%%%%%%%%%%%%%%%%%%%%%%%%%%%%%%%%%%%%%%%%%%%%%%%%%%%%%%%%%%%
% Bibliography
%%%%%%%%%%%%%%%%%%%%%%%%%%%%%%%%%%%%%%%%%%%%%%%%%%%%%%%%%%%%%%%%%%%%%%%%%%%%%%

%\end{document}

\addtocounter{chapter}{1}
\providecommand{\tabularnewline}{\\}
\providecommand{\boldsymbol}[1]{\mbox{\boldmath $#1$}}

\def\beq{\begin{equation}}
\def\eeq{\end{equation}}
\def\beqn{\begin{eqnarray}}
\def\eeqn{\end{eqnarray}}
\def\ba{\begin{eqnarray}}
\def\ea{\end{eqnarray}}
\def\atp{\frac{\alpha_s(Q^2)}{2\pi}}
\def\afp{\frac{\alpha_s(Q^2)}{4\pi}}

%%%%%%%%%%%%%%%%%%%%%%%%%%%%%%%%%%%%%%%%%%%%%%%%%%%%%%%%%%
%\documentstyle [12pt] {report}   % Specifies the document style.
%\RequirePackage{graphics}
%%%%%%%%%%%%%%%% two floating figures, side by side %%%%%%%%%%%%%%%
\newenvironment{2figures}[1]{\begin{figure}[#1] 
  \begin{center}
    \begin{tabular}{p{.47\textwidth}p{.47\textwidth}} }
 {  \end{tabular}
  \end{center} 
 \end{figure}
}
%%%%%%%%%%%%%%%%%%%%%%%%%%%%%%%%%%%%%%%%%%%%%%%%%%%%%%%%%%%%%%%%%%%
%\begin{document}
%\begin{center}
%\vspace{2.cm}
\mchapter{NNLO Evolution of the Pdf's and their Errors: 
 Benchmarks and Predictions for Drell-Yan}
{Alessandro Cafarella, Claudio Corian\`{o},
Marco Guzzi}

%\section*{Abstract}

\begin{center}
\section*{Abstract}
\end{center}
We quantify the impact of the 
next-to-next-to-leading order evolution on the Drell-Yan total cross section and on the corresponding rapidity distributions of the lepton pair and compute the corresponding errors. 
We base our analysis on \textsc{Candia}, a program that solves the DGLAP equations using the method of the $x$-space iterates. 
%\newpage

\section{Introduction}

In the search for new physics at the LHC we need high precision in the determinations 
of the QCD background, possibly at next-to-next-to-leading order 
(NNLO) in the strong coupling constant $\alpha_s$. While 
it is expected that only a few processes 
will be computed in the near future at this order of accuracy in QCD, for 
Drell-Yan lepton pair production, some of these corrections - for instance those involving the 
invariant mass distributions - have been available for some time \cite{Van_Neerven1}. The study of this process will be essential both in the search of extra neutral interactions and for partonometry, where the impact of the perturbative 
resummation \cite{Nason} can be studied in detail given the large number of events expected at the LHC.
More recently, following the computation of the invariant mass distribution $d\sigma/d Q$, 
where $Q^2$ is the invariant 
mass of the lepton pair, also the rapidity distributions $d\sigma/(dY d Q)$, implemented in VRAP \cite{Anastasiou} have been computed. A fully exclusive 
numerical computation has also been presented \cite{Melnikov}. 
Next-to-leading order (NLO) analysis of the forward-backward asymmetries 
$A_{FB}$ on the Z resonance and NNLO 
charge asymmetries have also been determined \cite{Anastasiou}. 
The computation of the hard scatterings for some of these processes 
has been performed much before that the analytical computation of 
the NNLO evolution kernels needed for a consistent extraction of 
NNLO parton distributions (pdf's) were available. Following the computation of the kernels 
\cite{vogt1}, some 
benchmarks for the NNLO evolution have been presented \cite{LesHouches02}, followed by a later update \cite{LesHouches05}. 

Testing the benchmarks by using independent 
approaches that solve the equations is not only a demanding numerical problem, but involves subtle 
issues concerning the types of solutions that are implemented in a given numerical algorithm. Specifically, we have shown in \cite{candia,CCG2} that the selection of 
a given ansatz - either in Mellin space or in x-space - in the 
solution of the DGLAP involves a specific arrangement of the 
logarithmic expansion that solves iteratively the equations. 
For instance, ansatze for the exact solutions perform automatically 
a resummation of the contributions identified by the simplest logarithmic 
ansatz (also called ``truncated solutions''). These involve logarithms 
of the ratio of two couplings $\log(\alpha_s(Q^2)/\alpha_s(Q_0^2))$ at two different scales $Q$ and $Q_0$. Exact solutions, instead, replace these logarithmic expansions with more complicated functions of $\alpha_s$. Details can be found in \cite{candia,CCG2,CCG1}. 

These expansions apply generically 
both to forward and non-forward twist-2 operators, and 
converge to the solutions of the evolution equations with very high precision.
Concerning the structure of the evolution codes, these 
are usually based either on the numerical Mellin inversion 
(using an ansatz in moment space) or on ``brute force methods''. In this second case the numerical solution is built by a discretization 
of the equations, reduced to a stable finite-difference scheme. The 
theoretical indetermination coming from the various approaches, as we have pointed out in \cite{CCG2}, has to do with the the selected accuracy of the 
solution.

\section{The choice of the solution and the theoretical indetermination}

We have re-analized the issue of the initial state dependence of the predictions for the $d\sigma/d Q$ and $d\sigma/(dy d Q)$ cross sections in \cite{CCG2,CCG1}, focusing our attention near the $Z_0$ peak. 

Being the 
predictions coming from the inclusions of the NNLO corrections quite small, it is natural to worry 
whether the errors coming from the statistical fits in the determination of the pdf's to the experimental data and the indeterminations due to 
the treatment of the evolution can be compared so that the size of the errors claimed in the hadronic cross sections are quantified consistently. We point out that the specific choice 
of the solution of the evolution equation brings in an intrinsic 
indetermination which is comparable with the error coming from the fits to the pdf's. This indetermination is of theoretical origin. 

In our analysis we will omit most of the theoretical details that can be found in \cite{CCG1} 
and come to a discussion of the numerical results, since these can be of interest for actual 
experimental searches. We also present some benchmarks for the NNLO evolution that can be useful 
for a consistent comparison with other codes. An extensive numerical analysis of the predictions on the 
Z resonance that illustrates the points summarized here can be found in \cite{CCG2}.  

As we are going to see, the variations induced by the choice of the solution of the DGLAP
induce variations on the cross section of the order of $1 \%$ or so at NNLO, and clearly affect also the 
NNLO $K$-factor for the total cross section. We have used for this study 
\textsc{Candia} \cite{candia}, which implements the truncated and exact solutions of the DGLAP built without the numerical Mellin inversions. On the countrary of other programs that need initial conditions of a specific functional form, 
\textsc{Candia} allows any initial conditions to be studied, being based on x-space algorithms. An 
extended version of \textsc{Candia}, called $\textsc{Candia}_{\textsc{DY}}$ allows to study the invariant mass distribution to 
NNLO. An interface with VRAP is also under development.

\begin{figure}
\subfigure[Alekhin Evolution]{%
\includegraphics[width=7cm,angle=-90]{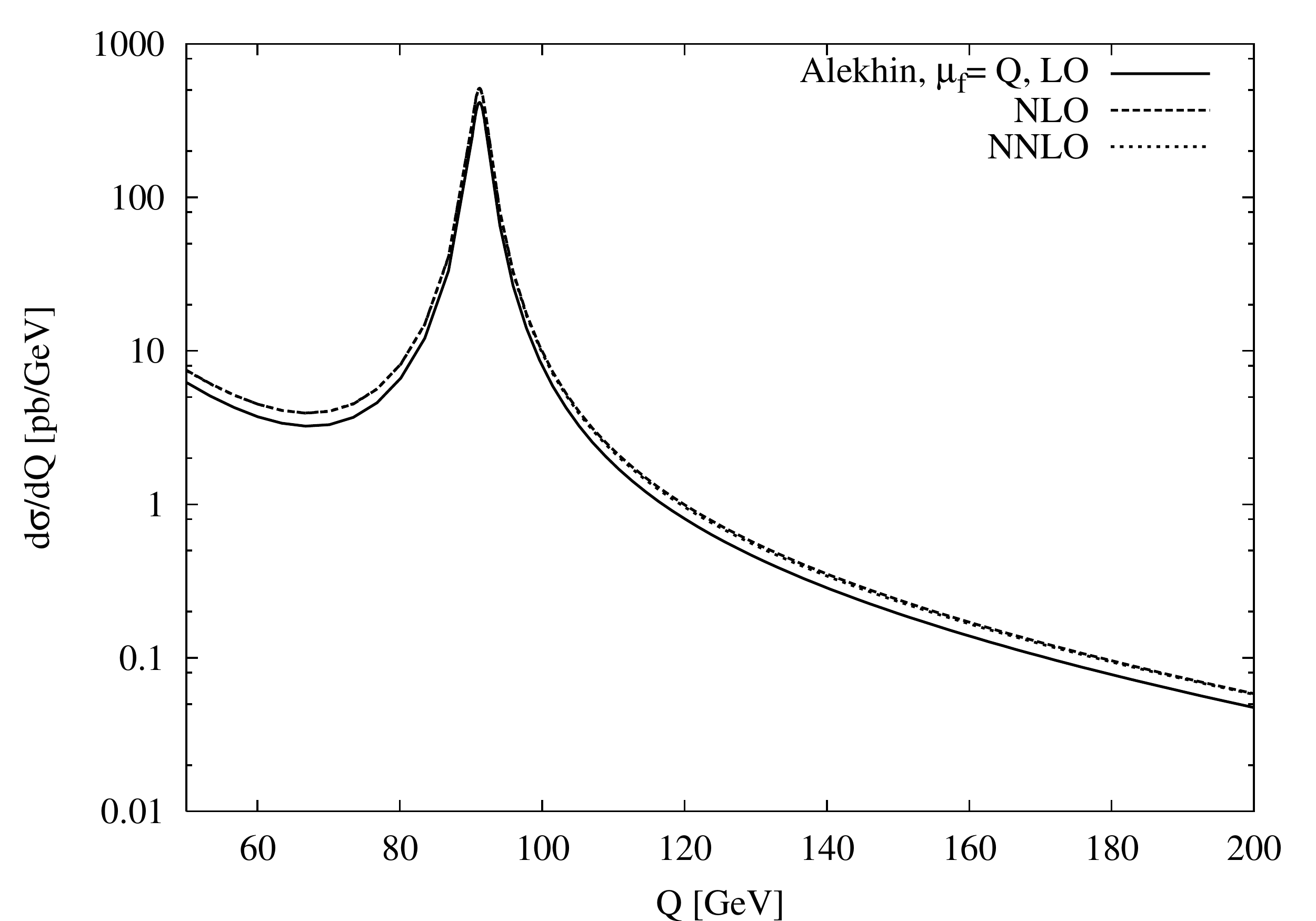}}
\subfigure[MRST Evolution]{%
\includegraphics[width=7cm,angle=-90]{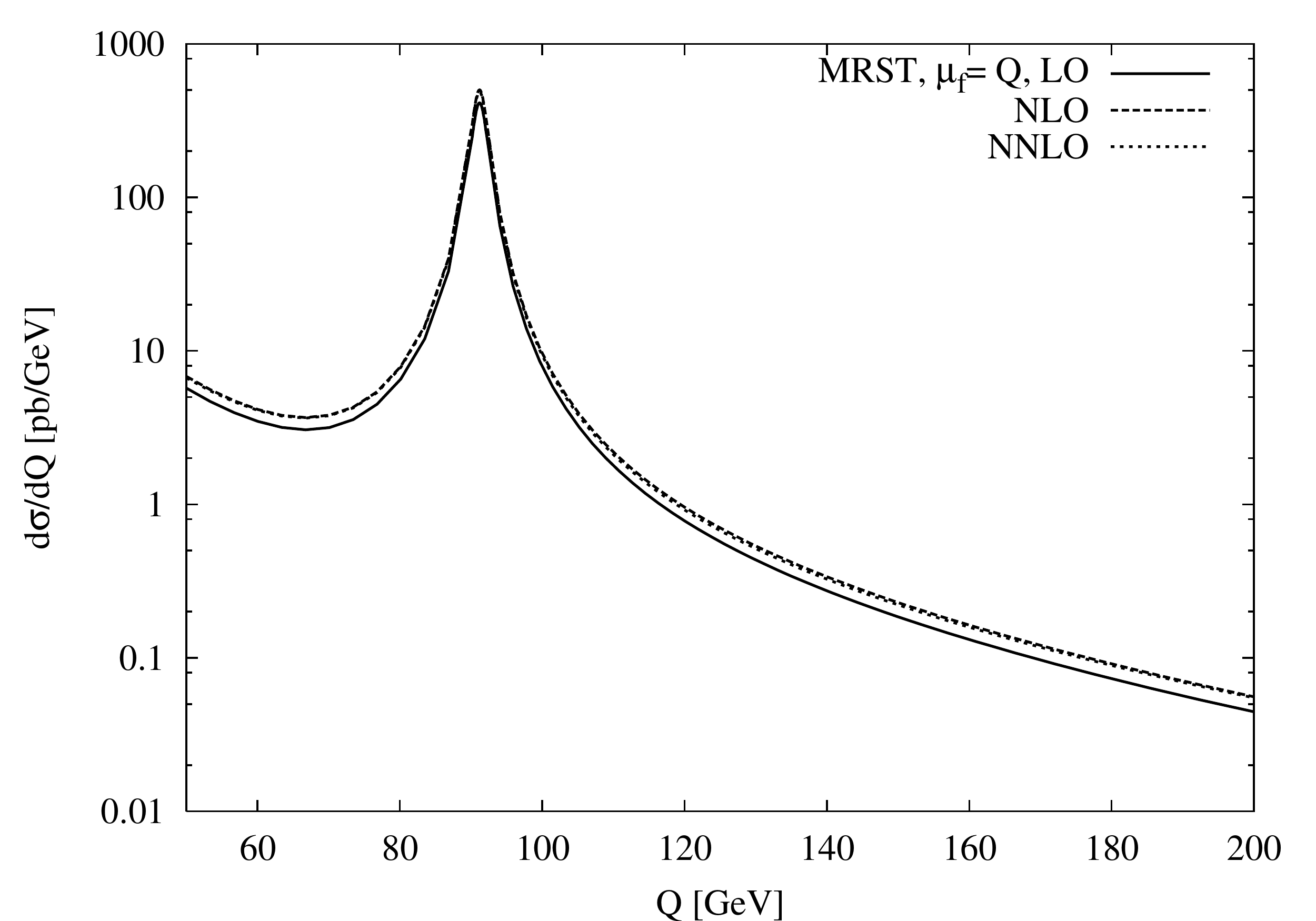}}
\caption{Cross Sections in the region of the peak of the Z
boson at LO, NLO, and NNLO obtained using the
luminosities evolved respectively by Alekhin and MRST}
\label{Cross1}
\end{figure}

\section{Benchmarks and Numerical Results}
Now, we come to illustrate a comparison between the NNLO evolution performed by 
\textsc{Candia} and \textsc{Pegasus} \cite{Pegasus}, using the initial conditions of the Les Houches model. This model works as a benchmark and allows to compare various evolution codes. 
We show the behaviour of the gluon distribution and the valence up quark distribution (non-singlet).
We observe that the differences in the singlet case are around $0.4-0.3\%$ or less, 
while for the valence up quark they are around $3\%$ at $x=10^{-5}$ and then decrease to $0.1-0.02\%$ at $x=0.1$ (see Table \ref{comp5}). 
We have denoted with $x\delta f(x)$ the relative differences normalized to the \textsc{Pegasus} determination, i.e. 
$x\delta f(x)\equiv (x f(x)_{\textsc{Pegasus}} - x f(x)_{\textsc{Candia}})/x f(x)_{\textsc{Pegasus}}$. 
We have set the factorization and renormalization scales to coincide and equal to $Q=100$ GeV. In 
\textsc{Candia} we have used the ``asymptotic solutions'', 
which are similar to those obtained by PEGASUS in one of its 3 running 
modes. The two algorithms and their implementations compare very well 
in the x-region relevant for the LHC (around $0.1 \%$ at NNLO).
Using the same benchmark we have calculated the NNLO cross sections using the two evolutions (see Tab. \ref{comp6}), and we
observe that the differences between the two methods in the kinematical region that we are considering 
are around $0.2-0.3\%$.
 
\begin{table}
\begin{footnotesize}
\begin{center}
\begin{tabular}{|c||c|c|c|c|c|c|}
\hline
\multicolumn{7}{|c|}{\textsc{Candia} vs \textsc{Pegasus} PDFs at NNLO, Les Houches input, VFN scheme, $Q=\mu_{F}=\mu_{R}=100$ GeV}
\tabularnewline
\hline
$ x $&
$xg(x)^{\textsc{Candia}}_{asymp}$     &
$xg(x)^{\textsc{Pegasus}}$&
$\delta xg(x) $         &
$xu_v(x)^{\textsc{Candia}}_{asymp}$     &
$xu_v(x)^{\textsc{Pegasus}}$&
$\delta xu_v(x) $\tabularnewline
\hline
\hline
$1e-05$&
$2.1922\cdot10^{+2}$&
$2.2012\cdot10^{+2}$&
$4.1108\cdot10^{-3}$&
$3.0823\cdot10^{-3}$&
$3.1907\cdot10^{-3}$&
$3.3962\cdot10^{-2}$
\tabularnewline
\hline
$0.0001$&
$8.8486\cdot10^{+1}$&
$8.8804\cdot10^{+1}$&
$3.5856\cdot10^{-3}$&
$1.3871\cdot10^{-2}$&
$1.4023\cdot10^{-2}$&
$1.0811\cdot10^{-2}$
\tabularnewline
\hline
$0.001$&
$3.0319\cdot10^{+1}$&
$3.0404\cdot10^{+1}$&
$2.8106\cdot10^{-3}$&
$6.0060\cdot10^{-2}$&
$6.0019\cdot10^{-2}$&
$6.9117\cdot10^{-4}$
\tabularnewline
\hline
$0.01$&
$7.7785\cdot10^{+0}$&
$7.7912\cdot10^{+0}$&
$1.6326\cdot10^{-3}$&
$2.3287\cdot10^{-1}$&
$2.3244\cdot10^{-1}$&
$1.8584\cdot10^{-3}$
\tabularnewline
\hline
$0.1$&
$8.5284\cdot10^{-1}$&
$8.5266\cdot10^{-1}$&
$2.1595\cdot10^{-4}$&
$5.4977\cdot10^{-1}$&
$5.4993\cdot10^{-1}$&
$2.9526\cdot10^{-4}$
\tabularnewline
\hline
$0.2$&
$2.4183\cdot10^{-1}$&
$2.4161\cdot10^{-1}$&
$9.1195\cdot10^{-4}$&
$4.8313\cdot10^{-1}$&
$4.8323\cdot10^{-1}$&
$2.0148\cdot10^{-4}$
\tabularnewline
\hline
$0.3$&
$7.9005\cdot10^{-2}$&
$7.8898\cdot10^{-2}$&
$1.3515\cdot10^{-3}$&
$3.4629\cdot10^{-1}$&
$3.4622\cdot10^{-1}$&
$1.9857\cdot10^{-4}$
\tabularnewline
\hline
$0.4$&
$2.5636\cdot10^{-2}$&
$2.5594\cdot10^{-2}$&
$1.6452\cdot10^{-3}$&
$2.1711\cdot10^{-1}$&
$2.1696\cdot10^{-1}$&
$6.7488\cdot10^{-4}$
\tabularnewline
\hline
$0.5$&
$7.6538\cdot10^{-3}$&
$7.6398\cdot10^{-3}$&
$1.8314\cdot10^{-3}$&
$1.1883\cdot10^{-1}$&
$1.1868\cdot10^{-1}$&
$1.2434\cdot10^{-3}$
\tabularnewline
\hline
$0.6$&
$1.9439\cdot10^{-3}$&
$1.9401\cdot10^{-3}$&
$1.9844\cdot10^{-3}$&
$5.4753\cdot10^{-2}$&
$5.4652\cdot10^{-2}$&
$1.8520\cdot10^{-3}$
\tabularnewline
\hline
$0.7$&
$3.7162\cdot10^{-4}$&
$3.7080\cdot10^{-4}$&
$2.2059\cdot10^{-3}$&
$1.9537\cdot10^{-2}$&
$1.9486\cdot10^{-2}$&
$2.6105\cdot10^{-3}$
\tabularnewline
\hline
$0.8$&
$4.1248\cdot10^{-5}$&
$4.1141\cdot10^{-5}$&
$2.5990\cdot10^{-3}$&
$4.4306\cdot10^{-3}$&
$4.4148\cdot10^{-3}$&
$3.5750\cdot10^{-3}$
\tabularnewline
\hline
$0.9$&
$1.1766\cdot10^{-6}$&
$1.1722\cdot10^{-6}$&
$3.7723\cdot10^{-3}$&
$3.3696\cdot10^{-4}$&
$3.3522\cdot10^{-4}$&
$5.1816\cdot10^{-3}$
\tabularnewline
\hline
\end{tabular}
\end{center}
\caption{NNLO pdf's determined with \textsc{Candia} and \textsc{Pegasus} using the Les Houches model.}
\label{comp5}
\end{footnotesize}
\end{table}

\begin{table}
\begin{center}
\begin{tabular}{|c||c|c|c|}
\hline
\multicolumn{4}{|c|}{$\textrm{d}\sigma_{NNLO}/\textrm{d}Q$ [pb/GeV]. \textsc{Candia} vs \textsc{Pegasus} with Les Houches input.}
\tabularnewline
\hline
$Q ~[\textrm{GeV}]$&
$\sigma_{NNLO}^{\textsc{Candia}}$&
$\sigma_{NNLO}^{\textsc{Pegasus}}$  &
$\delta\sigma_{NNLO}  $ \tabularnewline
\hline
\hline
$50.0000$&
$8.0734\cdot10^{+0}$&
$8.1044\cdot10^{+0}$&
$3.8288\cdot10^{-3}$
\tabularnewline
\hline
$60.0469$&
$4.8771\cdot10^{+0}$&
$4.8948\cdot10^{+0}$&
$3.6106\cdot10^{-3}$
\tabularnewline
\hline
$70.0938$&
$4.4033\cdot10^{+0}$&
$4.4184\cdot10^{+0}$&
$3.4110\cdot10^{-3}$
\tabularnewline
\hline
$80.1407$&
$8.9241\cdot10^{+0}$&
$8.9527\cdot10^{+0}$&
$3.1936\cdot10^{-3}$
\tabularnewline
\hline
$90.1876$&
$3.3570\cdot10^{+2}$&
$3.3669\cdot10^{+2}$&
$2.9388\cdot10^{-3}$
\tabularnewline
\hline
$91.1876$&
$5.4905\cdot10^{+2}$&
$5.5067\cdot10^{+2}$&
$2.9299\cdot10^{-3}$
\tabularnewline
\hline
$92.1876$&
$3.3344\cdot10^{+2}$&
$3.3441\cdot10^{+2}$&
$2.8919\cdot10^{-3}$
\tabularnewline
\hline
$120.0701$&
$1.0249\cdot10^{+0}$&
$1.0274\cdot10^{+0}$&
$2.4285\cdot10^{-3}$
\tabularnewline
\hline
$146.0938$&
$2.8527\cdot10^{-1}$&
$2.8590\cdot10^{-1}$&
$2.1826\cdot10^{-3}$
\tabularnewline
\hline
$172.1175$&
$1.2295\cdot10^{-1}$&
$1.2319\cdot10^{-1}$&
$1.9887\cdot10^{-3}$
\tabularnewline
\hline
$200.0000$&
$6.0923\cdot10^{-2}$&
$6.1029\cdot10^{-2}$&
$1.7369\cdot10^{-3}$
\tabularnewline
\hline
\end{tabular}
\caption{NNLO cross sections in the two evolution methods.}
\label{comp6}
\end{center}
\end{table}

\begin{figure}
\subfigure[$K=\sigma_{NNLO}/\sigma_{NLO}$]{%
\includegraphics[width=5cm,angle=-90]{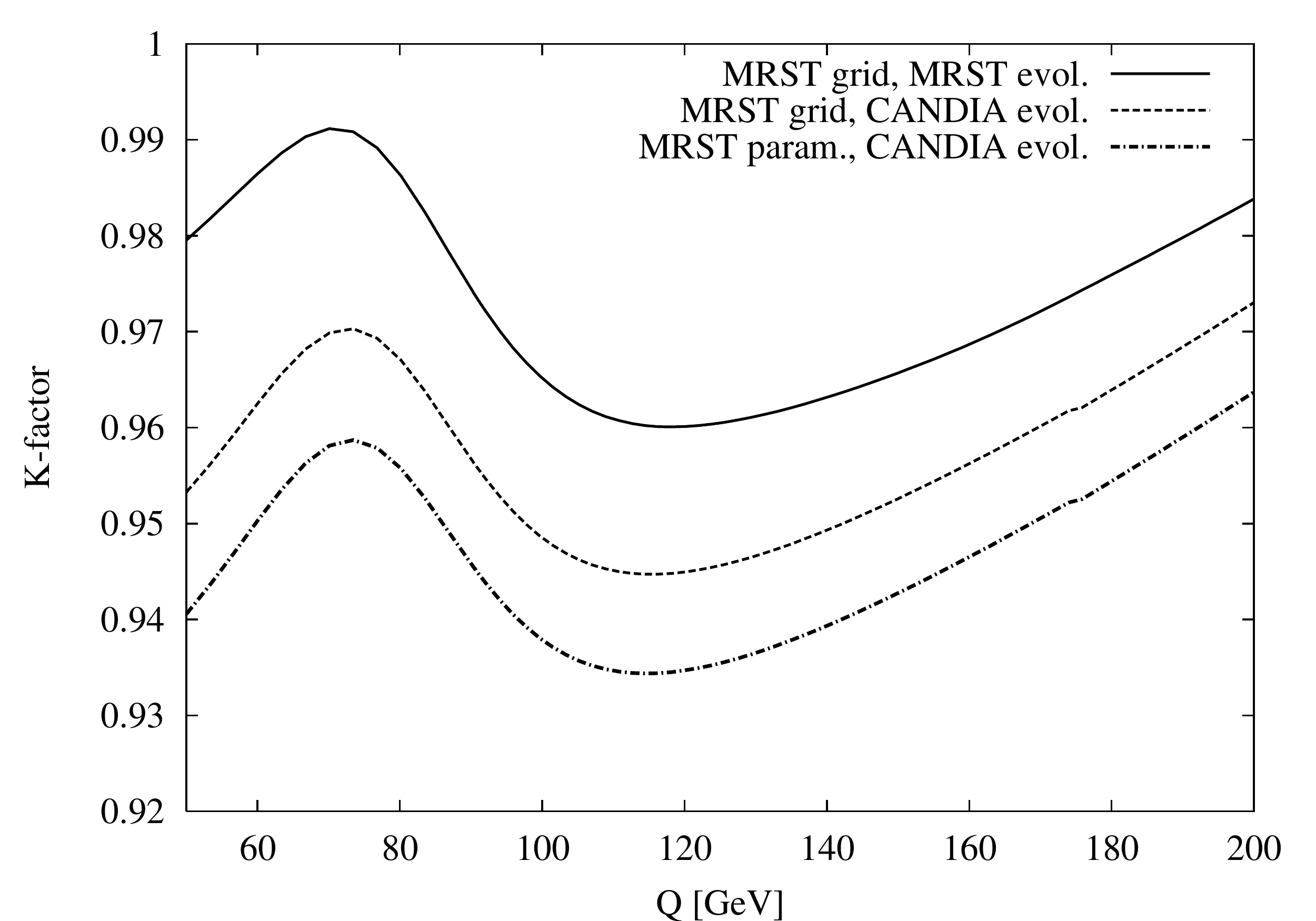}}
\subfigure[$K_1=\sigma_{NLO}/\sigma_{LO}$  and $K_2=\sigma_{NNLO}/\sigma_{LO}$]{%
\includegraphics[width=5cm,angle=-90]{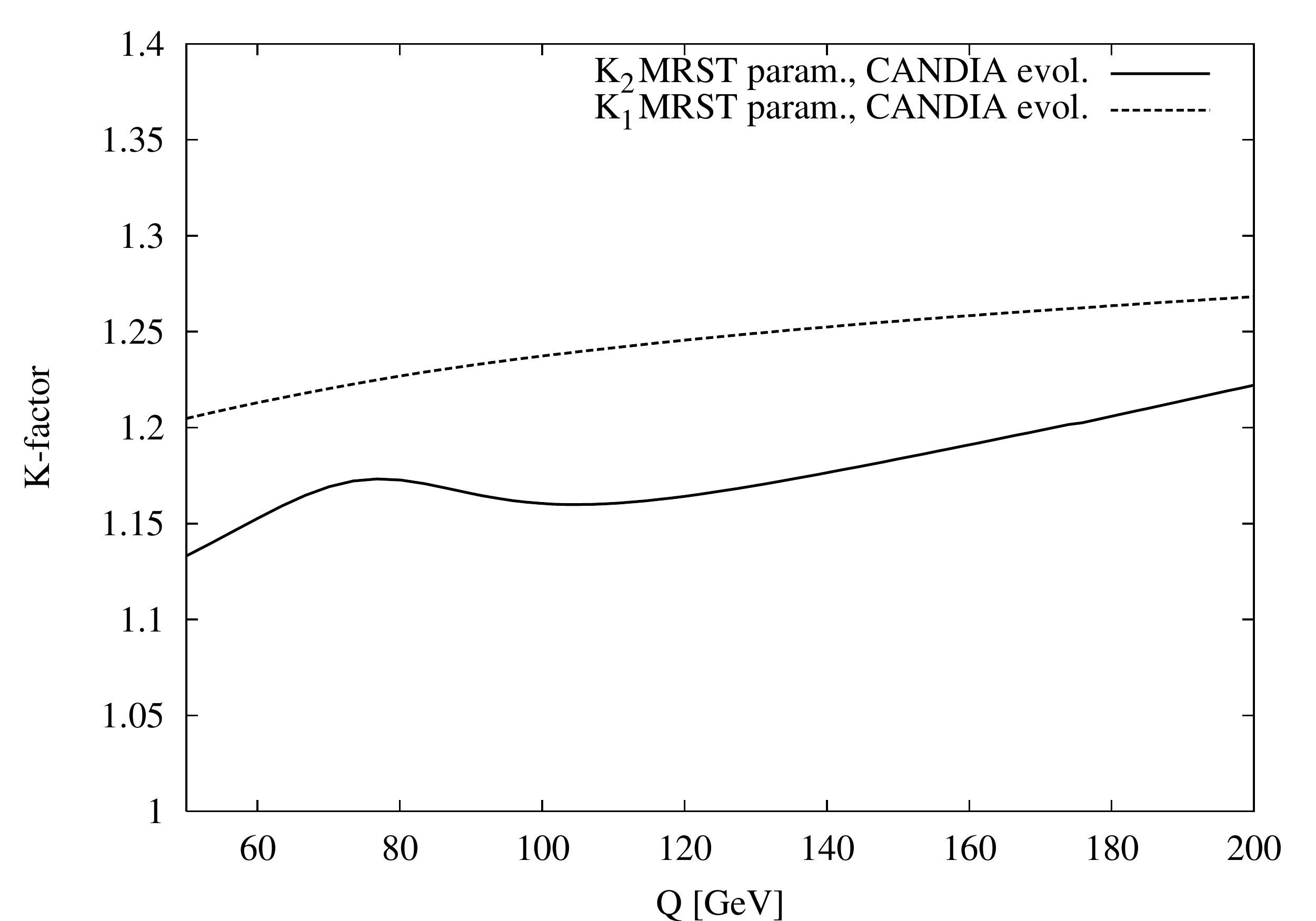}}
\subfigure[$K_1=\sigma_{NLO}/\sigma_{LO}$  and $K_2=\sigma_{NNLO}/\sigma_{LO}$]{%
\includegraphics[width=5cm,angle=-90]{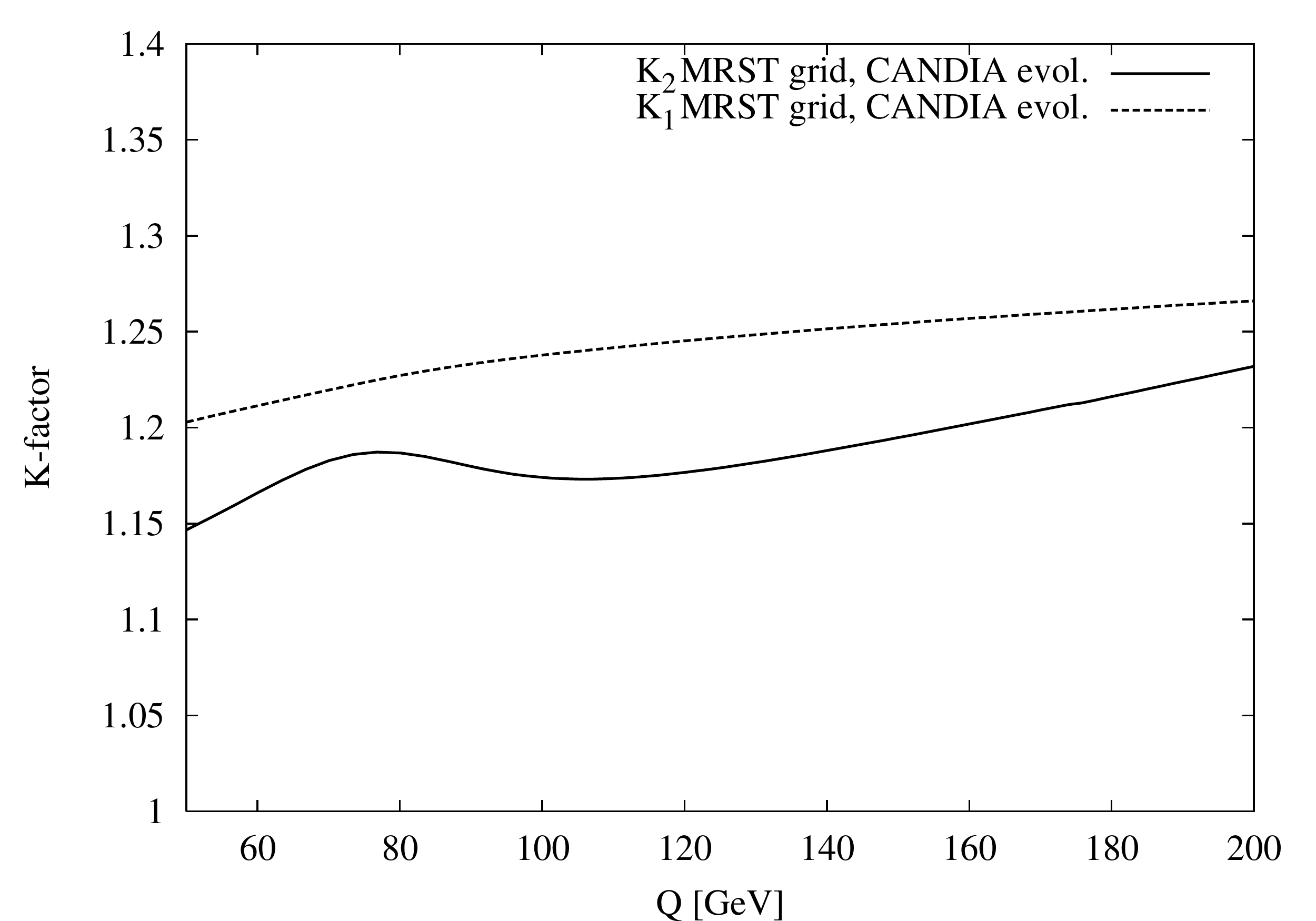}}
\subfigure[$K_1=\sigma_{NLO}/\sigma_{LO}$ and $K_2=\sigma_{NNLO}/\sigma_{LO}$]{%
\includegraphics[width=5cm,angle=-90]{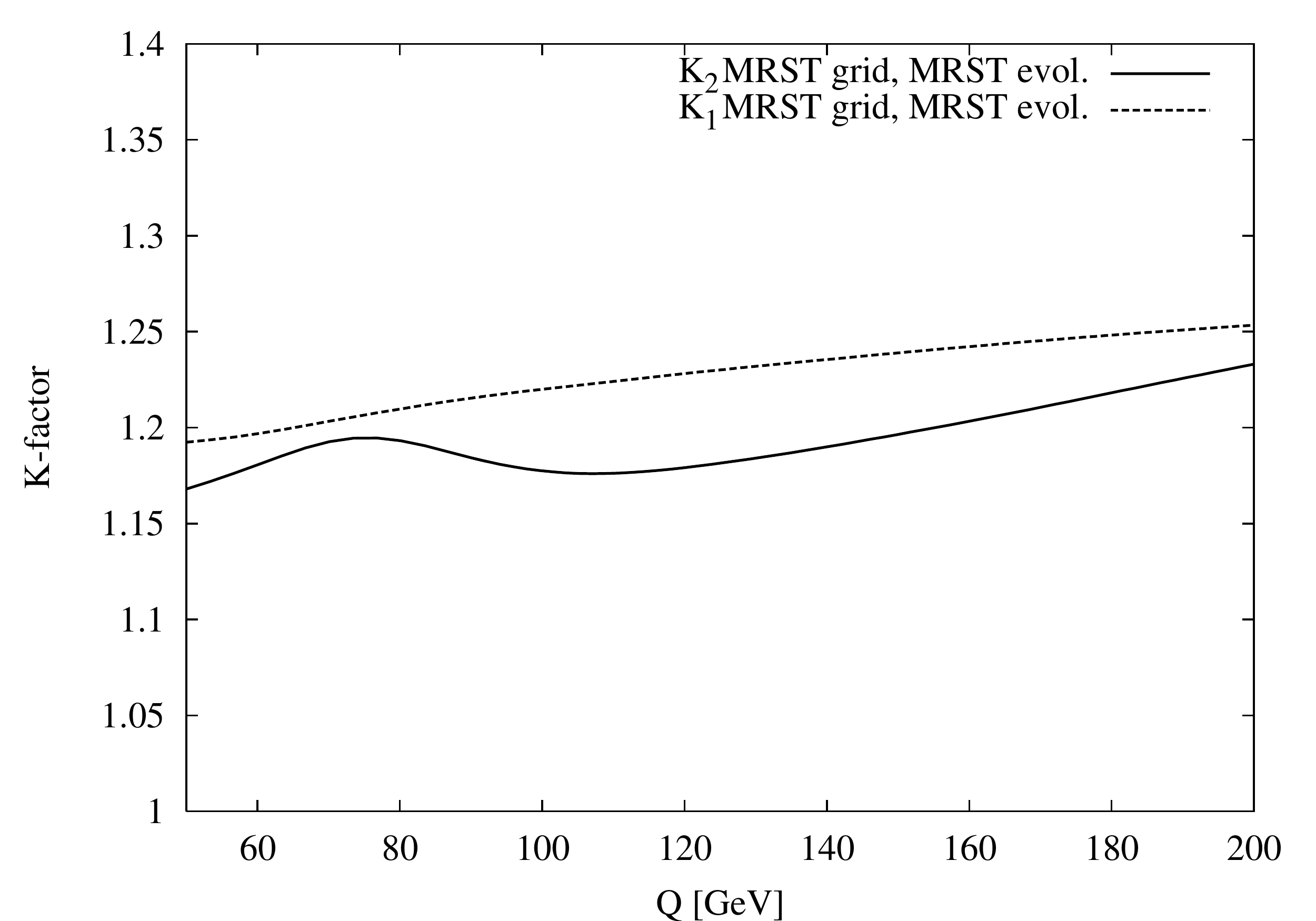}}
\caption{Various $K$-factors obtained with the evolution performed by \textsc{Candia} and MRST.}
\label{Kfact}
\end{figure}

Coming to a description of the NNLO cross section obtained from a 
realistic model, 
we show in Tab. \ref{alek1} results for the invariant mass distributions 
at $\sqrt{S}=14$ TeV according to Alekhin's model \cite{Alekhin}. Shown are also the 
errors which have been computed as discussed in \cite{CCG2}. A similar 
analysis has been performed using the MRST set \cite{MRST1} 
(see Tab. \ref{mmrst1}), 
for which the errors 
can be obtained, at this time, only at NLO. 

The differences between our prediction and the MRST result for the total cross sections are 
around 1 per cent or below at LO, vary from  $0.02 \%$ to $0.3 \%$ 
at NLO and are $2.6 \%$ and below at NNLO (see Tab. \ref{comp9}). In this case the maximum 
difference has been found for $Q=50$ GeV. These differences, clearly, affect the values
of the $K$-factors, as we are going to discuss below, which in our evolution are
larger compared to those of MRST.

\begin{table}
\begin{center}
\begin{tabular}{|c||c|c|c|c|}
\hline
\multicolumn{4}{|c|}{$\textrm{d}\sigma/\textrm{d}Q$ in [pb/GeV] for Alekhin with $Q^2=\mu_{F}^2=\mu_{R}^2$, $\sqrt{S}=14$ TeV}
\tabularnewline
\hline
$Q ~[\textrm{GeV}]$&
$\sigma_{LO}$&
$\sigma_{NLO}$&
$\sigma_{NNLO}$\tabularnewline
\hline
\hline
$        50$&
$         6.22$ $\pm$ $         0.27$&
$         7.48$ $\pm$ $         0.24$&
$         7.43$ $\pm$ $         0.21$
\tabularnewline
\hline
$        60.04$&
$         3.72$ $\pm$ $         0.15$&
$         4.50$ $\pm$ $         0.13$&
$         4.49$ $\pm$ $         0.12$
\tabularnewline
\hline
$        70.1$&
$         3.30$ $\pm$ $         0.12$&
$         4.03$ $\pm$ $         0.11$&
$         4.05$ $\pm$ $         0.10$
\tabularnewline
\hline
$        80.1$&
$         6.65$ $\pm$ $         0.24$&
$         8.20$ $\pm$ $         0.24$&
$         8.19$ $\pm$ $         0.23$
\tabularnewline
\hline
$        90.19$&
$       253$ $\pm$ $         8$&
$       313$ $\pm$ $         9$&
$       309$ $\pm$ $         8$
\tabularnewline
\hline
$        91.19$&
$       415$ $\pm$ $        14$&
$       514$ $\pm$ $        15$&
$       506$ $\pm$ $        15$
\tabularnewline
\hline
$       120.07$&
$         0.80$ $\pm$ $         0.02$&
$         0.99$ $\pm$ $         0.02$&
$         0.96$ $\pm$ $         0.03$
\tabularnewline
\hline
$       146.1$&
$         0.225$ $\pm$ $         0.006$&
$         0.277$ $\pm$ $         0.007$&
$         0.269$ $\pm$ $         0.007$
\tabularnewline
\hline
$       172.1$&
$         0.097$ $\pm$ $         0.002$&
$         0.119$ $\pm$ $         0.003$&
$         0.117$ $\pm$ $         0.003$
\tabularnewline
\hline
$        200$&
$         0.047$ $\pm$ $         0.001$&
$         0.058$ $\pm$ $         0.001$&
$         0.058$ $\pm$ $         0.001$
\tabularnewline
\hline
\end{tabular}
\caption{Cross sections derived from the best fits for the 3 orders with their errors for the
set by Alekhin.}
\label{alek1}
\end{center}
\end{table}
%%%%%%%%%%%%%%%%%%%%%%%%%%%%%%%%%%%%%%%%%%%%

\begin{table}
\begin{center}
\begin{tabular}{|c||c|c|c|}
\hline
\multicolumn{4}{|c|}{$\textrm{d}\sigma_{NNLO}/\textrm{d}Q$ [pb/GeV]. \textsc{Candia} vs MRST evolution with MRST input, $\mu_0^2=1.25$ GeV$^2$}
\tabularnewline
\hline
$Q ~[\textrm{GeV}]$&
$\sigma_{NNLO}^{\textsc{Candia}}$&
$\sigma_{NNLO}^{MRST}$  &
$\delta\sigma_{NNLO}  $ \tabularnewline
\hline
\hline
$50.0000$&
$6.4935\cdot10^{+0}$&
$6.6707\cdot10^{+0}$&
$2.6560\cdot10^{-2}$
\tabularnewline
\hline
$60.0469$&
$3.9997\cdot10^{+0}$&
$4.0961\cdot10^{+0}$&
$2.3534\cdot10^{-2}$
\tabularnewline
\hline
$70.0938$&
$3.6962\cdot10^{+0}$&
$3.7743\cdot10^{+0}$&
$2.0678\cdot10^{-2}$
\tabularnewline
\hline
$80.1407$&
$7.6755\cdot10^{+0}$&
$7.8198\cdot10^{+0}$&
$1.8455\cdot10^{-2}$
\tabularnewline
\hline
$90.1876$&
$2.9325\cdot10^{+2}$&
$2.9827\cdot10^{+2}$&
$1.6834\cdot10^{-2}$
\tabularnewline
\hline
$91.1876$&
$4.8006\cdot10^{+2}$&
$4.8822\cdot10^{+2}$&
$1.6702\cdot10^{-2}$
\tabularnewline
\hline
$92.1876$&
$2.9179\cdot10^{+2}$&
$2.9671\cdot10^{+2}$&
$1.6575\cdot10^{-2}$
\tabularnewline
\hline
$120.0701$&
$9.0411\cdot10^{-1}$&
$9.1687\cdot10^{-1}$&
$1.3918\cdot10^{-2}$
\tabularnewline
\hline
$146.0938$&
$2.5267\cdot10^{-1}$&
$2.5567\cdot10^{-1}$&
$1.1714\cdot10^{-2}$
\tabularnewline
\hline
$172.1175$&
$1.0938\cdot10^{-1}$&
$1.1049\cdot10^{-1}$&
$1.0028\cdot10^{-2}$
\tabularnewline
\hline
$200.0000$&
$5.4431\cdot10^{-2}$&
$5.4876\cdot10^{-2}$&
$8.1092\cdot10^{-3}$
\tabularnewline
\hline
\end{tabular}
\caption{NNLO cross section for Drell-Yan obtained by \textsc{Candia} using the MRST input and the evolved MRST pdf's}
\label{comp9}
\end{center}
\end{table}
%%%%%%%%%%%%%%%%%%%%%%%%%%%%%%%%%

%%%%%%%%%%%%%%%%%%%%%%%%%%%%%%%%%%%%%%%%%%%%
\begin{table}
\begin{center}
\begin{tabular}{|c||c|c|}
\hline
\multicolumn{2}{|c|}{$\textrm{d}\sigma/\textrm{d}Q$ in [pb/GeV] for MRST with $Q^2=\mu_{F}^2=\mu_{R}^2$, $\sqrt{S}=14$ TeV}
\tabularnewline
\hline
$Q ~[\textrm{GeV}]$&
$\sigma_{NLO}$\tabularnewline
\hline
\hline
$        50$&
$         6.77$ $\pm$ $         0.19$
\tabularnewline
\hline
$        60.04$&
$         4.13$ $\pm$ $         0.10$
\tabularnewline
\hline
$        70.1$&
$         3.79$ $\pm$ $         0.08$
\tabularnewline
\hline
$        80.1$&
$         7.90$ $\pm$ $         0.14$
\tabularnewline
\hline
$        90.19$&
$       305$ $\pm$ $         5$
\tabularnewline
\hline
$        91.19$&
$       499$ $\pm$ $         8$
\tabularnewline
\hline
$       120.1$&
$         0.952$ $\pm$ $         0.014$
\tabularnewline
\hline
$       146.1$&
$         0.264$ $\pm$ $         0.003$
\tabularnewline
\hline
$       172.1$&
$         0.113$ $\pm$ $         0.001$
\tabularnewline
\hline
$       200$&
$         0.0556$ $\pm$ $         0.0007$
\tabularnewline
\hline
\end{tabular}
\caption{Cross sections derived from the best fits at NLO with the errors for the
MRST set.}
\label{mmrst1}
\end{center}
\end{table}

We have summarized in Fig.~\ref{Kfact} four plots of the behavior of the 3 $K$-factors
$K=\sigma_{NNLO}/\sigma_{NLO}$, $K_1=\sigma_{NLO}/\sigma_{LO}$ and $K_2=\sigma_{NNLO}/\sigma_{LO}$
obtained using \textsc{Candia} and the MRST evolution.
These are shown as a function of $Q$, and evaluated at the center of mass energy of $\sqrt{S}=14$
TeV. The dependence of the results on
the evolution is significant. In fact, from Fig.~\ref{Kfact} it is evident that while the shapes of the
plots of the $K$-factors are similar, there are
variations of the order $2\%$, in the results using the two different evolutions.
Both in the evolution performed with \textsc{Candia} and in the MRST evolution we use the same
MRST input, choosing the initial scale $\mu_0^2=1.25$ GeV$^2$, and the same treatment of the heavy flavors. On the Z resonance we get
\ba
&&K(M_Z)=(\hat{\sigma}_{NNLO}\otimes\Phi^{NNLO}_{MRST})/(\hat{\sigma}_{NLO}\otimes\Phi^{NLO}_{MRST})=0.97\nonumber\\
&&K(M_Z)=(\hat{\sigma}_{NNLO}\otimes\Phi^{NNLO}_{\textsc{Candia}})/(\hat{\sigma}_{NLO}\otimes\Phi^{NLO}_{\textsc{Candia}})=0.95 \nonumber \\
&& K(M_Z)=(\hat{\sigma}_{NNLO}\otimes\Phi^{NNLO}_{Alekhin})/(\hat{\sigma}_{NLO}\otimes\Phi^{NLO}_{Alekhin})=0.98
\ea
which corresponds to a reduction by $2.7 \%$ of the NNLO cross section compared to the NLO result, 
(MRST evolution) and larger for the \textsc{Candia} evolution, 
$4.4 \% $, while for Alekhin is $1.5 \%$. From the analysis of the errors on the pdf's to NNLO, for instance for the Alekhin's set, the differences among these determinations are still 
compatible, being the variations on the $K$-factors of the order of $4 \%$. 
\section{The rapidity distributions}
In this case the QCD cross section is given by
\begin{eqnarray}
&&\frac{d\sigma^{Z}}{dY}=\sum_{ab}\int_{\sqrt{\tau}e^{Y}}^{1}
\int_{\sqrt{\tau}e^{-Y}}^{1}dx_1 dx_2
f_a^{h_1}(x_1,Q^2/\mu_F^2,\mu_R^2/\mu_F^2)f_b^{h_2}(x_2,Q^2/\mu_F^2,\mu_R^2/\mu_F^2)
\times \nonumber\\
&&\frac{d\sigma^{Z}_{ab}}{dY}(x_1,x_2,Q^2/\mu_F^2,\mu_R^2/\mu_F^2).
\label{rapid}
\end{eqnarray}
Notice that the evolution implemented in \textsc{Candia} allows to analyze the
renormalization/factorization scale dependence also in the evolution, 
which is not present in the MRST parameterizations. We have made explicit this dependences in (\ref{rapid}).

% \begin{figure}
% \subfigure[Pdf's Error bands in the Alekhin model with $Q=\mu_F=\mu_R$ and $\sqrt{S}=14$ TeV]{\includegraphics[%
%   width=8.5cm,
%   angle=-90]{Zerror_A_11196e+6_100_102_colour}}
% \subfigure[MRST error bands with $Q=\mu_F=\mu_R$ and $\sqrt{S}=14$ TeV]{\includegraphics[%
%   width=8.5cm,
%   angle=-90]{Zerror_M_11196e+6_100_102_colour}}
% \caption{Errors of the pdf's on the cross sections at LHC. Zoom in the region of $100$ GeV.}
% \label{error1}
% \end{figure}
% 
% 
% 
\begin{figure}
\subfigure[Alekhin's model]{%
\includegraphics[width=8cm,angle=-90]{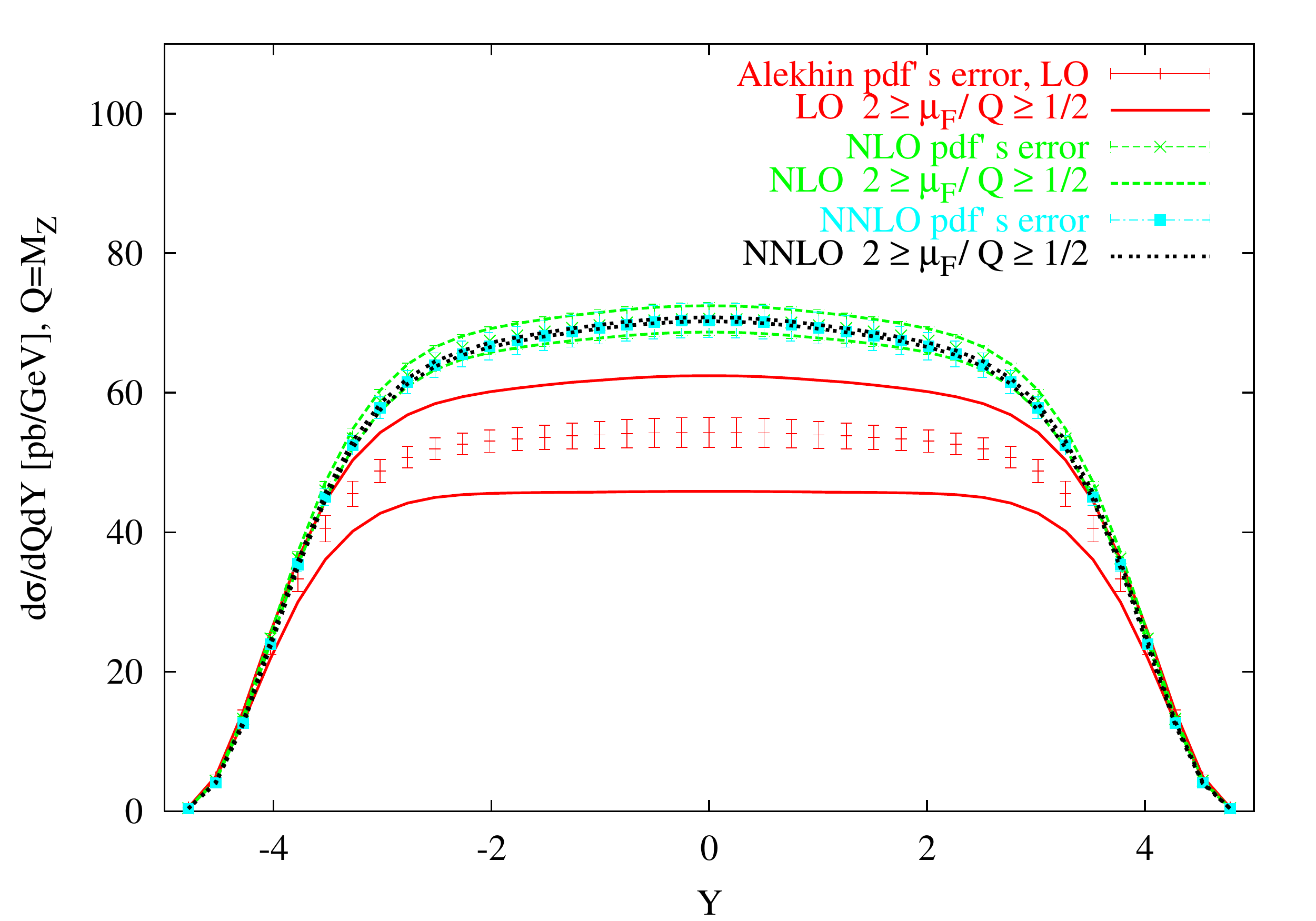}}
\subfigure[MRST model]{%
\includegraphics[width=8cm,angle=-90]{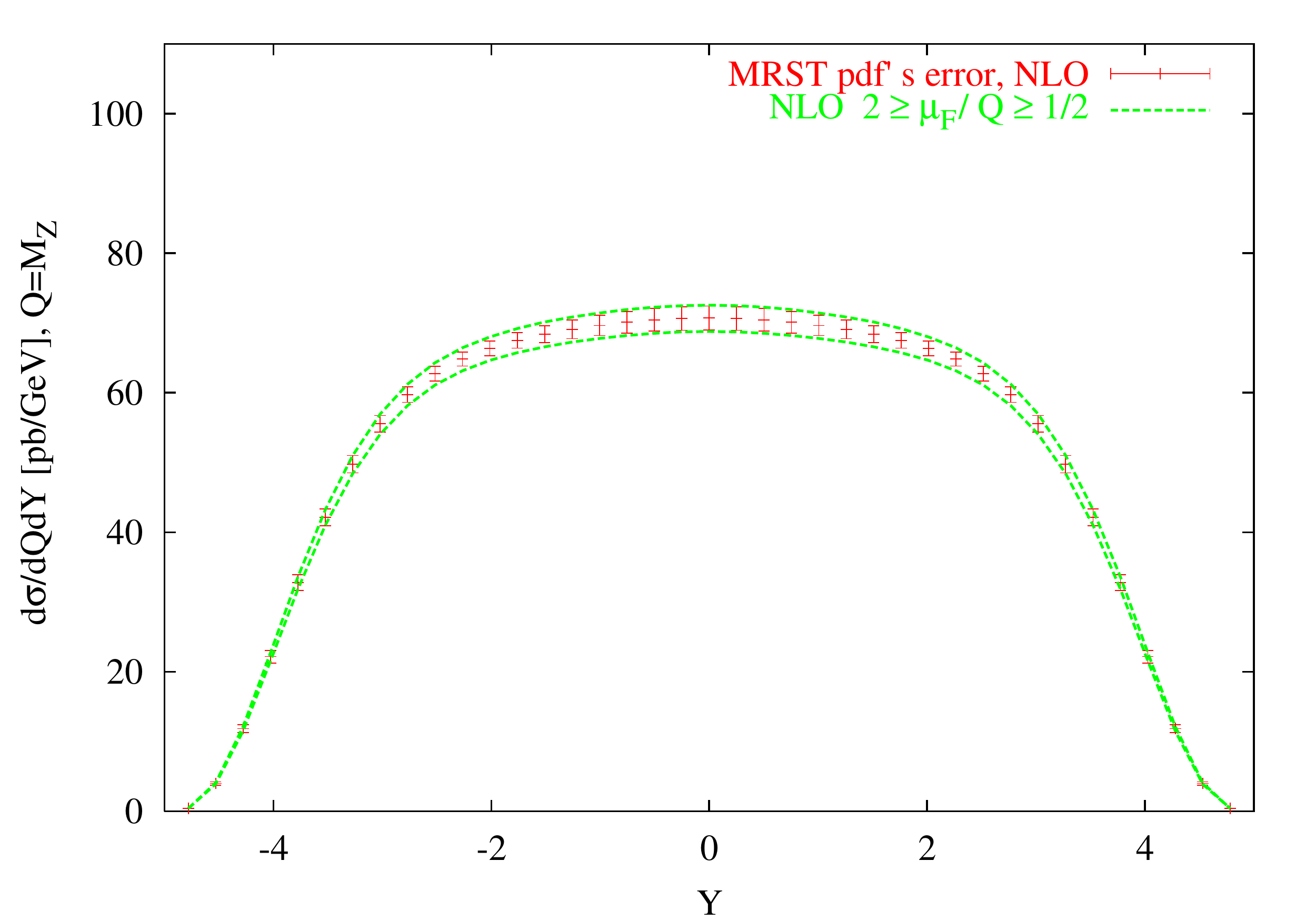}}
\caption{Plot of the rapidity distributions at LO, NLO and NNLO for Alekhin's model and MRST.
Shown are also the bands due to the variation of the $\mu_F$ scale,
and the errors on the cross sections at the corresponding orders.}
\label{Rapex}
\end{figure}
We have set the scales to be equal, $\mu_F=\mu_R$ and varied $\mu_F$ in the interval $1/2 Q \leq\mu_F\leq 2 Q$, obtaining results which differ from
those obtained in \cite{Anastasiou} by $2\%$ due to the different implementation of the 
evolution. Using \textsc{Candia} and as initial condition the 
MRST grid input with $\mu_0^2=1.25$ GeV$^2$ the NNLO band and the NLO one are resolved
separately. We have also found that with the inclusion of the $\mu_R^2/\mu_F^2$ effects in the pdf's evolution, the dependence on $\mu_R$ is quite sizeable at NLO, but is reduced at NNLO \cite{CCG2}. We show in Fig.~(\ref{Rapex}) the plots of the
variations of the rapidity distributions at the three orders and the corresponding errors on the pdf's for Alekhin's model and for MRST for $Q=M_Z$. In both cases the reduction of the variation of the cross sections
as we move toward higher orders is quite evident. We report also the errors
on these distributions obtained in both models, which get systematically smaller as the accuracy of the calculation increases.

\section{Conclusions}
In the search for extra neutral currents precise theoretical determinations of lepton pair 
production via the Drell-Yan mechanism are going to play a very relevant role 
(see Ref. \cite{pap2,pap3,pap4,pap5,pap7,pap8}).

In our determination, the change
in the value of the cross section from NLO to NNLO is around $4 \%$ on the Z 
peak, while the MRST and the Alekhin determinations are $2.6 \%$ 
and about $1.5 \%$ respectively. While these variations appear to be more modest compared to 
the analogous ones at a lower order (which are of the order of $20 \%$ or so), they are nevertheless important 
for the discovery of extra neutral currents at large invariant mass of the lepton pair in DY, given the fast falling cross section at those large values. 
The errors on the pdf's induce percentile variations of the cross section 
as we move from NLO to NNLO of the order of $4 \%$ around
the best-fit result, reducing the NNLO cross section compared to the
NLO prediction and rendering these results compatible.

\addtocounter{chapter}{1}
%\documentclass[a4paper,12pt,twoside]{report}
%\usepackage{epsfig}
%\usepackage{amssymb}
%\usepackage{lineno}
%\usepackage{setspace}
%%%%%%%%%%%%%%%%%%%%%%%%%%%%%%%%%%%%%%%%%%%%%%%%%%%%%%%%%%
%\begin{document}
%%%%%%%%%%%%%%%%%%%%%%%%%%%%%%%%%%%%%%%%%%%%%%%
% Toggle line numbering
% Won't work with the PRD revtex4 !
%\pagewiselinenumbers
% uncomment if you want doublespace
%\doublespace
%%%%%%%%%%%%%%%%%%%%%%%%%%%%%%%%%%%%%%%%%%%%%%%
%\chapter{Introduction} {\it C. Mariotti, E. Migliore and P. Nason}
%%%%%%%%%%%%%%%%%%%%%%%%%%%%%%%%%%%%%%%%%%%%%%%%%%%%%%%%%

\mchapter{Combination of QCD and electroweak corrections to Drell-Yan processes} 
{G. Balossini, C.M. Carloni Calame, G. Montagna, M. Moretti, O. Nicrosini, F. Piccinini, 
M. Treccani and A. Vicini}

%%%%%%%%%%%%%%%%%%%%%%%%%%%%%%%%%%%%
\section*{Introduction}
%%%%%%%%%%%%%%%%%%%%%%%%%%%%%%%%%%%%
%%\subsubsection*{Physics motivations}

Precision measurements of electroweak (EW) gauge boson production and properties will be a
crucial goal of the physics program of proton-proton collisions at the LHC. $W$ and $Z$
bosons will be produced copiously and careful measurements of their observables will be
important in testing the Standard Model (SM) and uncovering signs of new physics 
\cite{UB,acfgi}.

Thanks to the high luminosity achievable at the LHC, the systematic errors will play a dominant 
role in determining the accuracy of the measurements, implying, in particular, that the theoretical predictions will have to be of the highest standard as possible. For Drell-Yan 
(D-Y) processes, this amounts
to make available calculations of $W$ and $Z$ production processes including simultaneously 
higher-order corrections coming from the EW and QCD sector of the SM. 
Actually, in spite of a detailed knowledge of EW and QCD corrections 
separately, the combination of their effects have been addressed only recently 
\cite{cy,ward2007,jadach2007} and need to be deeply scrutinized in view of the anticipated
experimental accuracy.

In this contribution, after a review of existing calculations and codes, we present the results of a study aiming at combining EW and QCD radiative corrections to D-Y processes consistently.
We do not include in our analysis uncertainties due to factorization/renormalization scale variations, 
as well as uncertainties in the Parton Distribution Functions arising from diverse experimental and
theoretical sources, which are left to a future publication. Some results 
already available in this direction can be 
found in \cite{pdf}.

\section*{Status of theoretical predictions and codes}

Concerning QCD calculations and tools, the present situation reveals quite a rich structure, 
that includes 
next-to-leading-order (NLO) and next-to-next-to-leading-order (NNLO) 
corrections to $W/Z$ total production rate \cite{AEM,HvNM}, 
NLO calculations for $W, Z + 1, 2 \, \, {\rm jets}$ 
signatures \cite{GGK,MCFM} 
(available in the codes DYRAD and MCFM), resummation of 
leading and next-to-leading logarithms due to soft gluon 
radiation \cite{BY,resbos} (implemented 
in the Monte Carlo ResBos),
NLO corrections merged with QCD Parton Shower (PS) 
evolution (in the event generators
MC@NLO~\cite{MC@NLO} and POWHEG~\cite{POWHEG}), 
 NNLO corrections to $W/Z$ production in fully differential 
form~\cite{mp,mp1} 
 (available in the Monte Carlo program FEWZ),  
 as well as leading-order multi-parton matrix elements 
generators matched with vetoed PS, such as, for instance, 
ALPGEN~\cite{Alpgen}, MADEVENT \cite{MadEvent}, HELAC \cite{Helac} and 
SHERPA \cite{Sherpa}.

As far as complete ${\cal O}(\alpha)$ EW corrections 
to D-Y processes
are concerned, they have  been computed independently by various 
authors in \cite{dk, bw,ZYK,SANC,CMNV} for $W$ production and 
in \cite{zgrad2, Zykunov2007,HORACEZ,SANCZ} for $Z$ production.
Electroweak tools implementing exact NLO corrections to $W$ 
production are DK~\cite{dk}, WGRAD2 \cite{bw}, SANC \cite{SANC} 
and HORACE 
\cite{CMNV}, while ZGRAD2~\cite{zgrad2}, HORACE~\cite{HORACEZ}  
and SANC~\cite{SANCZ} include the full set of 
${\cal O}(\alpha)$ EW
corrections to $Z$ production. The predictions of a subset of such 
calculations have been 
compared, at the level of same input parameters and cuts, 
in the proceedings 
of the Les Houches 2005~\cite{LH} and TEV4LHC~\cite{tev4lhc} workshops for $W$ production, 
finding a very satisfactory agreement between the various, 
independent calculations. 
A first set of tuned comparisons for the $Z$ production process has been recently performed and is available in~\cite{LH2007}. 

From the calculations above, it turns out that NLO EW 
corrections are dominated, 
in the resonant region, by final-state QED radiation containing 
large collinear logarithms
of the form  $\log(\hat{s}/m_l^2)$, where $\hat{s}$ is the squared 
partonic centre-of-mass (c.m.) energy
and $m_l$ is the lepton mass. Since these corrections amount to 
several per cents around the
jacobian peak of the $W$ transverse mass and lepton transverse momentum 
distributions and cause a
significant shift (of the order of 100-200~MeV) in the extraction of the 
$W$ mass $M_W$ at the Tevatron, 
the contribution of higher-order corrections due to multiple photon 
radiation from the 
final-state leptons must be taken into account in the theoretical 
predictions, in view of
the expected precision (at the level of 15-20 MeV) in the $M_W$ 
measurement at the LHC.  The contribution
 due to multiple photon radiation has been computed, by means of a QED PS
 approach, in \cite{CMNTW} for $W$ production and
 in \cite{CMNTZ} for $Z$ production, and implemented in the event generator HORACE. 
 Higher-order QED contributions to $W$ production have been calculated independently
 in \cite{winhac} using the YFS exponentiation, and are available in the generator
 WINHAC. They have been also computed in the collinear approximation, within the structure functions approach, in~\cite{DK2007}. 

A further important phenomenological feature of EW corrections is that, in the 
region important for new physics searches (i.e. where the $W$ transverse mass is much
larger than the $W$ mass or the invariant mass of the final state leptons is much larger
than the $Z$ mass), the NLO EW effects become large (of the order of 
20-30\%) and negative, due to the appearance of EW Sudakov logarithms 
$\propto - (\alpha/\pi) \log^2 ({\hat s}/M_V^2)$, $V = W,Z$ \cite{dk,bw,CMNV,zgrad2,Zykunov2007,HORACEZ}. 
Furthermore, in this region, weak boson emission processes 
(e.g. $pp\to e^+\nu_eV + X$), 
that contribute at the same order in perturbation theory, can partially cancel the large Sudakov 
corrections, when the weak boson $V$ decays into unobserved $\nu\bar{\nu}$ or
jet pairs, as recently shown in \cite{baurw}.

%%%%%%%%%%%%%%%%%%%%%%%%%%%%%%%%%%%%
\section*{Theoretical approach}
%%%%%%%%%%%%%%%%%%%%%%%%%%%%%%%%%%%%

A first strategy  for the combination of EW and QCD 
corrections consists in the 
following formula

\begin{eqnarray}
\left[\frac{d\sigma}{d\cal O}\right]_{{\rm QCD} \& {\rm EW}} = 
\left\{\frac{d\sigma}{d\cal O}\right\}_{{\rm MC@NLO}}
+\left\{\left[\frac{d\sigma}{d\cal O}\right]_{{\rm EW}} - 
\left[\frac{d\sigma}{d\cal O}\right]_{{\rm Born}} \right\}_{{\rm HERWIG\, \,  PS}}
\label{eq:qcd-ew}
\end{eqnarray}
where ${d\sigma/d\cal O}_{{\rm MC@NLO}}$ stands for the prediction of the 
observable ${d\sigma/d\cal O}$ 
as  obtained  by means of MC@NLO, 
${d\sigma/d\cal O}_{{\rm EW}}$ is the HORACE
prediction for the EW corrections to the ${d\sigma/d\cal O}$ observable, 
and ${d\sigma/d\cal O}_{{\rm Born}}$ is the lowest-order result 
for the observable of interest. The label {\rm HERWIG PS} in the second term in r.h.s. 
of eq. (\ref{eq:qcd-ew}) means that EW corrections are convoluted with QCD PS 
evolution through the HERWIG event generator, in order to (approximately) include
mixed ${\cal O}(\alpha \alpha_s)$ corrections and to obtain a more realistic 
description of the observables under study. However, it is worth noting that the convolution 
of NLO EW corrections with QCD PS implies that the contributions of the order of 
$\alpha \alpha_s$ are not reliable when hard non-collinear  QCD radiation 
turns out to be relevant, e.g. for the lepton and vector boson transverse momentum distributions 
in the absence of severe cuts able to exclude resonant $W/Z$ production. In this case, a full  ${\cal O}(\alpha \alpha_s)$ calculation would be 
needed for a sound evaluation of mixed EW and QCD corrections. Full ${\cal O}(\alpha)$ 
EW corrections to the exclusive process $pp \to W + j$ (where $j$ stands for jet) have been
recently computed, in the approximation of real $W$ bosons, in \cite{hkk,kkp}, while 
one-loop weak corrections to $Z$ hadro-production have been computed, for on-shell 
$Z$ bosons, in \cite{mmr}. It is also 
worth stressing that in eq.~(\ref{eq:qcd-ew}) the infrared part of QCD corrections is factorized, whereas the infrared-safe matrix element residue is included in  an additive  form. It is otherwise possible to implement a fully factorized combination (valid for infra-red safe observables) as follows: 

\begin{eqnarray}
\left[\frac{d\sigma}{d\cal O}\right]_{{\rm QCD} \otimes {\rm EW}} = & &
\left( 1 + \frac{
\left[{d\sigma} / {d\cal O}\right]_{\rm MC@NLO} - \left[{d\sigma}/{d\cal O}\right]_{\rm HERWIG\, \, PS}
}{
\left[{d\sigma}/{d\cal O}\right]_{\rm Born}
}
\right) \times \nonumber \\
& & \times 
\left\{ 
\frac{d\sigma}{{d\cal O}_{\rm EW}}
\right\}_{{\rm HERWIG\, \,  PS}} , 
\label{eq:qcd-ew-factor}
\end{eqnarray}
where the ingredients are the same as in eq.~(\ref{eq:qcd-ew}) but also the QCD matrix element residue in now factorized.  Eqs.~(\ref{eq:qcd-ew}) and ~(\ref{eq:qcd-ew-factor}) have the very same ${\cal O}(\alpha)$ and ${\cal O}(\alpha_s)$ content, differing by terms of the order of 
$\alpha \alpha_s$. Their relative difference has been checked to be of the order of a few per cent in the
resonance region around the $W$/$Z$ mass, and can be taken as an estimate of the uncertainty of QCD and EW combination. 

\section*{Numerical results: $W$ and $Z$ production}

In order to assess the phenomenological relevance of the combination of QCD and EW 
corrections, we study, for definiteness, the charged-current process $p p \to W^\pm \to 
\mu^\pm + X$ at the LHC, 
imposing the following selection criteria 
\begin{eqnarray}
&& {\rm a.} \quad p_{\perp}^{\mu} \geq~25~{\rm GeV},~~~~\rlap{\slash}{\! E_T}  \geq~25~{\rm GeV},~~~~| \eta_\mu|< 2.5,
\nonumber\\
&& {\rm b.} \quad {\rm the \, \, cuts \, \, as \, \, above} \, \, \oplus \, \, M_\perp^W \geq 1~{\rm TeV},
\label{eq:cutsw}
\end{eqnarray}
where  $p_{\perp}^{\mu}$ and
$\eta_\mu$ are the transverse momentum and the pseudorapidity of the muon, $\rlap{\slash}{\! E_T} $
is the missing transverse energy, which we identify with the transverse momentum of 
the neutrino, as typically done in several phenomenological studies. For 
set up b., a severe cut on the $W$ transverse mass $M_\perp^W$ is 
superimposed to the cuts of set up a., in order to isolate the region 
of the high tail of $M_T^W$, which is interesting for new physics searches. 
We also consider the neutral-current reaction  $p p \to \gamma, Z \to e^+ e^- + X$, selecting the
events according to the cuts
\begin{equation}
p_\perp^{e^\pm} \geq 25~ {\rm GeV},~~~~|\eta^{e^\pm}| < 2.5,~~~~
M_{e^+e^-} \geq 200~{\rm GeV}.
\label{eq:cutsz}
\end{equation}
The granularity of the detectors and the size of the electromagnetic
showers in the calorimeter  
make it difficult to discriminate between electrons and photons with a
small opening angle. We adopt the following procedure to select the event:
we recombine the four-momentum vectors of the electron and
photon into an effective electron four-momentum vector if, defining
\begin{equation}
\Delta R(e,\gamma) = \sqrt{
  \Delta\eta(e,\gamma)^2+\Delta\phi(e,\gamma)^2 },
\end{equation}
$\Delta R(e,\gamma)<0.1$ (with $\Delta\eta,\Delta\phi$ the distances
of electrons and photons along the longitudinal and azimuthal directions). 
We do not recombine electrons and photons if $\eta_{\gamma}>2.5$ (with
$\eta_\gamma$ the photon pseudo-rapidity).
We apply the event selection cuts as in Eq.~(\ref{eq:cutsz}) only after the recombination procedure.

The parton distribution function (PDF) set MRST2004QED \cite{mrst04qed}
has been used to describe the proton partonic content. The QCD 
factorization/renormalization scale and the analogous QED scale (present 
in the PDF set MRST2004QED) are chosen to be equal, as usually done in the literature \cite{dk,bw,CMNV,zgrad2,HORACEZ}, and fixed at $\mu_R = \mu_F = 
\sqrt{\left(p_{\perp}^{W}\right)^2 +  M_{\mu \nu_\mu}^2}$ (for the charged-current
case), where $M_{\mu\nu_\mu}$ is the $\mu\nu_\mu$ invariant mass, and at 
$\mu_R = \mu_F=\sqrt{\left(p_{\perp}^{Z}\right)^2+
M^2_{e^+e^-}}$ (for the neutral-current case),  where $M_{e^+e^-}$ 
is the invariant mass of the lepton pair. 

In order to avoid systematics theoretical  effects, all the generators used in our study have been
properly tuned at the level of input parameters, PDF set and scale to give the same LO/NLO results. 
The tuning procedure validates the interpretation of the various relative effects as due to the
radiative corrections and not to a mismatch in the setups of the codes under consideration.

\begin{center}
\begin{figure}[h]
\hskip 12pt\includegraphics[width=6.cm]{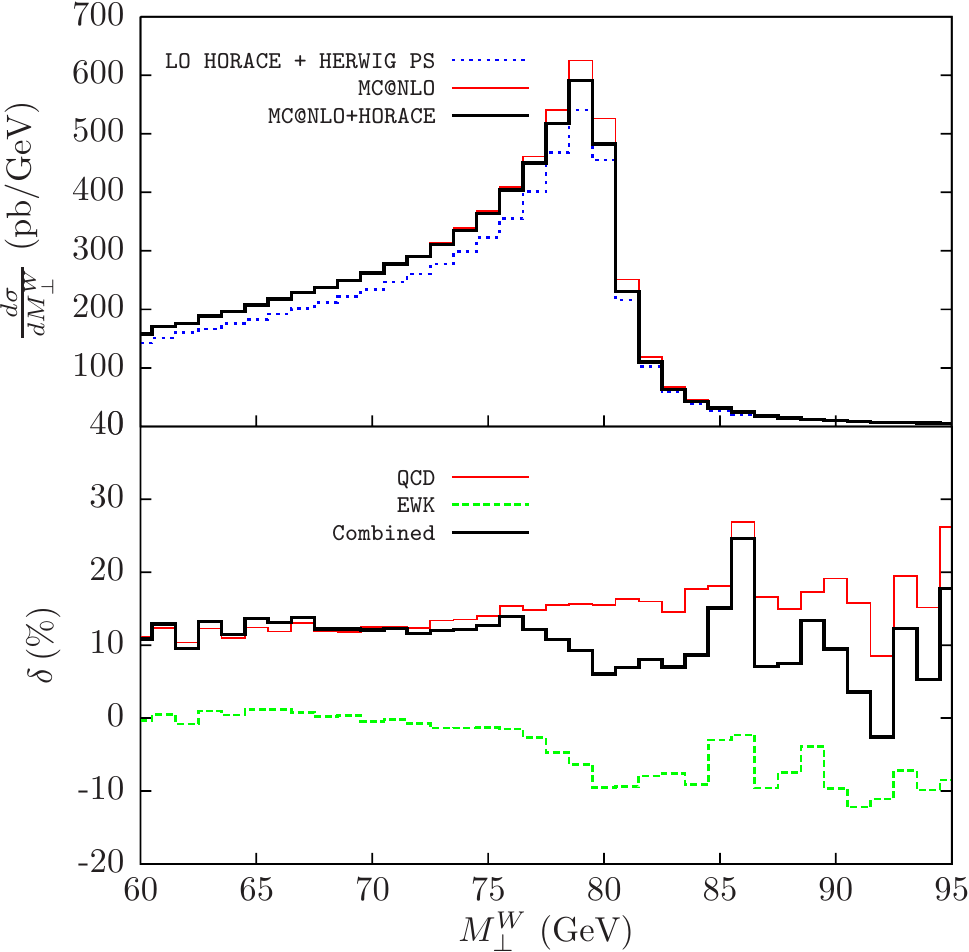}%
~\hskip 8pt\includegraphics[width=6.cm]{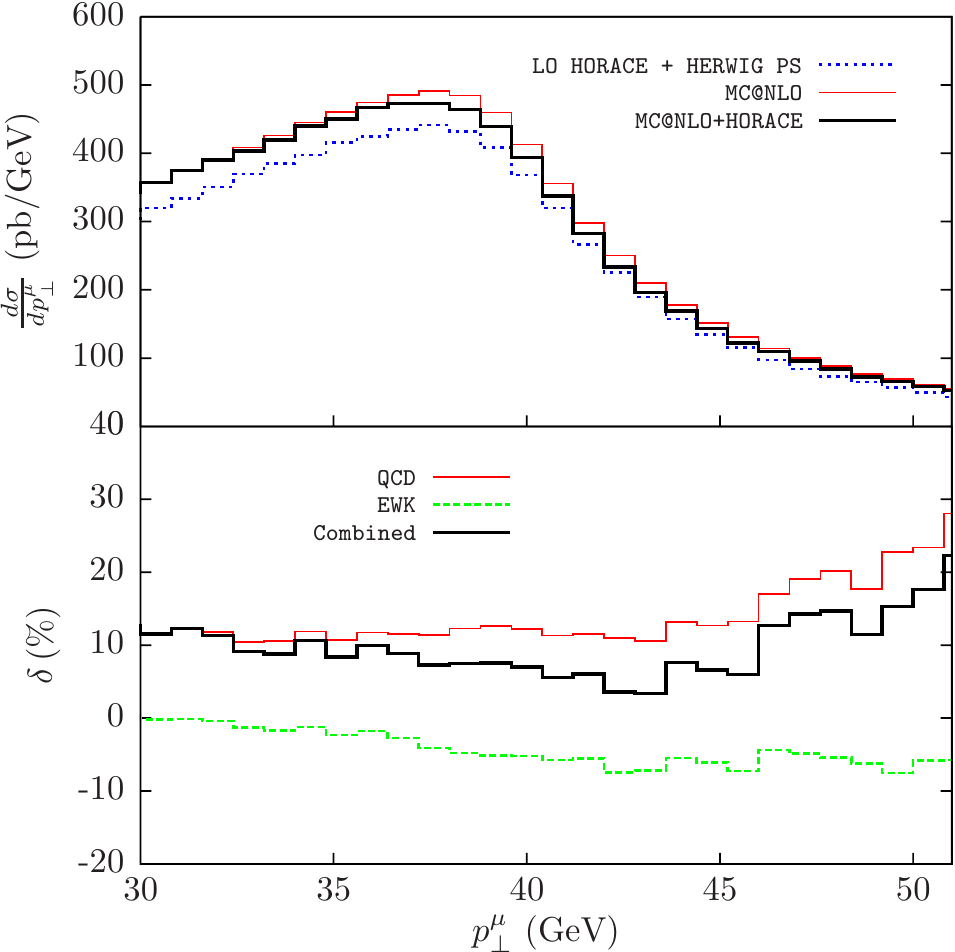}
\caption{Upper panel: predictions of MC@NLO, MC@NLO+HORACE and leading-order  HORACE+HERWIG 
PS for the $M_\perp^W$ (left) and $p_\perp^\mu$ (right) distributions at the LHC, according to the cuts of set up a. of Eq.~(\ref{eq:cutsw}). Lower panel: relative effect of QCD and EW corrections, and their sum, for the 
corresponding observables in the upper panel.}
\label{peak}
\end{figure}
\end{center}

\begin{center}
\begin{figure}[h]
\hskip 4pt\includegraphics[width=6.cm]{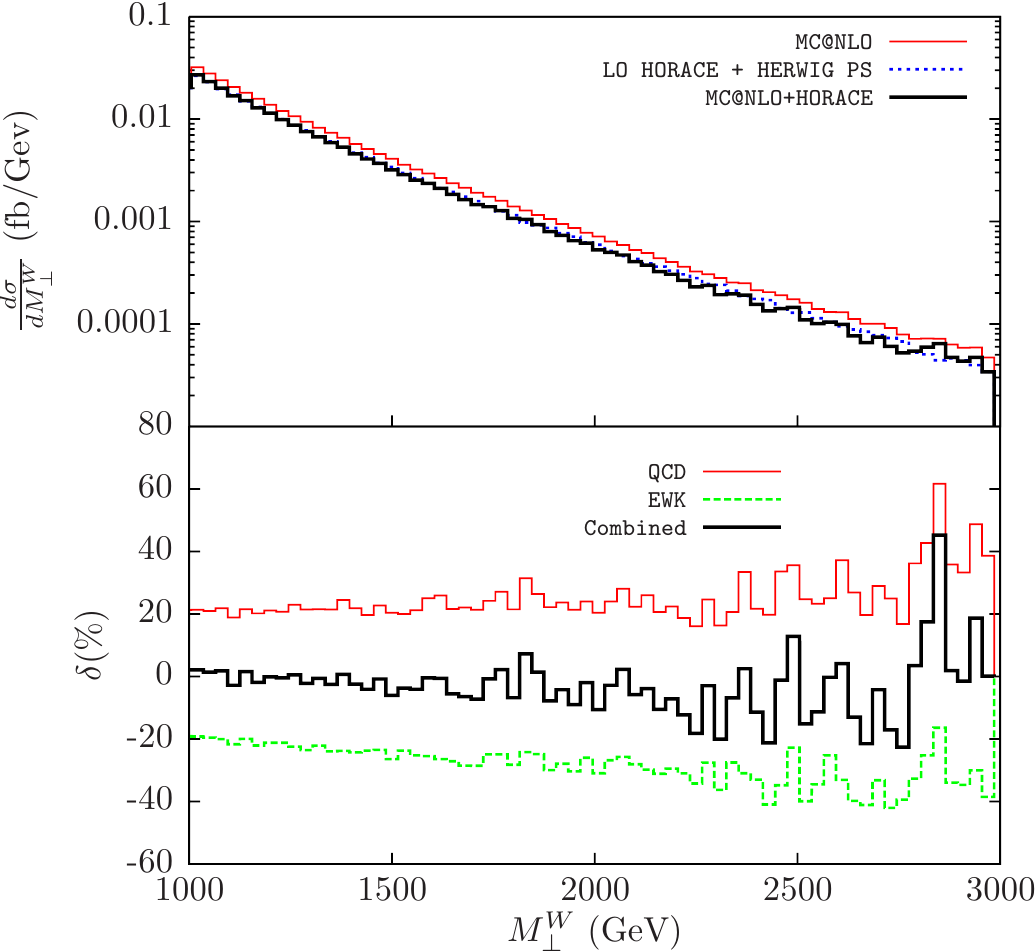}%
~\hskip 8pt\includegraphics[width=6.cm]{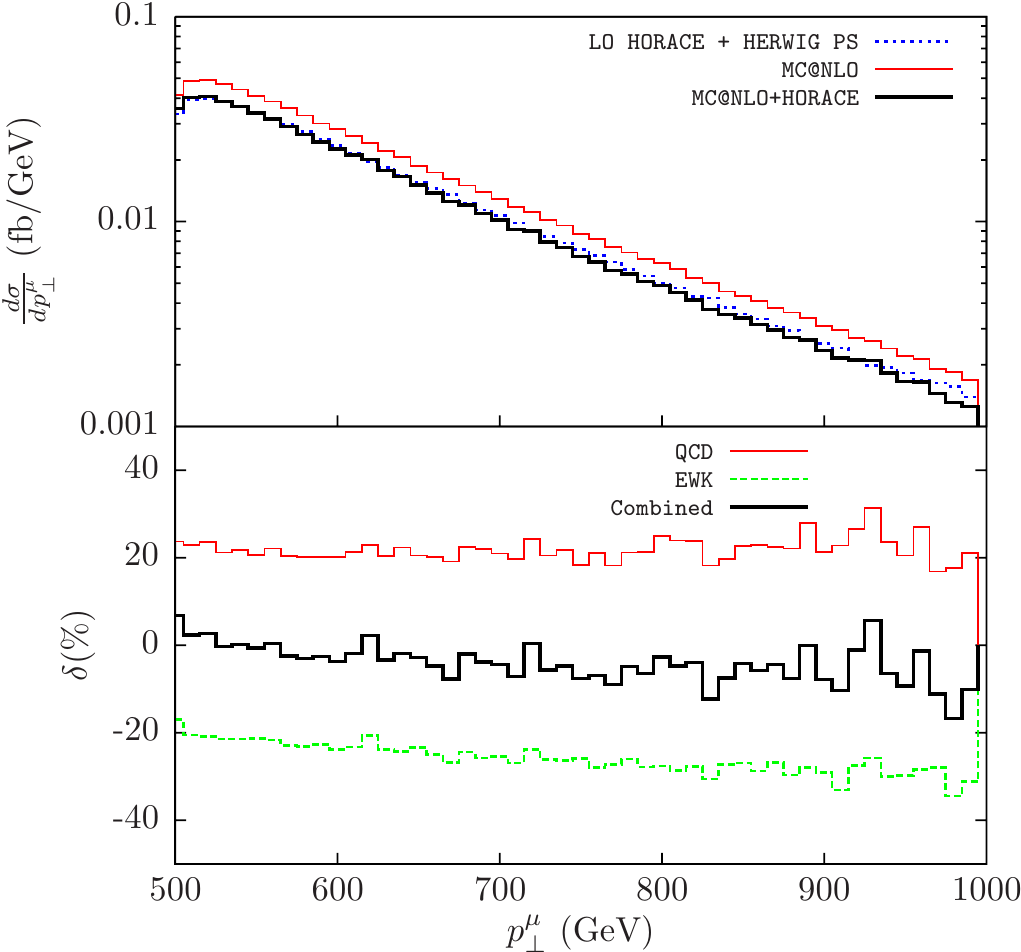}
\caption{The same as Fig. \ref{peak} according to the cuts of set up b. of Eq.~(\ref{eq:cutsw}).}
\label{Woffpeak}
\end{figure}
\end{center}

A sample of our numerical results is shown in Fig. \ref{peak} for the $W$ transverse mass  $M_\perp^W$ and muon transverse momentum  $p_{\perp}^{\mu}$ distributions according to set up a. 
of Eq. (\ref{eq:cutsw}), and in Fig. \ref{Woffpeak} for the
same distributions according to set up b. In Fig. \ref{peak} and Fig. \ref{Woffpeak}, the upper panels show the 
predictions of the generators MC@NLO and MC@NLO + HORACE interfaced to 
HERWIG PS (according to eq.~(\ref{eq:qcd-ew})), in 
comparison with the leading-order result by HORACE convoluted with HERWIG
shower evolution. The lower panels illustrate the relative effects of  the matrix element residue of NLO QCD and of full EW corrections, as well as their sum, that can be obtained by appropriate 
combinations of the results shown  in the upper panels. More precisely, the percentage corrections shown have been defined as
$\delta = \left(\sigma_{\rm NLO}-\sigma_{\rm Born+ HERWIG \, PS}\right)/\sigma_{\rm Born+ HERWIG \, PS}$, where $\sigma_{\rm NLO}$ stands for the predictions of the generators including
exact NLO corrections matched with QCD PS.

From Fig. \ref{peak} it can be seen that  the QCD corrections are positive around the 
$W$ jacobian peak, of about 10-20\%, and tend to compensate the negative effect due to 
EW corrections. Therefore, 
their interplay is crucial for a precise $M_W$ extraction at the LHC and their combined 
contribution can not be accounted for in terms of a pure QCD PS approach, 
as it can be inferred from the comparison of the predictions of MC@NLO versus 
the leading-order result by HORACE convoluted with HERWIG PS. It is also worth noting 
that the convolution of NLO corrections with the QCD PS broadens the sharply peaked shape of 
the fixed-order NLO QCD and EW effects.
 
The interplay between QCD and EW corrections to $W$ production in the region interesting for 
new physics searches, i.e. in the high tail of $M_\perp^W$ and $p_\perp^\mu$ distributions,
is shown in Fig. \ref{Woffpeak}. For both $M_\perp^W$ and $p_\perp^\mu$,  
the QCD corrections are positive and largely cancel the 
negative EW Sudakov logarithms. Therefore, a precise
normalization of the SM background to new physics searches 
necessarily requires the simultaneous control of QCD and EW corrections.

\begin{center}
\begin{figure}[h]
\hskip 4pt\includegraphics[width=6.5cm]{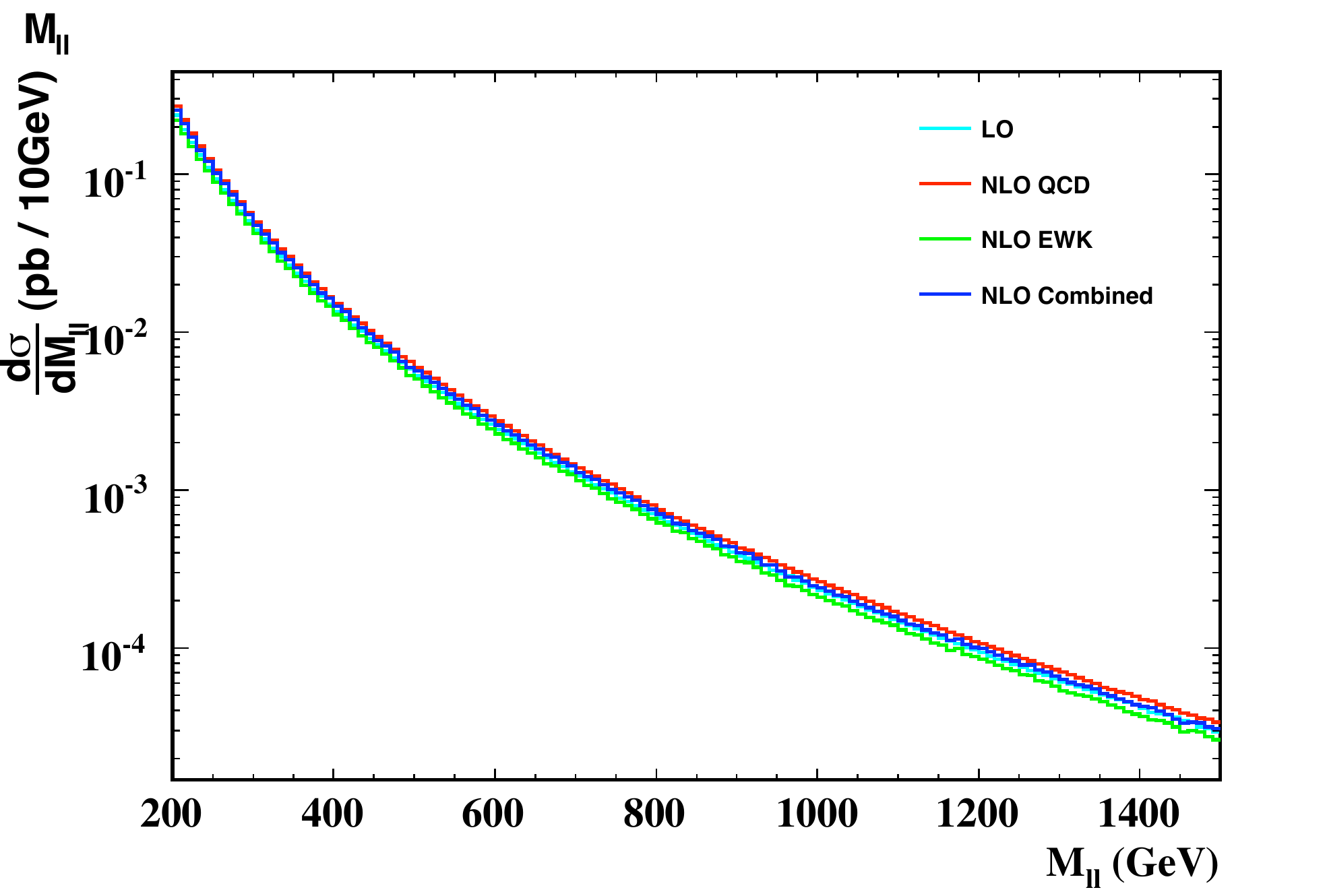}%
~\includegraphics[width=6.5cm]{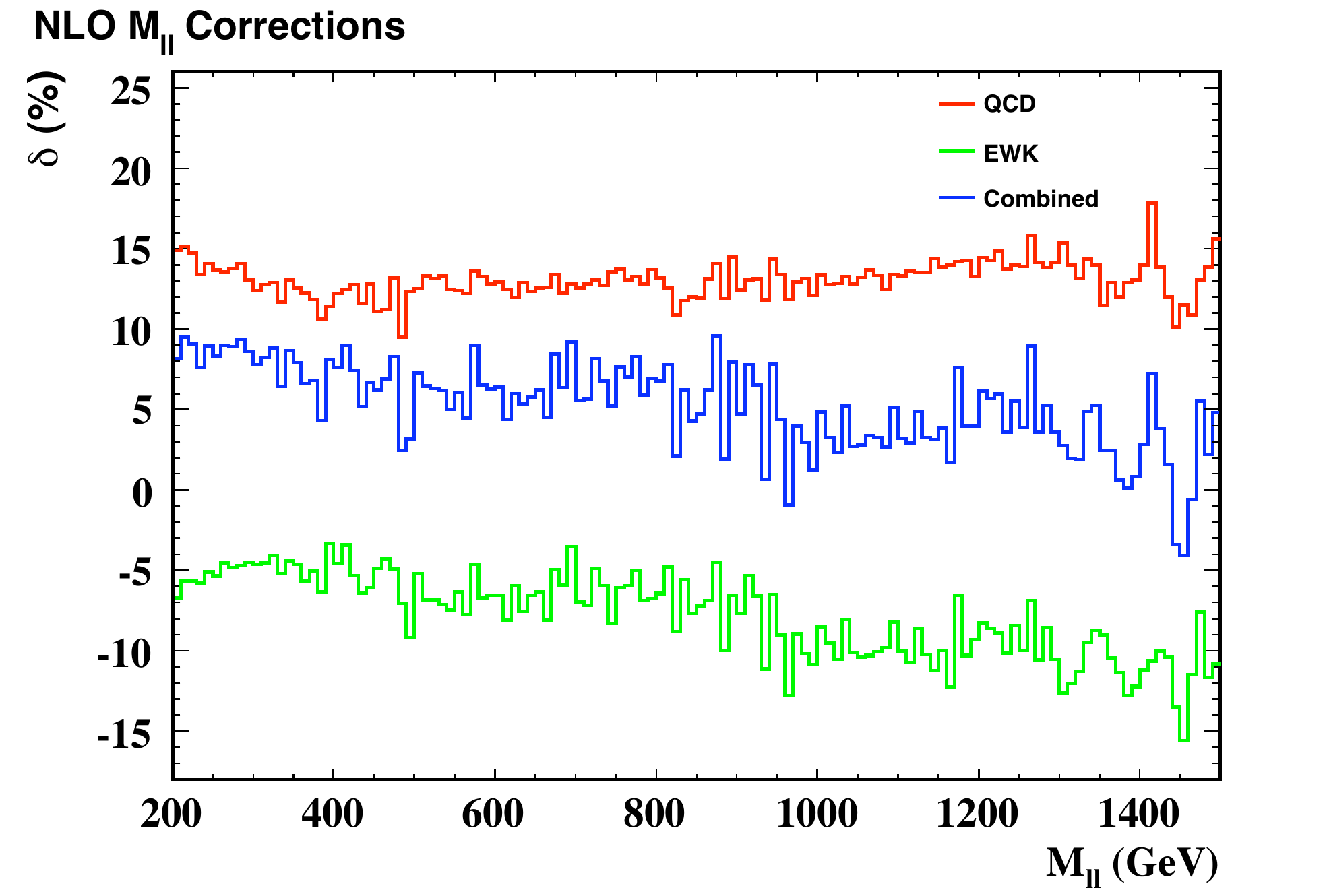}
\caption{Left panel: the di-electron invariant mass distribution according to the leading-order 
and NLO EW predictions of HORACE, of MC@NLO and of MC@NLO+HORACE at the LHC, 
using the cuts of Eq.~(\ref{eq:cutsz}). Right panel: relative effect of QCD and 
EW corrections, and their combination.}
\label{Zoffpeak}
\end{figure}
\end{center}

Figure~\ref{Zoffpeak} shows the combination of QCD and 
EW  corrections for the di-lepton invariant mass in the neutral-current D-Y process $p p \to \gamma, Z \to e^+ e^- + X$, according to the cuts of Eq.~(\ref{eq:cutsz})~\cite{zlh}.
The QCD corrections are quite flat and positive 
with a value of about $15\%$ over the mass range 200--1500~GeV. The
EW corrections are negative and vary from about $-5\%$ to
$-10\%$ and thus partially cancel the QCD contribution. Therefore, as for the charged-current channel, the
search for new physics in di-lepton final states needs a careful combination of EW and
QCD effects.

\section*{Conclusions}

During the last few years, there has been a big effort towards high-precision
predictions for D-Y-like processes, addressing the calculation of higher-order QCD and 
EW corrections. Correspondingly, precision computational tools have been developed to keep under control theoretical systematics in view of the future measurements at the LHC.

 We presented some original results about the combination of EW and 
QCD corrections to a sample of observables of $W$  and $Z$ production processes at the LHC. 
Our investigation shows that a high-precision knowledge of QCD and a careful combination of 
EW and strong contributions is mandatory in view of the anticipated experimental accuracy. We plan, however,
to perform a more complete and detailed phenomenological study, including the predictions of other 
QCD generators and considering further observables of interest for the many facets of the
$W/Z$ physics program at the LHC.  

\vskip 24pt\noindent
\leftline{\bf Acknowledgements}
C.M. Carloni Calame is supported by a 
INFN postdoc Fellowship and thanks the School of Physics\&Astronomy, 
University of Southampton, for hospitality. We acknowledge useful discussions with P. Nason and
other participants of the workshop.
We are also grateful to the colleagues of the common 
paper~\cite{zlh} for precious collaboration.

%\end{document}

\addtocounter{chapter}{1}
%\documentclass[a4paper,12pt,twoside]{report}
%\usepackage{epsfig}
%\usepackage{amssymb}
%\usepackage{lineno}
%\usepackage{setspace}
%\usepackage{amsmath} 
%%%%%%%%%%%%%%%%%%%%%%%%%%%%%%%%%%%%%%%%%%%%%%%%%%%%%%%%%%
%\begin{document}
%%%%%%%%%%%%%%%%%%%%%%%%%%%%%%%%%%%%%%%%%%%%%%%
% Toggle line numbering
% Won't work with the PRD revtex4 !

%\pagewiselinenumbers

% uncomment if you want doublespace
%\doublespace
%%%%%%%%%%%%%%%%%%%%%%%%%%%%%%%%%%%%%%%%%%%%%%%
\mchapter{Subtraction at NNLO and Higgs boson production at hadron colliders}%
{M. Grazzini}
%%%%%%%%%%%%%%%%%%%%%%%%%%%%%%%%%%%%%%%%%%%%%%%%%%%%%%%%%
\section{Introduction}

\newcommand\as{\alpha_{\mathrm{S}}} 
\def\ptmin{p_{T{\rm min}}}
\def\ptmax{p_{T{\rm max}}}
\newcommand\f[2]{\frac{#1}{#2}} 
\def\ltap{\raisebox{-.6ex}{\rlap{$\,\sim\,$}} \raisebox{.4ex}{$\,<\,$}} 
\def\gtap{\raisebox{-.6ex}{\rlap{$\,\sim\,$}} \raisebox{.4ex}{$\,>\,$}} 
\def\ptveto{p_T^{\rm veto}}

The dynamics of hard scattering processes
involving hadrons
is nowadays remarkably 
well described by perturbative QCD predictions.
Thanks to asymptotic freedom, the cross section for sufficiently 
inclusive reactions can be computed as a series expansion
in the QCD coupling $\as$.
Until few years ago, the standard for
such calculations was next-to-leading order (NLO) accuracy.
Next-to-next-to-leading order (NNLO)
results were known only for 
few
highly-inclusive reactions
(see e.g. Refs.~\cite{sigmatotgr,Hamberg:1990np,Higgstot}).

The extension from NLO to NNLO accuracy is important
to improve QCD predictions and to better assess
their uncertainties. In particular, this extension is essential in 
two cases: in those processes whose NLO corrections are comparable to the
leading order (LO) contribution; in those `benchmark' processes that are
measured with high experimental precision.
Such a task, however, implies finding methods and techniques to
cancel the infrared (IR) divergences that appear at 
intermediate steps of the calculations.

Recently, a new general method \cite{Anastasiou:2003gr},
based on sector decomposition \cite{sector}, has been proposed
and applied to the NNLO QCD calculations of $e^+e^-\to 2~{\rm jets}$
\cite{Anastasiou:2004qd}, Higgs \cite{Hdiff}
and vector \cite{DYdiff} boson production in hadron collisions, and to the NNLO
QED calculation of the electron energy spectrum in muon decay
\cite{Anastasiou:2005pn}.
The calculations of Refs.~\cite{Hdiff,DYdiff}
allow us to compute the corresponding cross sections with arbitrary 
cuts on the momenta of the partons produced in the final state.

The traditional approach to
perform NLO computations
is based on the introduction of auxiliary cross sections that are obtained
by approximating the QCD scattering amplitudes in the relevant IR (soft and
collinear) limits.
This strategy led to the proposal
of the {\em subtraction} \cite{Ellis:1980wv}
and {\em slicing} \cite{Fabricius:1981sx} methods.
Exploiting the universality properties of soft and collinear emission,
these methods were later developed in the form of 
general algorithms \cite{Giele:1991vf,Frixione:1995ms,Catani:1996vz}.
that make possible
to perform NLO calculations in a (relatively) straightforward manner,
once the corresponding QCD amplitudes are available.
In recent years, several research groups have been working on general NNLO
extensions of the subtraction method 
\cite{Kosower,Weinzierl,Frixione:2004is,GGG,ST}. 
Results have been obtained in some specific processes:
$e^+e^-\to 2~{\rm jets}$ \cite{Gehrmann-DeRidder:2004tv,Weinzierl:2006ij}
and, more recently, $e^+e^-\to 3~{\rm jets}$ \cite{Gehrmann-DeRidder:2006ez,GehrmannDeRidder:2007hr}.

In Ref.~\cite{Catani:2007vq} we proposed an extension of the subtraction method to NNLO for a specific, though important, class of 
processes: the production of colourless high-mass systems in hadron collisions.
We presented a formulation of the subtraction method for this class
of processes, and we applied
it to the NNLO calculation of Higgs boson production via the gluon fusion
subprocess $gg\to H$. The calculation has now been implemented in the numerical program {\tt HNNLO},
which includes all the relevant decay modes of the Higgs boson for this production subprocess,
namely, $H\to\gamma\gamma$ \cite{Catani:2007vq}, $H\to WW\to l\nu l\nu$ and
$H\to ZZ\to 4l$ \cite{Grazzini:2008tf}.

This contribution is organized as follows. In Sect.~\ref{sec:method}
we discuss the version of the subtraction formalism we use.
In Sect.~\ref{sec:results} we present a selection of numerical results
that can be obtained by our program. In Sect.~\ref{sec:summa} we summarize our
results.

\section{The method}
\label{sec:method}

We consider the inclusive hard-scattering reaction
\begin{equation}
h_1+h_2\to F(Q)+X,
\end{equation}
where the collision of the two hadrons $h_1$ and $h_2$
%(with momenta $p_1$ and $p_2$, respectively)
produces the triggered final state $F$. The final 
state $F$ consists of one or more colourless particles (leptons, photons, vector
bosons, Higgs bosons, $\dots$) with momenta $q_i$ and total invariant mass $Q$
.
Note that, since $F$ is colourless, the LO partonic subprocess
is either $q{\bar q}$ annihilation, as in the case of the Drell--Yan process, 
or $gg$ fusion,
as in the case of Higgs boson production.

At NLO, two kinds of corrections contribute: i) {\em real} corrections,
where one parton recoils against $F$; ii) {\em one-loop virtual} corrections to
the LO subprocess. Both contributions are separately 
%infrared 
IR divergent, but 
the divergences cancel in the sum.
At NNLO, three kinds of corrections must be considered: i) {\em double real} 
contributions, where two partons recoil against $F$; ii) {\em real-virtual} 
corrections, where one parton recoils against $F$ at one-loop order; 
iii) {\em two-loop virtual} corrections to the LO subprocess.
The three contributions are still
separately divergent, and the calculation has
to be organized so as to explicitly achieve the cancellation of the 
IR divergences.

Our
method
is based on a generalization
of the procedure
used in the specific NNLO calculation of
Ref.~\cite{Catani:2001cr}.
We first note that, at LO, the transverse momentum 
${\bf q}_{\, T}$ of the triggered final state $F$ is exactly zero.
As a consequence, as long as $q_T\neq 0$, the (N)NLO contributions are actually given by the (N)LO 
contributions to the triggered final state $F+{\rm jet(s)}$.
Thus, we can write the 
cross section as
\begin{equation}
\label{Fplusjets}
d{\sigma}^{F}_{(N)NLO}|_{q_T\neq 0}=d{\sigma}^{F+{\rm jets}}_{(N)LO}
\;\; .
\end{equation}
This means that, when $q_T\neq 0$, the 
IR divergences in our NNLO calculation are those in 
$d{\sigma}^{F+{\rm jets}}_{NLO}$: they can be handled and
cancelled by using 
available NLO formulations of the subtraction method.
The only remaining singularities of NNLO type are associated to the limit 
$q_T\to 0$, and we treat them by an additional subtraction. 
Our key point is that the singular behaviour 
of $d{\sigma}^{F+{\rm jets}}_{(N)LO}$ when $q_T\to 0$ is well known:
it appears in the
resummation program
\cite{Parisi:1979se,Collins:1984kg,Catani:2000vq}
of logarithmically-enhanced contributions
to transverse-momentum distributions.
Then,
to perform the additional subtraction, we follow the formalism used in 
Ref.~\cite{qtresum,ww} to combine resummed and fixed-order calculations.

We use a shorthand notation that mimics the notation 
of Ref.~\cite{qtresum}. We define the subtraction 
counterterm\footnote{The symbol $\otimes$ understands convolutions over momentum
fractions and sum over flavour indeces of the partons.}
\begin{equation}
\label{ct}
d{\sigma}^{CT}
%\sim
=
d{\sigma}_{LO}^F\otimes\Sigma^F(q_T/Q)\, d^2{\bf q}_{\, T}.
\end{equation}
The function $\Sigma^F(q_T/Q)$ embodies the
singular behaviour of $d{\sigma}^{F+{\rm jets}}$
when $q_T\to 0$. In this limit it can be expressed as
follows in terms of $q_T$-independent coefficients $\Sigma^{F(n;k)}$:
\begin{equation}
\label{sigmalimit}
\Sigma^F(q_T/Q)
\xrightarrow[q_T\to 0]{}
\sum_{n=1}^\infty
\left(\f{\as}{\pi}\right)^n\sum_{k=1}^{2n}
\Sigma^{F(n;k)} \;\f{Q^2}{q_T^2}\ln^{k-1} \f{Q^2}{q_T^2}  \;\; .
\end{equation}
The extension of Eq.~(\ref{Fplusjets}) to include
the contribution at $q_T=0$ is finally:
%we can write:
\begin{equation}
\label{main}
d{\sigma}^{F}_{(N)NLO}={\cal H}^{F}_{(N)NLO}\otimes d{\sigma}^{F}_{LO}
%+\left(d{\sigma}^{F+{\rm jets}}_{(N)LO}-
%d{\sigma}^{CT}_{(N)LO}\right)\;\; .
+\left[ d{\sigma}^{F+{\rm jets}}_{(N)LO}-
d{\sigma}^{CT}_{(N)LO}\right]\;\; .
\end{equation}
%With respect to 
Comparing with the right-hand side of
Eq.~(\ref{Fplusjets}), we have subtracted
the truncation of Eq.~(\ref{ct}) at (N)LO
and added a contribution at $q_T=0$ needed to 
%restore the correct normalization.
%get
obtain the correct total cross section.
The coefficient ${\cal H}^{F}_{(N)NLO}$ does not depend on $q_T$
and is obtained by the (N)NLO truncation of the perturbative function
\begin{equation}
{\cal H}^{F}=1+\f{\as}{\pi}\,
{\cal H}^{F(1)}+\left(\f{\as}{\pi}\right)^2
{\cal H}^{F(2)}+ \dots \;\;.
\end{equation}
The counterterm of Eq.~(\ref{ct})
regularizes the singularity of $d{\sigma}^{F+{\rm jets}}$ when $q_T\to 0$:
the term in the square bracket on the right-hand side of Eq.~(\ref{main}) is 
thus IR finite (or, better, integrable over $q_T$). Note that, at NNLO,
$d{\sigma}^{CT}_{(N)LO}$ acts as a counterterm for the {\em sum} of the two
contributions to $d{\sigma}^{F+{\rm jets}}$: the 
{\em double real} plus {\em real-virtual} contributions.
We also note that the counterterm
function $\Sigma^F(q_T/Q)$ can be defined 
in different ways:
the only property we require is that
in the small $q_T$ limit it must take the form given in Eq.~(\ref{sigmalimit}),
so as to match the singular behaviour of $d{\sigma}^{F+{\rm jets}}$.
Note that the perturbative coefficients $\Sigma^{F(n;k)}$ are 
universal: more precisely, the NNLO coefficients $\Sigma^{F(2;1)}$
and $\Sigma^{F(2;2)}$
have a non-universal contribution that, nonetheless, is proportional
to the NLO coefficient ${\cal H}^{F(1)}$.
The above coefficients
only depend on the type of partons (quarks or gluon) involved in the
LO partonic subprocess ($q{\bar q}$ annihilation or $gg$ fusion).
We finally note that
the simplicity of the LO subprocess is such that final-state partons 
actually appear only in 
the term $d{\sigma}^{F+{\rm jets}}$ on the
right-hand side of Eq.~(\ref{main}).
Therefore,
arbitrary IR-safe cuts on the jets at (N)NLO can effectively be accounted for 
through a (N)LO computation. Owing to this feature,
our NNLO extension of the subtraction formalism is observable-independent.

At NLO (NNLO), the physical information of the {\em one-loop (two-loop)
virtual} correction to the LO subprocess is 
contained in the coefficients 
${\cal H}^{(1)}$ (${\cal H}^{(2)}$). 
Once an explicit form of Eq.~(\ref{ct}) is chosen, 
the hard coefficients ${\cal H}^{F (n)}$ are uniquely 
identified (a different choice would correspond to different ${\cal H}^{F (n)}$).
According to Eq.~(\ref{main}), the NLO calculation  of $d{\sigma}^{F}$ 
requires the knowledge
of ${\cal H}^{F(1)}$ and the LO calculation of $d{\sigma}^{F+{\rm jets}}$.
The general (process-independent) form of 
the coefficient 
${\cal H}^{F(1)}$ is basically known: the precise relation between 
${\cal H}^{F(1)}$ and the IR finite part of the
one-loop correction to a generic LO subprocess is explicitly derived in 
Ref.~\cite{deFlorian:2000pr}.

%At NNLO, the coefficient ${\cal H}^{F(2)}$ is also needed, together with the
%NLO calculation for $c {\bar c}\to F+{\rm jet(s)}$.
At NNLO, the coefficient ${\cal H}^{F(2)}$ is also needed, together with the
NLO calculation of $d{\sigma}^{F+{\rm jets}}$.
The coefficients
${\cal H}^{H(2)}$ for Higgs boson production in
the large-$M_{top}$ limit have been computed \cite{inprep}.
Since the NLO corrections to $gg\to H+{\rm jet(s)}$ are available 
\cite{deFlorian:1999zd} 
in the same limit,
we are able to apply Eq.~(\ref{main})
at NNLO. We have 
encoded our computation in a parton level
Monte Carlo program, in which
we can implement arbitrary IR-safe cuts on the final state.

\section{Results}
\label{sec:results}

In the following
we present numerical results for Higgs boson production at the LHC.
We use the MRST2004 parton distributions \cite{Martin:2004ir},
with densities and $\as$ evaluated at each corresponding order
(i.e., we use $(n+1)$-loop $\as$ at N$^n$LO, with $n=0,1,2$). The 
renormalization and factorization scales are fixed to the value 
$\mu_R=\mu_F=M_H$, where $M_H$ is the mass of the Higgs boson.

\subsection{$H\to \gamma\gamma$}
\label{gammagamma}

We consider the Higgs boson decay in the
$H\to \gamma\gamma$ channel and follow
Ref.~\cite{CMStdr} to apply 
cuts on the photons.
For each event, we classify the photon transverse momenta according to their
minimum and maximum value,  
$\ptmin$ and $\ptmax$. The photons are required to be in the
central rapidity region, $|\eta|<2.5$, with  $\ptmin>35$~GeV
and $\ptmax>40$~GeV. We also require the photons to be isolated:
the hadronic (partonic) transverse energy in a cone of radius $R=0.3$ along the
photon direction 
has to be smaller 
than 6~GeV. When $M_H=125$~GeV, by applying these cuts
the impact of the NNLO corrections on the NLO total cross section
is reduced from 19\% to 11\%.
 
%%====================================
\begin{figure}[htb]
\begin{center}
\begin{tabular}{c}
\epsfxsize=12truecm
\epsffile{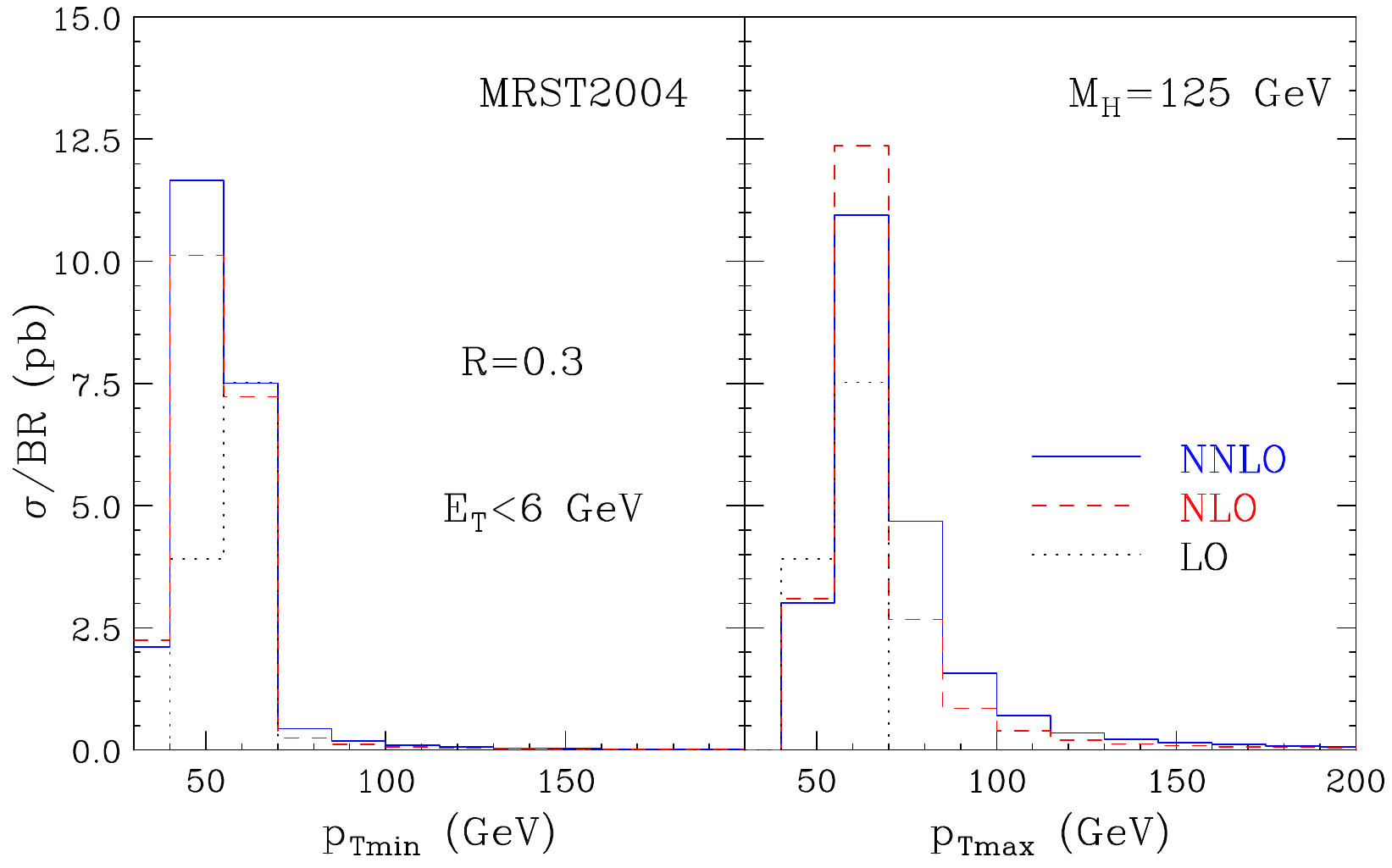}\\
\end{tabular}
\end{center}
\caption{\label{fig:isol}
{\em Distributions in $\ptmin$ and $\ptmax$ for the diphoton signal at 
the LHC. The cross section is divided by the branching ratio in two photons.}}
\end{figure}
%%====================================

In Fig.~\ref{fig:isol} we plot
the 
distributions in $\ptmin$ and $\ptmax$
for the $gg\to H\to\gamma\gamma$ signal.
We note that the shape of these distributions sizeably
differs when going from LO to NLO and to NNLO.
The origin of these perturbative instabilities is well known 
\cite{Catani:1997xc}.
Since the LO spectra
are kinematically bounded by $p_T\leq M_H/2$,
each higher-order perturbative contribution produces
(integrable) logarithmic singularities in the vicinity of
that boundary. More detailed studies are necessary to assess
the theoretical uncertainties of these fixed-order results
and the relevance of all-order resummed calculations.

%==========================================
\begin{figure}
\begin{center}
\includegraphics[width=0.5\textwidth]{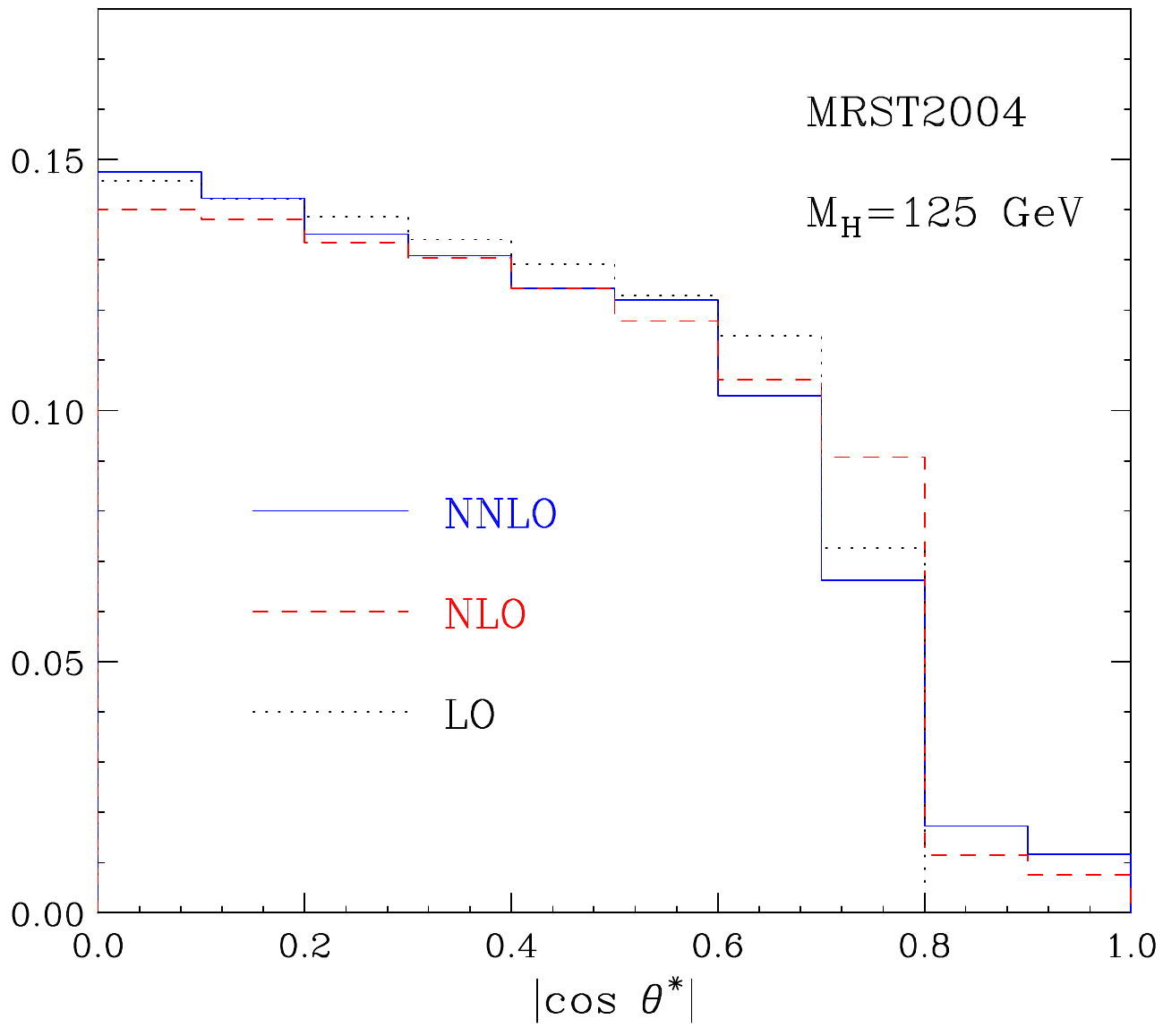}
 \caption{Normalized distribution in the variable $\cos\theta^*$.}
\label{fig:ctheta}
\end{center}
\end{figure}
%==========================================

In Fig.~\ref{fig:ctheta} we consider the (normalized) distribution in
the variable $\cos\theta^*$, where $\theta^*$ is the polar angle of one of the photons
in the rest frame of the Higgs boson.
At small values of $\cos\theta^*$ the distribution
is quite stable with respect to higher order QCD corrections.
We also note that the LO distribution vanishes beyond the value $\cos\theta^*_{\rm max}<1$.
The upper bound $\cos\theta^*_{\rm max}$ is due to the fact that the photons
are required to have a minimum $p_T$ of $35$ GeV.
As in the case of Fig.~\ref{fig:isol}, in the vicinity of this LO kinematical boundary
there is an instability of the
perturbative results beyond LO.

\subsection{$H\to WW\to l\nu l\nu$}

We now consider the production of a Higgs boson with mass $M_H=165$ GeV in the decay mode $H\to WW\to l\nu l\nu$ \cite{Grazzini:2008tf}.
We apply a set of selection cuts taken from the study of Ref.~\cite{Davatz:2004zg}. The charged leptons are classified according to their minimum and maximum $p_T$. The $\ptmin$ should be larger than 25 GeV, and $\ptmax$ should be
between 35 and 50 GeV. The charged lepton rapidity should fullfil $|\eta|<2$. The missing $p_T$ of the event is
required to be larger than $20$ GeV and the invariant mass of the charged
leptons is smaller than $35$ GeV.
The azimuthal separation of the charged leptons in the
transverse plane ($\Delta\phi$) is smaller than $45^o$.
Finally, there should be no jet with $p_T^{\rm jet}$ larger than $\ptveto$
\footnote{Jets are reconstructed with the $k_T$ algorithm \cite{ktalg}
with jet size $D=0.4$.}.

In Table \ref{tab:wwsel} we report the corresponding
cross sections in the case of $\ptveto=30$ GeV.

\begin{table}[htbp]
\begin{center}
\begin{tabular}{|c|c|c|c|}
\hline
$\sigma$ (fb)& LO & NLO & NNLO\\
\hline
\hline
$\mu_F=\mu_R=M_H/2$ & $17.36\pm 0.02$ & $18.11\pm 0.08$ & $15.70\pm 0.32$\\
\hline
$\mu_F=\mu_R=M_H$ & $14.39\pm 0.02$ & $17.07\pm 0.06$ & $15.99\pm 0.23$ \\
\hline
$\mu_F=\mu_R=2M_H$ & $12.00 \pm 0.02$ & $15.94\pm 0.05$ & $15.68\pm 0.20$\\
\hline
\end{tabular}
\end{center}
\caption{{\em Cross sections for $pp\to H+X\to WW+X\to l\nu l\nu+X$ at the LHC
when selection cuts are applied and $\ptveto=30$ GeV.}}
\label{tab:wwsel}
\end{table}

The cuts are quite hard, 
the efficiency being $8\%$ at NLO and $6\%$ at NNLO.
The scale dependence of the result is strongly reduced at NNLO,
being of the order of the error from the numerical integration.
The impact of higher order corrections is also
drastically changed. The $K$-factor is now 1.19 at NLO and
1.11 at NNLO.
As expected, the jet veto tends to stabilize the
perturbative expansion, and the NNLO cross section turns out to be smaller than the NLO one.

\subsection{$H\to ZZ\to e^+e^-e^+e^-$}

We now consider the production of
a Higgs boson with mass $M_H=200$ GeV \cite{Grazzini:2008tf}.
In this mass region the dominant decay mode is $H\to ZZ\to 4$ leptons,
providing a clean four lepton signature.
In the following we
consider the decay of the Higgs boson in two identical lepton pairs.
When no cuts are applied,
the NLO $K$-factor is $K=1.87$ whereas at NNLO we have $K=2.26$.
We find that the interference contribution is smaller than $1\%$ in this
region of Higgs boson masses.

We consider the following cuts \cite{CMStdr}:
\begin{enumerate}
\item For each event, we order the transverse momenta of the leptons from the largest ($p_{T1}$) to the smallest ($p_{T4}$). They are required to fulfil the following
thresholds:\\ $p_{T1}>30~{\rm GeV}~~~~p_{T2}>25~{\rm GeV}~~~~p_{T3}>15~{\rm GeV}~~~~p_{T4}>7~{\rm GeV}$\,;
\item Leptons should be central: $|y|< 2.5$;
\item Leptons should be isolated: the total transverse energy $E_T$ in a cone of radius 0.2 around each lepton should fulfil $E_T< 0.05~p_T$;
\item For each possible $e^+e^-$ pair, the closest ($m_1$)
and next-to-closest ($m_2$) to $M_Z$ are found.
Then $m_1$ and $m_2$ are required to be $81$ GeV $< m_1 < 101$ GeV and $40$ GeV $< m_2 < 110$ GeV.
\end{enumerate}
These cuts are designed
to maximize the statistical significance for an early discovery,
but to keep the possibility for a more detailed analysis of
the properties of the Higgs boson.
The corresponding cross sections are reported in Table \ref{tab:cuts}.
\begin{table}[htbp]
\begin{center}
\begin{tabular}{|c|c|c|c|}
\hline
$\sigma$ (fb)& LO & NLO & NNLO\\
\hline
\hline
$\mu_F=\mu_R=M_H/2$ & $1.541 \pm 0.002$ & $2.764\pm 0.005$ & $ 3.013\pm 0.023$\\
\hline
$\mu_F=\mu_R=M_H$ & $1.264\pm 0.001$ & $2.360\pm 0.003$ & $2.805\pm 0.015$\\
\hline
$\mu_F=\mu_R=2M_H$ & $1.047\pm 0.001$ & $2.044 \pm 0.003$ & $2.585\pm 0.010$\\
\hline
\end{tabular}
\end{center}
\caption{{\em Cross sections for $pp\to H+X\to ZZ+X\to e^+e^-e^+e^-+X$ at the LHC
when cuts are applied.}}
\label{tab:cuts}
\end{table}

Contrary to what happens in the $H\to WW\to l\nu l\nu$ decay mode,
the cuts are quite mild, the efficiency being $63\%$ at NLO and $62\%$ at NNLO.
The NLO and NNLO $K$-factors are $1.87$ and $2.22$, respectively.
Comparing with the inclusive case, we
conclude that these cuts do not change significantly
the impact of QCD radiative corrections.
We also find that the effect of lepton isolation is mild: at NNLO it reduces the accepted cross section by about $4\%$.

%%====================================
\begin{figure}[htb]
\begin{center}
\begin{tabular}{c}
\epsfxsize=12truecm
\epsffile{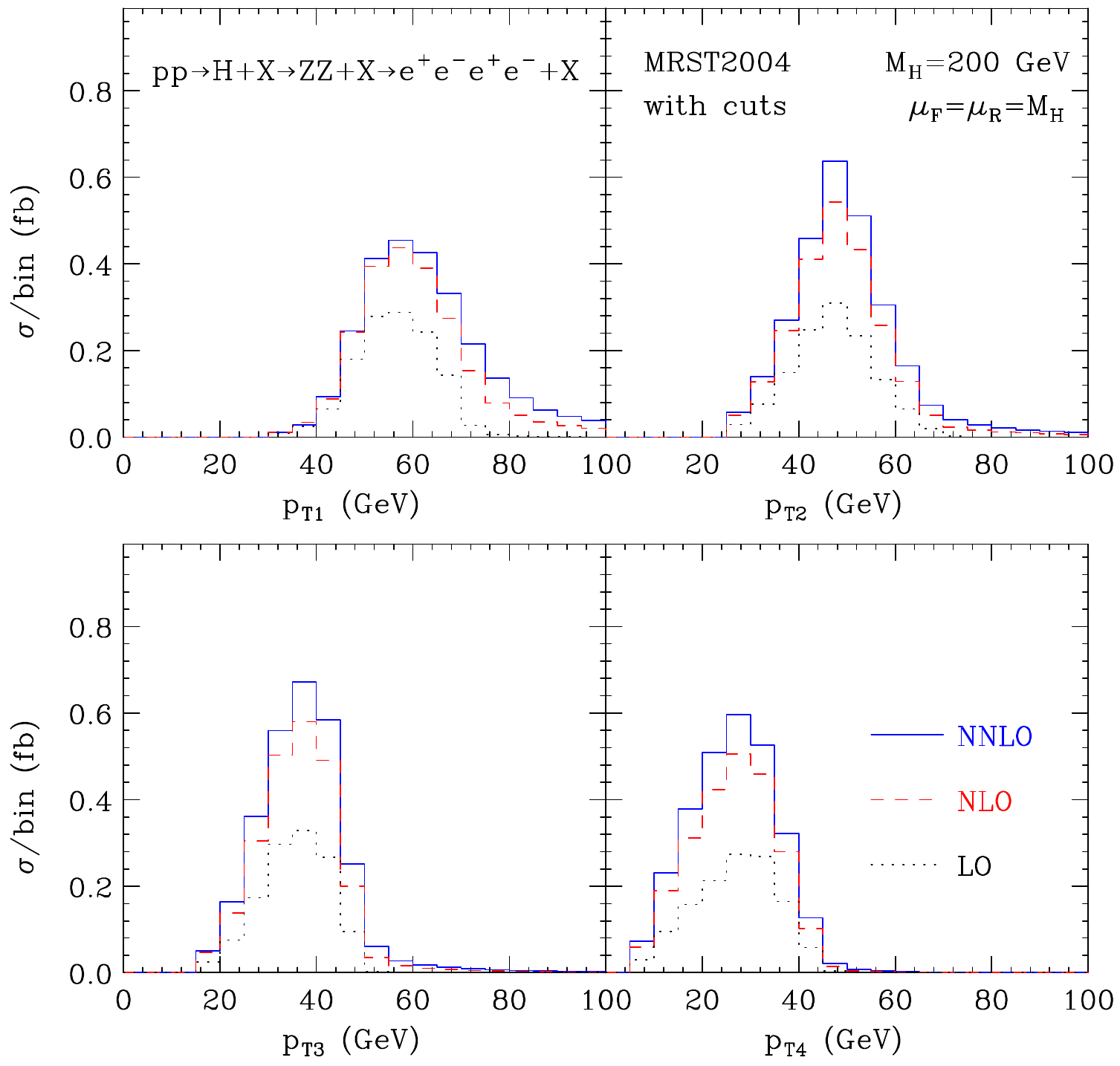}\\
\end{tabular}
\end{center}
\caption{\label{fig:ptlept}
{\em Tranverse momentum spectra of the final state leptons
for $pp\to H+X\to ZZ+X\to e^+e^-e^+e^-+X$,
ordered according to decreasing $p_T$,
at LO (dotted), NLO (dashed), NNLO (solid).}}
\end{figure}
%%====================================

In Fig.~\ref{fig:ptlept} we report the $p_T$ spectra of the charged leptons.
We note that at LO, without cuts, the $p_{T1}$ and $p_{T2}$ are kinematically bounded by $M_H/2$, whereas $p_{T3}< M_H/3$ and $p_{T4}<M_H/4$.
Contrary to what happens in the $H\to\gamma\gamma$ decay mode
(see Sect.~\ref{gammagamma}) the distributions smoothly reach the kinematical boundary,
and no perturbative instability is observed beyond LO.

\section{Summary}
\label{sec:summa}

We have illustrated 
an extension of the subtraction formalism to
compute NNLO QCD corrections to
the production of high-mass systems in hadron collisions. We have
considered an explicit application of
our method to the NNLO computation of $gg\to H$
at the LHC, including the decay of the Higgs boson in all the relevant decay modes, namely, $H\to\gamma\gamma$, $H\to WW\to l\nu l\nu$
and $H\to ZZ\to 4$ leptons. We have presented few selected results, including
kinematical
cuts on the final state. In the case of the $H\to\gamma\gamma$ and $H\to WW\to l\nu l\nu$ decay modes, our computation parallels the one of
Refs.~\cite{Hdiff,Anastasiou:2007mz}, but it is performed with a completely independent method. 
In the quantitative studies that we have carried out, the two computations
give results in numerical agreement. 
In our approach the calculation is
directly implemented in a parton level event generator.
This feature makes it particularly suitable for practical applications
to the computation of distributions in the form of bin histograms.
The calculation is implemented in the numerical program {\tt HNNLO}, which can
be downloaded from \cite{hnnloweb}.

%\end{document}

\addtocounter{chapter}{1}
%\documentclass[a4paper,12pt,twoside]{report}
%\usepackage{epsfig}
%\usepackage{amssymb}
%\usepackage{lineno}
%\usepackage{setspace}
%%%%%%%%%%%%%%%%%%%%%%%%%%%%%%%%%%%%%%%%%%%%%%%%%%%%%%%%%%
%\begin{document}
%%%%%%%%%%%%%%%%%%%%%%%%%%%%%%%%%%%%%%%%%%%%%%%
% Toggle line numbering
% Won't work with the PRD revtex4 !
%\pagewiselinenumbers
% uncomment if you want doublespace
%\doublespace
%%%%%%%%%%%%%%%%%%%%%%%%%%%%%%%%%%%%%%%%%%%%%%%
\mchapter{QCD final states: resummation and Monte Carlo simulations}
{A. Banfi}

%%%%%%%%%%%%%%%%%%%%%%%%%%%%%%%%%%%%%%%%%%%%%%%%%%%%%%%%%
\section{Introduction}
Any short distance dominated cross section $d\sigma$ in QCD can be
written as a formal series in the QCD coupling $\alpha_s$:
\begin{equation}
  \label{eq:series}
  d\sigma = d\sigma_0 + \alpha_s \cdot d\sigma_1 + \alpha_s^2 \cdot 
  d\sigma_2 + \dots  
\end{equation}
where $d\sigma_0$ is the leading order (LO) or Born contribution,
$d\sigma_1$ the next-to-leading order (NLO) contribution,
$d\sigma_2$ the next-to-next-to-leading order (NNLO) and so on. In
spite of the smallness of $\alpha_s$, the coefficients of the
expansion may be large. This happens typically when a process is
characterised by two widely separated scales $Q$ and $Q_0$, where $Q$
represents the hard scale of the process and $Q_0$ an energy
resolution. In this case large logarithms $L=\ln Q/Q_0$ arise at any
order in the perturbative (PT) series eq.~(\ref{eq:series}), and only
after an all-order resummation can one give meaning to the PT
expansion. In many cases resummation makes it possible to rewrite
$d\sigma$ as an exponent
\begin{equation}
  \label{eq:resummation}
  d\sigma = C(\alpha_s)\exp\left(Lg_1(\alpha_s L)+g_2(\alpha_s L)+
    \alpha_s g_3(\alpha_s L)\right) + \mbox{suppressed terms}\>,   
\end{equation}
where $g_1$ resums the leading logarithms (LL, $\alpha_s^n L^{n+1}$),
$g_2$ the next-to-leading logarithms (NLL, $\alpha_s^n L^{n}$), $g_3$
the next-to-next-to-leading logarithms (NNLL, $\alpha_s^n L^{n-1}$), and so on. 

The physical origin of large logarithms is the incomplete cancellation
of soft and collinear (SC) singularities between real and virtual
contributions. In particular we distinguish between double logarithms
$\alpha_s L^2$ arising from soft and collinear emissions, and single
logarithms $\alpha_s L$ from hard collinear or soft large-angle
emissions. SC singularities factorise~\cite{CSS} from hard matrix
elements and build up the exponent in eq.~(\ref{eq:resummation}).
Finite virtual corrections and the exact treatment of the phase space
in the SC limit give the multiplicative constant $C(\alpha_s)$, while
hard emission contributions are suppressed by powers of $Q_0/Q$.

\begin{figure}[htbp]
  \centering
  \epsfig{file=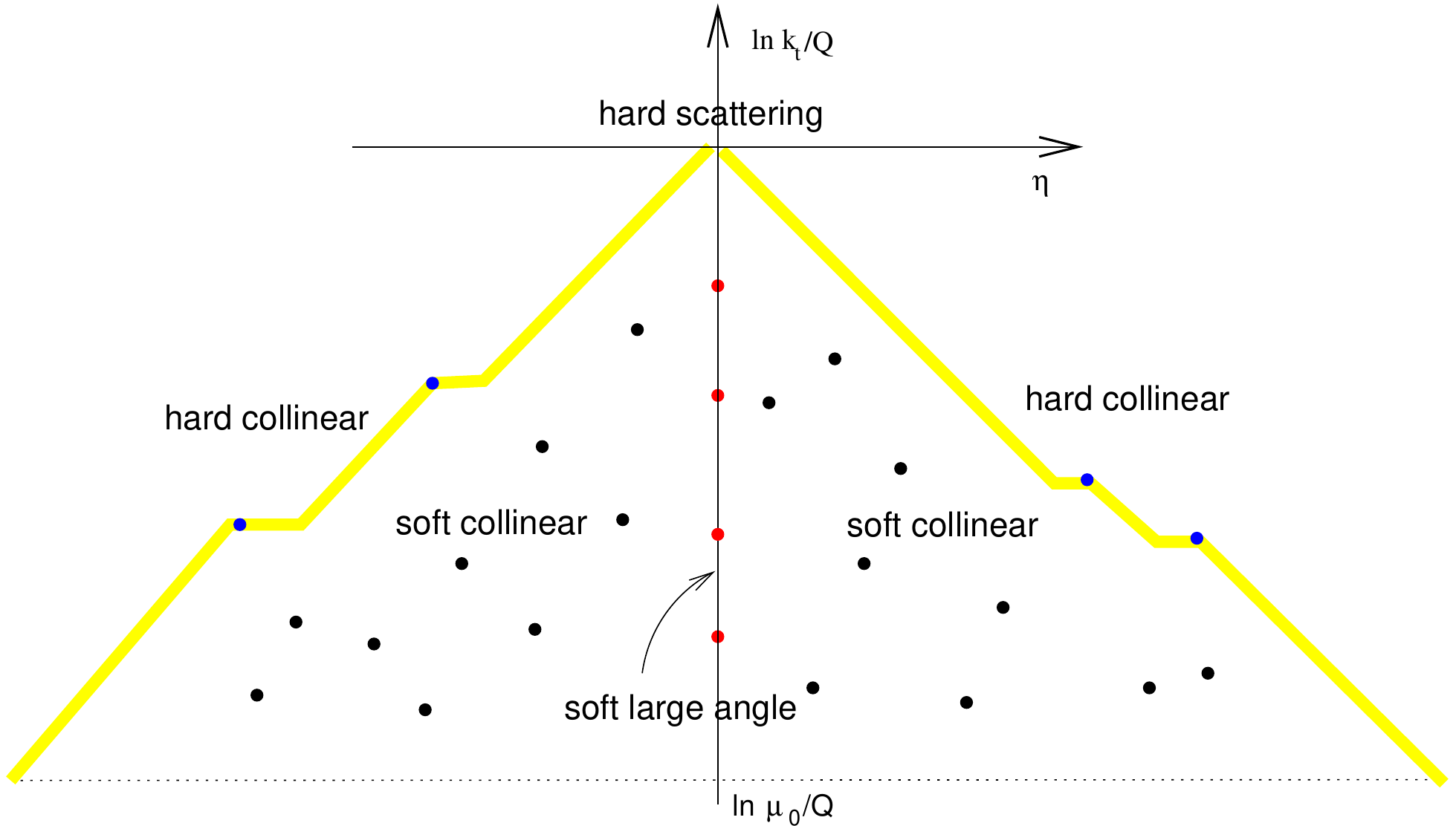, width=.7\textwidth}  
  \caption{The Lund diagram for a generic contribution to the total
    cross section for the production of a vector or Higgs boson at
    hadronic colliders. Here $\mu_0$ is an infrared cut-off and the
    hard scale $Q$ is the boson mass.}
  \label{fig:lozenge-DY}
\end{figure}

The above discussion can be visualised with the help of Lund diagrams.
Consider for instance the well-known example of vector or Higgs boson
production in hadron-hadron collisions. A generic contribution to the
total cross section for this process in the SC limit is illustrated in
fig.~\ref{fig:lozenge-DY}.  Each dot represents an emitted parton,
identified via its rapidity ($\eta$, on the $x$ axis) and transverse
momentum ($\ln k_t/Q$, on the $y$ axis) with respect to the beam. The
hard vertex is the origin of the axes. The yellow bands represent the
collinear limit $\eta < \ln (2E_\ell/k_t)$, where $E_\ell$ is the
energy of emitting hard parton (leg) $\ell=1,2$.  Hard collinear
emissions (blue) are kinks on the two bands, since they reduce the
emitting parton energy by a significant fraction. Soft large-angle
emissions (red) are along the line $\eta=0$, while all remaining black
dots are soft and collinear emissions.  Since for fixed $\alpha_s$
emissions are distributed uniformly in $\ln k_t/Q$ and $\eta$, an area
in the picture corresponds to a double logarithmic contribution, a
line to a single logarithm, while points represent ${\cal
  O}(\alpha_s)$ corrections.

Virtual corrections are universal and can be shown to
exponentiate~\cite{exponentiation}.  The contribution of real
emissions is instead observable dependent and can be represented as a
vetoed region, where real emissions are forbidden, and only virtual
contributions survive. According to the way the veto condition is
imposed one can distinguish between
\begin{enumerate}
\item \emph{inclusive} observables: no hadrons are directly observed,
  QCD radiation is restricted via energy-momentum conservation. In
  this case, after an integral transform, real contributions
  exponentiate to all (logarithmic) orders.  An example of an
  inclusive observable is the cross section for the production of a
  non-QCD particle, for instance a Higgs boson~\cite{inclusive};
\item \emph{final-state} observables: one measures final-state hadron
  momenta, a typical example being event shape distributions and jet
  rates. In this case there is no general statement concerning the
  level of accuracy at which exponentiation holds.
\end{enumerate}
Inclusive observables have been discussed during the workshop by Massimiliano
Grazzini. Here I will concentrate on final-state observables,
introducing specific classes of observables and discussing to what
extent a resummation for those observables is feasible.

\section{Global observables}
\label{sec:global}
Let us consider a global variable that measures hadrons everywhere,
for instance the thrust in $e^+e^-$ annihilation, defined as
\begin{equation}
  \label{eq:thrust}
  T=\max_{\vec n}\frac{\sum_i |\vec p_i \cdot \vec n|}{\sum_i |\vec p_i|}\>.
\end{equation}
The thrust, as all other event shapes, is a measure of the geometrical
properties of hadron energy-momentum flow. For instance, for
pencil-like events we have $T\! \simeq \!1$, for planar events $T\!
\simeq\! 2/3$, while for spherical events $T\!\simeq\! 1/2$. This
variable has the property of infrared and collinear (IRC) safety,
i.e.~its value does not change after emission of extremely soft
particles and/or quasi-collinear splittings. IRC safety is precisely
the condition that ensures that observables related to the thrust,
i.e.~distributions and mean values, can be computed using the
quark-gluon language, in spite of the fact that the variable
definition involves hadrons, and the difference between parton and
hadron level predictions is suppressed by powers of the hard scale
$Q$.

The basic quantity we are interested in is $\Sigma(\tau)$, the
probability that $1\!-\!T\!<\!\tau$, the differential distribution
$\sigma^{-1}d\sigma/d\tau$ being the derivative of $\Sigma(\tau)$.
Fixed order QCD predictions are reliable as long as $\tau$ is large,
but fail as soon as events approach the Born limit $\tau\!=\!0$. In
the small $\tau$ region one needs to resum logarithmic enhanced
contributions to all orders, and the resulting resummed distribution
has the same shape as the data. However, to get on top of the data,
one needs to add a further correction that can be interpreted as the
difference between parton and hadron level. This hadronisation
correction can be estimated using Monte Carlo (MC) event generators
like \textsc{herwig}~\cite{nnherwig},
\textsc{pythia}~\cite{pythia-old,pythia-new} or
\textsc{ariadne}~\cite{ariadne}, taking the ratio of the distributions
obtained with MC's before and after hadronisation, and estimating
hadronisation uncertainties using different event generators. This has
lead to a successful description of IRC safe event shape distributions
and jet rates in $e^+e^-$ annihilation, giving one of the most
accurate measurements of $\alpha_s$ (see~\cite{DS-review} for a recent
review). The validity of this procedure relies strongly on the fact
that MC event generators contain the physics that is needed to
describe the main features of final-state observables.
%and that the characteristic scale of
%hadronisation corrections is much lower than that of PT radiation.
This statement is in general true for variables whose LL
exponentiate, as we shall see in the following.

%This is somewhat confirmed by the fact that the biggest
%uncertainties do not come from hadronisation corrections, but rather
%from missing PT higher orders. This picture breaks down at extremely
%small values of $\tau$, where hadronisation effects dominate, which is
%reflected by large hadronisation uncertainties.
%
%Given the above discussion, it is a crucial issue to determine whether
%a variable distribution can be resummed or not.  There are in fact
%variables that have the very same expression with a single amission
%but have completely different resummation properties. For instance,
%double logarithms do exponentiate for the thrust distribution, but
%they do not for the three-jet rate with the JADE
%algorithm~\cite{JADE}, although they are both IRC safe and the value
%of both variables after a single emission is the same.

\begin{figure}[htbp]
  \begin{minipage}[l]{0.5\linewidth}
     \begin{center}
        \epsfig{file=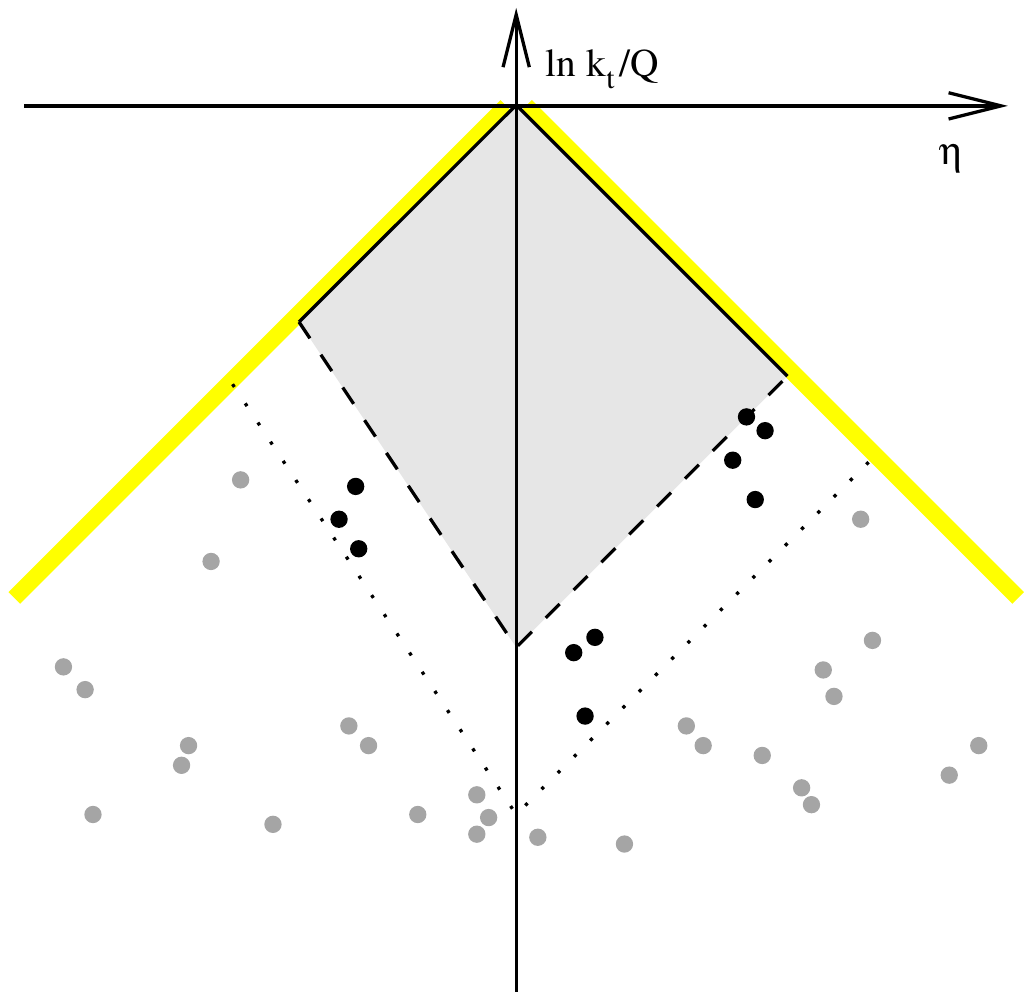,width=\textwidth}    
      \end{center}
  \end{minipage}
  \begin{minipage}[r]{0.5\linewidth}
     \begin{center}
        \epsfig{file=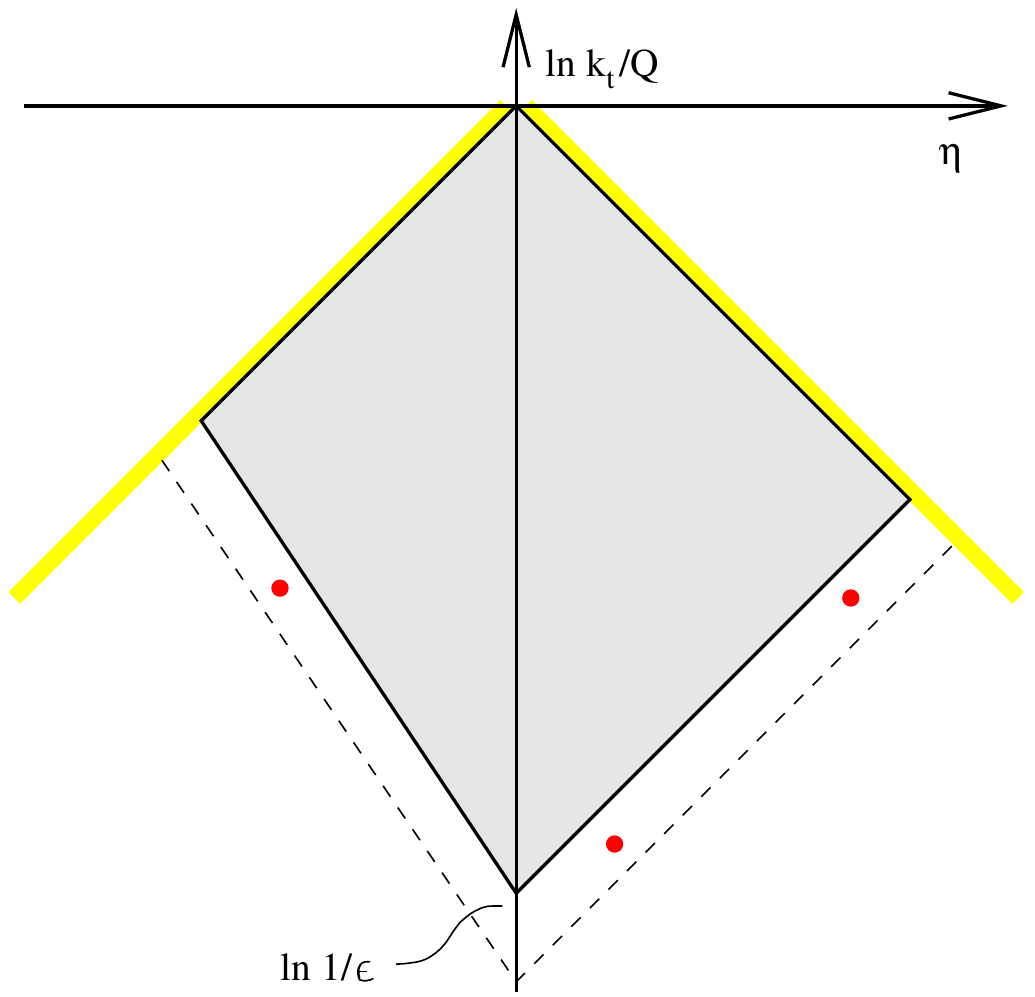,width=\textwidth}    
      \end{center}
  \end{minipage}
  \caption{The Lund diagram for emissions contributing to a generic
    final-state observable (left) and its simplified version in case
    of a rIRC safe observable (right).}
  \label{fig:mc-ff}
\end{figure}

Consider then a generic final-state variable $V(\{k_i\})$, a function
of final-state momenta $\{k_i\}$, and its rate $\Sigma(v)$,
the probability that $V(\{k_i\})\!<\!v$. A generic contribution to
$\Sigma(v)$ can be represented by the Lund diagram on the left hand
side of fig.~\ref{fig:mc-ff}. The grey area corresponds to the vetoed
region where no real emissions are allowed, and only virtual
corrections survive, and one can write in general:
\begin{equation}
  \label{eq:Sigma}
  \Sigma(v) = e^{-R(v)} {\cal F}(v)\>,
\end{equation}
where $R(v)$ is the exponent representing virtual corrections up to
the scale $vQ$, while real emission outside the vetoed region and the
remaining virtual corrections build up the function ${\cal F}(v)$. The
variable $V$ is said to exponentiate if all leading (double)
logarithms are contained in the exponent $R(v)$ and ${\cal F}(v)$ is a
pure NLL function, usually denoted by ${\cal F}(R')$, with
$R'(v)=-vdR/dv$.  There are two basic conditions for this to happen,
which go under the name of recursive infrared and collinear (rIRC)
safety conditions~\cite{caesar}:
\begin{enumerate}
\item the variable must scale in the same fashion with multiple
  emissions as with a single emission. Formally, parametrising
  the momentum $k_i$ of each emission in terms of $V(k_i)$,\footnote{
    For simplicity, we will always write $V(k_1,\dots,k_n)$ instead of
    the more correct form $V(\{\tilde p\},k_1,\dots,k_n)$, where
    $\{\tilde p\}$ denotes final-state hard parton momenta.} the value
  the variable $V$ would have if only emission $k_i$ were present, and
  defining $V(k_i)=\bar v \zeta_i$, the first rIRC safety condition
  states that the following limit:
  \begin{equation}
    \label{eq:rIRC-1}
    \lim_{\bar v\to 0} \frac{V(k_1(\bar v \zeta_1),\dots,k_n(\bar v \zeta_n))}
    {\bar v}
  \end{equation}
  has to be finite and non-zero. This ensures that the boundary of
  the vetoed region in fig.~\ref{fig:mc-ff} does not change
  substantially whatever is the number of emissions;
\item the variable's scaling property~(\ref{eq:rIRC-1}) must not be
  altered by the addition of extra-soft particles or by
  quasi-collinear splittings, formally:
  \begin{equation}
    \label{eq:rIRC-2}
    [\lim_{\zeta_{n+1}\to 0},\lim_{\bar v\to 0}] 
    \frac{V(k_1(\bar v \zeta_1),\dots,k_n(\bar v \zeta_n),
      k_{n+1}(\bar v \zeta_{n+1}))}
    {\bar v} = 0\>,
%\lim_{\bar v\to 0} 
%    \frac{V(k_1(\bar v \zeta_1),\dots,k_n(\bar v \zeta_n))}
%    {\bar v} 
  \end{equation}
  where the only non-trivial part of the commutator is the one where
  one takes the limit $\zeta_{n+1}\!\to\! 0$ after the limit $\bar v\!\to
  \!0$, the other part being equal to the limit in eq.~(\ref{eq:rIRC-1})
  due to IRC safety. An analogous condition should hold also for
  collinear splittings.
%  Notice the order of the limits, first one rescales all emissions,
%  and then one eliminates the softest emission.
  This implies that in fig.~\ref{fig:mc-ff} one can eliminate all
  emissions in grey, far from the boundary of the vetoed region,
  without altering the value of $V$, and for all emissions with
  $V(k_i)\sim V(k_1,\ldots,k_n)$ one can replace clusters of emissions
  close in rapidity with a single emission having the total momentum
  of the cluster.
\end{enumerate}
In the end, for a rIRC safe variable, a generic contribution to
$\Sigma(v)$ can be represented by a Lund diagram like the one on the
right hand side of fig.~\ref{fig:mc-ff}, where one has a vetoed area,
giving rise to the LL exponent $R(v)$, and real emissions contributing
to ${\cal F}(R')$ at NLL accuracy are both soft and collinear, well
separated in rapidity and confined in a narrow region of width $\ln
1/\epsilon \ll \ln 1/v$ close to the boundary of the vetoed area. The
fact that emissions are well separated in rapidity makes it possible
to exploit QCD coherence, and consider soft gluons as radiated
independently (like in QED) from the hard legs.  This simplification
of multi-gluon soft matrix elements makes it possible to compute
${\cal F}(R')$ with a MC procedure, where emissions are ordered in
$V(k_i)=\bar v \zeta_i$, with $\zeta_i<\zeta_{i-1}$, $V(k_1)$ is fixed
at $\bar v$, and the probability of emission of gluon $k_i$ collinear
to leg $\ell$ with rapidity $\eta_i$ and azimuth $\phi_i$ (with
respect to leg $\ell$) is
\begin{equation}
  \label{eq:dPell}
  dP(k_i(\zeta_i,\eta_i, \phi_i,\ell))=
  R'_\ell\>\frac{d\eta_i}{\Delta
    \eta_i}\>\frac{d\phi_i}{2\pi}\>
  \frac{d\zeta_i}{\zeta_i}
  \left(\frac{\zeta_i}{\zeta_{i-1}}\right)^{R'}\,,
  \qquad 
  \sum_\ell R'_\ell = R'.
\end{equation}
The function ${\cal F}(R')$ can then be computed as the following average:
\begin{equation}
  \label{eq:F(R')}
  {\cal F}(R')= 
  \left\langle\lim_{\bar v\to 0}
    \left(\frac{V(k_1, \ldots, k_n)}{V(k_1)}\right)^{-R'}
  \right\rangle\>,
\end{equation}
where the limit $\bar v\to 0$ ensures that the result contains no
NNLL contributions. Eq.~(\ref{eq:F(R')}) is an example of
application of MC techniques used in parton shower event generators to
obtain exact QCD results, and is one of the building blocks of the
automated resummation program \textsc{caesar}~\cite{caesar}.

We can now discuss what level of accuracy can be achieved by parton
shower event generators. MC parton showers produce emissions in the
whole of the phase space (for instance all emissions in the diagram on
the left hand side of fig.~\ref{fig:mc-ff}), with approximated matrix
elements that are exact in the collinear limit but mistreat the
soft large-angle region, both because they do not have full
interference terms, and because they are correct only in the
large-$N_c$ limit. However, for rIRC safe observables, LL and NLL
contributions are determined only by emissions that are well separated
in rapidity and close to the boundary of the vetoed region. MC event
generators correctly describe such emissions, so that one expects
that they reproduce not only LL, but most NLL contributions to rIRC
safe observables.

\section{Non-global observables}
\label{sec:non-global}
\begin{figure}[htbp]
  \centering \epsfig{file=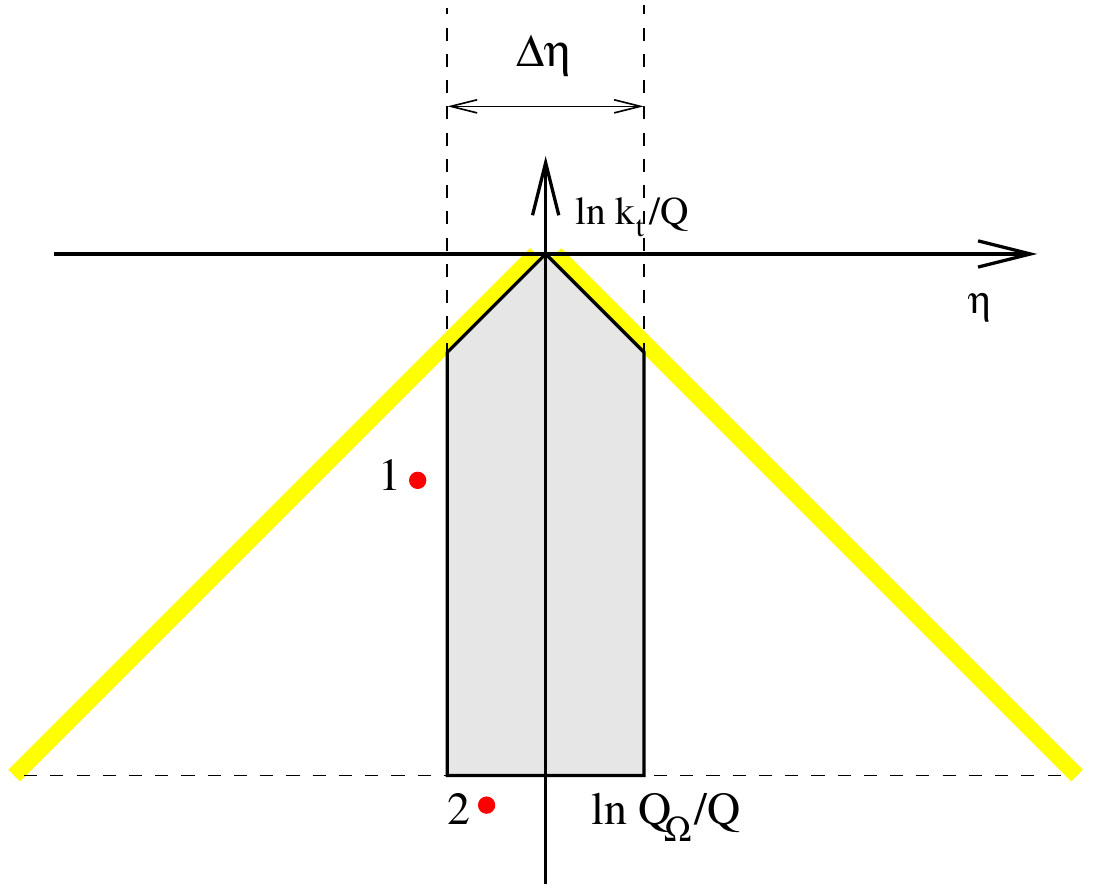,width=0.46\textwidth}
  \caption{A configuration giving rise to non-global logarithms. The
    region $\Omega$ corresponds to a rapidity gap of width
    $\Delta\eta$ between the jets.  }
\label{fig:nonglobal}
\end{figure}

Non-global variables measure hadrons in a restricted part of the phase
space. The most relevant example is the hadron transverse energy flow
in a region $\Omega$ away from the hard jets, defined as~\cite{etflow}
\begin{equation}
  \label{eq:etflow}
  E_t = \sum_{i\in\Omega} E_{ti}\>,\qquad\Sigma(Q_\Omega) = 
  \int_0^{Q_\Omega} \!\!\! dE_t \frac{1}{\sigma}\frac{d\sigma}{dE_t}\>.
\end{equation}
The distribution $\Sigma(Q_\Omega)$ is sensitive only to soft
emissions at large angles, so that LL are single logarithms
$\alpha_s^n L^n$, with $L=\ln Q/Q_{\Omega}$. To achieve a LL
resummation for $\Sigma(Q_\Omega)$ one cannot rely on an independent
emission picture of QCD radiation, because configurations like the one
in fig.~\ref{fig:nonglobal}, where one has two emissions close in
rapidity and the $k_t$ of the harder emission can take any value from
$Q$ to $Q_\Omega$, give a single logarithm~\cite{etflow}.
These non-global logarithms are ruled by the correlated part of soft
gluon emission matrix elements, and can be resummed only in the
large-$N_c$ limit.  Furthermore, due to the fact that one needs to
take into account the exact form of $\Omega$ for an arbitrary number
of emissions, resummation can be performed only numerically. In the MC
procedure of ref.~\cite{etflow}, emissions are ordered in the variable
$t=\int_{k_t}^Q \frac{dk}{k}\frac{\alpha_s(k)}{2\pi}\sim \alpha_s\ln
Q/k_t$, with $t_i>t_{i-1}$, and, given an ensemble of $n\!-\!1$ colour
connected soft gluons forming $n$ dipoles, a softer gluon $k_i$ is
emitted from dipole $j$ with probability:
\begin{equation}
  \label{eq:dPsoft}
  dP(k_i(t_i,\eta_i,\phi_i,j)) =  2 C_A d\eta_i\>
  \Theta\left(\frac{\Delta \eta_j}{2}-|\eta_i|\right)\>
  \frac{d\phi_i}{2\pi} \>dt_i\>
  e^{-2 C_A \Delta\eta_{\mathrm{tot}}(t_i-t_{i-1})}\>,
\end{equation}
where $\eta_i$ and $\phi_i$ are the rapidity and azimuth of emission
$k_i$ in the frame in which partons forming dipole $j$ are
back-to-back, $\Delta \eta_j$ represents a collinear cutoff and
$\Delta \eta_{\mathrm{tot}}\!=\!\sum_j \Delta\eta_j$. The MC starts
from a quark-antiquark dipole, and continues emitting gluons as long
as one emission falls into $\Omega$. If one bins the value of $t$
corresponding to this last emission, when the MC stops one has
reconstructed $d\Sigma(t)/dt$ at LL accuracy, with $t\!\sim\! \alpha_s\ln
Q/Q_\Omega$.

Given the fact that the away-from-jet $E_t$ flow is used to tune MC
models of the underlying event in hadron-hadron collisions, it is
crucial to understand whether MC parton shower generators are able to
describe $\Sigma(Q_\Omega)$. If this is not the case, missing PT
contributions, which are non-universal, will be included by tuning
hadronisation and underlying event parameters, which are supposed to
be universal. In ref.~\cite{etflow-mc} it was first investigated the
difference between a full treatment of soft radiation as in
eq.~(\ref{eq:dPsoft}), and an angular ordering (AO) approximation
resulting from a free azimuthal average of $dP(k_i)$ (see
~\cite{etflow-mc} for details).  The difference between the full and
the AO distributions was found to be less than 10\% in the whole range
of values of $t$ relevant for phenomenology ($t\!<\!0.15$,
corresponding to $Q_\Omega\!>\!1$GeV at the LEP1 energy), thus
supporting the idea that MC generators implementing angular ordering
should be able to give a reasonable description of $\Sigma(Q_\Omega)$.
Then a direct comparison of a LL resummation for $\sigma^{-1}
d\sigma/dE_t$ in $e^+e^-$ annihilation and the corresponding
predictions obtained with \textsc{herwig} and \textsc{pythia} at
parton level was performed, the away-from-jet region $\Omega$ being a
rapidity gap of width $\Delta \eta$ between the two jets.  The
comparison showed that \textsc{herwig}, whose evolution variable is
the angle of each branching, is in good agreement with the LL
resummation, discrepancies being less than 10\%, both for large and
small gaps.  On the contrary, the old \textsc{pythia}
shower~\cite{pythia-old}, which uses the virtuality as an evolution
variable and rejects a posteriori configurations not respecting AO,
lies below the resummation, about 50\% less at $t=0.15$. The new
\textsc{pythia} shower~\cite{pythia-new} is in agreement with the
resummation for small gaps, $\Delta\eta=1$, while for large gaps,
$\Delta\eta=3$, the two distributions have different shapes, and the
point where they start to deviate seems to be exponentially related to
the gap size.  While it is known that the AO requirement in the old
\textsc{pythia} shower places too strong a veto on soft emissions,
this problem should be overcome with the new \textsc{pythia} shower.
Therefore the origin of the disagreement is unclear and needs further
investigation.

Unlike rIRC safe global observables, resummation of non-global
observables cannot rely on general approximations for soft radiation.
For instance, the distribution for the away-from-jet hadron energy
flow in eq.~(\ref{eq:etflow}) has the form $\Sigma(t) =
\exp[-R(t)]S(t)$, where $S(t)$ contains non-global logarithms. Real
emissions from the hard legs cancel with virtual corrections at scales
larger than $t$, leaving only the exponent $R(t)$. Therefore within LL
accuracy there is no analogous of the function ${\cal F}(R')$ of
eq.~(\ref{eq:F(R')}). This might be argued from the fact that for
global observables ${\cal F}(R')$ is sensitive only to soft and
collinear emissions, while here all relevant emissions are soft and at
large angles. This however is not the case if one considers the
away-from-jet energy flow of mini-jets, defined by replacing the sum
over hadron with the sum over jets. This observable was introduced to
reduce the impact of non-global logarithms, because emissions close in
rapidity tend to be clustered together by jet
algorithms~\cite{etflow-AS}.  Unfortunately, recombination spoils the
complete real-virtual cancellation of primary emissions, so that one
needs to introduce a new correction factor $C(t)$~\cite{etflow-BD}.
Also the statement that inter-jet energy flows are sensitive only to
soft large-angle emissions might not be completely true. In
ref.~\cite{superleading} it has been shown that, in hadron-hadron
collisions, if one assumes that $k_t$ is the ordering variable for
virtual corrections, a gluon outside the gap collinear to one of the
incoming legs and emitting a softer gluon inside the gap gives a
double-logarithmic contribution if it is accompanied by two-loop
non-cancelling Coulomb phases. This higher order contribution to
$\Sigma(t)$ is $\alpha_s^4 \pi^2 \Delta\eta L^5$, representing thus a
super-leading logarithm. The presence of super-leading logarithms
depends on the ordering variable used for virtual corrections, so that
at the moment one cannot clearly state whether these logarithms are
actually present, and in case they are how they can be resummed to all
orders.

\section{Conclusions and outlook}
\label{sec:conclusion}
Final-state observables are very sensitive to QCD radiation. They are
therefore extremely valuable tools to test our understanding of QCD
dynamics. In particular, comparison of data, resummed analytical
calculations and MC simulations can be used to improve our description
of multi-soft gluon radiation. We hope that such a study, which was
very successful at LEP, will continue at hadron colliders.

One research direction could be then trying to use global rIRC safe
observables for the tuning of models of the underlying event. The
advantage of this class of observables is that they are correctly
described by MC event generators at LL accuracy. This is not the case
for the variable that is traditionally used for this purpose, the
away-from-jet energy flow, which is instead non-global.  However,
globalness may be a problem at hadron colliders, since one cannot
measure hadrons too close to the beam pipe. This problem has been
addressed in ref.~\cite{hh-shapes}, where one can find a list of
global rIRC safe event-shapes and jet resolution parameters in hadronic
dijet production.  We look forward to experimental investigations in
this direction.

%Once it has been established what kind of observables can be
%reasonably describes by MC event generators, one can even try to
%investigate event-shape sensitivity to new physics. For instance, one
%naively expects different distributions in the heavy-jet mass
%according to whether one has a pure QCD event or the production of a
%heavy supersymmetric particle.

Concerning non-global logarithms, in hadronic collisions they appear
in a variety of contexts.  First of all, they contribute significantly
to the distribution of the $E_t$ flow away from the hard jets, for
which even a LL calculation is missing.
Furthermore, non-global logarithms give NLL effects in distributions
of non-global event shapes such as the ones defined in~\cite{nagy},
which measure emissions only in a central rapidity region.  These
variables are preferred from an experimental point of view, since
hadron momenta can be measured by combining central tracker and
calorimeter information, thus considerably reducing systematic
uncertainties.
Non-global logarithms will appear also in more inclusive
distributions, like jet transverse momentum spectra or dijet azimuthal
correlations. Theoretical studies in these directions are in progress.

\addtocounter{chapter}{1}
\newcommand{\msbar}{$\overline{\mbox{MS}}$ }
%\usepackage{setspace}
%%%%%%%%%%%%%%%%%%%%%%%%%%%%%%%%%%%%%%%%%%%%%%%%%%%%%%%%%%
%\begin{document}
%%%%%%%%%%%%%%%%%%%%%%%%%%%%%%%%%%%%%%%%%%%%%%%
% Toggle line numbering
% Won't work with the PRD revtex4 !
%\pagewiselinenumbers
% uncomment if you want doublespace
%\doublespace
%%%%%%%%%%%%%%%%%%%%%%%%%%%%%%%%%%%%%%%%%%%%%%%
\mchapter{Bottom-quark fragmentation: resummations and Monte Carlo
simulations}{G. Corcella}

%%%%%%%%%%%%%%%%%%%%%%%%%%%%%%%%%%%%%%%%%%%%%%%%%%%%%%%%%
Heavy-quark phenomenology is currently one of the main 
fields of investigation in theoretical and experimental
particle physics.
In the following, we shall study $B$-hadron production in
$e^+e^-\to b\bar b$ annihilation, top ($t\to bW$) and Higgs
($H\to b\bar b$) decays. 
We will describe $b$-quark production using resummed calculations as well
as Monte Carlo generators, and get non-perturbative information by
tuning hadronization models to experimental data from 
SLD \cite{sld} and LEP \cite{aleph,opal,delphi}.
We shall also use a recently proposed non-perturbative model 
\cite{shirkov,ugo,gianca}, which includes power corrections in an effective
strong coupling constant and does not introduce any further 
tunable parameter.
We first consider $b\bar b$ 
pair production at the $Z^0$ pole in the 
next-to-leading order (NLO) approximation,
\begin{equation}
e^+e^-\to Z^0(q) \to b(p_b) \bar b(p_{\bar b}) \left( g(p_g)\right),
\end{equation}
and define the $b$-quark energy fraction
\begin{equation}
x_b={{2p_b\cdot q}\over{q^2}}.
\label{xb}
\end{equation}
The energy spectrum of a massive $b$ quark is given by:
 \begin{equation}
{1\over{\sigma_0}}{{d\sigma}\over {dx_b}}=\delta(1-x_b)
+{{\alpha_S}\over{2\pi}}
\left[P_{qq}(x_b)\ln{{m_Z^2}\over{m_b^2}}+A(x_b)\right] +
{\cal O} \left( {{m_b^2}\over{m_Z^2}}\right)^p  ,
\label{massb}
\end{equation}
where $\sigma_0$ is the cross section of the Born process $e^+e^-\to q\bar q$,
$p\geq 1$,
$A(x_b)$ is a function independent of the $b$ mass, 
$P_{qq}(x_b)$ is the Altarelli--Parisi (AP) splitting function.
%\begin{equation}
%P_{qq}(x_b)=C_F\left( {{1+x_b^2}\over {1-x_b}}\right)_+.
%\label{pqq}
%\end{equation} 
The large logarithm $\sim\alpha_S\ln(m_Z^2/m_b^2)$ can be resummed 
by the use of the perturbative fragmentation formalism.

Following \cite{mele}, 
the  $b$ spectrum is expressed as the convolution of a coefficient
function, corresponding to the emission from a massless parton, 
and a perturbative fragmentation function $D(m_b,\mu_F)$,
associated with the transition of a massless parton into a heavy $b$:
\begin{eqnarray}
{1\over {\sigma_0}} {{d\sigma_b}\over{dx_b}} (x_b,m_Z,m_b) &=&
\sum_i\int_{x_b}^1
{{{dz}\over z}\left[{1\over{\sigma_0}}
{{d\hat\sigma_i}\over {dz}}(z,Q,\mu,\mu_F)
\right]^{\overline{\mathrm{MS}}}
D_i^{\overline{\mathrm{MS}}}\left({x_b\over z},\mu_F,m_b \right)} \nonumber \\
&+& {\cal O}\left((m_b/m_Z)^p\right) \; .
\label{pff}
\end{eqnarray}
In Eq.~(\ref{pff}), $d\hat\sigma_i /dz$ is the coefficient function for the
production of a massless parton $i$, 
after subtracting the collinear singularity in the \msbar factorization
scheme. Neglecting $g\to b\bar b$ splitting, 
$i=b$ and $D_b^{\overline{\mathrm{MS}}}$ expresses the fragmentation
of a massless $b$ into a massive $b$.
%The NLO $e^+e^-\to q\bar q$ \msbar coefficient function can be read from the
%formulas in \cite{mele}.

The perturbative fragmentation function follows the 
DGLAP evolution equations
\cite{ap,dgl} and its value at a  scale $\mu_F$ can be obtained once
an initial condition at $\mu_{0F}$
is given. In \cite{mele} the NLO initial condition
$D_b^{\rm ini}(x_b,\mu_{0F},m_b)$ was calculated and its process-independence 
was established in \cite{cc}.
%It is given at NLO by:
%\begin{equation}
%D_b^{\rm ini}(x_b,\mu_{0F},m_b)=\delta(1-x_b)+
%{{\alpha_S(\mu_0^2)C_F}\over{2\pi}}
%\left[{{1+x_b^2}\over{1-x_b}}\left(\ln {{\mu_{0F}^2}\over{m_b^2}}-
%2\ln (1-x_b)-1\right)\right]_+.
%\label{dbb}
%\end{equation}
Solving the DGLAP equations for an evolution
from $\mu_{0F}$ to $\mu_F$, with a NLO kernel, allows one to resum leading (LL)
$\alpha_S^n\ln^n(\mu_F^2/\mu_{0F}^2)$ and next-to-leading (NLL) 
$\alpha_S^n \ln^{n-1}(\mu_F^2/\mu_{0F}^2)$ logarithms (collinear
resummation).
%The explicit expression for the solution of the DGLAP equations can be found,
%for instance, in Ref.~\cite{mele}.
Setting $\mu_{0F}\simeq m_b$ and $\mu_F\simeq m_Z$, one succeeds in resumming
the logarithms
$\ln(m_Z^2/m_b^2)$ appearing in the massive spectrum (\ref{massb}).

%The next-to-next-to-leading order
%(NNLO) initial condition of the perturbative fragmentation function
%was calculated in Refs.~\cite{mitov1}; following \cite{cc,ugo},
%we shall however use it in the NLO approximation and delay the inclusion
%of NNLO corrections to future work. Moreover, the recent calculation of
%the NNLO time-like splitting functions \cite{mitov2}, 
%entering in the kernel of the
%DGLAP equations, may also allow NNLL collinear resummation.

Furthermore, both initial condition and coefficient 
function \cite{mele} present terms, $\sim 1/(1-x_b)_+$ and
$\sim [\ln(1-x_b)/(1-x_b)]_+$, which become large for $x_b\to 1$.
The large-$x_b$ limit corresponds to soft-
or collinear-gluon radiation. Such contributions are usually resummed
in Mellin moment space, where they correspond, at ${\cal O}(\alpha_S)$, to
single ($\sim\alpha_S\ln N$) and double ($\sim\alpha_S\ln^2 N$) logarithms of
the Mellin variable $N$ (soft or threshold resummation).
In \cite{cc} threshold resummation was implemented in the
NLL approximation; in \cite{ugo} even large-$N$ NNLL contributions
were resummed.
To NNLL accuracy, terms $\sim \alpha_S^n\ln^{n+1}N$  (LL),
$\sim \alpha_S^n\ln^nN$ (NLL) and  $\sim \alpha_S^n\ln^{n-1}N$
are kept in the resummed exponent.
%In Refs.~\cite{cc,ugo},
%soft and collinear resummations 
%are matched to the exact NLO results, in order to 
%deal with a reliable prediction over the full $x_b$ ($N$) range. This implies
%that the total widths are NLO as well.

As for Monte Carlo parton shower algorithms, implemented in 
event generators such as HERWIG \cite{her} and PYTHIA \cite{nnpythia},
they rely on the universality of the elementary branching
probability for soft or collinear radiation. 
Referring, e.g., to parton cascades in $e^+e^-\to q\bar q$ processes,
the probability of soft or collinear 
emission reads:
\begin{equation}
  \label{elem}
  dP={{\alpha_S}\over{2\pi}}{{dQ^2}\over{Q^2}}\ 
  P(z)\  dz\ 
 {{ \Delta_S(Q^2_{\mathrm{max}},Q^2)}\over {\Delta_S(Q^2,Q_0^2)}}.
\end{equation}
In (\ref{elem}) $P(z)$ is still the AP splitting
function,
$z$ is the energy fraction of the emitted parton, 
$Q^2$ is the shower ordering variable.

In HERWIG, $Q^2$ is an energy-weighted angle, equivalent 
to angular ordering in soft limit \cite{nmarweb}. 
In PYTHIA \cite{nnpythia}, $Q^2$ is the momentum squared of
the radiating parton, with an option to veto branchings that do not
fulfil angular ordering.
Moreover, the latest PYTHIA version
offers, as an alternative, the possibility to order 
final-state showers according to the transverse momentum of the emitted
parton with respect to the emitter
\cite{skands}. It was found out \cite{banfi}
that the PYTHIA transverse-momentum-ordered 
showers yield a better treatment of angular ordering, although its 
implementation is still not as accurate as it is in HERWIG.
Hereafter,
we shall use PYTHIA 6.220, whose cascades are ordered in virtuality
with the option to 
reject non-angular-ordered showers turned on,
and the version 6.506 of HERWIG.
In (\ref{elem}) $\Delta_S(Q_1^2,Q_2^2)$ is the Sudakov form
factor, expressing the probability of evolution from $Q_1^2$ to $Q_2^2$ 
with no resolvable emission. 
%The ratio of form factors
%in (\ref{elem}) represents the probability that the considered emission
%is the first, i.e. there is no emission
%between $Q^2$ and $Q^2_{\mathrm{max}}$, where $Q^2_{\mathrm{max}}$ is set by
%the hard-scattering process. $Q_0^2$ is the shower cutoff, i.e.
%value of $Q^2$ at which the evolution is terminated.
%In diagrammatic terms, the Sudakov form factor
%sums up all virtual and unresolved real emissions to all orders.

For multiple emissions,
iterating the branching probability (\ref{elem}) allows one to resum
soft- and collinear-enhanced radiation: as discussed in \cite{ncmw},
parton shower algorithms resum leading logarithms in the
Sudakov exponent, and include a class of subleading NLLs as well.

%Moreover, both HERWIG and PYTHIA 
%yield the leading-order total cross section for all implemented processes. 
%The more recent `Monte Carlo
%at next-to-leading order' program (MC@NLO) \cite{mcnlo} 
%implements both real
%and virtual corrections to the hard interaction, in such
%a way that predicted observables, including total cross sections,
%are correct to NLO accuracy.

Calculations based on the perturbative fragmentation formalism are
supplemented by non-perturbative fragmentation functions 
to yield hadron spectra.
Up to power corrections, the $B$-hadron spectrum reads:
\begin{equation}
{1\over {\sigma}} {{d\sigma_B}\over{dx_B}} (x_B,Q,m_b)={1\over{\sigma}}
\int_{x_B}^1 {{{dz}\over z}{{d\sigma_b}\over {dz}}(z,Q,m_b)
D^{\mathrm{np}}\left({x_B\over z}\right)},
\label{npff}
\end{equation}
where $x_B$ is the $B$ energy fraction and $D^{\rm np}$ the
non-perturbative fragmentation function.
In the following, we shall use the the NLO/NLL perturbative calculation
in Ref.~\cite{cc} along with Kartvelishvili model \cite{kart}:
\begin{equation}
D^{\mathrm{np}}(x;\gamma)=(1+\gamma)(2+\gamma) (1-x) x^\gamma,
\label{kk}
\end{equation}
and fit $\gamma$ to experimental data. 

We shall also use a non-perturbative model, 
based on an extension of \cite{shirkov},
consisting in including power corrections in an
effective coupling constant.
Such a model was presented in
detail in Refs.~\cite{ugo,gianca}; here we
just point out that it employs the following effective coupling constant:
\begin{equation}
\tilde\alpha_S(k^2) = \frac{i}{2 \pi} \, \int_0^{k^2} d s 
\ {\rm Disc}_s\  \frac{ \bar\alpha_S(-s) }{ s },
\label{time}
\end{equation}
where in the integrand function one sets:
\begin{equation}
\bar\alpha_S(k^2)= 
\frac{1}{2\pi i}
\int_0^{\infty}  \, \frac{ds}{s+k^2} \, 
{\rm Disc}_s \, \alpha_S(-s),
\label{space}
\end{equation}
with $\alpha_S$ being the standard coupling constant. $\tilde\alpha_S(k^2)$
and $\bar\alpha_S(k^2)$ are usually called \cite{ugo} time-like and
space-like effective coupling constants, respectively.
As discussed in \cite{ugo,gianca}, one can prove that 
the effective coupling $\tilde\alpha_S(k^2)$
is free from the Landau pole and that 
at small momenta it includes power-suppressed
effects. 
Also, it is remarkable that the model based on
Eq.~(\ref{time}) does not introduce any extra tunable parameter.
As in \cite{ugo}, the effective coupling
will be implemented in the NNLO approximation, and used
along with a calculation based on the perturbative
fragmentation formalism, with NLO coefficient function and initial
condition, NLL DGLAP evolution and NNLL threshold resummation.

As far as HERWIG and PYTHIA are concerned, their parton showers 
terminate when a scale $Q_0$, of the order
of 1 GeV, is reached. The hadronization is simulated according to 
the cluster \cite{ncluster} and string \cite{nstring} models, respectively.

We shall now consider data on $B$-hadron spectra at the $Z^0$ pole, collected
by the SLD \cite{sld}, OPAL \cite{opal} and ALEPH \cite{aleph} 
collaborations. The ALEPH data contain only $b$-flavoured mesons, the
OPAL and SLD ones a small fraction of baryons as well.
As in \cite{ugo,cv}, when using resummed calculations, we limit the
comparison to $x_B\leq 0.92$, in order to avoid very large-$x_B$ data,
where our computation is still not completely reliable and the 
spectra become negative or oscillate.
We convolute the NLO/NLL calculation of \cite{cc} with the Kartvelishvili
model (\ref{kk}) and find that, in the considered range, the best fit 
is obtained for $\gamma=17.178\pm 0.303$,
with $\chi^2/\mathrm{dof}=46.2/53$.

The effective-coupling model does not have any free parameter, 
but nonetheless in \cite{ugo} all quantities entering in the perturbative 
computation were varied within conventional ranges, in order to
gauge the theoretical uncertainty on the prediction.
For the sake of brevity, we do not present here all the 
plots shown in \cite{ugo};
we just point out that the best comparison with the data is obtained 
for $\mu_{0F}=m_b/2$, where $\mu_{0F}$ is the factorization scale 
in the initial condition of the perturbative fragmentation function.
We obtain $\chi^2/\mathrm{dof}= 103.0/54$, which is
a quite reasonable value, since we are not tuning any parameter.
It was also shown in Ref.~\cite{ugo} that  setting
$\mu_{0F}=m_b$ and $m_b=5.3$~GeV, a mass value characteristic of
a $B$ meson, leads to an excellent description of the ALEPH data, with
$\chi^2/\mathrm{dof}=11.9/16$.

\begin{figure}
\centerline{\resizebox{0.6\textwidth}{!}{\includegraphics{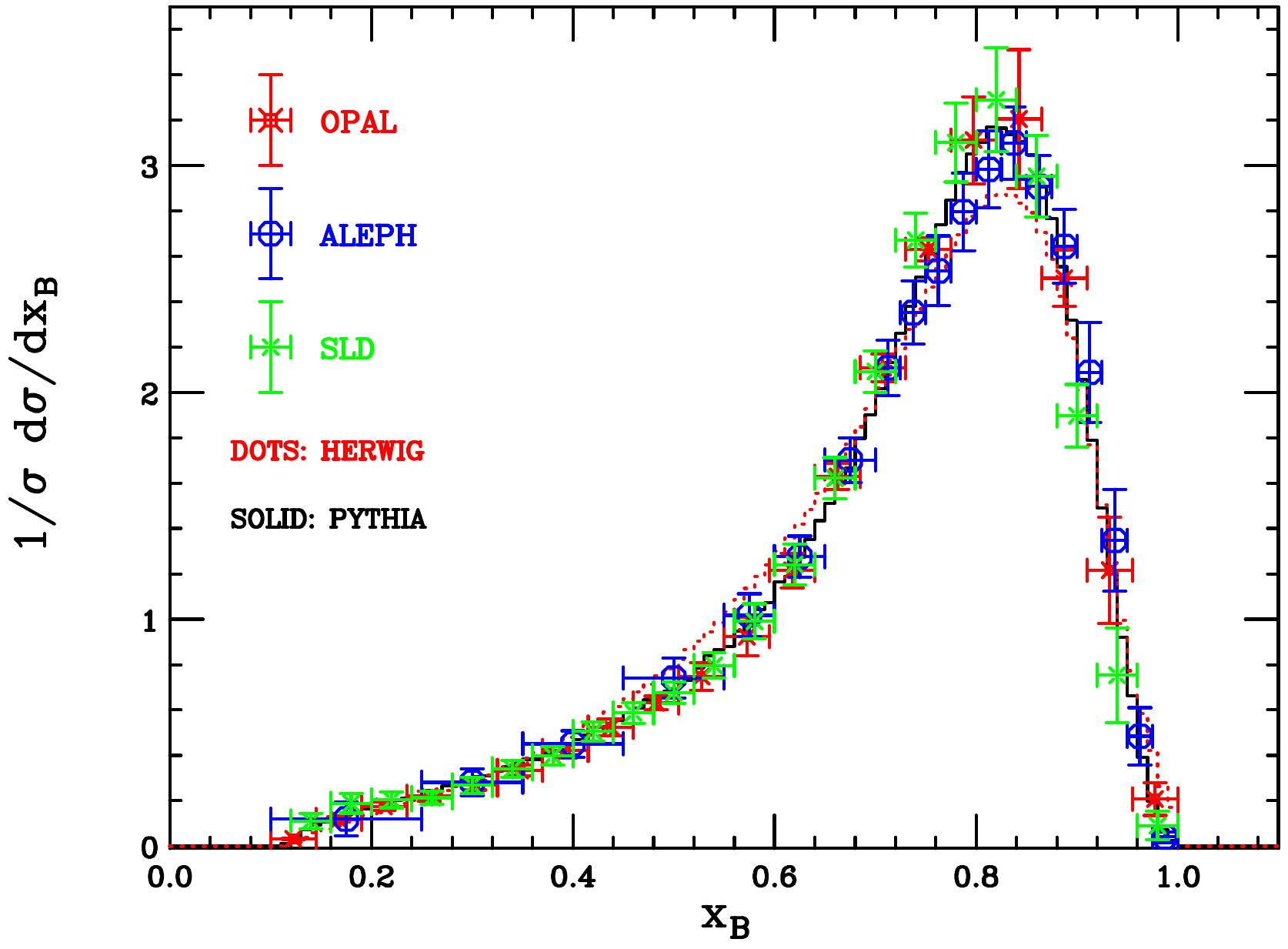}}}
\caption{\small $B$-hadron spectrum measured by ALEPH, OPAL and SLD
experiments, along with the HERWIG and PYTHIA predictions,
after fitting the hadronization parameters to the data.}
\label{hp}
\end{figure}

\begin{figure}
\centerline{\resizebox{0.6\textwidth}{!}{\includegraphics{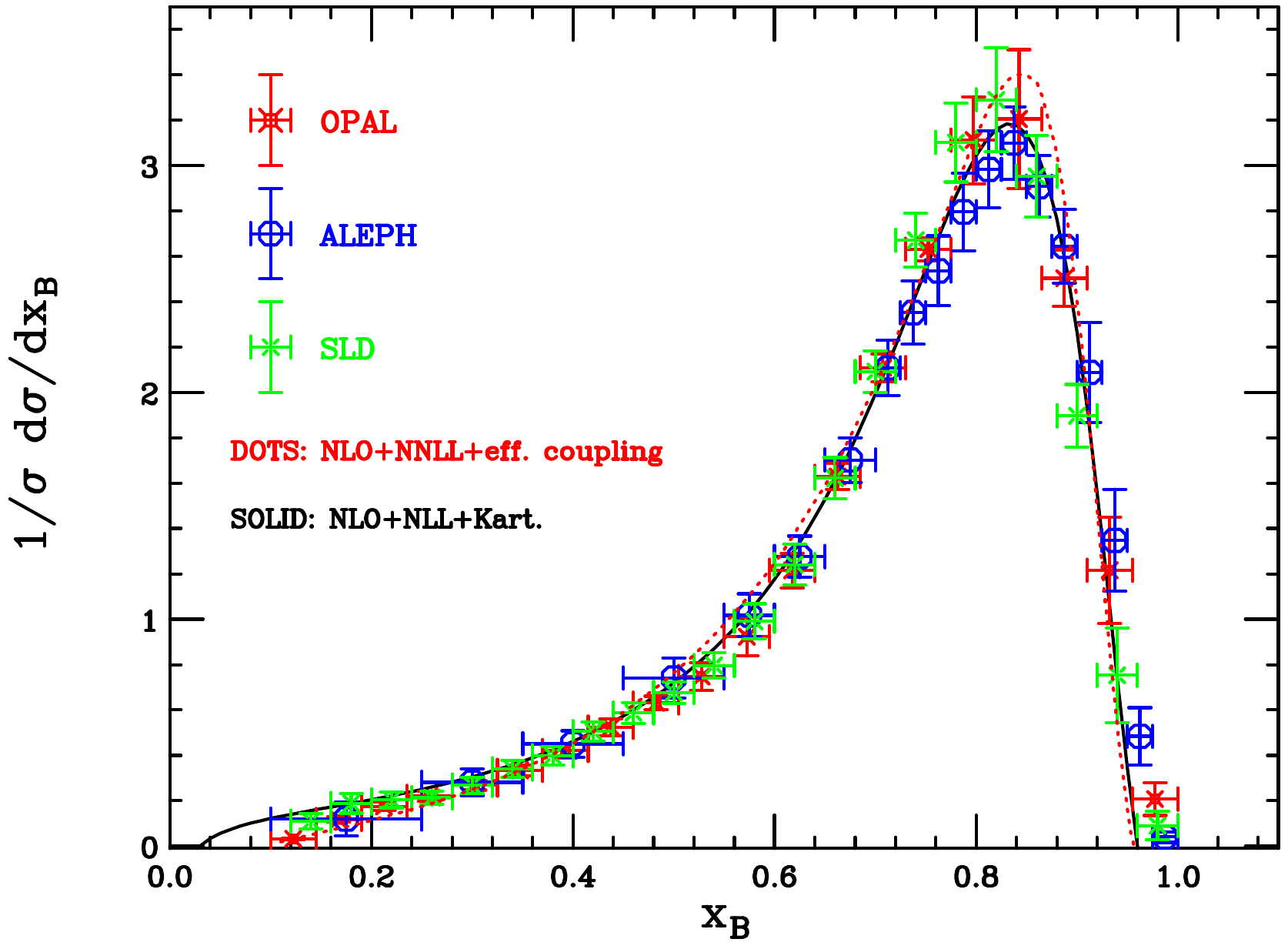}}}
\caption{\small As in Fig.~\ref{hp}, but comparing the data with
the NLO/NLL calculation using the Kartvelishvili hadronization
model and the NLO/NNLL computation with the effective strong
coupling constant.}
\label{keff}
\end{figure}

For the purpose of HERWIG and PYTHIA, as in \cite{cv}, we fit only the
non-perturbative parameters of the cluster and string models; in fact,
the default parametrizations yield rather poor fits of the
$b$-fragmentation data, as we obtain $\chi^2/\mathrm{dof}=739.4/61$ 
for HERWIG and $\chi^2/\mathrm{dof}=467.9/61$ 
for PYTHIA.
In HERWIG, we change CLSMR(1) and CLSMR(2), ruling
the Gaussian smearing of the hadron direction with respect to
the original constituent quarks; PLSPLT(2), which determines the
mass distribution of $b$-flavoured cluster decays;
DECWT, affecting the
relative weight of decuplet and octet baryons; and CLPOW, to which
the heavy-cluster yield and the baryon/meson ratio are sensitive.
The fitted values are: CLSMR(1) = 0.4 (default 0), CLSMR(2) = 0.3 (0),
PSPLT(2) = 0.33 (1), DECWT = 0.7 (1), CLPOW = 2.1 (2). 
After the tuning, the agreement with the data is still not very good,
but it is much better than with the default parametrization:
$\chi^2/\mathrm{dof}=222.4/61$. 
In PYTHIA, we modify the values of the fragmentation
parameters
PARJ(41) and PARJ(42), which control the
$a$ and $b$ parameters of the Lund symmetric fragmentation function, and
PARJ(46), which modifies the endpoint of the Lund function 
according to the Bowler hadronization model \cite{bowler}.
Our tuning gives: 
PARJ(41) = 0.85 (default value 0.3), PARJ(42) = 1.03 (0.58), PARJ(46) = 
0.85 (1).
After our tuning, PYTHIA matches the $e^+e^-$ data very well, 
and we obtain $\chi^2/\mathrm{dof}$ = 45.7/61 from the fit.
We have checked that our tuning works well also for the new model
implemented in PYTHIA 6.3, which orders parton showers in transverse momentum;
we found $\chi^2/\mathrm{dof}=46.0/61$
from the comparison with the $x_B$ data.
%\begin{table}[ht!]
%\small
%\caption{\label{para} Parameters of HERWIG and PYTHIA
%hadronization models that we have tuned to improve the agreement
%with $e^+e^-$ data, along with the $\chi^2$ per degree of freedom.}
%\begin{center}
%\begin{tabular}{|c|c|}\hline
%HERWIG & PYTHIA \\
%\hline\hline
%CLSMR(1) = 0.4  &                 \\
% \hline
%CLSMR(2) = 0.3  & PARJ(41) = 0.85 \\
%\hline
%DECWT = 0.7     & PARJ(42) = 1.03 \\
%\hline
%CLPOW = 2.1     & PARJ(46) = 0.85 \\
%\hline
%PSPLT(2) = 0.33 &                \\
%\hline
%\hline
%$\chi^2/\mathrm{dof}$ = 222.4/61 & $\chi^2/\mathrm{dof}$ = 45.7/61 \\
%\hline
%\end{tabular}
%\end{center}
%%\end{table}
In Fig.~\ref{hp} we compare LEP and SLD data with HERWIG and PYTHIA;
in Fig.~\ref{keff}
we present the experimental spectra along with the NLO/NLL calculation
using the Kartvelishvili model and the NLO/NNLL one with the
analytic coupling constant.
All approaches use the best-fit parametrizations.

\par We note in Fig.~\ref{hp} 
that PYTHIA, after the tuning, gives a good description of
the experimental spectra, while HERWIG's distribution is broader,   
below the data around the peak and above them at small $x_B$.
From Fig.~\ref{keff}, we learn that the NLO/NLL calculation using the
Kartvelishvili model reproduces the data quite well, but it becomes
negative at large $x_B$. The plot relying on the effective-coupling model
lies above the data around the peak and approaches zero more 
rapidly at large $x_B$. In any case, even this result is acceptable,
considering that, when modelling power corrections by means of the
effective coupling constant, we are not tuning any free parameter to the data.

\begin{figure}
\centerline{\resizebox{0.6\textwidth}{!}{\includegraphics{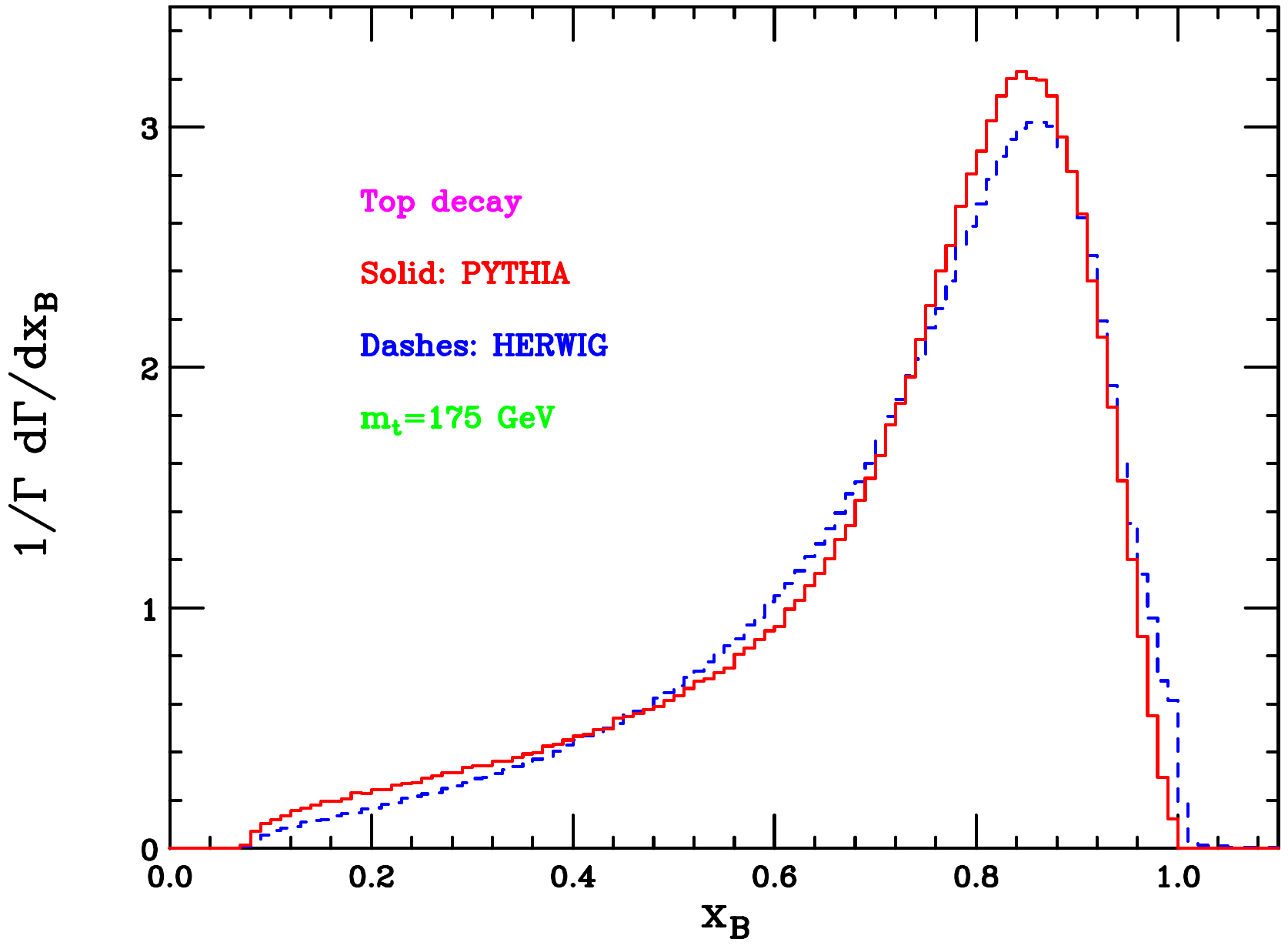}}}
\caption{\small $B$-hadron spectrum in top quark decay according to
HERWIG, PYTHIA and the NLO/NLL calculation which includes non-perturbative
corrections via the Kartvelishvili hadronization model.}
\label{tb}
\end{figure}

\begin{figure}
\centerline{\resizebox{0.6\textwidth}{!}{\includegraphics{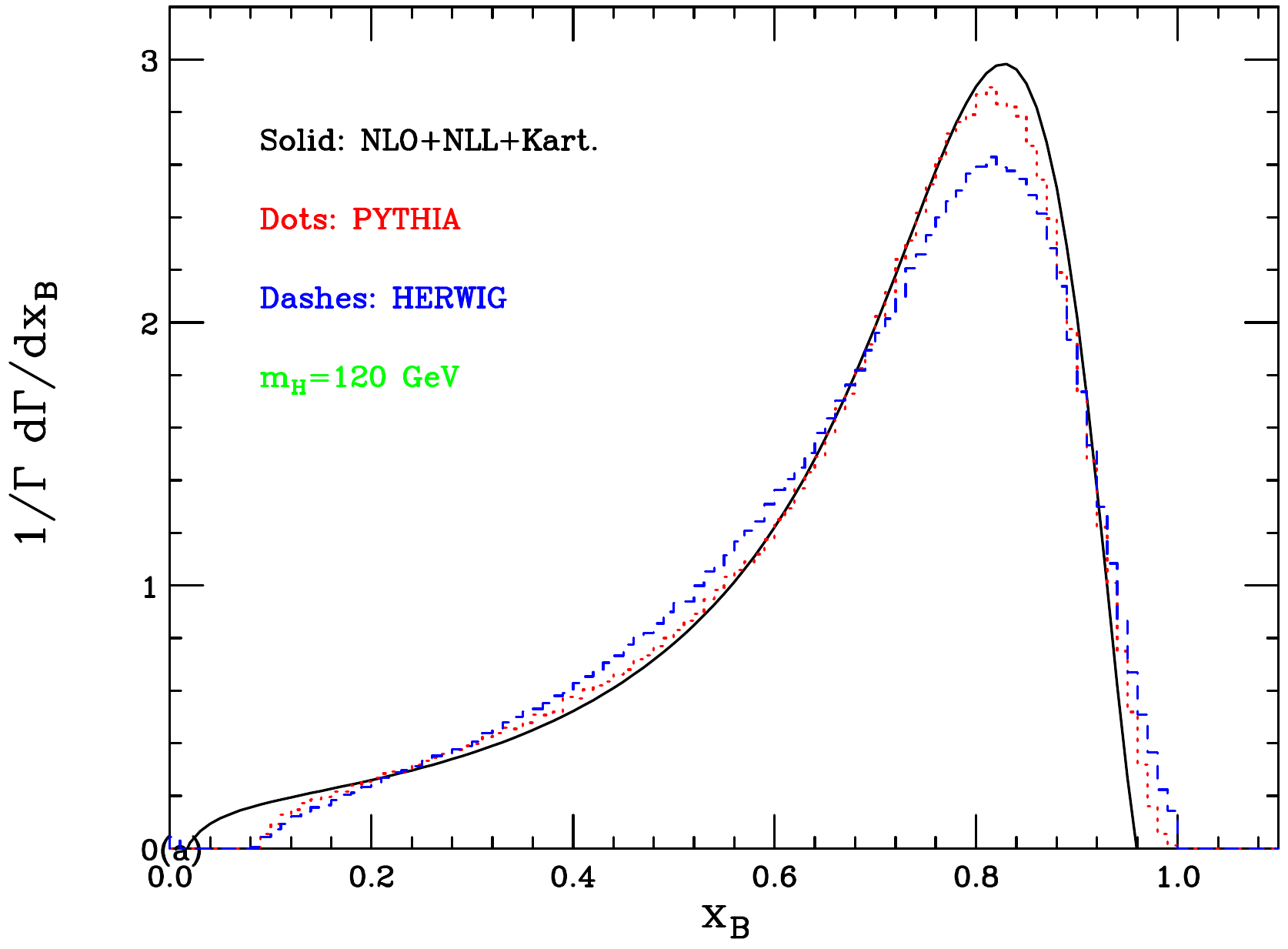}}}
\caption{\small As in Fig.~\ref{tb}, but for
$b$-flavoured hadron production in the decay of the 
Standard Model Higgs $H\to b\bar b$.}
\label{hig}
\end{figure}

Using the fits to LEP and SLD data, we can predict the $B$-hadron
spectrum in other processes, such as top-quark decay ($t\to bW$) and
the decay of the Standard Model  Higgs boson $H\to b\bar b$.
In Figs.~\ref{tb} and \ref{hig} we show the predictions yielded by
HERWIG, PYTHIA and the resummed
calculation based on the perturbative fragmentation approach for 
$B$-hadron production in
top ($t\to bW$) and  Higgs ($H\to b\bar b$) decays.
We parametrize cluster, string and Kartvelishvili models using the best
fits to LEP and SLD data. As for the resummation, we use the NLO/NLL
calculations in Refs.~\cite{top,top1} for top decay and in Ref.~\cite{hbb}
for $H\to b\bar b$ processes. 
In our plots we 
have set the top and Higgs masses to the values $m_t=175$~GeV and 
$m_H=120$~GeV. 
The results in Figs.~\ref{tb} and \ref{hig} exhibit similar features to
the comparison presented in Figs.~\ref{hp} and \ref{keff}: 
PYTHIA and the NLO/NLL calculation using the Kartvelishvili model are
in good agreement, while the spectra yielded by HERWIG show some
discrepancy, as they are broader than the other two predictions and  lie
below them at small $x_B$ and above at large $x_B$.
\par Finally, we present the same comparison in moment space,
where the moments of the differential cross section are defined 
as follows: 
\begin{equation}
\sigma_N  =\int_0^1 {dz  \ z^{N-1}
{1\over{\sigma}}{{d\sigma}\over{dz}}(z) }.
\label{moment}
\end{equation}
In Ref.~\cite{delphi}, the DELPHI collaboration presented the
first five moments
for $B$ production in $e^+e^-$ annihilation.
From the point of view of resummed calculations, 
working in moment space \cite{canas} presents several advantages.
In $N$-space, convolutions become ordinary products, and the relation
between parton- and hadron-level cross sections becomes:
$\sigma_N^B=\sigma_N^bD_N^{\mathrm{np}}$,
where $\sigma_N^b$ and $\sigma_N^B$ are the moments of the $b$ and $B$ cross 
sections, and $D_N^{\mathrm{np}}$ is the $N$-space counterpart of the 
non-perturbative fragmentation function.
Therefore, there is no need to assume any functional form for the 
non-perturbative fragmentation function in $x_B$-space.
\begin{table}
\caption{\label{mom}\small  Moments
$\sigma^B_N$ from
DELPHI~\protect\cite{delphi}, and $N$-space results 
in $e^+e^-$ annihilation, Higgs ($H$) and top ($t$) decay, 
according to NLO/NLL calculations, HERWIG (HW) and PYTHIA (PY).
Also presented are the $N$-space $e^+e^-$ results obtained using the
effective coupling constant $\tilde\alpha_S$ along with the
theoretical errors \cite{ugo}.}
\small\begin{tabular}{| c | c c c c |}
\hline
& $\langle x\rangle$ & $\langle x^2\rangle$ & $\langle x^3\rangle$
& $\langle x^4\rangle$ \\
\hline
\hline
$e^+e^-$ data $\sigma_N^B$&0.715$\pm$0.005 &0.540$\pm$0.006 &
0.424$\pm$0.007 &0.341$\pm$0.006  \\
\hline
\hline
$e^+e^-$ NLL $\sigma_N^b$   & 0.780 & 0.644 & 0.548 & 0.476  \\
\hline
$D^{\mathrm{np}}_N$ & 0.917 & 0.839 & 0.773 & 0.716 \\
\hline
$e^+e^-$ HW  & {0.711} & 
{0.535} & {0.418} & {0.335}  \\
\hline
$e^+e^-$ PY  & 0.716
& 0.541 & {0.424} & {0.340}\\
\hline
$e^+e^-$ NNLL+$\tilde\alpha_S$  & $0.687 \pm 0.040$ 
& $0.5019 \pm 0.047$ & $0.381 \pm 0.046$& 
$0.298 \pm 0.046 $\\
\hline
\hline
$t$-dec. NLL & 0.723 & 0.556 & 0.443 & 0.363 \\
\hline
$t$-dec. HW & 0.733 
& 0.570 & 0.461 & 0.381\\
\hline
$t$-dec. PY & 0.722 & 
0.559 & 0.449 & 0.369\\
\hline
\hline
$H$-dec. NLL
& 0.695 & 0.517 & 0.402 & 0.321 \\
\hline
$H$-dec. HW & 0.684 & 0.504 & 0.388 & 0.308 \\
\hline
$H$-dec. PY & 0.688 & 
0.508 & 0.391 & 0.310 \\
\hline\hline
\end{tabular}
\end{table}
The results of our $N$-space analysis are summarized in Table~\ref{mom}. 
We note that, after the fits to LEP and SLD data, HERWIG and PYTHIA 
agree with
the DELPHI moments, within the experimental uncertainties.
As for the calculation based on the effective coupling constant, it is
able to reproduce the experimental moments within the theoretical errors which
were calculated in \cite{ugo} and quoted in Table~\ref{mom}.
As done for the $x_B$-space analysis, we also present in Table~\ref{mom}
the moments
of the differential width for the production of $B$ hadron in top
or Higgs decays, using the moments $D_N^{\mathrm{np}}$ taken from 
the fits to the DELPHI data. 

In summary, we studied bottom-quark fragmentation in $e^+e^-$ annihilation,
top and Higgs decays using resummed calculations and the HERWIG and 
PYTHIA parton shower models. We fitted the Kartvelishvili, cluster and
string models to $B$-hadron spectra measured at LEP and SLD, and then 
predicted the $B$-energy distribution in other processes. We also presented
the results yielded by a model which incorporates non-perturbative
power corrections via an effective strong coupling constant. 
The analysis was finally extended to Mellin moment space.
We believe that the study here presented can be a useful starting point 
to address $b$-quark fragmentation at present and future colliders, 
as it sets some benchmarks for the hadronization models which are typically 
used along with Monte Carlo generators and resummed computations.

%\end{document}

\addtocounter{chapter}{1}
%\documentclass[a4paper,12pt,twoside]{report}
%\usepackage{epsfig}
%\usepackage{amssymb}
%\usepackage{lineno}
%\usepackage{setspace}
%%%%%%%%%%%%%%%%%%%%%%%%%%%%%%%%%%%%%%%%%%%%%%%%%%%%%%%%%%
% \documentclass[paper]{JHEP3}
% \usepackage{epsfig}
% \usepackage{axodraw}
% \jot 7pt
% \def    \eqnum          #1{(\ref{#1})}       %equation # in round parenthesis
% \def    \scite          #1{$^{\cite{#1}}$}     %superscript biblio ref
% %--------------------------------------------
% %       SET PAGE SIZE
% %       \evensidemargin 0.0in
% %        \oddsidemargin -1cm
%         \textwidth 16.5cm
%         \textheight 25cm
%         \hoffset=-3cm
%         \headsep 0in
%         \newdimen\eqskip
%         \newdimen\txtskip
%         \eqskip=25pt
%         \txtskip=25pt
%         \baselineskip=\txtskip
%         \parskip 5pt plus 1pt
%         \floatsep 0cm
%         \textfloatsep 0.2cm
%         \renewcommand{\textfraction}{0.3}
%         \renewcommand{\topfraction}{0.7}
%         \newdimen\mysep                
%         \newdimen\hmysep
%         \mysep=-0.4cm
%         \hmysep=-0.4cm
% %        \mysep=0cm
%\begin{document}
       
  \newcommand{\ccaption}[2]{
    \begin{center}
    \parbox{0.85\textwidth}{
      \caption[#1]{\small{{#2}}}
      }
    \end{center}
    }
% \newcommand{\BS}{\bigskip}
% MATH SYMBOLS
\def    \be             {\begin{equation}}
\def    \ee             {\end{equation}}
\def    \ba             {\begin{eqnarray}}
\def    \ea             {\end{eqnarray}}
\def    \nn             {\nonumber}
\def    \=              {\;=\;}
\def    \frac           #1#2{{#1 \over #2}}
\def    \ret            {\\[\eqskip]}
\def    \ie             {{\em i.e.\/} }
\def    \eg             {{\em e.g.\/} }
\def \lsim{\mathrel{\vcenter
     {\hbox{$<$}\nointerlineskip\hbox{$\sim$}}}}
\def \gtrsim{\mathrel{\vcenter
     {\hbox{$>$}\nointerlineskip\hbox{$\sim$}}}}
\def    \bentarrow      {\:\raisebox{1.1ex}{\rlap{$\vert$}}\!\rightarrow}
\def    \rd             {{\mathrm d}}    
\def    \Im             {{\mathrm{Im}}}  
\def    \bra#1          {\mbox{$\langle #1 |$}}
\def    \ket#1          {\mbox{$| #1 \rangle$}}

% UNITS                 
\def    \kev            {\mbox{$\mathrm{keV}$}}
\def    \mev            {\mbox{$\mathrm{MeV}$}}
\def    \gev            {\mbox{$\mathrm{GeV}$}}

% PARTICLE SYMBOLS
\def    \nubar   {\bar{\nu}}
\def    \ubar   {\bar{u}}
\def    \dbar   {\bar{d}}
\def    \qbar   {\bar{q}}
\def    \bbar   {\bar{b}}
\def    \cbar   {\bar{c}}
\def    \cpbar   {\bar{c}'}
\def    \tbar   {\bar{t}}
\def    \Qbar   {\overline{Q}}
\def \sss {\scriptscriptstyle}
% KINEMATICAL VARIABLES 
\def    \mq             {\mbox{$m_Q$}}  
\def    \mZ             {\mbox{$m_Z$} }
\def    \mZsq             {\mbox{$m_Z^2$} }
\def    \mW             {\mbox{$m_W$} }
\def    \mWsq             {\mbox{$m_W^2$} }
\def    \mH             {\mbox{$m_H$} }
\def    \mHsq             {\mbox{$m_H^2$} }
\def    \mt             {\mbox{$m_{\sss{top}}$}}  
\def    \mtsq             {\mbox{$m_{\sss{top}}^2$}}
\def    \mb             {\mbox{$m_b$}}  
\def    \mbsq             {\mbox{$m_b^2$}}  
\def    \mqq            {\mbox{$m_{Q\bar Q}$}}
\def    \mqqsq          {\mbox{$m^2_{Q\bar Q}$}}
\def    \pt             {\mbox{$p_{\sss T}$}}
\def    \ptjet             {\mbox{$p^{\sss{jet}}_{\sss T}$}}
\def    \ptpart             {\mbox{$p^{\sss{part}}_{\sss T}$}}
\def    \ptjet#1         {\mbox{$p_{\sss T, #1}$}}
\def    \etajet             {\mbox{$\eta_{\sss{jet}}$}}
\def    \etapart             {\mbox{$\eta_{\sss{part}}$}}
\def    \etamax         {\mbox{$\eta_{\sss{max}}$}}
\def    \pttop             {\mbox{$p^{\sss{top}}_{\sss T}$}}
\def    \pttopsq             {\mbox{$p^2_{\sss t, \sss T}$}}
\def    \pttbar             {\mbox{$p_{\bar{\sss t},\sss T}$}}
\def    \pttbarsq             {\mbox{$p^2_{\overline{\sss t},\sss T}$}}
\def    \ptpair             {\mbox{$p^{\sss t\bar{\sss t}}_{\sss T}$}}
\def    \ptmin             {\mbox{$p_{\sss T}^{\sss{min}}$}}
\def    \Rmin             {\mbox{$R_{\sss{min}}$}}
\def    \dRjj             {\mbox{$\Delta R_{\sss{jj}}$}}
\def    \drjp           {\mbox{$\Delta R_{\sss{jp}}$}} 
\def    \dR             {\mbox{$\Delta R$}}
\def    \dphitt        {\mbox{$\Delta \phi^{\sss t \overline{\sss t}}$}}
\def    \dphil        {\mbox{$\Delta \phi^{\sss{lept}}$}}
\def    \dphi             {\mbox{$\Delta \phi$}}
\def    \et             {\mbox{$E_T$}}
\def	\etmin		{\mbox{$E_T^{\sss{min}}$}}
\def    \etclus         {\mbox{$E^{\sss{clus}}_{\sss T}$}}
\def    \etaclmax       {\mbox{$\eta^{\sss{clus}}_{\sss{max}}$}}
\def    \rjet           {\mbox{$R_{\sss{jet}}$}}
\def    \rjj            {\mbox{$R_{\sss{jj}}$}}
\def    \rpp            {\mbox{$R_{\sss{pp}}$}}
\def    \Rsep           {\mbox{$R_{\sss{sep}}$}}
\def    \Rmat           {\mbox{$R_{\sss{match}}$}}
\def    \rmin           {\mbox{$R_{\sss{min}}$}}
\def    \rcone          {\mbox{$R_{\sss{cone}}$}}
\def    \rclus          {\mbox{$R_{\sss{clus}}$}}
\def    \cclus          {\mbox{$C_{\sss{clus}}$}}
\def    \njet           {\mbox{$N_{\sss{jet}}$}} 
\def    \nclus           {\mbox{$N_{\sss{clus}}$}} 
\def    \npart          {\mbox{$N_{\sss{part}}$}}
\def    \exc            {\mbox{$_{\sss{exc}}$}}
\def    \inc            {\mbox{$_{\sss{inc}}$}}
\newcommand{\pv}{\mathbf p}
\newcommand {\ptr}{\pt}  %\mbox { $ p_{\sss T} $} }
\newcommand {\ptja}{\mbox {$p_{\sss T,1}$}}
\newcommand {\ptjb}{\mbox {$p_{\sss T}^{{\sss j_2}}$}}
\newcommand {\yjet}{\mbox {$y_{{\sss 1}}$}}
\newcommand {\ul}{a) (upper left panel) }
\newcommand {\ur}{b) (upper right panel) }
\newcommand {\dlf}{c) (lower left panel) }
\newcommand {\drg}{d) (lower right panel) }
\newcommand {\pttb} {\pttbar} %\mbox{$p_{\sss T}^{{\sss t}\bar{\sss t}}$ }  }
\newcommand {\ptop} {\pttop}%\mbox {$  p_{\sss T}{{\sss t }} $ } }
\newcommand {\ytop}{\mbox{$ y_{\sss {top}}$}}
\newcommand {\phitt}{\dphitt} %\mbox {$ \Delta \Phi^{\sss t} $ } }
        
% QCD PARAMETERS                                      
\newcommand     \MSB            {\ifmmode {\overline{\rm MS}} \else 
                                 $\overline{\rm MS}$  \fi}
\def    \muf            {\mbox{$\mu_{\rm F}$}}
\def    \mug            {\mbox{$\mu_\gamma$}}
\def    \mufsq          {\mbox{$\mu^2_{\rm F}$}}
\def    \mur            {{\mbox{$\mu_{\rm R}$}}}
\def    \mursq          {\mbox{$\mu^2_{\rm R}$}}
\def    \mul            {{\mu_\Lambda}}
\def    \mulsq          {\mbox{$\mu^2_\Lambda$}}

\def    \bzero          {\mbox{$b_0$}}
\def    \as             {\ifmmode \alpha_s \else $\alpha_s$ \fi}
\def    \aem             {\mbox{$\alpha_{em}(\mZ)$}}
\def    \asb            {\mbox{$\alpha_s^{(b)}$}}
\def    \assq           {\mbox{$\alpha_s^2$}}
\def \oacube {\mbox{$ {\cal O}(\alpha_s^3)$}}
\def \oaemacube {\mbox{$ {\cal O}(\alpha\alpha_s^3)$}}
\def \oafour {\mbox{$ {\cal O} (\alpha_s^4)$}}
\def \oatwo {\mbox{$ {\cal O} (\alpha_s^2)$}}
\def \oaematwo {\mbox{$ {\cal O}(\alpha \alpha_s^2)$}}
\def \oaemas {\mbox{$ {\cal O}(\alpha \alpha_s)$}} 
\def \oas   {\mbox{$ {\cal O}(\alpha_s)$}}
\def\slash#1{{#1\!\!\!/}}
\def\rt1{\raisebox{-1ex}{\rlap{$\; \rho \to 1 \;\;$}}
\raisebox{.4ex}{$\;\; \;\;\simeq \;\;\;\;$}}
\def\ltap{\raisebox{-.5ex}{\rlap{$\,\sim\,$}} \raisebox{.5ex}{$\,<\,$}}
\def\gtap{\raisebox{-.5ex}{\rlap{$\,\sim\,$}} \raisebox{.5ex}{$\,>\,$}} 

\newcommand\LambdaQCD{\Lambda_{\scriptscriptstyle \rm QCD}}

\def\naive{na\"{\i}ve}
\def\asp{{\alpha_s}\over{\pi}}
\def\half{\frac{1}{2}}
\def\herwig{{\small HERWIG}}
\def\herwigs{{\small HERWIG} \ }
\def\isajet{{\small ISAJET}}
\def\pythia{{\small PYTHIA}}
\def\grace{{\small GRACE}}
\def\amegic{{\small AMEGIC++}}
\def\vecbos{{\small VECBOS}}
\def\madgraph{{\small MADGRAPH}}
\def\comphep{{\small CompHEP}}
\def\ALPHA{{\small ALPHA}}
\def\ALPGEN{{\small ALPGEN}}
\def\ALPGENs{{\small ALPGEN} \ }
\def\alpgen{{\small ALPGEN}}
\def\alpgens{{\small ALPGEN} \ }
\def\mcnlo{{\small MC@NLO}}
\def\mcnlos{{\small MC@NLO} \ }
\def\ppbar{\mbox{$p \bar{p}$}}
\def\ttbar{\mbox{$t \bar{t}$}}
\def\ttzero{$0_{\sss exc}$}
\def\ttone{{\tt tt1}}
\def\met{$\rlap{\kern.2em/}E_T$}

\newcommand{\ben}{\begin{enumerate}}
\newcommand{\een}{\end{enumerate}}
\newcommand{\bit}{\begin{itemize}}
\newcommand{\eit}{\end{itemize}}
%\begin{document}
%%%%%%%%%%%%%%%%%%%%%%%%%%%%%%%%%%%%%%%%%%%%%%%
% Toggle line numbering
% Won't work with the PRD revtex4 !
%\pagewiselinenumbers
% uncomment if you want doublespace
%\doublespace
%%%%%%%%%%%%%%%%%%%%%%%%%%%%%%%%%%%%%%%%%%%%%%%
% \chapter{Introduction} {\it C. Mariotti, E. Migliore and P. Nason}
%%%%%%%%%%%%%%%%%%%%%%%%%%%%%%%%%%%%%%%%%%%%%%%%%%%%%%%%%
% \section{Physics at the Large Hadron Collider (LHC)} {\it (C. Mariotti and E. Migliore)}
%%%%%%%%%%%%%%%%%%%%%%%%%%%
% \subsection{Why LHC?}
% \begin{document}
 \mchapter{Monte Carlo simulations of 
       top-quark pair production in hadronic collisions}{M.Treccani}
\section{Introduction}
 \label{sec:intro}
In view of the starting of the LHC and the accumulated statistics at Tevatron,
there appears the need for further improvement in the accuracy of theoretical predictions.
One of the most interesting fields refer to the class of events with multiple final states, giving rise to multiple jets with complicated topologies.
There exists different strategies to tackle this problem, with distinct features and points of strength.
The main problem is how to consistently compose the contributions due to Matrix Element (ME) calculations with the contributions of the Monte Carlo showering codes (MC), in order to exploit their complementariety and avoid at the same time the so-called double counting phenomenon \cite{Hoche:2006ph}.\\
One of these strategies, known as \mcnlo, put the emphasis on achieving the next-to-leading-order (NLO)
accuracy in the description of the inclusive rates for a given final
state $F$, accompanied by the exact leading-order (LO) description of the emission of
one extra jet ($F+$jet). For a detailed explanation of this approach and its implementation in several cases, see \cite{Frixione:2006he,Frixione:2002ik,Frixione:2003ei}.\\
One alternative approach relies on a consistent leading-logarithmic (LL) accuracy in the prediction of a final state $F$ accompanied by a varying number of
extra jets. The removal of double counting of jet is achieved by the so-called {\em matching algorithm} for matrix elements and parton shower.
It is understood that the 
matching algorithm approach cannot improve the intrinsic LL accuracy of the predictions; however it will give a better accuracy in the prediction of the observables more sensible to the production of two or more jets in addition to $F$.\\
In this note, we study in detail the so-called {\em \small MLM}
matching~\cite{mlmfnal,Mangano:2006rw} embedded in the the ME generator \alpgen\ \cite{Mangano:2002ea} in the \ttbar\ pair production at hadron colliders.
First we will address its stability w.r.t its internal parameters, and after we will perform detailed numerical comparison between {\em \small MLM} matching and the \mcnlo\ code.\\
In particular, in  Section~2 we will perform some robusteness test on the \alpgen\ calculations, comparing predictions 
obtained with different parameters and discussing the related uncertainties. Section~3
covers the detailed comparison between \alpgen\ and \mcnlo predictions, and in Section~4 we will present our conclusions.
\section{Consistency studies of the matching algorithm}
\label{sec:sanity}

In this section we study the overall consistency of
the matching algorithm applied to the case of \ttbar\ final states. 
We shall consider \ttbar\ production at the Tevatron
(\ppbar\ collisions at $\sqrt{S}=1.96$~TeV) and at the LHC ($pp$
collisions at $\sqrt{S}=14$~TeV).\\
The generation parameters for the
light partons are defined by the following kinematical cuts: the default values for the event
samples at the Tevatron (LHC) are given by: \,\ptmin=20 (30) \gev and
\Rmin=0.7 (0.7), while they are considered only in the geometrical region defined by $\eta \le 4 (5)$.\\
The top particle is assumed to be stable, and
therefore all jets coming from the decay of top quarks are neglected.
For the shower evolution we use \herwig, version
6.510~\cite{Marchesini:1988cf}. We stopped the evolution after the perturbative phase, in order to drop down all the common systematics that could smooth out any possible discrepancy between the various simulations.
For all generations we chose the parton distribution function set
{\small MRST2001J}\cite{Martin:2001es}, with renormalization and
factorization scales squared set equal to:
$\mursq=\mufsq=\sum_{i=t,\bar{t},\mathrm{jets}} \; [m_i^2+(p_{\sss T}^i)^2]$.
Jet observables are built out of the  partons emerging form the shower in the rapidity range
 $\vert\eta\vert \le 6$ and adopting the cone algorithm {\small GETJET}\cite{getjet}. The jet cone size is set to
$\rcone=0.7$ and the minimum transverse momentum to define a jet at the Tevatron(LHC) is
15(20)~\gev\ .\\
Having defined the environmental parameters of these studies, we then explore the systematic uncertainties due to the
variation of the internal parameters. These uncertainties
reflect the underlying fact that this approach relies on the
LO evaluation of the hard ME and on the LL accuracy in the removal of double counting and in
the description of the shower evolution. In this section we shall show that the size of the resulting
uncertainties is  consistent with what can be expected in
such a LL approach in the case of \ttbar\ production.\\

To our analysis, the important feature of the whole procedure is the presence of two set of parameters: the generation cuts and the matching cuts (see \cite{mlmfnal,Mangano:2006rw}).
The first set is necessary to avoid the Infra-Red (IR) and collinear singularities:$ \ptmin$ , the minimum transverse momentum of the extra-parton(s) to be generated, and $\Rmin$, the minimum separation between extra-partons in the $(\eta,\phi)$ plane.
Along with these parameters, there exist an analogous set, but with slightly different meaninings : the matching cuts $\etclus$ and $\Rmat$.
It's worth to stress that the latter parameters are necessary to effectively separate the phase space, but the prediction should be stable against (slight) modifications of them, together with the choice of the particular cone jet algorithm adopted in the matching procedure.

In our examples here we consider two independent variations of the
generation and of two of the matching cuts, as in table~\ref{tab:cuts_syst}, keeping fixed our
definition of the physical objects (the jets) and of the
observables.

\begin{figure}
\begin{center}
\includegraphics[height=0.22\textheight,clip]{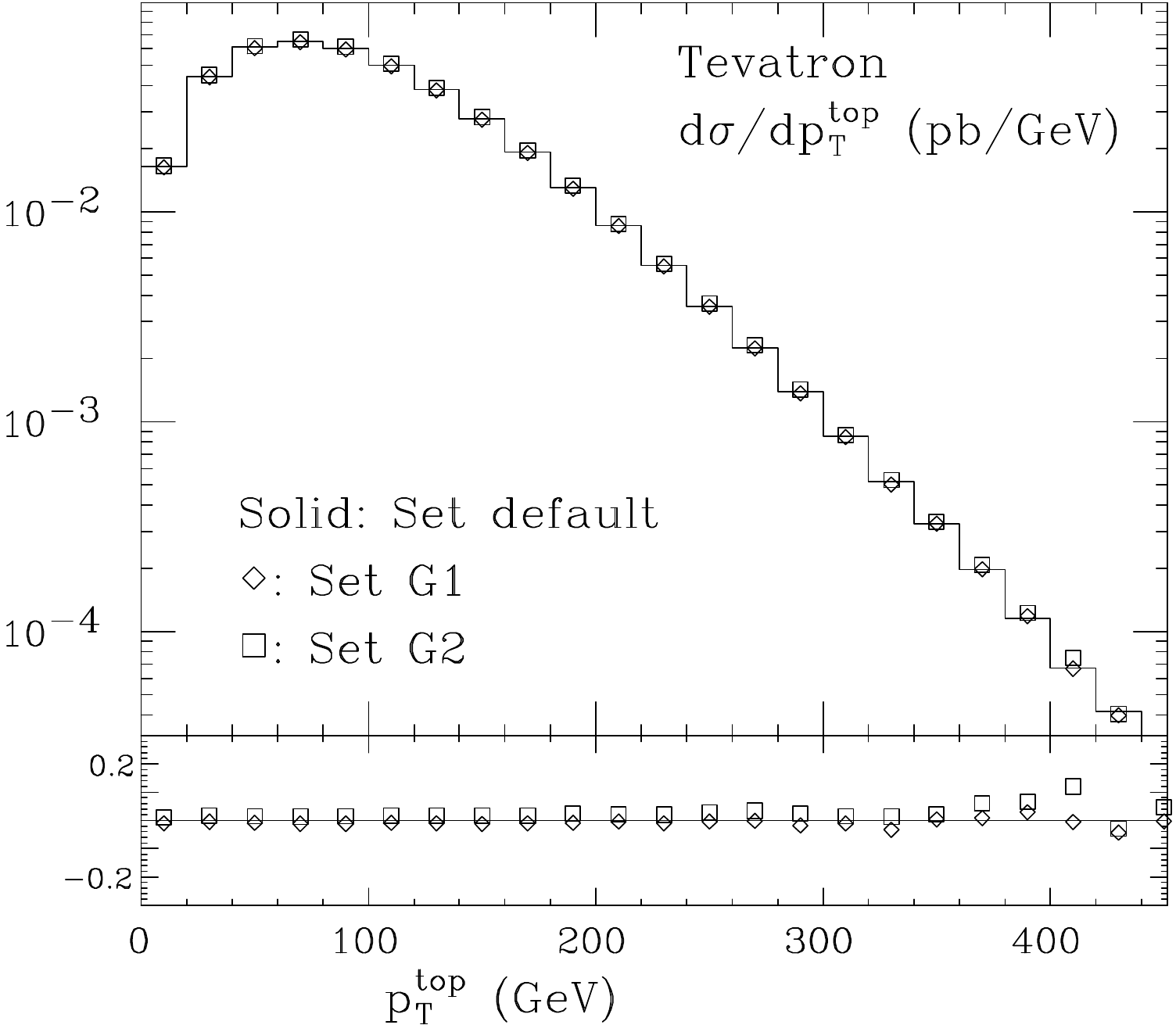}
\hspace*{0.5cm}
\includegraphics[height=0.22\textheight,clip]{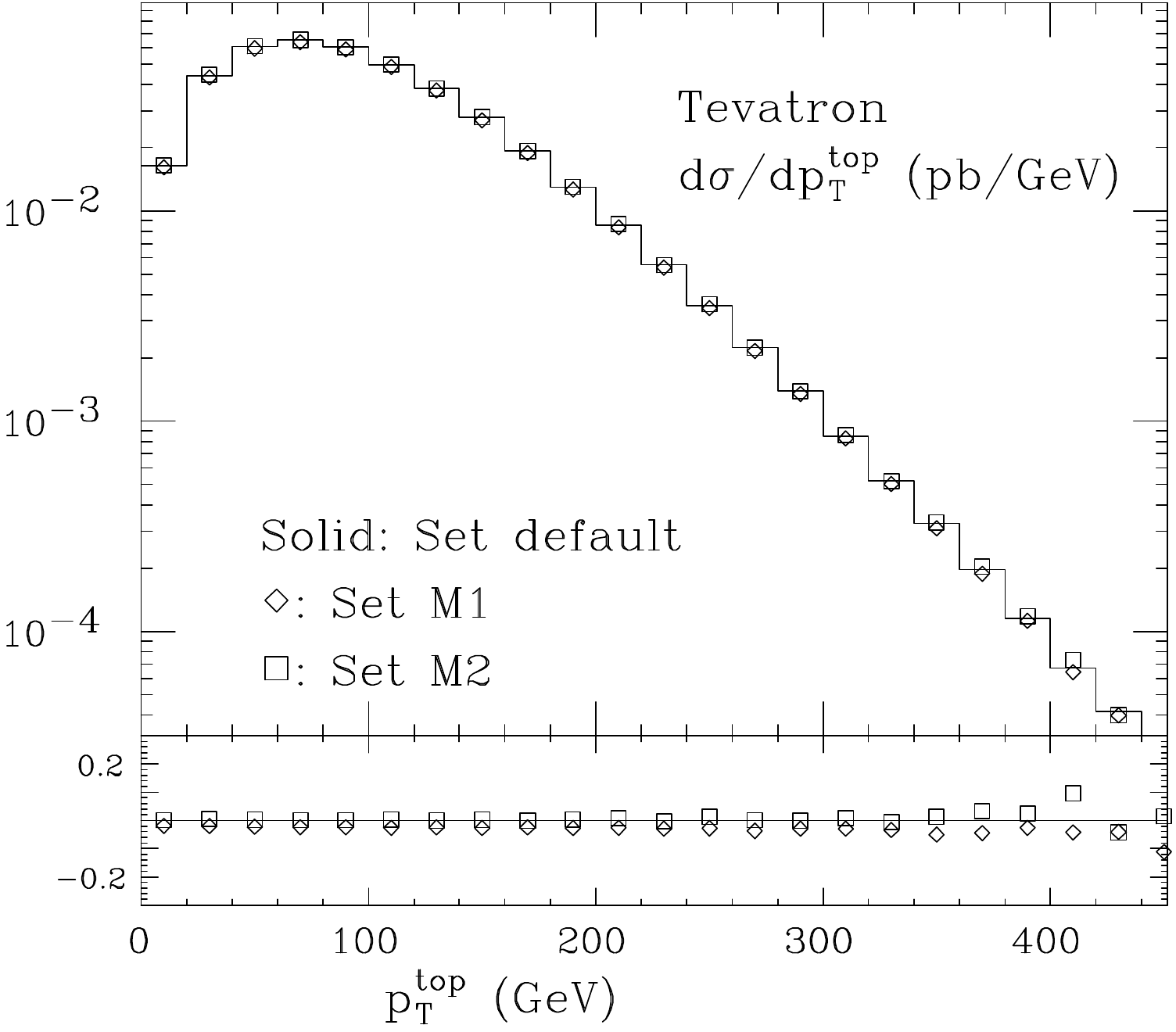}
\\
\includegraphics[height=0.22\textheight,clip]{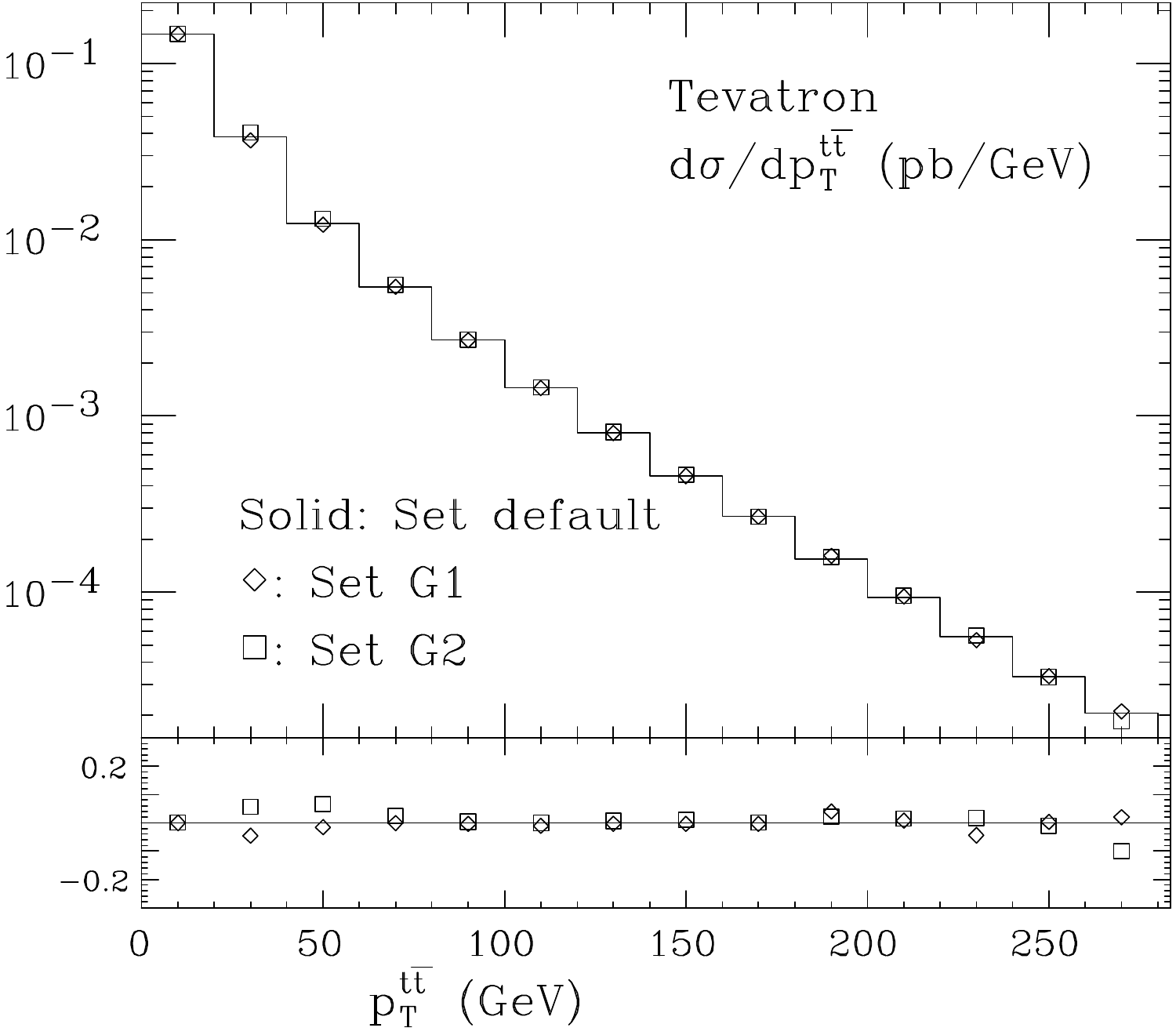}
\hspace*{0.5cm}
\includegraphics[height=0.22\textheight,clip]{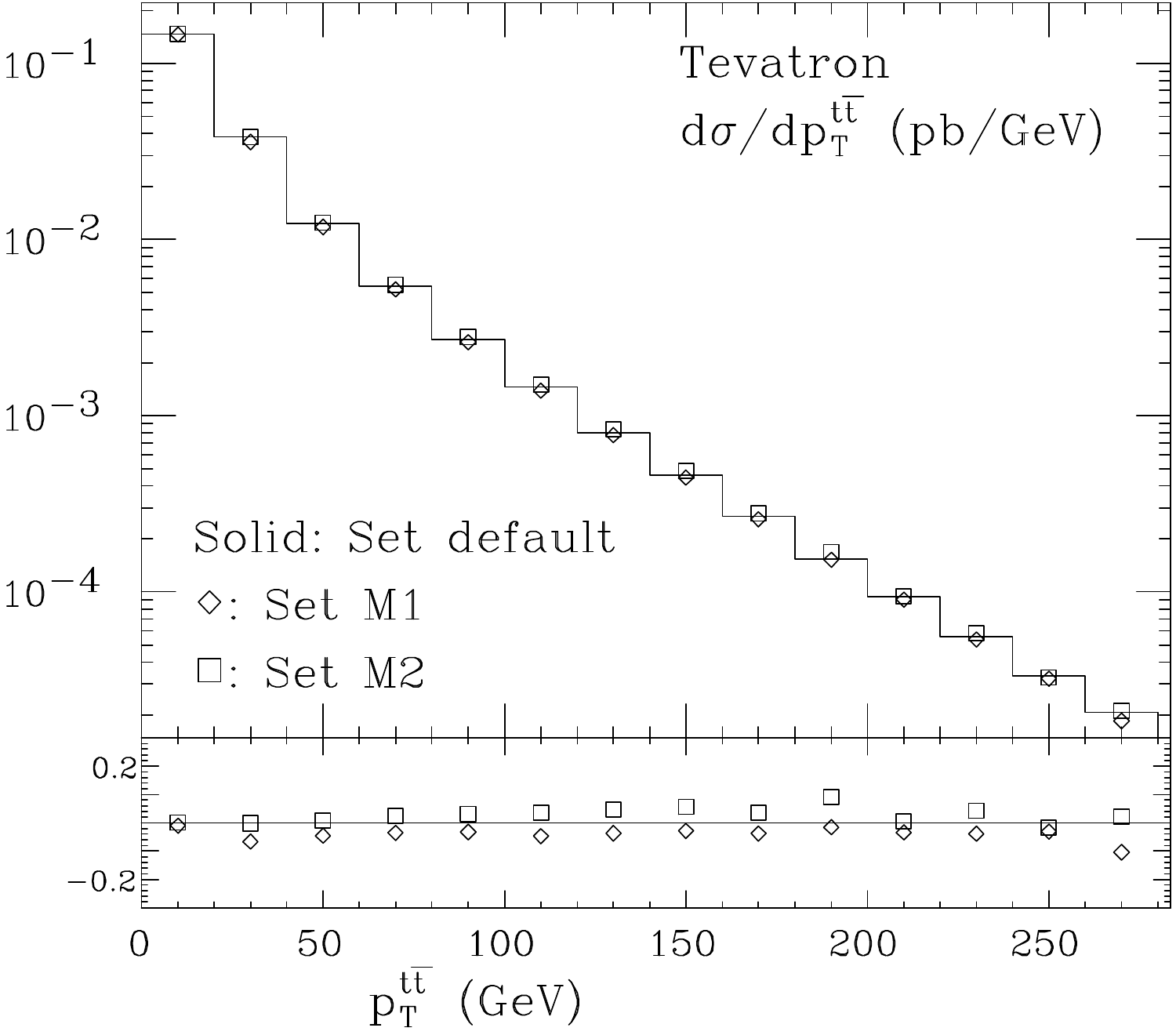}
\\
\includegraphics[height=0.22\textheight,clip]{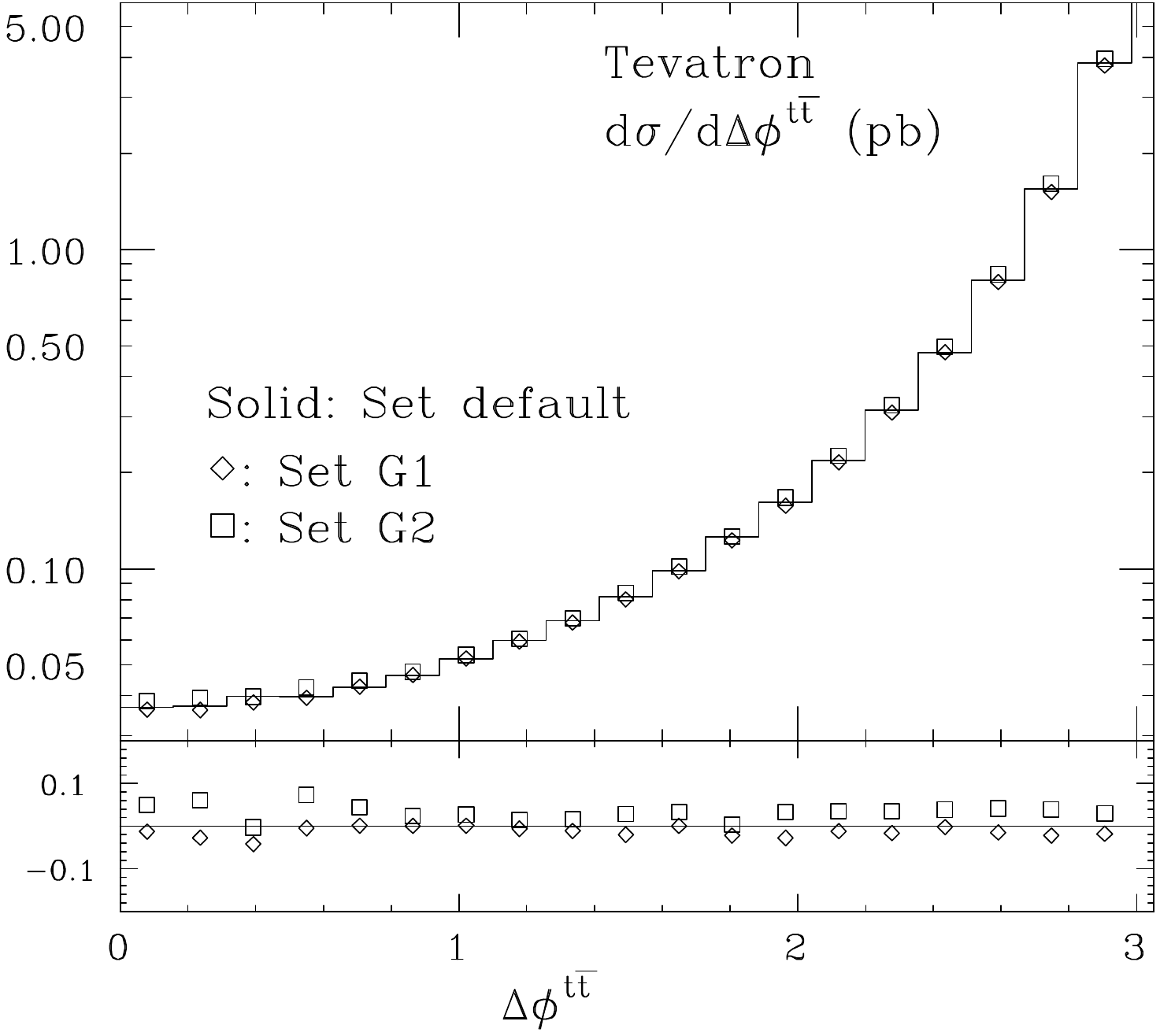}
\hspace*{0.5cm}
\includegraphics[height=0.22\textheight,clip]{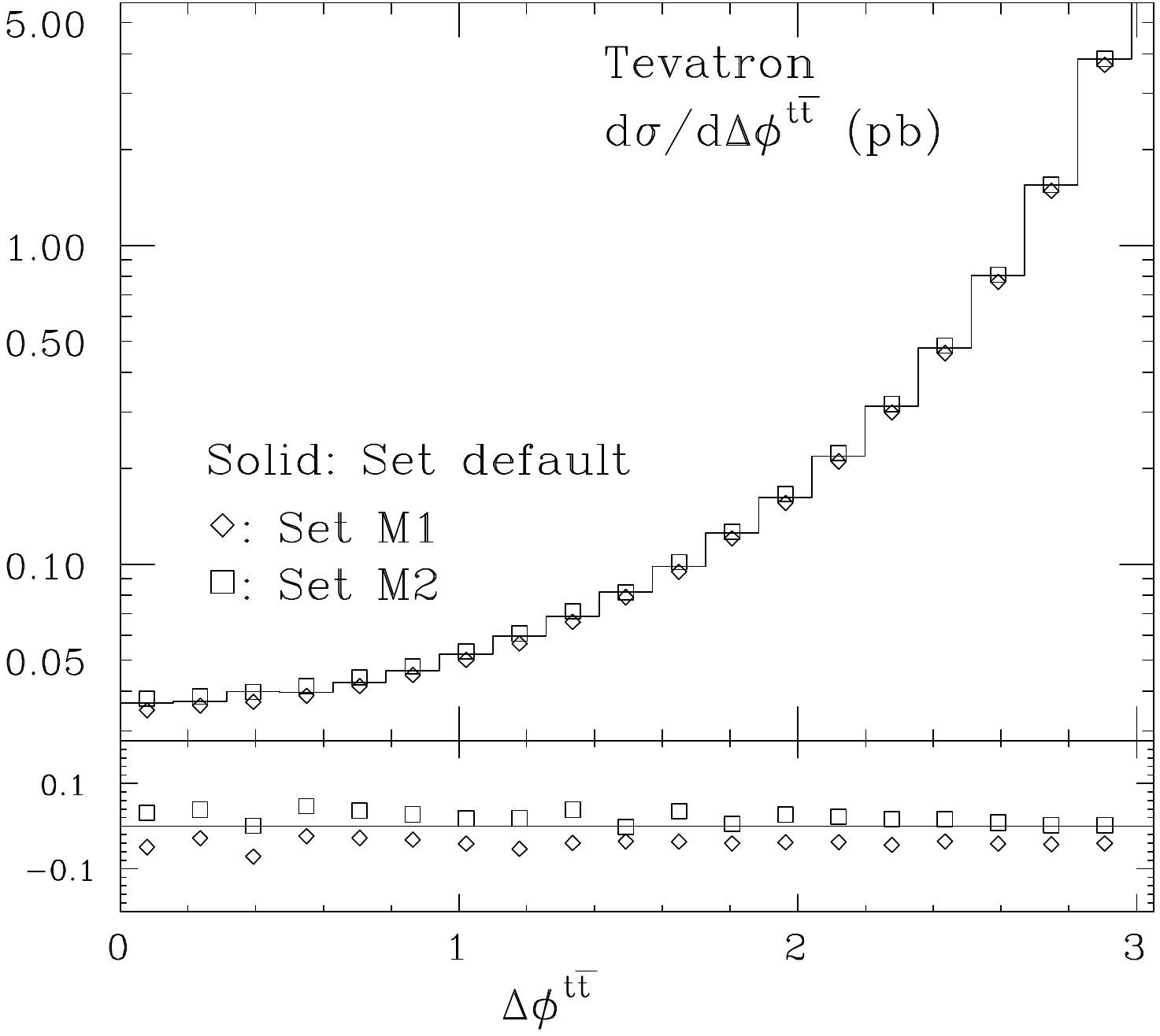}
\\
\includegraphics[height=0.22\textheight,clip]{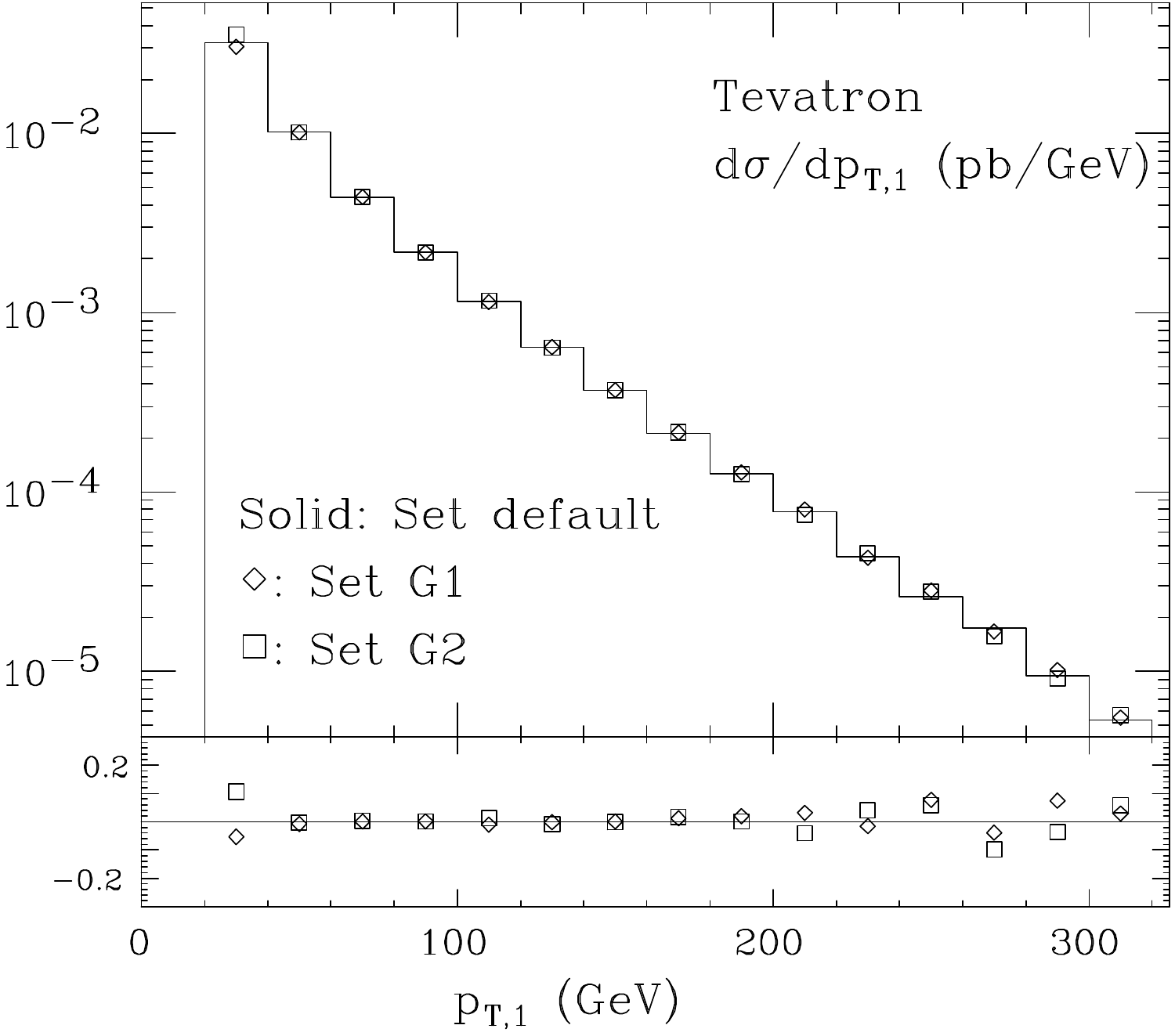}
\hspace*{0.5cm}
\includegraphics[height=0.22\textheight,clip]{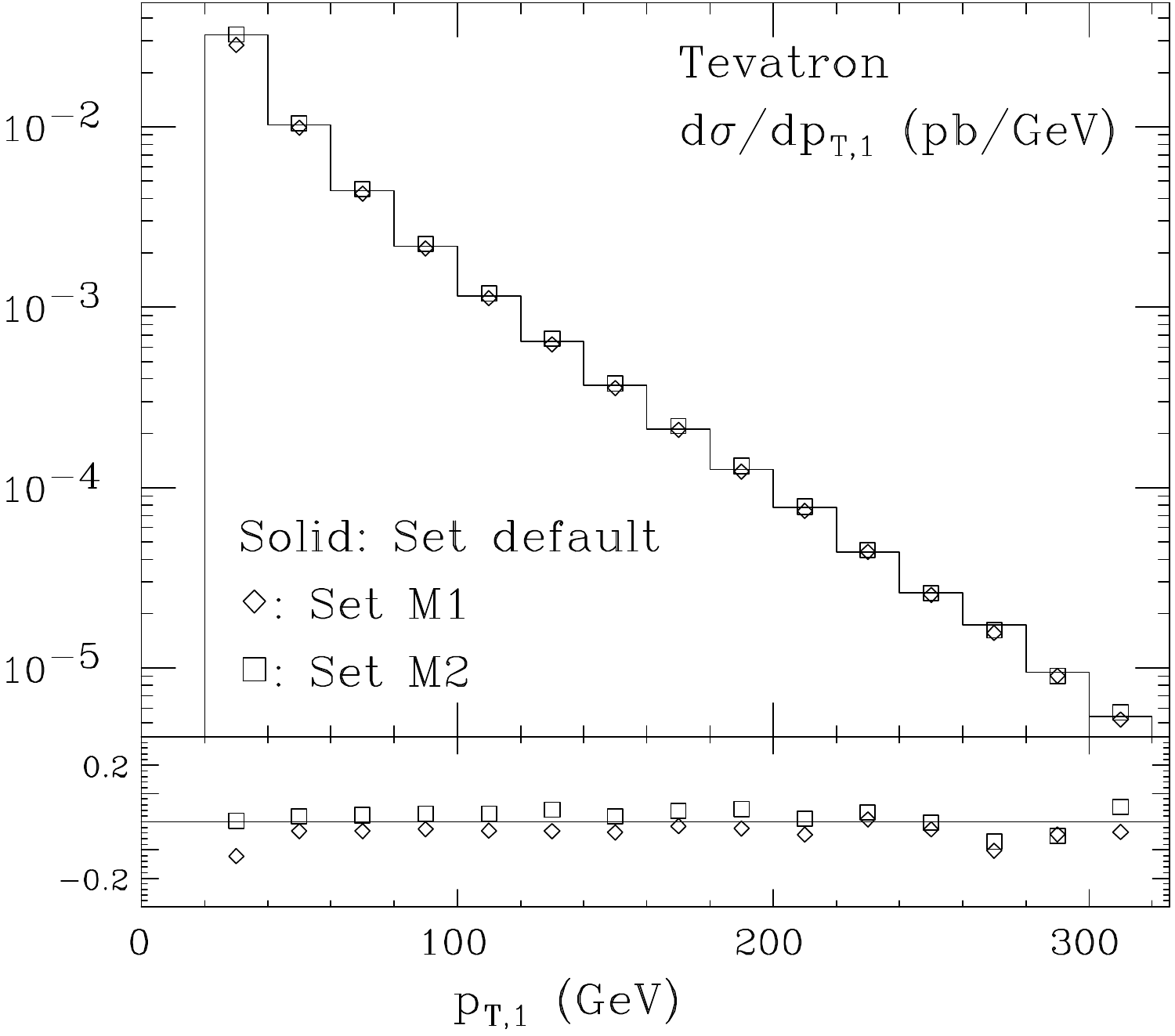} \\
\ccaption{}{\label{fig:pt_GM} Comparison between the three
alternative sets of generation (left) and matching (right) parameters
given in table~\ref{tab:cuts_syst}, at the Tevatron.}
\end{center}
\end{figure}

\begin{figure}
\begin{center}
\includegraphics[height=0.22\textheight,clip]{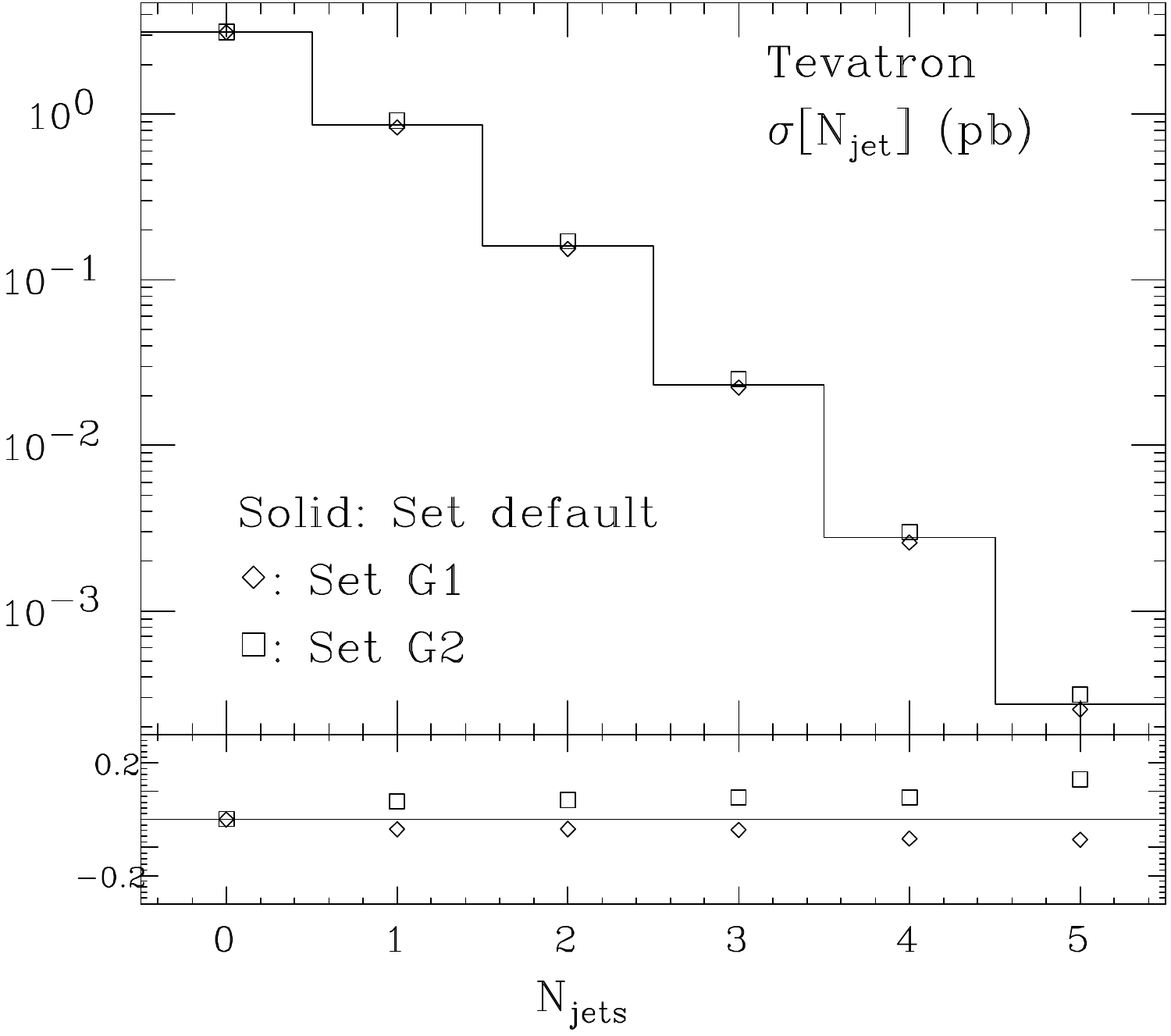}
\hspace*{0.5cm}
\includegraphics[height=0.22\textheight,clip]{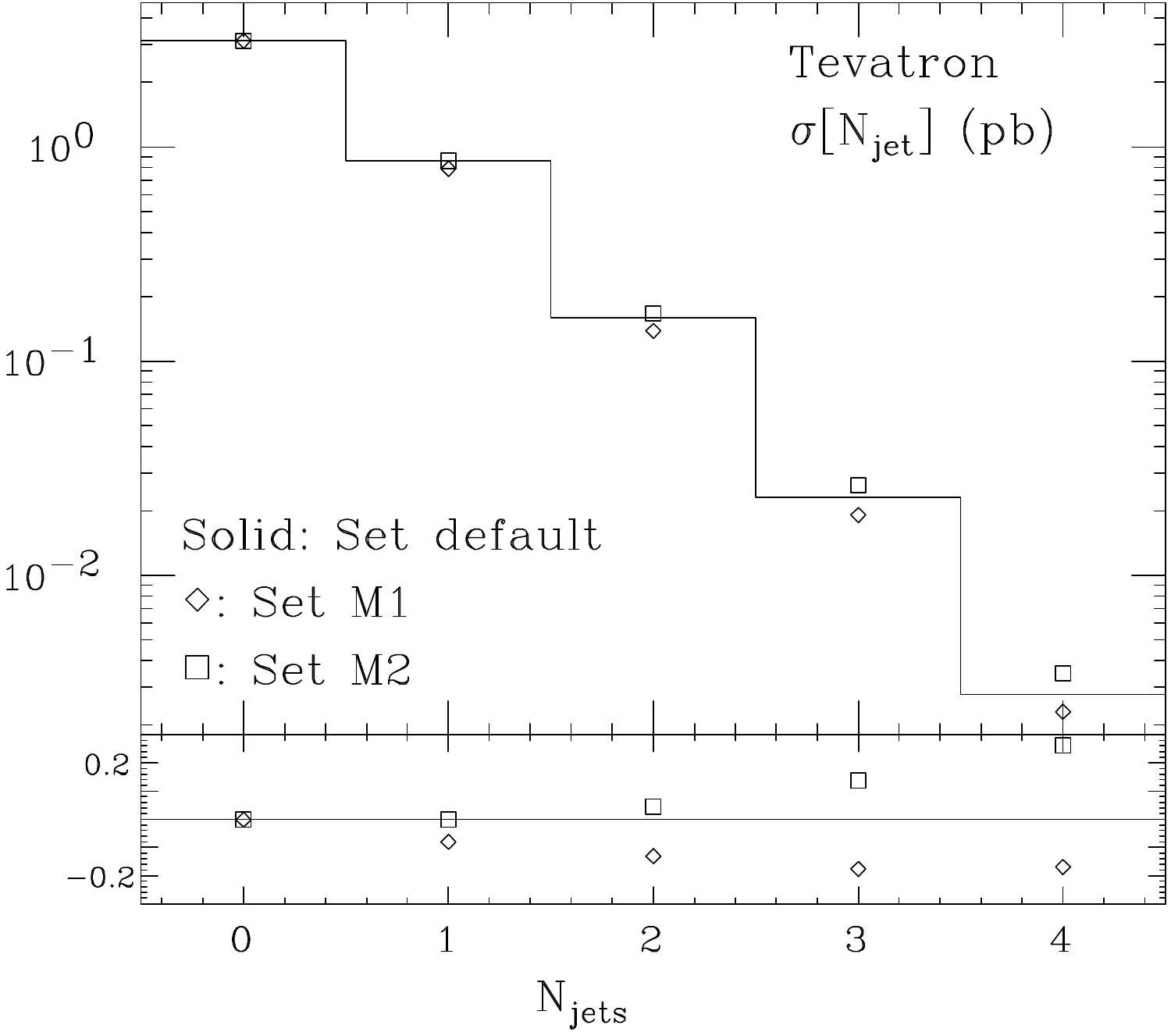}
\\
\includegraphics[height=0.22\textheight,clip]{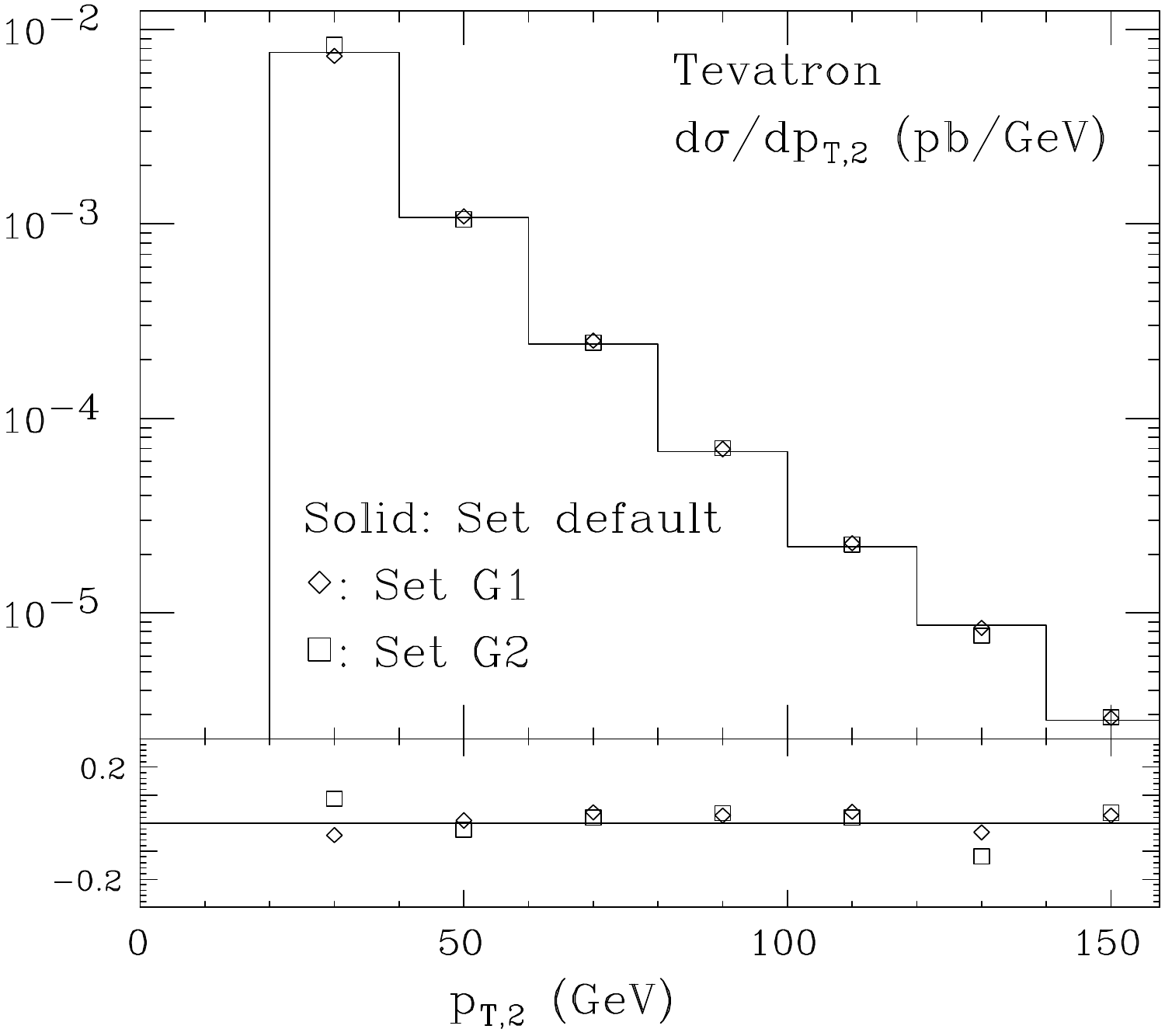}
\hspace*{0.5cm}
\includegraphics[height=0.22\textheight,clip]{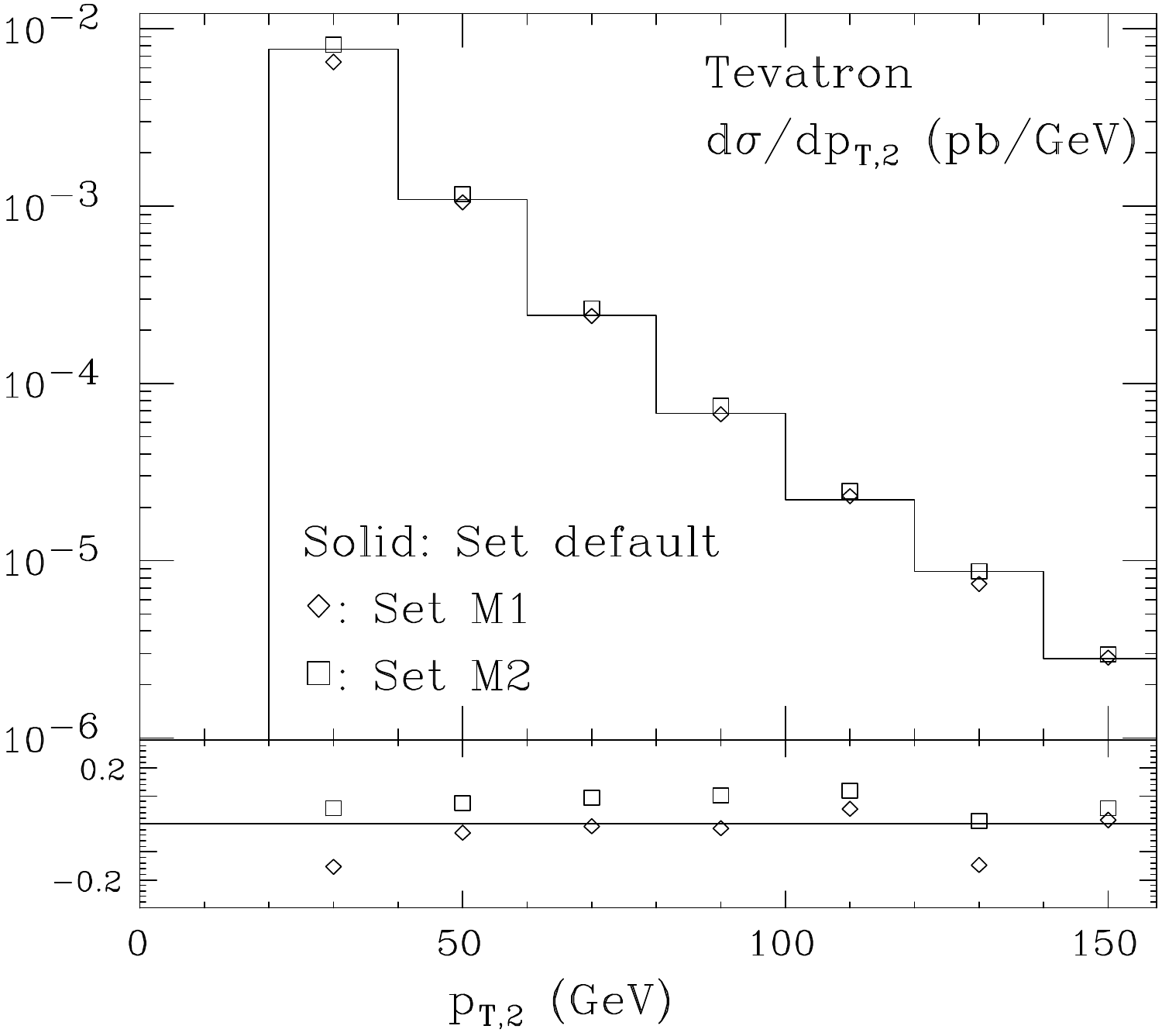}
\\
\includegraphics[height=0.22\textheight,clip]{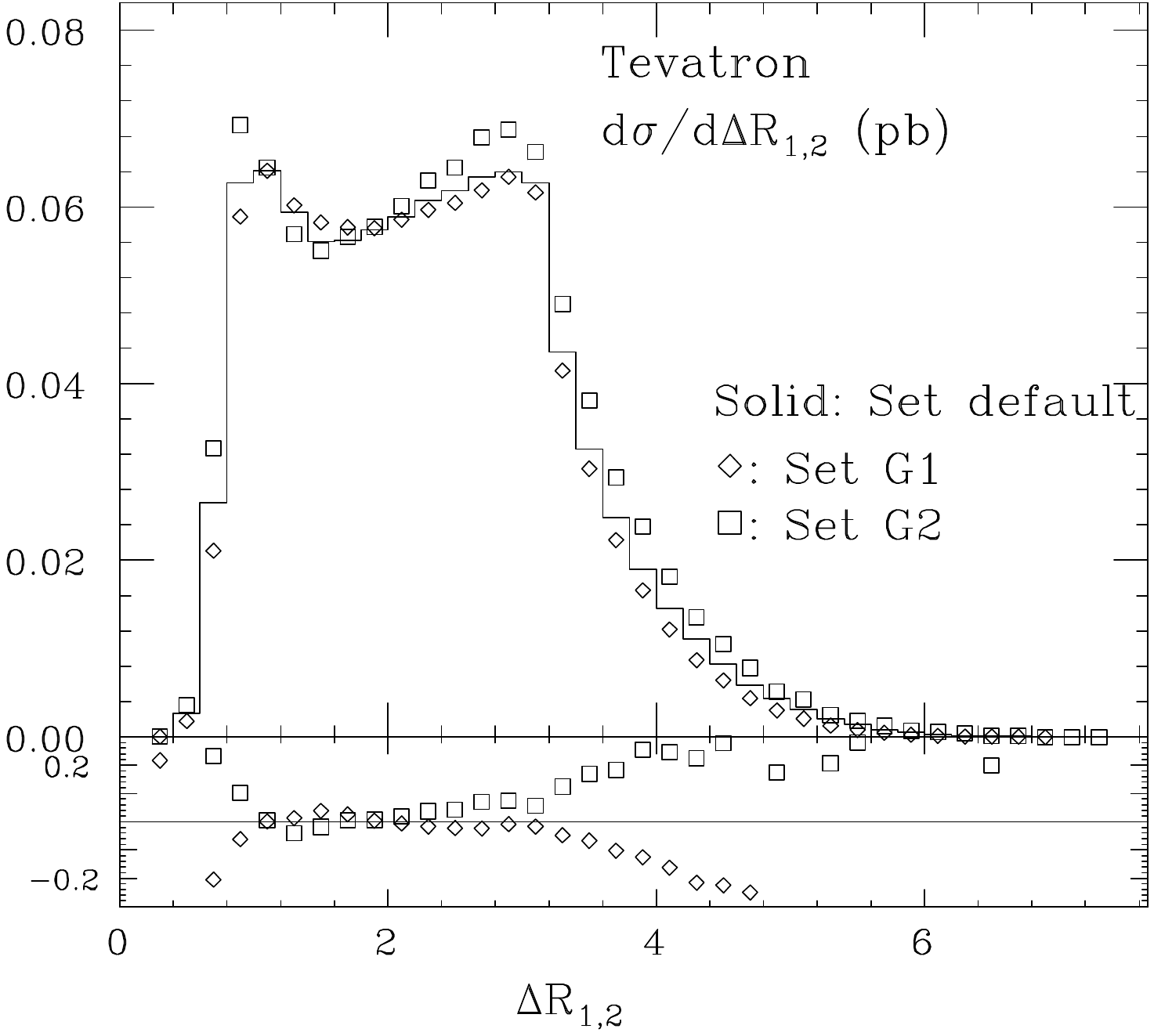}
\hspace*{0.5cm}
\includegraphics[height=0.22\textheight,clip]{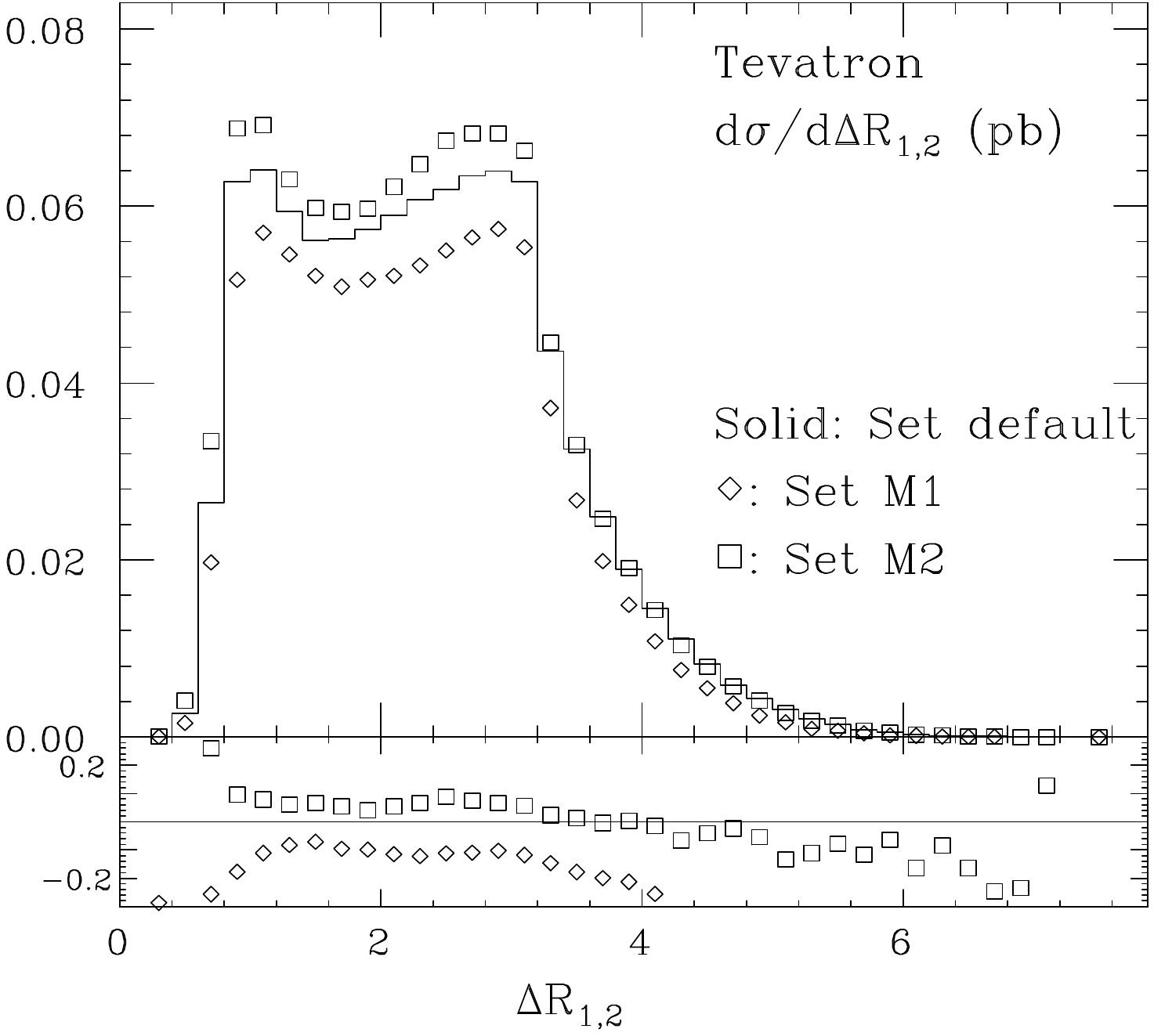}
\\
\includegraphics[height=0.22\textheight,clip]{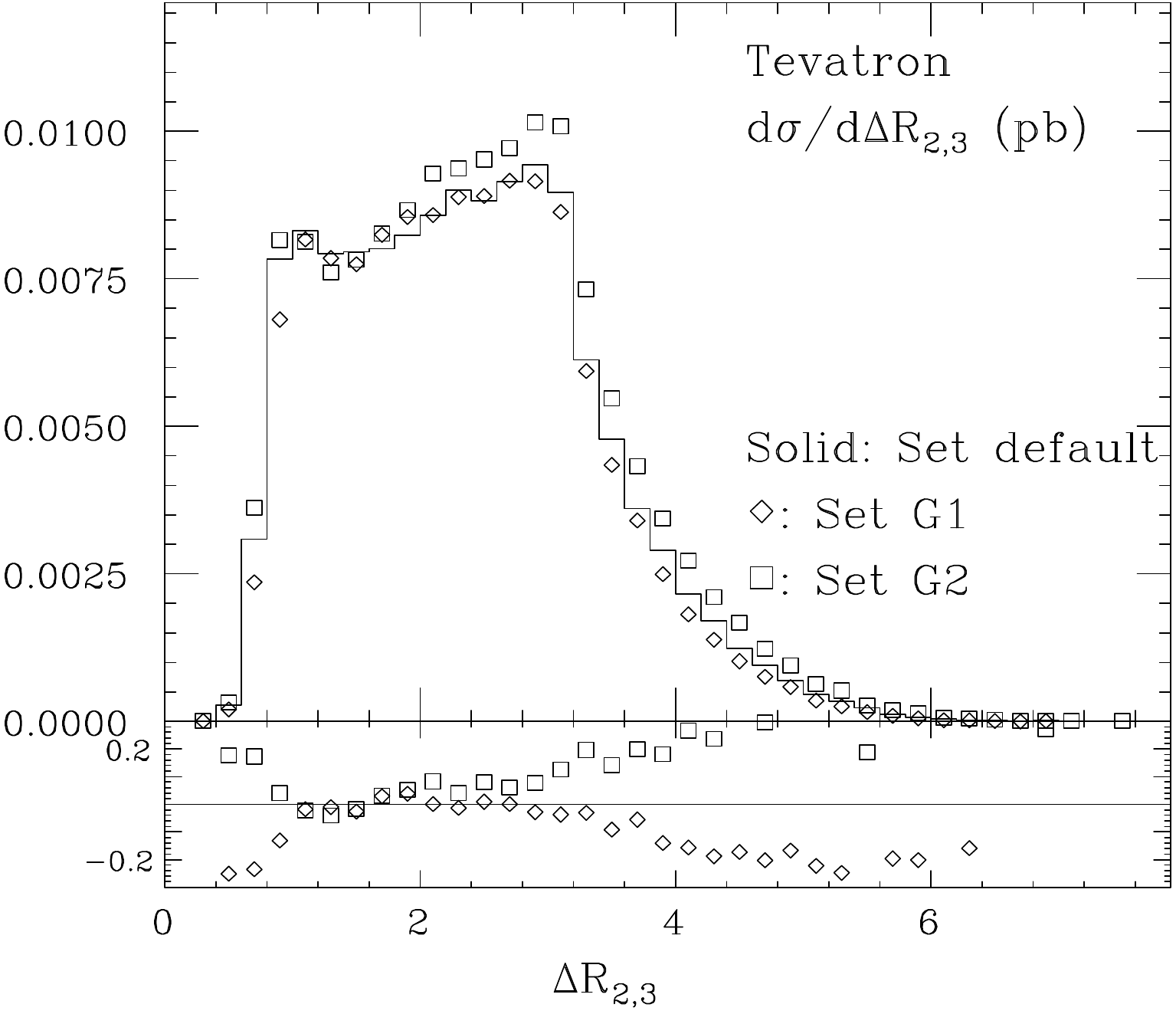}
\hspace*{0.5cm}
\includegraphics[height=0.22\textheight,clip]{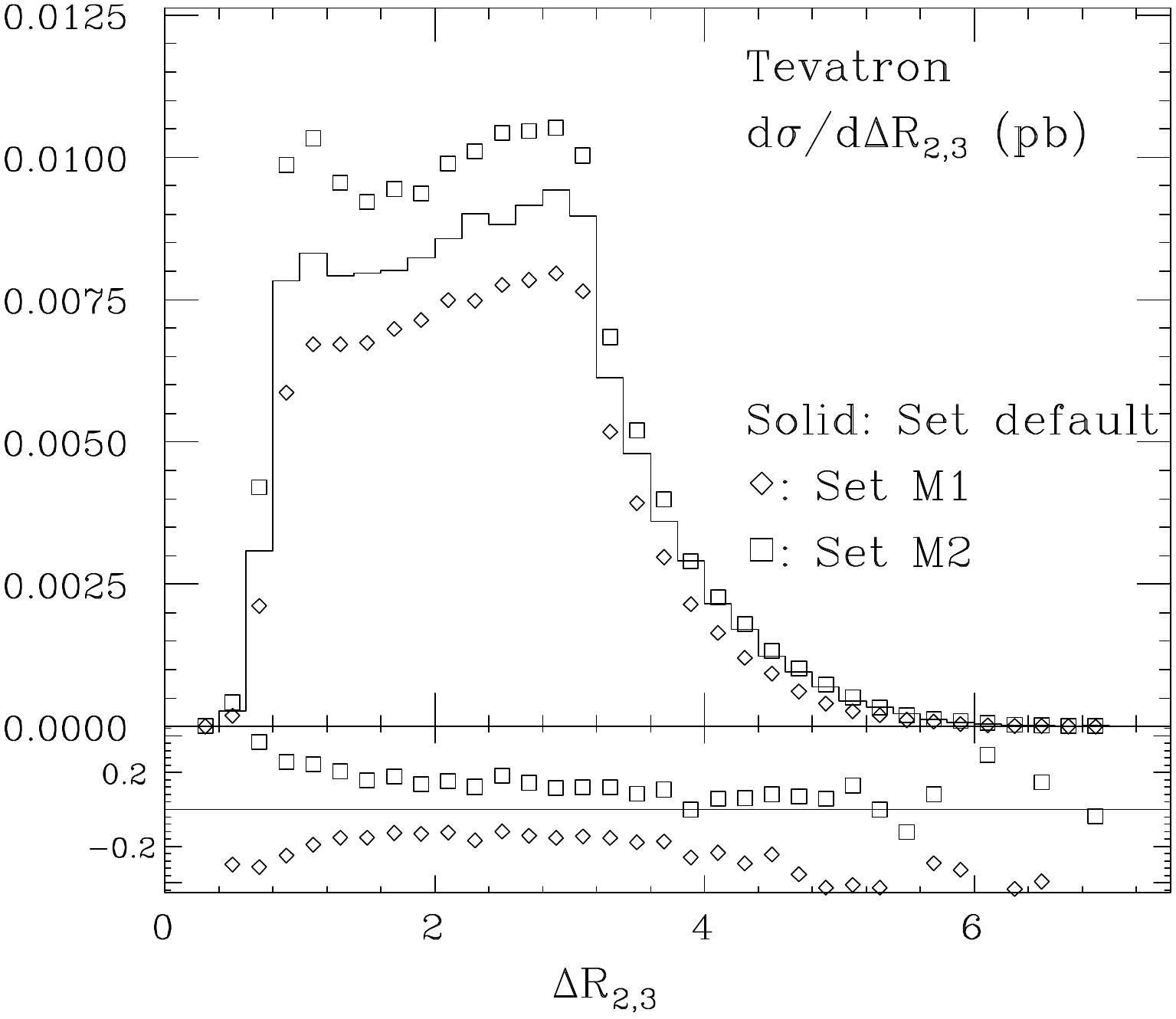}
\\
\ccaption{}{\label{fig:Njet_GM} Comparison between the three
  alternative sets of generation and matching parameters given in
  table~\ref{tab:cuts_syst}, for multijet distributions at the Tevatron.}
\end{center}
\end{figure}

Then we proceed to study some distributions for the Tevatron, showing the observables dominated by contributions with up to 1 hard parton in fig.~\ref{fig:pt_GM}, and those relative
to multijet final states in fig.~\ref{fig:Njet_GM}.
We find that these distribution are stable against reasonable variations of the internal parameters, with relative differencies confined well below few percents, both in matching and generation parameter variations.
Angular observables, such as $\Delta R$, are more sensible, since they are directly related to the matching variables, and their agreement is within 10\%.
\begin{table}
\begin{center}
\begin{tabular}{|l|cc|cc|}
\hline
&   \multicolumn{2}{c|}{Generation parameters} &
       \multicolumn{2}{c|}{Matching parameters} \\
Param set & \ptmin & \Rmin & min \etclus & \Rmat\\
\hline
Tevatron, default  & 20 &  0.7 & 25 & 1.5 $\times$ 0.7 \\
Tevatron, Set G1   & 15 &  0.7 & 20 & 1.5 $\times$ 0.7 \\
Tevatron, Set G2   & 30 &  0.7 & 36 & 1.5 $\times$ 0.7 \\
Tevatron, Set M1   & 20 &  0.7 & 20 & 1.5 $\times$ 0.7 \\
Tevatron, Set M2   & 20 &  0.7 & 25 & 1.5 $\times$ 1.0 \\
\hline
LHC, default  & 30 &  0.7 & 36 & 1.5 $\times$ 0.7 \\
LHC, Set G1   & 25 &  0.7 & 30 & 1.5 $\times$ 0.7 \\
LHC, Set G2   & 40 &  0.7 & 48 & 1.5 $\times$ 0.7 \\
LHC, Set M1   & 30 &  0.7 & 30 & 1.5 $\times$ 0.7 \\
LHC, Set M2   & 30 &  0.7 & 36 & 1.5 $\times$ 1.0 \\
\hline
\end{tabular}
\ccaption{}{\label{tab:cuts_syst}
Variations of  the generation and matching parameters used for the
study of the systematics.}
\end{center}
\end{table}

The analysis at the LHC, which will not be shown here, leads to qualitatively and quantitatively
similar results.

\section{Comparisons with MC@NLO}
\label{sec:MC@NLO}

We shall now compare in detail the description of \ttbar\ events as
provided by \ALPGENs and \mcnlo.  For consistency with the \mcnlo\
approach, where only the \oacube\ ME effects are included, we use
\alpgens samples obtained by stopping the ME contributions only to 1 extra-parton besides the \ttbar\ pair.
This strategy allow to highlight the different features of the two alternative approaches applied to same set of contributions. It is understood that a homogeneous comparison can only be done through the introduction of a proper K factor, determined by the ratio of the total rates of the two predictions.
We adopt the same simulation setup, modifying only the same factorization and renormalization scale in order to match \mcnlo's default:
 \\
$\mursq=\mufsq=\sum_{i=t,\bar{t}} \;\frac{1}{2}[m_i^2+(p_{\sss T}^i)^2].$

\begin{figure}
\begin{center}
\includegraphics[height=0.22\textheight,clip]{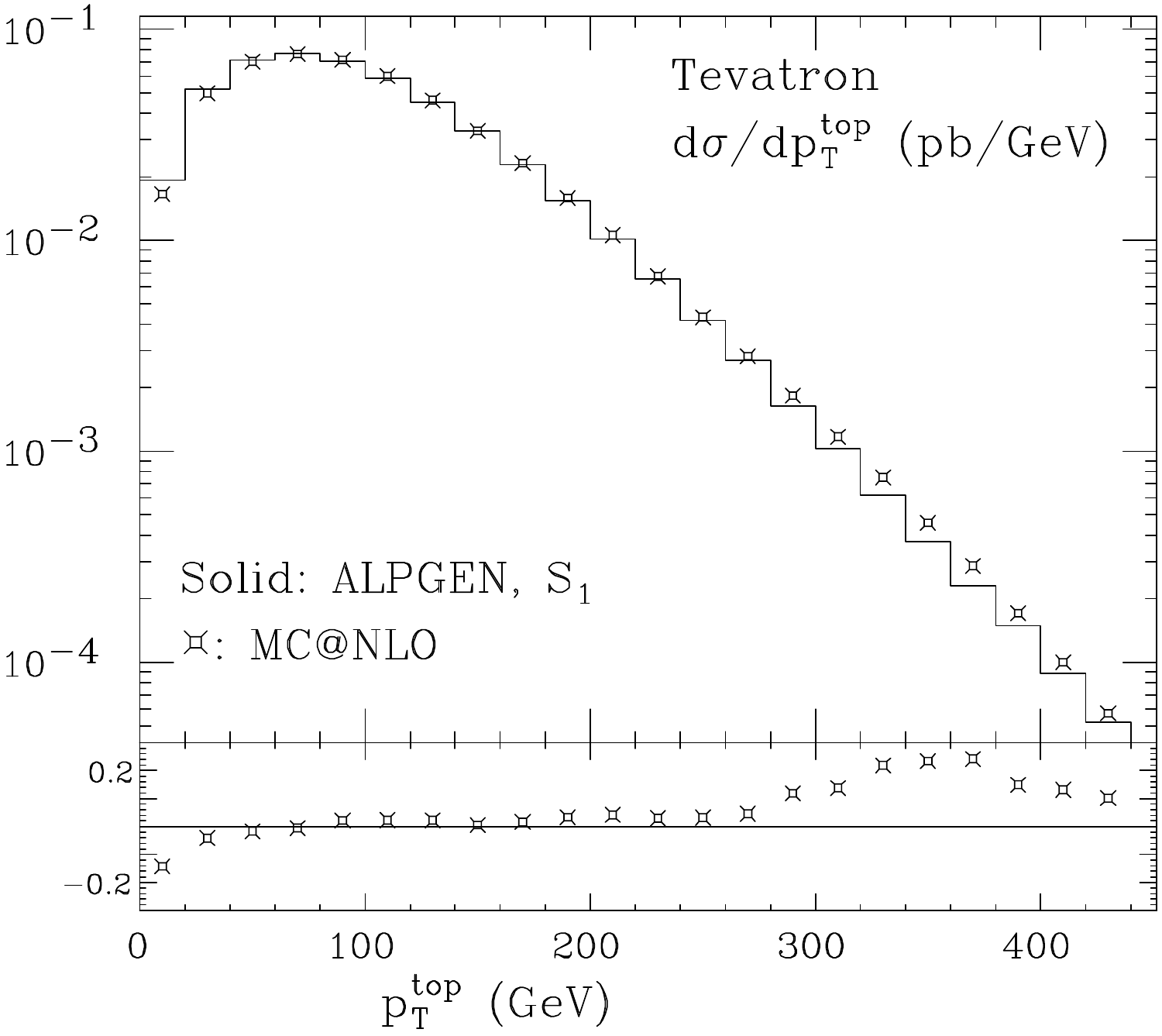}
\hspace*{0.5cm}
\includegraphics[height=0.22\textheight,clip]{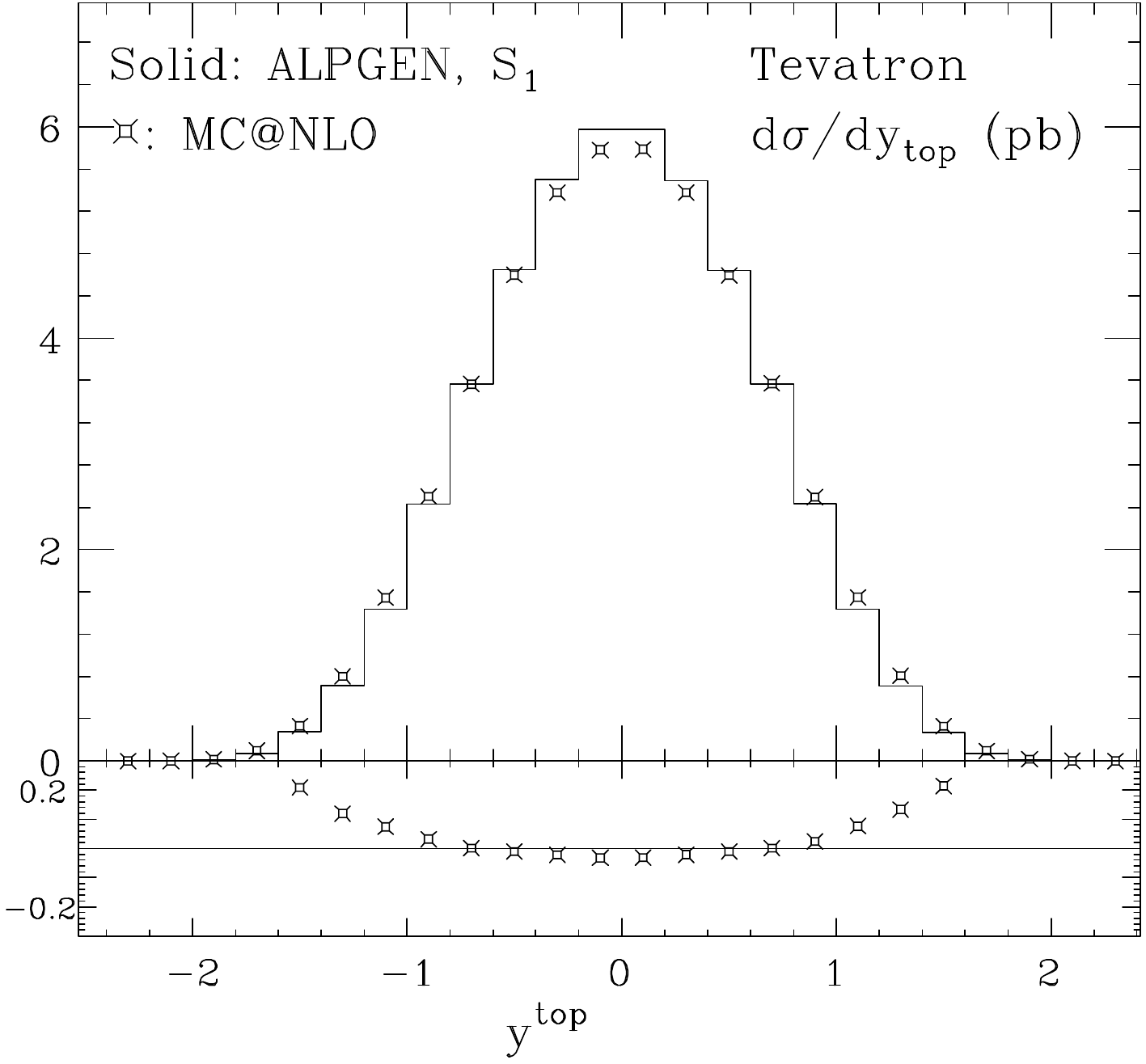}
\\
\includegraphics[height=0.22\textheight,clip]{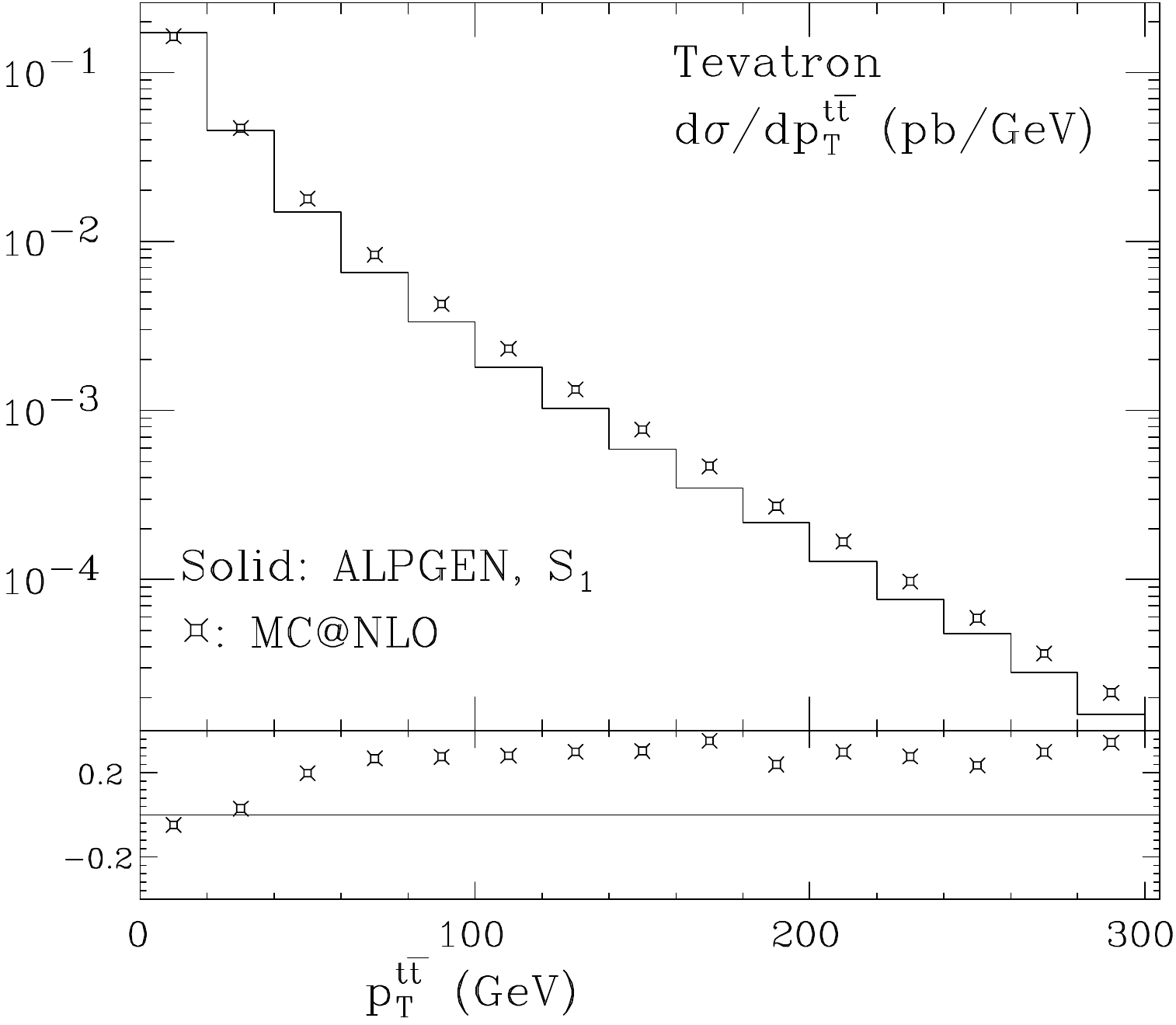}
\hspace*{0.5cm}
\includegraphics[height=0.22\textheight,clip]{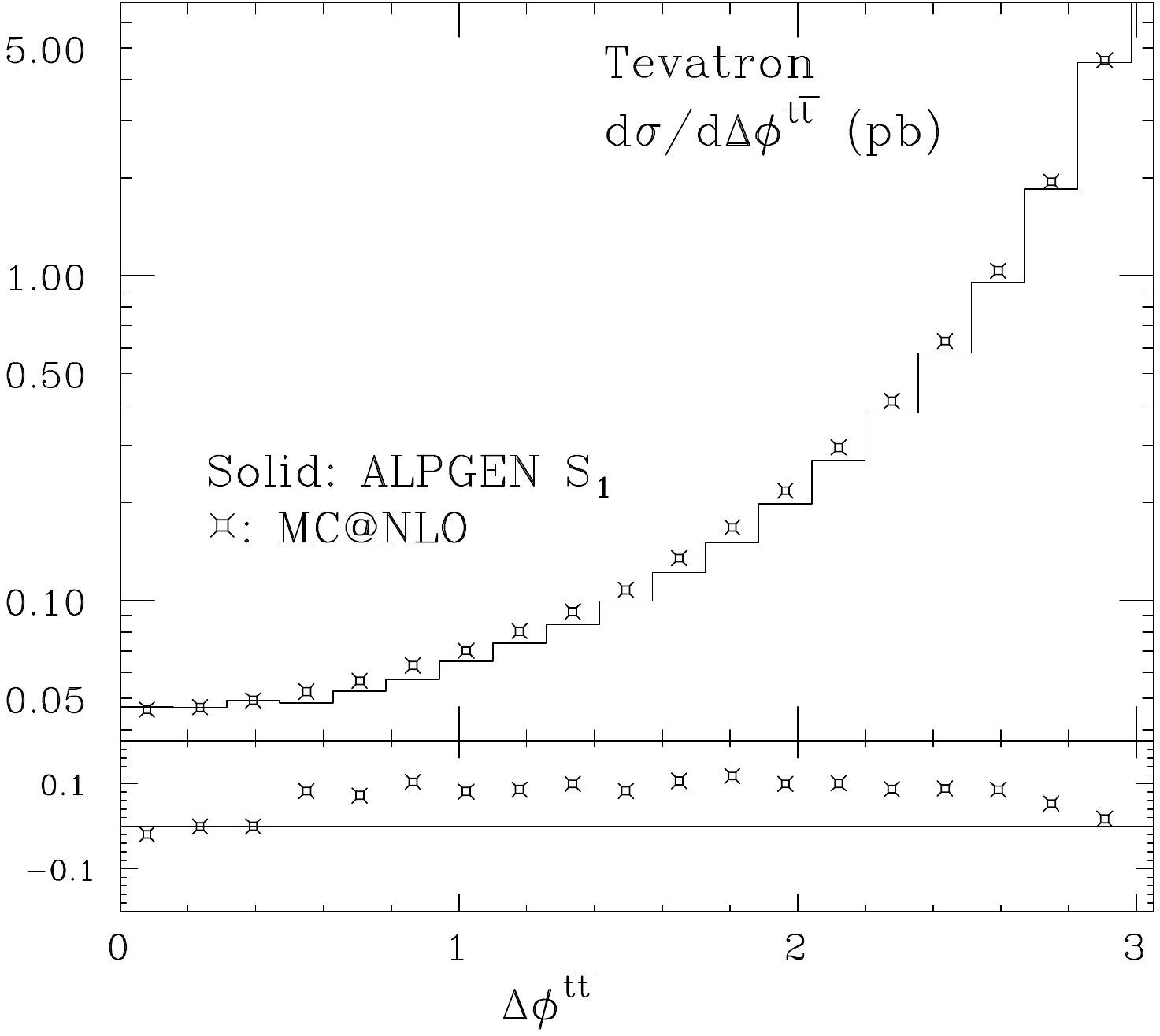}
\\ 
\includegraphics[height=0.22\textheight,clip]{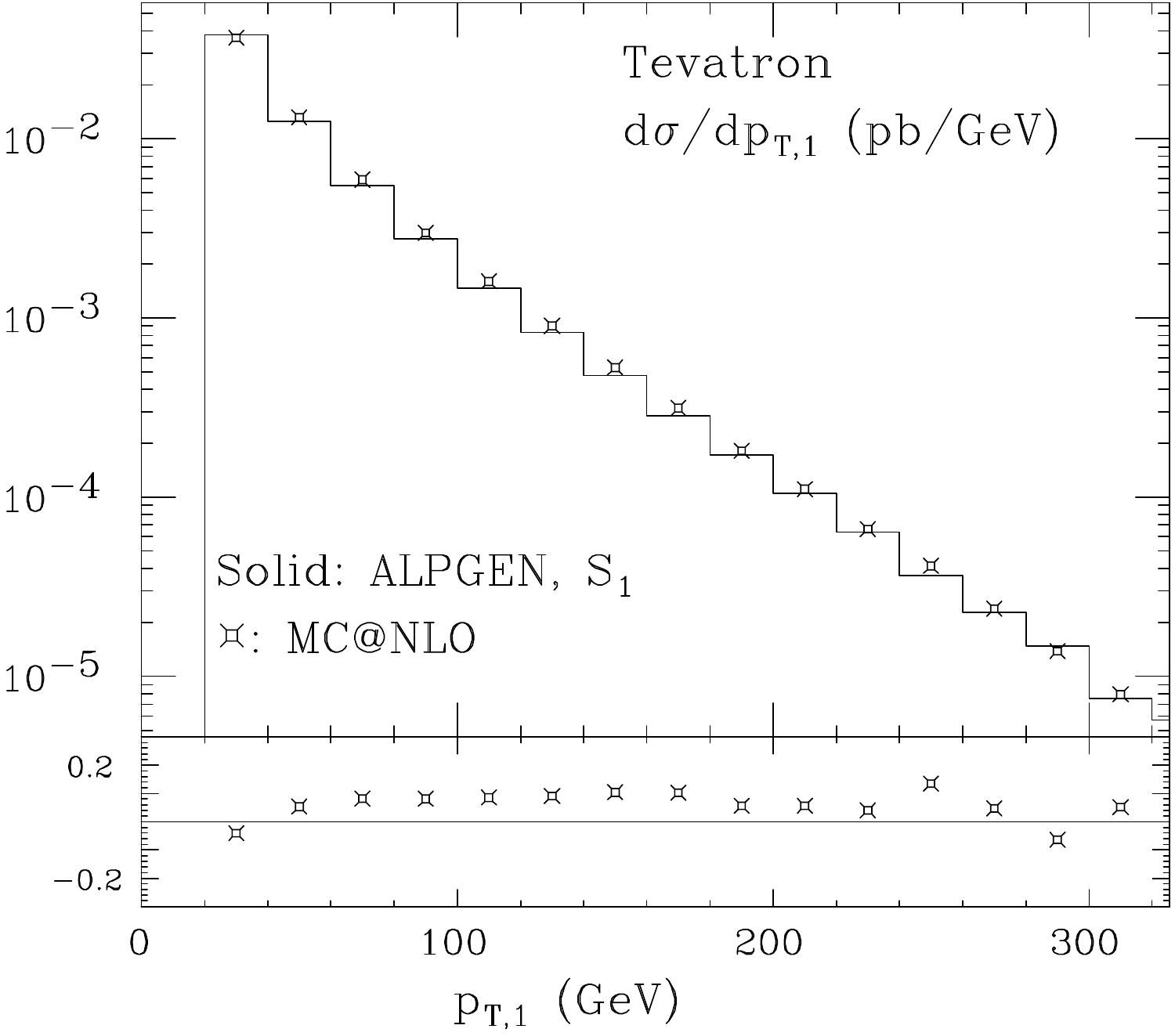}
\hspace*{0.5cm}
\includegraphics[height=0.22\textheight,clip]{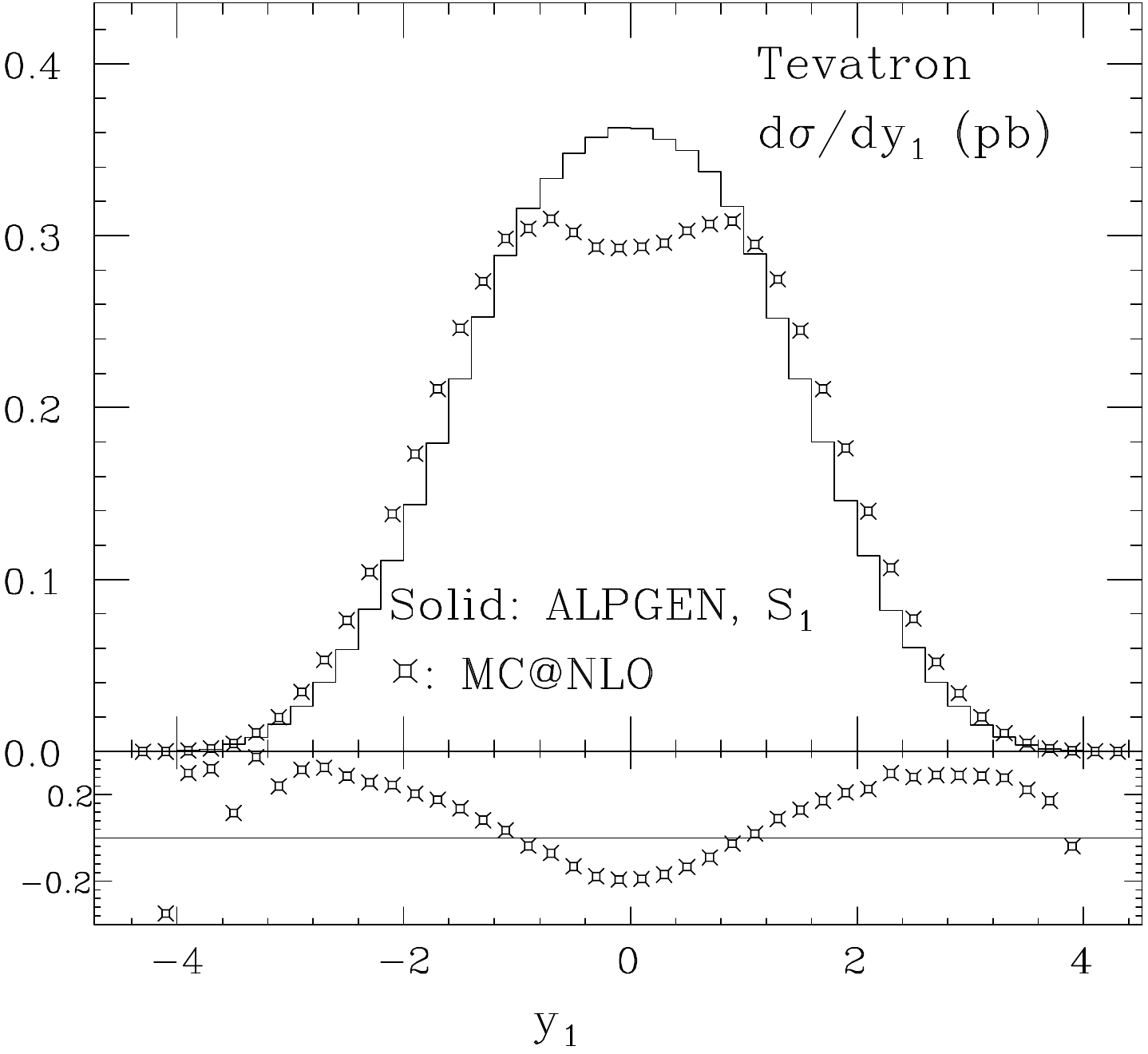} 
\\
\includegraphics[height=0.22\textheight,clip]{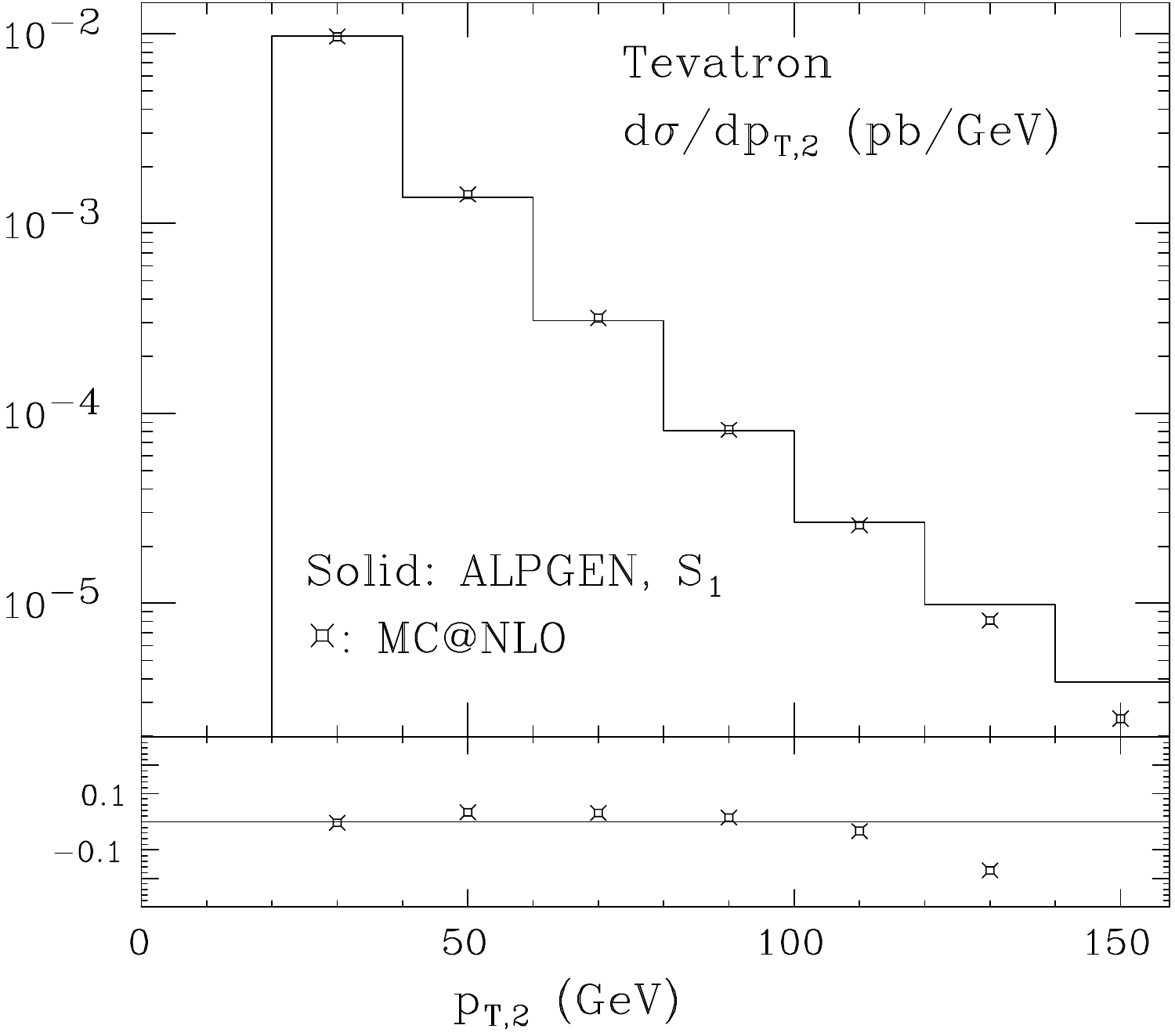}
\hspace*{0.5cm}
\includegraphics[height=0.22\textheight,clip]{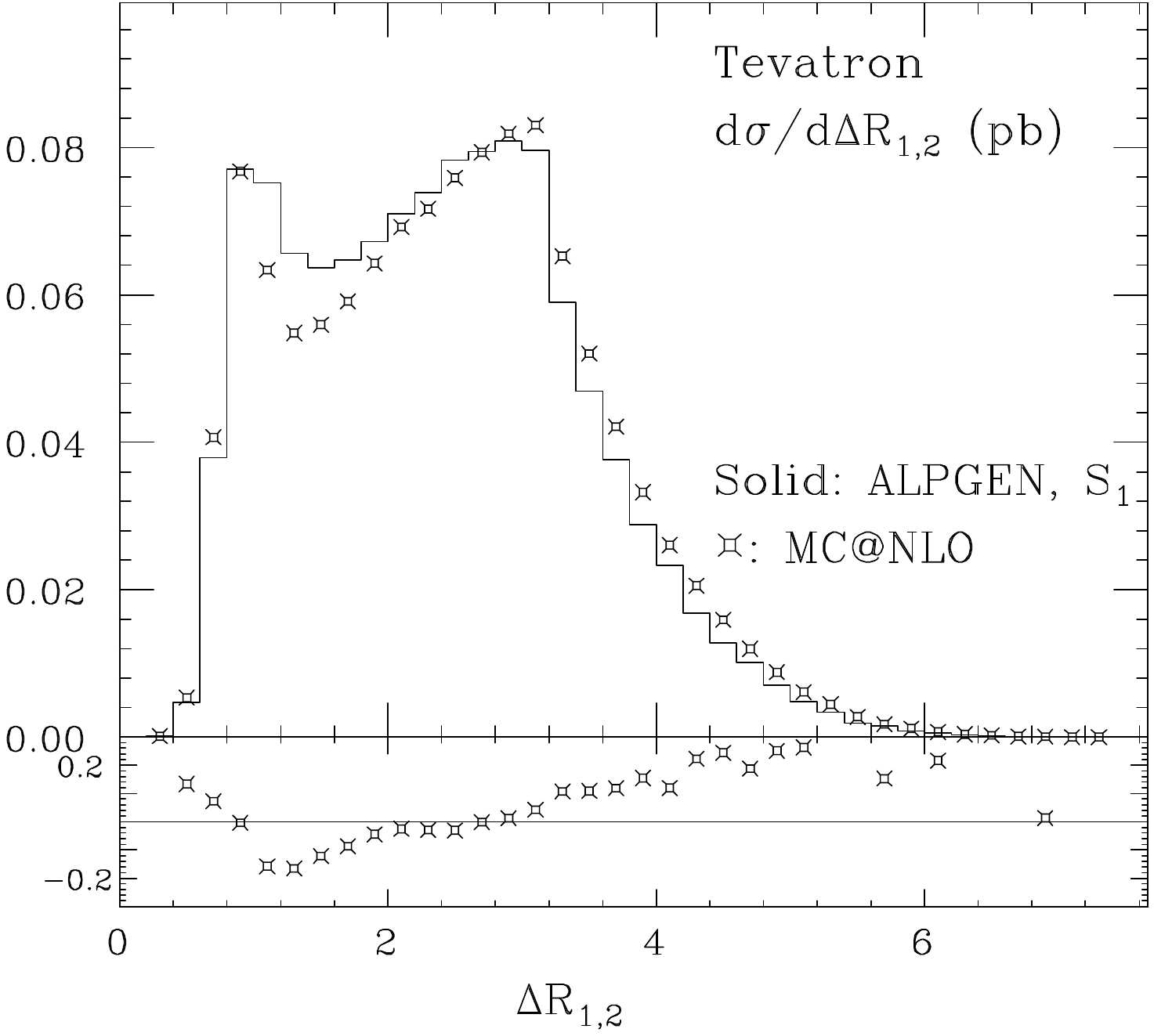}
\ccaption{}{\label{fig:inclusiveTEV} Comparison of \alpgen\
(histogram) and \mcnlo\ (plot) distributions, at the Tevatron. The
\alpgen\ results are rescaled to \mcnlo, using the K factor of 1.36.
The relative difference (\mcnlo-\alpgen)/\alpgen) is shown at the
bottom of each plot.  }
\end{center}
\end{figure}

% \begin{figure}
% \begin{center}
% \includegraphics[height=0.22\textheight,clip]{pt_MC_LHC}
% \hspace*{0.5cm}
% \includegraphics[height=0.22\textheight,clip]{ytop_MC_LHC}
% \\
% \includegraphics[height=0.22\textheight,clip]{ptt_MC_LHC}
% \hspace*{0.5cm}
% \includegraphics[height=0.22\textheight,clip]{dphi_MC_LHC}
% \\ 
% \includegraphics[height=0.22\textheight,clip]{pt1_MC_LHC}
% \hspace*{0.5cm}
% \includegraphics[height=0.22\textheight,clip]{y1_MC_LHC} 
% \\
% \includegraphics[height=0.22\textheight,clip]{pt2_MC_LHC}
% \hspace*{0.5cm}
% \includegraphics[height=0.22\textheight,clip]{Dr12_MC_LHC}
% \ccaption{}{\label{fig:inclusiveLHC} Same as
%   fig.~\ref{fig:inclusiveTEV} for the LHC, using the K
%   factor of 1.51.}
% \end{center}
% \end{figure}

% \[
% \mu^2 =\frac{1}{2} \left ({\pttopsq+\mtsq} + 
% {\pttbarsq+\mtsq} \right )
% \]

The upper two rows of 
plots in figs.~\ref{fig:inclusiveTEV} 
refer to inclusive properties of the \ttbar\ system, 
namely the transverse momentum and rapidity
of the top and anti-top quark, the transverse momentum of the 
\ttbar\ pair, and the azimuthal angle $\phitt$
between the top and anti-top quark.
The overall agreement is good, once \alpgens is corrected with the proper
K-factor (1.36 for the Tevatron, and 1.51 for the LHC), 
and no large  discrepancy is seen between the two 
descriptions of the chosen distributions.
The most significant differencies (10 to 20\%) are seen in the 
$\pttop$ distribution, \ALPGEN's one being slightly softer.

The study of jet quantities reveals instead 
one important difference: the rapidity of the
leading jet, \yjet, is different in the two descriptions, where \mcnlos
exhibits a dip at $\yjet=0$. This difference is particularly
marked at the Tevatron, but is very visible also at the LHC.
This is shown in the right figure of the third row in
fig.~\ref{fig:inclusiveTEV}
 Visible differences are also present 
in the distribution of the 1st and 2nd jet separation in 
$(\eta,\phi)$ space, $\Delta R_{1,2}$.
% \noindent
% \begin{figure}
% \begin{center}
% \includegraphics[height=0.25\textheight,clip]{Njet_MC_TEV}
% \hspace*{0.5cm}
% \includegraphics[height=0.25\textheight,clip]{Njet_MC_LHC}
% \ccaption{}{ \label{fig:Njets}
% Jet multiplicity from \alpgen\ and \mcnlo, at the Tevatron (left) and
% at the LHC (right).
% The relative difference (\mcnlo-\alpgen)/\alpgen\
% is shown at the bottom of each plot.}
% \end{center}
% \end{figure}

\begin{figure}
\begin{center}
\includegraphics[height=0.25\textheight,clip]{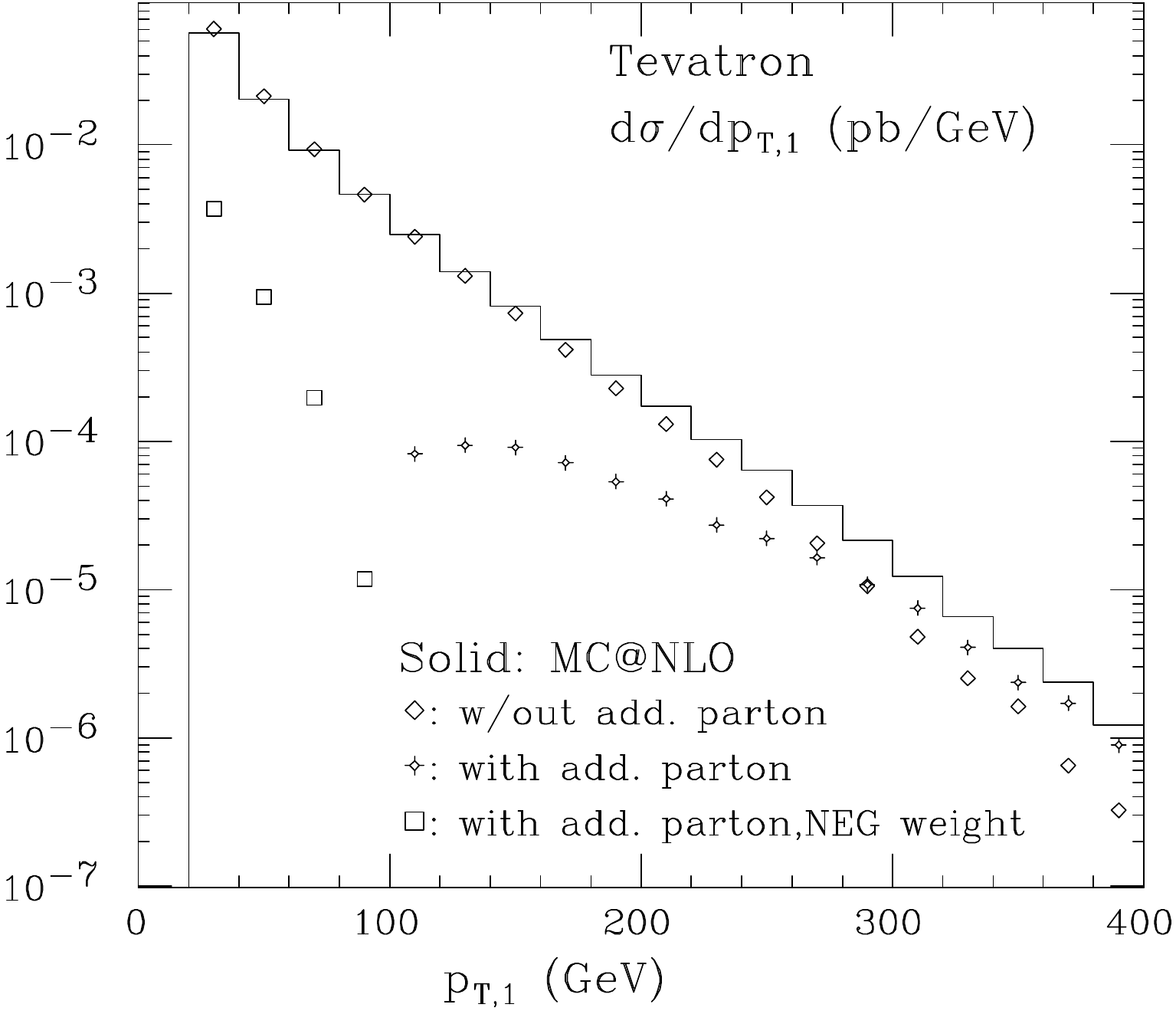}
\hspace*{0.5cm}
\includegraphics[height=0.25\textheight,clip]{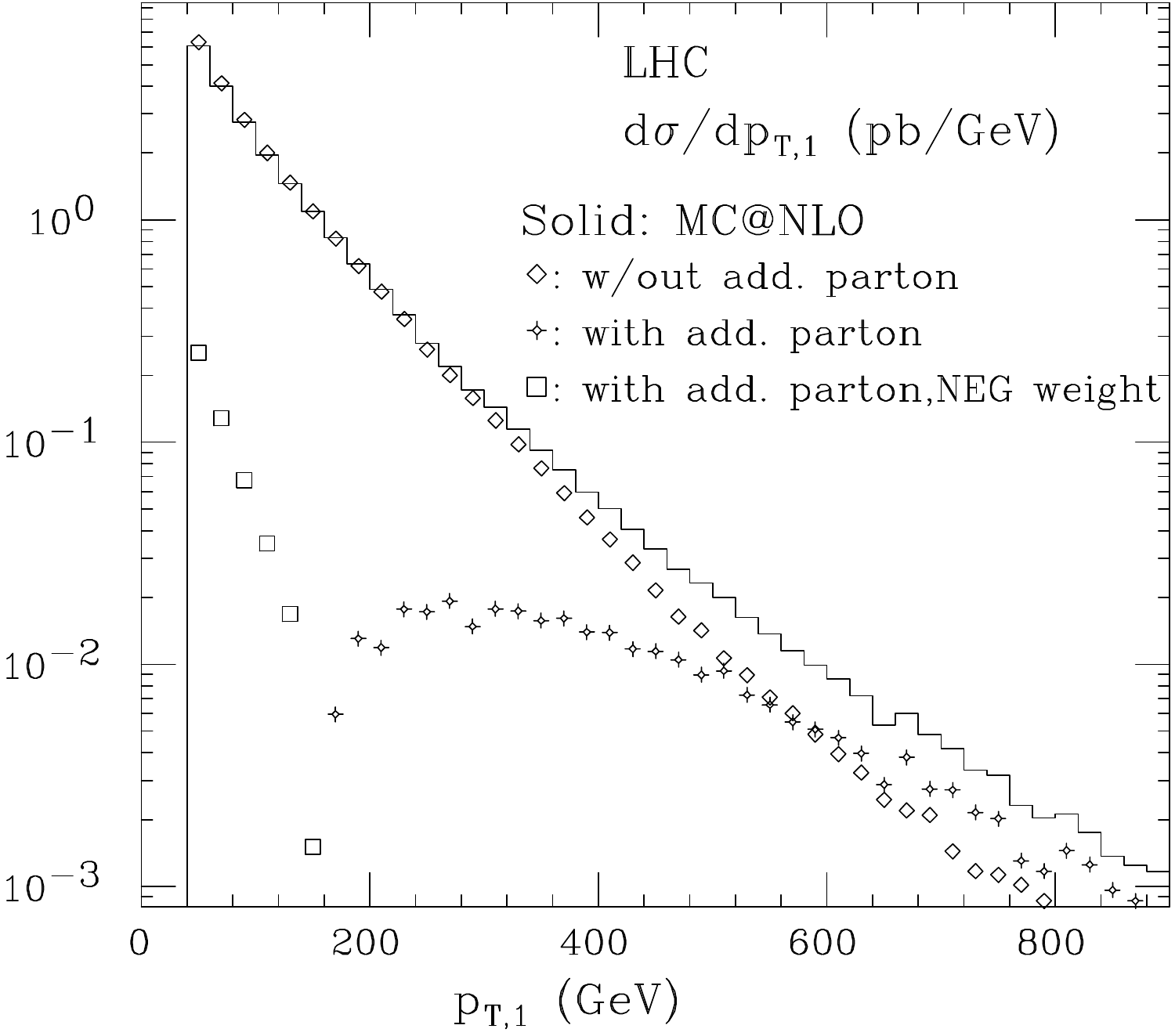}
\ccaption{}{\label{fig:01mcnlo} Contributions to the transverse
  momentum of the leading jet in \mcnlo. 
Tevatron (left) and LHC (right). }
\end{center}
\end{figure}

\begin{figure}
\begin{center}
\includegraphics[width=0.28\textwidth,clip,]{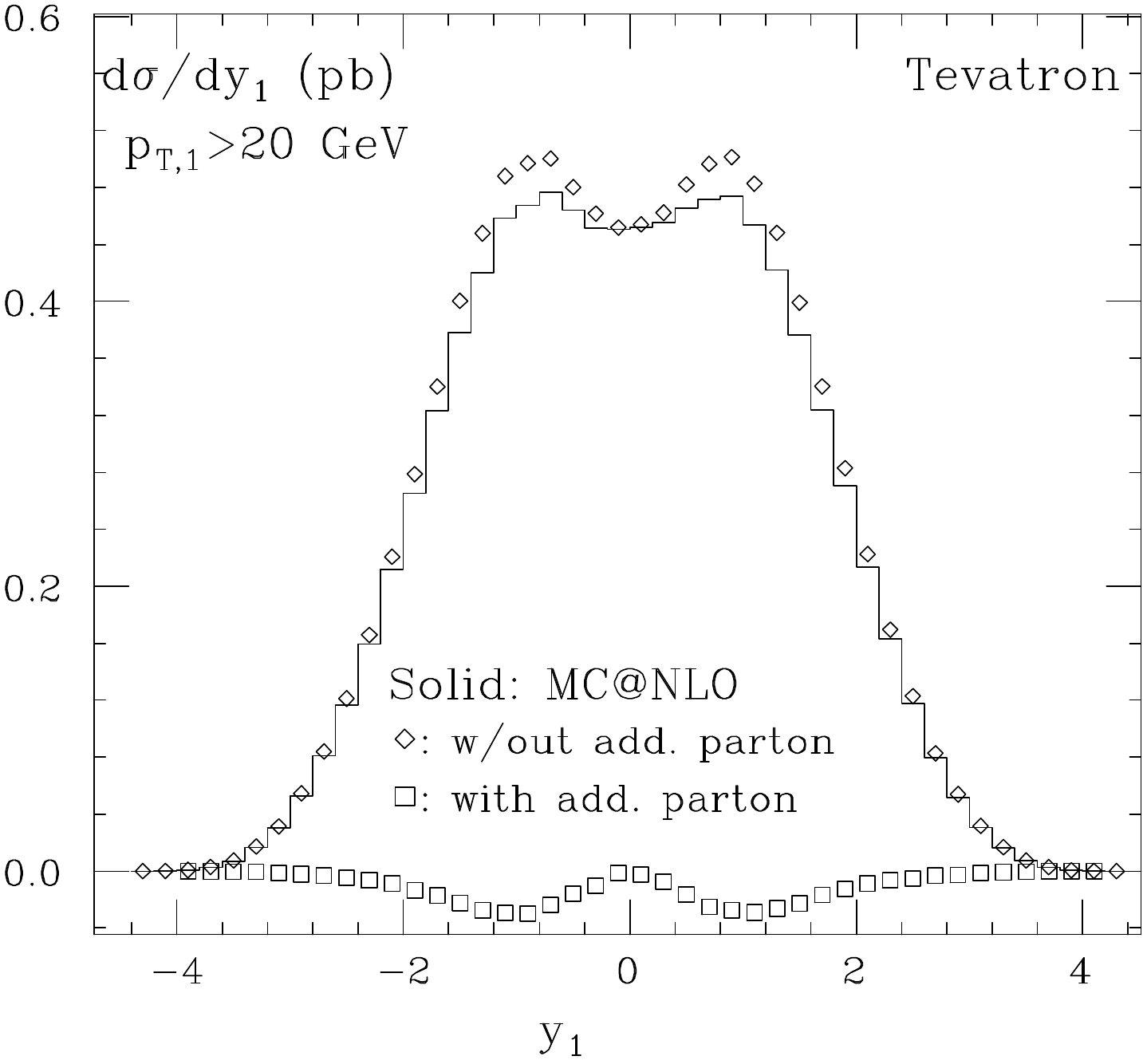}
\hspace*{0.5cm}
\includegraphics[width=0.28\textwidth,clip,]{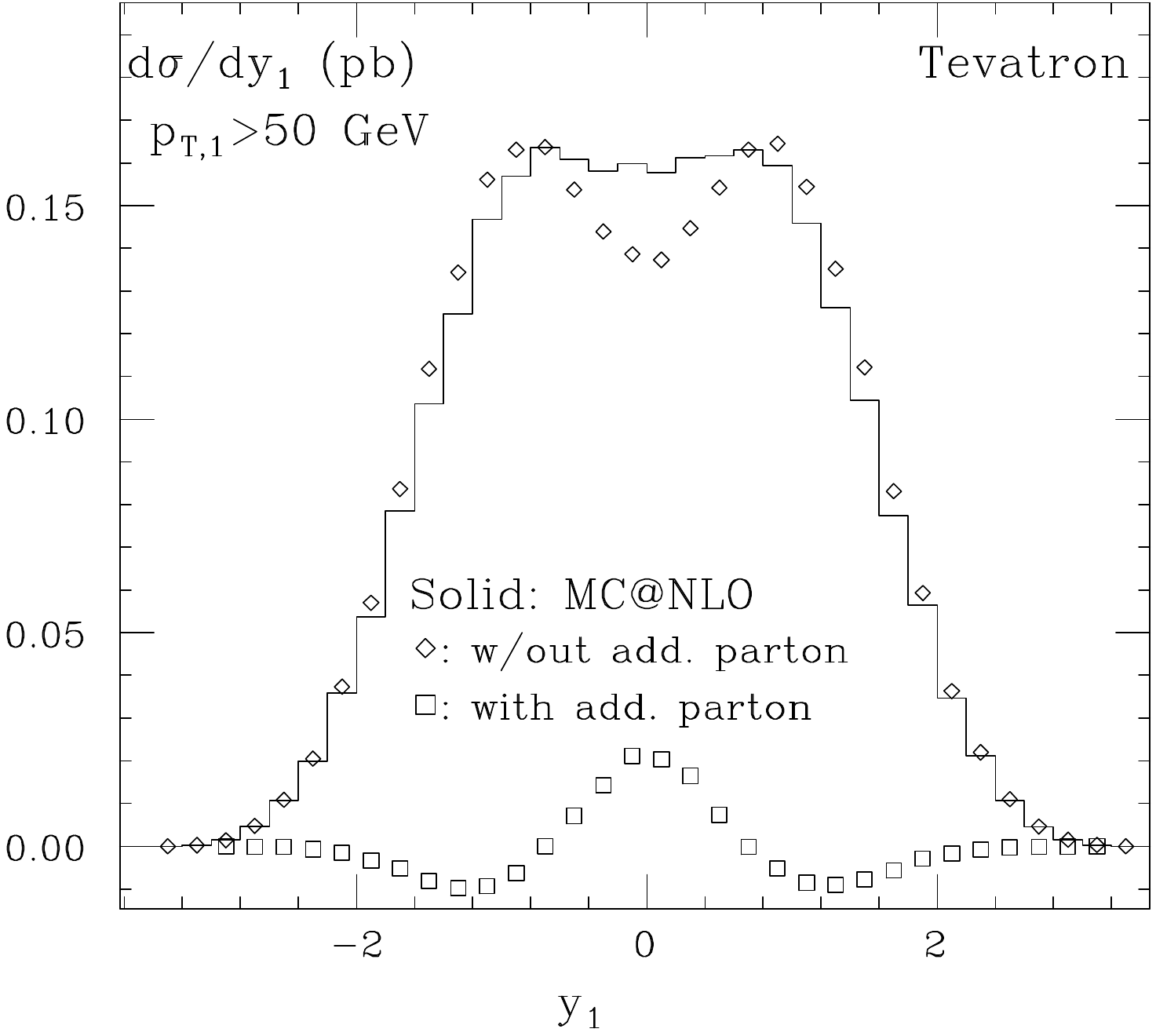}
\hspace*{0.5cm}
\includegraphics[width=0.28\textwidth,clip,]{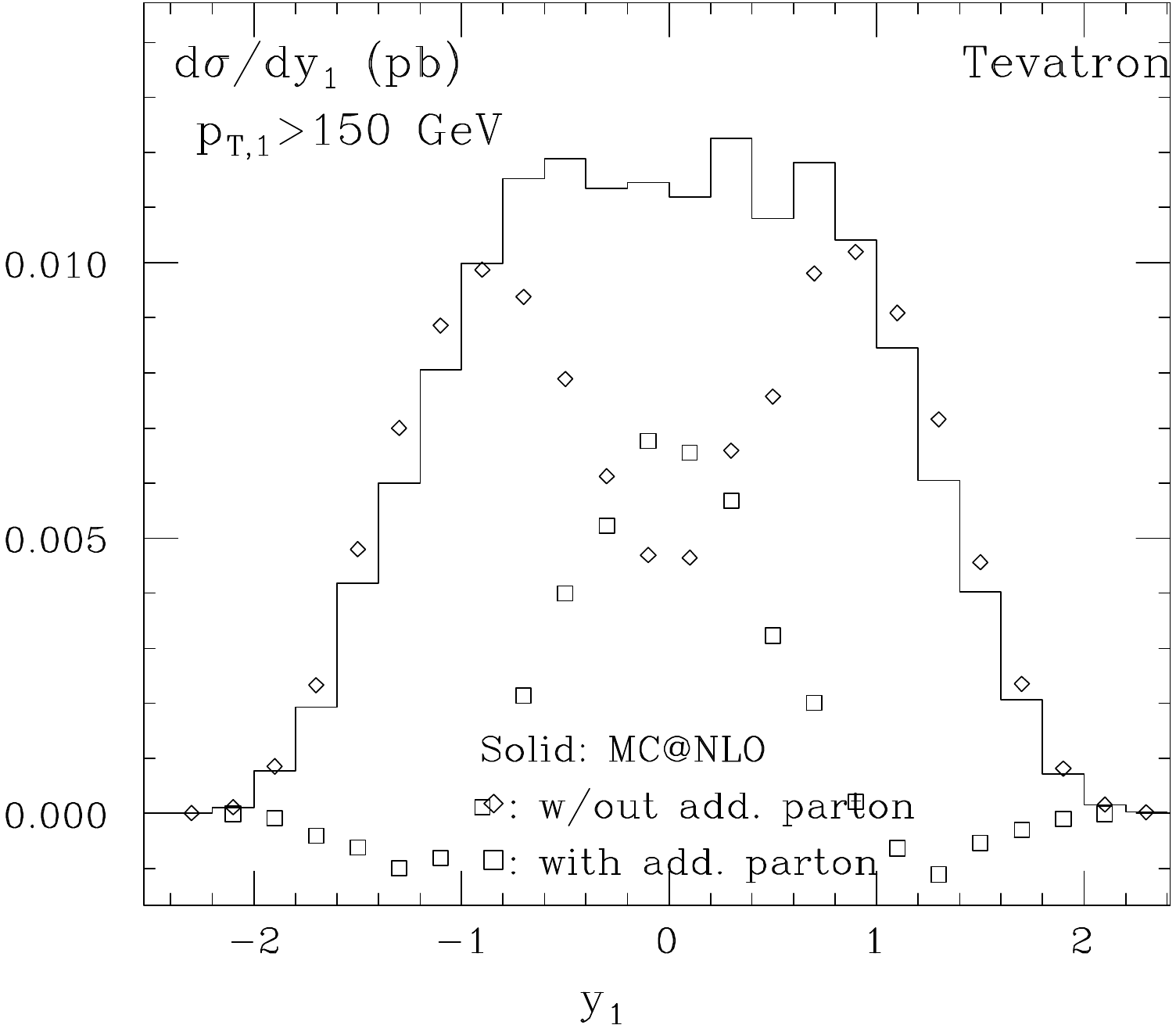}
\\
\includegraphics[width=0.28\textwidth,clip,]{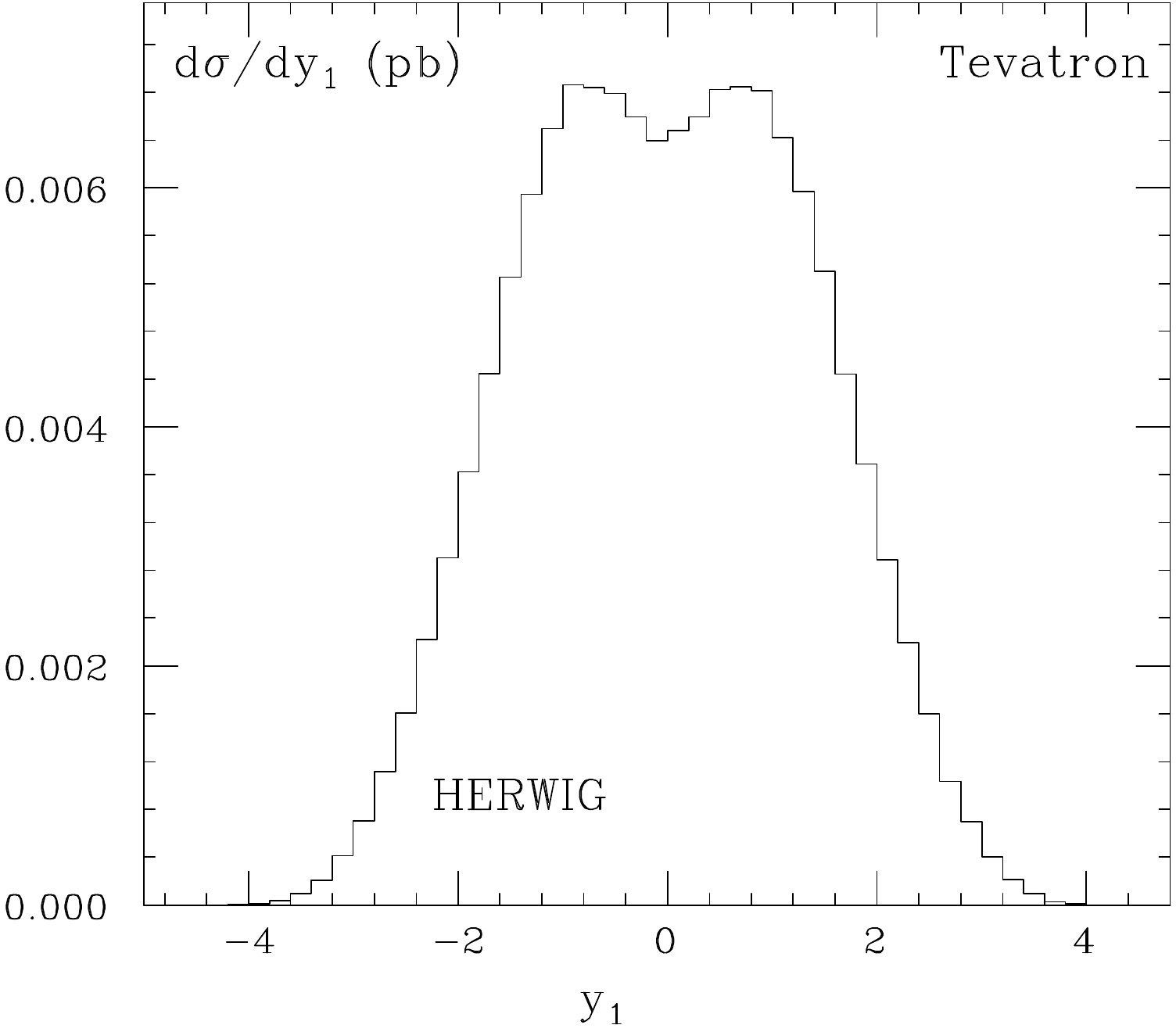}
\hspace*{0.5cm}
\includegraphics[width=0.28\textwidth,clip,]{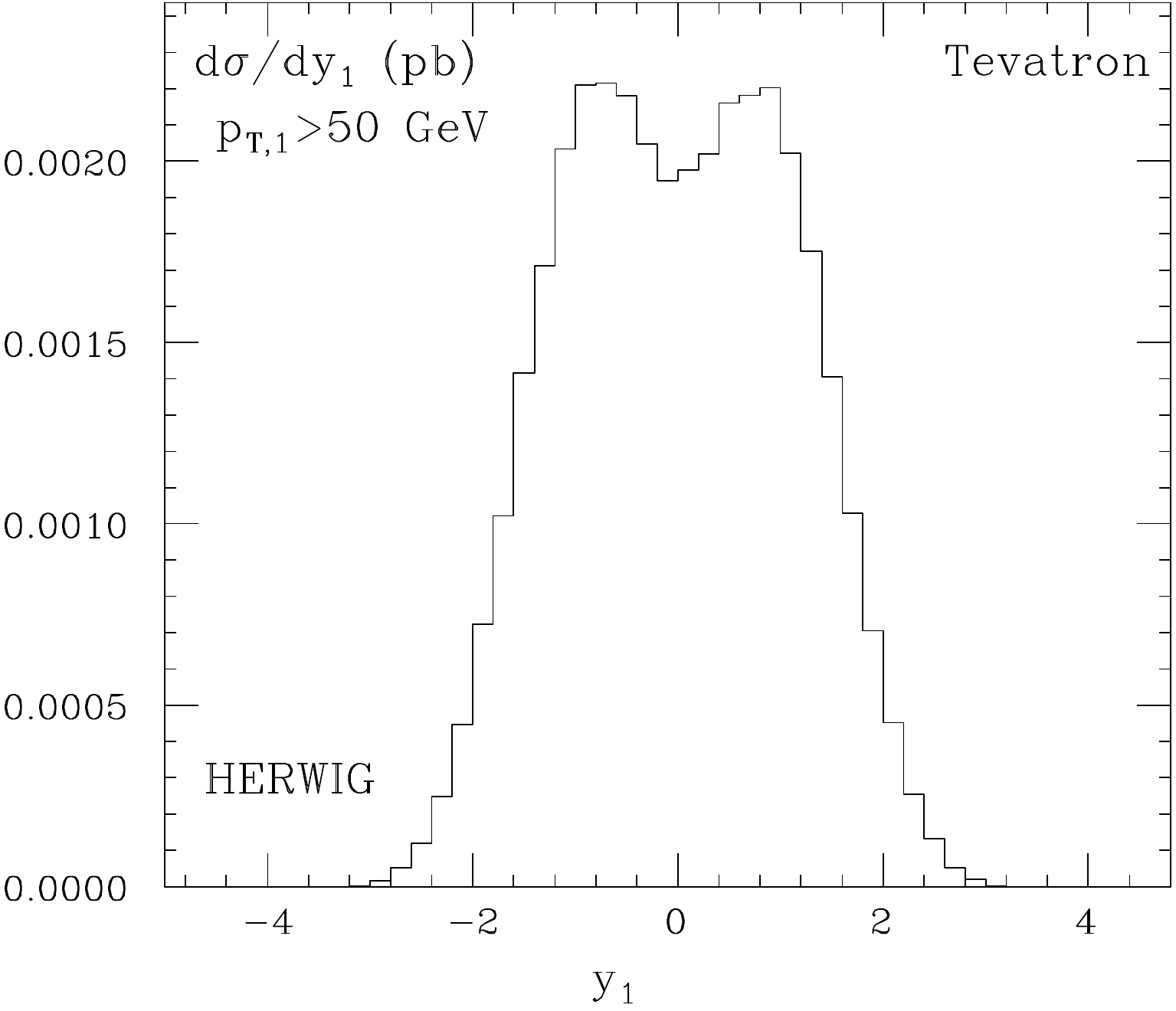}
\hspace*{0.5cm}
\includegraphics[width=0.28\textwidth,clip,]{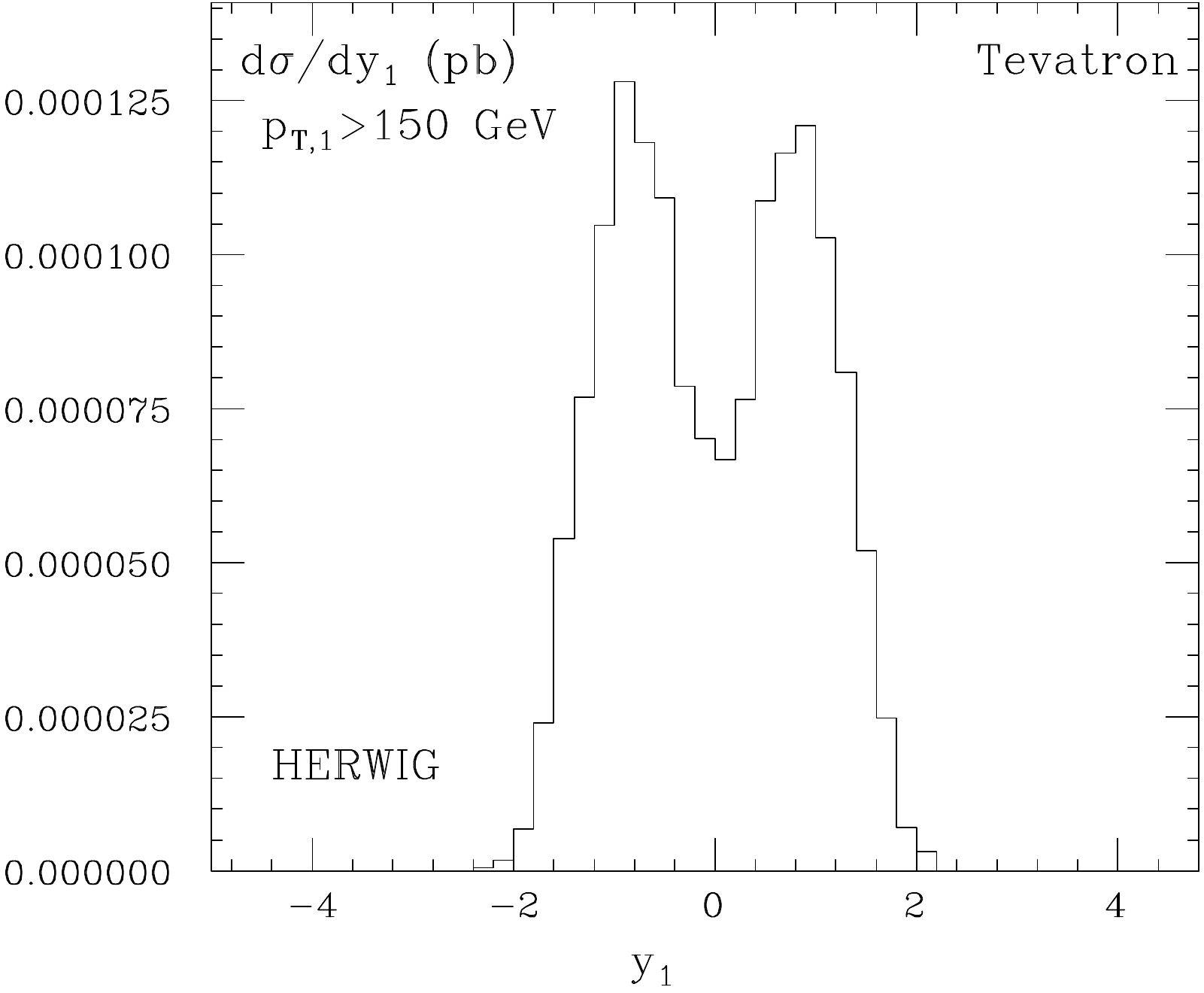}
\\
\includegraphics[width=0.28\textwidth,clip,]{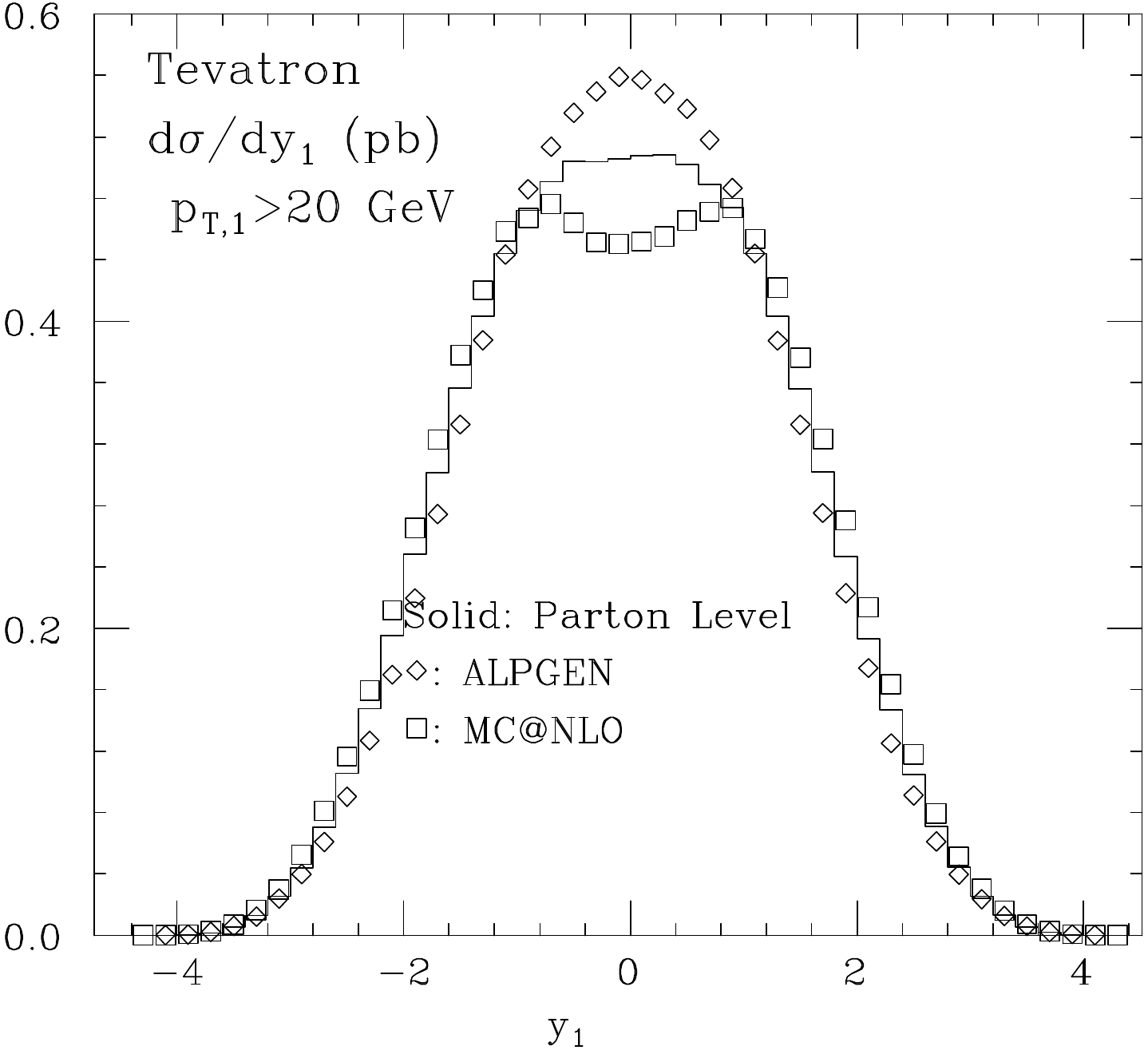}
\hspace*{0.5cm}
\includegraphics[width=0.28\textwidth,clip,]{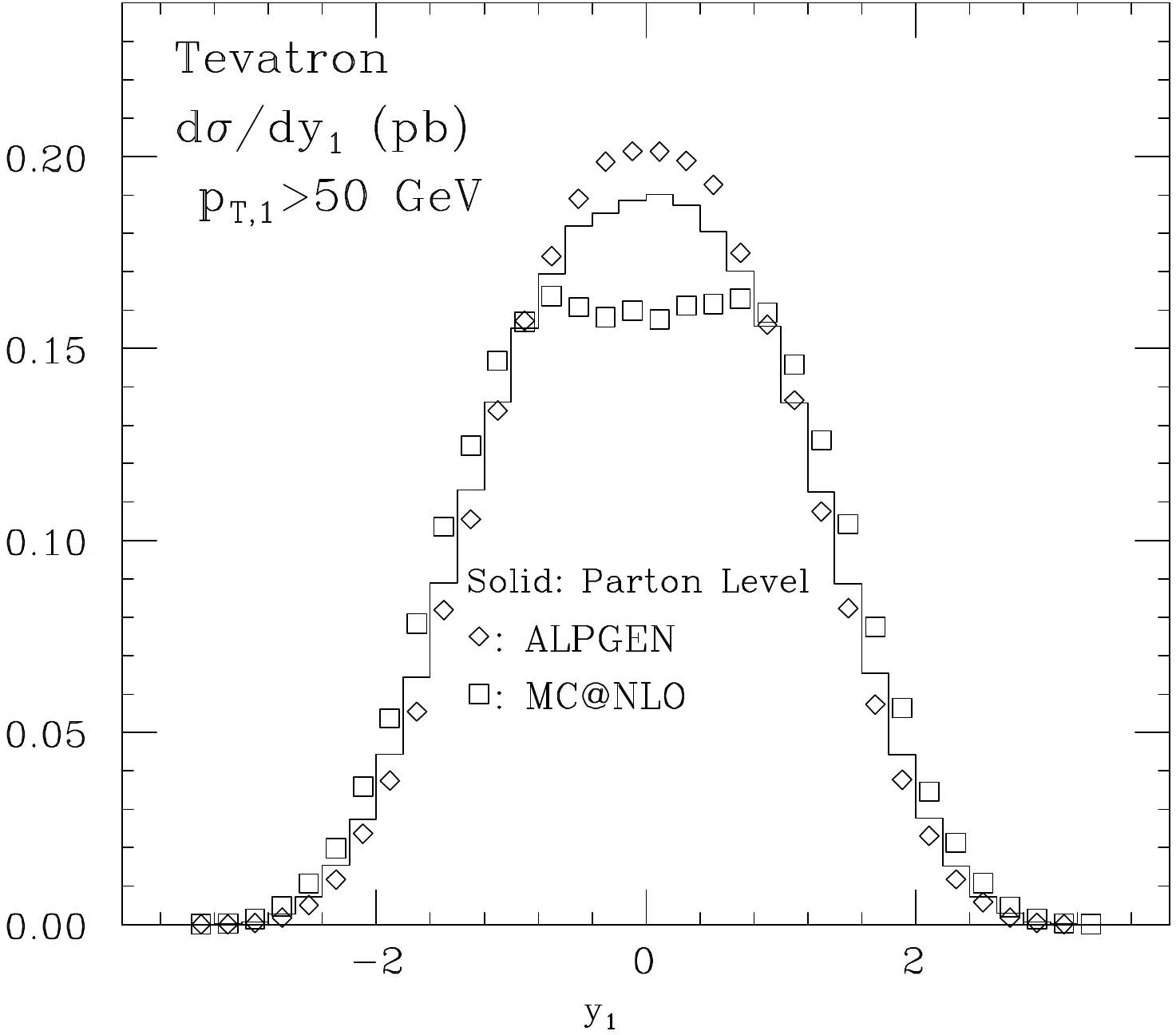}
\hspace*{0.5cm}
\includegraphics[width=0.28\textwidth,clip,]{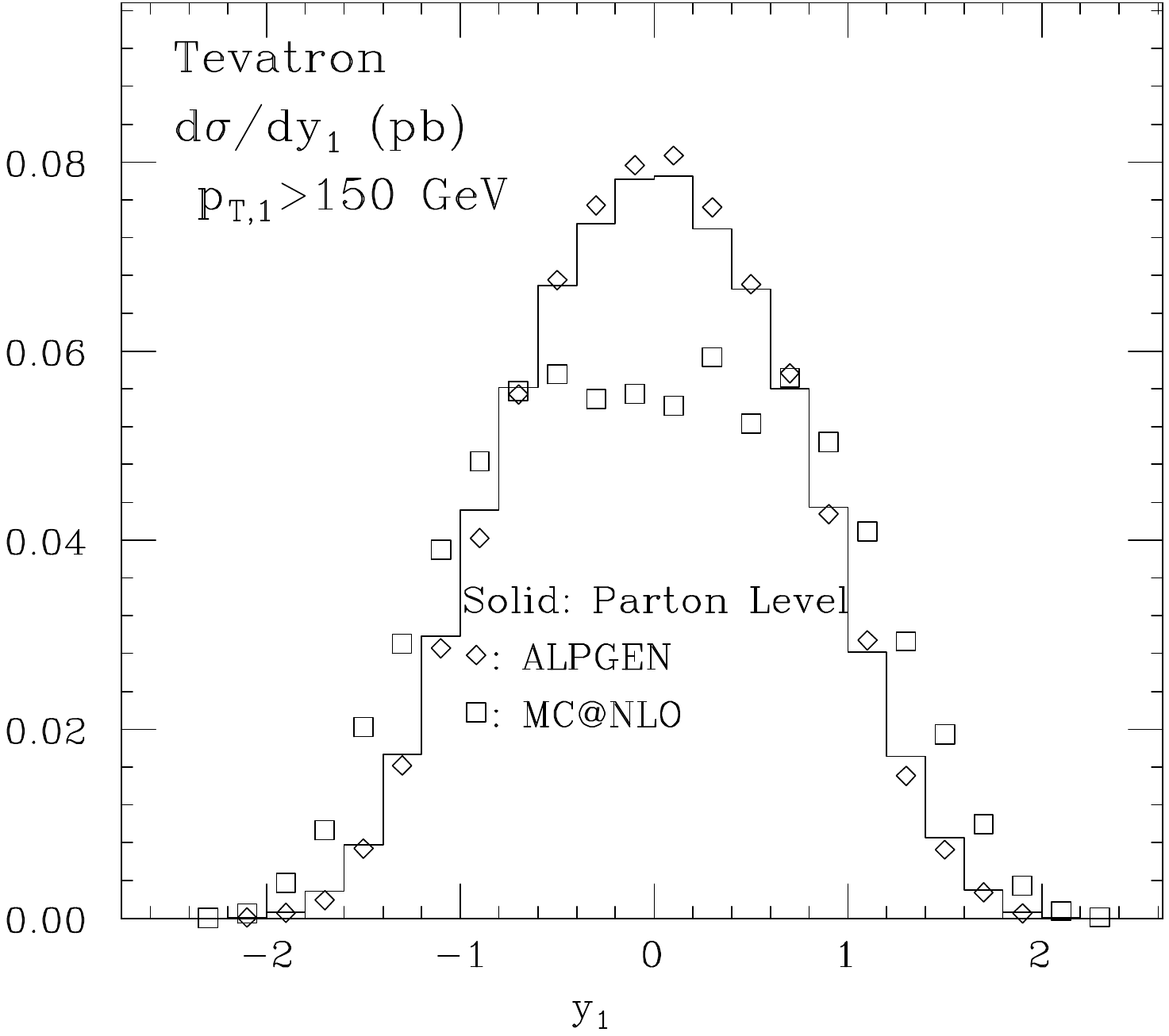}
 \\

%\includegraphics[width=0.25\textwidth,clip,]{y1gt20_MC_partial_LHC}
%\hspace*{0.5cm}
%\includegraphics[width=0.25\textwidth,clip,]{y1gt100_MC_partial_LHC}
%\hspace*{0.5cm}
%\includegraphics[width=0.25\textwidth,clip,]{y1gt300_MC_partial_LHC}
%\\
\ccaption{}{\label{fig:01y} 
Rapidity of the leading jet $\yjet$ at Tevatron for for various jet \pt\ thresholds. 
Upper set: \mcnlo, with partial contributions. Central set: \herwig.
Lower set: comparison between \alpgen, \mcnlo, and the parton
 level predictions }
\end{center}
\end{figure}

% \begin{figure}
% \begin{center}
% \includegraphics[width=0.25\textwidth,clip,]{y1_herwig_TEV}
% \hspace*{0.5cm}
% \includegraphics[width=0.25\textwidth,clip,]{y1_gt50_herwig_TEV}
% \hspace*{0.5cm}
% \includegraphics[width=0.25\textwidth,clip,]{y1_gt150_herwig_TEV}
% %\\
% %\includegraphics[width=0.25\textwidth,clip,]{y1_herwig_LHC}
% %\hspace*{0.5cm}
% %\includegraphics[width=0.25\textwidth,clip,]{y1_gt100_herwig_LHC}
% %\hspace*{0.5cm}
% %\includegraphics[width=0.25\textwidth,clip,]{y1_gt300_herwig_LHC}
% \\
% \ccaption{}{\label{fig:01y-her} 
% Rapidity of the leading jet $\yjet$
% as described by \herwigs  at TeVatron}
% %as described by \herwig. The plots show the results
% %for various jet \pt\ thresholds. Upper set: Tevatron, lower set: LHC} 
% \end{center}
% \end{figure}
% %

To understand the difference in the rapidity distribution,
we look in more detail in
fig.~\ref{fig:01mcnlo} at some
features in the \mcnlo\ description of the leading jet.
For the \pt\ of the leading jet, \ptja,  we plot separately the contribution from the various components
of the \mcnlo\ generation: 
events in which the shower is initiated by the LO \ttbar\ hard process,
and  events in which the shower is initiated by 
a $t \bar t + q (g)$ hard process. In this last case, we separate the
contribution of positive- and negative-weight events, where the
distribution of negative events is shown in absolute value.
The plots show that for \mcnlos
the contribution of the $t \bar t + q (g)$ hard process is 
almost negligible over most of the relevant range and becomes appreciable
only for very large values of $\ptja$.
This hierarchy is stronger at the LHC than at the Tevatron. 

Upper set of fig.~\ref{fig:01y} shows the various contributions to the rapidity
distribution \yjet\ for different
jet \pt\ thresholds. It
appears that the \yjet\ distribution resulting from the shower evolution
of the \ttbar\ events in \mcnlo\ has a strong dip at \yjet=0, a dip that
cannot be compensated by the more central distributions of the jet
from the  $t \bar t + q (g)$ hard process, given its marginal role in
the overall jet rate. 

That the dip at \yjet=0 is a feature typical of jet emission from the \ttbar\
state in \herwig\ is shown in central set of fig.~\ref{fig:01y}, obtained from
the standard \herwig\ code rather than from \mcnlo. 
We speculate that this feature is a consequence of the dead-cone
description of hard emission from heavy quarks implemented in the
\herwig\ shower algorithm.
To complete our analysis, we show in lower set of fig.~\ref{fig:01y} the
comparison between the \alpgen, \mcnlo\ and the parton-level \yjet\
spectra, for different jet \pt\ thresholds. We notice that at large
\pt, where the Sudakov effects that induce potential differences
between the shower and the PL results have vanished, the \alpgen\
result reproduces well the PL result, while still differing
significantly from the \mcnlo\ distributions.
\section {Conclusions}
The study presented in this paper examines the predictions of \alpgen\ and
its matching algorithm for the description of \ttbar+jets
events. Several checks of the algorithm have
shown its internal consistency, and indicate a rather mild dependence
of the results on the parameters that define it. 
The consistency of the approach is confirmed by the comparison with
\mcnlo. In particular, inclusive variables show excellent
agreement, once the NLO/LO $K$ factor is included. \\
Instead we found a rather surprising difference between
the predictions of two codes for the rapidity distribution of the
leading jet accompanying the \ttbar\ pair. In view of the relevance of
this variable for the study at the LHC of new physics signals, it is important to
further pursue the origin of this discrepancy, with independent
calculations, and with a direct comparison with data. Preliminary
results obtained with the new positive-weight NLO shower MC introduced
in~\cite{Nason:2004rx,Nason:2006hf} and presented in this Meeting \cite{Nason:2006XX}, appear to support the
distributions predicted by \alpgen.
%\section*{Acknowledgements}
%We wish to thank the organizers for 
%their effort in this very active and fruitful Workshop 

%\end{document}
----------------------------------------------------------------------

\addtocounter{chapter}{1}
%\documentclass[a4paper,12pt,twoside]{report}
%\usepackage{epsfig}
%\usepackage{amssymb}
%\usepackage{lineno}
%\usepackage{setspace}
%%%%%%%%%%%%%%%%%%%%%%%%%%%%%%%%%%%%%%%%%%%%%%%%%%%%%%%%%%
%\begin{document}
%%%%%%%%%%%%%%%%%%%%%%%%%%%%%%%%%%%%%%%%%%%%%%%
% Toggle line numbering
% Won't work with the PRD revtex4 !
%\pagewiselinenumbers
% uncomment if you want doublespace
%\doublespace
%%%%%%%%%%%%%%%%%%%%%%%%%%%%%%%%%%%%%%%%%%%%%%%
\mchapter{Phenomenology of the Standard Model Higgs boson
at the LHC}{G. Corcella and D. Rebuzzi}

%%%%%%%%%%%%%%%%%%%%%%%%%%%%%%%%%%%%%%%%%%%%%%%%%%%%%%%%%
The Higgs boson plays a crucial role in the Standard Model (SM)
of electroweak
interactions, as it is responsible of the mechanism of mass generation.
However, this particle has not yet been experimentally discovered.
Searches for the Higgs boson will be one of the main goals of the Large
Hadron Collider, which will be capable of 
exploring the Higgs mass spectrum from 100 GeV to about 1 TeV.
In order to accurately perform such searches, the use of precise
QCD calculations and reliable Monte Carlo event generators will be 
mandatory.

We study several observables related to the phenomenology of the
SM Higgs boson and compare the predictions yielded by
HERWIG
\cite{nherwig} and PYTHIA \cite{npythia}, the two most popular event generators,
as well as QCD computations resumming the large logarithms 
appearing in the
Higgs transverse momentum spectrum \cite{hqt}.
In fact, some differences between HERWIG and PYTHIA are to be expected,
since they implement parton showers \cite{marweb,beng}, 
matrix-element matching \cite{mike,miu}
and hadronization \cite{cluster,string}
in a different fashion (see \cite{cor} for some discussions
and comparison between the two Monte Carlo codes).
Hereafter we shall use the HERWIG 6.510 and PYTHIA 6.403 versions.
As parton distribution functions (PDFs), 
the leading-order (LO) CTEQ6L1 set \cite{cteq} will be employed.

We shall consider Higgs production via gluon-gluon fusion (GGF), which
is the dominant channel at the LHC, and through vector-boson fusion
(VBF).
In HERWIG, the same user-defined process simulates both $gg\to H$
and $q\bar q\to H$, 
while in PYTHIA the two subprocesses can be run separately.
In the following, for the sake of comparison,
we shall use a modified version of HERWIG, with the 
$q\bar q\to H$ subprocess turned off.
We checked that, with the PDF set \cite{cteq}, HERWIG simulates 
about 6--7\% of events according to $q\bar q\to H$.

Indeed, if we compare the HERWIG and PYTHIA total cross sections for $gg\to H$,
using the default parametrizations,
we find meaningful discrepancies, about 15--20\% for a Higgs mass 
$110~\mathrm{GeV}<m_H<190~\mathrm{GeV}$.
We investigated the possible causes determining such differences and 
understood that they are mostly due to the value of 
the strong coupling constant $\alpha_S(m_H)$
implemented in the two programs.
Both codes tuned the QCD parameter $\Lambda$ to LEP data, along with
other quantities, such as the shower cutoff, quark and gluon effective masses
and the hadronization non-perturbative parameters. However, 
such fits led to pretty different results for the
strong coupling constant at the $Z$ mass: $\alpha_S(m_Z)\simeq
0.116$ in HERWIG and $\alpha_S(m_Z)\simeq 0.127$ in PYTHIA. 
While the value of HERWIG is consistent with the world average,
i.e. $\alpha_S(m_Z)=0.118\pm 0.002$ \cite{pdg}, PYTHIA uses a somewhat 
higher value. 
Different values of $\alpha_S(m_Z)$ clearly do not affect the total LO 
$e^+e^-\to q\bar q$ cross section, but do have an impact on the LO
$gg\to H$ one, which is ${\cal O}(\alpha_S^2(m_H))$. 
Another difference between the two codes is the implementation of
the QCD beta function, which is at two loops in HERWIG and at one loop in 
PYTHIA.
At hadron colliders, while HERWIG still uses its own value
of $\alpha_S(m_Z)$, PYTHIA employs, for the hard process and the initial-state
parton cascade, the same value of $\alpha_S(m_Z)$ as the one
in the chosen PDF.
The LO CTEQ6L1 set uses $\alpha_S(m_Z)=0.130$, even larger than the
PYTHIA default, and much above the world average.
In any case, we point out that, since
HERWIG and PYTHIA yield only 
LO rates, they should not be used to calculate the 
Higgs production total cross section, which is currently available up to 
next-to-next-to-leading order (NNLO) \cite{nnlo}.
Parton showers predict instead more reliably differential 
distributions, which exhibit milder dependence on $\alpha_S(m_Z)$, since
they are equivalent to resummations (see 
\cite{cmw} for some comparison between Monte Carlo algorithms and
resummed computations). Throughout our analysis we shall nonetheless 
use the same value of $\alpha_S(m_H)$ in both HERWIG and PYTHIA.
For example, we can employ, as a reference point, the value
$\alpha_S(m_Z)=0.130$, as in the parton density \cite{cteq},
leading to $\alpha_S(m_H)\simeq 0.123$ using one-loop evolution
from $m_Z$ to $m_H$.
In HERWIG, in order to increase $\alpha_S(m_Z)$, we shall have to increase
the parameter QCDLAM (default value 0.18 GeV), which roughly corresponds to
QCD quantity $\Lambda$ in the $\overline{\mathrm{MS}}$ renormalization scheme
at high momentum fractions \cite{cmw}.
However, QCDLAM cannot be made arbitrarily 
large without modifying other
HERWIG non-perturbative parameters, such as the shower cutoffs  
for gluons and quarks, namely VGCUT (default 0.10 GeV) and VQCUT
(default 0.48 GeV). We found that a possible combination yielding
$\alpha_S(m_H)=0.123$ in HERWIG is the following:
\begin{equation}
\rm{QCDLAM}=0.378\ \rm{GeV}\  ;\  VGCUT=1.00\  \rm{GeV}\  ;\  
VGCUT=1.50\  \rm{GeV}.\label{para}
\end{equation}
It can be seen that using the parametrization (\ref{para}) in HERWIG the
two programs give approximately the same rates, as shown in
Table~\ref{tabgg}, where we quote the total cross section
given by default PYTHIA (PY) and HERWIG (HW), and by HERWIG tuned according
to Eq.~(\ref{para}) (HW$^*$).  
Of course, one can also modify the PYTHIA parameters in order to
have the same $\alpha_S(m_H)$ as in HERWIG.
Before moving on, we point out that we are aware that 
changing the default parameters of a
Monte Carlo program can be dangerous, as this may likely spoil the
agreement with the data taken into account in the fits.
In fact, we are not recommending that one should
use a different parametrization, such as Eq.~(\ref{para}), but 
just trying to understand the reason of the discrepancy and 
whether it is possible to improve the agreement
between HERWIG and PYTHIA. 
\begin{table}[ht!]
\caption{\label{tabgg} Cross sections for Higgs production in the
gluon-fusion channel, according to PYTHIA (PY) and HERWIG (HW), using their
default parametrization, and according to HERWIG, tuned as in Eq.~(\ref{para})
(HW$^*$).}\begin{center}
\begin{tabular}{|c|c|c|c|}\hline
$m_H$ & $\sigma$ (PY)  & $\sigma$ (HW) & $\sigma$ (HW$^*$) \\
\hline\hline
110 GeV & 20.7 pb & 16.6 pb & 20.2 pb\\
 \hline
130 GeV  & 15.5 pb & 13.2 pb & 15.5 pb\\
\hline
150 GeV  & 12.2 pb & 10.2 pb & 12.2 pb\\
\hline
170 GeV  & 10.3 pb & 7.9 pb & 10.7 pb\\
\hline
190 GeV & 7.9 pb & 6.6  pb & 8.1 pb\\
\hline
\hline
\end{tabular}
\end{center}
\end{table}
\begin{table}[ht!]
\caption{\label{totwid} 
Total Higgs decay width according to PYTHIA and HERWIG, using its default
parameters and the tuning (\ref{para}).}
\begin{center}
\begin{tabular}{|c|c|c|c|}\hline
$m_H$ & $\Gamma$ (PY)  & $\Gamma$ (HW) & $\Gamma$ (HW$^*$) \\
\hline\hline
110 GeV & 2.5 MeV & 3.5 MeV &  3.0 MeV\\
 \hline
130 GeV  & 4.4 MeV & 5.4 MeV & 4.8 MeV\\
\hline
150 GeV  &  15.8 MeV & 16.4 GeV & 15.6 GeV\\
\hline
170 GeV  & 355.2 MeV & 328.6 MeV & 337.3 MeV\\
\hline
190 GeV  &  981.6 MeV & 919.3 MeV & 919.6 MeV\\
\hline
\hline
\end{tabular}
\end{center}
\end{table}
\par
As far as the decays of the Higgs 
boson are concerned, we investigated both the total width as well
as the partial rates into a few given channels..
The total rates yielded by 
HERWIG and PYTHIA, even after tuning HERWIG as in Eq.~(\ref{para}),
are a bit different, as quoted in Table~\ref{totwid}.
Several issues contribute to the discrepancy exhibited by Table~\ref{totwid}. 
%First, unlike HERWIG, PYTHIA implements also the decays $H\to gg$ and 
%$H\to \gamma Z^*$. 
For example, the treatment of the decay $H\to ZZ$, on which our study 
will be later on mostly concentrated, is quite different in the two codes.
While PYTHIA implements the general 
case, where both $Z$ bosons are allowed to be off-shell, in HERWIG at least one
$Z$ is forced to be on-shell. 
We checked that if we selected events where both $Z$'s are
on the mass shell, e.g. within five widths, 
HERWIG and PYTHIA rates would agree up to a good accuracy level.

We present in Table~\ref{hzz} the $H\to ZZ$ rates according to PYTHIA, HERWIG
and HDECAY, a code, based on Ref.~\cite{hdec}, computing the
total and partial Higgs decay rates, possibly including 
higher-order radiative corrections.
We note reasonable agreement between PYTHIA and HDECAY, which also
permits that both $Z$'s are off-shell, while HERWIG yields
slightly lower widths.

Among the other Higgs decay modes, major differences between HERWIG and PYTHIA 
are present especially in the channels into heavy 
quarks, such as $H\to c\bar c$ or 
$H\to b\bar b$, where the discrepancies between the two 
default codes can
be up to $\sim 50\%$.
Considering, e.g., the decay $H\to b\bar b$, both HERWIG and PYTHIA 
implement the $b$-quark $\overline{\mathrm{MS}}$ mass $\bar m_b(m_H)$, 
which is an appropriate mass definition
for $b\bar b$ production at the Higgs mass scale.
However HERWIG, unlike PYTHIA, also
includes the leading-logarithmic (LL) resummation of the large
contributions $\sim\alpha_S^n(m_H)\ln^n(m_H/m_b)$ and NLO corrections to 
$\bar m_b(m_H)$. Removing such higher-order corrections, and still using
the tuning (\ref{para}), we expect that HERWIG and PYTHIA should agree.
The partial rates of the two Monte Carlo generators can again 
be compared with the results of the HDECAY code.
In the $H\to b\bar b$ mode, HDECAY includes the NNLO 
corrections to the $\overline{\mathrm{MS}}$ $b$-quark mass, the NLO ones
to the massive rate $\Gamma(H\to b\bar b)$ near threshold, 
and even the NNNLO ones, in the massless approximation
$m_b\ll m_H$, far above threshold.
We present in Table~\ref{hbb} the widths for
$H\to b\bar b$ according to HERWIG,  
PYTHIA and HDECAY for few values of $m_H$ and using  
$\alpha_S(m_H)=0.123$ everywhere.
HERWIG yields the largest rate, even above the HDECAY result; hence,
we may conclude that some of the higher-order corrections that HDECAY
implements, whereas HERWIG does not, have negative sign. 
We finally remark 
that, unlike HERWIG, PYTHIA includes the modes $H\to gg$ and 
$H\to \gamma Z^*$.
\begin{table}[ht!]
\caption{\label{hzz} Width $\Gamma(H\to Z^{(*)}Z^{(*)})$ for 
different values of
$m_H$ according to PYTHIA, HERWIG with the parametrization (\ref{para}) 
and HDECAY,
for several values of the Higgs boson mass}
\begin{center}
\begin{tabular}{|c|c|c|c|}\hline
$m_H$ & $\Gamma$ (PY)  & $\Gamma$ (HW$^*$) & $\Gamma$ (HDECAY) \\
\hline\hline
110 GeV & 0.012 MeV & 0.011 MeV & 0.012 MeV \\
\hline
120 GeV  & 0.053 MeV & 0.051 MeV & 0.054 MeV\\
\hline
130 GeV  & 0.186 MeV & 0.173 MeV & 0.189 MeV \\
\hline
140 GeV  & 0.530 MeV & 0.503 MeV & 0.541 MeV\\
\hline
150 GeV &  1.357 MeV & 1.311 MeV & 1.374 MeV\\
\hline
\hline
\end{tabular}
\end{center}
\end{table}
\begin{table}[ht!]
\caption{\label{hbb} As in Table~\ref{hzz}, but for the
decay $H\to b\bar b$.}\begin{center}
\begin{tabular}{|c|c|c|c|}\hline
$m_H$ & $\Gamma$ (PY)  & $\Gamma$ (HW$^*$) & $\Gamma$ (HDECAY) \\
\hline\hline
110 GeV & 1.88 MeV & 2.46 MeV & 2.23 MeV \\
 \hline
130 GeV  & 2.17 MeV & 2.82 MeV & 2.55 MeV\\
\hline
150 GeV  & 2.46 MeV & 3.18 MeV & 2.85 MeV \\
\hline
170 GeV  & 2.74 MeV & 3.51 MeV & 3.15 MeV\\
\hline
190 GeV & 3.02 MeV & 3.84 MeV & 3.44 MeV\\
\hline
\hline
\end{tabular}
\end{center}
\end{table}
\par
\begin{figure}[ht!]\begin{center}
\epsfig{file=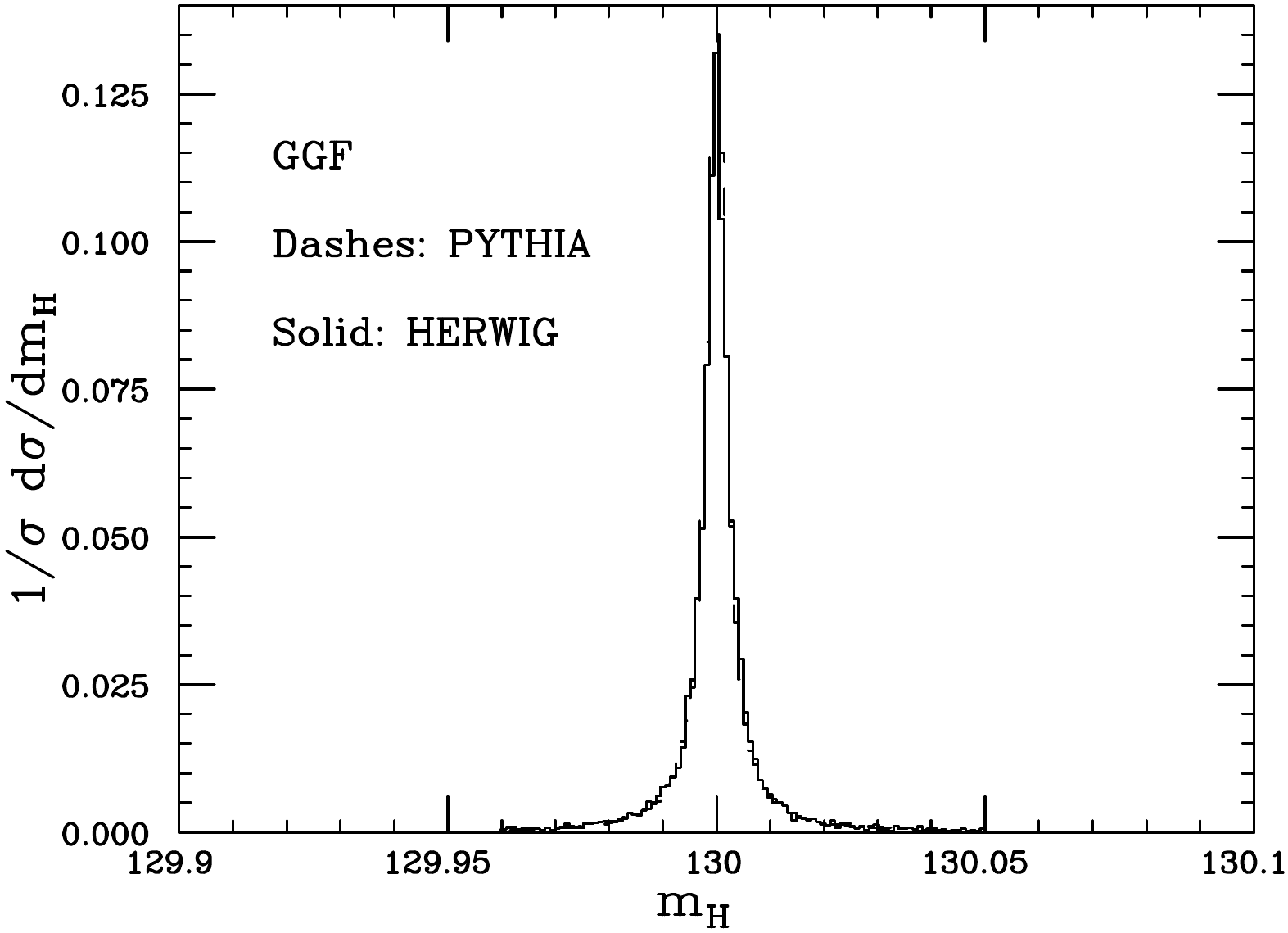,width=0.495\textwidth}
\epsfig{file=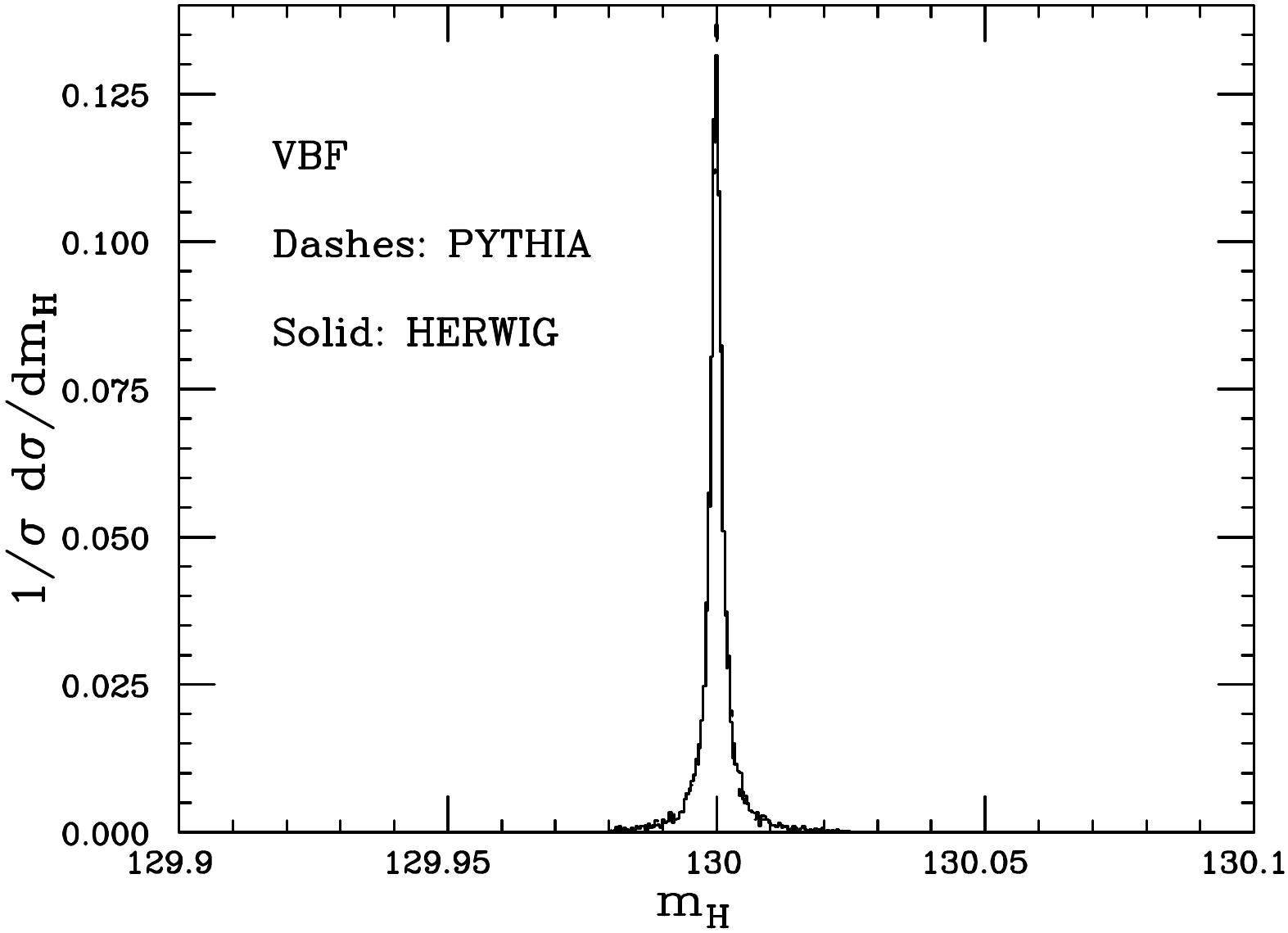,width=0.495\textwidth}
\caption{Higgs invariant-mass distribution for $H\to ZZ$ decays and
Higgs production in gluon-gluon (left) and vector-boson (right) fusion,
according to HERWIG (solid line) and PYTHIA (dashes).}
\label{hm}
\end{center}
\end{figure}
\begin{figure}[ht!]\begin{center}
\epsfig{file=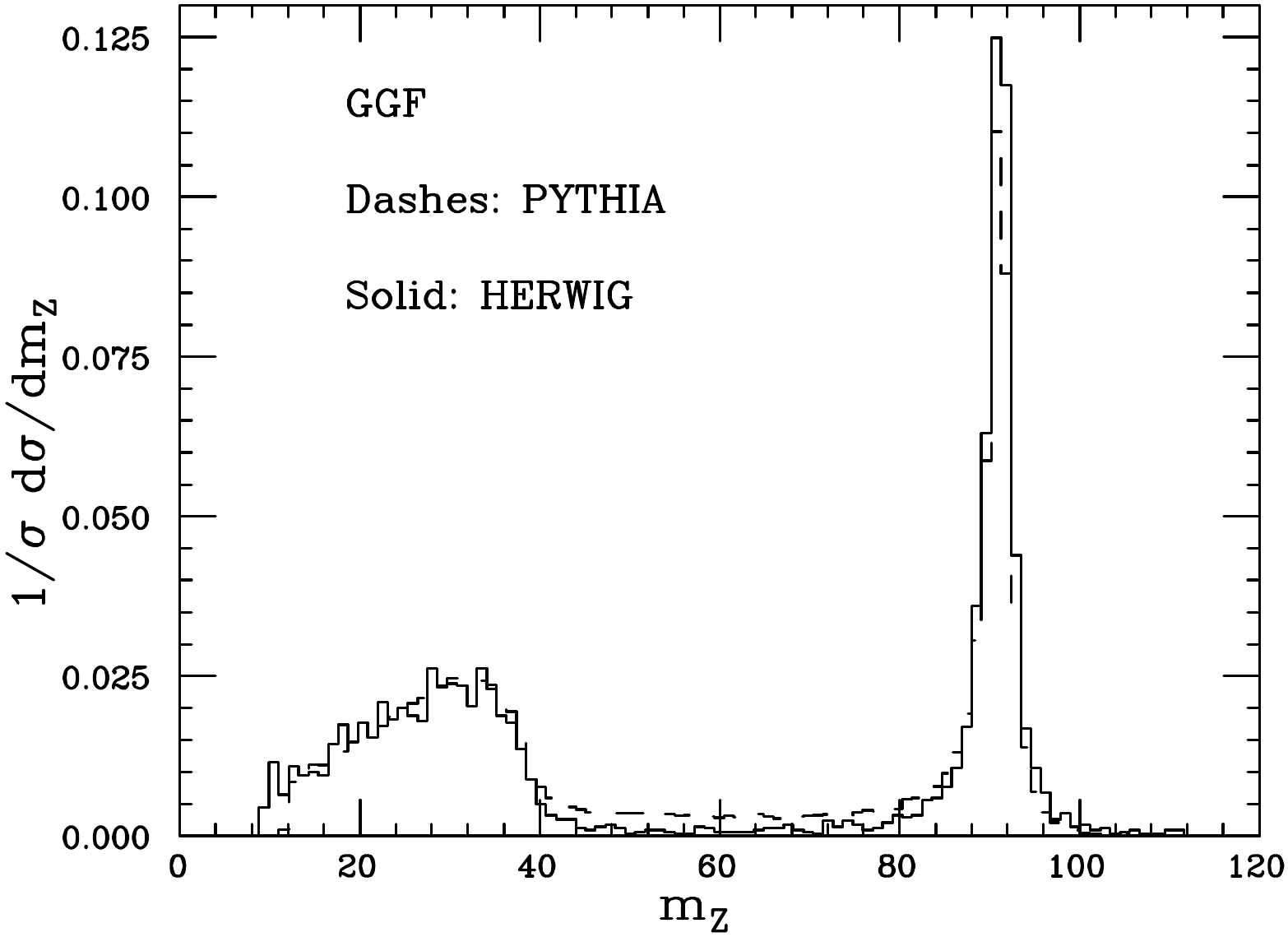,width=0.495\textwidth}
\epsfig{file=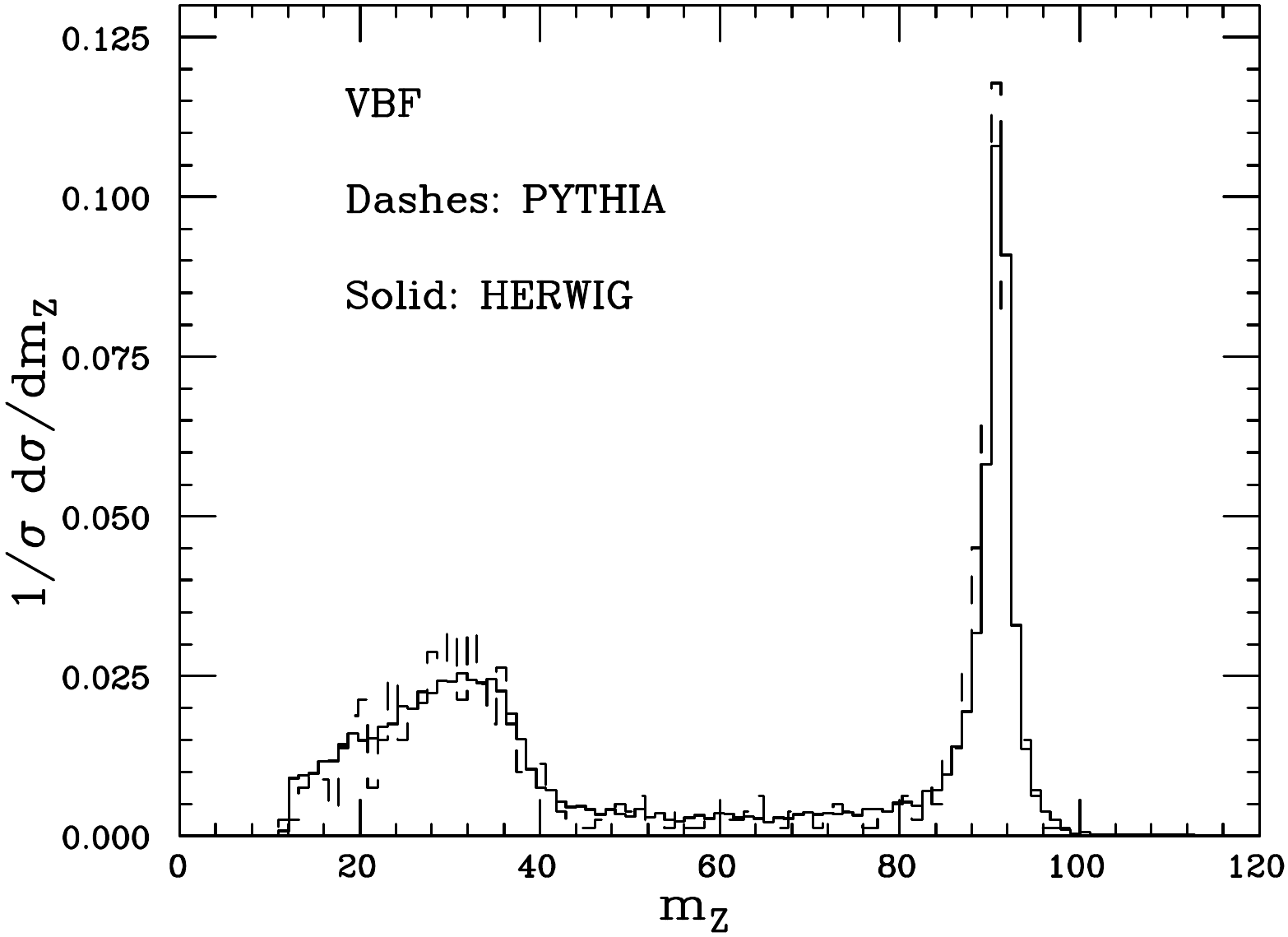,width=0.495\textwidth}
\caption{As in Fig.~\ref{hm}, but showing the $Z$-mass distribution.}
\label{zm}
\end{center}
\end{figure}
\begin{figure}[ht!]\begin{center}
\epsfig{file=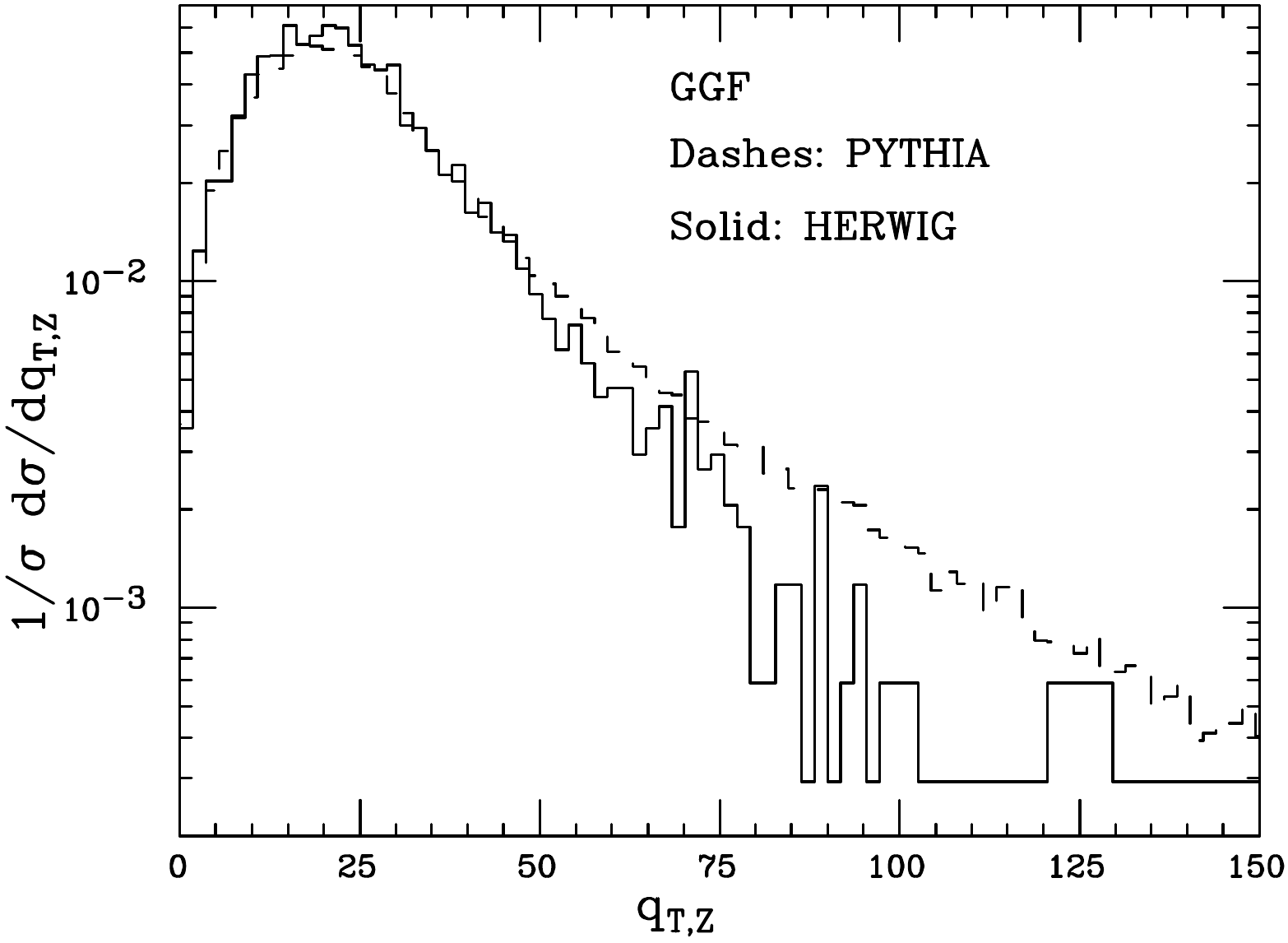,width=0.495\textwidth}
\epsfig{file=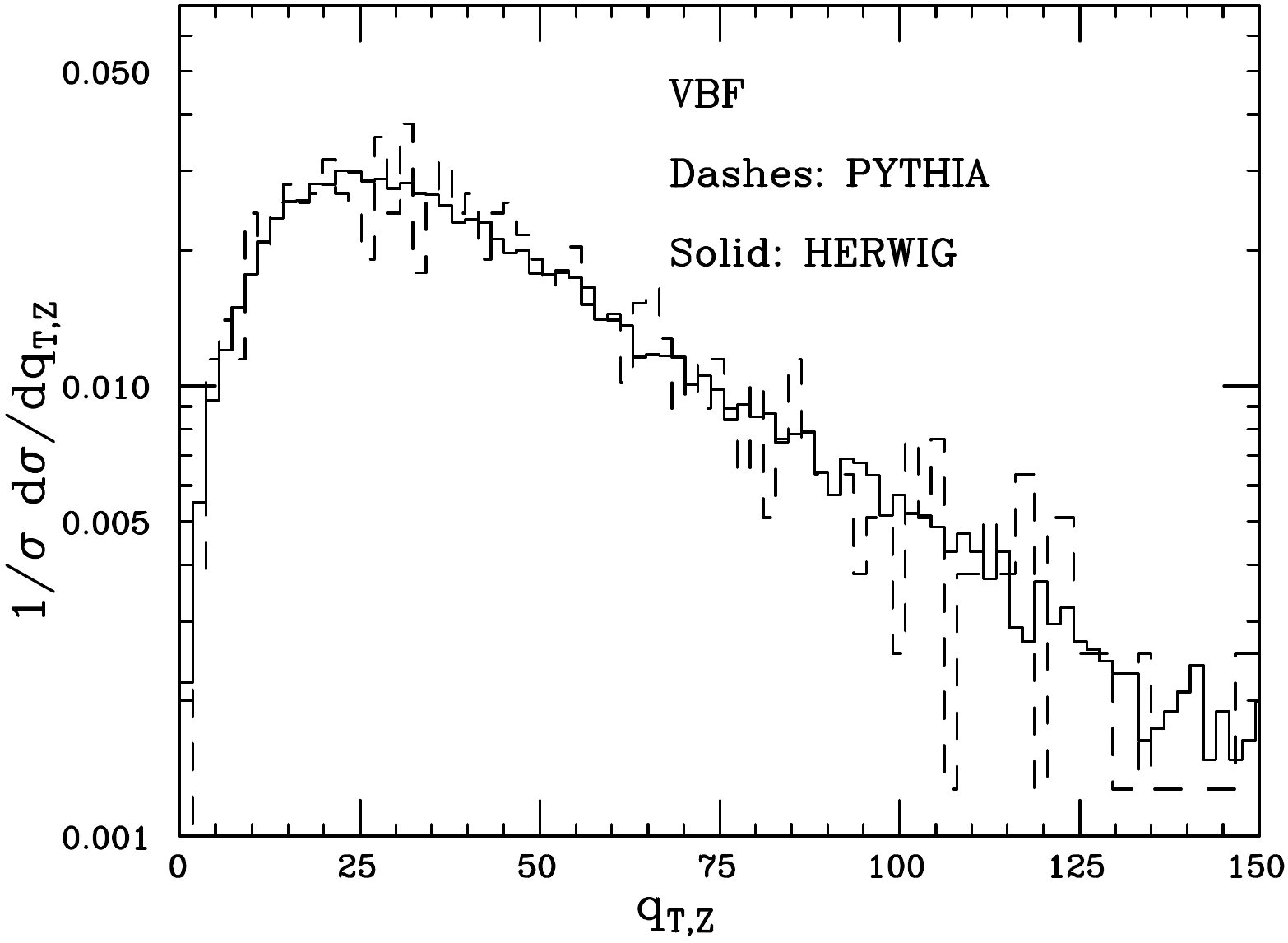,width=0.495\textwidth}
\caption{$Z$ transverse momentum spectrum in GGF (left) and VBF (right).}
\label{zp}
\end{center}
\end{figure}
\begin{figure}[ht!]\begin{center}
\epsfig{file=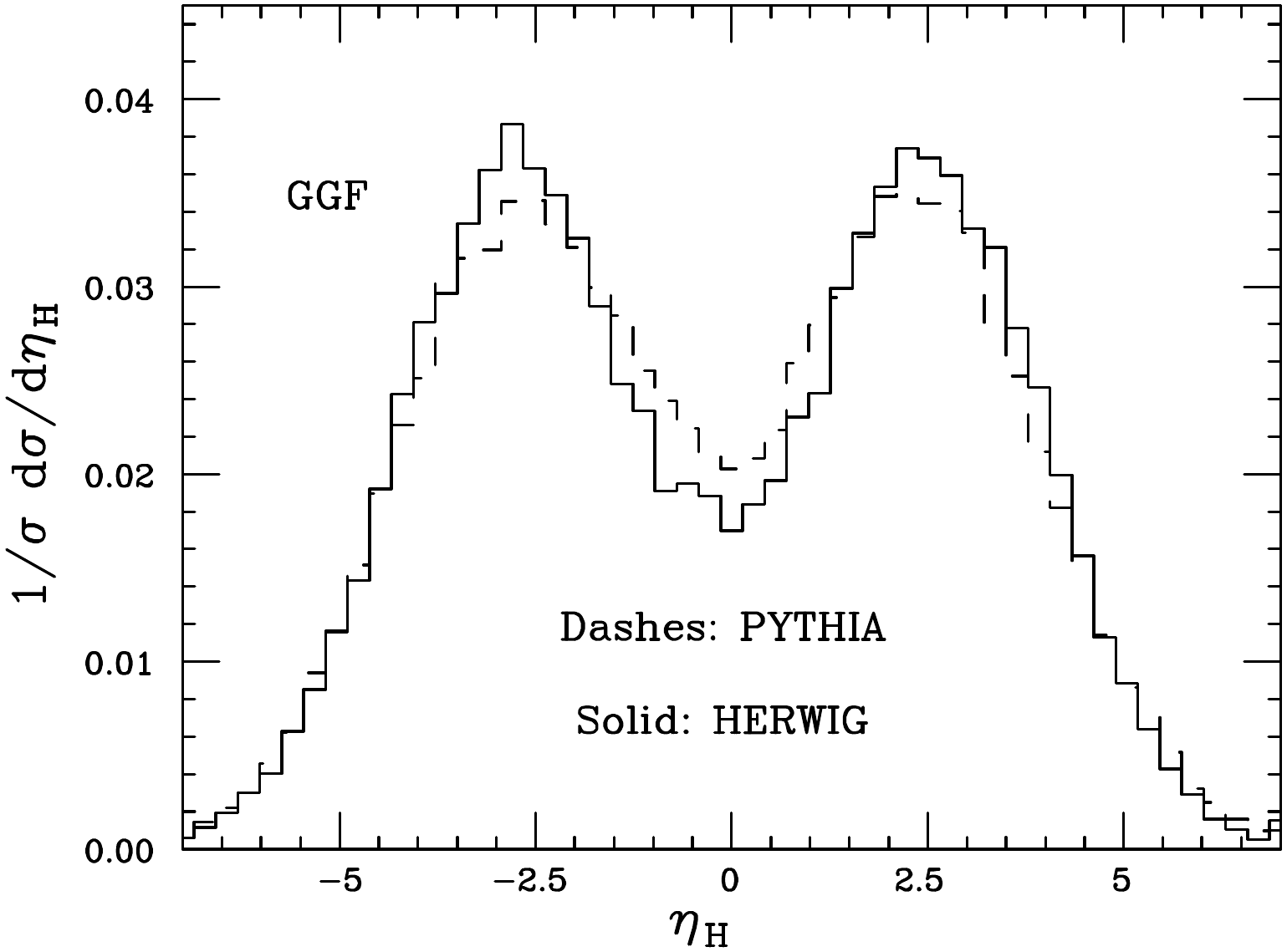,width=0.495\textwidth}
\epsfig{file=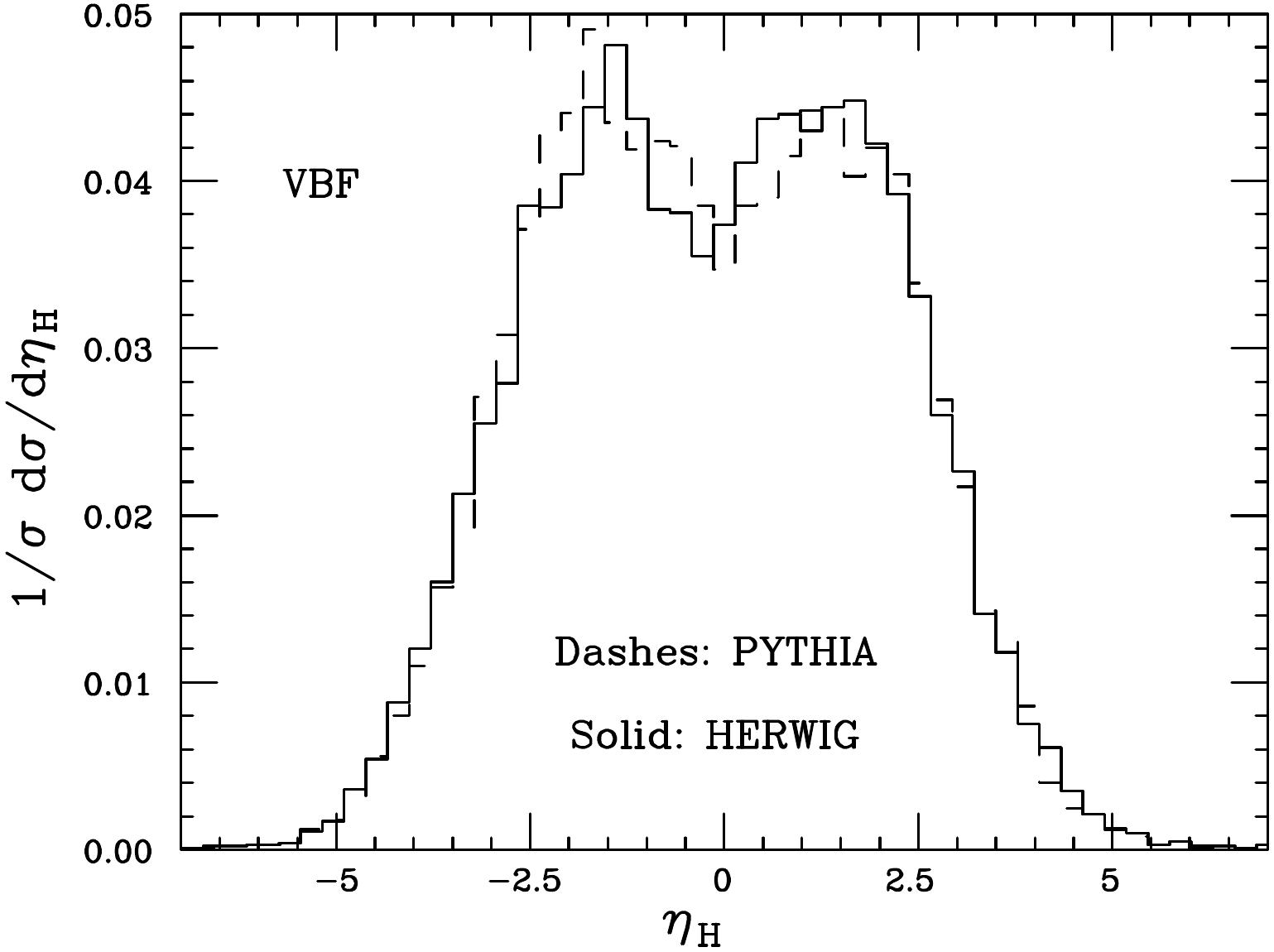,width=0.495\textwidth}
\caption{Higgs pseudorapidity distribution in GGF (left) and
VBF (right).}
\label{het}
\end{center}
\end{figure}
\begin{figure}[ht!]\begin{center}
\epsfig{file=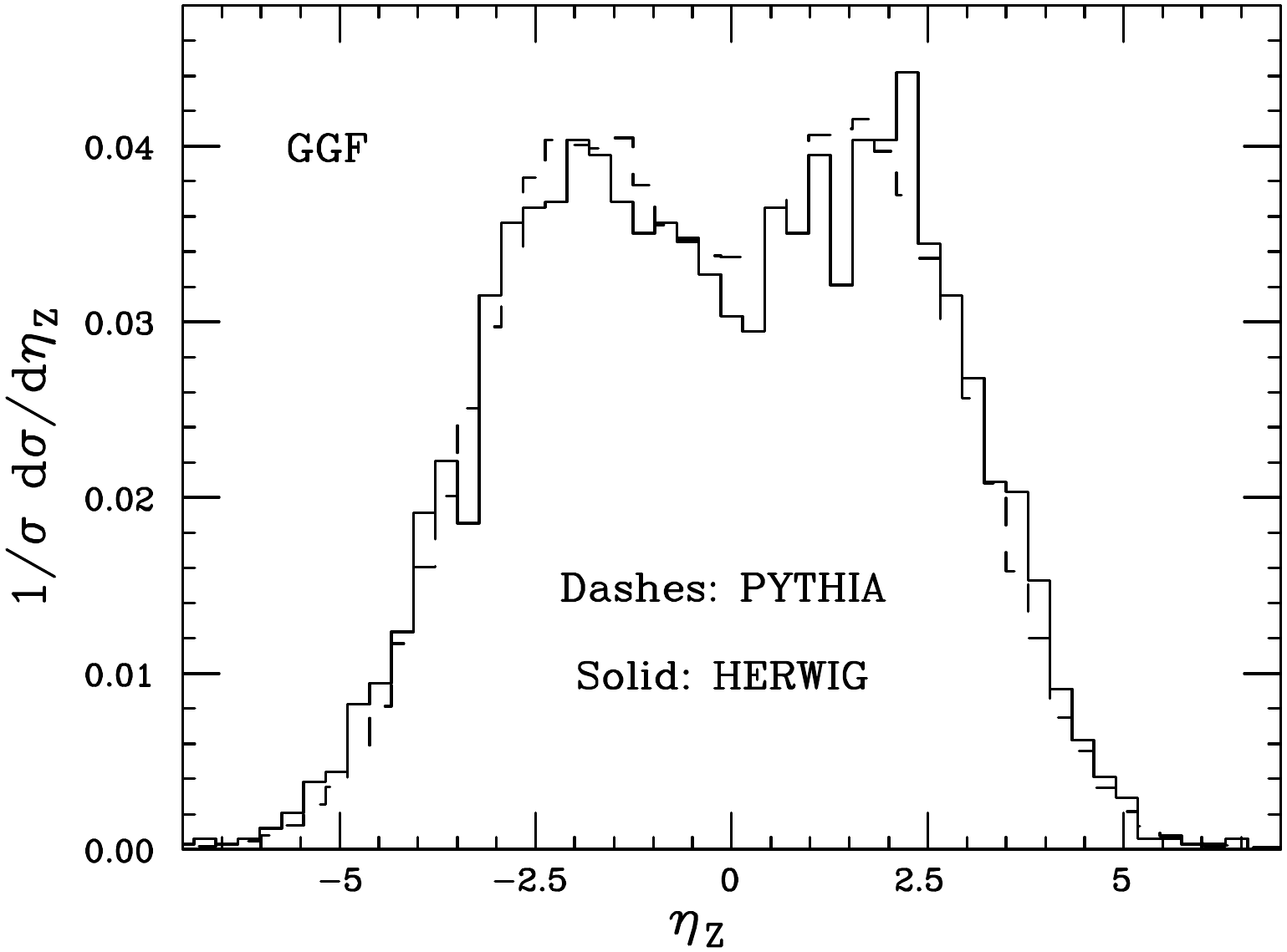,width=0.495\textwidth}
\epsfig{file=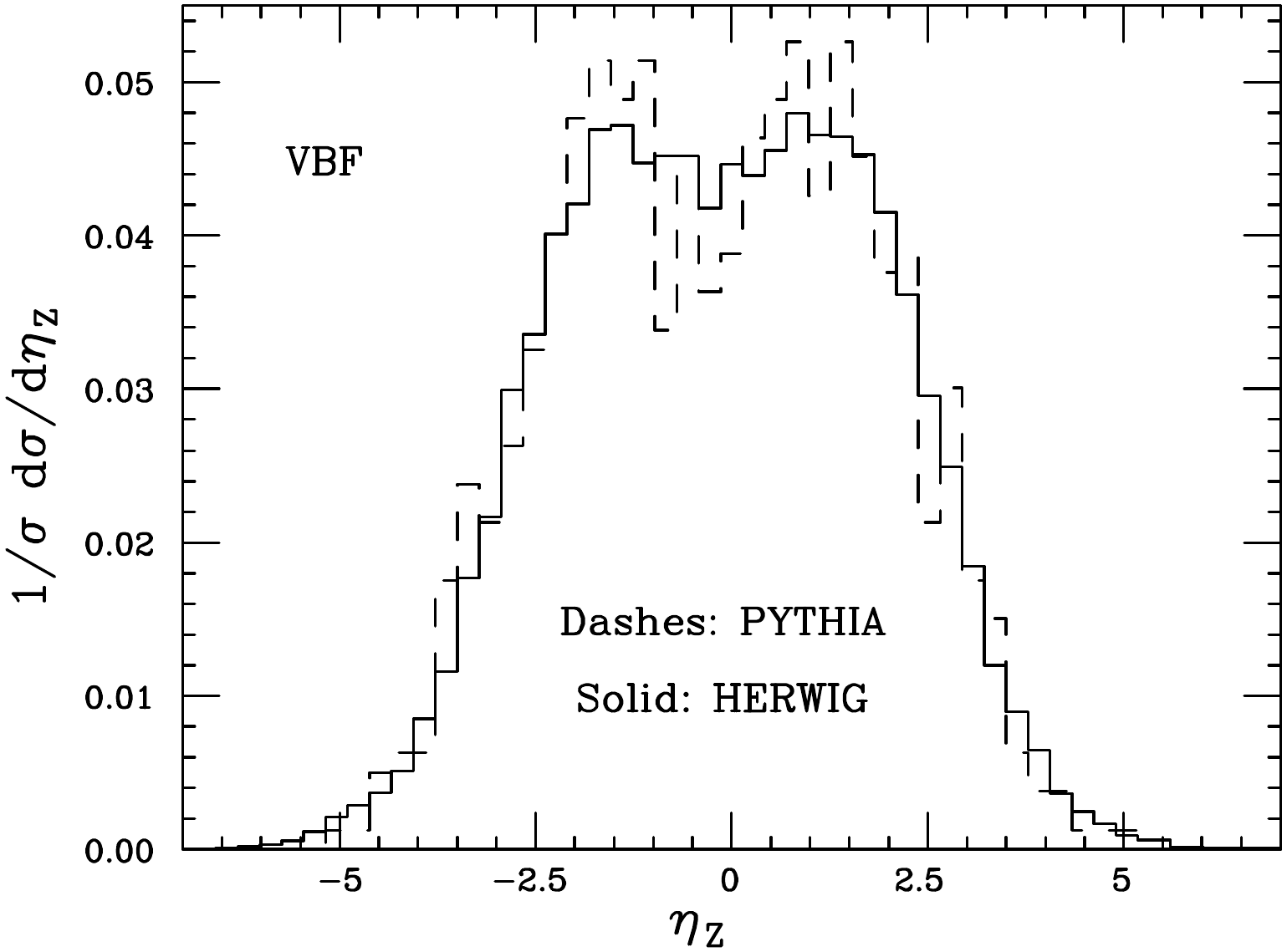,width=0.495\textwidth}
\caption{As in Fig.~\ref{het}, but presenting the $Z$ pseudorapidity.}
\label{zet}
\end{center}
\end{figure}

As anticipated, in our phenomenological analysis 
we shall consider SM Higgs production in GGF and VBF and
concentrate ourselves mostly on the decay channel
$H\to ZZ\to 4\ell$. 
We shall study the following distributions:
the $Z$- and $H$-boson mass spectrum, the Higgs and $Z$ 
transverse momentum ($q_{T,H}$ and $q_{T,Z}$) and 
pseudorapidity ($\eta_H$
and $\eta_Z$).
Such spectra are presented in Figs.~\ref{hm}--\ref{zet}, for a Higgs mass
$m_H=130$~GeV.

The $m_H$ distributions of HERWIG and PYTHIA (Fig.~\ref{hm}) look compatible. 
On the contrary, the fact that PYTHIA allows both $Z$'s to be off-shell has 
an evident impact on Fig.~\ref{zm}: in the
intermediate $Z$-mass range, say 40--80 GeV, where both $Z$'s are
off-shell, and PYTHIA yields more events.
Once again, if we set a filter allowing only $Z$'s near the mass shell, 
the discrepancies in the intermediate $m_Z$ range will disappear.
In Fig.~\ref{zp} we instead compare the $Z$ transverse momentum distributions:
in GGF PYTHIA predicts more events than HERWIG at
large $q_T$, while in VBF the two codes roughly agree.
As we shall discuss later in more detail, for the time being, the
default version of PYTHIA includes matrix-element corrections to Higgs 
production in GGF, 
while HERWIG does not. Such corrections are responsible of the
simulation of a few events with a Higgs of large transverse momentum,
whose decays still yield $Z$'s at large $q_T$.
In Fig.~\ref{het} we instead present the Higgs pseudorapidity distributions: we
clearly note an asymmetry, about 5\%, for VBF in HERWIG,
with more events simulated at positive rather than negative $\eta_H$.
PYTHIA yields instead a symmetric spectrum.
This asymmetry exhibited by HERWIG is currently under investigation
\cite{sey} and should be clarified in a forthcoming publication \cite{correb}.
As for the $\eta_H$ distribution in GGF, both 
HERWIG and PYTHIA spectra are symmetric, although some discrepancy is
still present, with PYTHIA leading to more events around $\eta_H=0$.
The HERWIG and PYTHIA
$Z$-pseudorapidity spectra, presented in Fig.~\ref{zet}, are instead rather
similar.
\par
We finally wish to present results on 
the Higgs transverse momentum distribution, which 
has been thoroughly investigated using Monte Carlo generators as well as
resummed calculations. In fact, the NLO 
Higgs $q_T$ spectrum exhibits 
contributions, $\sim\alpha_S^n \ln^k(m_H^2/q_T^2)$, with $k\leq n+1$,
which are large for small
values of the Higgs transverse momentum.
Small values of $q_T$ correspond to initial-state soft or collinear 
parton radiation.
Such logarithms have been resummed in \cite{hqt} up to 
next-to-next-to-next-to-leading logarithmic
accuracy (NNLL) in a Sudakov-like exponential
factor.
The authors of Ref.~\cite{hqt} also released a computing code, named
HqT, implementing numerically their resummation.

In detail, the LLs are $\sim\alpha_S^n\ln^{n+1}(m_H^2/q_T^2)$, the NLLs
$\sim\alpha_S^n\ln^n(m_H^2/q_T^2)$, the NNLLs 
$\sim\alpha_S^n\ln^{n-1}(m_H^2/q_T^2)$ and so forth. 
While at small $q_T$ the resummation works fine, at large $q_T$ one
should rely on fixed-order calculations. The HqT code allows one to 
match the resummation with the fixed-order spectrum up to NNLO accuracy
\footnote{Throughout this paper, by NLO we always mean corrections  
of ${\cal O} (\alpha\alpha_S^3)$, which are NLO for the total $H$-production
cross section. 
For the $q_T$ spectrum, such contributions are instead LO.}.
As for HERWIG and PYTHIA, their
standard algorithms \cite{marweb,beng} are reliable 
for soft or collinear parton radiation, i.e. for Higgs production at
small $q_T$. In fact, it can be shown \cite{cmw} that parton showers 
include all soft/collinear leading logarithms, plus some NLLs as well.
Higgs production at large $q_T$ corresponds instead to hard or large-angle 
initial-state radiation: therefore, it cannot be simulated by 
standard parton shower algorithms, but must be described by the
use of the exact tree-level NLO matrix-element \cite{baur,spira}.

In order to allow hard and large-angle radiation, 
HERWIG and PYTHIA have been provided with matrix-element
corrections \cite{mike,miu},
although their actual implementation is indeed somewhat
different.
HERWIG splits the phase space into two regions: a region corresponding to
soft/collinear emission, where one uses the parton shower approximation,
and a region associated with hard or wide-angle radiation, the so-called
`dead zone' of the standard algorithm, where the exact 
tree-level amplitude is used. Moreover, the exact matrix element is 
also employed to correct the radiation in the HERWIG parton-shower region
any time an emission is capable of being the hardest so far, i.e. it
has the largest $q_T$ with respect to the radiating parton.
Unlike HERWIG, PYTHIA uses instead the parton-shower approximation in all
physical phase space and corrects with the exact amplitude only the first
branching \cite{miu}. As discussed in \cite{mike}, correcting 
only the first emission will however lead to
an unphysical dependence of the hard-emission
probability on the infrared cutoff, appearing in the Sudakov form factor.

\begin{figure}[ht!]
\centerline{\resizebox{0.7\textwidth}{!}{\includegraphics{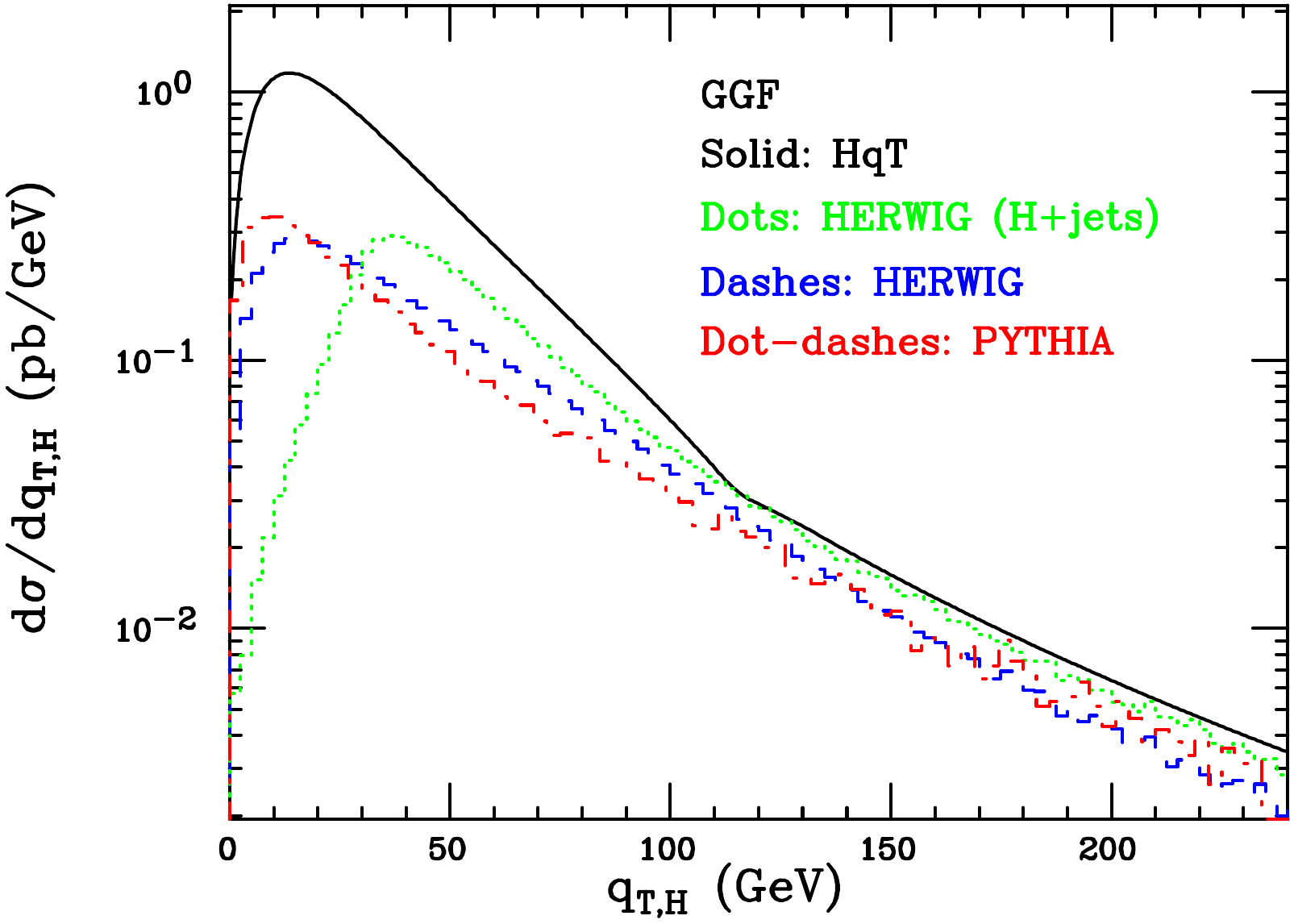}}}
\caption{\small Transverse momentum distribution of the Higgs boson, produced
in gluon fusion, according to HqT (solid line), HERWIG (dashed),
PYTHIA (dot-dashed) and the HERWIG `Higgs+jet' process (dotted).}
\label{qtqt}
\end{figure}

The latest version of PYTHIA \cite{npythia} does include matrix-element 
corrections to Higgs production in gluon-gluon fusion, though treating the
top quark in the loop in the infinite-mass limit. 
The official version of HERWIG \cite{nherwig} does not include yet 
the corrections to Higgs production, although a preliminary code is available,
based on the paper \cite{cormor}, which extends the earlier work 
for Drell--Yan processes \cite{corsey}.
With respect to PYTHIA, Ref.~\cite{cormor} 
fully includes top-quark mass effects and 
possibly corrects even to $q\bar q\to H$, whenever
this subprocess is turned on.
We also compare the Monte Carlo spectra with the HqT code, which we run in the
NLL approximation, matched to the NLO total cross section in the
point $q_T=m_H$.
The results are presented in Fig~\ref{qtqt}, where we also show the 
result yielded by the HERWIG `Higgs+jet' process, where the hard-scattering
process is always generated according to one of the tree-level
corrections to $gg\to H$, i.e. $gg\to Hg$, $qg\to Hq$, etc.
We set in all codes $\alpha_S(m_H)$ to the same value, however, at large
$q_T$ still different scales $\mu$ for the strong coupling are used.
In fact, by default,
HERWIG uses the Higgs transverse mass $\mu=\sqrt{q_T^2+m_H^2}$,
which is a reasonable scale since the dead zone 
corresponds to $q_T\geq m_H$; PYTHIA uses instead $\mu=q_T$;
HqT sets $\mu=m_H$.
For the sake of comparison, we set $\mu=q_T$ in all codes for $q_T\geq m_H$: 
this way, all spectra should roughly agree at large transverse momentum.
From Fig.~\ref{qtqt} we learn that 
the Higgs+jet spectrum roughly agrees with HqT at large $q_T$. The small
discrepancy can be due to the fact that HqT also includes 
bottom quarks in the loop and to different choices of the PDF 
factorization scale, 
which is set to $\mu_F=m_H$ in HqT and $\mu_F=q_T$ in
HERWIG and PYTHIA.
The HERWIG and PYTHIA  $q_T$ spectra, though in agreement with each other, 
are a bit below the other two curves, even though the same 
tree-level NLO matrix element has been used.
Such a discrepancy obviously needs further investigation \cite{correb}.
\par
In summary, we performed a study on SM Higgs boson production and decay at the
LHC, mainly using PYTHIA and HERWIG, the two most popular Monte Carlo
generators. We found several differences between the two codes: we 
understood the causes of most discrepancies, whereas a few are still
under investigation. In a forthcoming work \cite{correb} 
we shall present an even more detailed analysis of Higgs boson 
phenomenology, where we shall also study the impact of the NLO
corrections to Higgs production in GGF, implemented in the
MC@NLO code \cite{mcnlo}.

\addtocounter{chapter}{1}

\newcommand{\GeVcc}{~{\rm GeV/}c^{\rm 2}~}
\newcommand{\TeVcc}{~{\rm TeV/}c^{\rm 2}~}
\newcommand{\fbinv} {~\rm fb^{-1}~}

%%%%%%%%%%%%%%%%%%%%%%%%%%%%%%%%%%%%%%%%%%%%%%%%%%%%%%%%%%
%\begin{document}
%%%%%%%%%%%%%%%%%%%%%%%%%%%%%%%%%%%%%%%%%%%%%%%
% Toggle line numbering
% Won't work with the PRD revtex4 !
%\pagewiselinenumbers
% uncomment if you want doublespace
%\doublespace
%%%%%%%%%%%%%%%%%%%%%%%%%%%%%%%%%%%%%%%%%%%%%%%
\mchapter{MSSM Higgs Searches with CMS}{G. Masetti}
%%%%%%%%%%%%%%%%%%%%%%%%%%%%%%%%%%%%%%%%%%%%%%%%%%%%%%%%%
\section{Introduction}

%%The LHC (\textbf{L}arge \textbf{H}adron \textbf{C}ollider) project 
%The Large Hadron Collider (LHC) project
%is the proton-proton accelerator currently under construction at CERN.
%The start-up of the collider is foreseen for year 2007 and the 
%center-of-mass energy will be 14 TeV.
%The Compact Muon Solenoid (CMS) is one of the two general purpose 
%detector designed to optimize the discovery potential of LHC. 
%In particular, one of the main physics goals is the discovery of the
%Higgs boson.

According to the Minimal Supersymmetric Standard Model (MSSM), two isospin
Higgs doublets have to be introduced.
After electroweak symmetry breaking, five Higgs scalar mass eigenstates remain:
one CP-odd neutral scalar boson A, two charged scalars
$H^{\pm}$, and two CP-even neutral scalars h and H.
At the tree level, the Higgs sector is completely defined by only two
parameters: they are usually chosen as 
the ratio of the vacuum expectation values ($\tan\beta = v_{2}/v_{1}$) 
and the Higgs boson A mass ($M_{A}$).
The tree level hierarchies ($M_{h} < M_{Z}$, $M_{A} < M_{H}$ and $M_{W} < M_{H}$)
are modified by large radiative corrections: the leading one-loop correction
is proportional to $m_{t}^{4}$ and the upper bound of $M_{h}$ is 
shifted to $M_{h} \leq 135 \GeVcc$.

Varying $M_{A}$ in the range $91 \GeVcc < M_{A} < 1 \TeVcc$ we can distinguish
three different regimes:
\begin{itemize}
	\item \textbf{Decoupling regime.} If $M_{A} \gg M^{max}_{h}$ then the Higgs 
	bosons H, A and $H^{\pm}$ are very heavy and almost degenerate in mass, while 
	h has a mass very close to $M^{max}_{h}$ and becomes SM-like. H and A, besides 
	their masses, are degenerate also in width and cross section.
	\item \textbf{Low $M_{A}$ regime.} If $M_{A} < M^{max}_{h}$ the behavior of the 
	two CP-even neutral Higgs bosons h and H is swapped with respect to the decoupling 
	regime: h is almost degenerate in mass, width and cross section with A, and
	H is the SM-like Higgs, with a mass close to $M^{max}_{h}$.
	\item \textbf{Intense coupling regime.} This occurs for $M_{A} \sim M^{max}_{h}$
	and high $\tan\beta$ \cite{Boos:2002,Boos:2004} and it leads to similar, 
	but not degenerate, masses for the 
	three neutral Higgs bosons. This property, in principle, allows to detect the three 
	neutral Higgs separately.
\end{itemize}

\begin{figure}
  \includegraphics[height=.3\textheight,angle=270]{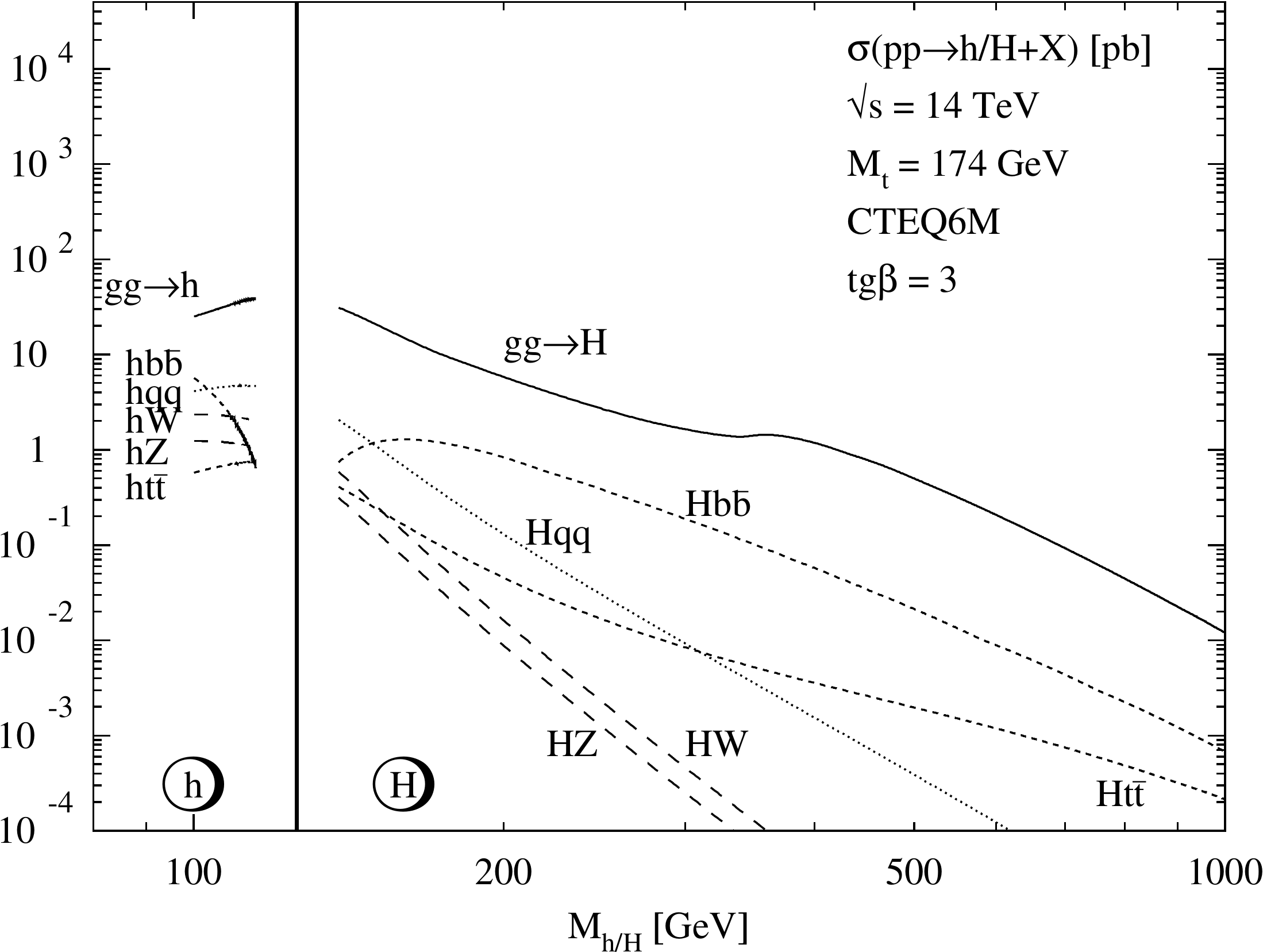}
  \includegraphics[height=.3\textheight,angle=270]{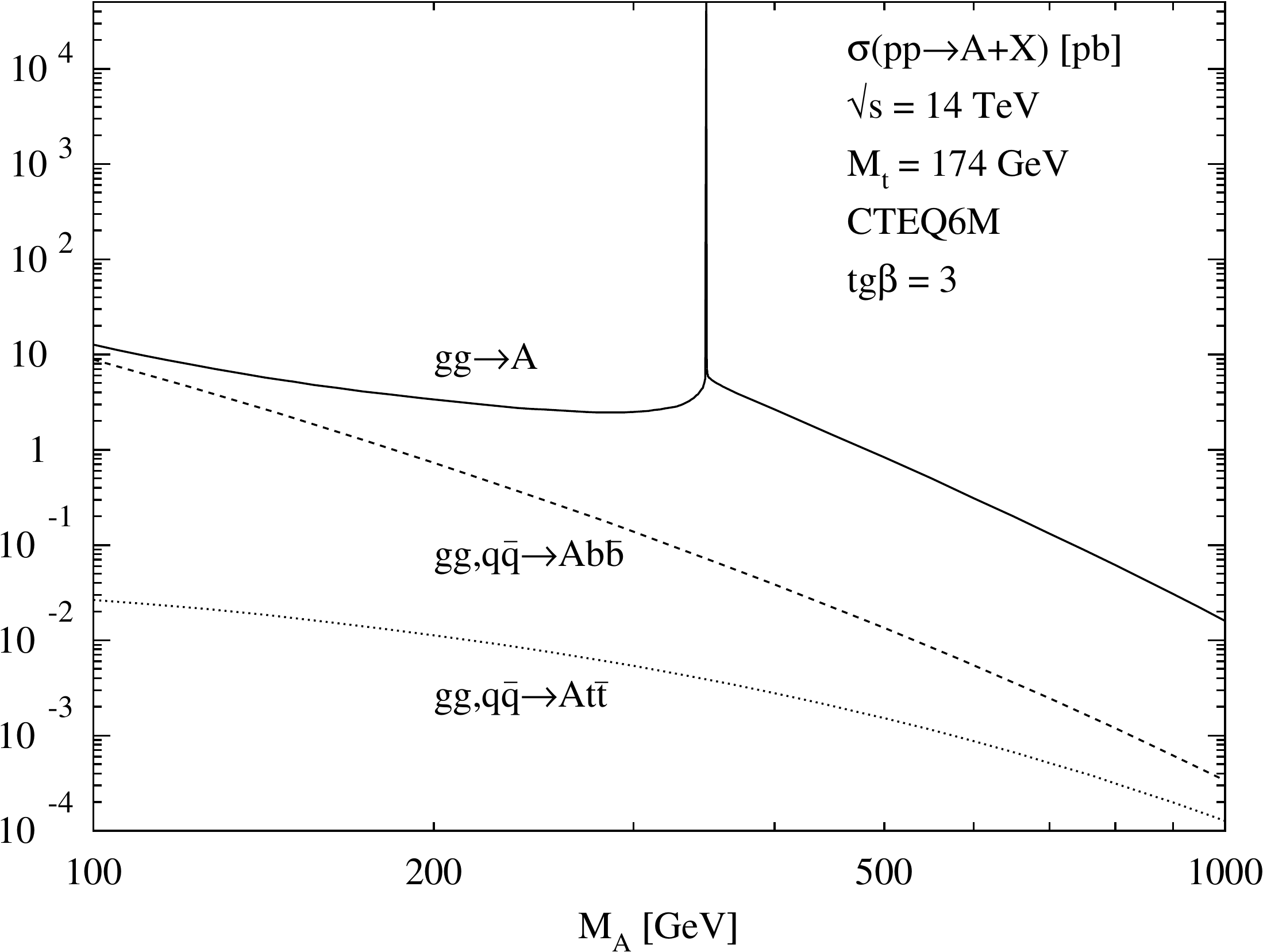}
  \newline
  \includegraphics[height=.3\textheight,angle=270]{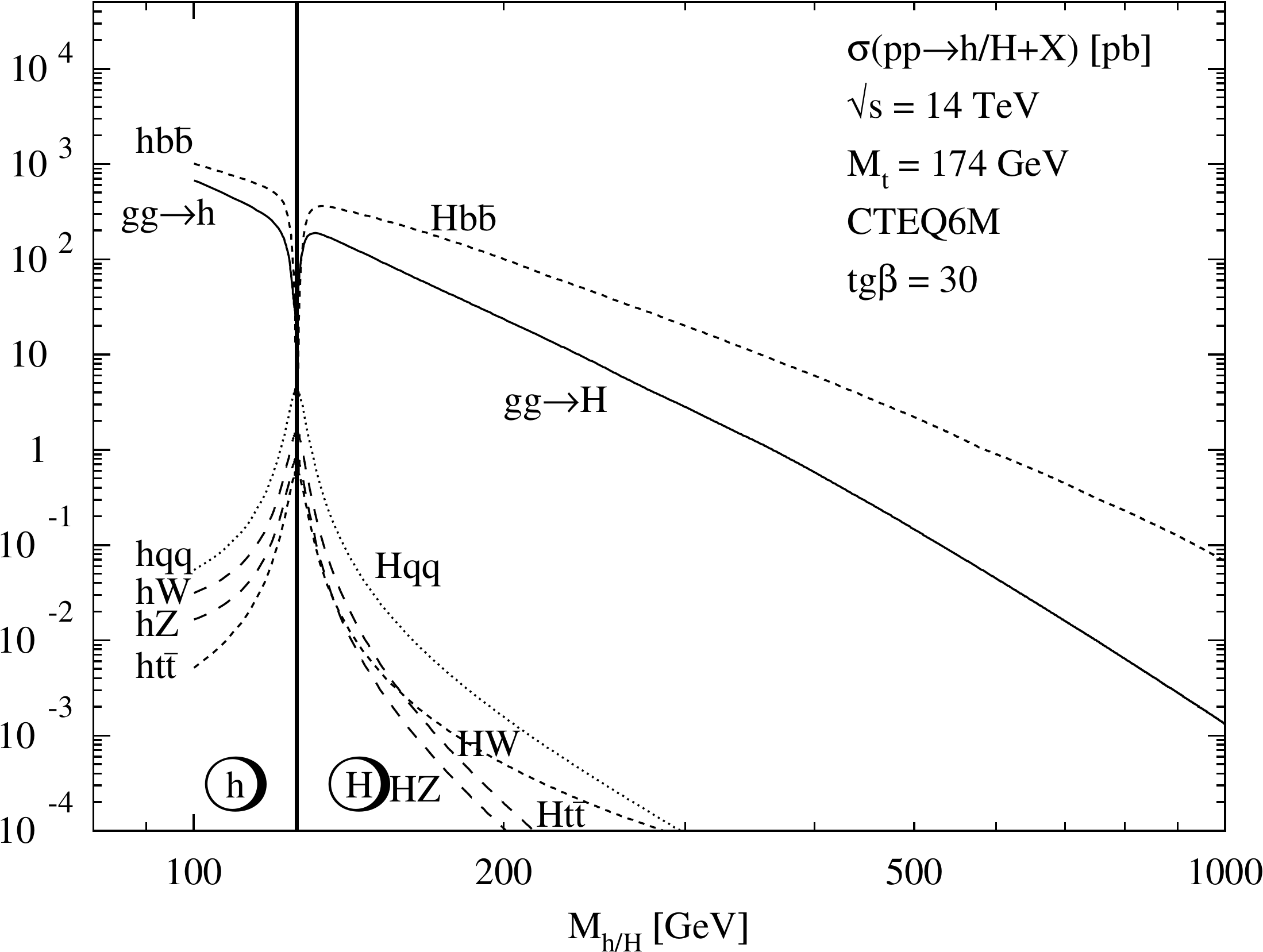}
  \includegraphics[height=.3\textheight,angle=270]{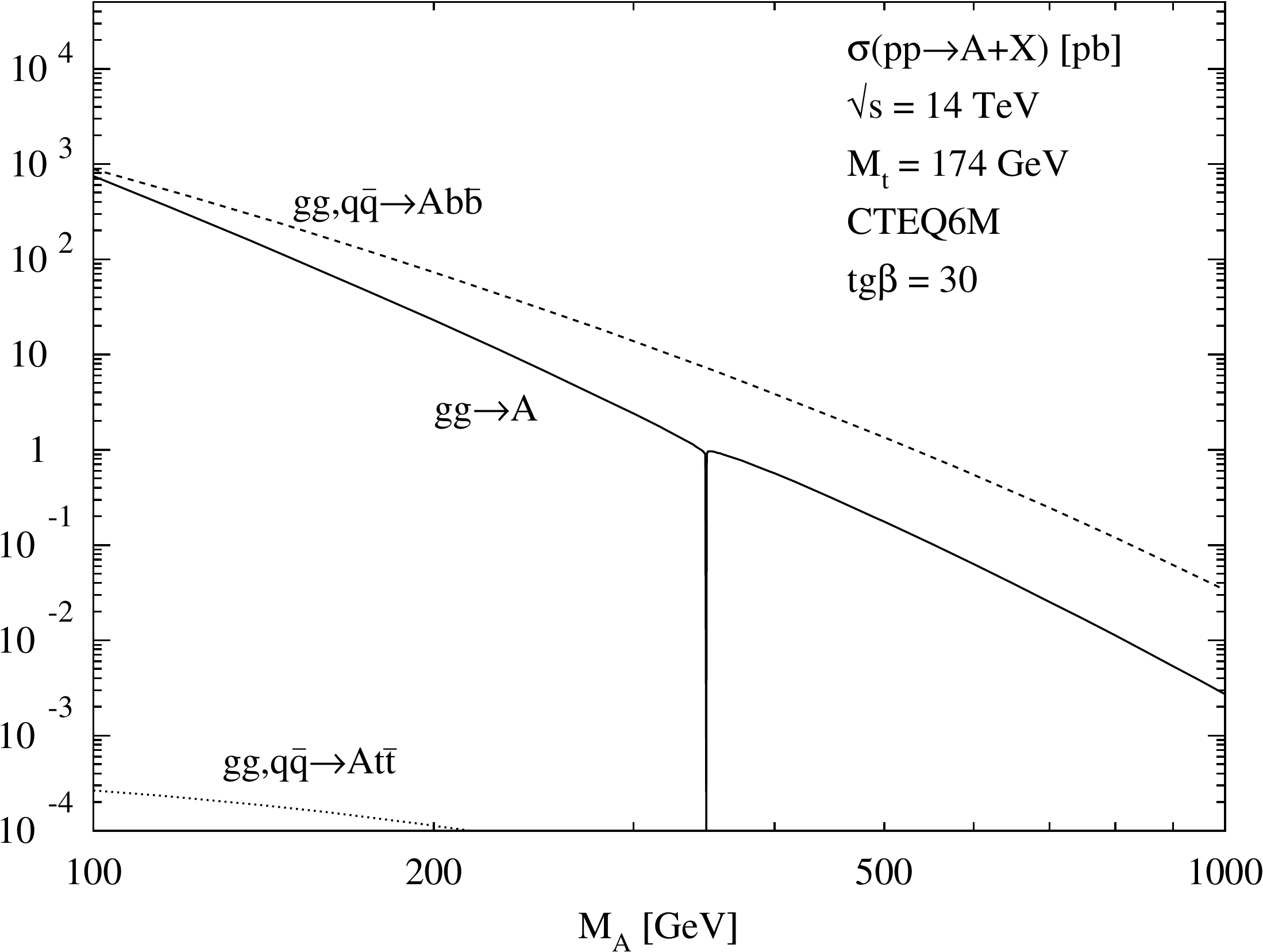}     
  \caption{
  Neutral MSSM Higgs production cross sections at the LHC.
 % The 5$\sigma$ discovery regions for the neutral Higgs bosons $\phi$ ($\phi$=h, H, A)
%      produced in the association with b quarks
%      $\rm p \rm p \rightarrow \rm b \bar{\rm b} \phi$ with the
%      $\phi \rightarrow \mu \mu$ and $\phi \rightarrow \tau \tau$ decay modes
%      (left plot) and for the light, neutral
%     	Higgs boson h from the inclusive $\rm p \rm p \rightarrow \rm h$+X production
%     	with the $\rm h \rightarrow \gamma \gamma$ decay and for the light and heavy scalar
%     	Higgs bosons, h and H, produced in the vector boson fusion
%     	$\rm q \rm q \rightarrow \rm q \rm q \rm h (\rm H)$ with the
%     	$\rm h (\rm H) \rightarrow \tau \tau \rightarrow \ell$+jet decay (right plot).
%     	The $\rm m_{\rm h}^{\rm max}$ scenario is used.
     	}
  \label{fig:cross_section}
\end{figure}

The LEP experiments have excluded the Higgs masses $M_{A} < 91.9 \GeVcc$, 
$M_{h,H} < 91 \GeVcc$ \cite{LEP1} and $M_{H^{\pm}} < 78.6 \GeVcc$ \cite{LEP2}.

In this report the discovery potential of the MSSM neutral Higgs boson
with the CMS detector at LHC is presented. These analysis are also 
described in \cite{PTDR2_Higgs}.

\section{Neutral Higgs bosons searches}

The production cross-section for the MSSM neutral Higgs bosons is strongly dependent
on the value of the $\tan\beta$ parameter.
All neutral MSSM Higgs production cross sections including NLO QCD corrections 
are shown in Fig.\ref{fig:cross_section}.

\subsection{Large $\tan\beta$}

In the region with large $\tan\beta$ values ($>$ 15)
%In case of large $\tan\beta$ ($>$ 15), the neutral
the neutral Higgs bosons are mainly produced in association with b-quarks:
$pp \rightarrow q\bar{q}/gg \rightarrow h/A/H + b\bar{b}$.
The presence of a $b\bar{b}$ pair is important to suppress the very 
large background from Drell-Yan processes. The Higgs bosons mainly
decay in a $b\bar{b}$ pair (90$\%$) and in a $\tau\bar{\tau}$ pair (10$\%$). 
In CMS six channels have been studied:
%taken into account to study the discovery potential of the detector:
\begin{itemize}
\item
  $A/H \rightarrow \mu\mu$
\item
 	$A/H \rightarrow \tau\tau \rightarrow e + jet + X$ %\cite{CMS_NOTE_2006-075}
\item
 	$A/H \rightarrow \tau\tau \rightarrow \mu + jet + X$ %\cite{CMS_NOTE_2006-105}
\item
 	$A/H \rightarrow \tau\tau \rightarrow jet + jet + X$ %\cite{CMS_NOTE_2006-126}
\item
 	$A/H \rightarrow \tau\tau \rightarrow e + \mu + X$ %\cite{CMS_NOTE_2006-101} 	
\item
  $A/H \rightarrow bb$ %\cite{bbAbb}
\end{itemize}

The muon final state, with respect to the other channels, has a much lower
branching ratio ($\approx 3 \times 10^{-4}$), but the event is very clean 
and Higgs masses and widths can be reconstructed precisely. Moreover
it is possible to exploit the theoretical relation between the Higgs decay
width and $\tan\beta$ ($\Gamma_{H} \propto \tan^{2}\beta$) 
to perform a direct measurement of this latter quantity.

\begin{figure}
  \includegraphics[height=.25\textheight]{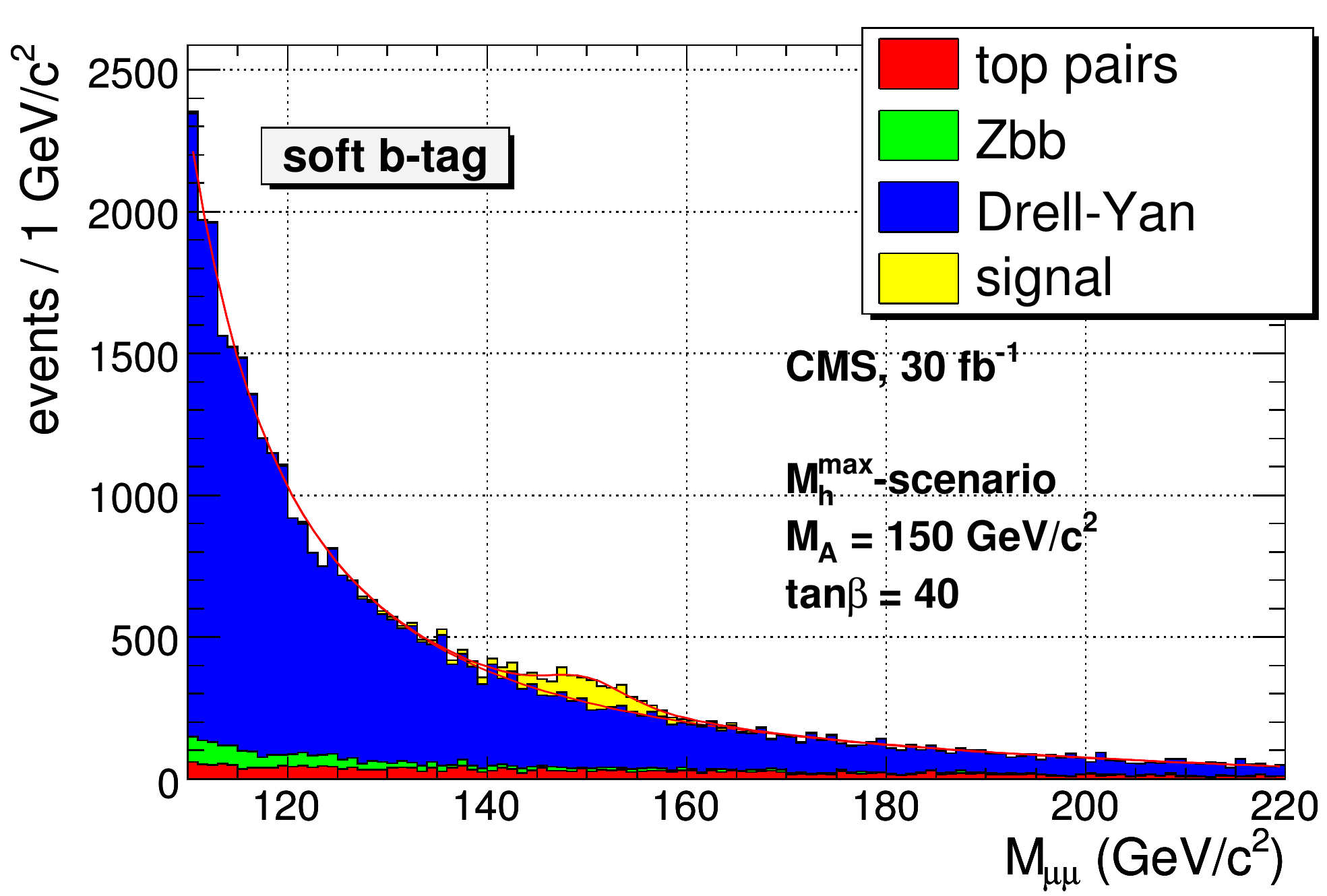}
  \includegraphics[height=.25\textheight]{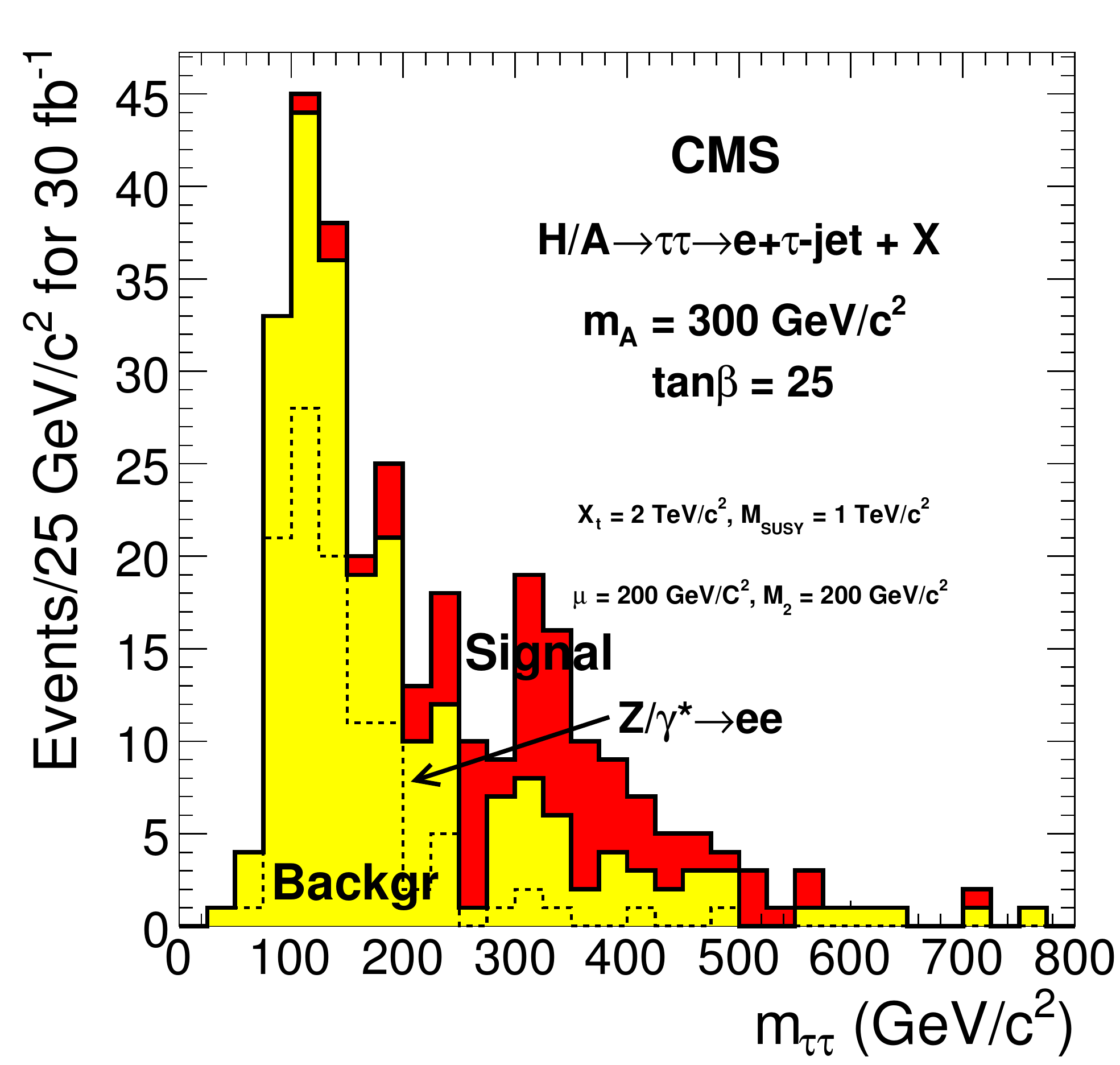}
  \caption{
  %The 5$\sigma$ discovery contours for 30 and 60$\fbinv$ integrated luminosity 
  %for the $A \rightarrow Zh$ channel (left plot). 
  Reconstructed dimuon mass for the main background and for
  the signal sample with $M_{A}$ = 150$\GeVcc$ and $\tan\beta$ = 40 
  for the $A/H \rightarrow \mu\mu$ channel (left) and with 
  $M_{A}$ = 300$\GeVcc$ and $\tan\beta$ = 25 
  for the $A/H \rightarrow \tau\tau \rightarrow e + jet$ channel (right). 
  %The Higgs boson H mass is degenerate with the A mass.
  }
  \label{fig:lowtb}
\end{figure}

The rejection strategy is mainly based on identification of isolated muons and on b-tagging. 
This latter selection is particular important: b jets from signal events 
are mainly produced in the forward region with lower $p_{T}$ with
respect to the b jets coming from $t\bar{t}$ background.
Figure \ref{fig:lowtb} (left) shows the reconstructed dimuon invariant mass
for signal and background.

The tau channels, on the other hands, have a better signal to background ratio
and can reach 
larger discovery region in the plane ($M_{A},\tan\beta$),
as can be seen in Fig. \ref{fig:hightb} (left).
Fig. \ref{fig:lowtb} (right) shows the reconstructed invariant mass for signal 
and background that can be obtained with the 
$A/H \rightarrow \tau\tau \rightarrow e + jet$ channel. 
Indeed, despite the escaping neutrino, the Higgs boson mass can be 
reconstructed also for these channels, exploiting the collinearity approximation: the 
neutrino is assumed to be emitted along the $\tau$ direction.

Finally, the bb channel must take into account the huge QCD background and,
to perform a discovery, one needs to know in advance masses and widths of the Higgs bosons.
Thus this channel can be considered mainly as a cross-check for the discovery.

%\begin{figure}
%  \includegraphics[height=.3\textheight]{AH_discovery}
%  \includegraphics[height=.3\textheight]{h_discovery}
%  \caption{The 5$\sigma$ discovery regions for the neutral Higgs bosons $\phi$ ($\phi$=h, H, A)
%      produced in the association with b quarks
%      $\rm p \rm p \rightarrow \rm b \bar{\rm b} \phi$ with the
%      $\phi \rightarrow \mu \mu$ and $\phi \rightarrow \tau \tau$ decay modes
%      (left plot) and for the light, neutral
%     	Higgs boson h from the inclusive $\rm p \rm p \rightarrow \rm h$+X production
%     	with the $\rm h \rightarrow \gamma \gamma$ decay and for the light and heavy scalar
%     	Higgs bosons, h and H, produced in the vector boson fusion
%     	$\rm q \rm q \rightarrow \rm q \rm q \rm h (\rm H)$ with the
%     	$\rm h (\rm H) \rightarrow \tau \tau \rightarrow \ell$+jet decay (right plot).
%     	The $\rm m_{\rm h}^{\rm max}$ scenario is used.}
%  \label{fig:hightb}
%\end{figure}

\subsection{Small $\tan\beta$}

Concerning low $\tan\beta$ values, the dominant neutral MSSM Higgs production
mechanism is the gluon fusion $gg \rightarrow h/A/H$, which can be mediated by top and botton
loops (as in the SM case), but also by stop and sbottom (Fig.\ref{fig:cross_section}).

In CMS two channels have been investigated:

\begin{itemize}
\item
  $A \rightarrow Zh \rightarrow \ell^{+}\ell^{-} b\bar{b}$
\item
 	$A/H \rightarrow \tilde{\chi}^{0}_{2} \tilde{\chi}^{0}_{2} \rightarrow 4\ell + E^{miss}_{T}$
\end{itemize}

\begin{figure}
  \includegraphics[height=.3\textheight]{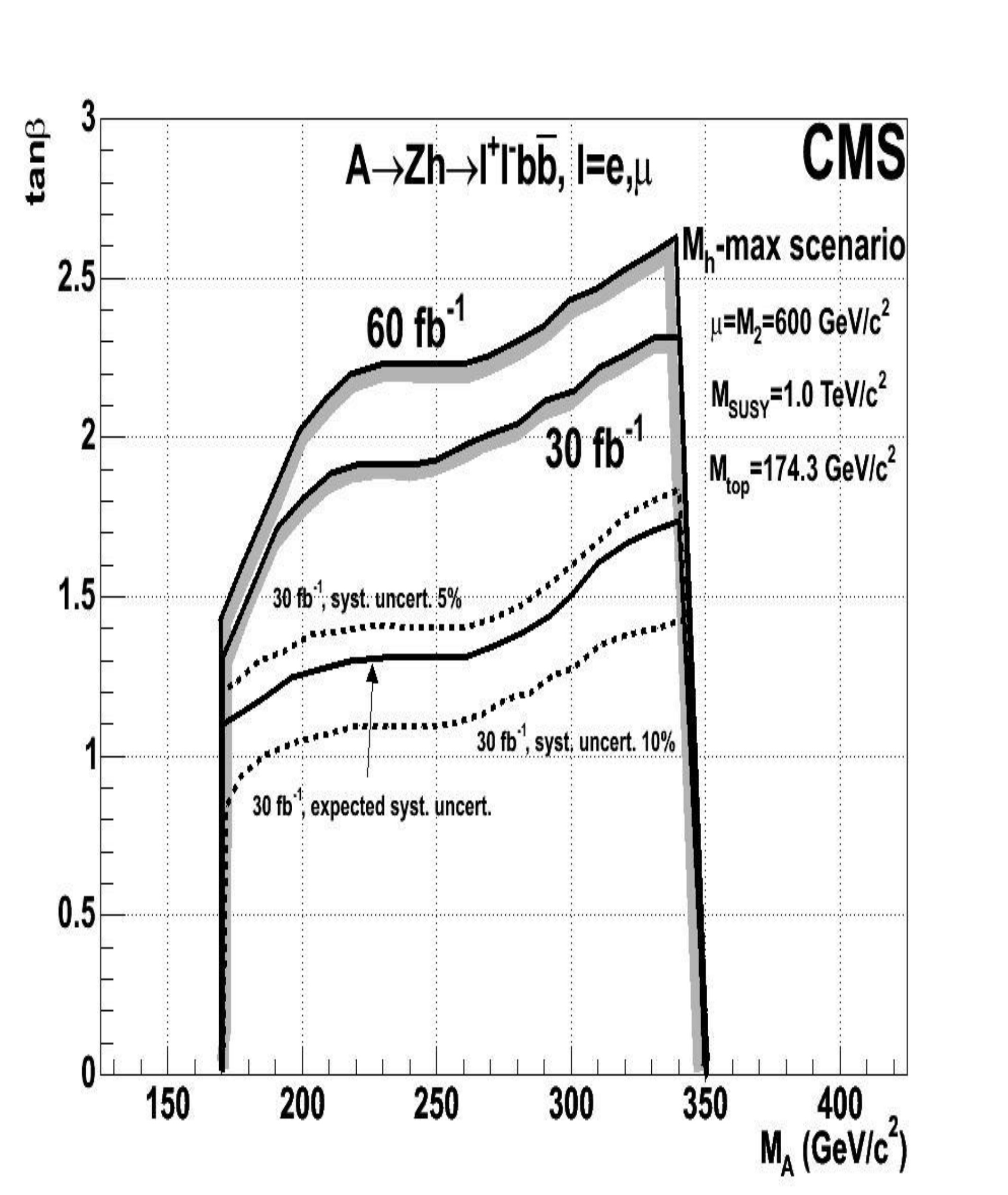}
  \includegraphics[height=.3\textheight]{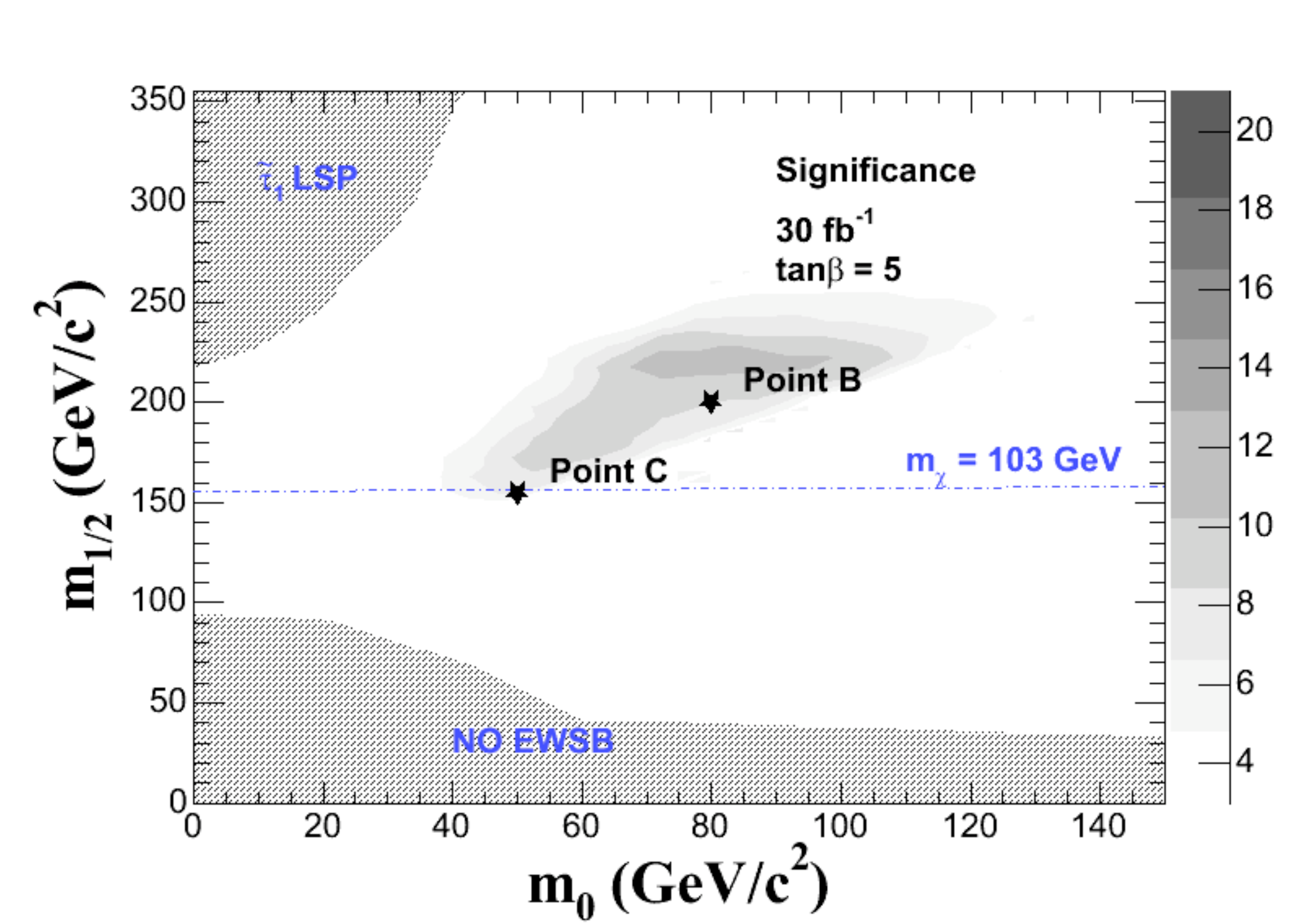}
  \caption{(left) The $5\sigma$ discovery region contours for 30 and 60 $\fbinv$ integrated
  luminosity for the $A \rightarrow Zh$ channel. The effect of underestimation or overestimation
  of the background systematic uncertainty can be seen in the courve of 30 $\fbinv$.
  (right)The $5\sigma$ discovery region contours for 30 $\fbinv$ integrated
  luminosity for the $A/H \rightarrow \tilde{\chi}^{0}_{2} \tilde{\chi}^{0}_{2}$ channel.}
  \label{fig:atoh}
\end{figure}

The first channel provides an interesting way to detect A and h simultaneously.
The cross section increases with decreasing of $\tan\beta$, while the 
mass range is $m_{Z}+m_{h} \leq m_{A} \leq 2m_{top}$.
However results are strongly dependent on the MSSM parameters $\mu$ and $M_{2}$, 
because the Higgs boson decay $A \rightarrow \tilde{\chi}^{0}_{1} \tilde{\chi}^{0}_{1}$ 
may become dominant
(the best results being obtained for large values of $\mu$ and $M_{2}$).
Fig. \ref{fig:atoh} (left) shows the discovery region for this channel
for 30 and 60$\fbinv$ of integrated luminosity.

To increase the discovery region in the low and intermediate region of $\tan\beta$,
the second channel has been studied which takes into account the decay modes of the
neutral Higgs bosons to supersymmetric particles.
The final state studied by this channel is particular clean (four leptons plus 
missing transverse energy). The analysis is performed in three benchmark points of 
the minimal Super Gravity constrained version of the MSSM (mSUGRA): these points 
are obtained varying the parameters $m_{0}$ and $m_{\frac{1}{2}}$ for $\tan\beta$ = 5, 10,
$sign(\mu)$ = + and $A_{0}$ = 0.
Fig.\ref{fig:atoh}(right) shows the discovery region in the ($m_{0},m_{\frac{1}{2}}$) plane, 
for an integrated luminosity of 30$\fbinv$.
% (figure \ref{fig:char}(left)).
%For the $A/H \rightarrow \chi^{0}_{2} \chi^{0}_{2}$ channel three mSUGRA benchmark points have been considered. 

%\begin{figure}
%%  \includegraphics[height=.3\textheight]{Atoh_discovery}
%  \includegraphics[height=.3\textheight]{Hmumu_mass}
%%  \includegraphics[height=.3\textheight]{Atochi_discovery}
%  \caption{
%  %The 5$\sigma$ discovery contours for 30 and 60$\fbinv$ integrated luminosity 
%  %for the $A \rightarrow Zh$ channel (left plot). 
%  Dimuon reconstruction mass for the main background and for
%  the signal sample with $M_{A}$ = 150$\GeVcc$ and $\tan\beta$ = 40 
%  for the $A/H \rightarrow \mu\mu$ channel 
%  %(right plot)
%  .}
%  \label{fig:lowtb}
%\end{figure}

\begin{figure}
  \includegraphics[height=.3\textheight]{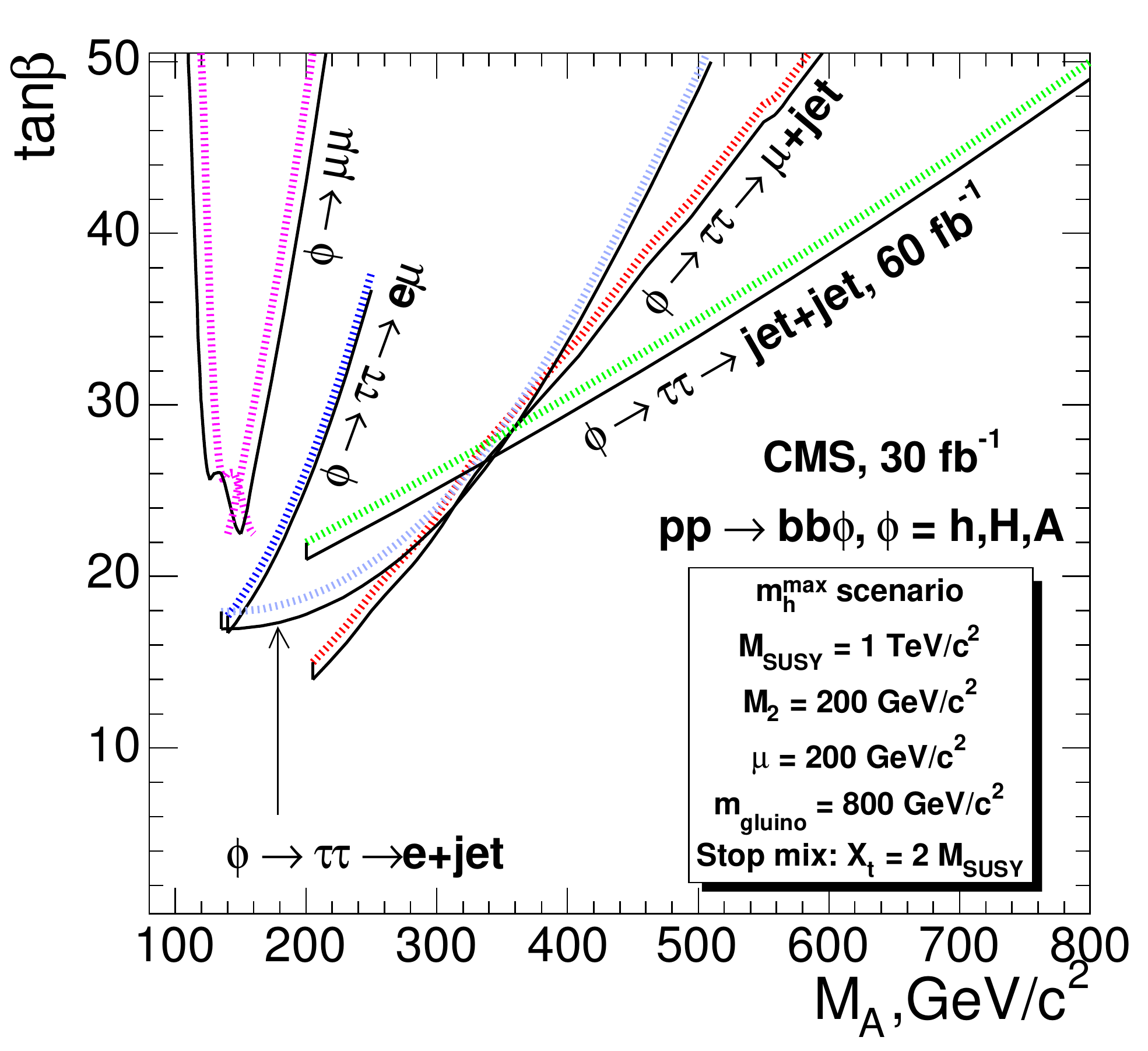}
  \includegraphics[height=.3\textheight]{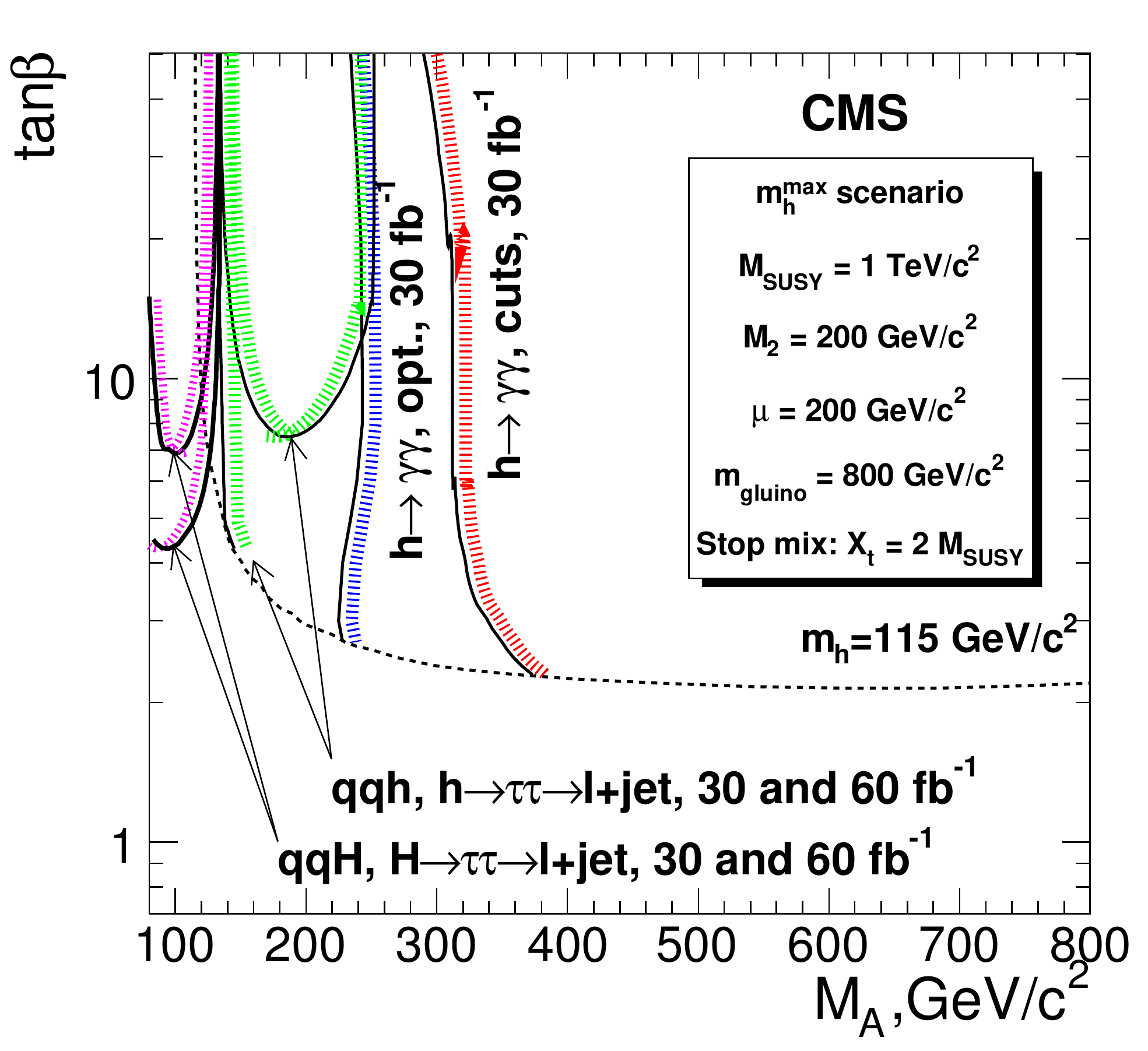}
  \caption{The 5$\sigma$ discovery regions for the neutral Higgs bosons $\phi$ ($\phi$=h, H, A)
      produced in the association with b quarks
      $\rm p \rm p \rightarrow \rm b \bar{\rm b} \phi$ with the
      $\phi \rightarrow \mu \mu$ and $\phi \rightarrow \tau \tau$ decay modes
      (left plot) and for the light, neutral
     	Higgs boson h from the inclusive $\rm p \rm p \rightarrow \rm h$+X production
     	with the $\rm h \rightarrow \gamma \gamma$ decay and for the light and heavy scalar
     	Higgs bosons, h and H, produced in the vector boson fusion
     	$\rm q \rm q \rightarrow \rm q \rm q \rm h (\rm H)$ with the
     	$\rm h (\rm H) \rightarrow \tau \tau \rightarrow \ell$+jet decay (right plot).
     	The $\rm m_{\rm h}^{\rm max}$ scenario is used.}
  \label{fig:hightb}
\end{figure}

Fig.\ref{fig:hightb} summarize the $5\sigma$ discovery region that can be obtained 
with 30$\fbinv$ of integrated luminosity.

\section{Charged Higgs bosons searches}

Three channels, depending on the final state and on the Higgs boson mass,
have been investigated for charged Higgs bosons:

\begin{itemize}
\item
  $H^{\pm} \rightarrow \tau\nu_{\tau}$, with $M_{H} < M_{top}$ %\cite{CMS_NOTE_2006-056}
\item
 	$H^{\pm} \rightarrow \tau\nu_{\tau}$, with $M_{H} > M_{top}$ %\cite{CMS_NOTE_2006-100}
\item
  $H^{\pm} \rightarrow tb$, with $M_{H} > M_{top}$ %\cite{CMS_NOTE_2006-109}
\end{itemize}

\begin{figure}
  \includegraphics[height=.3\textheight]{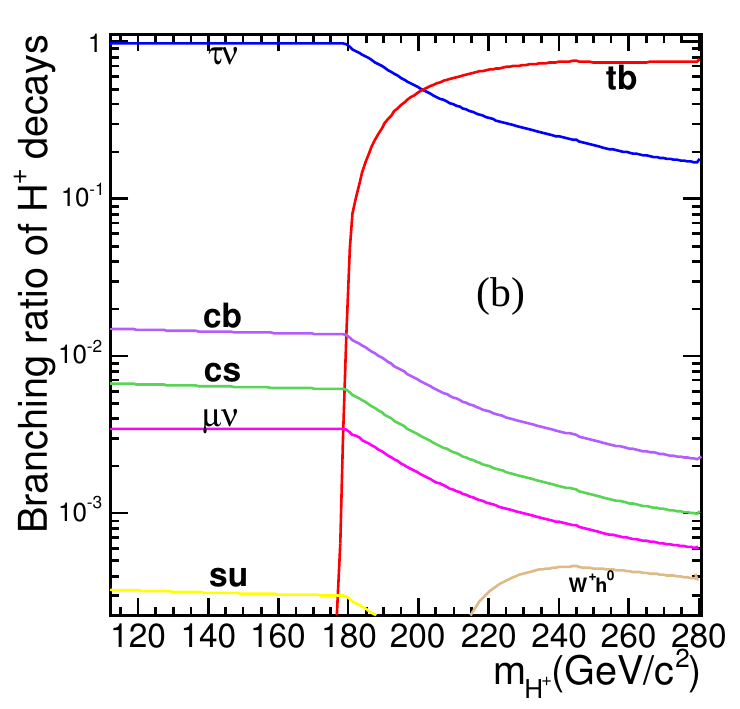}
  \includegraphics[height=.3\textheight]{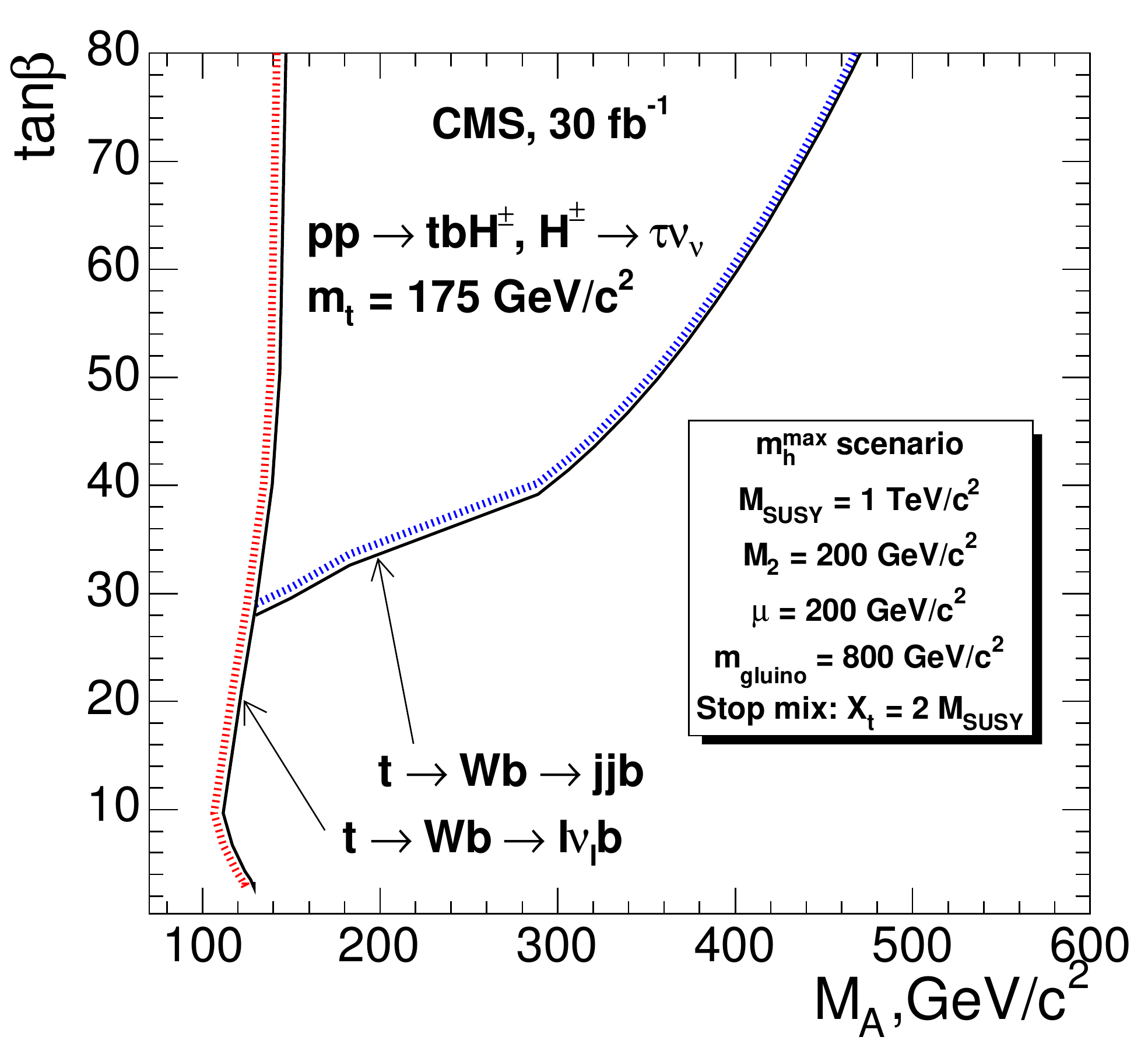}
  \caption{Branching ratios for charged Higgs boson decaying to different 
  final states for $\tan\beta$ = 20 (left).
%  For integrated luminosity of  $30\fbinv$ 
  The $5\sigma$-discovery regions for
%	$\mathrm{A}^{0}/\mathrm{H}^{0} \to \chi^0_2\chi^0_2 \to  
%	4\ell~+~\mathrm{E}_\mathrm{T}^\mathrm{miss}$ channel in the 
%	$(m_0,m_{1/2})$ plane for fixed $A_0 = 0$, sign$(\mu) = +$ and  $\tan\beta = 5$ (left plot) and
%	for 
	the charged Higgs boson with the $\tau \nu$ decay mode 
	with 30$\fbinv$ of integrated luminosity (right).}
  \label{fig:char}
\end{figure}

For Higgs boson masses below $M_{top}$, the main production mechanism is
through top decay, $t \rightarrow H^{+}b$, and 
the branching ratio of the $\tau\nu$ channel is about 98\% (Fig.\ref{fig:char}(left)). 
The study is performed considering the leptonic decay of the W: 
$t\bar{t} \rightarrow H^{\pm}W^{\mp}b\bar{b} \rightarrow \tau\nu_{\tau}\ell\nu_{\ell}b\bar{b}$, 
$\tau \rightarrow $ hadrons.
The discovery region, as can be seen in 
figure~\ref{fig:char}(right), covers almost the entire allowed region in the 
($M_{A},\tan\beta$) plane.

If $M_{H} > M_{top}$, the charged Higgs bosons are mainly produced 
% the main production mechanism for the charged Higgs bosons is 
in association with a top-bottom pair, $gg \rightarrow tbH^{\pm}$. 
The final state for this channel is very clean
($H^{\pm} \rightarrow \tau^{\pm}\nu$, $r \rightarrow hadrons+\nu$ and $W^{\mp} \rightarrow jj$)
and, after the selection cuts, almost background free. 
The characteristics for this channel
is the presence of large missing transverse energy and 
the $\tau$ helicity correlations favouring the $H^{\pm} \rightarrow \tau^{\pm}\nu$ decay
over the $W^{\pm} \rightarrow \tau^{\pm}\nu$ decay.
A large sector 
of the ($M_{A},\tan\beta$) plane can be covered (Fig. ~\ref{fig:char}(right)).
%Figure~\ref{fig:char}(right) shows the discovery region that can be obtained at CMS for charged 
%Higgs bosons.

Finally, for masses above $M_{top} + M_{bottom}$, the channel $H^{\pm} \rightarrow tb$ opens up.
Two production mechanism are considered:
\begin{itemize}
\item
  $gb \rightarrow tH^{\pm} \rightarrow ttb \rightarrow W^{+}W^{-}bbb \rightarrow qq'\mu\nu_{\mu}bbb$ 
\item
  $gg \rightarrow tH^{\pm}b \rightarrow ttbb \rightarrow W^{+}W^{-}bbbb \rightarrow qq'\mu\nu_{\mu}bbbb$   
\end{itemize}

Unfortunately no sensitivity is obtained in the MSSM parameter space with this analysis, 
due to the large background
and the resulting large effects of systematic uncertainties on its knowledge.

Fig.\ref{fig:char}(right) summarize the $5\sigma$ discovery region that can be obtained 
with 30$\fbinv$ of integrated luminosity.

\section{Conclusions}
Many channels have been studied to estimate the discovery potential of 
MSSM Higgs bosons at CMS. A large area in the ($M_{A},\tan\beta$) plane 
will be explored: the most promising channels are $\phi \rightarrow \tau\tau$ 
($\phi$ = h,A,H) and $H^{\pm} \rightarrow \tau\nu$.

%\end{document}

\addtocounter{chapter}{1}

\newcommand{\ordEW}{\mathcal{O}(\alpha_{\scriptscriptstyle EM}^6)}
\newcommand{\ordQCD}{\mathcal{O}(\alpha_{\scriptscriptstyle EM}^4
  \alpha_{\scriptscriptstyle S}^2)}

%\title {Physics studies at the LHC with PHANTOM}
%\author{Giuseppe Bevilacqua  \\
%Dipartimento di Fisica Teorica, Università di Torino}
%\date{}

%\begin{document}
%%%%%%%%%%%%%%%%%%%%%%%%%%%%%%%%%%%%%%%%%%%%%%%
% Toggle line numbering
% Won't work with the PRD revtex4 !
%\pagewiselinenumbers
% uncomment if you want doublespace
%\doublespace
%%%%%%%%%%%%%%%%%%%%%%%%%%%%%%%%%%%%%%%%%%%%%%%

%\maketitle

\mchapter{Physics studies at the LHC with PHANTOM}{G. Bevilacqua}

\section{Introduction}
To shed light on the nature of Electroweak Symmetry Breaking (EWSB) is one 
among the main purposes of the LHC experiments. The Standard Model provides the
most economical explanation in terms of the Higgs mechanism. The remarkable
agreement between its predictions and precision electroweak data is consistent 
with a light Higgs boson with a mass lower than 200 GeV\cite{EWWG1},
while direct searches have established a lower bound of 114.4 GeV\cite{EWWG2}
at 95\% CL.

The Higgs mechanism is essential to ensure the renormalizability of the Standard
Model. Moreover, without a Higgs the theory is not self-consistent and violates
perturbative unitarity at high energies, which implies that for several 
processes the amplitude grows indefinitely with energy.
This unphysical behaviour is particularly manifest in the scattering of
longitudinally polarized vector bosons, whose cross section goes over the 
unitarity limit at about one TeV.
If the Higgs hypothesis is not realized, effects of new physics are 
expected to restore unitarity at this scale.

On the other side, if a massive Higgs boson exists, a resonance will be observed
in the $VV$ invariant mass spectrum in correspondence of the Higgs mass.
Vector Boson Fusion represents the second most important contribution to the 
cross section for Higgs production at LHC, moreover $H \rightarrow VV$ is the 
preferred decay channel above about 150 GeV.
Final states in which the vector bosons decay into four leptons or two leptons 
plus two jets provide one of the cleanest signatures for Higgs detection.

It should be mentioned that even if a light Higgs will be discovered, the 
gauge boson scattering could nonetheless deviate from the Standard Model 
predictions in the high energy domain. Several models have been proposed, 
whereby the Higgs is a composite Goldstone boson associated to some new strong 
dynamics active at the TeV scale\cite{Rattazzi:2005di,Giudice:2007fh}.
The $VV$-scattering cross section is expected to be enhanced at high $M_{VV}$ as
a consequence of the new strongly interacting sector.

In brief, Boson-Boson Scattering has a great potential for probing the 
mechanism of EWSB at LHC, independently of its particular realization.
Unfortunately, no beam of on-shell bosons will be available.
The only consistent way to extract some information from data is to rely upon 
a complete calculation of final states with six fermions.

A Monte Carlo event generator, \texttt{PHANTOM}\cite{Phantom}, has
been recently developed to provide a complete description at $\ordEW+\ordQCD$ 
of all six-parton final states in the Standard Model, including for the first 
time all EW and mixed EW+QCD contributions at this perturbative order. 
Exact matrix elements at tree level are evaluated efficiently. A good 
coverage of phase space is achieved thanks to a new 
approach\cite{Accomando:2005cc} which combines the best features of multichannel
with the adaptivity of the \texttt{VEGAS} algorithm\cite{VEGAS}.
Apart from Boson-Boson Scattering and Higgs search, the program gives access to
the full tree-level description of $t\bar{t}$ production which is important 
both in itself and as a background in connection with the previous topics.

In this note we present the first parton-level studies based on 
\texttt{PHANTOM}. The analysis is focused on Boson-Boson Scattering and 
Higgs search via Vector Boson Fusion in the semi-leptonic channels,
that is final states characterized by four quarks and two charged leptons or one
charged lepton plus one neutrino.
A light Higgs scenario is compared with no-Higgs ($M_H \rightarrow \infty$) 
results, to be intended at this stage as a minimal benchmark in the search for 
new physics beyond the Standard Model.

\section{Boson-Boson Scattering signature and its irreducible background}
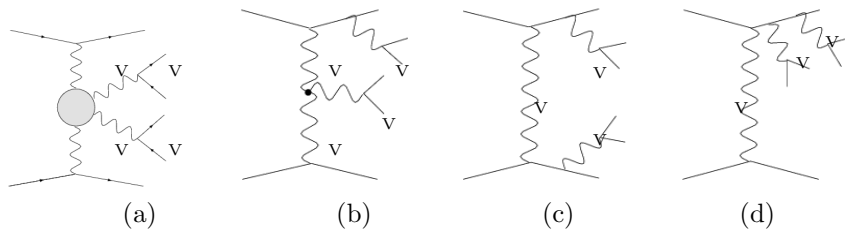
\begin{figure}[htb]
\centering
\epsfig{file=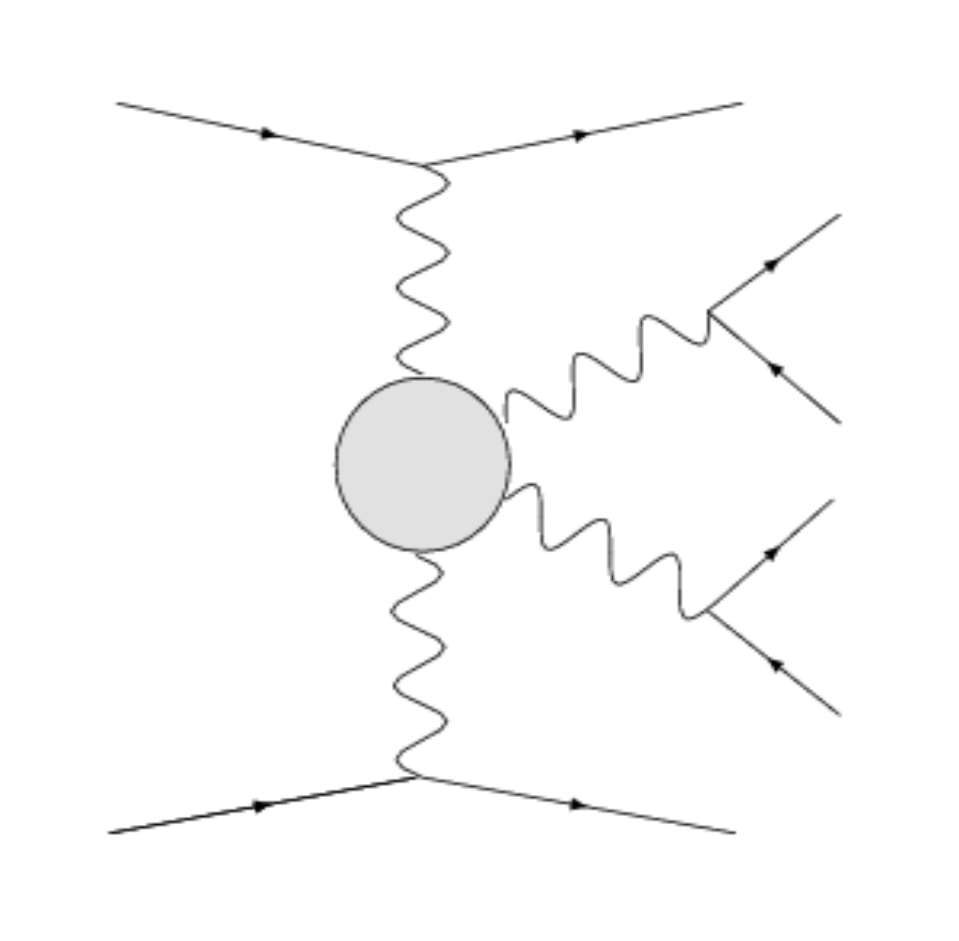,width=0.18\textwidth,height=2.65cm}
\hspace{0.1cm}
\epsfig{file=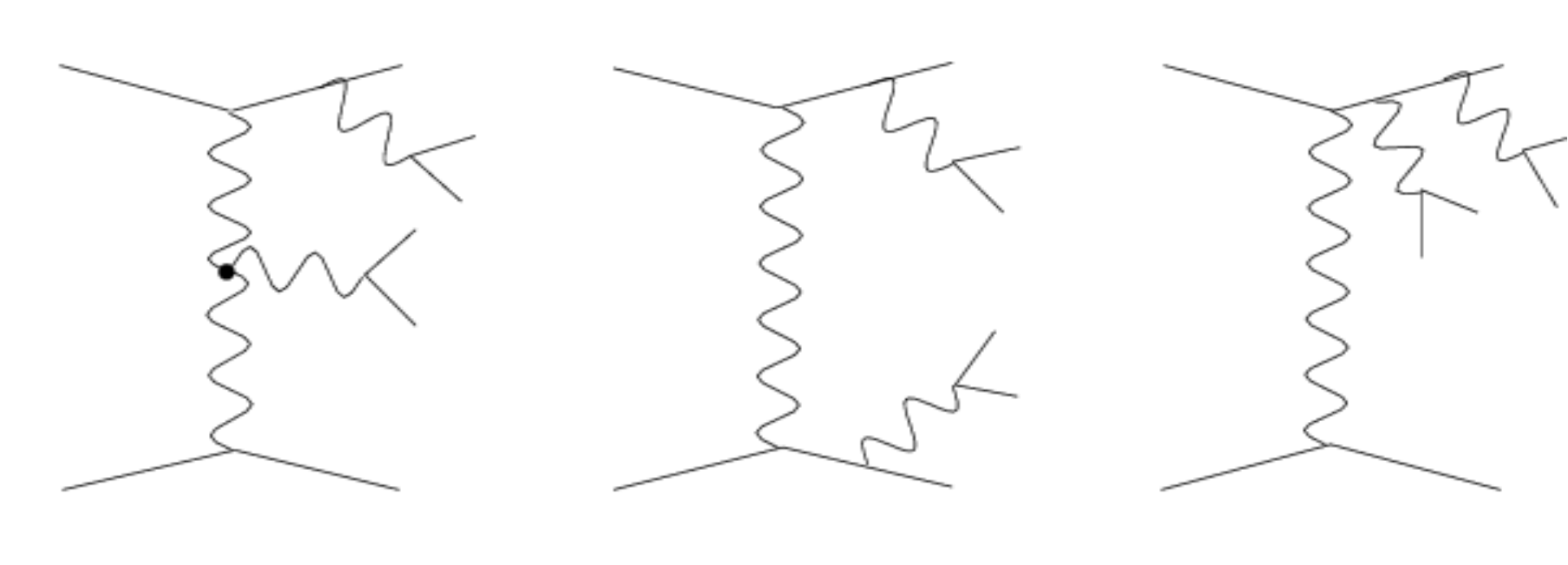,width=0.55\textwidth}
\begin{picture}(0,0) (0,0)
  \put(-274,-5) {\small{(a)}}
  \put(-277,50) {\tiny{V}}
  \put(-277,20) {\tiny{V}}
  \put(-257,50) {\tiny{V}}
  \put(-257,20) {\tiny{V}}
  \put(-194,-5) {\small{(b)}}
  \put(-197,50) {\tiny{V}}
  \put(-197,20) {\tiny{V}}
  \put(-172,50) {\tiny{V}}
  \put(-177,30) {\tiny{V}}
  \put(-117,-5) {\small{(c)}}
  \put(-120,36) {\tiny{V}}
  \put(-98,49) {\tiny{V}}
  \put(-98,24) {\tiny{V}}
  \put(-43,-5) {\small{(d)}}
  \put(-45,36) {\tiny{V}}
  \put(-22,53) {\tiny{V}}
  \put(-11,61) {\tiny{V}}
\end{picture}
\caption{The Boson-Boson Scattering topology (a) and other typical contributions
to the amplitude of a six-fermion final state (b,c,d).}
\label{fig:VBSbckgr_EW}
\end{figure}

The problem of how to define a signal for Boson-Boson Scattering has been 
debated since a long time. Ideally, one should isolate the contribution of
diagrams like the one depicted in Fig.\ref{fig:VBSbckgr_EW}(a), in which the 
incoming partons emit two space-like vector bosons which scatter among 
themselves and finally decay into fermions. 
The resulting amplitude should then be deconvoluted from the PDF's and finally 
projected to on-shell bosons.
In practice, this procedure turns out to be problematic due to gauge invariance 
and cancellation effects. In fact the subset of boson-boson fusion diagrams is 
not separately gauge invariant and there are relevant negative 
interferences\cite{Kleiss:1986xp} with the rest of the diagrams, some
of which are illustrated in Fig.\ref{fig:VBSbckgr_EW}(b,c,d).
Previous studies have evidentiated substantial differences among results based 
on this approximation and exact tree-level predictions on the $VV$ mass 
spectrum\cite{Accomando:2006mc,Accomando:2005hz}, showing that 
a complete evaluation of all contributions at $\ordEW$ is necessary.

In the two-step approach we follow, we first study the characteristics of 
Boson-Boson Scattering from a six-fermion point of view. 
As no contribution to $VV$ scattering can be found outside $\ordEW$, it is 
convenient to start with a complete study of all processes at this 
perturbative order. In our intention, this represents a benchmark for physics
analyses and does not claim to be fully realistic.
With the help of appropriate selection cuts, we isolate a sample of events 
which best exemplifies the signature of the physical process under 
investigation. This means that candidates for top or three-boson 
production are subtracted to the advantage of events characterized by two tag 
jets widely separated in pseudorapidity and two couples of jets and leptons 
resonant at $W$ or $Z$ mass. Subsequently, further procedures are examined
to enhance the contribution of longitudinally polarized vector bosons, which are
most sensitive to effects of EWSB.
The inclusion of the relevant QCD background represents the step forward which 
is essential to get meaningful results at parton level and completes our 
analysis.

\begin{figure}[h!tb]
\centering
\epsfig{file=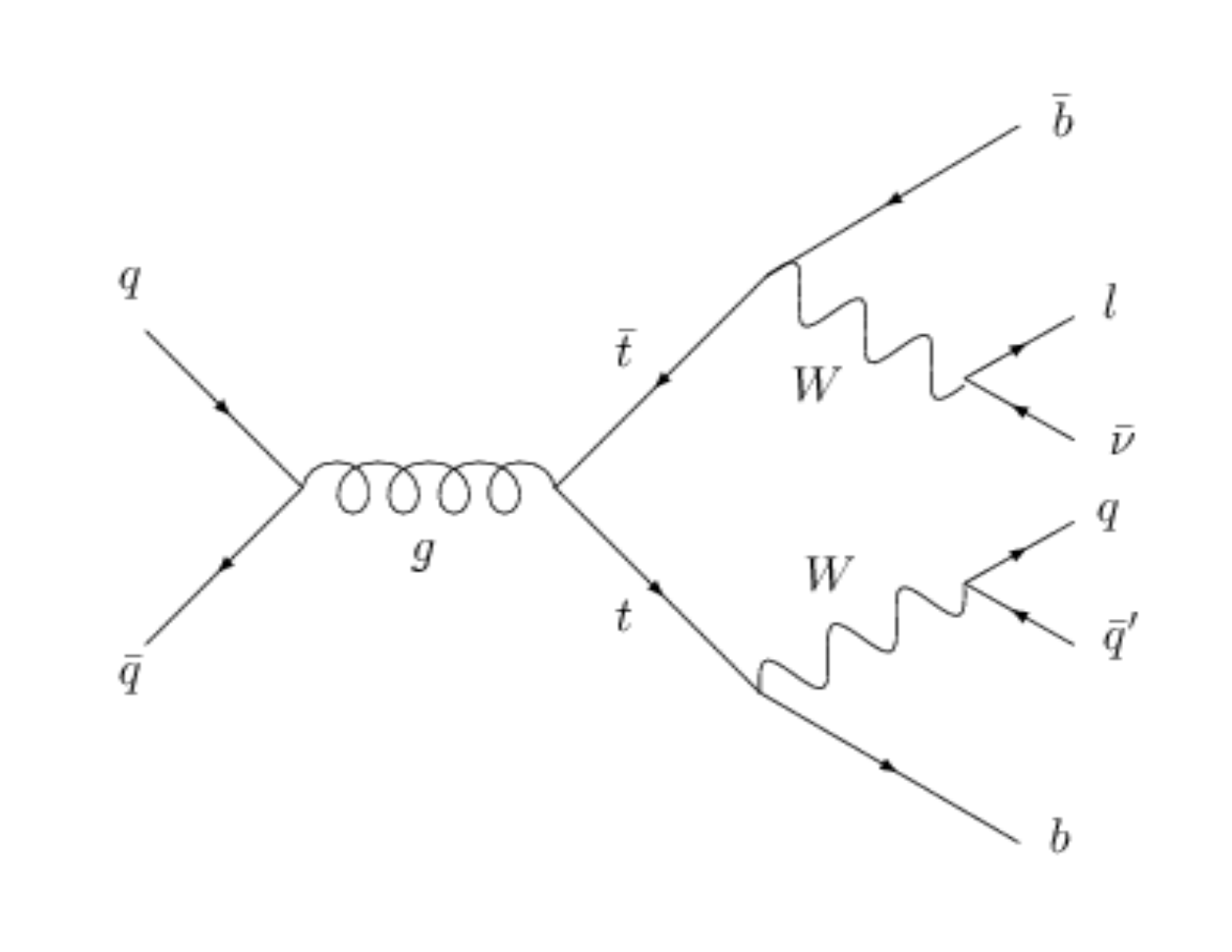,width=0.26\textwidth,height=2.8cm}
\epsfig{file=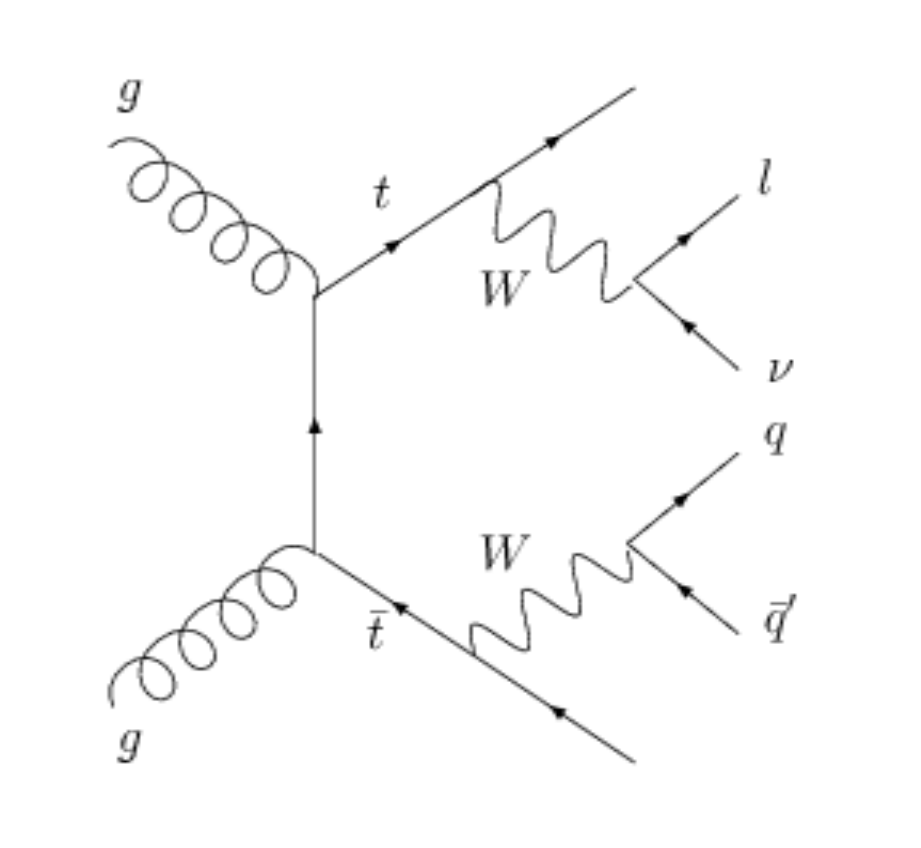,width=0.26\textwidth,height=2.8cm}
\epsfig{file=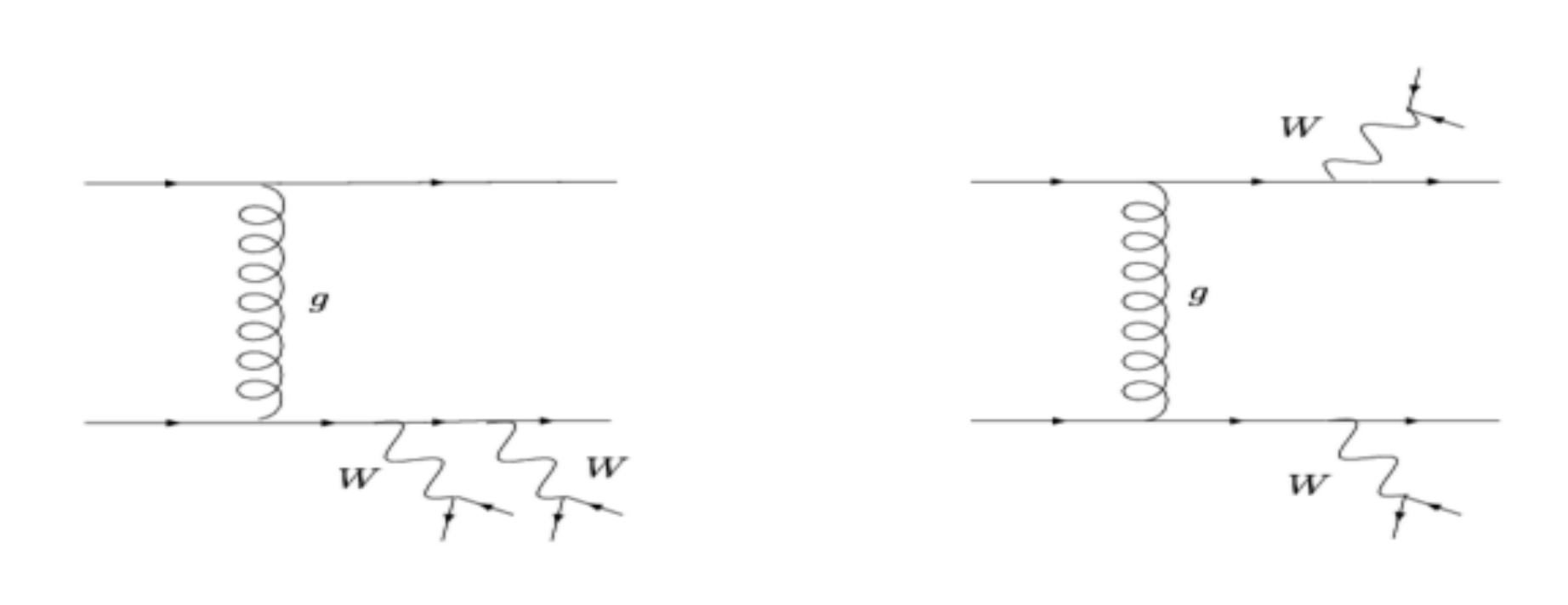,width=0.4\textwidth,height=3.2cm}
\begin{picture}(0,0) (0,0)
  \put(-324,-5) {\small{(a)}}
  \put(-224,-5) {\small{(b)}}
  \put(-129,-5) {\small{(c)}}
  \put(-43,-5) {\small{(d)}}
\end{picture}
\caption{Examples of contributions to the QCD irreducible background: $t\bar{t}$
production (a,b) and $VV+2\,jets$ (c,d)}
\label{fig:VBSbckgr_QCD}
\end{figure}
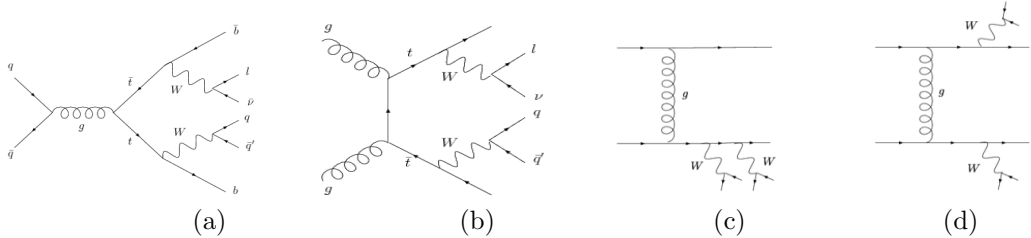

Some examples of contributions to the QCD irreducible background are 
illustrated in Fig.\ref{fig:VBSbckgr_QCD}.
Top production, either from single $t$ or $t\bar{t}$, is a huge source of
background for this kind of studies. In particular $t\bar{t}$ is dominated 
by $\ordQCD$ contributions due to diagrams shown in 
Fig.\ref{fig:VBSbckgr_QCD}(a,b).
$VV+2\,jets$ at $\ordQCD$ is another important background with a signature 
characterized by two outgoing vector bosons, quite similar to the scattering one
and therefore particularly challenging to subract.

In the case of totally leptonic channels, i.e. final states with four leptons,
$\ordQCD$ encompasses the full QCD background at tree level for
six-fermion analyses.
On the contrary, semileptonic channels get additional contributions from 
$\mathcal{O}(\alpha_{\scriptscriptstyle EM}^2\alpha_{\scriptscriptstyle S}^4)$ 
diagrams which are responsible for the $V+4 \,jets$ background. 
The latter must be included as well in a complete analysis and can be covered by
other Monte Carlo event generators, such as \texttt{AlpGen}\cite{nMangano:2002ea}
or \texttt{MadEvent}\cite{Maltoni:2002qb}.
It should be noticed nevertheless that these contributions have quite 
different kinematical features with respect to the scattering signature, 
therefore they are expected to be easily suppressed by means of appropriate 
selection cuts.
Of course, the final word is left to the results of a complete study.

\section{$\ordEW$ results in the semileptonic $\mu^+\mu^-$ channel}
This section is devoted to the analysis at $\ordEW$ of all the processes of type
$qq \rightarrow qqqq\mu^+\mu^-$. 
All our results have been obtained using the CTEQ5L\cite{CTEQ5L} PDF set with 
scale
\begin{equation}
\label{PDFScale1}
Q^2 = M_W^2 + \frac{1}{6}\,\sum_{i=1}^6 p_{Ti}^2,
\end{equation}
where $p_{Ti}$ denotes the transverse momentum of the $i$-th final state 
particle.

\begin{figure}[htb]
\centering
\epsfig{file=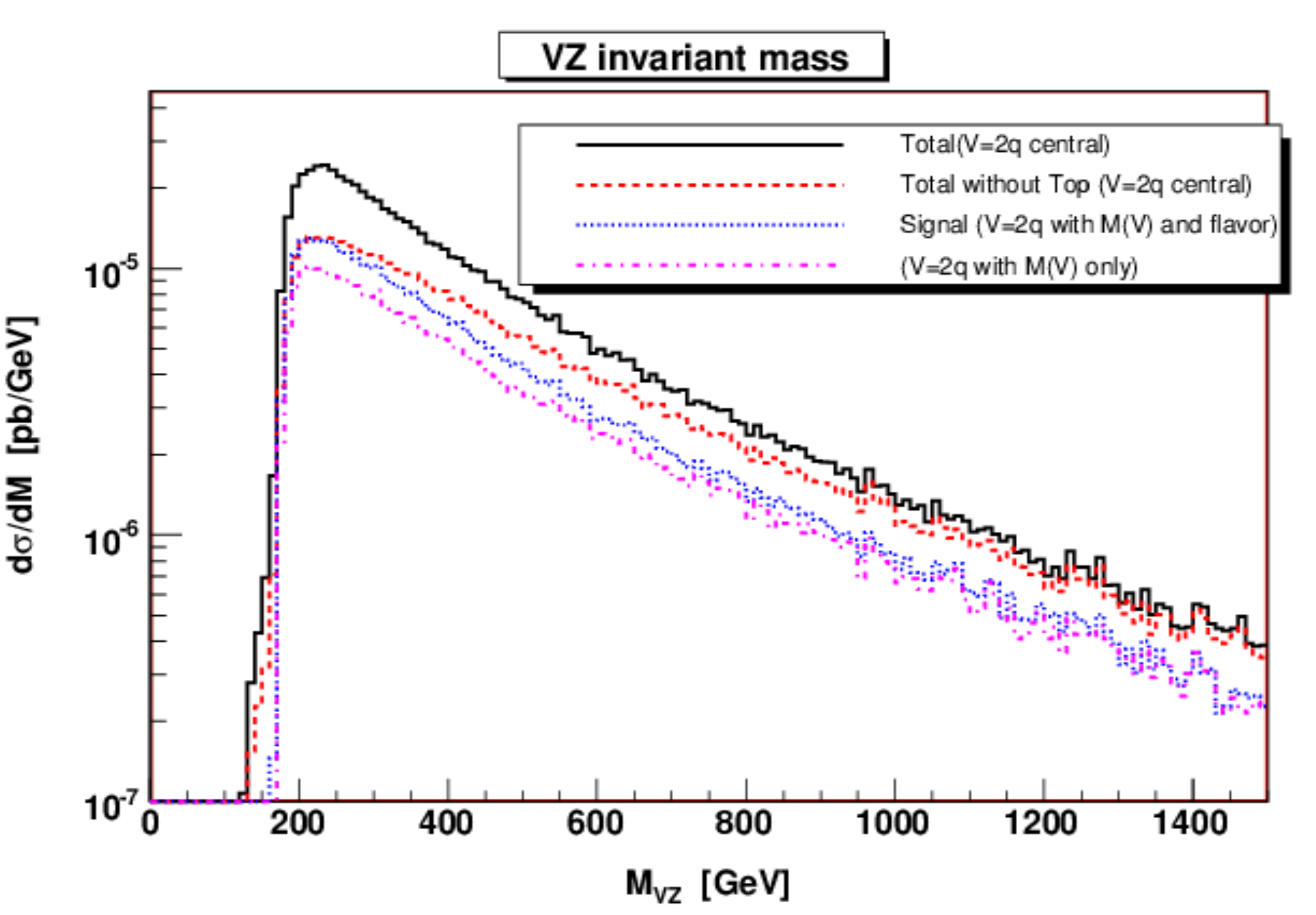,width=0.6\textwidth,height=6.5cm}
\caption{Invariant mass distribution of the lepton pair and the two jets from
boson decay for the no-Higgs case.
The solid(dashed) line is obtained identifying the two most central jets as the 
vector boson decay products before(after) top vetoing.
The dotted line is obtained requiring the correct flavour content for the jets
identified as decay products of both the vector boson and the top.
The dot--dashed lines is produced using solely invariant mass informations
to identify the vector boson and the top decay products.}
\label{fig:cuts_sensitivity}
\end{figure}

As already mentioned, the choice to start from pure EW results is motivated by 
the fact that no QCD diagram contributes to the scattering topology.
At this stage we are mainly concerned with isolating as much as possible the 
$VV$ scattering signature in the spirit of the discussion of the previous 
section.
Following this approach, the selection procedure we adopt in this section
makes use of flavour information and is not fully realistic.
Results are found anyway to be not too sensitive to the details of the selection
cuts, as shown in Fig.\ref{fig:cuts_sensitivity}. We compare different methods 
of reconstructing the $VZ$ invariant mass.
Results based on flavour selection and predictions obtained using cuts on the 
invariant mass only are shown.
The two distributions differ by about 20\% at small invariant masses but agree 
quite nicely above 800 GeV, showing that results based on flavour information 
are not seriously degraded when selection procedures closer to the actual 
experimental practice are adopted.

\begin{figure}[h!tb]
\centering
\epsfig{file=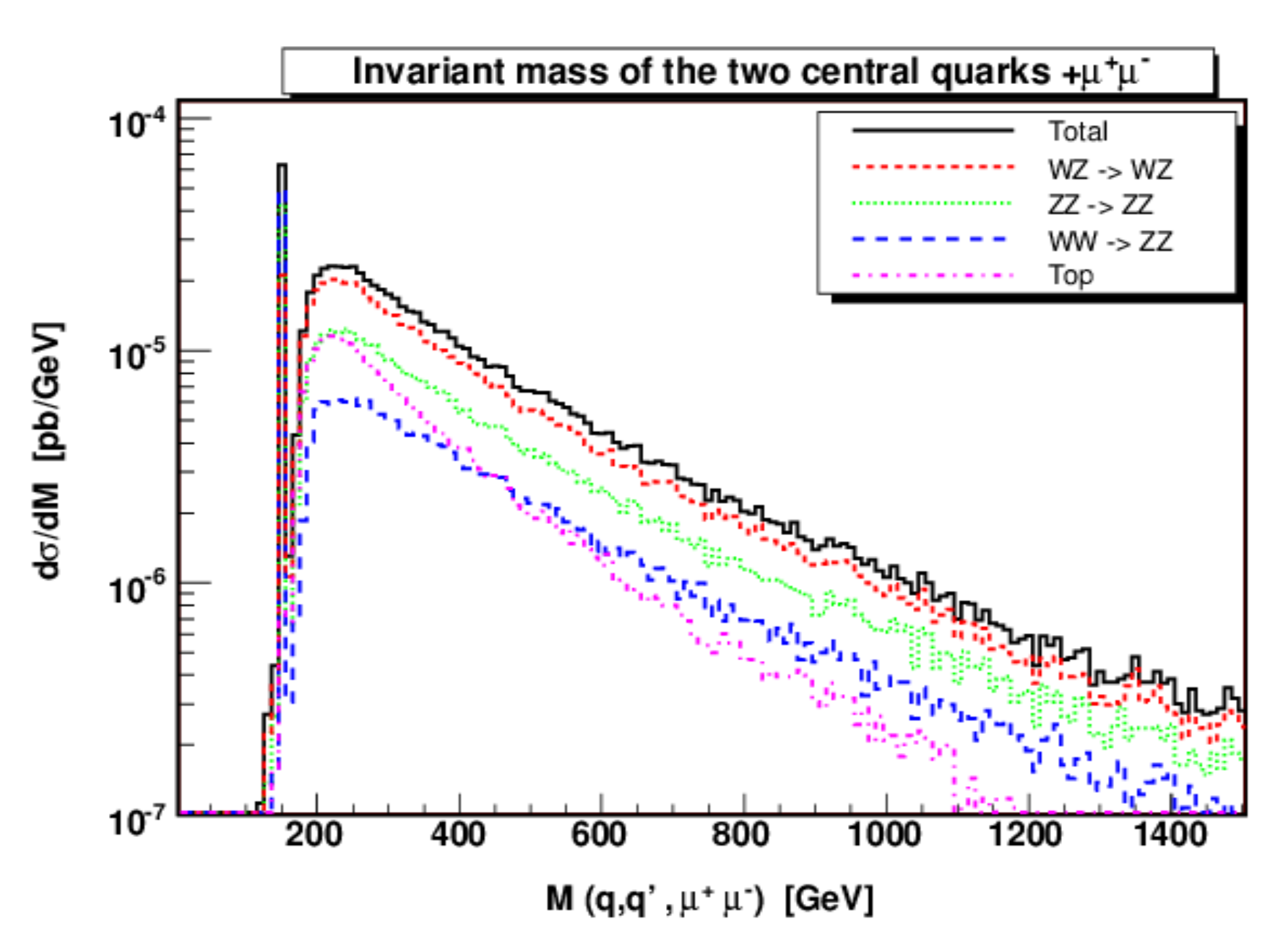,width=0.48\textwidth,height=5.7cm}
\epsfig{file=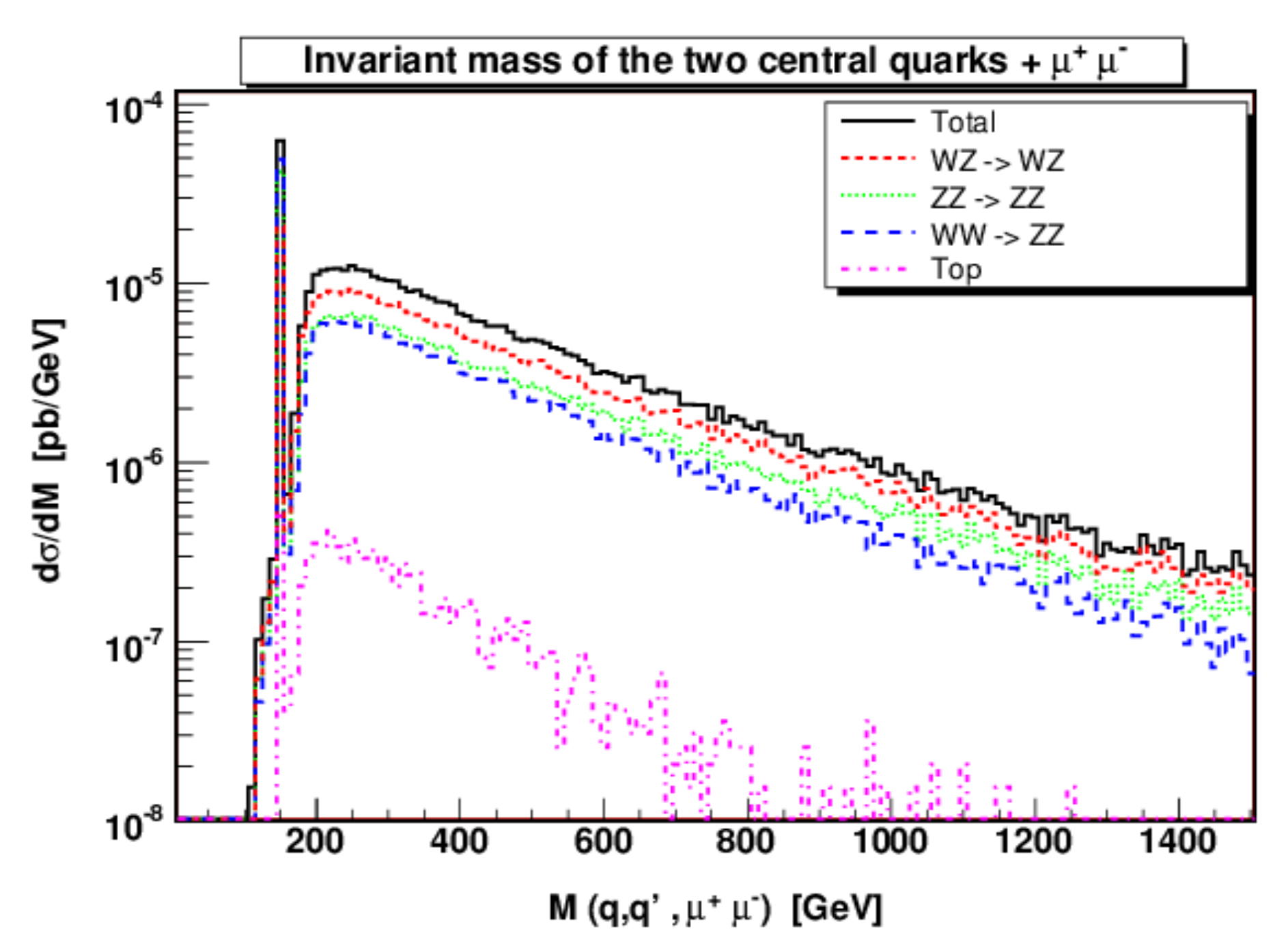,width=0.48\textwidth,height=5.7cm}
\caption{Invariant mass distribution of the two charged leptons and the two 
most central quarks, for different sets of processes.
The plot on the left includes the set of acceptance cuts described in 
the left part of Tab.\ref{tab:standard-cuts}.
In the plot on the right a further cut for vetoing top production is applied.}
\label{fig:scatt_subproc}
\end{figure}

It is not possible to separate the contribution of the various subprocesses 
($ZW \rightarrow ZW$, $WW \rightarrow ZZ$, $ZZ \rightarrow ZZ$, $t\bar{t}$, 
single $t$) to a given six-fermion final state.
Still, it is possible to isolate different subgroups of reactions including 
different subsets of diagrams. For example, only the subprocess 
$ZZ \rightarrow ZZ$ can be identified inside the scattering set of the diagrams 
describing the reaction $uu \rightarrow uus\bar{s}\mu^+\mu^-$, hence the latter 
will be unambigously assigned to the $ZZ \rightarrow ZZ$ subgroup.
If more than one subprocess contributes to a given reaction, the same will 
appear in different subgroups. This is the case, for instance, with 
$ud \rightarrow uds\bar{s}\mu^+\mu^-$ ($ZZ \rightarrow ZZ$, 
$WW \rightarrow ZZ$).
The plot on the left in Fig.\ref{fig:scatt_subproc} shows the invariant mass 
distribution of the two most central quarks and of the two leptons for all
reactions which contain the different subprocesses as well as the distribution
for the complete set of processes. A light Higgs with $M_H=150$ GeV has been 
assumed.
In order to comply with typical acceptance and trigger requirements, the 
standard acceptance cuts in Tab.\ref{tab:standard-cuts} have been applied.
It should be clear that the total cross section in Fig.\ref{fig:scatt_subproc}
is smaller than the sum of the cross sections for the various groups as a 
consequence of double counting.

\begin{table}[htb]
\begin{center}
\begin{tabular}{|c|c|} 
\cline{1-1} \cline{2-2} 
\textbf{Acceptance cuts} & \textbf{Selection cuts} \\
\cline{1-1} \cline{2-2}
$p_T(\ell^\pm)> 20$ GeV  & $\vert M(bW) - M_{top} \vert > 15$ GeV \\
\cline{1-1} \cline{2-2}
$\vert\eta(\ell^\pm)\vert<3$ & $\vert M(\ell^+\ell^-) - M_Z \vert < 10$ GeV \\
\cline{1-1} \cline{2-2}
$E(q)>30$ GeV & $\vert M(q_1q_2\mbox{ from W}) - M_W \vert < 10$ GeV \\
\cline{1-1} \cline{2-2}
$p_T(q)>20$ GeV &$\vert M(q_1q_2\mbox{ from Z}) - M_Z \vert < 10$ GeV  \\
\cline{1-1} \cline{2-2}
$\vert\eta(q)\vert<5$ & $\vert M(q_3q_4) - M_W \vert > 10$ GeV \\
\cline{1-1} \cline{2-2}
$M(\ell^+\ell^-) > 20$ GeV & $\vert M(q_3q_4) - M_Z \vert > 10$ GeV \\
\cline{1-1} \cline{2-2}
$M(qq) > 60$ GeV  &  \multicolumn{1}{|c}{} \\
\cline{1-1}
$\vert \Delta \eta(\mbox{tag-quarks}) \vert > 3.8$ & \multicolumn{1}{|c}{} \\
\cline{1-1}
\end{tabular}
\caption{Standard acceptance and selection cuts applied in results on the 
$\mu^+\mu^-$ channel.
$M(q_1q_2\mbox{ from W(Z)})$ is the invariant mass of the two quarks with the 
correct flavour content to be produced in a $W(Z)$ decay. If more than one 
combination of two quarks satisfies this requirement, the one closest to the 
corresponding central mass value is selected. $M(q_3q_4)$ denotes the invariant
mass of the two remaining quarks.}
\label{tab:standard-cuts}
\end{center}
\end{table}

For the selected type of final states, top background enters only through
single $t$ production. Top candidates are rejected requiring a $b\:(\bar{b})$ 
quark in the final state together with two other quarks of the right flavour 
combination to be associated to a $W^+\:(W^-)$ decay, with an invariant mass 
$M(bW)$ between 160 and 190 GeV. 
The plot on the right in Fig.\ref{fig:scatt_subproc} shows the results after 
top subtraction. The subprocess $ZW \rightarrow ZW$ provides the most relevant 
contribution.

\begin{figure}[htb]
\centering
\epsfig{file=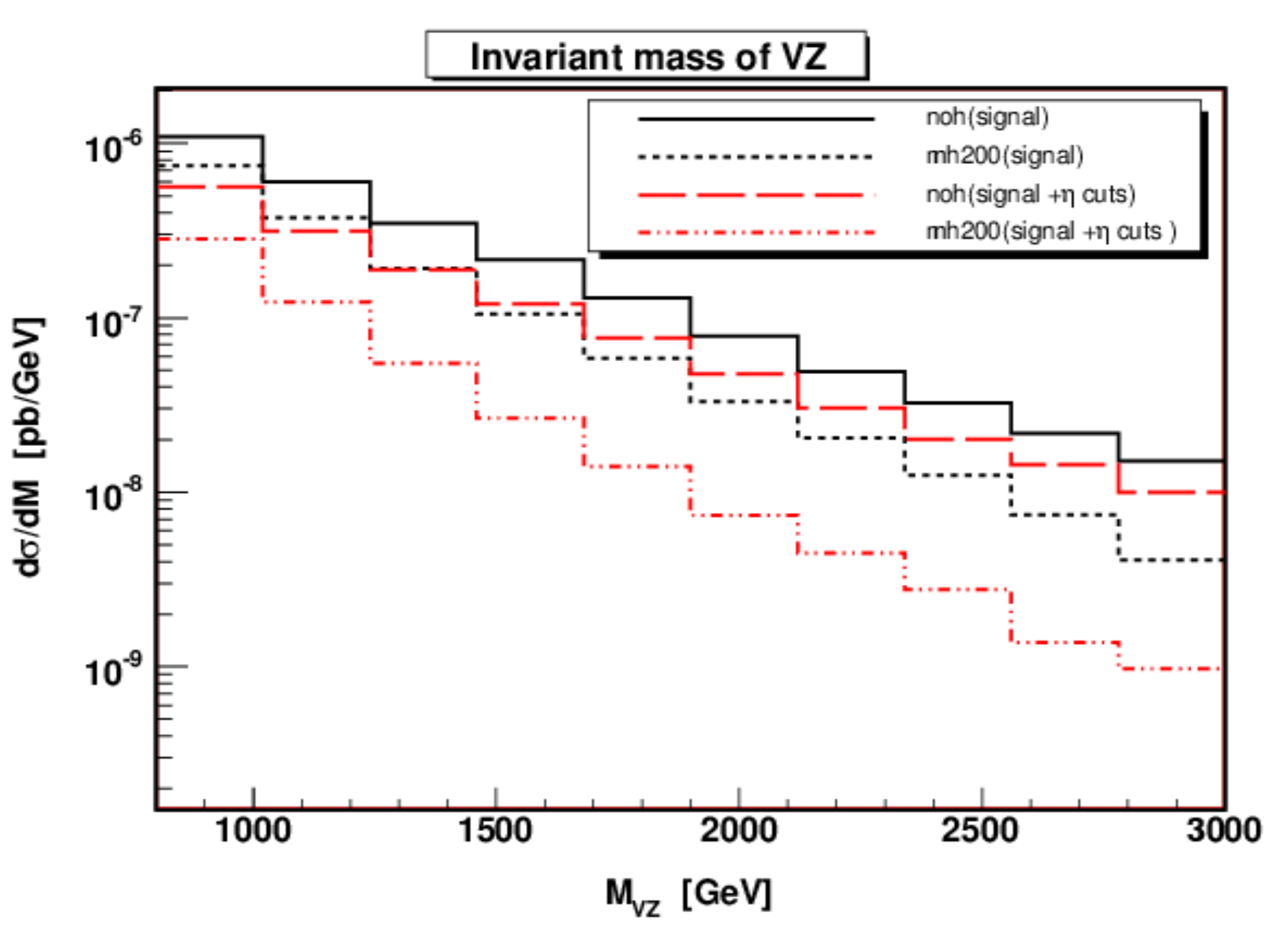,width=0.48\textwidth,height=5.7cm}
\epsfig{file=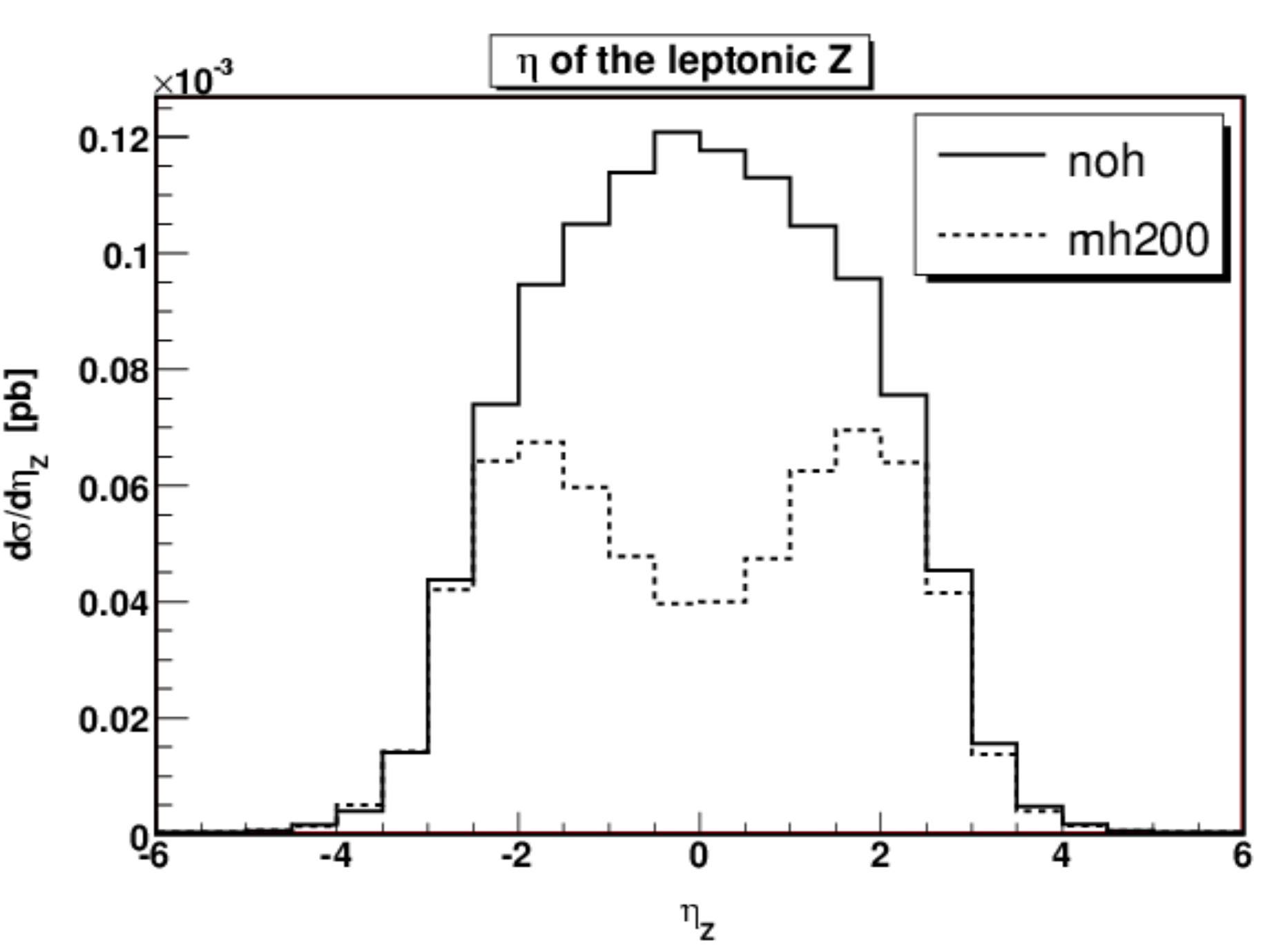,width=0.48\textwidth,height=5.7cm}
\caption{On the left: invariant mass distribution  for $M(VZ)>800$ GeV.
The full and long--dashed line refer to the no-Higgs case, the short--dashed 
and dot--dashed ones to $M_H=200$ GeV. 
All results satisfy the standard acceptance and selection cuts reported in 
Tab.\ref{tab:standard-cuts}.
For the long--dashed and dot--dashed histograms we have further required 
$\vert\eta(\ell^+\ell^-)\vert<\ $2 and $\vert\eta(qq\mbox{ from } V)\vert<\ $2.
On the right: pseudorapidity distribution of the $Z$ reconstructed from leptons.
}
\label{fig:high-MVV}
\end{figure}

We intend to investigate the differences between
the Standard Model and alternative scenarios in the high invariant mass region.
In this analysis we have considered the Standard Model without Higgs sector,
hereafter denoted no-Higgs scenario, which acts as an upper limit for new 
physics effects in models like those mentioned in Ref.\cite{Giudice:2007fh}.
Despite longitudinally polarized gauge bosons become strongly interacting in
absence of the Higgs particle, detecting any deviation still remains a 
challenge at LHC because the rise of the cross section is masked by the decrease
of parton luminosities at high energies.
An analysis of selection cuts capable to enhance the difference between the 
no-Higgs and light-Higgs case could provide some guidance in the search for 
signals of new physics. As shown in Fig.\ref{fig:high-MVV}, simple requirements 
of centrality of the final-state bosons achieve this result at $\ordEW$.

The procedure we adopted for separating jets imposes a minimum invariant mass 
of 60 GeV for each pair of quarks. An alternative approach consists in 
requiring a minimum $\Delta R$ separation among coloured particles and is 
discussed in Tab.\ref{tab:DeltaR}.
We find out that the no-Higgs case is more sensitive than light-Higgs to cuts 
based on $\Delta R$. Indeed, in the model without Higgs the outgoing vector 
bosons are favoured to be more central and they have a larger $p_T$. 
As a consequence of Lorentz boost they decay producing two quarks with small 
relative angle which are most likely to merge into one jet unless a sufficiently
small $\Delta R$ is selected.
The larger is the minimum separation required, the smaller the number of 
expected events. This is an indication that low $\Delta R$ cuts should be 
studied to evidentiate new physics.

On the other hand, alternative procedures for jet reconstruction based on the 
$K_\perp$ algorithm have been recently proposed\cite{Butterworth:2007ke}, which
may prove useful in connection with this kind of studies as they lead to
encouraging results in identifying hadronic decays of heavy bosons via a cut on 
the sub-jet separation scale.

\begin{table}[hbt]
\begin{center}
\begin{tabular}{|c|c c|c c|c|}
\hline
$M_{cut}$& \multicolumn{2}{c|}{\hspace*{6.5mm}no Higgs \hspace*{6.5mm}}
& \multicolumn{2}{c|}{$M_H=200$ GeV }& Ratio \\
\hline
800 GeV &\hspace*{2.mm}31 & (14,17) &\hspace*{2.mm}12 & (7,5) & 2.59\\
900 GeV &\hspace*{2.mm}25 & (12,13) &\hspace*{2.mm} 8 & (5,3) & 3.12\\
1.0 TeV &\hspace*{2.mm}19 &  (9,10) &\hspace*{2.mm} 6 & (4,2) & 3.16\\
\hline
\hline
\multicolumn{6}{|c|}{$\Delta R = 0.4$} \\
\hline
$M_{cut}$& \multicolumn{2}{c|}{\hspace*{6.5mm}no Higgs \hspace*{6.5mm}}
& \multicolumn{2}{c|}{$M_H=200$ GeV }& Ratio \\
\hline
800 GeV &\hspace*{2.mm}18 & (8,10)&\hspace*{2.mm}10 & (6,4) & 1.80\\
900 GeV &\hspace*{2.mm}12 &  (5,7)&\hspace*{2.mm} 6 & (4,2) & 2.00\\
1.0 TeV &\hspace*{2.mm} 8 &  (4,4)&\hspace*{2.mm} 4 & (2,2) & 2.00\\
\hline
\hline
\multicolumn{6}{|c|}{$\Delta R = 0.5$} \\
\hline
$M_{cut}$& \multicolumn{2}{c|}{\hspace*{6.5mm}no Higgs \hspace*{6.5mm}}
& \multicolumn{2}{c|}{$M_H=200$ GeV }& Ratio \\
\hline
800 GeV &\hspace*{2.mm}12 & (5,7)&\hspace*{2.mm}8 & (5,3) & 1.50\\
900 GeV &\hspace*{2.mm} 8 & (4,4)&\hspace*{2.mm}5 & (4,2)& 1.60\\
1.0 TeV &\hspace*{2.mm} 5 & (2,3)&\hspace*{2.mm}3 & (2,1)& 1.60\\
\hline
\end{tabular}
\end{center}
\caption{Number of events as a function of the minimum
invariant mass of the $ZV \rightarrow \mu^+\mu^-jj$ pair for 
$\mathcal{L}=100\;\mbox{fb}^{-1}$, having applied the cuts in 
Tab.\ref{tab:standard-cuts}\cite{Accomando:2006vj}.
All events satisfy $\vert\eta(\ell^+\ell^-)\vert < 2$ and 
$\vert\eta(qq\mbox{ from } V)\vert < 2$. 
In brackets we show the contribution of the ($ZW$, $ZZ$) final 
states.}
\label{tab:DeltaR}
\end{table}

\section{$\ordEW+\ordQCD$ results in the semileptonic $\mu \nu_\mu$ channel}
We will now investigate how much the sensitivity to the $VV$ scattering signal 
is affected by the QCD irreducible background, which is expected to considerably
dilute the differences between the light-Higgs and no-Higgs scenarios 
evidentiated in the previous section. 
To this purpose, we consider the full set of parton-level processes involved at 
$\ordEW+\ordQCD$,
\begin{eqnarray}
& q q \rightarrow q q q q \mu \bar{\nu}_{\mu} \nonumber  \hspace{0.8cm}
g g \rightarrow q q q q \mu \bar{\nu}_{\mu} \nonumber & \\
& g q \rightarrow g q q q \mu \bar{\nu}_{\mu} \nonumber  \hspace{0.8cm}
q q \rightarrow g g q q \mu \bar{\nu}_{\mu} \nonumber & ,
\end{eqnarray}
together with a selection procedure as close as possible to the actual 
experimental practice, without resorting to any flavour information other than
the one which a typical $b$-tagging algorithm is able to provide.
The $p_Z$ of the neutrino is approximately reconstructed by imposing the 
invariant mass of the two leptons equal to the $W$ boson nominal mass:
\begin{equation}
\label{eq:nu_reco_equation}
(p^{\mu}+p^{\nu})^2 = M_W^2 .
\end{equation}
This equation has two solutions,
\begin{equation}
\label{eq:nu_reco}
p_{z}^{\nu} = \frac{\alpha p_z^\mu \pm \sqrt{\alpha^2 p_z^{\mu 2} - 
 (E^{\mu 2} - p_z^{\mu 2})(E^{\mu 2} p_T^{\nu 2} - \alpha^2)}}
  {E^{\mu 2} - p_z^{\mu 2}}  \; ,
\end{equation}
where
\begin{equation}
\alpha = \frac{M_W^2}{2} + p_x^{\mu}p_x^{\nu} + p_y^{\mu}p_y^{\nu}  \; .
\end{equation}
The event is rejected if the discriminant of Eq.(\ref{eq:nu_reco}) is negative, 
otherwise both the solutions are required to pass the selection cuts.

All the results presented in this section have been obtained using the CTEQ5L 
PDF set and the QCD coupling constant running at the scale
\begin{equation}
\label{PDFScale2}
Q^2 = M_{top}^2 + p_T(top)^2,
\end{equation}
where $p_T(top)$ is the transverse momentum of the reconstructed top, for all
processes in which a $t$ or $\bar{t}$ can be produced. For all the other 
processes the scale has been evaluated as in Eq.(\ref{PDFScale1}).

\begin{table}[h!tb]
\begin{center}
\begin{tabular}{|c|c|c|}
\cline{1-3}
 \multicolumn{1}{|c|}{\scriptsize\boldmath$M_H = 200$ \textsl{\textbf{GeV}}} & $\sigma_{\scriptscriptstyle{{EW}}}$
 & $\sigma_{\scriptscriptstyle{EW + QCD}}$ \\
\hline
all events      & 0.89 pb  &  80.8 pb  \\
\hline
top events      & 0.52 pb  &  71.6 pb  \\
\hline
ratio top/all   & 0.58     &  0.89     \\
\hline
\end{tabular}
\end{center}
\caption{Contribution of $t\bar{t}$/single $t$ to the total cross section with 
standard acceptance cuts only (see the left part of Tab.\ref{tab:cuts_munuqcd}).
Comparison between results at $\ordEW$ (EW) and $\ordEW+\ordQCD$ (EW+QCD).
Interferences between the two perturbative orders are neglected.}
\label{tab:xsec_EW_vs_EW+QCD}
\end{table}

It should be clear from the results shown in Tab.\ref{tab:xsec_EW_vs_EW+QCD} 
that suppressing the top background is the primary objective to achieve. 
In this analysis we assume the possibility to tag $b$-jets in the central region
with 0.8 efficiency for $\vert \eta \vert < 1.5$, which allows to discard part 
of the events containing $b$ quarks in the final state.
We impose additional cuts against top on the invariant mass of triplets of type 
$\{jjj\}$ and $\{j\mu\nu\}$, where $j$ denotes any final-state quark or gluon. 
In order to isolate two vector boson production, kinematical cuts are
applied on the invariant mass of the two most central jets, which are associated
in our analysis to a $W$ or $Z$ decaying hadronically. 
The $VV$ fusion signature is further isolated by requiring a minimum 
$\Delta \eta$ separation between the two forward/backward jets. 

At this stage, however, any attempt to appreciate differences between
Higgs and no-Higgs scenarios at high invariant masses would still be vain.
This is essentially due to the fact that the contribution of the 
QCD diagrams depicted in Fig.\ref{fig:VBSbckgr_QCD}(c,d) is not substantially 
affected by the above-mentioned selection criteria.
Investigating the differences between the kinematics of $VV$ scattering and 
$VV+2\,jets$, we have identified additional cuts that serve our purpose.
As the background dominates in the phase space regions characterized by one 
vector boson and the forward/backward jet produced with small invariant mass, 
a viable method of taming $VV+2\,jets$ consists in applying cuts on the $p_T$ 
and $\eta$ of the $W$ reconstructed from leptons as well as on the invariant 
mass of the $W$ plus one of the two tag jets.
All details about the selection cuts applied are reported in 
Tab.\ref{tab:cuts_munuqcd}.

\begin{table}[h!tb]
\begin{center}
\begin{tabular}{|c|c|} 
\cline{1-1} \cline{2-2} 
\textbf{Acceptance cuts} & \textbf{Selection cuts} \\
\cline{1-1} \cline{2-2}
$p_T(\ell^\pm,j) > 10$ GeV & $b$-tagging for $\vert \eta \vert < 1.5$ 
 (80\% efficiency) \\
\cline{1-1} \cline{2-2}
$E(\ell^\pm,j) > 20$ GeV & $\vert M(jjj;j\ell^\pm\nu_{rec}) - M_{top} \vert >15$ GeV\\
\cline{1-1} \cline{2-2}
$\vert \eta(\ell^\pm) \vert < 3$ & $70 \mbox{GeV} < M(j_cj_c) < 100$ GeV \\
\cline{1-1} \cline{2-2}
$\vert \eta(j) \vert < 6.5$ & $M(j_fj_b)<70$ GeV ; $M(j_fj_b)>100$ GeV \\
\cline{1-1} \cline{2-2}
\multicolumn{1}{c|}{} & $M(jj) > 60$ GeV  \\
\cline{2-2} 
\multicolumn{1}{c|}{} & $\Delta\eta(j_fj_b) > 4$ \\
\cline{2-2}
 \multicolumn{1}{c|}{} & $ p_T(\ell^\pm\nu_{rec}) > 100$ GeV \\
\cline{2-2}
 \multicolumn{1}{c|}{} & $ \eta(\ell^\pm\nu_{rec}) < 2$ \\
\cline{2-2}
\multicolumn{1}{c|}{} & $M(j_{f/b}\ell^\pm\nu_{rec}) > 250$ GeV \\
\cline{2-2} 
\end{tabular}
\caption{List of kinematical cuts applied in all results on the 
$\mu\nu_{\mu}$ channel. $j$ denotes any final-state quark or gluon, while 
$\ell^\pm$ is the charged lepton. The subfixes $c$,$f$,$b$ mean 
\textit{central}, \textit{forward}, \textit{backward} respectively.
$\nu_{rec}$ is the neutrino reconstructed following the prescription of 
Eq.(\ref{eq:nu_reco_equation})}
\label{tab:cuts_munuqcd}
\end{center}
\end{table}

\begin{table}[h!tb]
\begin{center}
\begin{tabular}{|c|c|c|c|c|c|} 
\cline{1-5}
\multirow{2}{*}{{\footnotesize\boldmath$\ordEW$}} & \multicolumn{2}{|c|}{no Higgs} & \multicolumn{2}{|c|}{$M_H = 200$ GeV} & \multicolumn{1}{c}{}\\
\cline{2-6}
 & $\sigma$ & $\mathcal{L}$=$100\,\mbox{fb}^{-1}$
 & $\sigma$ & $\mathcal{L}$=$100\,\mbox{fb}^{-1}$ & ratio  \\
\hline
all events & 12.46 fb & 1246 $\pm$ 35 & 13.57 fb & 1357 $\pm$ 37 & 0.918 \\
\hline
$M_{cut}=0.8$ TeV & 3.19 fb & 319 $\pm$ 18 & 1.45 fb & 145 $\pm$ 12 & 2.200 \\
\hline
$M_{cut}=1.2$ TeV & 1.28 fb & 128 $\pm$ 11 & 0.41 fb & 41 $\pm$ 6 & 3.122 \\
\hline
$M_{cut}=1.6$ TeV & 0.60 fb & 60 $\pm$ 8 & 0.14 fb & 14 $\pm$ 4 & 4.286 \\
\hline
\end{tabular}
\caption{Integrated $\ordEW$ cross section for 
$M(j_cj_cl\nu)>M_{cut}$ and number of expected events after one year at high 
luminosity having applied the cuts listed in Tab.\ref{tab:cuts_munuqcd}.}
\label{tab:xsec_EW}
\end{center}
\end{table}

\begin{table}[h!tb]
\begin{center}
\begin{tabular}{|c|c|c|c|c|c|} 
\cline{1-5}
\multicolumn{1}{|c|}{\footnotesize\boldmath$\ordEW \;+$} & \multicolumn{2}{|c|}{no Higgs} & \multicolumn{2}{|c|}{$M_H = 200$ GeV} & \multicolumn{1}{c}{}\\
\cline{2-6}
\multicolumn{1}{|c|}{\footnotesize\boldmath$\ordQCD$} & $\sigma$ & $\mathcal{L}$=$100\,\mbox{ fb}^{-1}$
 & $\sigma$ & $\mathcal{L}$=$100\,\mbox{ fb}^{-1}$ & ratio  \\
\hline
all events & 40.70 fb & 4070 $\pm$ 64 & 40.73 fb & 4073 $\pm$ 64 & 0.999 \\
\hline
$M_{cut}=0.8$ TeV & 7.61 fb & 761 $\pm$ 28 & 5.14 fb & 514 $\pm$ 23 & 1.481 \\
\hline
$M_{cut}=1.2$ TeV & 2.53 fb & 253 $\pm$ 16 & 1.73 fb & 173 $\pm$ 13 & 1.462 \\
\hline
$M_{cut}=1.6$ TeV & 1.00 fb & 100 $\pm$ 10 & 0.55 fb & 55 $\pm$ 7 & 1.818 \\
\hline
\end{tabular}
\caption{Integrated $\ordEW+\ordQCD$ cross section for 
$M(j_cj_cl\nu)> M_{cut}$ and number of expected events after one year at high 
luminosity having applied the cuts listed in Tab.\ref{tab:cuts_munuqcd}.
Interferences between the two perturbative orders are neglected.}
\label{tab:xsec_EW+QCD}
\end{center}
\end{table}

\begin{figure}[h!tb]
\centering
\epsfig{file=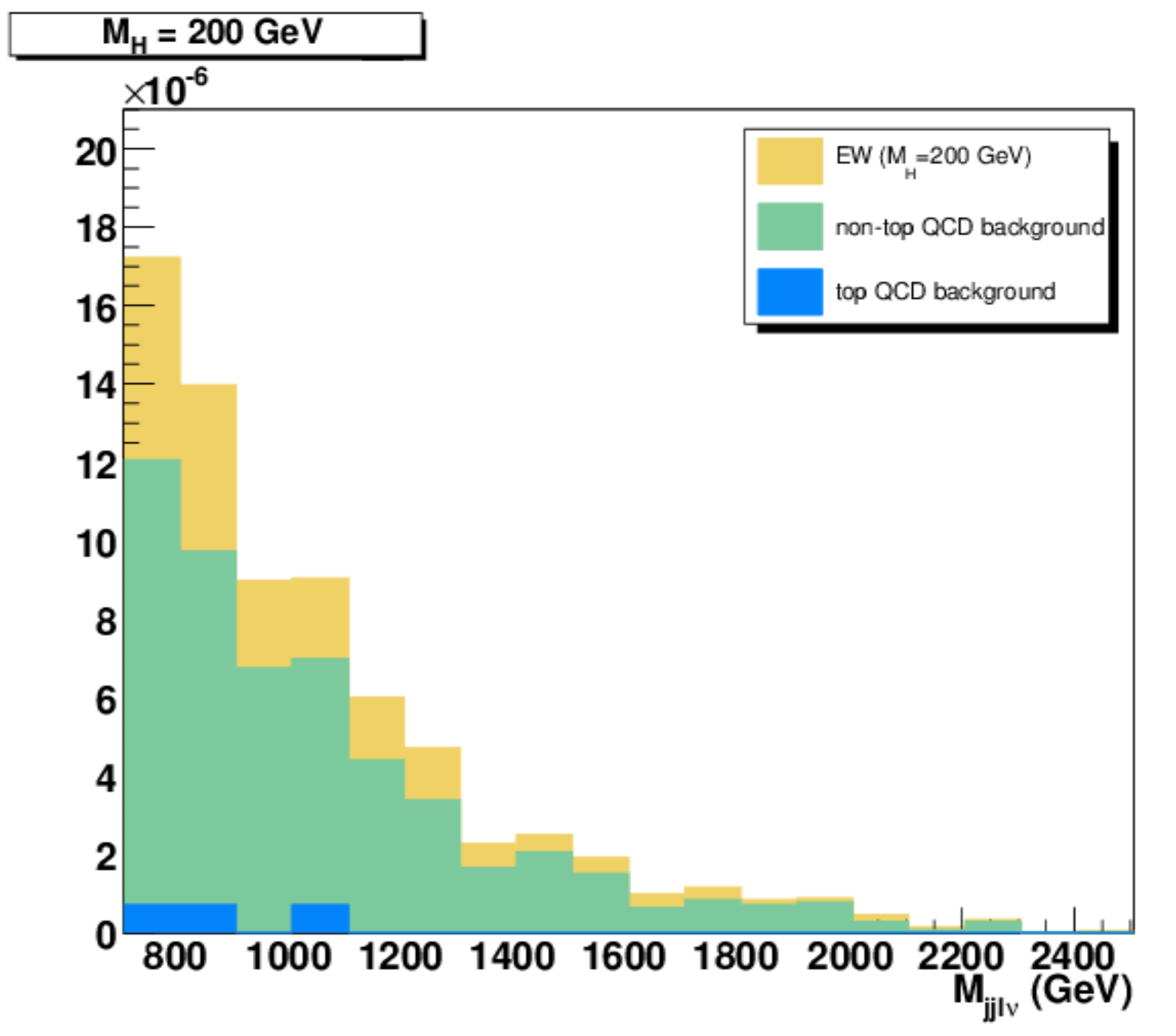,width=0.49\textwidth,height=6.4cm}
\epsfig{file=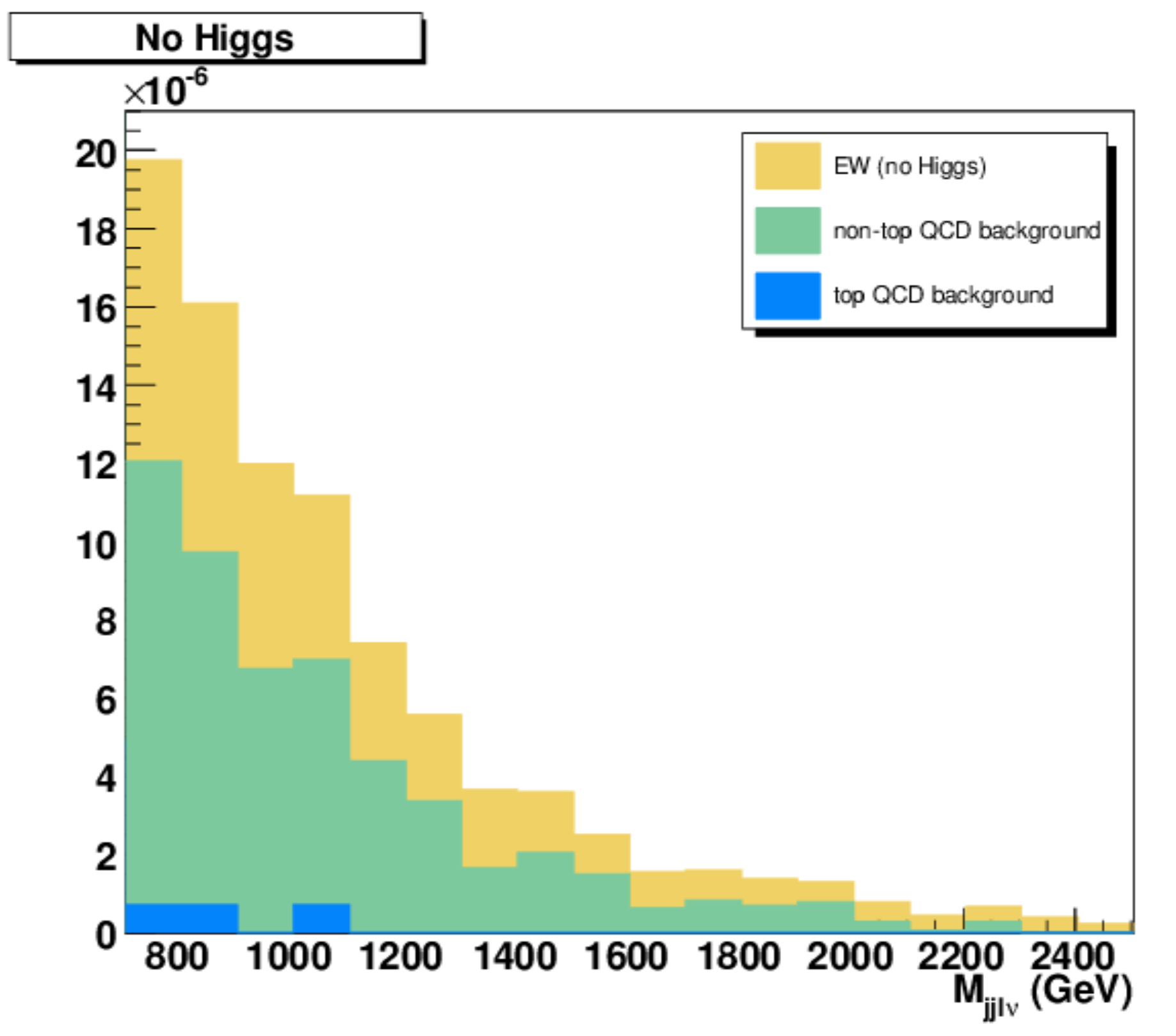,width=0.49\textwidth,height=6.4cm} \\
\vspace{0.6cm} 
\epsfig{file=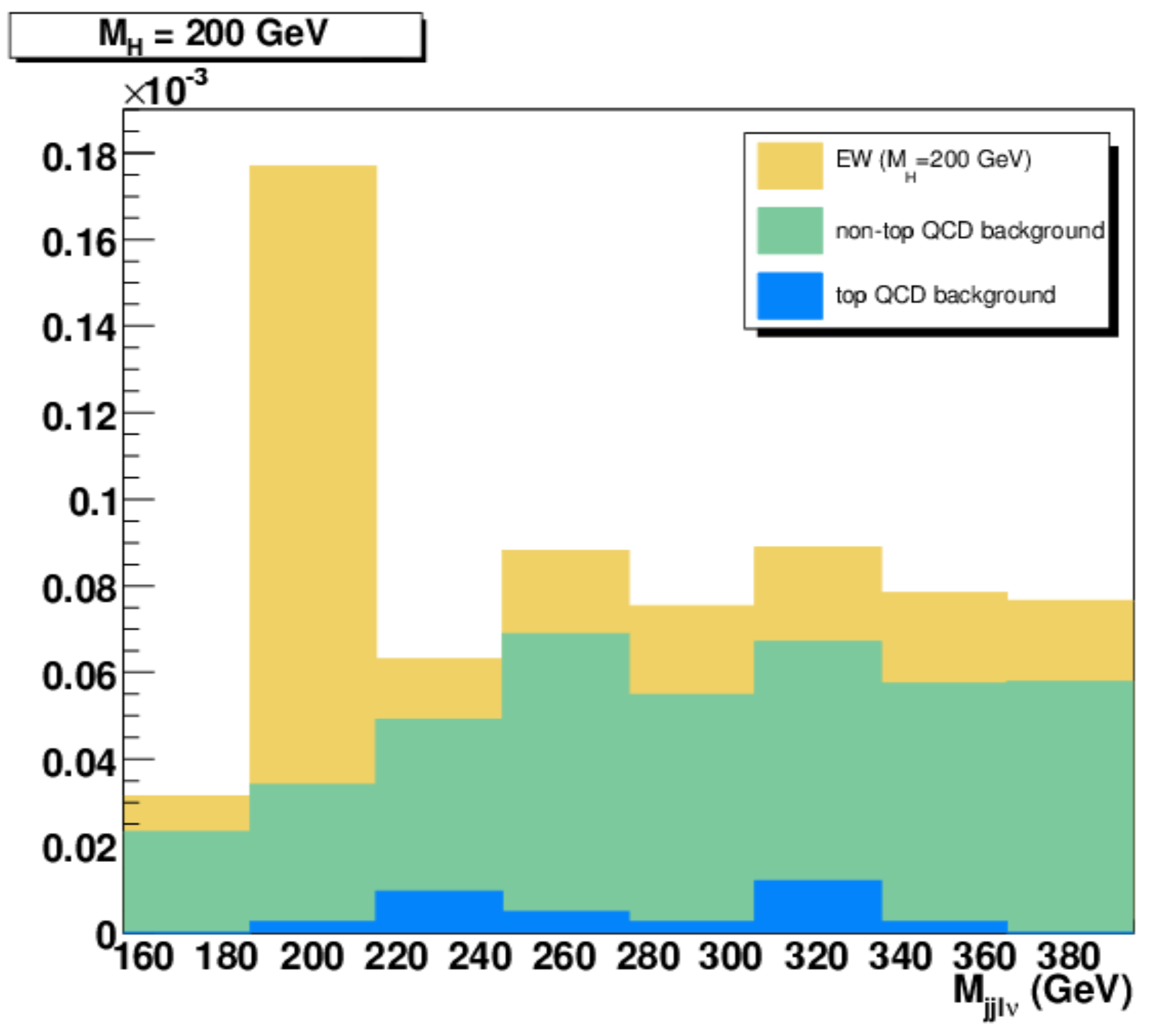,width=0.49\textwidth,height=6.4cm}
\epsfig{file=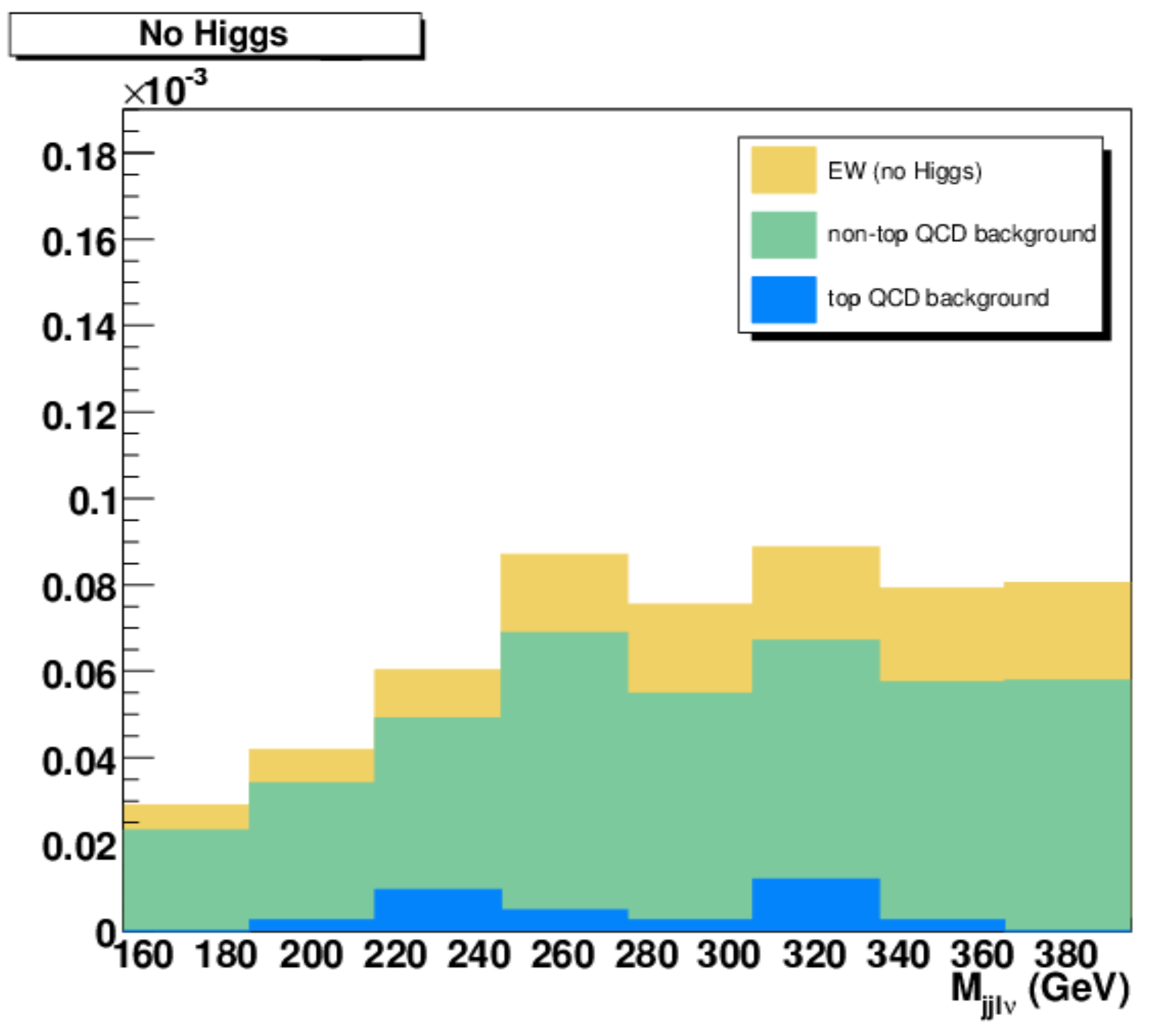,width=0.49\textwidth,height=6.4cm}
\caption{ Invariant mass distribution of the two leptons and the two most 
central jets in the Standard Model with a light Higgs (on the left) and in the 
no-Higgs scenario (on the right). The cuts applied are listed in 
Tab.\ref{tab:cuts_munuqcd}.
$\ordEW$ (EW) and $\ordQCD$ (QCD) contributions to the differential cross 
section have been isolated and are shown separately. The QCD contributions are
further split into \textit{top background} (in blue) and $VV+2\,jets$ 
(in green).}
\label{fig:munuqcd_compare_signal_bckg}
\end{figure}

\begin{figure}[h!tb]
\centering
\epsfig{file=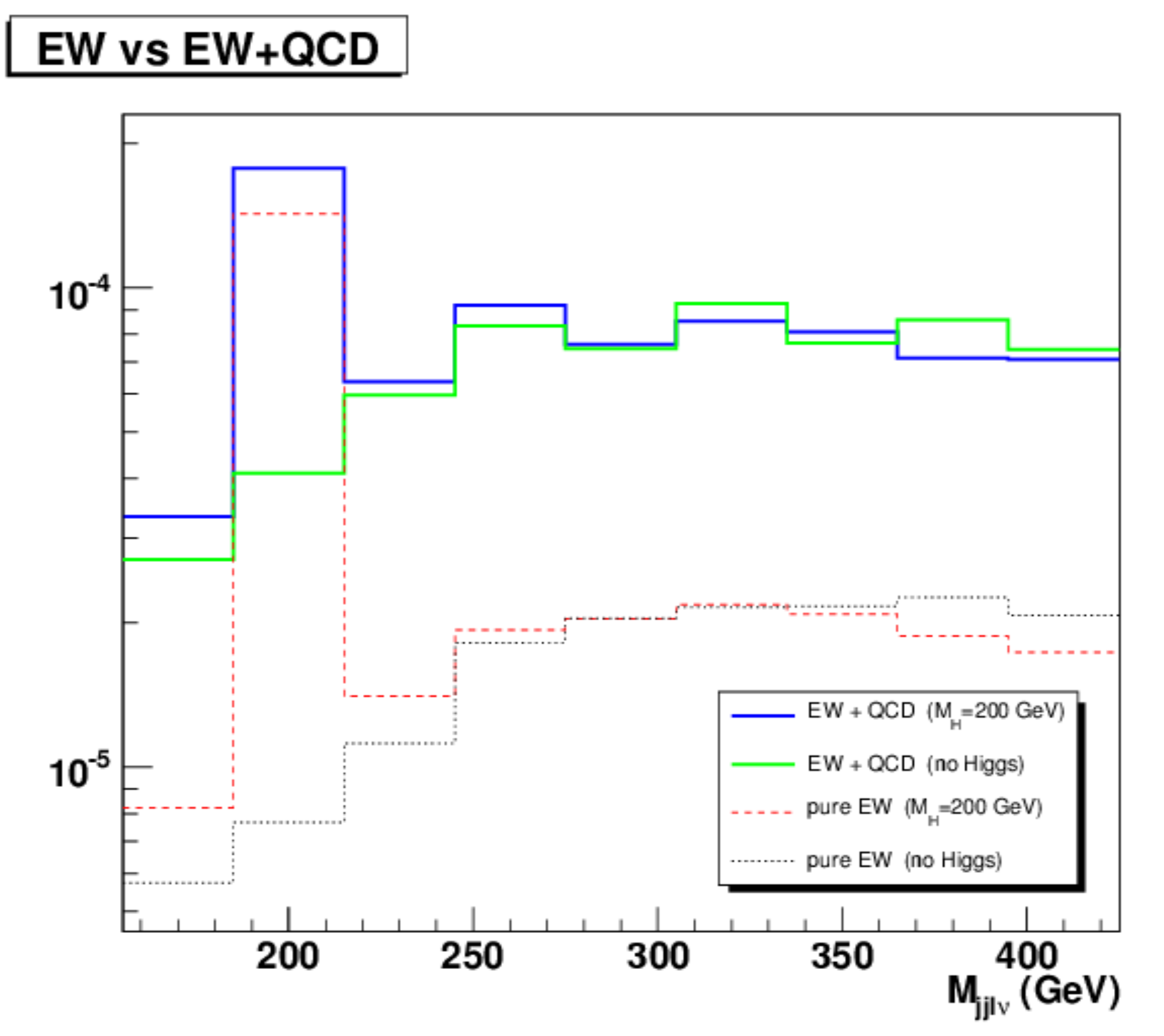,width=0.49\textwidth,height=6.3cm}
\epsfig{file=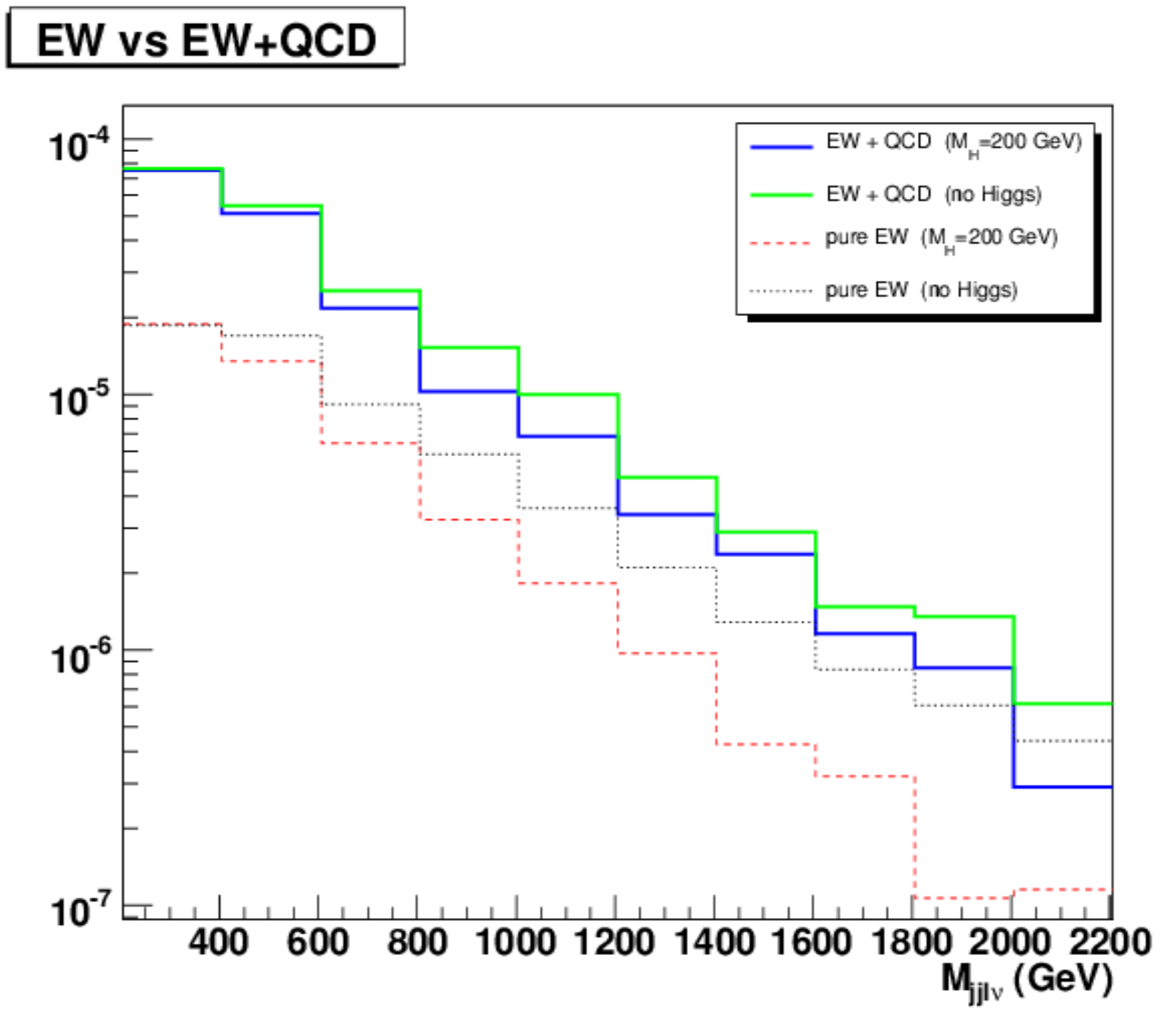,width=0.49\textwidth,height=6.3cm}
\caption{Invariant-mass distribution of the two leptons plus the two most
central jets in the pure EW ($\ordEW$) and EW+QCD ($\ordEW+\ordQCD$) case.
Comparison between light-Higgs ($M_H=200$ GeV) and no-Higgs results.
Interferences between the two different perturbative orders are neglected. 
The cuts applied are listed in Tab.\ref{tab:cuts_munuqcd}}
\label{fig:munuqcd_compare_EW_EW+QCD}
\end{figure}

Fig.\ref{fig:munuqcd_compare_signal_bckg},\ref{fig:munuqcd_compare_EW_EW+QCD} 
illustrate the final results of our analysis, showing that the top 
background is basically under control. $VV+2\,jets$ still provides a 
non-negligible contribution over the whole invariant mass spectrum, 
nevertheless differences between light-Higgs and no-Higgs can be appreciated.
In Tab.\ref{tab:xsec_EW},\ref{tab:xsec_EW+QCD} we show the integrated 
cross section at high energies as a function of the minimum invariant mass,
comparing results of the pure EW and EW+QCD cases.
Despite reducing the ratio between no-Higgs and light-Higgs cross sections, the
inclusion of QCD background seems not to compromise the possibility of finding 
signals of EWSB at LHC.
We find that about 500 events are expected above 800 GeV after one year of high
luminosity running ($\mathcal{L}=100\mbox{ fb}^{-1}$) in case of a Higgs boson 
with mass 200 GeV. The Higgsless model predicts about 250 more events in 
accordance with the enhancement of the $VV$ differential cross section at high 
energies. These numbers refer to the muon channel only, and are obviously
improved by summing up the muon and electron channels.
It should nevertheless be noticed that imposing a minimum $\Delta R$ separation
among coloured particles could degrade these preliminary results and requires 
further investigations.

\section{Conclusions}
The results presented in this note should be intended as the starting point of a
larger complete study which aims at providing an up-to-date and hopefully more 
reliable answer on the sensitivity of Boson-Boson Scattering as a probe of the 
symmetry breaking mechanism at LHC.
In spite of theoretical uncertaintes tipically estimated of $\mathcal{O}(10)$\%,
these results put a benchmark at parton level to the possibility of extracting
the $VV$ scattering signal from its irreducible background.
There are clear indications that, with a complete calculation and appropriate 
kinematical cuts, the Higgs and no-Higgs cases show appreciable differences
even after including the most relevant QCD contributions. 
We find in our analysis that the number of events expected in the Higgsless 
scenario is about 50\% larger than the one predicted by the Standard Model with 
a 200 GeV Higgs boson.
More refined selection criteria are under study in order to enhance further the 
Higgs signal as well as any possible deviation from the Standard Model 
expectations in the high invariant mass region.
As an evolution for the near future, the analysis will be extended to cover 
$V+4\,jets$ contributions so that a full estimate of the QCD background in the 
semileptonic channel will be available.

%\end{document}

\addtocounter{chapter}{1}

\providecommand{\tabularnewline}{\\}
\providecommand{\boldsymbol}[1]{\mbox{\boldmath $#1$}}

\def\beq{\begin{equation}}
\def\eeq{\end{equation}}
\def\beqn{\begin{eqnarray}}
\def\eeqn{\end{eqnarray}}
\def\ba{\begin{eqnarray}}
\def\ea{\end{eqnarray}}
\def\atp{\frac{\alpha_s(Q^2)}{2\pi}}
\def\afp{\frac{\alpha_s(Q^2)}{4\pi}}

%\begin{document}
%%%%%%%%%%%%%%%%%%%%%%%%%%%%%%%%%%%%%%%%%%%%%%%
% Toggle line numbering
% Won't work with the PRD revtex4 !
%\pagewiselinenumbers
% uncomment if you want doublespace
%\doublespace

%\begin{center}
\mchapter{Searching for Extra Neutral Interactions at the LHC}
{Roberta Armillis, Claudio Corian\`{o}, Alon E. Faraggi, 
Marco Guzzi, Nikos Irges}

%\begin{center}
\section*{Abstract}
%\end{center}

We present a brief overview of some aspects of the theory and 
phenomenology of models containing extra
neutral currents in abelian extensions of the Standard Model. 
We illustrate the mechanism of anomaly cancellation as a way to infer the 
charge assignments of the fermion spectrum and then briefly discuss 
a variant of this approach where an effective action is rendered anomaly-free by a 
mechanism involving an axion via a Wess-Zumino term. Measurements at the 
LHC on the Z resonance in leptoproduction will be able to exclude a class of these models for variations of the cross section at the level of $4 \%$, which is obtained at larger values of the anomalous coupling $(g_z \approx 1)$. The anomalous nature is unlikely to be resolved with an inclusive NNLO analysis.

%\end{abstract}
%\end{center}
%\newpage

%%%%%%%%%%%%%%%%%%%%%%%%%%%%
\section{Introduction}
%%%%%%%%%%%%%%%%%%%%%%%%%%%%
The search for extra abelian gauge interactions will surely be an important 
component of the experimental program at the LHC. In fact, extra neutral interactions, or ``extra $Z'$'', 
as they are commonly known, are predicted by several theoretical
constructions such as Grand Unified Theories, Superstring Theory,
and, more recently, by a class of models characterized by intersecting 
branes, just to mention a few (see \cite{Langacker}, \cite{e6zprime},\cite{Kiritsis}). 
One of the most relevant channels useful for the search of extra neutral
currents is lepton pair production, at an invariant mass of the
lepton pair not too large compared to the Z mass,
since the invariant mass distribution $d \sigma/d M^2$ is rapidly falling, so
to allow to gather enough statistics and separate the signal from the
Standard Model background. Being the mass of the extra gauge boson (or bosons)
undetermined, as is the coupling constant of the extra neutral current, a scan
of the entire high energy tail of the Drell-Yan distribution is required. The
width of these resonances is also undetermined
by the theoretical models and may change considerably depending on the
underlying assumptions of each theoretical construction. 
We will summarize below some of the main motivations of these searches discussing both anomaly-free models 
and a class of models where the anomaly is cancelled via the inclusion of  
a pseudoscalar, the Axi-Higgs \cite{CorianoIrgesKiritsis}, which is part 
of the Higgs sector. In models of this type the phenomenology of the extra $Z'$ shows some very distinctive 
features at the level of trilinear gauge interactions. 
A more detailed analysis of the topics addressed in this contribution can be found in 
\cite{npap2,CorianoIrgesMorelli,npap3,npap4,npap5,npap7,npap8}. 
The numerical study that we present is performed on the abelian extension 
of \cite{CarenaetAl}, whose analysis is extended here, in part, 
to next-to-next-to-leading order in QCD. 
Other recent analysis addressing a general family-blind $U(1)$ 
is in \cite{Appelquist} and briefly highlighted in the next section. 
A more detailed analysis of some of the issues addressed here 
are presented in \cite{npap5}.

\section{Non Anomalous $U(1)$'s}
The interaction of an extra neutral gauge boson $U(1)$ with the fermions of the
Standard Model requires a suitable definition of their charges with respect to
the new additional neutral interactions.
Some of the most useful and powerful constraints in fixing the charges of these
models is the
requirement that they are free of anomalies. We will bring
in a simple example to illustrate how the
cancellation works for the case of one extra gauge factor, $U(1)_z$, assuming a
chiral spectrum
that includes also a set of extra fermions $\nu_k$, with $k=1,2,...,n$
indicating a certain number of
right-handed neutrinos. For simplicity we will assume that the interaction is
family universal,
which means that the cancellations take place generation by generation. We
consider a model with the gauge symmetry $SU(3)\times SU(2) \times U(1)\times
U(1)_z$. Breaking the gauge symmetry down to
$SU(3)\times U(1)_{em}$ requires an extra scalar sector compared to the SM.
This can be achieved in different ways, for instance by including, beside the 
Higgs doublet, a $SU(2)_W$ singlet $\phi$, whose
vacuum expectation value is $v_\phi$. In a more general framework, the
constraints on the
interactions of the $U(1)_z$ gauge boson with the fermions of the SM are
relaxed if two Higgs
doublets, $H_1$ and $H_2$, together with an extra scalar $\phi$,
are introduced (see \cite{CarenaetAl}).

\section{Phenomenological Models: An Example}

Following \cite{Appelquist} we consider  three generations of quarks, $q_L^i,
u_R^i, d_R^i$, and leptons, $l_L^i, e_R^i$, $i=1,2,3$, and a number $n$
of right-handed neutrinos, $\nu_R^k$, $k=1, ... , n$, which are singlets
under $SU(3)_C \times SU(2)_W$. We label the $U(1)_z$ charges as $z_q, z_u, z_d, z_l, z_e$,
for the standard model fermions,
and $z_k$ are the charges of the right-handed neutrinos. The Higgs sector 
of the Standard Model is enlarged with an extra singlet $\phi$, as shown in Table 1. 
$H$ is the electroweak Higgs.
The cancellation of the anomalous interactions, for instance those involving 
the $[SU(2)_W]^2 U(1)_z$ and $[SU(3)_C]^2 U(1)_z$ gauge currents requires that
\ba
&&z_l = -3z_q\nonumber\\
&&z_d = 2 z_q - z_u .
\label{first}
\ea
Similarly, the $[U(1)_Y]^2 U(1)_z$ anomaly cancellation then implies that
\ba
z_e = -2z_q - z_u.
\label{second}
\ea
Eqs.~(\ref{first}) and (\ref{second}) together lead
to the conclusion that only two independent real parameters,
$z_q$ and $z_u$,
describe the allowed $U(1)_z$ charges of the quarks
and $U(1)_Y$-charged leptons. Equivalently,
the $U(1)_z$ charges may be expressed as a linear combination
of $Y$ and $B-L$: $( z_{u}- z_{q} ) Y + ( 4z_{q} - z_{u} ) ( B - L )$.

\begin{table}[t]
\centering
 \renewcommand{\arraystretch}{1.5}
\begin{tabular}{|c| |c|c|c|c|}\hline
& $SU(3)_C$ & $SU(2)_W$ & $U(1)_Y$ & $U(1)_z$ \\ \hline\hline
$q_L^i$ &  3 & 2 & $1/3$ & $z_q$\\ \hline
$u_R^i$ &  3 & 1 & $4/3$ & $z_u$\\ \hline
$d_R^i$ &  3 & 1 & $-2/3$ & $2z_q - z_u$\\ \hline
$l_L^i$ &  1 & 2 & $-1$ & $-3z_q$\\ \hline
$e_R^i$ &  1 & 1 & $-2$ & $-2z_q - z_u$\\ \hline
$\nu_R^k$ , $k=1, ... , n$ &  1 & 1 & 0 & $z_k$\\ \hline\hline
$H$ &  1 & 2 & $+1$ & $-z_q + z_u$\\ \hline
$\varphi$ &  1 & 1 & $0$ & $1$\\ \hline
\end{tabular}
\caption{Charge assignement for the Appelquist model}
\label{table1}
\end{table}

Additional restrictions on the $U(1)_z$ charges are imposed by the
mixed gravitational-$U(1)_z$ and $[U(1)_z]^3$
anomaly cancellation conditions
\ba
&& \frac{1}{3} \sum_{k=1}^n z_k = -4 z_q + z_u ~, \label{sum} \\
&& \left( \sum_{k=1}^n z_k \right)^{\! 3} = 9 \sum_{k=1}^n z_k^3 ~.
\ea

For $n=1$ or $2$, the charge assignments compatible with 
these equations constrain $U(1)_z $ to be proportional to the 
hypercharge, giving a solution termed ``sequential'', while more general solutions can be found already 
for $n=3$, which can be expressed in terms of a free parameter. 
The analysis of the possible neutrino mass terms compatible with a given 
charge assignment can also be used as a way to select the most 
interesting solutions of these equations. For instance, in the case 
$n=3$ both Dirac and Majorana mass terms are possible in this minimal 
model (see \cite{Appelquist} for more details). It is understood that a modification of the Higgs sector renders the study more involved \cite{CarenaetAl}.

%%%%%%%%%%%%%%%%%%%%%%%%%%%%
\section{Heterotic--string inspired $Z'$}
%%%%%%%%%%%%%%%%%%%%%%%%%%%%
The heterotic--string gives rise to effective field theories
that descend from the $E_8\times E_8$ or $SO(32)$ groups of the ten
dimensional theories. The first case gives rise to additional $Z^\prime$s
that arise in the $SO(10)$ and $E_6$ extensions of the Standard Model.
A basis for the extra $Z^\prime$ arising in these models is
formed by the two groups $U(1)_\chi$ and $U(1)_\psi$ via the decomposition
$E_6\rightarrow SO(10)\times U(1)_\psi$ and $SO(10)\rightarrow SU(5)\times
U(1)_\chi$ \cite{e6zprime}.
Additional, flavor non--universal $U(1)$'s, may arise in
heterotic $E_8\times E_8$ string models from the $U(1)$ currents in the Cartan
subalgebra of the four dimensional gauge group, that are external to $E_6$.
Non--universal $Z^\prime$s typically must be beyond the LHC reach, to
avoid conflict with Flavor Changing Neutral Currents (FCNC) constraints. 
Recently \cite{npap2} a novel $Z^\prime$ in quasi--realistic
string models that do not descend from the $E_8\times E_8$ has been identified.
Under the new $U(1)$
symmetry left--handed components and right--handed components in
the 16 spinorial $SO(10)$ representation, of each Standard Model generation,
have charge 
$-1/2$ and $+1/2$, respectively. As a consequence, the extra $U(1)$ is
family universal and anomaly free. It arises in left-right symmetric
string models, in which the $SO(10)$ symmetry is broken directly 
at the string level to $SU(3)\times U(1)_{B-L}\times SU(2)_L\times 
SU(2)_R\times U(1)_{Z^\prime}\times U(1)^n\times {\rm hidden}$
\cite{cfs}. The $U(1)^n$ are flavor dependent $U(1)$s that are
broken near the string scale. The Standard Model matter states are
neutral under the hidden sector gauge group, which in these
string models is typically a rank eight group. It is important to 
note that the fact that the spectrum is derived from a string vacuum 
that satisfies the modular invariance constraints, establishes
that the model is free from gauge and gravitational anomalies.
The pattern of $U(1)_{Z^\prime}$
charges in the quasi--realistic string models of ref. \cite{cfs}
does not arise in related string models in which the $SO(10)$ symmetry
is broken to a different subgroup.

The important
function of this $Z^\prime$ is that it forbids dimension
four, five and six proton decay mediating operators \cite{lowscalesprime}. 
The extra $U(1)$ is anomaly free and family universal. It allows
the fermions Yukawa couplings to the Higgs field and the 
generation of small neutrino masses via a seesaw mechanism.
The existence of an 
extra $Z^\prime$ at low energies is motivated by proton longevity,
and the suppression of the proton decay mediating operators.
String models contain several $U(1)$ symmetries that suppress
the proton decay mediating operators. However, these
are typically non--family universal. They constrain the fermion
mass terms and hence must be broken at a high scale. 
Thus, the existence of a $U(1)$ symmetry that can remain unbroken
down to low energies is highly nontrivial. The $U(1)$ symmetry
in ref. \cite{cfs, npap2} satisfies all of these requirements.
Furthermore, as the generation of small neutrino masses in the
string models arises from the breaking of the $B-L$ current,
the extra $U(1)$ allows lepton number violating terms, but forbids
the baryon number violating terms. Hence, it predicts that
$R$--parity is violated and its phenomenological implications
for SUSY collider searches differ substantially from models
in which $R$--parity is preserved. 

\section{Anomalous $U(1)'s$:
Cancelling the anomalies via higher dimensional operators }
The simplest effective theory that allows a cancellation of the gauge
anomalies includes a Wess-Zumino term with a shifting axion. The role of the axion 
has been a matter of debate in the past \cite{Preskill}, since with a suitable 
gauge choice one can set the axion 
to vanish. Undoubtly, the axion can also be interpreted as the phase of an additional 
(second) Higgs when either large Yukawa couplings or a large vev of this additional Higgs is responsible for the decoupling of one or more chiral fermions \cite{CorianoIrgesMorelli}. According to this picture, an effective anomalous theory is then the result of the partial 
decoupling of part of the fermion spectrum, leaving the left-over fermions in a reducible representation.   
The induced Wess-Zumino term is of the
form $b/M_1 F \wedge F$, a dimension-5 operator, where $b$ is an axion and $F$ is
the field
strength of a gauge interaction. The $1/M_1$ suppression of the axion-gauge interaction 
is directly related to the vev of the 
Higgs and $b$ is its pseudoscalar phase \cite{CorianoIrgesMorelli}. The scale $M_1$ is completely 
unrelated to the mechanism held responsible for the generation of 
a mass for the axion and remains a free parameter (see \cite{Roncadelli}). 
The mechanism of partial decoupling is a generic feature of effective anomalous theories 
that brings into the effective action a gauged axion. 

In general, models based on intersecting branes predict similar 
structures, though the axion of the 
effective theory is not a consequence of partial decoupling. The Standard Model, in this case, is enlarged with several extra anomalous 
$U(1)$'s, one combination of which is anomaly free and is identified with the 
hypercharge, while the remaining U(1)'s are accompanied by axions. These models predict a single physical axion (the Axi-Higgs) \cite{CorianoIrgesKiritsis} and contain generalized 
Chern-Simons terms in the effective action \cite{CorianoIrgesKiritsis}, \cite{Pascal}, 
\cite{CorianoIrgesMorelli}. These terms are necessary in order to render the abelian interactions anomaly-free. On the contrary of non-anomalous $U(1)$ models, which have similar 
properties and difficult to discern at the LHC, anomalous U(1)'s show some interesting features 
both in Drell-Yan at NNLO \cite{npap8}, due to non-cancelling anomalous contribution 
and to the possibility of an axion exchange. We refer to \cite{CorianoIrgesMorelli,npap3} for 
a discussion of the detection of modified trilinear interaction in the neutral sector 
( $Z \,\gamma\, \gamma$ vertex).

\section{The detection of extra $Z$-primes at the LHC} 

We quantify below the rates for the invariant mass distributions in the case of the model of 
\cite{CarenaetAl}. The approach follows closely \cite{CCG,npap1,npap6}. 
Our convention for the couplings of the ${\cal Z}=Z,Z^{\prime}$ to the quarks and leptons are written below
\ba
&& {g_V}^{Z,j}=2 \left[c_w^2 (T_3^{L,j}+T_3^{R,j})-s_w^2(\frac{\hat{Y}^{j}_L}{2}+\frac{\hat{Y}^{j}_R}{2})
+\varepsilon \frac{g_z}{g} c_w (\frac{\hat{z}_{L,j}}{2}+\frac{\hat{z}_{R,j}}{2})\right]
\nonumber\\
&& {g_A}^{Z,j}=2\left[c_w^2 (T_3^{R,j}-T_3^{L,j})
-s_w^2(\frac{\hat{Y}^{j}_R}{2}-\frac{\hat{Y}^{j}_L}{2})
+\varepsilon \frac{g_z}{g} c_w (\frac{\hat{z}_{R,j}}{2}-\frac{\hat{z}_{L,j}}{2})\right]
\nonumber\\
&& {g_V}^{Z^{\prime},j}=
2\left[ -\varepsilon c_w^2 (T_3^{L,j}+T_3^{R,j})
+\varepsilon s_w^2(\frac{\hat{Y}^{j}_L}{2}+\frac{\hat{Y}^{j}_R}{2})
+ \frac{g_z}{g}c_w(\frac{\hat{z}_{L,j}}{2}+\frac{\hat{z}_{R,j}}{2})\right]
\nonumber\\
&& {g_A}^{Z^{\prime},j}= 2\left[ -\varepsilon c_w^2 (T_3^{R,j}-T_3^{L,j})
+\varepsilon s_w^2(\frac{\hat{Y}^{j}_R}{2}-\frac{\hat{Y}^{j}_L}{2})
+ \frac{g_z}{g}c_w(\frac{\hat{z}_{R,j}}{2}-\frac{\hat{z}_{L,j}}{2})\right]
\nonumber\\
\ea
where $j$ is an index that runs over the quarks and the lepton, 
and $\sin\theta_W=s_w,\cos\theta_W=c_w$ for brevity. The mixing 
in the neutral sector appears through the parameter $\varepsilon$, defined as 
\ba
&&\varepsilon=\frac{\delta M^2_{Z Z^{\prime}}}{M^2_{Z^{\prime}}-M^2_{Z}}\nonumber\\
&&M_Z^2=\frac{g^2}{4 \cos^2\theta_W}(v_{H_1}^2+v_{H_2}^2)\left[1+O(\varepsilon^2)\right]
\nonumber\\
&&M_{Z^{\prime}}^2=\frac{g_z^2}{4}(z_{H_1}^2 v_{H_1}^2+z_{H_2}^2v_{H_2}^2+z_{\phi}^2 v_{\phi}^2)\left[1+O(\varepsilon^2)\right]
\nonumber\\
&&\delta M^2_{Z Z^{\prime}}=-\frac{g g_z}{4\cos\theta_W}(z_{H_1}^2 v_{H_1}^2+z_{H_2}^2v_{H_2}^2),\,
\ea
where we have chosen \cite{CarenaetAl} $z_{H_1}=z_{H_2}=0$, $v_{H_2}=246$ GeV and $\tan{\beta}=1$ and have defined $g={e}/{\sin\theta_W}, g_Y={e}/{\cos\theta_W}$. Precision studies at LEP 
constrain $\varepsilon$ to be smaller than $10^{-3}$.

The decay rates into leptons for the $Z$ and the $Z^{\prime}$ are
universal and are given by
\begin{footnotesize}
\ba
&&\Gamma({\cal Z}\rightarrow l\bar{l})=\frac{g^2}{192\pi c_w^2}
M_{{\cal Z}}\left[(g_{V}^{{\cal Z},l})^2+(g_{A}^{{\cal Z},l})^2\right]=
\frac{\alpha_{em}}{48 s_w^2 c_w^2}M_{{\cal Z}}\left[(g_{V}^{{\cal Z},l})^2+
(g_{A}^{{\cal Z},l})^2\right]\,,
\nonumber\\
&&\Gamma({\cal Z}\rightarrow \psi_i\bar{\psi_i})=\frac{N_c\alpha_{em}}{48
s_w^2 c_w^2}
M_{{\cal Z}}\left[(g_{V}^{{\cal Z},\psi_i})^2+(g_{A}^{{\cal
Z},\psi_i})^2\right]
\times\nonumber\\
&&\hspace{3cm}
\left[1+ \frac{\alpha_s(M_{{\cal Z}})}{\pi}
+1.409\frac{\alpha_s^2(M_{{\cal
Z}})}{\pi^2}-12.77\frac{\alpha_s^3(M_{\cal Z})}{\pi^3}\right],\,
\ea
\end{footnotesize}
where $i=u,d,c,s$ and ${\cal Z}=Z,Z^{\prime}$.
For the $Z^{\prime}$ and $Z$ decays into heavy quarks we 
have used (see Ref.\cite{Kuhn:1985ps,Kniehl:1989qu})
\begin{footnotesize}
\ba
&&\Gamma({\cal Z}\rightarrow b\bar{b})=\frac{N_c\alpha_{em}}{48 s_w^2 c_w^2}
M_{{\cal Z}}
%\times
%\nonumber\\
%&&
\left\{(g_{V}^{{\cal Z},b})^2\left[ c_0 +c_1 \frac{\alpha_s(M_{\cal Z})}{\pi}+\dots\right]
+(g_{A}^{{\cal Z},b})^2\left[ d_0 + d_1 \frac{\alpha_s(M_{\cal Z})}{\pi}+\dots\right]
\right\},
\nonumber\\
&&\Gamma({\cal Z}\rightarrow t\bar{t})=\frac{N_c\alpha_{em}}{48 s_w^2 c_w^2}
M_{{\cal Z}}\beta
\left\{(g_{V}^{{\cal Z},t})^2 \frac{(3-\beta^2)}{2} A_1 
+(g_{A}^{{\cal Z},t})^2 A_2 \right\},
\ea
\end{footnotesize}
where the coefficients $c_0,c_1,d_0,d_1$ are defined in Eq.~(4.1) of 
\cite{Kniehl:1989qu} and where as in \cite{Kuhn:1985ps} we have defined  
\begin{footnotesize}
\ba
&&\beta=\sqrt{1 - 4 \frac{m_t^2}{M_{{\cal Z}}^{2}} },
\hspace{1cm}
A_1=\left\{1+\frac{4}{3}\alpha_s(M_{\cal Z}) \left[\frac{\pi}{2\beta}
-\frac{3+\beta}{4} \left(\frac{\pi}{2}-\frac{3}{4\pi} \right)  \right] \right\},
\nonumber\\
&&A_2=\left\{1+\frac{4}{3}\alpha_s(M_{\cal Z}) \left[\frac{\pi}{2\beta} 
-\left(\frac{19}{10} -\frac{22\beta}{5}+\frac{7\beta^2}{2}\right)\left(\frac{\pi}{2}-\frac{3}{4\pi} \right)
\right]\right\}.
\ea
\end{footnotesize}
Then, the total decay rate for the $Z$ and $Z^{\prime}$ 
is obtained by summing over each fermionic contribution, for instance
\ba
&&\Gamma_{Z^{\prime}}=\sum_{i=u,d,c,s}\Gamma(Z^{\prime}\rightarrow \psi_i\bar{\psi_i})+\Gamma(Z^{\prime}\rightarrow b\bar{b})
+\Gamma(Z^{\prime}\rightarrow t\bar{t})+3\Gamma(Z^{\prime}\rightarrow l\bar{l})
+3\Gamma(Z^{\prime}\rightarrow \nu_l\bar{\nu_l}). \nonumber \\
\ea

\subsection{Calculation of the point-like cross section in the $Z^{\prime}$ case}

We come now briefly to discuss the quantification of the invariant 
mass distributions around the resonances in leptoproduction.

The colour-averaged inclusive differential cross section
for the reaction $H_1 +H_2 \rightarrow l_1 +l_2 +X $, is given by \cite{VN}
\ba
\frac{d\sigma}{dQ^2}=\tau \sigma_{V}(Q^2,M_V^2) W_{V}(\tau,Q^2)\hspace{1cm} \tau=\frac{Q^2}{S},
\ea
where $V$ refers to the generic vector bosons $Z,\gamma,Z'$ and  
where all the hadronic initial state information is contained in the hadronic structure function
which is defined as
\ba
W_{V}(\tau,Q^2)=\sum_{i,j} \int_{0}^{1}dx_1 \int_0^1 dx_2 \int_{0}^{1}dx \delta(\tau-x x_1 x_2)
PD_{i,j}(x_1,x_2,\mu_F^2)\Delta_{i,j}(x,Q^2,\mu_F^2)\,,
\nonumber\\
\ea
where the quantity $PD_{i,j}(x_1,x_2,\mu_F^2)$ contains all the information about the parton
distribution functions and their evolution to the $\mu_F^2$ scale and the 
$\Delta$'s are the flavour-dependent $(i,j)$ hard scatterings.

The point-like cross sections for the case of the $Z^{\prime}$ boson are given by
\ba
&&\sigma_{{Z^{\prime}}}(Q^2)=\frac{\pi\alpha_{em}}{4 M_{{Z^{\prime}}}\sin^2\theta_W \cos^2\theta_W N_c}
\frac{\Gamma_{{Z^{\prime}}\rightarrow \bar{l} l}}{(Q^2-M_{Z^{\prime}}^2)^2 + M_{Z^{\prime}}^2 \Gamma_{Z^{\prime}}^2}
\nonumber\\
&&\sigma_{{Z^{\prime}},\gamma}(Q^2)=\frac{\pi\alpha_{em}^2}{6 N_c} \frac{g_V^{Z^{\prime},l}g_V^{\gamma,l}}{\sin^2{\theta_W}\cos^2{\theta_W}}
\frac{(Q^2-M_{Z^{\prime}}^2)}{Q^2(Q^2-M_{Z^{\prime}}^2)^2+M_{Z^{\prime}}^2\Gamma_{Z^{\prime}}^2},\nonumber\\
&&\sigma_{{Z^{\prime}},Z}(Q^2)=\frac{\pi\alpha_{em}^2}{96}
\frac{\left[g_V^{Z^{\prime},l}g_V^{Z,l}+g_A^{Z^{\prime},l}g_A^{Z,l}\right]}{\sin^4{\theta_W}\cos^4{\theta_W}N_c}\nonumber \\
&& \qquad \qquad \qquad\times \frac{(Q^2-M_Z^2)(Q^2-M^2_{Z^{\prime}}) +M_Z\Gamma_{Z}M_{Z^{\prime}}\Gamma_{Z^{\prime}}} {\left[(Q^2-M_{Z^{\prime}}^2)^2 + M_{Z^{\prime}}^2\Gamma_{Z^{\prime}}^2\right]
\left[(Q^2-M_Z^2)^2 + M_Z^2\Gamma_{Z}^2\right]}.
\nonumber\\
\ea
In Table~\ref{tablethird} we have shown the numerical results
for the total cross sections on the $Z$ peak in the different models 
considered in our analysis. In the first
line of each column we show the results for
the total cross section in [fb], in the second line the total width
$\Gamma_{Z^{\prime}}$, expressed in GeV, and
in the third line the observable $\sigma_{tot}\times BR(Z'\rightarrow l\bar{l})$, where
$BR(Z'\rightarrow l\bar{l})=\Gamma_{Z^{\prime}\rightarrow l\bar{l}}/\Gamma_{Z^{\prime}}$.
All these quantities refer to the value of the coupling constant $g_z$ listed in the first column.

In Table~\ref{table08} we have shown a comparison between 
the Drell-Yan NNLO invariant mass distribution \cite{npap1} 
for the Free fermionic model (anomaly free) in the TeV region and the 
SM background for different values of the coupling $g_z$ (see Ref. \cite{npap5}). 
We have performed the PDF evolution with CANDIA \cite{npap6}, and we have 
chosen the MRST \cite{MRST} set as input distributions.
The mass of the extra Z-prime has been taken 
$M_{Z'}=2.5$ TeV, while $\tan{\beta}=40$.
We observe that in correspondence of the peak value of the extra resonance, the cross
section is enhanced of about 2 order of magnitude with respect to the SM, while moving 
away from the peak, the value of the Free fermionic cross section decreases rapidly
and the difference with respect to the background is around 1-2\%.

\subsection{Results for anomaly-free and anomalous models}

\begin{figure}
\begin{center}
%\subfigure[]
{\includegraphics[width=7cm,angle=-90]{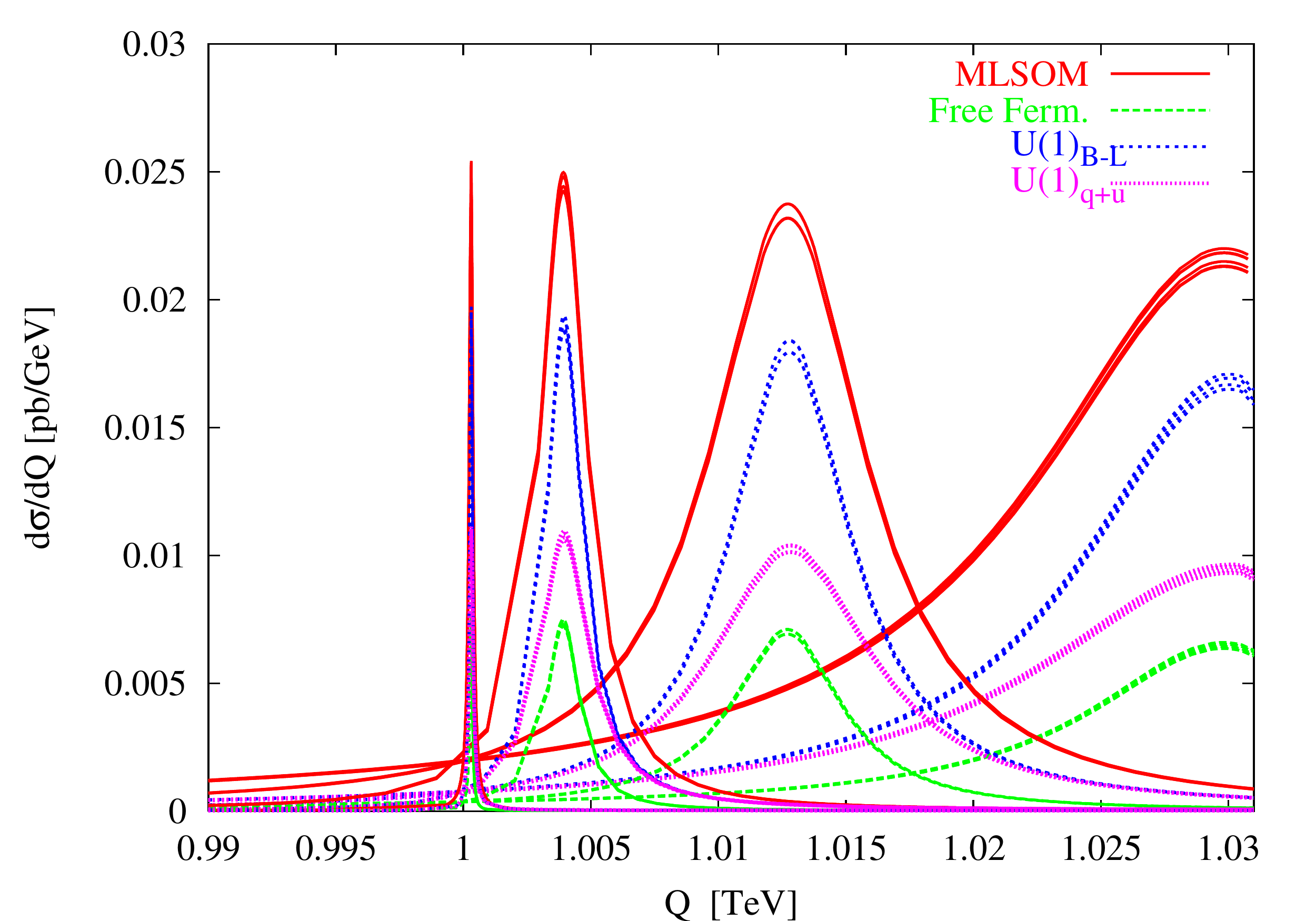}}
\caption{Comparisons among anomalous Drell-Yan in the MLSOM versus several anomaly-free models.}
\label{cross1}
\end{center}
\end{figure}

\begin{figure}
\begin{center}
\subfigure[ Overlaps at NLO/NNLO ]{%
\includegraphics[width=6.5cm,angle=-90]{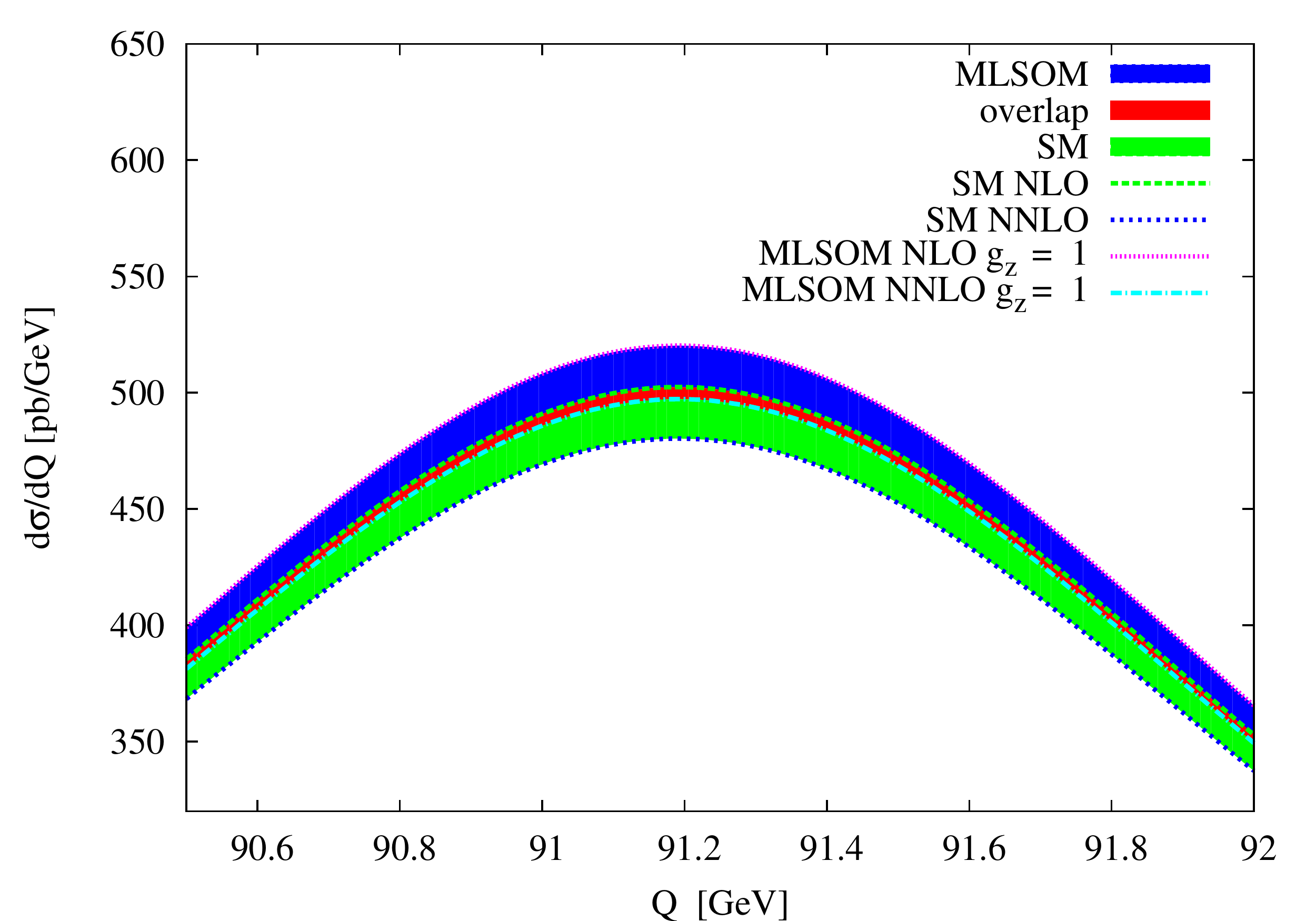}}
\caption{\small Zoom on the $Z$ resonance for anomalous Drell-Yan in
the $\mu_F=\mu_R=Q$ at NLO/NNLO for all the models.}
\label{compare}
\end{center}
\end{figure}

An analysis of the anomalous effects in Drell-Yan and in double prompt photon 
can be found in \cite{npap7}, where several comparisons between anomaly-free and anomalous models 
(the MLSOM) are reported. 
In Fig.~\ref{cross1} we show a comparison between
the MLSOM and the anomaly-free extensions. We have 
included the $\mu_R/\mu_F$ scale dependence,
which appears as a band, and the variations with respect to $g_z$.
As shown in this figure, the red lines correspond to the MLSOM, the 
green ones to the free fermionic model, the blue ones to the $U(1)_{B-L}$ model
and the purple ones to the  $U(1)_{q+u}$ model. The first peak
corresponds to $g_z=0.1$, the second to $g_z=g_Y$, 
the third to $g_z=g_2$ and the fourth to $g_z=1$.
The width of each peak gets larger as $g_z$ grows, but the 
peak-value of the cross section decreases.
Different choices of $g_z$ correspond to slightly different values of
the mass of the extra $Z^{\prime}$ because of the relation between the
St\"uckelberg mass $M_1$ and $M_{Z^{\prime}}$ given in \cite{CorianoIrgesMorelli}.
For a fixed value of the coupling, the effects due to the variations
of the scales $\mu_R$ and $\mu_F$ become visible only
for $g_z=1$ and in this case they are around 2-3\%.
In the case $g_z=1$ (red line), the uppermost lines correspond
to the choice $\mu_F=2 Q$, $\mu_R=1/2 \mu_F$ and $\mu_R= 2 \mu_F$,
while the lowermost lines correspond to the choice
$\mu_F=Q$, $\mu_R=1/2 \mu_F$ and $\mu_R= 2 \mu_F$. The peak-value obtained for the anomalous model is the largest one, with a cross section which is
around $0.022$ [pb/GeV], while the free fermionic model
appears to be the smallest with a value around $0.006$ [pb/GeV]. A sizeable coupling of the extra $Z$ prime affects also the width and the height of the $Z$ resonance. We show in Fig. \ref{compare} that there is an
overlap (red band) between the theoretical uncertainty for the invariant mass distribution in Drell-Yan on the $Z$ resonance, due to the change of the perturbative order in the SM (green band) and in the MLSOM (blue band). 
From this figure it is evident that for larger values of the anomalous coupling ($g_z>1$) 
a class of anomalous models based on intersecting branes can be excluded, 
due to the small overlap for $g_z\approx1$ \cite{npap7}.

\section{Conclusions} 

\begin{table}[ht]
\begin{center}
\begin{footnotesize}
\begin{tabular}{|c||c|c|c|c|}
\hline
\multicolumn{5}{|c|}{$\sigma_{tot}^{nnlo}$ [fb], $\sqrt{S}=14$ TeV, $M_{1}=1$ TeV, $\tan\beta=40$}
\tabularnewline
\hline
$g_z$&
MLSOM&
$U(1)_{B-L}$&
$U(1)_{q+u}$&
$Free Ferm.$
\tabularnewline
\hline
\hline
$0.1$ & $5.982$& $3.575$& $2.701$& $1.274$  \\
       & $0.173$& $0.133$& $0.177$& $0.122$  \\
       & $0.277$& $0.445$& $0.252$& $0.017$
\tabularnewline
\hline
$0.36$ & $106.674$& $105.567$& $53.410$& $42.872$  \\
       & $2.248$& $1.733$& $2.308$& $1.583$  \\
       & $4.937$& $13.138$& $4.991$& $0.586$
\tabularnewline
\hline
$0.65$ & $240.484$& $143.455$& $108.344$& $51.155$  \\
       & $7.396$& $5.700$& $7.592$& $5.205$  \\
       & $11.127$& $17.853$& $10.124$& $0.699$
\tabularnewline
\hline
$1$ & $532.719$& $317.328$& $239.401$& $113.453$  \\
       & $17.810$& $13.720$& $18.274$& $12.530$  \\
       & $24.639$& $39.491$& $22.370$& $1.550$
\tabularnewline
\hline
\end{tabular}
\end{footnotesize}
\end{center}
\caption{\small Total cross sections,  widths and
$\sigma_{tot}\times BR(Z'\rightarrow l\bar{l})$, where
$BR(Z'\rightarrow l\bar{l})=\Gamma_{Z^{\prime}\rightarrow l\bar{l}}/\Gamma_{Z^{\prime}}$,  for the MLSOM and three anomaly-free extensions of the SM.}
\label{tablethird}
\end{table}
We have presented a brief discussion of the search of extra neutral interactions 
at the LHC. While the anomaly-free construction are generated  quite 
automatically in the context of GUT's and low energy string models, 
the possibility of detecting an anomalous gauge 
interaction can not be excluded. In this second case the anomaly cancellation procedure 
of the theory requires the introduction of a Wess-Zumino counterterm and the effective action contains 
a physical axion as a fingerprint of partial decoupling of part of the fermion spectrum. 

We have presented some numerical predictions for the NNLO invariant mass 
distributions in leptoproduction using different anomaly free models 
as a case study, and shown that the cross sections 
are rather sensitive to the masses of the extra neutral gauge bosons and to the coupling 
$g_z$. In this first analysis, the masses of the extra gauge bosons
and the extra coupling can be considered in general, free parameters of the theory. In the anomalous case, anomaly effects in Drell-Yan have been found to be small. However, while the nature (whether anomalous or not) of the extra $Z'$ will be difficult to resolve at the LHC in Drell-Yan or in double photon, precision study on the $Z$ resonance can be used to set exclusion limits on many of these models especially 
at larger values of the anomalous coupling ($g_z\approx 1$).

\begin{table}[hb]
\begin{center}
\begin{footnotesize}
\begin{tabular}{|c||c|c|c|c|c|c|}
\hline
\multicolumn{7}{|c|}{$d\sigma^{nnlo}/dQ$ [pb/GeV] for the FF model with $M_{Z^{\prime}}=2.5$ TeV, $\tan\beta=40$, Candia evol.}
\tabularnewline
\hline
$Q ~[\textrm{TeV}]$      &
$g_z=0.1$                 &
$g_z=0.4$                 &
$g_z=0.6$                 &
$g_z=0.8$                 &
$g_z=1$                   &
$\sigma_{nnlo}^{SM}(Q)$\tabularnewline
\hline
\hline
$2.400$&
$2.6475\cdot10^{-7}$&
$3.3941\cdot10^{-7}$&
$5.0947\cdot10^{-7}$&
$8.7995\cdot10^{-7}$&
$1.5720\cdot10^{-6}$&
$2.6141\cdot10^{-7}$
\tabularnewline
\hline
$2.423$&
$2.4961\cdot10^{-7}$&
$3.5212\cdot10^{-7}$&
$6.0291\cdot10^{-7}$&
$1.1654\cdot10^{-6}$&
$2.2223\cdot10^{-6}$&
$2.4543\cdot10^{-7}$
\tabularnewline
\hline
$2.446$&
$2.3629\cdot10^{-7}$&
$4.0068\cdot10^{-7}$&
$8.4077\cdot10^{-7}$&
$1.8529\cdot10^{-6}$&
$3.7317\cdot10^{-6}$&
$2.3050\cdot10^{-7}$
\tabularnewline
\hline
$2.469$&
$2.2656\cdot10^{-7}$&
$6.0047\cdot10^{-7}$&
$1.7162\cdot10^{-6}$&
$4.2536\cdot10^{-6}$&
$8.5322\cdot10^{-6}$&
$2.1654\cdot10^{-7}$
\tabularnewline
\hline
$2.492$&
$2.4932\cdot10^{-7}$&
$3.7446\cdot10^{-6}$&
$1.2697\cdot10^{-5}$&
$2.3281\cdot10^{-5}$&
$3.0409\cdot10^{-5}$&
$2.0349\cdot10^{-7}$
\tabularnewline
\hline
$2.5000$&
$3.7947\cdot10^{-5}$&
$3.7947\cdot10^{-5}$&
$3.7947\cdot10^{-5}$&
$3.7947\cdot10^{-5}$&
$3.7947\cdot10^{-5}$&
$1.9900\cdot10^{-7}$
\tabularnewline
\hline
$2.5003$&
$8.5271\cdot10^{-6}$&
$3.7283\cdot10^{-5}$&
$3.7757\cdot10^{-5}$&
$3.7858\cdot10^{-5}$&
$3.7892\cdot10^{-5}$&
$1.9886\cdot10^{-7}$
\tabularnewline
\hline
$2.5005$&
$2.7949\cdot10^{-6}$&
$3.5983\cdot10^{-5}$&
$3.7438\cdot10^{-5}$&
$3.7730\cdot10^{-5}$&
$3.7824\cdot10^{-5}$&
$1.9873\cdot10^{-7}$
\tabularnewline
\hline
$2.5770$&
$1.5907\cdot10^{-7}$&
$1.4769\cdot10^{-7}$&
$2.2368\cdot10^{-7}$&
$5.0120\cdot10^{-7}$&
$1.1340\cdot10^{-6}$&
$1.6192\cdot10^{-7}$
\tabularnewline
\hline
$2.636$&
$1.3692\cdot10^{-7}$&
$1.2412\cdot10^{-7}$&
$1.3364\cdot10^{-7}$&
$1.9772\cdot10^{-7}$&
$3.6561\cdot10^{-7}$&
$1.3839\cdot10^{-7}$
\tabularnewline
\hline
$2.700$&
$1.1628\cdot10^{-7}$&
$1.0680\cdot10^{-7}$&
$1.0536\cdot10^{-7}$&
$1.2481\cdot10^{-7}$&
$1.8637\cdot10^{-7}$&
$1.1718\cdot10^{-7}$
\tabularnewline
\hline
\end{tabular}
\end{footnotesize}
\end{center}
\caption{NNLO cross sections for the Free Fermionic model with a
$M_{Z^{\prime}}=2.5$ TeV for values of the coupling constant $g_z$ larger than $g_z=0.1$}
\label{table08}
\end{table}

%\end{document}


\begin{thebibliography}{99}

%  The ALICE detector

\bibitem{ALICE-PPRVol1} ALICE Collaboration, ALICE Physics Performance Report Volume I, J. Phys. G: Nucl. Part. Phys. {\bf 30} (2004) 1517-1763; CERN/LHCC 2003-049.

\bibitem{ALICE-PPRVol2} ALICE Collaboration, ALICE Physics Performance Report Volume II, J. Phys. G: Nucl. Part. Phys. {\bf 32} (2006) 1295-2040; CERN/LHCC 2005-030.

\bibitem{ppDay1} P. Giubellino et al, ``Day-one proton--proton physics with the ALICE central detector'', ALICE Internal Note, ALICE-PHY-2000-28. 

\bibitem{Dainese_MCWS} A. Dainese, ``Measurement of heavy-flavour production with ALICE'', in these proceedings.

\bibitem{Stocco_MCWS} D. Stocco, ``Quarkonia detection with the ALICE Muon Spectrometer in pp collisions at 14 TeV and PDF sensitivity in the low x region'', in these proceedings.

\bibitem{ALICE_EMCal} ALICE Collaboration, \emph{Addendum to the Technical Proposal: Electromagnetic Calorimeter}, CERN/LHCC/2006-014.

\bibitem{ALICE_EMCal_TDR} ALICE Collaboration, \emph{Electromagnetic Calorimeter Technical Design Report}, ALICE-TDR-014, CERN/LHCC/2008-014, 1 September 2008. 

\bibitem{MultipSPD} R. Caliandro, R. A. Fini and T. Virgili, 2002, ALICE Internal Note, ALICE-INT-2002-43. 

\bibitem{MultipSPD2006} T. Virgili, 2006, ALICE Internal Note, ALICE-INT-2006-013.  

\bibitem{MBtrigger} J. Conrad, J. G. Contreras and C. E. Jorgensen, 2005, ALICE Internal Note, ALICE-INT-2005-30.  

%  First Physics measurements with ALICE

\bibitem{Feynman} R.~Feynman, Phys. Rev. Lett. {\bf 23}, 1415 (1969).

\bibitem{UA5_dNdeta} K.~Alpg\r{a}rd {\it et al.} (UA5 Collaboration), Phys. Lett. {\bf B107}, 310 (1981); G.~J.~Alner {\it et al.} (UA5 Collaboration), Z. Phys. {\bf C33}, 1 (1986).

\bibitem{UA1_dNdeta} G.~Arnison {\it et al.} (UA1 Collaboration), Phys. Lett. {\bf B123}, 108 (1982).
 
\bibitem{CDF_dNdeta} F. Abe {\it et al.} (CDF Collaboration), Phys. Rev.  {\bf D41}, 2300 (1990).

\bibitem{QGSM}  A.B.~Kaidalov, Phys. Lett. {\bf B116} (1982) 459; A.B.~Kaidalov
and K.A.~Ter-Martirosyan, Phys. Lett. {\bf B117} (1982) 247; A.~B.~Kaidalov and O.~I.~Piskunova, Z. Phys. {\bf C 30} 145 (1986); A.~B.~Kaidalov, Nucl. Phys. {\bf A 525}, 39c (1991).

\bibitem{sigmatot} K.A. Ter-Martirosyan, Yad. Fiz. {\bf 44} (1986) 1257; 
Sov. J. Nucl. Phys. {\bf 44} (1986) 817. 

\bibitem{Kaidalov} A.~B.~Kaidalov, Physics-Uspekhi, 46 (11) 1121-1136 (2003). 
 
\bibitem{CDF_tuning} R.~Field (for the CDF Collaboration), ``Min-bias and the underlying event in Run 2 at CDF'', Acta Phys. Polon. {\bf B 36} (2005) 167. 

\bibitem{ATLAS_tuning} C.M.Buttar, D.Clements, I. Dawson and A.Moraes, ``Simulations of minimum bias events and the underlying event, MC tuning and predictions for the LHC'', Acta Phys. Polon. {\bf B 35} (2004) 433; ATLAS Note ATLAS-PHYS-PUB-2005-007.

\bibitem{CMS_tuning} D.Acosta {\it et al.}, ``The underlying event at the LHC'', CMS Note 2006-067.   

\bibitem{PYTHIA} T. Sjostrand, P. Eden, C. Friberg, L.Lonnblad, G. Miu, S. Mrenna and E. Norrbin, Comput. Phys. Commun. {\bf 135} (2001) 238; arXiv:hep-ph/0010017.

\bibitem{JIMMY} J. M. Butterworth, J.R. Forshaw and M.H.Seymour, Z.Phys.{\bf C72} (1996) 637; arXiv:hep-ph/9601371.  

\bibitem{SHERPA} T. Gleisberg, S. Hoche, F. Krauss, A. Schalicke. S. Schumann and J. C. Winter,  JHEP {\bf 0402} (2004) 056; arXiv:hep-ph/0311263.

\bibitem{HERWIG++} M. Bahr, S. Gieseke and M.H. Seymour, arXiv:hep-ph/0806.4250

\bibitem{PHOJET} R. Engel,  Z.Phys.{\bf C66} (1995) 203. 

\bibitem{JFGO1} J.F. Grosse-Oetringhaus and C. E. Jorgensen, 2007, ALICE Internal Note, ALICE-INT-2007-05.  
 
\bibitem{ISR_mult} W.~Thome {\it et al.}, Nucl. Phys. {\bf B 125}, 365 (1977).

\bibitem{KNO} Z.~Koba, H.~B.~Nielsen and P.~Olesen, Nucl. Phys. {\bf B 40}, 317 (1972).

\bibitem{UA5_multip} G.~J.~Alner {\it et al.} (UA5 Collaboration), Phys. Lett. {\bf B138}, 304 (1984).

\bibitem{GiovannUgocc_1} A.~Giovannini and R.~Ugoccioni, Phys. Rev. {\bf D59} 094020 (1999).

\bibitem{GiovannUgocc_2} A.~Giovannini and R.~Ugoccioni, Phys. Rev. {\bf D60} 074027 (1999).

\bibitem{two-comp} J.~Dias de Deus and R.~Ugoccioni, Phys. Lett.  {\bf B469} 243 (1999).

\bibitem{Walker} W.~D.~Walker, Phys. Rev. {\bf D69} 034007 (2004). 

\bibitem{Matinyan} S.~G.~Matinyan and W.~D.~Walker, Phys. Rev. {\bf D59} 034022 (1999).

\bibitem{E735_98} T.~Alexopoulos {\it et al} [E735 Collaboration], Phys. Lett. 
{\bf B435}, 453 (1998).

\bibitem{Anykeev} V.~Anykeev {\it et al}, Nucl. Instr. Meth. {\bf A303} 350 (1991).

\bibitem{Dagostini} G.D'Agostini, DESY 94-099, June 1994. 

\bibitem{UA5_multfit} R~.E.~Ansorge {\it et al} (UA5 Collaboration), Z. Phys. 
{\bf C43} 357 (1989).

\bibitem{JFGO2} J.F. Grosse-Oetringhaus, 2008, ALICE Internal Note, ALICE-INT-2008-022.  

\bibitem{VanHove} L.~Van Hove, Phys. Lett. {\bf B118} 138 (1982).
 
\bibitem{collider_pt} B.Alper {\it et al}, Nucl. Phys. {\bf B 100} 237 (1975); C.~Albajar {\it et al}, Nucl. Phys. {\bf B 335} 261 (1990); C.~Abe {\it et al}, Phys. Rev. Lett. {\bf 61} 1818 (1998).

\bibitem{Kalman} P.~Billoir, Nucl. Instr. Meth. {\bf A225} 352 (1984). 

\bibitem{UA1_avept} G.~Arnison {\it et al} (UA1 Collaboration) Phys. Lett 
{\bf B 118} 173 (1982); G.~Bocquet {\it et al} (UA1 Collaboration) Phys. Lett. {\bf B 366} 434 (1996).

\bibitem{ISR_avept} A.~Breakstone {\it et al} Phys. Lett.{\bf B 132} 458 (1983); A.~Breakstone {\it et al} Phys. Lett.{\bf B 132} 463 (1983); A.~Breakstone {\it et al} Phys. Lett.{\bf B 183} 227 (1987).

\bibitem{E735_avept}  T.~Alexopoulos T {\it et al} (E735 Collaboration) Phys. 
Rev. Lett. {\bf 64} 991 (1990); T.~Alexopoulos {\it et al} (E735 Collaboration) Phys. Rev. {\bf D 48} 984 (1993).

\bibitem{CDF_soft} D.~Acosta {\it et al} (CDF Collaboration) Phys. Rev 
{\bf D 65} 072005 (2002).

\bibitem{Lat80} C.~M.G.~Lattes {\it et al}, Phys. Rep. {\bf 65} 151 (1980).

\bibitem{Wan87} X.~N.~Wang and R.~C.~Hwa, Phys.Rev. {\bf D 39} 187 (1987); X.~N.~Wang and M.~Gyulassy, Phys. Lett. {\bf B 282} 466 (1992).

\bibitem{HERWIG} G. Corcella {\it et al}, JHEP {\bf 0101} (2001) 010; arXiv:hep-ph/0011363.
 
\bibitem{CDF_UE} T.~Affolder {\it et al} (CDF Collaboration], Phys. Rev. 
{\bf D 65} 092002 (2002); D.~Acosta {\it et al} (CDF Collaboration], Phys. 
Rev. {\bf D 70} 072002 (2004).

%  Strange particle measurements

\bibitem{Wroblewski} A.~Wroblewski, Acta. Phys. Polon. B16 (1985) 379.

\bibitem{lambda_s_models}
V. V. Anisovich and M. N. Kobrinski, Phys.\ Lett.\ {\bf B52} (1974) 217;
V. M. Shekhter and L. M. Scheglova, Sov.\ J. Nucl.\ Phys.\ {\bf 27} (1978) 567.

\bibitem{Becattini_strange} F. Becattini {\em et al.}, Phys. Rev. {\bf C64} (2001) 024901.

\bibitem{Becattini_analyses}
F. Becattini, U. Heinz, Z. Phys.\ {\bf C76} (1997) 269;
F. Becattini, M. Gazdzicki, J. Sollfrank, Eur.\ Phys.\ J. {\bf C5} (1998) 143;
F. Becattini and G. Passaleva, Eur.\ Phys.\ J. {\bf C23} (2002) 551.

\bibitem{Alexopoulos_2002} T.~Alexopoulos {\em et al.} (E735 Collab.)
Phys. Lett. {\bf B528} (2002) 43. 

\bibitem{STAR_ptspectra} B.I.~Abelev et al (STAR Collaboration) 
Phys. Rev. {\bf C75} (2007) 064901.

\bibitem{KKP} B.~A.~Kniehl, G.~Kramer and B.~Potter, Nucl. Phys. {\bf B597} (2001) 337. 

\bibitem{DSV} D.~DeFlorian, M.~Stratmann and W.~Vogelsang,  
Phys. Rev. {\bf D57} (1998) 5811.

\bibitem{AKK} S.~Albino, B.~A.~Kniehl and G.~Kramer, 
Phys. Lett. {\bf B725} (2005) 181. 

\bibitem{OPAL_flavFF} G.~Abbiendi {\em et al.} (OPAL Collab.), Eur. Phys.J. 
{\bf C16} (2000) 407.

\bibitem{Alexopoulos_1993}
{Alexopoulos, T. et al.\ (E735 Collaboration)}, Phys.\ Rev.\ {\bf D48}  (1993)
  984.

\bibitem{VanHove_ptQGP} L. Van Hove, Phys.\ Lett.\ {\bf B118}  (1982) 138.

\bibitem{Vernet} R. Vernet, ``Prospects for strangeness measurements in 
ALICE'', Proceedings of the Workshop on Relativistic Nuclear Physics (WRNP07), 
June 2007, Kiev (Ukraine), Phys. Atom. Nucl. {\bf 71} (2008) 1523-1534; arXiv:nucl-ex/0802.0095v1. 

%  Baryon Flow measurements

\bibitem{DPM} A.~Capella, U.~Sukhatme, C.I.~Tan and J.~Tran Thanh Van, 
Phys.  Rep.  {\bf 236} (1994) 225.

\bibitem{ISR_baryons} B.~Alper {\it et al.}, Nucl. Phys. {\bf B100} (1975) 
237; T.~Akesson {\it et al.}, Nucl. Phys. {\bf B228} (1983) 409; 
L.~Camilleri {\it et al.}, Phys. Rep. {\bf 53} (1987) 144.

\bibitem{rv} G.C.~Rossi, G.~Veneziano, Nucl. Phys. {\bf B123} (1977)
507; Phys.  Rep.  {\bf 63} (1980) 149.

\bibitem{kz89} B.Z.~Kopeliovich and B.G.~Zakharov, Z. Phys. {\bf C43}
(1989) 241.

\bibitem{kharz96} D.~Kharzeev, Phys. Lett. {\bf B378} (1996) 238. 

\bibitem{kp97} B.Z.~Kopeliovich and B.~Povh, Z. Phys. {\bf C75} (1997)
693. 

\bibitem{H1baryons} H1 Collaboration, C. Adloff {\it et al.}, ``Measurement of the baryon-antibaryon asymmetry in photoproduction at HERA'', Proceedings of 29th Int.Conference on High Energy Physics ICHEP98, Vancouver, Canada, July 1998. 

\bibitem{kp99} B.Z.~Kopeliovich and B.~Povh, Phys. Lett. {\bf B446} (1999) 321.

\bibitem{BRAHMS_ratios} I.G. Bearden {\it et al.} (BRAHMS Collaboration), Phys. Lett. {\bf B607} (2005) 42;   arXiv:nucl-ex/0409002.

\bibitem{Christag_ppbar} P. Christakoglou and M.Oldenburg, Private Communication, to be published as an ALICE Internal Note.

%  Correlations and EbyE fluctuations

\bibitem{UA1_HBT} B. Buschbeck and H.C.Eggers, Nucl.Phys. (Proc.Suppl.) {\bf B92} (2001) 235; B. Buschbeck, H.C.Eggers and P.Lipa, Phys. Lett. {\bf B481} (2000) 187.  

\bibitem{E735_HBT} C. Lindsey (for the E735 Collaboration), FERMILAB-Conf-91/336

\bibitem{STAR_HBT} Z. Chajecki (for the STAR Collaboration), Proceedings of the conference Quark Matter 2005, Budapest, Hungary, August 4-9 2005, Nucl. Phys. 
{\bf A774} (2006) 599-602; arXiv:nucl-ex/0510014.

\bibitem{NA22_HBT} N.M.Agababyan {\it et al} (NA22 COLLAB.), Z.Phys.{\bf C59} (1993) 195. 

\bibitem{BudaLund_STAR} T.Cs\"{o}rg\"{o}. M.Csan\'{a}d, B.L\"{o}rstad and A.Ster, 
arXiv:hep-ph/0406042.

\bibitem{Humanic_HBT} T. Humanic, Phys. Rev. {\bf C76} (2007) 025205.

\bibitem{Kawrakow} I. Kawrakow,  Phys. Rev. {\bf D49} (1994) 2275. 

\bibitem{Berger} E. Berger, Nucl. Phys. {\bf B85} (1975) 61.

\bibitem{Ranft_clusters} J. Ranft, Fortschr. Phys. {\bf 23} (1975) 467.

\bibitem{GiovanniniVanhove} A. Giovannini and L. Van Hove,  Z.Phys.{\bf C30} (1986) 391.

\bibitem{ChouYang} T.T. Chou and C.N.Yang, Phys. Lett. {\bf B135} (1984) 175.

\bibitem{Lim} S.L. Lim {\it et al}, Z.Phys.{\bf C43} (1989) 621; Z.Phys.{\bf C54} (1992) 107.

\bibitem{GiovanniniUgoccioni_FB} A. Giovannini and R. Ugoccioni, Phys. Lett. {\bf B558} (2003) 59. 

\bibitem{Wang} X.N. Wang,  Phys. Rev. {\bf D47} (1993) 2754. 

\bibitem{STAR_whitebook} J. Adams {\it et al} (STAR Collaboration) Nucl. Phys. {\bf A757} (2005) 102.

\bibitem{PorterTrainor} R.J. Porter and A.T. Trainor (for the STAR Collaboration), J.Phys.Conf.Ser. {\bf 27} (2005) 98; Acta Phys. Polonica {\bf B36} (2005) 353. 

\bibitem{JACEE} T.H. Burnett {\it et al} (JACEE COLLAB.), Phys. Rev. Lett. {\bf 50} (1983) 2062. 

\bibitem{NA22} M. Adamus {\it et al} (NA22 COLLAB.), Phys. Lett. {\bf B185} (1987) 200. 

%  Diffractive physics

\bibitem{Schicker} R.~Schicker, ``Diffractive physics in ALICE'', Proceedings 
of 2th International Conference on Elastic and Diffractive Scattering 
``Forward Physics and QCD'', (DESY, Hamburg, 21 - 25 May 2007), 
DESY-PROC-2007-02; R.~Schicker, ``Low-mass diffractive systems at LHC'', Proc. of the conference ``New Trends in high energy physics'', (Yalta, 15-22 Sep. 2007).

\bibitem{Bzdak} A.~Bzdak, L.~Motyka, L.~Szymanowski,J.~R.~Cudell, Phys. Rev. 
{\bf D 75} (2007) 094023.

\bibitem{Goncalves1} V.~P.~Goncalves, M.~V.~Machado, Phys. Rev. {\bf D 71} 
(2005) 014025.

\bibitem{Goncalves2} V.~P.~Goncalves, M.~V.~Machado, Phys. Rev. {\bf D 75} 
(2007) 031502.

%  Jets

\bibitem{Gyulassy} M.~Gyulassy, I.~Vitev, X.~N.~Wang and B.~W.~Zhang, 
{\it Quark Gluon Plasma 3}, edited by R.~C.~Hwa and X.~N.~Wang  
(World Scientific, Singapore, 2003), p.123; nucl-th/0302077.

\bibitem{BDMS} R.~Baier, Yu.~L.~Dokshitzer, A.~J.~Mueller and D.~Schiff, 
Phys. Rev. {\bf C58}, 1706 (1998).  

\bibitem{BSZ} R.~Baier, D.~Schiff and B.~G.~Zakharov, Ann. Rev. Nucl. Part.
Sci.{\bf 50}, 37 (2000).  

\bibitem{Kovner} A.~Kovner and U.~A.~Wiedemann in Ref.\cite{Gyulassy}, 
p.192; hep-ph/0304151.

\bibitem{Bjorken_1982} J.~D.~Bjorken, preprint FERMILAB-PUB-82-059-THY.

\bibitem{Mustafa} M.~G.~Mustafa, Phys. Rev. {\bf C 72}, 014905 (2005).

\bibitem{Dutt} A.~K.~Dutt-Mazumder, J.~Alam, P.~Roy and B.~Sinha, 
Phys. Rev. {\bf D 71}, 094016 (2005).

\bibitem{Djordjevic} M.~Djordjevic, Phys. Rev. {\bf C 74}, 064907 (2006).

\bibitem{Djordjevic_2} M.~Djordjevic, M.~Gyulassy, R.~Vogt and S.~Wicks, 
Phys. Lett. {\bf B 632}, 81 (2006).

\bibitem{Morsch:2005sv} A.~Morsch, J.\ Phys.\ {\bf G31} (2005) S597.

\bibitem{Heinz_STAR_HP08} M. Heinz (for the STAR Collaboration),
``Systematic studies of fragmentation functions in inclusive jets from 
p+p collisions at 200 GeV by STAR'', Hard Probes Conference 2008 (HP08), 
Illa de Toxa, Galicia, Spain, 8-14 June 2008. 

\bibitem{Armesto_jets} N.~Armesto, ``Jet rates in p+A collisions at the LHC'', Workshop on Proton-Nucleus Collisions at the LHC (CERN, 25-27 May 2005),  
A.~Accardi {\it et al.}, \emph{PDFs, shadowing and pA collisions}, 
``Hard Probes in heavy-ion collisions at the LHC'', CERN Yellow Report 
{\bf 2004-009}, hep-ph/0308248.  

\bibitem{Loizides_2005PhDThesis} C.~A.~Loizides. Ph.~D.~Thesis, Univ.~of 
Frankfurt am Main, Germany, arXiv:nucl-ex/0501017. 

% Photons

\bibitem{Tornbull} R. M. Tornbull, J.\ Phys.\ {\bf G14}, 135 (1988).

\bibitem{Aurenche_1999} P. Aurenche, M. Fontannaz, J.-Ph. Guillet, B. Kniehl, 
E. Pilon and M. Werlen, Eur. Phys. J. {\bf C9}, 107 (1999).

\bibitem{Aurenche_2000} P. Aurenche, M. Fontannaz, J.-Ph. Guillet, B. Kniehl, 
E. Pilon and M. Werlen, Eur. Phys. J. {\bf C13}, 347 (2000); hep-ph/9910252.

\bibitem{Binoth2000} T. Binoth {\it et al.}, Eur. Phys. J. {\bf C16}, 311 
(2000).

\bibitem{Catani2002} S. Catani {\it et al.}, J. High Energy Phys. {\bf 05}, 
028 (2002).

\bibitem{D0_2006} V. M. Abazov {\it et al.}, (D0 Collaboration), Phys. Lett. 
{\bf B639}, 151 (2006).

\bibitem{PHENIX_2006} S. S. Adler {\it et al.} (PHENIX Collaboration), Phys. 
Rev. {\bf D71}, 071102 (2005).

\bibitem{Aurenche_2006} P. Aurenche, M. Fontannaz, J.-Ph. Guillet, E. Pilon 
and M. Werlen, Phys. Rev. {\bf D73}, 094007 (2006).

\bibitem{E706_2004} L. Apanasievich {\it et al.}, (E706 Collaboration), 
Phys. Rev.{\bf D70}, 092009 (2004).

\bibitem{recoil} E.~Laenen, G.~Sterman and W.~Vogelsang, Phys. Rev. Lett. 
{\bf 84}, 4296 (2000); Phys. Rev. {\bf D63} 114018 (2001).

\bibitem{kT} H.-L.~Lai and H.-N.~Li, Phys. Rev. {\bf D58}, 114020 (1998).

\bibitem{Gordon93}
L.E. Gordon and W. Vogelsang, Phys. Rev. {\bf D 48} 3136 (1993).

\bibitem{GTagged} F. Arleo {\it et al.}, J. High Energy Phys. {\bf 11}, 
009 (2004).

\bibitem{YellowReport_2003_photons} F.~Arleo  {\it et al.},
\emph{Photon Physics in heavy-ion collisions at the LHC}, CERN Yellow Report 
{\bf 2004-009}; hep-ph/0311131.  

\bibitem{DPMJET} J. Ranft, Phys. Rev. {\bf D51}, 64 (1995); R. Engel, Z. Phys. 
{\bf C66}, 203 (1995); R. Engel and J. Ranft, Phys. Rev. {\bf D54}, 4244 
(1996); S. Roesler, R. Engel and J. Ranft,  hep-ph/0012252. 

\bibitem{frag-phot} L. Bourhis, M. Fontannaz and J.-Ph. Guillet, Eur. Phys. J. 
{\bf C2}, 529 (1998) 529.

\bibitem{SCINT} G. Conesa {\it et al.}, Nucl.~Inst.~Meth.{\bf A 537} 363 
(2005).

\bibitem{GAMMAJET} G. Conesa {\it et al.}, ALICE-INT-2005-014, (2005).

\bibitem{kniehl} B.A.~Kniehl, G.~Kramer and B.~Potter, Nucl. Phys. {\bf B582}, 
514 (2000).

\bibitem{Jalilian00} J. Jalilian-Marian, K. Orginos and I. Sarcevic, Phys. 
Rev. {\bf C 63} 041901 (2001); hep-ph/0010230. 

\bibitem{Wong98} C. Wong and H. Wang, Phys. Rev. {\bf C 58} 376 (1998).

\bibitem{Papp99} G. Papp, P. L\'evai, and G. Fai, Phys. Rev. {\bf C 61}, 
0219021 (1999). 
 
\bibitem{Zakharov04} B.G. Zakharov, JETP Lett. \textbf{80}, 1 (2004). 

\bibitem{SalgadoWiedemann} C.~A.~Salgado and U.~A.~Wiedemann, Phys. Rev. Lett. 
{\bf 93} 042301 (2004).

% Exotica (mini BH)

\bibitem{Humanic} T.~Humanic, ALICE Internal Report, ALICE-INT-2005-017; 

\bibitem{Humanic2} T.~J.~Humanic, B.~Koch and H.~Stocker, arXiv:hep-ph/0607097.

\bibitem{CHARYBDIS} C.~M.~Harris, P.~Richardson and B.~R.~Webber, arXiv:hep-ph/0409309.

\end{thebibliography}

\begin{thebibliography}{100}

\bibitem{alicePPR1}
  ALICE Collaboration, Physics Performance Report Vol.~I,  
  CERN/LHCC 2003-049 and J.~Phys.~G~{\bf 30}, 1517 (2003).

\bibitem{notehvq}
  N.~Carrer and A.~Dainese, ALICE Internal Note, ALICE-INT-2003-019 (2003), 
   arXiv:hep-ph/0311225. 

\bibitem{hvqmnr} M.L.~Mangano, P.~Nason and G.~Ridolfi, Nucl. Phys. 
    B~{\bf 373}, 295 (1992).

\bibitem{EKS98} 
  K.J.~Eskola, V.J.~Kolhinen, C.A.~Salgado, Eur.~Phys.~J.~C~{\bf 9}, 61
  (1999).

\bibitem{monteno}
  M.~Monteno, these proceedings.

\bibitem{stocco}
  D.~Stocco, these proceedings.

\bibitem{alicePPR2}
  ALICE Collaboration, Physics Performance Report Vol.~II,
  CERN/LHCC 2005-030
  and J.~Phys.~G~{\bf 32} 1295, (2006).

\bibitem{fonll}
  M.~Cacciari, S.~Frixione, M.L.~Mangano, P.~Nason and G.~Ridolfi,
  JHEP {\bf 0407}, 033 (2004).

\bibitem{cacciari}
  M.~Cacciari, private communication.

\bibitem{ua1Bextraction}
  C.~Albajar {\it et al.}, UA1 Collaboration, 
 Phys.~Lett.~B~{\bf 213}, 405 (1988); Phys.~Lett.~B~{\bf 256}, 121 (1991).

\bibitem{hotquarks06}
  A.~Dainese, arXiv:nucl-ex/0608005.  

\bibitem{heralhc}
  J.~Baines {\it et al.},
  arXiv:hep-ph/0601164, in CERN-2005-014 / DESY-PROC-2005-01.

\end{thebibliography}

\begin{thebibliography}{9}
  \bibitem{TDR} ALICE Collaboration,
	LHCC 99-22 / ALICE TDR 5.
  \bibitem{CEM} V. D. Barger and W. Y. Keung and R. J. Phillips,
	\PL{B91}{1980}{253}.
  \bibitem{CEMvogt} M. Bedjidian et al.,
	CERN YellowReport, CERN-2004-009, [arXiv:hep-ph/0311048].
  \bibitem{Martin:1998sq}
        A.~D.~Martin, R.~G.~Roberts, W.~J.~Stirling and R.~S.~Thorne,
        Eur.\ Phys.\ J.\  C {\bf 4} (1998) 463
        [arXiv:hep-ph/9803445].
  \bibitem{Acosta:2004yw}
        D.~Acosta {\it et al.}  [CDF Collaboration],
        Phys.\ Rev.\  D {\bf 71} (2005) 032001
        [arXiv:hep-ex/0412071].
  \bibitem{Acosta:2001gv}
        D.~Acosta {\it et al.}  [CDF Collaboration],
        Phys.\ Rev.\ Lett.\  {\bf 88}, 161802 (2002).
   \bibitem{carrerDainese} N. Carrer and A. Dainese,
	\ALI{2003}{019}, [arXiv:hep-ph/0311225].
   \bibitem{PPR2} ALICE Collaboration,
	ALICE Physics Performance Report Volume II,
        J. Phys. G: Nucl. Part. Phys. 32 1295-2040.
   \bibitem{PPR1} ALICE Collaboration,
	ALICE Physics Performance Report Volume I,
	J. Phys. G: Nucl. Part. Phys. 30 1517-1763.
   \bibitem{Martin:2002dr}
        A.~D.~Martin, R.~G.~Roberts, W.~J.~Stirling and R.~S.~Thorne,
        Phys.\ Lett.\  B {\bf 531} (2002) 216
        [arXiv:hep-ph/0201127].
   \bibitem{Lai:1999wy}
        H.~L.~Lai {\it et al.}  [CTEQ Collaboration],
        Eur.\ Phys.\ J.\  C {\bf 12}, 375 (2000)
        [arXiv:hep-ph/9903282].
   \bibitem{Pumplin:2002vw}
        J.~Pumplin, D.~R.~Stump, J.~Huston, H.~L.~Lai, P.~Nadolsky and W.~K.~Tung,
        JHEP {\bf 0207} (2002) 012
        [arXiv:hep-ph/0201195].
\end{thebibliography}

\begin{thebibliography}{99}

\bibitem{LHCPrimer}
J.M. Campbell, J.W. Huston, W.J. Stirling, 
{\em  Hard Interactions of Quarks and Gluons: A Primer for LHC Physics.\/}
Rept. Prog. Phys. 70:89 (2007).

\bibitem{Devenish_Cooper-Sarkar}
R. Devenish, A. Cooper-Sarkar
{\em Deep Inelastic Scattering.\/}
Oxford University Press, Oxford (2004).

\bibitem{PDFError_ED}
S. Ferrag, 
{\em Proton structure impact on sensitivity to extra-dimensions at LHC.\/} 
hep-ph/0407303 (2004).

\bibitem{PDFError_Higgs}
A. Djouadi, S. Ferrag,
{\em PDF uncertainties in Higgs production at hadron colliders.\/} 
Phys. Lett. {\bfseries B 586} 345-352 (2004). %[hep-ph/0310209]

\bibitem{Florian}
  D.Phil. thesis by F. Heinemann (Oxford) 2007.

\bibitem{Tricoli_Photon05}
A. Tricoli for the ATLAS Coll.
{\em  Structure function measurements at the LHC.\/}
Acta Phys. Polon. {\bfseries B 37} 711-714 (2006).
 
\bibitem{Hollins}
  Ph.D. thesis by I. Hollins (Birmingham) 2006.

\bibitem{Zb_VerducciEtAl}
S. Diglio, A. Tonazzo and M. Verducci (RomaTre) 2005.

\bibitem{CSCNote}
ATLAS CSC Note {\em  Electroweak boson cross-section measurements with ATLAS\/} (2008).

\bibitem{NNLO_WZ_y}
C. Anastasiou, L. Dixon, K. Melnikov, F. Petriello,
{\em High precision QCD at hadron colliders: Electroweak gauge boson rapidity distributions at NNLO.\/}
Phys. Rev. {\bfseries D 69} 094008 (2004).

\bibitem{CarloniPolesello}
Thanks to C. Carloni and G. Polesello.

\bibitem{TricoliCooper-Sarkar}
A. Cooper-Sarkar (Oxford) and A. Tricoli (RAL) 2006.

\bibitem{Tricoli}
D.Phil. thesis by A. Tricoli (Oxford) 2006.

\bibitem{HERAtoLHC}
A. Tricoli, A. Cooper-Sarkar, C. Gwenlan,
{\em Uncertainties on W and Z production at the LHC.\/}
CERN-2005-014, FERMILAB-CONF-05-586-E, IFUM-853-FT,
hep-ex/0509002 (2005). 

\bibitem{Mandy}
A. Cooper-Sarkar (Oxford) private communication, 2005.

\bibitem{Carli-Salam-Siegert}
T. Carli, G. P. Salam, F. Siegert, 
{\em A posteriori inclusion of PDFs in NLO QCD final-state calculations.\/}
hep-ph/0510324 (2005). 

\bibitem{Clements-et-al}
T. Carli, D. Clements, Cooper-Sarkar, C. Gwenlan, A. G. P. Salam, P. Starovoitov, M Sutton, ``PDF4LHC'' meeting (2008).

\end{thebibliography}

\begin{thebibliography}{99}
\bibitem{Van_Neerven1} R. Hamberg, W.L. van Neerven, T. Matsuura,
Nucl.Phys. {\bf B 359} 343, (1991), Erratum-ibid.{\bf B 644} 403, (2002).
W. Van Neerven and A. Vogt, Nucl.Phys.\, {\bf B 603}, 42, (2001); Nucl.Phys. {\bf B 588}, 345, (2000).
 \bibitem{Nason} S. Catani, D. de Florian, M. Grazzini, P. Nason, JHEP {\bf 0307} 028, 
(2003); G. Corcella and L. Magnea, Phys.Rev.{\bf 72},074017, (2005);
V. Ravindran, J. Smith, W.L. van Neerven,{\bf hep-ph/0608308}.
\bibitem{Anastasiou} C. Anastasiou, L. J. Dixon, K. Melnikov and F. Petriello
Phys.Rev. \textbf{D 69}, 094008, (2004).
\bibitem{Melnikov} K. Melnikov and F. Petriello, Phys.Rev.{\bf D74}114017, (2006).  
\bibitem{vogt1}S. Moch, J. Vermaseren and A. Vogt, Nucl. Phys.\, {\bf B 688}, 101, (2004);
Nucl. Phys.\, {\bf B 691}, 129, (2004).
\bibitem{LesHouches02} The QCD / SM working group: Summary report.
W. Giele et al,  hep-ph/0204316, (2002)
\bibitem{LesHouches05} Working Group I: Parton distributions:
Summary report for the HERA LHC Workshop Proceedings. M. Dittmar et. al., hep-ph/0511119, (2005)
\bibitem{candia} A. Cafarella,C. Corian\`o and M. Guzzi Comput.Phys.Commun. {\bf 179}, 665 (2008), arXiv:0803.0462 [hep-ph]
\bibitem{CCG2} A. Cafarella, C. Corian\`o, M. Guzzi, JHEP 0708, 030 (2007),  hep-ph/0702244.
\bibitem{CCG1} A. Cafarella, C. Corian\`o, M. Guzzi Nucl.Phys. {\bf B 748}, 253 (2006), hep-ph/0512358; 
A. Cafarella and C. Corian\`o, Comput.Phys.Commun.\,{\bf 160}, 213 (2004), hep-ph/0311313.
\bibitem{Pegasus} A. Vogt, Comput.Phys.Commun. {\bf 170} 65, (2005).
\bibitem{Alekhin} S.I.~Alekhin, Phys.Rev.{\bf D 68}, 014002, (2003);
S.I.~Alekhin, Eur.Phys.J. {\bf C 10}, 395, (1999);
\bibitem{MRST1} A.D. Martin, R.G. Roberts, W.J. Stirling and R.S. Thorne, Eur.Phys.J. {\bf C 23}, 73 (2002);
Phys.Lett.{\bf B 531}, 216 (2002).
\bibitem{pap2} C. Corian\`o, A. E. Faraggi and M. Guzzi, Eur.Phys.J.C53, 421 (2008), arXiv:0704.1256 [hep-ph].
\bibitem{pap3} R. Armillis, C. Corian\`o and M. Guzzi, JHEP 0805, 015 (2008), arXiv:0711.3424 [hep-ph] 
\bibitem{pap4} C. Corian\`o, M. Guzzi and S. Morelli, Eur.Phys.J.C 55, 629 (2008), arXiv:0801.2949 [hep-ph] 
\bibitem{pap5} C. Corian\`o, A.E. Faraggi and M. Guzzi, Phys.Rev.D 78, 015012 (2008), arXiv:0802.1792 [hep-ph] 
\bibitem{pap7} R. Armillis, C. Corian\`o, M. Guzzi and S. Morelli, JHEP 0810, 034 (2008), arXiv:0808.1882 [hep-ph] 
\bibitem{pap8} R. Armillis, C. Corian\`o, M. Guzzi and S. Morelli, arXiv:0809.3772 [hep-ph]
\end{thebibliography}

\begin{thebibliography}{99}

\bibitem{UB} U. Baur, Electroweak physics at the Tevatron and LHC: Theoretical status and perspectives, arXiv:hep-ph/0511064

\bibitem{acfgi} R. Armilis, C. Corian\`{o}, A. E. Faraggi, M. Guzzi, N. Irges, 
Extra neutral interactions at the LHC, these proceedings

\bibitem{cy} Q.-H. Cao and C.-P. Yuan, Phys. Rev. Lett. {\bf 93} (2004) 
042001; arXiv:hep-ph/0401171

\bibitem{ward2007} B.F.L. Ward and S.A. Yost, Acta Phys. Polon. {\bf B38} (2007) 2395

\bibitem{jadach2007} S. Jadach, W. Placzek, M. Skrzypek, P. Stephens and Z. Was, Acta Phys. Polon. { \bf B38} (2007) 2305

\bibitem{pdf} A. Tricoli, Parton densities at the LHC, these proceedings;\\
A. Cafarella, C. Corian\`{o} and M. Guzzi, NNLO evolution of Pdf's, these proceedings;\\
N.E. Adam, V. Halyo and S.A. Yost, arXiv:0802.3251 [hep-ph];\\
S. Frixione and M.L. Mangano,  JHEP {\bf 0405} (2004) 056

\bibitem{AEM}
  G.~Altarelli, R.~K.~Ellis and G.~Martinelli, Nucl. Phys. {\bf B157} (1979) 461
  
\bibitem{HvNM}
  R.~Hamberg, W.~L.~van Neerven and T.~Matsuura, Nucl. Phys. {\bf B359} (1991) 343
  [Erratum Nucl. Phys. {\bf B644} (2002) 403]
 
\bibitem{GGK} W.T. Giele, E.W.N. Glover and D.A. Kosower, 
Nucl. Phys. {\bf B403} (1993) 633 

\bibitem{MCFM}  J.M. Campbell and R.K. Ellis,  Phys. Rev. {\bf D65}  (2002) 113007

\bibitem{BY}
  C.~Balazs and C.~P.~Yuan, Phys. Rev. {\bf D56} (1997) 5558
  
\bibitem{resbos} F. Landry, R. Brock, P.M. Nadolsky and C.-P. Yuan, 
Phys. Rev. {\bf D67} (2003) 073016

\bibitem{MC@NLO}
  S.~Frixione and B.~R.~Webber, JHEP {\bf 0206} (2002) 029
  
\bibitem{POWHEG}  S.~Frixione, P. Nason and C. Oleari, JHEP {\bf 0711} (2007) 070
    
\bibitem{mp} K.~Melnikov and F.~Petriello, Phys. Rev. Lett. {\bf 96} (2006) 231803

\bibitem{mp1} K.~Melnikov and F.~Petriello, Phys. Rev.  {\bf D74} (2006) 114017

\bibitem{Alpgen}
M.L. Mangano, M. Moretti, F. Piccinini, R. Pittau and A.D. Polosa, JHEP {\bf 0307} (2003) 001

\bibitem{MadEvent} T. Stelzer and W.F. Long, 
Comp. Phys. Commun. {\bf 81} (1994) 357; 
F. Maltoni and T. Stelzer, JHEP {\bf 02} (2003) 027

\bibitem{Helac} A. Kanaki and C.G. Papadopoulos, Comput. Phys. Commun. {\bf 132} (2000) 
306; C.G.~Papadopoulos and M.~Worek, Eur. Phys. J. {\bf C50} (2007) 843; 
A. Cafarella, C.G. Papadopoulos and M. Worek, arXiv:0710.2427 [hep-ph].


\bibitem{Sherpa} T.~Gleisberg, S.~H\"oche, F.~Krauss, A.~Sch\"alicke, 
S.~Schumann and J.~Winter, JHEP {\bf 0402} (2004) 056

\bibitem{dk} S. Dittmaier and M. Kr\"amer, Phys. Rev. {\bf D65} (2002) 0703007

\bibitem{bw} U. Baur and D. Wackeroth, Phys. Rev. {\bf D70} (2004) 073015

\bibitem{ZYK}
  V.~A.~Zykunov, Eur. Phys. J. Direct {\bf C3} (2001) 9; Phys. Atom. Nucl. 
   {\bf 69} (2006) 1522

\bibitem{SANC}
  A.~Arbuzov, D.~Bardin, S.~Bondarenko, P.~Christova, L.~Kalinovskaya, G.~Nanava and R.~Sadykov, Eur. Phys. J. {\bf C46} (2006) 407
  
 \bibitem{CMNV}
 C.M. Carloni Calame, G. Montagna, O. Nicrosini and A. Vicini, JHEP {\bf 12} (2006) 016
 
\bibitem{zgrad2} U.~Baur, O.~Brein, 
W.~Hollik, C.~Schappacher and D. Wackeroth, Phys. Rev. {\bf D65} (2002) 033007

\bibitem{Zykunov2007} V.A. Zykunov, Phys. Rev. {\bf D75} (2007) 073019

\bibitem{HORACEZ} C.M. Carloni Calame, G. Montagna, O. Nicrosini and A. Vicini, JHEP {\bf 10} (2007) 109

\bibitem{SANCZ} A.~Arbuzov, D.~Bardin, S.~Bondarenko, P.~Christova, L.~Kalinovskaya, G.~Nanava and R.~Sadykov, arXiv:0711.0625 [hep-ph]

\bibitem{LH} C. Buttar {\it et al.}, arXiv:hep-ph/0604120

\bibitem{tev4lhc} C.E. Gerber {\it et al.},  FERMILAB-CONF-07-052, arXiv:0705.3251 [hep-ph]

\bibitem{LH2007} C. Buttar {\it et al.}, arXiv:0803.0678 [hep-ph]

\bibitem{CMNTW} C.M. Carloni Calame, G.~Montagna, O.~Nicrosini and 
M.~Treccani,  Phys. Rev. {\bf D69} (2004) 037301

\bibitem{CMNTZ} C.M. Carloni Calame G.~Montagna, O.~Nicrosini and 
M.~Treccani,  JHEP {\bf 05} (2005) 019

\bibitem{winhac} S. Jadach and W. P\l{}aczek, Eur. Phys. J.  {\bf C29} (2003) 325

\bibitem{DK2007} S. Brensing, S. Dittmaier, M. Kr\"amer and A. Muck, arXiv:0710.3309 [hep-ph]

\bibitem{baurw} U. Baur, Phys. Rev. {\bf D75} (2007) 013005

\bibitem{hkk} W. Hollik, T. Kasprzik and B.A. Kniehl, Nucl. Phys. {\bf B790} (2008) 138

\bibitem{kkp} J.H. K\"uhn, A. Kulesza and S. Pozzorini, Nucl. Phys. {\bf B797} (2008) 27

\bibitem{mmr} E. Maina, S. Moretti and D.A. Ross,  Phys. Lett. {\bf 593} (2004) 143, 
Erratum ibid. {\bf 614} (2005) 216.

\bibitem{mrst04qed} A.D. Martin, R.G. Roberts, W.J. Stirling and 
R.S. Thorne, Eur. Phys. J. {\bf C39} (2005) 155

\bibitem{zlh} U. Baur {\it et al.}, The neutral-current Drell-Yan process in the high invariant mass region,
in \cite{LH2007}

\end{thebibliography}

\begin{thebibliography}{99}

% Total rates

%\cite{sigmatotgr}
\bibitem{sigmatotgr}
  %\cite{Gorishnii:1990vf}
  S.~G.~Gorishnii, A.~L.~Kataev and S.~A.~Larin,
  %``The O (alpha-s**3) corrections to sigma-tot (e+ e- $\to$ hadrons) 
  %and Gamma (tau- $\to$ tau-neutrino + hadrons) in QCD,''
  Phys.\ Lett.\  B {\bf 259} (1991) 144;
  %%CITATION = PHLTA,B259,144;%%%\cite{Surguladze:1990tg}
  %\cite{Surguladze:1990tg}
  L.~R.~Surguladze and M.~A.~Samuel,
  %``Total Hadronic Cross-Section In E+ E- Annihilation At The Four Loop Level
  %Of Perturbative QCD,''
  Phys.\ Rev.\ Lett.\  {\bf 66} (1991) 560
  [Erratum-ibid.\  {\bf 66} (1991) 2416];
  %%CITATION = PRLTA,66,560;%%
  %\cite{Chetyrkin:1996ez}
  K.~G.~Chetyrkin,
  %``Corrections of order alpha(s)**3 to R(had) in pQCD with light gluinos,''
  Phys.\ Lett.\  B {\bf 391} (1997) 402.
  %[arXiv:hep-ph/9608480].
  %%CITATION = PHLTA,B391,402;%%

%\cite{Hamberg:1990np}
\bibitem{Hamberg:1990np}
R.~Hamberg, W.~L.~van Neerven and T.~Matsuura,
%``A Complete Calculation Of The Order Alpha-S**2 Correction To The Drell-Yan K
%Factor,''
Nucl.\ Phys.\ B {\bf 359} (1991) 343
[Erratum-ibid.\ B {\bf 644} (2002) 403];
%%CITATION = NUPHA,B359,343;%%
%
%%\cite{Zijlstra:1992qd}
%\bibitem{Zijlstra:1992qd}
E.~B.~Zijlstra and W.~L.~van Neerven,
%``Order alpha-s**2 QCD corrections to the deep inelastic proton structure
%functions F2 and F(L),''
Nucl.\ Phys.\ B {\bf 383} (1992) 525;
%%CITATION = NUPHA,B383,525;%%
%
%%\cite{Zijlstra:1992kj}
%\bibitem{Zijlstra:1992kj}
E.~B.~Zijlstra and W.~L.~van Neerven,
%``Order alpha-s**2 correction to the structure function F3 (x, Q**2) in deep
%inelastic neutrino - hadron scattering,''
Phys.\ Lett.\ B {\bf 297} (1992) 377.
%%CITATION = PHLTA,B297,377;%%

%\cite{Harlander:2002wh}
\bibitem{Higgstot}
R.~V.~Harlander and W.~B.~Kilgore,
  %``Next-to-next-to-leading order Higgs production at hadron colliders,''
  Phys.\ Rev.\ Lett.\  {\bf 88} (2002) 201801;
%[arXiv:hep-ph/0201206].
%%CITATION = HEP-PH 0201206;%%
%\cite{Anastasiou:2002yz}
%\bibitem{Anastasiou:2002yz}
C.~Anastasiou and K.~Melnikov,
%``Higgs boson production at hadron colliders in NNLO QCD,''
Nucl.\ Phys.\ B {\bf 646} (2002) 220;
%[arXiv:hep-ph/0207004].
%%CITATION = HEP-PH 0207004;%%
%\cite{Ravindran:2003um}
%\bibitem{Ravindran:2003um}
V.~Ravindran, J.~Smith and W.~L.~van Neerven,
%``NNLO corrections to the total cross section for Higgs boson production  in
%hadron hadron collisions,''
Nucl.\ Phys.\ B {\bf 665} (2003) 325.
%[arXiv:hep-ph/0302135].
%%CITATION = HEP-PH 0302135;%%


%%% Sector decomposition

%\cite{Anastasiou:2003gr}
\bibitem{Anastasiou:2003gr}
C.~Anastasiou, K.~Melnikov and F.~Petriello,
%``A new method for real radiation at NNLO,''
Phys.\ Rev.\ D {\bf 69} (2004) 076010.
%[arXiv:hep-ph/0311311].
%%CITATION = HEP-PH 0311311;%%

%\cite{sector}
\bibitem{sector}
  T.~Binoth and G.~Heinrich,
  %``An automatized algorithm to compute infrared divergent multi-loop
  %integrals,''
  Nucl.\ Phys.\ B {\bf 585} (2000) 741,
%[arXiv:hep-ph/0004013];
%%CITATION = HEP-PH 0004013;%%
%\cite{Binoth:2004jv}
%\bibitem{Binoth:2004jv}
%T.~Binoth and G.~Heinrich,
%``Numerical evaluation of phase space integrals by sector decomposition,''
Nucl.\ Phys.\  B {\bf 693} (2004) 134;
%[arXiv:hep-ph/0402265].
%%CITATION = NUPHA,B693,134;%%
%\cite{Hepp:1966eg}
%\bibitem{Hepp:1966eg}
  K.~Hepp,
  %``Proof Of The Bogolyubov-Parasiuk Theorem On Renormalization,''
  Commun.\ Math.\ Phys.\  {\bf 2} (1966) 301.
  %%CITATION = CMPHA,2,301;%%





%%% e+e-->2jets, Higgs and DY at NNLO

%\cite{Anastasiou:2004qd}
\bibitem{Anastasiou:2004qd}
  C.~Anastasiou, K.~Melnikov and F.~Petriello,
  %``Real radiation at NNLO: e+ e- --> 2jets through O(alpha(s)**2),''
  Phys.\ Rev.\ Lett.\  {\bf 93} (2004) 032002.
%  [arXiv:hep-ph/0402280].
  %%CITATION = HEP-PH 0402280;%%

%\cite{Anastasiou:2004xq}
\bibitem{Hdiff}
  C.~Anastasiou, K.~Melnikov and F.~Petriello,
  %``Higgs boson production at hadron colliders: Differential cross sections
  %through next-to-next-to-leading order,''
  Phys.\ Rev.\ Lett.\  {\bf 93} (2004) 262002,
%[arXiv:hep-ph/0409088].
  %%CITATION = HEP-PH 0409088;%%
%\cite{Anastasiou:2005qj}
%\bibitem{Anastasiou:2005qj}
%C.~Anastasiou, K.~Melnikov and F.~Petriello,
  %``Fully differential Higgs boson production and the di-photon signal through
  %next-to-next-to-leading order,''
Nucl.\ Phys.\ B {\bf 724} (2005) 197.
%  [arXiv:hep-ph/0501130].
  %%CITATION = HEP-PH 0501130;%%


%\cite{Melnikov:2006di}
\bibitem{DYdiff}
  K.~Melnikov and F.~Petriello,
  %``The W boson production cross section at the LHC through O(alpha(s)**2),''
  Phys.\ Rev.\ Lett.\  {\bf 96} (2006) 231803,
%  [arXiv:hep-ph/0603182].
%%CITATION = HEP-PH 0603182;%%
%\cite{Melnikov:2006kv}
%\bibitem{Melnikov:2006kv}
%  K.~Melnikov and F.~Petriello,
  %``Electroweak gauge boson production at hadron colliders through
  %O(alpha(s)**2),''
Phys.\ Rev.\ D {\bf 74} (2006) 114017.
%[arXiv:hep-ph/0609070].
%%CITATION = HEP-PH 0609070;%%
%\cite{Anastasiou:2005pn}

\bibitem{Anastasiou:2005pn}
  C.~Anastasiou, K.~Melnikov and F.~Petriello,
  %``The electron energy spectrum in muon decay through O(alpha**2),''
  arXiv:hep-ph/0505069.
  %%CITATION = HEP-PH/0505069;%%

% Subtraction, early

%\cite{Ellis:1980wv}
\bibitem{Ellis:1980wv}
R.~K.~Ellis, D.~A.~Ross and A.~E.~Terrano,
%``The Perturbative Calculation Of Jet Structure In E+ E- Annihilation,''
Nucl.\ Phys.\ B {\bf 178} (1981) 421.
%%CITATION = NUPHA,B178,421;%%

% Slicing early

%\cite{Fabricius:1981sx}
\bibitem{Fabricius:1981sx}
K.~Fabricius, I.~Schmitt, G.~Kramer and G.~Schierholz,
%``Higher Order Perturbative QCD Calculation Of Jet Cross-Sections In E+ E-
%Annihilation,''
Z.\ Phys.\ C {\bf 11} (1981) 315.
%%CITATION = ZEPYA,C11,315;%%

% Modern versions of subtraction and slicing

%\cite{Giele:1991vf}
\bibitem{Giele:1991vf}
W.~T.~Giele and E.~W.~N.~Glover,
%``Higher order corrections to jet cross-sections in e+ e- annihilation,''
Phys.\ Rev.\ D {\bf 46} (1992) 1980;
%%CITATION = PHRVA,D46,1980;%%
%%\cite{Giele:1993dj}
%\bibitem{Giele:1993dj}
W.~T.~Giele, E.~W.~N.~Glover and D.~A.~Kosower,
%``Higher order corrections to jet cross-sections in hadron colliders,''
Nucl.\ Phys.\ B {\bf 403} (1993) 633.
%[arXiv:hep-ph/9302225].
%%CITATION = HEP-PH 9302225;%%

%\cite{Frixione:1995ms}
\bibitem{Frixione:1995ms}
S.~Frixione, Z.~Kunszt and A.~Signer,
%``Three-jet cross sections to next-to-leading order,''
Nucl.\ Phys.\ B {\bf 467} (1996) 399;
%[arXiv:hep-ph/9512328];
%%CITATION = HEP-PH 9512328;%%
%\cite{Frixione:1997np}
%\bibitem{Frixione:1997np}
S.~Frixione,
%``A general approach to jet cross sections in QCD,''
Nucl.\ Phys.\ B {\bf 507} (1997) 295.
%[arXiv:hep-ph/9706545].
%%CITATION = HEP-PH 9706545;%%

%\cite{Catani:1996vz}
\bibitem{Catani:1996vz}
S.~Catani and M.~H.~Seymour,
%``A general algorithm for calculating jet cross sections in NLO QCD,''
Nucl.\ Phys.\ B {\bf 485} (1997) 291
[Erratum-ibid.\ B {\bf 510} (1997) 503].
%[arXiv:hep-ph/9605323].
%%CITATION = HEP-PH 9605323;%%



%%% NNLO attempts

%\cite{Kosower:1997zr}
\bibitem{Kosower}
  D.~A.~Kosower,
%``Antenna factorization of gauge-theory amplitudes,''
  Phys.\ Rev.\ D {\bf 57} (1998) 5410,
%[arXiv:hep-ph/9710213].
%%CITATION = HEP-PH 9710213;%%
%\cite{Kosower:2002su}
%\bibitem{Kosower:2002su}
%D.~A.~Kosower,
%``Multiple singular emission in gauge theories,''
Phys.\ Rev.\ D {\bf 67} (2003) 116003,
%[arXiv:hep-ph/0212097].
%%CITATION = PHRVA,D67,116003;%%
%\cite{Kosower:2003bh}
%\bibitem{Kosower:2003bh}
%  D.~A.~Kosower,
%``Antenna factorization in strongly-ordered limits,''
Phys.\ Rev.\ D {\bf 71} (2005) 045016.
%[arXiv:hep-ph/0311272].
%%CITATION = HEP-PH 0311272;%%


%\cite{Weinzierl:2003fx}
\bibitem{Weinzierl}
S.~Weinzierl,
%``Subtraction terms at NNLO,''
JHEP {\bf 0303} (2003) 062,
%  [arXiv:hep-ph/0302180].
%%CITATION = HEP-PH 0302180;%%
%\cite{Weinzierl:2003ra}
%\bibitem{Weinzierl:2003ra}
%  S.~Weinzierl,
%``Subtraction terms for one-loop amplitudes with one unresolved parton,''
JHEP {\bf 0307} (2003) 052.
%[arXiv:hep-ph/0306248].
%%CITATION = HEP-PH 0306248;%%


%\cite{Frixione:2004is}
\bibitem{Frixione:2004is}
  S.~Frixione and M.~Grazzini,
  %``Subtraction at NNLO,''
  JHEP {\bf 0506} (2005) 010.
%[arXiv:hep-ph/0411399].
%%CITATION = HEP-PH 0411399;%%


%\cite{Gehrmann-DeRidder:2005hi}
\bibitem{GGG}
A.~Gehrmann-De Ridder, T.~Gehrmann and E.~W.~N.~Glover,
%``Quark-gluon antenna functions from neutralino decay,''
Phys.\ Lett.\ B {\bf 612} (2005) 36,
%[arXiv:hep-ph/0501291].
%%CITATION = HEP-PH 0501291;%%
%\cite{Gehrmann-DeRidder:2005aw}
%\bibitem{Gehrmann-DeRidder:2005aw}
%A.~Gehrmann-De Ridder, T.~Gehrmann and E.~W.~N.~Glover,
%``Gluon gluon antenna functions from Higgs boson decay,''
Phys.\ Lett.\ B {\bf 612} (2005) 49,
%[arXiv:hep-ph/0502110].
%%CITATION = HEP-PH 0502110;%%
%\cite{Gehrmann-DeRidder:2005cm}
%\bibitem{Gehrmann-DeRidder:2005cm}
%A.~Gehrmann-De Ridder, T.~Gehrmann and E.~W.~N.~Glover,
%``Antenna subtraction at NNLO,''
JHEP {\bf 0509} (2005) 056;
%[arXiv:hep-ph/0505111].
%%CITATION = HEP-PH 0505111;%%
%\cite{Daleo:2006xa}
%\bibitem{Daleo:2006xa}
A.~Daleo, T.~Gehrmann and D.~Maitre,
%``Antenna subtraction with hadronic initial states,''
hep-ph/0612257.
%%CITATION = HEP-PH 0612257;%%


%\cite{Somogyi:2005xz}
\bibitem{ST}
  G.~Somogyi, Z.~Trocsanyi and V.~Del Duca,
  %``Matching of singly- and doubly-unresolved limits of tree-level QCD  squared
  %matrix elements,''
  JHEP {\bf 0506} (2005) 024,
%[arXiv:hep-ph/0502226].
%%CITATION = JHEPA,0506,024;%%
%\cite{Somogyi:2006da}
%\bibitem{Somogyi:2006da}
%G.~Somogyi, Z.~Trocsanyi and V.~Del Duca,
%``A subtraction scheme for computing QCD jet cross sections at NNLO:
%Regularization of doubly-real emissions,''
JHEP {\bf 0701} (2007) 070;
%[arXiv:hep-ph/0609042].
%%CITATION = JHEPA,0701,070;%%
%\cite{Somogyi:2006db}
%\bibitem{Somogyi:2006db}
  G.~Somogyi and Z.~Trocsanyi,
  %``A subtraction scheme for computing QCD jet cross sections at NNLO:
  %Regularization of real-virtual emission,''
  JHEP {\bf 0701} (2007) 052.
%[arXiv:hep-ph/0609043].
%%CITATION = JHEPA,0701,052;%%



%%% 2 jets at NNLO

%\cite{Gehrmann-DeRidder:2004tv}
\bibitem{Gehrmann-DeRidder:2004tv}
  A.~Gehrmann-De Ridder, T.~Gehrmann and E.~W.~N.~Glover,
  %``Infrared structure of e+ e- --> 2jets at NNLO,''
  Nucl.\ Phys.\ B {\bf 691} (2004) 195.
%[arXiv:hep-ph/0403057].
%%CITATION = HEP-PH 0403057;%%

%\cite{Weinzierl:2006ij}
\bibitem{Weinzierl:2006ij}
  S.~Weinzierl,
  %``NNLO corrections to 2-jet observables in electron positron annihilation,''
  Phys.\ Rev.\ D {\bf 74} (2006) 014020.
%[arXiv:hep-ph/0606008].
%%CITATION = HEP-PH 0606008;%%

%%% 3 jets at NNLO



%\cite{Gehrmann-DeRidder:2006ez}
\bibitem{Gehrmann-DeRidder:2006ez}
A.~Gehrmann-De Ridder, T.~Gehrmann, E.~W.~N.~Glover and G.~Heinrich,
%``Infrared structure of e+ e- --> 3jets at NNLO: QED-type contributions,''
Nucl.\ Phys.\ Proc.\ Suppl.\  {\bf 160} (2006) 190.
%[arXiv:hep-ph/0607042].
%%CITATION = NUPHZ,160,190;%%

%\cite{GehrmannDeRidder:2007hr}
\bibitem{GehrmannDeRidder:2007hr}
  A.~Gehrmann-De Ridder, T.~Gehrmann, E.~W.~N.~Glover and G.~Heinrich,
  %``NNLO corrections to event shapes in $e^+e^-$ annihilation,''
  JHEP {\bf 0712} (2007) 094
  [arXiv:0711.4711 [hep-ph]].
  %%CITATION = JHEPA,0712,094;%%

%%%

%\cite{Catani:2007vq}
\bibitem{Catani:2007vq}
  S.~Catani and M.~Grazzini,
  %``An NNLO subtraction formalism in hadron collisions and its application   to
  %Higgs boson production at the LHC,''
  Phys.\ Rev.\ Lett.\  {\bf 98} (2007) 222002
  [arXiv:hep-ph/0703012].
  %%CITATION = PRLTA,98,222002;%%

%\cite{Grazzini:2008tf}
\bibitem{Grazzini:2008tf}
  M.~Grazzini,
  %``NNLO predictions for the Higgs boson signal in the H->WW->lnu lnu and
  %H->ZZ->4l decay channels,''
  JHEP {\bf 0802} (2008) 043
  [arXiv:0801.3232 [hep-ph]].
  %%CITATION = JHEPA,0802,043;%%

%%% Veto

%\cite{Catani:2001cr}
\bibitem{Catani:2001cr}
S.~Catani, D.~de Florian and M.~Grazzini,
  %``Direct Higgs production and jet veto at the Tevatron and the LHC in  NNLO
  %QCD,''
  JHEP {\bf 0201} (2002) 015.
%  [arXiv:hep-ph/0111164].
%%CITATION = HEP-PH 0111164;%%

%%% qt resummation

%\cite{Parisi:1979se}
\bibitem{Parisi:1979se}
  G.~Parisi and R.~Petronzio,
  %``Small Transverse Momentum Distributions In Hard Processes,''
  Nucl.\ Phys.\  B {\bf 154} (1979) 427.
  %%CITATION = NUPHA,B154,427;%%

%\cite{Collins:1984kg}
\bibitem{Collins:1984kg}
  J.~C.~Collins, D.~E.~Soper and G.~Sterman,
  %``Transverse Momentum Distribution In Drell-Yan Pair And W And Z Boson
  %Production,''
  Nucl.\ Phys.\  B {\bf 250} (1985) 199.
  %%CITATION = NUPHA,B250,199;%%

%\cite{Catani:2000vq}
\bibitem{Catani:2000vq}
  S.~Catani, D.~de Florian and M.~Grazzini,
  %``Universality of non-leading logarithmic contributions in transverse
  %momentum distributions,''
  Nucl.\ Phys.\  B {\bf 596} (2001) 299.
%  [arXiv:hep-ph/0008184].
  %%CITATION = NUPHA,B596,299;%%


%\cite{Bozzi:2003jy}
\bibitem{qtresum}
  G.~Bozzi, S.~Catani, D.~de Florian and M.~Grazzini,
  %``The q(T) spectrum of the Higgs boson at the LHC in QCD perturbation
  %theory,''
  Phys.\ Lett.\ B {\bf 564} (2003) 65,
%  [arXiv:hep-ph/0302104].
%%CITATION = HEP-PH 0302104;%%
%\cite{Bozzi:2005wk}
%\bibitem{Bozzi:2005wk}
%  G.~Bozzi, S.~Catani, D.~de Florian and M.~Grazzini,
  %``Transverse-momentum resummation and the spectrum of the Higgs boson at the
  %LHC,''
  Nucl.\ Phys.\ B {\bf 737} (2006) 73,
%  [arXiv:hep-ph/0508068].
%%CITATION = HEP-PH 0508068;%%
%\cite{Bozzi:2007pn}
%\bibitem{Bozzi:2007pn}
%  G.~Bozzi, S.~Catani, D.~de Florian and M.~Grazzini,
  %``Higgs boson production at the LHC: transverse-momentum resummation and
  %rapidity dependence,''
  Nucl.\ Phys.\  B {\bf 791} (2008) 1.
%  [arXiv:0705.3887 [hep-ph]].
  %%CITATION = NUPHA,B791,1;%%

% WW

\bibitem{ww}
  M.~Grazzini,
  %``Soft-gluon effects in W W production at hadron colliders,''
  JHEP {\bf 0601} (2006) 095.
%  [arXiv:hep-ph/0510337].
%%CITATION = JHEPA,0601,095;%%



%\cite{deFlorian:2000pr}
\bibitem{deFlorian:2000pr}
D.~de Florian and M.~Grazzini,
%``Next-to-next-to-leading logarithmic corrections at small transverse
%momentum in hadronic collisions,''
  Phys.\ Rev.\ Lett.\  {\bf 85} (2000) 4678,
%  [arXiv:hep-ph/0008152].
%%CITATION = HEP-PH 0008152;%%
%\cite{deFlorian:2001zd}
%\bibitem{deFlorian:2001zd}
%  D.~de Florian and M.~Grazzini,
%``The structure of large logarithmic corrections at small transverse
%momentum in hadronic collisions,''
  Nucl.\ Phys.\ B {\bf 616} (2001) 247.
%[arXiv:hep-ph/0108273].
%%CITATION = HEP-PH 0108273;%%

\bibitem{inprep}
S.~Catani, M. Grazzini, to appear.

% H+jet

%\cite{deFlorian:1999zd}
\bibitem{deFlorian:1999zd}
  D.~de Florian, M.~Grazzini and Z.~Kunszt,
  %``Higgs production with large transverse momentum in hadronic collisions  at
  %next-to-leading order,''
  Phys.\ Rev.\ Lett.\  {\bf 82} (1999) 5209;
%  [arXiv:hep-ph/9902483].
%%CITATION = PRLTA,82,5209;%%
see also J. Campbell and R.K. Ellis,
{\em MCFM - Monte Carlo for FeMtobarn processes}, {\tt http://mcfm.fnal.gov}.


%\cite{Martin:2004ir}
\bibitem{Martin:2004ir}
  A.~D.~Martin, R.~G.~Roberts, W.~J.~Stirling and R.~S.~Thorne,
  %``Physical gluons and high-E(T) jets,''
  Phys.\ Lett.\ B {\bf 604} (2004) 61.
%[arXiv:hep-ph/0410230].
%%CITATION = HEP-PH 0410230;%%

\bibitem{CMStdr}
CMS collaboration, {\em CMS Physics, Technical Design Report, Vol.~II 
Physics Performance},
report CERN/LHCC 2006-021.

%\cite{Catani:1997xc}
\bibitem{Catani:1997xc}
  S.~Catani and B.~R.~Webber,
  %``Infrared safe but infinite: Soft-gluon divergences inside the physical
  %region,''
  JHEP {\bf 9710} (1997) 005.
%[arXiv:hep-ph/9710333].
%%CITATION = JHEPA,9710,005;%%

%\cite{Davatz:2004zg}
\bibitem{Davatz:2004zg}
  G.~Davatz, G.~Dissertori, M.~Dittmar, M.~Grazzini and F.~Pauss,
  %``Effective K-factors for g g --> H --> W W --> l nu l nu at the LHC,''
  JHEP {\bf 0405} (2004) 009.
%[arXiv:hep-ph/0402218].
%%CITATION = JHEPA,0405,009;%%

\bibitem{ktalg}
%\cite{Catani:1993hr}
%\bibitem{Catani:1993hr}
  S.~Catani, Y.~L.~Dokshitzer, M.~H.~Seymour and B.~R.~Webber,
  %``Longitudinally Invariant K(T) Clustering Algorithms For Hadron Hadron
  %Collisions,''
  Nucl.\ Phys.\  B {\bf 406} (1993) 187;
  %%CITATION = NUPHA,B406,187;%%
%\cite{Ellis:1993tq}
%\bibitem{Ellis:1993tq}
  S.~D.~Ellis and D.~E.~Soper,
  %``Successive combination jet algorithm for hadron collisions,''
  Phys.\ Rev.\  D {\bf 48} (1993) 3160.
  %[arXiv:hep-ph/9305266].
  %%CITATION = PHRVA,D48,3160;%%  





%\cite{Anastasiou:2007mz}
\bibitem{Anastasiou:2007mz}
  C.~Anastasiou, G.~Dissertori and F.~Stockli,
  %``NNLO QCD predictions for the H -> WW -> l l nu nu signal at the LHC,''
  JHEP {\bf 0709} (2007) 018.
%[arXiv:0707.2373 [hep-ph]].
%%CITATION = JHEPA,0709,018;%%

\bibitem{hnnloweb}
{\tt http://theory.fi.infn.it/grazzini/codes.html}

\end{thebibliography}

\begin{thebibliography}{99}

\bibitem{CSS}
  J.~C.~Collins, D.~E.~Soper and G.~Sterman,
  % ``Factorization of Hard Processes in QCD,''
  Adv.\ Ser.\ Direct.\ High Energy Phys.\  {\bf 5} (1988) 1.
%  [hep-ph/0409313].
  %% CITATION = HEP-PH 0409313;%%

\bibitem{exponentiation}
  Y.~L.~Dokshitzer, D.~Diakonov and S.~I.~Troyan,
  %``Hard Processes In Quantum Chromodynamics,''
  Phys.\ Rept.\  {\bf 58} (1980) 269;
  %%CITATION = PRPLC,58,269;%%
  J.~G.~M.~Gatheral,
  %``Exponentiation Of Eikonal Cross-Sections In Nonabelian Gauge Theories,''
  Phys.\ Lett.\ B {\bf 133} (1983) 90; 
%
  A.~Bassetto, M.~Ciafaloni and G.~Marchesini,
  %``Jet Structure And Infrared Sensitive Quantities In Perturbative QCD,''
  Phys.\ Rept.\ {\bf 100} (1983) 201.
  %%CITATION = PRPLC,100,201;%%
  %%CITATION = PHLTA,B133,90;%%
  
\bibitem{inclusive}
  S.~Catani, D.~de Florian, M.~Grazzini and P.~Nason,
  %``Soft-gluon resummation for Higgs boson production at hadron colliders,''
  JHEP {\bf 0307} (2003) 028;
%  [hep-ph/0306211].
  %%CITATION = HEP-PH 0306211;%%
  G.~Bozzi, S.~Catani, D.~de Florian and M.~Grazzini,
%   ``Transverse-momentum resummation and the spectrum of the Higgs boson at the
  %LHC,''
  Nucl.\ Phys.\ B {\bf 737}(2006)73.
%  [hep-ph/0508068].
  %%CITATION = HEP-PH 0508068;%%

\bibitem{nnherwig}
  G.~Corcella {\it et al.},
  %``HERWIG 6: An event generator for hadron emission reactions with
  %interfering gluons (including supersymmetric processes),''
  JHEP {\bf 0101} (2001) 010.
%  [arXiv:hep-ph/0011363].
  %%CITATION = JHEPA,0101,010;%%

\bibitem{pythia-old}
  T.~Sjostrand, S.~Mrenna and P.~Skands,
  %``PYTHIA 6.4 physics and manual,''
  JHEP {\bf 0605} (2006) 026.
%  [hep-ph/0603175].
  %%CITATION = JHEPA,0605,026;%%

\bibitem{pythia-new}
  T.~Sjostrand and P.~Z.~Skands,
  %``Transverse-momentum-ordered showers and interleaved multiple
  %interactions,''
  Eur.\ Phys.\ J.\  C {\bf 39} (2005) 129.
%  [hep-ph/0408302].
  %%CITATION = EPHJA,C39,129;%%

\bibitem{ariadne}
  L.~Lonnblad,
  %``Ariadne Version 4: A Program For Simulation Of QCD Cascades Implementing
  %The Color Dipole Model,''
  Comput.\ Phys.\ Commun.\  {\bf 71} (1992) 15.
  %%CITATION = CPHCB,71,15;%%

\bibitem{DS-review}
  M.~Dasgupta and G.~P.~Salam,
  %``Event shapes in e+ e- annihilation and deep inelastic scattering,''
  J.\ Phys.\ G {\bf 30} (2004) R143.
%  [arXiv:hep-ph/0312283].
  %%CITATION = HEP-PH 0312283;%%

\bibitem{caesar}
  A.~Banfi, G.~P.~Salam and G.~Zanderighi,
%   ``Principles of general final-state resummation and automated
  %implementation,''
  JHEP {\bf 0503} (2005) 073.
%  [hep-ph/0407286].
  %%CITATION = HEP-PH 0407286;%%

\bibitem{etflow}
   M.~Dasgupta and G.~P.~Salam,
  %``Resummation of non-global QCD observables,''
  Phys.\ Lett.\  B {\bf 512} (2001) 323;\\
%  [arXiv:hep-ph/0104277].
  %%CITATION = PHLTA,B512,323;%%
% M.~Dasgupta and G.~P.~Salam,
  %``Accounting for coherence in interjet E(t) flow: A case study,''
  JHEP {\bf 0203} (2002) 017.
%  [hep-ph/0203009].
  %%CITATION = HEP-PH 0203009;%%
 
\bibitem{etflow-mc}
  A.~Banfi, G.~Corcella and M.~Dasgupta,
  %``Angular ordering and parton showers for non-global QCD observables,''
  JHEP {\bf 0703} (2007) 050.
%  [hep-ph/0612282].
  %%CITATION = JHEPA,0703,050;%%

\bibitem{etflow-AS}
  R.~B.~Appleby and M.~H.~Seymour,
%   ``Non-global logarithms in inter-jet energy flow with kt clustering
  %requirement,''
  JHEP {\bf 0212} (2002) 063.
%  [hep-ph/0211426];\\
  %%CITATION = HEP-PH 0211426;%%

\bibitem{etflow-BD}
 A.~Banfi and M.~Dasgupta,
  %``Problems in resumming interjet energy flows with k(t) clustering,''
  Phys.\ Lett.\ B {\bf 628} (2005) 49.
%  [hep-ph/0508159];\\
  %%CITATION = HEP-PH 0508159;%%


\bibitem{superleading}
  J.~R.~Forshaw, A.~Kyrieleis and M.~H.~Seymour,
  %``Super-leading logarithms in non-global observables in QCD,''
  JHEP {\bf 0608} (2006) 059.
%  [hep-ph/0604094].
  %%CITATION = HEP-PH 0604094;%%

\bibitem{hh-shapes}
  A.~Banfi, G.~P.~Salam and G.~Zanderighi,
  %``Resummed event shapes at hadron hadron colliders,''
  JHEP {\bf 0408} (2004) 062.
%  [hep-ph/0407287].
  %%CITATION = JHEPA,0408,062;%%

\bibitem{nagy}
  Z.~Nagy,
  %``Next-to-leading order calculation of three-jet observables in hadron
  %hadron collision,''
  Phys.\ Rev.\  D {\bf 68} (2003) 094002.
%  [hep-ph/0307268].
  %%CITATION = PHRVA,D68,094002;%%

\end{thebibliography}

\begin{thebibliography}{99}
\bibitem{sld}
SLD Collaboration, K. Abe et al., Phys. Rev. Lett. 84 (2000) 4300.

\bibitem{aleph}
ALEPH Collaboration, A. Heister et al., Phys. Lett. B512 (2001) 30.

\bibitem{opal}
OPAL Collaboration, G. Abbiendi et al., Eur. Phys. J. C29 (2003) 463.

\bibitem{delphi}
DELPHI Collaboration, G. Barker et al., DELPHI 2002-069, CONF 603.

\bibitem{shirkov}
D. Shirkov, Nucl. Phys. Proc. Suppl. 152 (2006) 51.


\bibitem{ugo}
U. Aglietti, G. Corcella and G. Ferrera, Nucl. Phys. B775 (2007) 162.


\bibitem{gianca} 
G. Corcella and G. Ferrera,  JHEP 0712 (2007) 029.

\bibitem{mele}
B. Mele and P. Nason, Nucl. Phys. B   361 (1991) 626.

\bibitem{ap}
G. Altarelli and G. Parisi, Nucl. Phys. B126 (1977) 298.

\bibitem{dgl}
V.N. Gribov and L.N. Lipatov, Sov. J. Nucl. Phys. 15 (1972) 438;\\
L.N. Lipatov, Sov. J. Nucl. Phys. 20 (1975) 95;\\
Yu.L. Dokshitzer, Sov. Phys. 46 (1977) 641.

\bibitem{cc}
M. Cacciari and S. Catani, Nucl.  Phys.  B617 (2001) 253.


\bibitem{her}
G. Corcella, I.G. Knowles, G. Marchesini, S. Moretti, K. Odagiri,
P. Richardson, M.H. Seymour, B.R. Webber, JHEP 0101 (2001) 010.

\bibitem{nnpythia}
 T.~Sjostrand, S.~Mrenna and P.~Skands, JHEP 0605 (2006) 026.

\bibitem{nmarweb}
G. Marchesini and B.R. Webber, Nucl. Phys. B238 (1984) 1;  
ibid. B310 (1988) 461.


\bibitem{kart}
V.G. Kartvelishvili, A.K. Likehoded and V.A. Petrov, Phys. Lett. B78 (1978)
615.



\bibitem{ncluster}
B.R. Webber, Nucl. Phys. B238 (1984) 492.

\bibitem{nstring}
B. Andersson, G. Gustafson, G. Ingelman, T. Sj\"ostrand,
Phys. Rept. 97 (1983) 31.






\bibitem{skands}
T. Sj\"ostrand and P.Z. Skands, Eur. Phys. J. C39 (2005) 129.

\bibitem{banfi}
A. Banfi, G. Corcella and M. Dasgupta, JHEP 0703 (2007) 050.

\bibitem{ncmw}
S. Catani, G. Marchesini and B.R. Webber, Nucl. Phys. B349 (1991) 635.

\bibitem{cv}
G. Corcella and V. Drollinger, Nucl. Phys. B730 (2005) 82.

\bibitem{bowler}
M.G. Bowler, Z. Phys. C11 (1981) 169.


\bibitem{top}
G. Corcella and A.D. Mitov, Nucl. Phys. B623 (2002) 247.

\bibitem{top1}
M. Cacciari, G. Corcella and A.D. Mitov, JHEP  0212  (2002) 015.

\bibitem{hbb}
G. Corcella, Nucl. Phys. B  705 (2005) 363
[Erratum-ibid. B 715 (2005) 609].

\bibitem{canas}
M. Cacciari and P. Nason, Phys. Rev. Lett. 89 (2002) 122003.

\end{thebibliography}

\begin{thebibliography}{99}                                            

\bibitem{Hoche:2006ph} S.~Hoche, F.~Krauss, N.~Lavesson, L.~Lonnblad,
  M.~Mangano, A.~Schalicke and S.~Schumann, 
%``Matching parton showers and matrix elements,'' 
arXiv:hep-ph/0602031.  
%%CITATION = HEP-PH 0602031;%%
\bibitem{Frixione:2006he}
  S.~Frixione and B.~R.~Webber,
  %``The MC@NLO 3.2 event generator,''
  arXiv:hep-ph/0601192.
  %%CITATION = HEP-PH/0601192;%%
%\cite{Frixione:2002ik}
\bibitem{Frixione:2002ik}
  S.~Frixione and B.~R.~Webber,
  %``Matching NLO QCD computations and parton shower simulations,''
  JHEP {\bf 0206} (2002) 029
  [arXiv:hep-ph/0204244].
  %%CITATION = HEP-PH 0204244;%%
%\cite{Nason:2004rx}
\bibitem{Frixione:2003ei}
  S.~Frixione, P.~Nason and B.~R.~Webber,
  %``Matching NLO QCD and parton showers in heavy flavour production,''
  JHEP {\bf 0308} (2003) 007
  [arXiv:hep-ph/0305252].
  %%CITATION = HEP-PH 0305252;%%
\bibitem{mlmfnal}
M.L. Mangano, presentation at the FNAL Matrix Element/Monte Carlo
Tuning Working Group, 15 Nov 2002,
http://www-cpd.fnal.gov/personal/mrenna/tuning/nov2002/mlm.pdf .
%\cite{Mangano:2006rw}
\bibitem{Mangano:2006rw}
  M.~L.~Mangano, M.~Moretti, F.~Piccinini and M.~Treccani,
  %``Matching matrix elements and shower evolution for top-quark production in
  %hadronic collisions,''
  JHEP {\bf 0701}, 013 (2007)
  [arXiv:hep-ph/0611129].
  %%CITATION = JHEPA,0701,013;%%
\bibitem{Mangano:2002ea}
  M.~L.~Mangano, M.~Moretti, F.~Piccinini, R.~Pittau and A.~D.~Polosa,
%   ``ALPGEN, a generator for hard multiparton processes in hadronic
%   collisions,''
  JHEP {\bf 0307} (2003) 001
  [arXiv:hep-ph/0206293];\\
%%CITATION = HEP-PH 0206293;%%
%\bibitem{Caravaglios:1999yr}
F.~Caravaglios, M.~L.~Mangano, M.~Moretti and R.~Pittau,
%``A new approach to multijet calculations in hadron collisions,''
Nucl.\ Phys.\ B {\bf 539} (1999) 215
[hep-ph/9807570].
%%CITATION = HEP-PH 9807570;%%
\bibitem{Marchesini:1988cf}
G.~Marchesini and B.~R.~Webber,
%``Monte Carlo Simulation Of General Hard Processes With Coherent QCD Radiation,''
Nucl.\ Phys.\ B {\bf 310} (1988) 461.
%%CITATION = NUPHA,B310,461;%%
%\cite{Marchesini:1992ch}
%\bibitem{Marchesini:1992ch}
G.~Marchesini, B.~R.~Webber, G.~Abbiendi, I.~G.~Knowles, M.~H.~Seymour and L.~Stanco,
%``HERWIG: A Monte Carlo event generator for simulating hadron emission reactions with interfering gluons. Version 5.1 - April 1991,''
Comput.\ Phys.\ Commun.\  {\bf 67} (1992) 465.
%%CITATION = CPHCB,67,465;%%
%\cite{Corcella:2001bw}
%\bibitem{Corcella:2001bw}
G.~Corcella {\it et al.},
%``HERWIG 6: An event generator for hadron emission reactions with  interfering gluons (including supersymmetric processes),''
JHEP {\bf 0101} (2001) 010
[hep-ph/0011363].  G.~Corcella {\it et al.},
  ``HERWIG 6.5 release note,''
  arXiv:hep-ph/0210213.
%%CITATION = HEP-PH 0011363;%%
%\cite{Corcella:2002jc}
%\bibitem{Corcella:2002jc}

%  %%CITATION = HEP-PH 0210213;%%

%\cite{Martin:2001es}
\bibitem{Martin:2001es}
  A.~D.~Martin, R.~G.~Roberts, W.~J.~Stirling and R.~S.~Thorne,
%``MRST2001: Partons and alpha(s) from precise deep inelastic scattering  and
%Tevatron jet data,''
  Eur.\ Phys.\ J.\ C {\bf 23} (2002) 73
  [arXiv:hep-ph/0110215].
  %%CITATION = HEP-PH 0110215;%%
%\cite{getjet}
\bibitem{getjet}
F.E.~Paige and S.D.~Protopopescu, in {\it Physics of the SSC}, Snowmass, 1986, 
Colorado, edited by R. Donaldson and J.~Marx.
\bibitem{Nason:2004rx}
  P.~Nason,
  %``A new method for combining NLO QCD with shower Monte Carlo algorithms,''
  JHEP {\bf 0411} (2004) 040
  [arXiv:hep-ph/0409146].
  %%CITATION = HEP-PH 0409146;%%
\bibitem{Nason:2006hf}
  P.~Nason and G.~Ridolfi,
%   ``A positive-weight next-to-leading-order Monte Carlo for Z pair
  %hadroproduction,''
  JHEP {\bf 0608} (2006) 077
  [arXiv:hep-ph/0606275].
  %%CITATION = HEP-PH 0606275;%%
%\cite{Frixione:2003ei}
\bibitem{Nason:2006XX}
  P.~Nason, presented in this Meeting
%  ``Workshop sui Monte Carlo, la fisica e le simulazioni a LHC'',
%  Frascati, 23-25 October 2006,
% {\tt  http://moby.mib.infn.it/\~nason/mcws3/Nason-10-2006.pdf}.
\end{thebibliography}

\begin{thebibliography}{99}

\bibitem{nherwig}
G. Corcella, I.G. Knowles, G. Marchesini, S. Moretti, K. Odagiri,
P. Richardson, M.H. Seymour, B.R. Webber, JHEP 0101 (2001) 010.

\bibitem{npythia}
T. Sjostrand, S. Mrenna and P. Skands, JHEP 0605 (2006) 026.


\bibitem{hqt}
G. Bozzi, S. Catani, D. de Florian and M. Grazzini, Phys. Lett. B564 
(2003) 65; Nucl. Phys. B 737 (2006) 73.


\bibitem{marweb}
G. Marchesini and B.R. Webber, Nucl. Phys. B238 (1984) 1;
Nucl. Phys. B310 (1988) 461.

\bibitem{beng}
H.U. Bengtsson and G. Ingelman, Comput. Phys. Commun. 34 (1985) 251.

\bibitem{mike}
M.H. Seymour, Comput. Phys. Commun. 90 (1995) 95.

\bibitem{miu}
G. Miu and T. Sjostrand, Phys. Lett. B449 (1999) 313.

\bibitem{cluster}
B.R. Webber, Nucl. Phys. B238 (1984) 492.

\bibitem{string}
B. Andersson, G. Gustafson, G. Ingelman, T. Sj\"ostrand,
Phys. Rept. 97 (1983) 31

\bibitem{cor}
G. Corcella, {\it Bottom-quark fragmentation: resummations and
Monte Carlo simulations}, these proceedings.

\bibitem{cteq}
J. Pumplin, D.R. Stump, J. Huston, H.L. Lai and P. Nadolsky, JHEP 0207 (2002) 
012.

\bibitem{pdg}
W.M. Yao et al., J. Phys. G33, (2006) 1.

\bibitem{nnlo}
C. Anastasiou and K. Melnikov, Nucl. Phys. B646 (2002) 220;\\
V. Ravindran, J. Smith and W.L. van Neerven, Nucl. Phys. B634 (2002) 247.


\bibitem{cmw}
S. Catani, G. Marchesini and B.R. Webber, Nucl. Phys. B349 (1991) 635.

\bibitem{hdec}
A. Djouadi, J. Kalinowski and M. Spira, Comput. Phys. Commun. 108
(1998) 56.

\bibitem{sey} 
M.H. Seymour, private communication.

\bibitem{correb} 
G. Corcella and D. Rebuzzi, work in progress.

\bibitem{baur}
U. Baur and E.W.N. Glover, Nucl. Phys. B339 (1990) 38.

\bibitem{spira}
M. Spira, A. Djouadi, D. Graudenz and P.M. Zerwas, Nucl. Phys. B453 (1995) 17..


\bibitem{cormor}
G. Corcella and S. Moretti, Phys. Lett. B590 (2004) 249.

\bibitem{corsey}
G. Corcella and M.H. Seymour, Nucl. Phys. B565 (2000) 227.

\bibitem{mcnlo}
S. Frixione and B.R. Webber, JHEP 0206 (2002) 029.



\end{thebibliography}

\begin{thebibliography}{99}
%\cite{Deile:2006tt}


\bibitem{Boos:2002}
  E.Boos, A.Djouadi, M.Muhlleitner, and A.Vologdin,
  ``The MSSM Higgs bosons in the intense-coupling regime,''
  Phys.Rev. D66(2002) 055004,
  arXiv:hep-ph/0205160.

\bibitem{Boos:2004}
  E.Boos, A.Djouadi, and A.Nikitenko,
  ``Detection of the MSSM Higgs bosons in the intense-coupling regime at the LHC,''
  Phys.Lett. B578(2004) 384-393,
  arXiv:hep-ph/0307079.

\bibitem{LEP1}
  ALEPH, DELPHI, L3, OPAL Collaboration, LEP Working Group for Higgs Boson Searches,
  ``Search for neutral MSSM Higgs bosons at LEP,''
  arXiv:hep-ex/0602042.

\bibitem{LEP2}
  LEP Working Group for Higgs Boson Searches,
  ``Search for Charged Higgs bosons: Preliminary combined results using LEP data 
  collected at energies up to 209 GeV,''
  arXiv:hep-ex/0107031.
  

\bibitem{PTDR2_Higgs}
  D.Acosta {\it et al.}  [CMS Collaboration],
  ``CMS Coll. Physics TDR Vol.II, Chapter 11,''
  CERN/LHCC 2006-021 (2006).
%  arXiv:hep-ph/0307079.
  
  %
%\bibitem{Deile:2006tt}
%  M.~Deile {\it et al.}  [TOTEM Collaboration],
% ``Diffraction and total cross-section at the Tevatron and the LHC,''
%  arXiv:hep-ex/0602021.
%  %%CITATION = HEP-EX 0602021;%%
%
%%\cite{Dobbs:2004bu}
%\bibitem{Dobbs:2004bu}
%  M.~Dobbs {\it et al.},
%  ``The QCD/SM working group: Summary report,''
%  arXiv:hep-ph/0403100.
%  %%CITATION = HEP-PH 0403100;%%\end{thebibliography}

\end{thebibliography}

\begin{thebibliography}{99}
\bibitem{EWWG1} 
  ALEPH, DELPHI, L3 and OPAL Collaborations and the LEP Electroweak Working 
  Group, \textit{A combination of preliminary Electroweak measurements and 
  constraints on the Standard Model}, [hep-ex/0612034] \\
%  \url{http://lepewwg.web.cern.ch/LEPEWWG}.
  {\tt http://lepewwg.web.cern.ch/LEPEWWG}
\bibitem{EWWG2} 
  ALEPH, DELPHI, L3 and OPAL Collaborations and the LEP Electroweak Working 
  Group for Higgs Boson Searches, \textit{Phys. Lett.} \textbf{B565} (2003) 61.
\bibitem{Rattazzi:2005di}
  R.~Rattazzi,
  %``Physics Beyond the Standard Model,''
  PoS {\bf HEP2005} (2006) 399,
  [hep-ph/0607058].
  %%CITATION = POSCI,HEP2005,399;%%
\bibitem{Giudice:2007fh}
  G.~F.~Giudice, C.~Grojean, A.~Pomarol and R.~Rattazzi,
  %``The strongly-interacting light Higgs,''
  hep-ph/0703164.
  %%CITATION = HEP-PH/0703164;%%
\bibitem{Phantom} 
  A.~Ballestrero, A.~Belhouari, G.~Bevilacqua and E.~Maina, in preparation.
\bibitem{Accomando:2005cc}
  E.~Accomando, A.~Ballestrero and E.~Maina,
  %``PHASE, a Monte Carlo event generator for six-fermion physics at the  LHC,''
  JHEP {\bf 0507} (2005) 016,
  [hep-ph/0504009].
  %%CITATION = JHEPA,0507,016;%%
\bibitem{VEGAS} 
  G.P.~Lepage, \textit{Jour. Comp. Phys.} 27 (1978) 192.
\bibitem{Kleiss:1986xp}
  R.~Kleiss and W.~J.~Stirling,
  %``ANOMALOUS HIGH-ENERGY BEHAVIOR IN BOSON FUSION,''
  Phys.\ Lett.\  B {\bf 182} (1986) 75.
  %%CITATION = PHLTA,B182,75;%%
\bibitem{Accomando:2006mc}
E.~Accomando, A.~Ballestrero, A.~Belhouari and E.~Maina,
  %``Isolating vector boson scattering at the LHC: Gauge cancellations and the
  %equivalent vector boson approximation vs complete calculations,''
  Phys.\ Rev.\  D {\bf 74} (2006) 073010,
  [hep-ph/0608019].
  %%CITATION = PHRVA,D74,073010;%%
\bibitem{Accomando:2005hz}
  E.~Accomando, A.~Ballestrero, S.~Bolognesi, E.~Maina and C.~Mariotti,
  %``Boson boson scattering and Higgs production at the LHC from a six fermion
  %point of view: Four jets + l nu processes at O(alpha(em)**6),''
  JHEP {\bf 0603} (2006) 093,
  [hep-ph/0512219].
  %%CITATION = JHEPA,0603,093;%%
\bibitem{nMangano:2002ea}
  M.~L.~Mangano, M.~Moretti, F.~Piccinini, R.~Pittau and A.~D.~Polosa,
  %``ALPGEN, a generator for hard multiparton processes in hadronic
  %collisions,''
  JHEP {\bf 0307} (2003) 001,
  [hep-ph/0206293].
  %%CITATION = JHEPA,0307,001;%%
\bibitem{Maltoni:2002qb}
  F.~Maltoni and T.~Stelzer,
  %``MadEvent: Automatic event generation with MadGraph,''
  JHEP {\bf 0302} (2003) 027,
  [hep-ph/0208156].
  %%CITATION = JHEPA,0302,027;%%
\bibitem{CTEQ5L} 
  CTEQ Coll.(H.L.~Lai {\it et al.}) {\it Eur. Phys. J.} C12 2000 375 .
\bibitem{Butterworth:2007ke}
  J.~M.~Butterworth, J.~R.~Ellis and A.~R.~Raklev,
  %``Reconstructing sparticle mass spectra using hadronic decays,''
  hep-ph/0702150.
  %%CITATION = HEP-PH/0702150;%%
\bibitem{Accomando:2006vj}
  E.~Accomando, A.~Ballestrero, A.~Belhouari and E.~Maina,
  %``Boson fusion and Higgs production at the LHC in six fermion final states
  %with one charged lepton pair,''
  hep-ph/0603167.
  %%CITATION = HEP-PH/0603167;%%

\end{thebibliography}

\begin{thebibliography}{99}
\bibitem{Langacker} P. Langacker, arXiv:0801.1345. 

\bibitem{e6zprime} See {\it e.g.}: A. Leike, Phys.Rep.317, 143 (1999).
\bibitem{Kiritsis} E. Kiritsis, Phys.Rept.421, 105 (2005).
\bibitem{GL} G. Leontaris and J. Rizos, Nucl.Phys.B 567, 32 (2000).
\bibitem{CorianoIrgesKiritsis} C. Corian\`o N. Irges and E. Kiritsis, Nucl.Phys.B 746, 77 (2006).
\bibitem{npap2} C. Corian\`o, A. E. Faraggi and M. Guzzi, Eur.Phys.J.C53, 421 (2008), arXiv:0704.1256 [hep-ph].
\bibitem{CorianoIrgesMorelli} 
C. Corian\`o N. Irges and S. Morelli, Nucl.Phys.B 789, 133 (2008), hep-ph/0703127;
C. Corian\`o N. Irges and S. Morelli, JHEP 0707, 008 (2007), hep-ph/0701010;
C. Corian\`o and N. Irges, Phys.Lett.B 651, 298 (2007) hep-ph/0612140.
\bibitem{npap3} R. Armillis, C. Corian\`o and M. Guzzi, JHEP 0805, 015 (2008), arXiv:0711.3424 [hep-ph] 
\bibitem{npap4} C. Corian\`o, M. Guzzi and S. Morelli, Eur.Phys.J.C 55, 629 (2008), arXiv:0801.2949 [hep-ph] 
\bibitem{npap5} C. Corian\`o, A.E. Faraggi and M. Guzzi, Phys.Rev.D 78, 015012 (2008), arXiv:0802.1792 [hep-ph] 
\bibitem{npap7} R. Armillis, C. Corian\`o, M. Guzzi and S. Morelli, JHEP 0810, 034 (2008), arXiv:0808.1882 [hep-ph] 
\bibitem{npap8} R. Armillis, C. Corian\`o, M. Guzzi and S. Morelli, arXiv:0809.3772 [hep-ph] 
\bibitem{CarenaetAl} M. Carena, A. Daleo, B. A. Dobrescu, T. M. P. Tait, Phys.Rev.D70 093009, (2004).
\bibitem{Appelquist} T. Appelquist, B.A. Dobrescu, A.R. Hopper. Phys.Rev.D 68 035012, (2003).
\bibitem{cfs} G.B.\ Cleaver, A.E.\ Faraggi and C.\ Savage, Phys.Rev.D 63 066001,(2001);\\
G.B.\ Cleaver, D.J.\ Clements and A.E.\ Faraggi, Phys.Rev.D 65 106003, (2002).
\bibitem{lowscalesprime} A.E.\ Faraggi and D.V. Nanopoulos, Mod.Phys.Lett A 6, 61 (1991);\\
J.C. Pati, Phys.Lett.B 388 532, (1996);\\
A.E.\ Faraggi, Phys.Lett.B 499 147, (2001).
\bibitem{Preskill} J. Preskill, Ann. of Phys. 210, 323 (1991).
\bibitem{Roncadelli} A. Dupays and M. Roncadelli, astro-ph/0612227;
E. Masso and J. Redondo, Phys.Rev.Lett. 97, 151802 (2006).
\bibitem{Pascal} P. Anastasopoulos, M. Bianchi, E. Dudas and E. Kiritsis, JHEP 0611, 057 (2006), hep-th/0605225.
\bibitem{CCG} A. Cafarella, C. Corian\`o and M. Guzzi, Nucl.Phys.B 748, 253 (2006), hep-ph/0512358.
\bibitem{npap1} A. Cafarella, C. Corian\`o and M. Guzzi, JHEP 0708, 030 (2007), hep-ph/0702244.
\bibitem{npap6} A. Cafarella, C. Corian\`o and M. Guzzi, Comput.Phys.Commun. 179, 665 (2008), arXiv:0803.0462 [hep-ph] 
\bibitem{Kuhn:1985ps} J.H. Kuhn, A. Reiter and P. M. Zerwas, Nucl.Phys.B 272, 560 (1986).
\bibitem{Kniehl:1989qu} B. A. Kniehl, J. H. Kuhn, Nucl.Phys.B 329, 547 (1990).
\bibitem{VN} R. Hamberg, W.L. van Neerven and T. Matsuura, Nucl.Phys.B 359, 343 (1991).
\bibitem{MRST} A.D. Martin, R.G. Roberts, W.J. Sterling and R.S. Thorne, Eur.Phys.J.C 23, 73 (2002); 
Phys.Lett.B 531, 216 (2002).
\end{thebibliography}
\end{document}